\documentclass[npb]{article}
\usepackage{axodraw}
\usepackage{graphicx}
\usepackage{amssymb}
\RequirePackage{mathptmx}
\RequirePackage[T1]{fontenc}
\usepackage{heppennames2}
\usepackage{lhchiggs}
\usepackage{cernunits}
\DeclareSymbolFont{forjmath}{OT1}{cmr}{m}{sl}
\DeclareMathSymbol{\Jmath}{\mathord}{forjmath}{'021}
\def\jmath{\Jmath}
\DeclareFontFamily{OT1}{cmr}{}
\DeclareFontFamily{OT1}{cmss}{}
\usepackage{pifont}
\usepackage{fancybox}
\usepackage{ctable}
\usepackage[center,small,bf]{caption}
\usepackage{parskip}
\usepackage{enumerate}
\usepackage{url}
\allowdisplaybreaks
\newcommand\pubdate{\today}

\def\csumb{Dipartimento di Fisica Teorica, Universit\`a di Torino, Italy\\
           INFN, Sezione di Torino, Italy}
\def\support{\footnote{Work supported by MIUR under contract
    2001023713$\_$006, by UniTo - Compagnia di San Paolo under contract ORTO11TPXK
and by the Executive Research Agency (REA) of the European Union under the Grant Agreement
PITN-GA-$2012$-$316704$ (HiggsTools).}}
\def\Title#1{\begin{center} {\Large\bf #1 } \end{center}}

\def\Author#1{\begin{center}{ \sc #1} \end{center}}
\def\Address#1{\begin{center}{ \it #1} \end{center}}

\newenvironment{Abstract}{\begin{quotation}  }{\end{quotation}}

\def\Acknowledgments{\bigskip  \bigskip \begin{center}
          \large\bf Acknowledgments\end{center}}
\def\email#1{\footnote{#1}}
\makeatletter
\def\section{\@startsection{section}{0}{\z@}{5.5ex plus .5ex minus
 1.5ex}{2.3ex plus .2ex}{\large\bf}}
\def\subsection{\@startsection{subsection}{1}{\z@}{3.5ex plus .5ex minus
 1.5ex}{1.3ex plus .2ex}{\normalsize\bf}}
\def\subsubsection{\@startsection{subsubsection}{2}{\z@}{-3.5ex plus
-1ex minus  -.2ex}{2.3ex plus .2ex}{\normalsize\sl}}

\renewcommand{\@makecaption}[2]{%
   \vskip 10pt
   \setbox\@tempboxa\hbox{\small #1: #2}
   \ifdim \wd\@tempboxa >\hsize     
       \small #1: #2\par          
     \else                        
       \hbox to\hsize{\hfil\box\@tempboxa\hfil}
   \fi}
 \def\citenum#1{{\def\@cite##1##2{##1}\cite{#1}}}
\def\citea#1{\@cite{#1}{}}
\newcount\@tempcntc
\def\@citex[#1]#2{\if@filesw\immediate\write\@auxout{\string\citation{#2}}\fi
  \@tempcnta\z@\@tempcntb\m@ne\def\@citea{}\@cite{\@for\@citeb:=#2\do
    {\@ifundefined
       {b@\@citeb}{\@citeo\@tempcntb\m@ne\@citea\def\@citea{,}{\bf }\@warning
       {Citation `\@citeb' on page \thepage \space undefined}}%
    {\setbox\z@\hbox{\global\@tempcntc0\csname b@\@citeb\endcsname\relax}%
     \ifnum\@tempcntc=\z@ \@citeo\@tempcntb\m@ne
       \@citea\def\@citea{,}\hbox{\csname b@\@citeb\endcsname}%
     \else
      \advance\@tempcntb\@ne
      \ifnum\@tempcntb=\@tempcntc
      \else\advance\@tempcntb\m@ne\@citeo
      \@tempcnta\@tempcntc\@tempcntb\@tempcntc\fi\fi}}\@citeo}{#1}}
\def\@citeo{\ifnum\@tempcnta>\@tempcntb\else\@citea\def\@citea{,}%
  \ifnum\@tempcnta=\@tempcntb\the\@tempcnta\else
  {\advance\@tempcnta\@ne\ifnum\@tempcnta=\@tempcntb \else\def\@citea{--}\fi
    \advance\@tempcnta\m@ne\the\@tempcnta\@citea\the\@tempcntb}\fi\fi}
\makeatother
\DeclareRobustCommand{\PA}{\HepParticle{A}{}{}\Xspace}
\DeclareRobustCommand{\PV}{\HepParticle{V}{}{}\Xspace}
\DeclareRobustCommand{\PX}{\HepParticle{X}{}{}\Xspace}
\DeclareRobustCommand{\PY}{\HepParticle{Y}{}{}\Xspace}
\DeclareRobustCommand{\Pf}{\HepParticle{f}{}{}\Xspace}

\DeclareRobustCommand{\PF}{\HepParticle{F}{}{}\Xspace}
\DeclareRobustCommand{\PI}{\HepParticle{I}{}{}\Xspace}
\DeclareRobustCommand{\PL}{\HepParticle{L}{}{}\Xspace}
\DeclareRobustCommand{\PG}{\HepParticle{G}{}{}\Xspace}
\DeclareRobustCommand{\PS}{\HepParticle{S}{}{}\Xspace}
\DeclareRobustCommand{\PGN}{\HepParticle{N}{}{}\Xspace}

\DeclareRobustCommand{\PQQ}{\HepParticle{Q}{}{}\Xspace}
\DeclareRobustCommand{\PQU}{\HepParticle{U}{}{}\Xspace}
\DeclareRobustCommand{\PQD}{\HepParticle{D}{}{}\Xspace}
\newcommand{\PHH}{\PH\PH}
\newcommand{\PAA}{\PA\PA}
\newcommand{\PAZ}{\PA\PZ}
\newcommand{\PZA}{\PZ\PA}
\newcommand{\PZZ}{\PZ\PZ}
\newcommand{\PWW}{\PW\PW}
\newcommand{\PVV}{\PV\PV}
\newcommand{\EFT}{\rm{\scriptscriptstyle{EFT}}}

\newcommand{\myLO}{\mathrm{\scriptscriptstyle{LO}}}
\newcommand{\myNLO}{\mathrm{\scriptscriptstyle{NLO}}}

\newcommand{\CT}{\mathrm{CT}}
\newcommand{\dZ}{\mathrm{dZ}}
\newcommand{\OS}{\mathrm{OS}}
\newcommand{\myQED}{\mathrm{QED}}

\newcommand{\mUV}{\mathrm{UV}}

\newcommand{\intf}{\mathrm{int}}
\newcommand{\mySM}{\rm{\scriptscriptstyle{SM}}}

\newcommand{\ssA}{{\mathrm{A}}}
\newcommand{\ssB}{{\mathrm{B}}}
\newcommand{\ssC}{{\mathrm{C}}}

\newcommand{\ssE}{{\mathrm{E}}}
\newcommand{\ssF}{{\mathrm{F}}}
\newcommand{\ssG}{{\mathrm{G}}}
\newcommand{\ssR}{{\mathrm{R}}}
\newcommand{\ssdR}{{\mathrm{dR}}}
\newcommand{\ssdZ}{{\mathrm{dZ}}}
\newcommand{\ssdCZ}{{\mathrm{d}{\mathcal{Z}}}}
\newcommand{\ssD}{{\mathrm{D}}}
\newcommand{\ssI}{{\mathrm{I}}}
\newcommand{\ssU}{{\mathrm{U}}}
\newcommand{\ssL}{{\mathrm{L}}}
\newcommand{\ssP}{{\mathrm{P}}}
\newcommand{\ssPR}{{\mathrm{PR}}}
\newcommand{\ssOS}{{\mathrm{OS}}}
\newcommand{\ssS}{{\mathrm{S}}}

\newcommand{\ssV}{{\mathrm{V}}}
\newcommand{\ssM}{{\mathrm{M}}}
\newcommand{\ssN}{{\mathrm{N}}}

\newcommand{\ssW}{{\mathrm{W}}}
\newcommand{\ssWF}{{\mathrm{WF}}}

\newcommand{\ssT}{{\mathrm{T}}}
\newcommand{\ssZ}{{\mathrm{Z}}}
\newcommand{\bqas}{\begin{eqnarray*}}
\newcommand{\eqas}{\end{eqnarray*}}
\newcommand{\nl}{\nonumber\\}

\newcommand{\lpar}{\left(}                            
\newcommand{\rpar}{\right)} 
\newcommand{\lbra}{\left[\Xspace}
\newcommand{\rbra}{\Xspace\right]}

\newcommand{\bq}{\begin{equation}}                    
\newcommand{\eq}{\end{equation}}
\newcommand{\bqa}{\arraycolsep 0.14em\begin{eqnarray}}
\newcommand{\eqa}{\end{eqnarray}}
\newcommand{\ba}[1]{\begin{array}{#1}}
\newcommand{\ea}{\end{array}}
\newcommand{\ben}{\begin{enumerate}}
\newcommand{\een}{\end{enumerate}}
\newcommand{\bei}{\begin{itemize}}
\newcommand{\eei}{\end{itemize}}
\newcommand{\eqn}[1]{Eq.(\ref{#1})}
\newcommand{\eqns}[2]{Eqs.(\ref{#1})--(\ref{#2})}

\newcommand{\tabn}[1]{Tab.~\ref{#1}}

\newcommand{\fig}[1]{Fig.~\ref{#1}}

\newcommand{\sect}[1]{Section~\ref{#1}}

\newcommand{\appendx}[1]{Appendix~\ref{#1}}

\newcommand{\bmid}{\Bigr|}

\newcommand{\oD}{{\overline \Delta}}
\newcommand{\hD}{{\hat{\Delta}}}

\newcommand{\Bref}[1]{Ref.~\cite{#1}}
\newcommand{\Brefs}[1]{Refs.~\cite{#1}}

\newcommand{\eg}{e.g.\xspace}
\newcommand{\ie}{i.e.\xspace}
\newcommand{\etc}{etc.\@\xspace}

\newcommand{\mw}{\mathswitch{M_{\PW}}}

\newcommand{\mz}{\mathswitch{M_{\PZ}}}

\newcommand{\mws}{\mathswitch{M^2_{\PW}}}
\newcommand{\mwc}{\mathswitch{M^3_{\PW}}}
\newcommand{\mwsOS}{\mathswitch{M^2_{\PW\,;\,\ssOS}}}
\newcommand{\mwq}{\mathswitch{M^4_{\PW}}}
\newcommand{\mwvi}{\mathswitch{M^6_{\PW}}}
\newcommand{\mzs}{\mathswitch{M^2_{\PZ}}}

\newcommand{\mzOS}{\mathswitch{M_{\PZ\,;\,\ssOS}}}
\newcommand{\mzsOS}{\mathswitch{M^2_{\PZ\,;\,\ssOS}}}
\newcommand{\mt}{\mathswitch{M_{\PQt}}}
\newcommand{\mb}{\mathswitch{M_{\PQb}}}
\newcommand{\mts}{\mathswitch{M^2_{\PQt}}}
\newcommand{\mbs}{\mathswitch{M^2_{\PQb}}}
\newcommand{\mtq}{\mathswitch{M^4_{\PQt}}}
\newcommand{\mbq}{\mathswitch{M^4_{\PQb}}}
\newcommand{\ml}{\mathswitch{M_{\Pl}}}
\newcommand{\mUs}{\mathswitch{M^2_{\PQU}}}
\newcommand{\mDs}{\mathswitch{M^2_{\PQD}}}

\newcommand{\mh}{\mathswitch{M_{\PH}}}
\newcommand{\mhOS}{\mathswitch{M_{\PH\,;\,\ssOS}}}
\newcommand{\mhs}{\mathswitch{M^2_{\PH}}}
\newcommand{\mhsOS}{\mathswitch{M^2_{\PH\,;\,\ssOS}}}
\newcommand{\mhq}{\mathswitch{M^4_{\PH}}}
\newcommand{\mhvi}{\mathswitch{M^6_{\PH}}}
\newcommand{\mqt}{\mathswitch{M_{\PQt}}}
\newcommand{\mqb}{\mathswitch{M_{\PQb}}}
\newcommand{\mqts}{\mathswitch{M^2_{\PQt}}}
\newcommand{\mqbs}{\mathswitch{M^2_{\PQb}}}

\newcommand{\sPF}{{\scriptscriptstyle{\PF}}}
\newcommand{\sPQQ}{{\scriptscriptstyle{\PQQ}}}
\newcommand{\sPL}{{\scriptscriptstyle{\PL}}}
\newcommand{\sPQU}{{\scriptscriptstyle{\PQU}}}
\newcommand{\sPQD}{{\scriptscriptstyle{\PQD}}}
\newcommand{\sPH}{{\scriptscriptstyle{\PH}}}
\newcommand{\sPS}{{\scriptscriptstyle{\PS}}}
\newcommand{\sPZ}{{\scriptscriptstyle{\PZ}}}
\newcommand{\sPW}{{\scriptscriptstyle{\PW}}}
\newcommand{\sPA}{{\scriptscriptstyle{\PA}}}
\newcommand{\sPB}{{\scriptscriptstyle{\PB}}}
\newcommand{\sPV}{{\scriptscriptstyle{\PV}}}
\newcommand{\sPVV}{{\scriptscriptstyle{\PV\PV}}}
\newcommand{\sPZZ}{{\scriptscriptstyle{\PZZ}}}
\newcommand{\sPWW}{{\scriptscriptstyle{\PWW}}}
\newcommand{\sPHAA}{{\scriptscriptstyle{\PH\PAA}}}
\newcommand{\sPHAZ}{{\scriptscriptstyle{\PH\PA\PZ}}}
\newcommand{\sPHZZ}{{\scriptscriptstyle{\PH\PZ\PZ}}}
\newcommand{\sPHWW}{{\scriptscriptstyle{\PH\PW\PW}}}
\newcommand{\sPHVV}{{\scriptscriptstyle{\PH\PV\PV}}}

\newcommand{\mqu}{\mathswitch{M_{\PQu}}}
\newcommand{\mqU}{\mathswitch{M_{\sPQU}}}

\newcommand{\mqd}{\mathswitch{M_{\PQd}}}
\newcommand{\mqD}{\mathswitch{M_{\sPQD}}}

\newcommand{\mle}{\mathswitch{M_{\Pl}}}
\newcommand{\mLe}{\mathswitch{M_{\sPL}}}
\newcommand{\mLes}{\mathswitch{M^2_{\sPL}}}

\newcommand{\muR}{\mathswitch{\mu_{\ssR}}}

\newcommand{\muRs}{\mathswitch{\mu^2_{\ssR}}}

\newcommand{\proc}{{\mbox{\scriptsize proc}}}

\newcommand{\off}{{\mbox{\scriptsize off}}}

\newcommand{\rest}{{\mbox{\scriptsize rest}}}
\newcommand{\fin}{{\mbox{\tiny fin}}}
\newcommand{\pole}{{\mbox{\tiny div}}}
\newcommand{\gen}{{\mbox{\scriptsize gen}}}
\newcommand{\ggen}{{\mbox{\tiny gen}}}
\newcommand{\ren}{{\mbox{\scriptsize ren}}}
\newcommand{\ct}{{\mbox{\scriptsize ct}}}

\newcommand{\ep}{\mathswitch \varepsilon}

\newcommand{\eph}{\mathswitch{\hat\varepsilon}}
\newcommand{\spro}[2]{{#1}\cdot{#2}}

\newcommand{\mrS}{{\mathrm{S}}}
\newcommand{\mrA}{{\mathrm{A}}}
\newcommand{\mrB}{{\mathrm{B}}}
\newcommand{\mrC}{{\mathrm{C}}}
\newcommand{\mrD}{{\mathrm{D}}}
\newcommand{\mrE}{{\mathrm{E}}}
\newcommand{\mrF}{{\mathrm{F}}}
\newcommand{\mrG}{{\mathrm{G}}}
\newcommand{\mrH}{{\mathrm{H}}}
\newcommand{\mrI}{{\mathrm{I}}}
\newcommand{\mrJ}{{\mathrm{J}}}
\newcommand{\mrK}{{\mathrm{K}}}
\newcommand{\mrL}{{\mathrm{L}}}
\newcommand{\mrM}{{\mathrm{M}}}
\newcommand{\mrN}{{\mathrm{N}}}
\newcommand{\mrO}{{\mathrm{O}}}
\newcommand{\mrP}{{\mathrm{P}}}
\newcommand{\mrQ}{{\mathrm{Q}}}
\newcommand{\mrR}{{\mathrm{R}}}
\newcommand{\mrT}{{\mathrm{T}}}
\newcommand{\mrU}{{\mathrm{U}}}
\newcommand{\mrV}{{\mathrm{V}}}
\newcommand{\mrW}{{\mathrm{W}}}
\newcommand{\mrX}{{\mathrm{X}}}
\newcommand{\mrZ}{{\mathrm{Z}}}

\newenvironment{definition}[1][Definition]{\begin{trivlist}
\item[\hskip \labelsep {\bfseries #1}]}{\end{trivlist}}

\newcommand{\Lag}{{\cal L}}
\newcommand{\Ope}{{\cal O}}

\newcommand{\srt}{\sqrt{2}}
\DeclareRobustCommand{\Ppp}{\HepParticle{\upphi}{}{+}\Xspace}
\DeclareRobustCommand{\Ppm}{\HepParticle{\upphi}{}{-}\Xspace}
\DeclareRobustCommand{\Pppm}{\HepParticle{\upphi}{}{\pm}\Xspace}
\DeclareRobustCommand{\Ppz}{\HepParticle{\upphi}{}{0}\Xspace}
\DeclareRobustCommand{\PAL}{\HepParticle{\mybar{\PL}}{}{}\Xspace}
\DeclareRobustCommand{\PpsiL}{\HepParticle{\uppsi}{L}{}\Xspace}
\DeclareRobustCommand{\PpsiR}{\HepParticle{\uppsi}{R}{}\Xspace}

\DeclareRobustCommand{\PQuR}{\HepParticle{\PQu}{R}{}\Xspace}
\DeclareRobustCommand{\PQdR}{\HepParticle{\PQd}{R}{}\Xspace}
\DeclareRobustCommand{\PQqL}{\HepParticle{\PQq}{L}{}\Xspace}

\DeclareRobustCommand{\PlR}{\HepParticle{\Pl}{R}{}\Xspace}
\DeclareRobustCommand{\PLL}{\HepParticle{\PL}{L}{}\Xspace}

\DeclareRobustCommand{\PAQqL}{\HepParticle{\PAQq}{L}{}\Xspace}

\DeclareRobustCommand{\PAlR}{\HepParticle{\PAl}{R}{}\Xspace}
\DeclareRobustCommand{\PALL}{\HepParticle{\PAL}{L}{}\Xspace}
\DeclareRobustCommand{\PAQuR}{\HepParticle{\PAQu}{R}{}\Xspace}
\DeclareRobustCommand{\PAQdR}{\HepParticle{\PAQd}{R}{}\Xspace}
\newcommand{\pdmu}{{\partial_{\mu}}}
\newcommand{\pdnu}{{\partial_{\nu}}}
\newcommand{\myGF}{G_{\sPF}}
\DeclareRobustCommand{\PK}{\HepParticle{\upPhi}{}{}\Xspace}
\DeclareRobustCommand{\Pphi}{\HepParticle{\upphi}{}{}\Xspace}
\DeclareRobustCommand{\PKdag}{\HepParticle{\PK}{}{\dagger}\Xspace}
\newcommand{\KdK}{\lpar \PKdag\,\PK\rpar }
\newcommand{\mrc}{\mathrm{c}}
\newcommand{\mrs}{\mathrm{s}}
\newcommand{\stw}{\mrs_{_{\theta}}}             
\newcommand{\ctw}{\mrc_{_{\theta}}}
\newcommand{\stws}{\mrs_{_{\theta}}^2}
\newcommand{\ctws}{\mrc_{_{\theta}}^2}

\newcommand{\ctwq}{\mrc_{_{\theta}}^4}

\newcommand{\ctwr}{\mrc^{\ren}_{_{\theta}}}
\newcommand{\stwr}{s^{\ren}_{_{\theta}}}
\newcommand{\stW}{\mrs_{_{\PW}}}             
\newcommand{\ctW}{\mrc_{_{\PW}}}
\newcommand{\stWs}{\mrs_{_{\PW}}^2}
\newcommand{\ctWs}{\mrc_{_{\PW}}^2}
\newcommand{\ctWc}{\mrc_{_{\PW}}^3}

\newcommand{\ctWq}{\mrc_{_{\PW}}^4}

\newcommand{\ctb}{\overline{\mrc}_{_{\theta}}}
\newcommand{\stb}{\overline{\mrs}_{_{\theta}}}
\newcommand{\ctbs}{\overline{\mrc}^2_{_{\theta}}}
\newcommand{\stbs}{\overline{\mrs}^2_{_{\theta}}}
\newcommand\mybar[1]{\ensuremath{\bar{#1}}}

\newcommand{\cth}{{\hat \mrc}_{_{\theta}}}
\newcommand{\sth}{{\hat \mrs}_{_{\theta}}}
\newcommand{\cths}{{\hat \mrc}^2_{_{\theta}}}
\newcommand{\cthc}{{\hat \mrc}^3_{_{\theta}}}
\newcommand{\sths}{{\hat \mrs}^2_{_{\theta}}}

\newcommand{\HSs}{{\Lambda^2}}

\DeclareRobustCommand{\PWpmmu}{\HepParticle{\PW}{\mu}{\pm}\Xspace}

\DeclareRobustCommand{\PZmu}{\HepParticle{\PZ}{\mu}{}\Xspace}
\DeclareRobustCommand{\PAmu}{\HepParticle{\PA}{\mu}{}\Xspace}

\DeclareRobustCommand{\PXpm}{\HepParticle{\PX}{}{\pm}\Xspace}

\DeclareRobustCommand{\PYz}{\HepParticle{\PY}{\PZ}{}\Xspace}
\DeclareRobustCommand{\PYa}{\HepParticle{\PY}{\PA}{}\Xspace}
\DeclareRobustCommand{\PHb}{\HepAntiParticle{\PH}{}{}\Xspace}
\DeclareRobustCommand{\Ppzb}{\HepAntiParticle{\upphi}{}{\,0}\Xspace}
\DeclareRobustCommand{\Pppmb}{\HepAntiParticle{\upphi}{}{\pm}\Xspace}

\DeclareRobustCommand{\PAbmu}{\HepAntiParticle{\PA}{}{\mu}\Xspace}
\DeclareRobustCommand{\PAbmud}{\HepAntiParticle{\PA}{\mu}{}\Xspace}

\DeclareRobustCommand{\PZbmu}{\HepAntiParticle{\PZ}{}{\mu}\Xspace}
\DeclareRobustCommand{\PZbmud}{\HepAntiParticle{\PZ}{\mu}{}\Xspace}

\DeclareRobustCommand{\PWpmbmud}{\HepAntiParticle{\PW}{\mu}{\pm}\Xspace}

\DeclareRobustCommand{\PXpmb}{\HepAntiParticle{\PX}{}{\pm}\Xspace}
\DeclareRobustCommand{\PYzb}{\HepAntiParticle{\PY}{\PZ}{}\Xspace}
\DeclareRobustCommand{\PYab}{\HepAntiParticle{\PY}{\PA}{}\Xspace}
\DeclareRobustCommand{\OMf}{\HepAntiParticle{\PM}{}{\Pf}\Xspace}
\DeclareRobustCommand{\OMs}{\HepAntiParticle{\PM}{}{\,2}\Xspace}

\DeclareRobustCommand{\OMHs}{\HepAntiParticle{\PM}{\,\PH}{\,2}\Xspace}

\newcommand{\myYM}{\rm{\scriptscriptstyle{YM}}}
\newcommand{\gfix}{{\mbox{\scriptsize gf}}}
\newcommand{\myFP}{\rm{\scriptscriptstyle{FP}}}

\newcommand{\nfact}{{\mbox{\scriptsize nfc}}}

\DeclareRobustCommand{\PM}{\HepParticle{M}{}{}\Xspace}

\newcommand{\bh}{\beta_{\mathrm{h}}}
\newcommand{\bhb}{\overline{\beta}_{\mathrm{h}}}
\newcommand{\RL}{\mathrm{R}_{\Lambda}}
\newcommand{\Ddmu}{\ssD^{(\leftrightarrow)}_{\mu}}
\newcommand{\Mbs}{M^2}

\newcommand{\Mbq}{M^4}
\newcommand{\mzb}{M_0}
\newcommand{\Mzbs}{M^2_0}

\newcommand{\apLWt}{a^{(3)}_{\upphi\,\sPL\,\scriptscriptstyle{\PW}}}
\newcommand{\apLldQ}{a_{\upphi\,\PL\Pl\PQd\PQQ}}
\newcommand{\apN}{a_{\upphi\,\PN}}
\newcommand{\aplVA}{a_{\upphi\,\Pl\,\scriptscriptstyle{\PV\PA}}}
\newcommand{\apLVA}{a_{\upphi\,\sPL\,\scriptscriptstyle{\PV\PA}}}
\newcommand{\apLA}{a_{\upphi\,\sPL\,\scriptscriptstyle{\PA}}}
\newcommand{\apLV}{a_{\upphi\,\sPL\,\scriptscriptstyle{\PV}}}
\newcommand{\apuVA}{a_{\upphi\,\PQu\,\scriptscriptstyle{\PV\PA}}}
\newcommand{\apdVA}{a_{\upphi\,\PQd\,\scriptscriptstyle{\PV\PA}}}
\newcommand{\apUVA}{a_{\upphi\,\sPQU\,\scriptscriptstyle{\PV\PA}}}
\newcommand{\apDVA}{a_{\upphi\,\sPQD\,\scriptscriptstyle{\PV\PA}}}
\newcommand{\apWAB}{a_{\upphi\,\scriptscriptstyle{\PW\PA\PB}}}
\newcommand{\apWAD}{a_{\upphi\,\scriptscriptstyle{\PW\PA\PD}}}
\newcommand{\alpBox}{a_{\Pl\,\upphi\,\scriptscriptstyle{\Box}}}

\newcommand{\aupBox}{a_{\PQu\,\upphi\,\scriptscriptstyle{\Box}}}
\newcommand{\adpBox}{a_{\PQd\,\upphi\,\scriptscriptstyle{\Box}}}
\newcommand{\aAA}{a_{\scriptscriptstyle{\PAA}}}
\newcommand{\aZZ}{a_{\scriptscriptstyle{\PZZ}}}
\newcommand{\aAZ}{a_{\scriptscriptstyle{\PAZ}}}
\newcommand{\ap}{a_{\upphi}}
\newcommand{\apBox}{a_{\upphi\,\scriptscriptstyle{\Box}}}
\newcommand{\apWB}{a_{\upphi\,\scriptscriptstyle{\PW\PB}}}
\newcommand{\apWBa}{a^{(a)}_{\upphi\,\scriptscriptstyle{\PW\PB}}}
\newcommand{\apWDpB}{a^{(+)}_{\upphi\,\scriptscriptstyle{\PW\PD\PB}}}
\newcommand{\apWDmB}{a^{(-)}_{\upphi\,\scriptscriptstyle{\PW\PD\PB}}}

\newcommand{\apWA}{a_{\upphi\,\scriptscriptstyle{\PW\PA}}}
\newcommand{\apWZ}{a_{\upphi\,\scriptscriptstyle{\PW\PZ}}}
\newcommand{\apB}{a_{\upphi\,\scriptscriptstyle{\PB}}}
\newcommand{\apW}{a_{\upphi\,\scriptscriptstyle{\PW}}}
\newcommand{\apD}{a_{\upphi\,\scriptscriptstyle{\PD}}}
\newcommand{\apDB}{a_{\upphi\,\scriptscriptstyle{\PD\PB}}}
\newcommand{\alp}{a_{\Pl\,\upphi}} 
\newcommand{\aup}{a_{\PQu\,\upphi}} 
\newcommand{\atp}{a_{\PQt\,\upphi}} 
\newcommand{\adp}{a_{\PQd\,\upphi}} 
\newcommand{\abp}{a_{\PQb\,\upphi}} 

\newcommand{\aLp}{a_{\sPL\,\upphi}} 
\newcommand{\aUp}{a_{\sPQU\,\upphi}} 
\newcommand{\aDp}{a_{\sPQD\,\upphi}} 
\newcommand{\apWDm}{a^{(-)}_{\upphi\,\scriptscriptstyle{\PW\PD}}}
\newcommand{\apWDp}{a^{(+)}_{\upphi\,\scriptscriptstyle{\PW\PD}}}
\newcommand{\auW}{a_{\PQu\,\scriptscriptstyle{\PW}}}
\newcommand{\auB}{a_{\PQu\,\scriptscriptstyle{\PB}}}
\newcommand{\auWB}{a_{\PQu\,\scriptscriptstyle{\PW\PB}}}
\newcommand{\auBW}{a_{\PQu\,\scriptscriptstyle{\PB\PW}}}
\newcommand{\atWB}{a_{\PQt\,\scriptscriptstyle{\PW\PB}}}
\newcommand{\atBW}{a_{\PQt\,\scriptscriptstyle{\PB\PW}}}
\newcommand{\aUW}{a_{\sPQU\,\scriptscriptstyle{\PW}}}
\newcommand{\aUB}{a_{\sPQU\,\scriptscriptstyle{\PB}}}
\newcommand{\aUWB}{a_{\sPQU\,\scriptscriptstyle{\PW\PB}}}
\newcommand{\aUBW}{a_{\sPQU\,\scriptscriptstyle{\PB\PW}}}
\newcommand{\adW}{a_{\PQd\,\scriptscriptstyle{\PW}}}
\newcommand{\adB}{a_{\PQd\,\scriptscriptstyle{\PB}}}
\newcommand{\adWB}{a_{\PQd\,\scriptscriptstyle{\PW\PB}}}
\newcommand{\adBW}{a_{\PQd\,\scriptscriptstyle{\PB\PW}}}
\newcommand{\abWB}{a_{\PQb\,\scriptscriptstyle{\PW\PB}}}
\newcommand{\abBW}{a_{\PQb\,\scriptscriptstyle{\PB\PW}}}
\newcommand{\aDW}{a_{\sPQD\,\scriptscriptstyle{\PW}}}
\newcommand{\aDB}{a_{\sPQD\,\scriptscriptstyle{\PB}}}
\newcommand{\aDWB}{a_{\sPQD\,\scriptscriptstyle{\PW\PB}}}
\newcommand{\aDBW}{a_{\sPQD\,\scriptscriptstyle{\PB\PW}}}
\newcommand{\alW}{a_{\Pl\,\scriptscriptstyle{\PW}}}
\newcommand{\alB}{a_{\Pl\,\scriptscriptstyle{\PB}}}
\newcommand{\alWB}{a_{\Pl\,\scriptscriptstyle{\PW\PB}}}
\newcommand{\alBW}{a_{\Pl\,\scriptscriptstyle{\PB\PW}}}
\newcommand{\aLW}{a_{\sPL\,\scriptscriptstyle{\PW}}}
\newcommand{\aLB}{a_{\sPL\,\scriptscriptstyle{\PB}}}
\newcommand{\aLWB}{a_{\sPL\,\scriptscriptstyle{\PW\PB}}}
\newcommand{\aLBW}{a_{\sPL\,\scriptscriptstyle{\PB\PW}}}
\newcommand{\aplWt}{a^{(3)}_{\upphi\,\Pl\,\scriptscriptstyle{\PW}}}
\newcommand{\apqWt}{a^{(3)}_{\upphi\,\PQq\,\scriptscriptstyle{\PW}}}
\newcommand{\apQWt}{a^{(3)}_{\upphi\,\sPQQ\,\scriptscriptstyle{\PW}}}
\newcommand{\aplV}{a_{\upphi\,\Pl\,\scriptscriptstyle{\PV}}}
\newcommand{\aplA}{a_{\upphi\,\Pl\,\scriptscriptstyle{\PA}}}
\newcommand{\apuV}{a_{\upphi\,\PQu\,\scriptscriptstyle{\PV}}}
\newcommand{\aptV}{a_{\upphi\,\PQt\,\scriptscriptstyle{\PV}}}
\newcommand{\apuA}{a_{\upphi\,\PQu\,\scriptscriptstyle{\PA}}}
\newcommand{\aptA}{a_{\upphi\,\PQt\,\scriptscriptstyle{\PA}}}
\newcommand{\apUV}{a_{\upphi\,\sPQU\,\scriptscriptstyle{\PV}}}
\newcommand{\apUA}{a_{\upphi\,\sPQU\,\scriptscriptstyle{\PA}}}
\newcommand{\apdV}{a_{\upphi\,\PQd\,\scriptscriptstyle{\PV}}}
\newcommand{\apbV}{a_{\upphi\,\PQb\,\scriptscriptstyle{\PV}}}
\newcommand{\apdA}{a_{\upphi\,\PQd\,\scriptscriptstyle{\PA}}}
\newcommand{\apbA}{a_{\upphi\,\PQb\,\scriptscriptstyle{\PA}}}
\newcommand{\apDV}{a_{\upphi\,\sPQD\,\scriptscriptstyle{\PV}}}
\newcommand{\apDA}{a_{\upphi\,\sPQD\,\scriptscriptstyle{\PA}}}
\newcommand{\apn}{a_{\upphi\,\PGn}}
\newcommand{\aplo}{a^{(1)}_{\upphi\,\Pl}}
\newcommand{\aplt}{a^{(3)}_{\upphi\,\Pl}}
\newcommand{\apqo}{a^{(1)}_{\upphi\,\PQq}}
\newcommand{\apQo}{a^{(1)}_{\upphi\,\sPQQ}}
\newcommand{\apqt}{a^{(3)}_{\upphi\,\PQq}}
\newcommand{\apLo}{a^{(1)}_{\upphi\,\sPL}}
\newcommand{\apLt}{a^{(3)}_{\upphi\,\sPL}}
\newcommand{\apQt}{a^{(3)}_{\upphi\,\sPQQ}}

\newcommand{\apu}{a_{\upphi\,\PQu}}
\newcommand{\apd}{a_{\upphi\,\PQd}}
\newcommand{\apl}{a_{\upphi\,\Pl}}

\newcommand{\apU}{a_{\upphi\,\sPQU}}

\newcommand{\apL}{a_{\upphi\,\sPL}}
\newcommand{\aLldQ}{a_{\sPL\,\Pl\,\PQd\,\sPQQ}}
\newcommand{\aoQuQd}{a^{(1)}_{\sPQQ\,\PQu\,\sPQQ\,\PQd}}
\newcommand{\aoLlQu}{a^{(1)}_{\sPL\,\Pl\,\sPQQ\,\PQu}}
\newcommand{\bfun}[3]{\mrB_0^{\fin}\lpar #1\,;\,#2\,,\,#3\rpar}
\newcommand{\sbfun}[2]{\mrB_0^{\fin}\lpar #1\,,\,#2\rpar}
\newcommand{\bfunp}[3]{\mrB_{0\mathrm{p}}^{\fin}\lpar #1\,;\,#2\,,\,#3\rpar}
\newcommand{\bfuns}[3]{\mrB_{0\mathrm{s}}^{\fin}\lpar #1\,;\,#2\,,\,#3\rpar}
\newcommand{\zbfunp}[2]{\mrB_{0\mathrm{p}}^{\fin}\lpar \,#1\,,\,#2\rpar}
\newcommand{\zbfuns}[2]{\mrB_{0\mathrm{s}}^{\fin}\lpar \,#1\,,\,#2\rpar}
\newcommand{\ssp}{\mathrm{p}}
\newcommand{\WC}[1]{\ssW_{#1}}
\newcommand{\WCr}[1]{\ssW^{\ren}_{#1}}
\newcommand{\vtqs}{\mathrm{v}^2_{\PQt}}
\newcommand{\vbqs}{\mathrm{v}^2_{\PQb}}
\newcommand{\vtq}{\mathrm{v}_{\PQt}}
\newcommand{\vbq}{\mathrm{v}_{\PQb}}
\newcommand{\vqu}{\mathrm{v}_{\PQu}}
\newcommand{\vqd}{\mathrm{v}_{\PQd}}
\newcommand{\vle}{\mathrm{v}_{\Pl}}
\newcommand{\vqus}{\mathrm{v}^2_{\PQu}}
\newcommand{\vqds}{\mathrm{v}^2_{\PQd}}
\newcommand{\vles}{\mathrm{v}^2_{\Pl}}
\newcommand{\vple}{\mathrm{v}^{+}_{\Pl}}

\newcommand{\vmle}{\mathrm{v}^{-}_{\Pl}}
\newcommand{\vmqu}{\mathrm{v}^{-}_{\PQu}}
\newcommand{\vmqd}{\mathrm{v}^{-}_{\PQd}}
\newcommand{\sla}[1]{/\!\!\!\!\!#1}
\newcommand{\OPI}{1\mathrm{PI}}
\newcommand{\OPR}{1\mathrm{PR}}
\newcommand{\IR}{1\mathrm{IR}}
\newcommand{\DUV}{\Delta_{\mathrm{UV}}}
\newcommand{\DIR}{\Delta_{\mathrm{IR}}}

\newcommand{\DUVM}{\Delta_{\mathrm{UV}}\lpar \mws \rpar}
\newcommand{\cfun}[6]{\ssC_0\lpar #1\,,\,#2\,,\,#3\,;\,#4\,,\,#5\,,\,#6\rpar}
\newcommand{\cfunf}[6]{\ssC^{\fin}_0\lpar #1\,,\,#2\,,\,#3\,;\,#4\,,\,#5\,,\,#6\rpar}

\newcommand{\afun}[1]{a^{\fin}_0\lpar #1 \rpar}
\newcommand{\xphvi}{x^6_{\sPH}}
\newcommand{\xphq}{x^4_{\sPH}}
\newcommand{\xpuq}{x^4_{\PQu}}
\newcommand{\xptq}{x^4_{\PQt}}
\newcommand{\xpdq}{x^4_{\PQd}}
\newcommand{\xpbq}{x^4_{\PQb}}
\newcommand{\xplq}{x^4_{\Pl}}
\newcommand{\xpss}{x^2_{\sPS}}
\newcommand{\xphs}{x^2_{\sPH}}
\newcommand{\xpus}{x^2_{\PQu}}
\newcommand{\xpts}{x^2_{\PQt}}
\newcommand{\xpds}{x^2_{\PQd}}
\newcommand{\xpbs}{x^2_{\PQb}}
\newcommand{\xpls}{x^2_{\Pl}}
\newcommand{\xpuc}{x^3_{\PQu}}
\newcommand{\xpdc}{x^3_{\PQd}}
\newcommand{\xplc}{x^3_{\Pl}}
\newcommand{\xpUs}{x^2_{\sPQU}}
\newcommand{\xpDs}{x^2_{\sPQD}}
\newcommand{\xpLs}{x^2_{\sPL}}
\newcommand{\xps}{x_{\sPS}}
\newcommand{\xph}{x_{\sPH}}
\newcommand{\xpu}{x_{\PQu}}
\newcommand{\xpd}{x_{\PQd}}
\newcommand{\xpl}{x_{\Pl}}
\DeclareRobustCommand{\xog}{x^{(1)}_{\gen}}
\DeclareRobustCommand{\xtg}{x^{(2)}_{\gen}}
\DeclareRobustCommand{\vog}{v^{(1)}_{\gen}}
\DeclareRobustCommand{\vtg}{v^{(2)}_{\gen}}
\newcommand{\sumg}{\sum_{\ggen}}
\newcommand{\xpUq}{x^4_{\sPQU}}
\newcommand{\xpDq}{x^4_{\sPQD}}

\newcommand{\xpLc}{x^3_{\sPL}}

\newcommand{\xpU}{x_{\sPQU}}
\newcommand{\xpD}{x_{\sPQD}}
\newcommand{\xpL}{x_{\sPL}}
\newcommand{\mcA}{\mathcal{A}}

\newcommand{\mcD}{\mathcal{D}}

\newcommand{\mcO}{\mathcal{O}}
\newcommand{\mcP}{\mathcal{P}}

\newcommand{\mcS}{\mathcal{S}}

\newcommand{\mcF}{\mathcal{F}}

\newcommand{\mcT}{\mathcal{T}}
\newcommand{\mcU}{\mathcal{U}}
\newcommand{\mcX}{\mathcal{X}}

\newcommand{\mcZ}{\mathcal{Z}}
\newcommand{\myNG}{{\mathrm N}_{\gen}}
\newcommand{\LR}{{\mathrm L}_{\ssR}}
\newcommand{\Lir}{{\mathrm L}_{\mathrm{ir}}}
\newcommand{\mrdim}{\mathrm{dim}}
\newcommand{\UVsl}{{\mathrm C}^{\sPL}_{\mathrm{UV}}}
\newcommand{\UVst}{{\mathrm C}^{\sPQU}_{\mathrm{UV}}}
\newcommand{\UVsb}{{\mathrm C}^{\sPQD}_{\mathrm{UV}}}
\newcommand{\lzs}{\lambda^2_{\sPZ}}
\newcommand{\lz}{\lambda_{\sPZ}}
\newcommand{\lws}{\lambda^2_{\sPW}}
\newcommand{\lw}{\lambda_{\sPW}}
\newcommand{\lazs}{\lambda^2_{\sPA\sPZ}}
\newcommand{\lzzs}{\lambda^2_{\sPZ\sPZ}}
\newcommand{\lwws}{\lambda^2_{\sPW\sPW}}
\newcommand{\laz}{\lambda_{\sPA\sPZ}}
\newcommand{\lzz}{\lambda_{\sPZ\sPZ}}
\newcommand{\lww}{\lambda_{\sPW\sPW}}
\newcommand{\WFHs}{\mrW^{(6)}_{\PH}}
\newcommand{\WFWs}{\mrW^{(6)}_{\PW}}
\newcommand{\degfs}{\mathrm{d}\mcZ^{(6)}_g}
\newcommand{\dMfs}{\mathrm{d}\mcZ^{(6)}_{\mw}}
\newcommand{\WFH}{\mrW^{(4)}_{\PH}}
\newcommand{\WFA}{\mrW^{(4)}_{\PA}}
\newcommand{\WFZ}{\mrW^{(4)}_{\PZ}}
\newcommand{\WFW}{\mrW^{(4)}_{\PW}}
\newcommand{\degf}{\mathrm{d}\mcZ^{(4)}_g}
\newcommand{\dMf}{\mathrm{d}\mcZ^{(4)}_{\mw}}
\newcommand{\dcthf}{\mathrm{d}\mcZ^{(4)}_c}
\newcommand{\WFZt}{\mrW^{(4)}_{\PZ\,\mid\,\PQt}}
\newcommand{\WFZb}{\mrW^{(4)}_{\PZ\,\mid\,\PQb}}
\newcommand{\WFZW}{\mrW^{(4)}_{\PZ\,\mid\,\PW}}
\newcommand{\WFZrf}{{\overline{\sum}}_{\gen}\,\mrW^{(4)}_{\PZ\,\mid\,\Pf}}
\newcommand{\WFHt}{\mrW^{(4)}_{\PH\,\mid\,\PQt}}
\newcommand{\WFHb}{\mrW^{(4)}_{\PH\,\mid\,\PQb}}
\newcommand{\WFHW}{\mrW^{(4)}_{\PH\,\mid\,\PW}}
\newcommand{\WFHtb}{\mrW^{(4)}_{\PH\,\mid\,\PQt,\PQb}}
\newcommand{\dMff}{\sum_{\gen}\,\mathrm{d}\mcZ^{(4)}_{M\,\mid\,\Pf}}
\newcommand{\dMfW}{\mathrm{d}\mcZ^{(4)}_{M\,\mid\,\PW}}
\newcommand{\dMftb}{\mathrm{d}\mcZ^{(4)}_{M\,\mid\,\PQt,\PQb}}
\newcommand{\dMfrf}{{\overline{\sum}}_{\gen}\,\mathrm{d}\mcZ^{(4)}_{M\,\mid\,\Pf}}
\newcommand{\degfrf}{\sum_{\gen}\,\mathrm{d}\mcZ^{(4)}_{g\,\mid\,\Pf}}
\newcommand{\degfW}{\mathrm{d}\mcZ^{(4)}_{g\,\mid\,\PW}}
\newcommand{\degftb}{\mathrm{d}\mcZ^{(4)}_{g\,\mid\,\PQt,\PQb}}
\newcommand{\degff}{\sum_{\gen}\,\mathrm{d}\mcZ^{(4)}_{g\,\mid\,\Pf}}
\newcommand{\dcthft}{\mathrm{d}\mcZ^{(4)}_{\ctw\,\mid\,\PQt}}
\newcommand{\dcthfb}{\mathrm{d}\mcZ^{(4)}_{\ctw\,\mid\,\PQb}}
\newcommand{\dcthfrf}{{\overline{\sum}}_{\gen}\,\mathrm{d}\mcZ^{(4)}_{\ctw\,\mid\,\Pf}}
\newcommand{\dcthfW}{\mathrm{d}\mcZ^{(4)}_{\ctw\,\mid\,\PW}}
\newcommand{\WFWtb}{\mrW^{(4)}_{\PW\,\mid\,\PQt,\PQb}}
\newcommand{\WFWW}{\mrW^{(4)}_{\PW\,\mid\,\PW}}
\newcommand{\WFWrf}{{\overline{\sum}}_{\gen}\,\mrW^{(4)}_{\PW\,\mid\,\Pf}}
\newcommand{\spc}{\,,}
\newcommand{\spp}{\,.}
\newcommand{\gds}{g_{_6}}
\begin{document}
\begin{titlepage}
\pubdate
\vfill
\def\thefootnote{\fnsymbol{footnote}}
\Title{\LARGE \sffamily \bfseries
NLO Higgs Effective Field Theory\\[0.1cm]
and $\upkappa\,$-framework\support}
\vspace{1.cm}
\Author{\normalsize \bfseries \sffamily
Margherita Ghezzi \email{margherita.ghezzi@to.infn.it},
Raquel Gomez-Ambrosio \email{raquel.gomez@to.infn.it},
Giampiero Passarino \email{giampiero@to.infn.it} and
Sandro Uccirati \email{uccirati@to.infn.it}}
\Address{\csumb}
\vspace{2.cm}
\begin{Abstract}
\noindent 
A consistent framework for studying Standard Model deviations is developed. It assumes that New 
Physics becomes relevant at some scale beyond the present experimental reach and uses the 
Effective Field Theory approach by adding higher-dimensional operators to the Standard Model 
Lagrangian and by computing relevant processes at the next-to-leading order, extending the
original $\upkappa\,$-framework.
\end{Abstract}
\vfill
\begin{center}
\end{center}
\end{titlepage}
\def\thefootnote{\arabic{footnote}}
\setcounter{footnote}{0}
\small
\thispagestyle{empty}
\tableofcontents
\normalsize
\clearpage
\setcounter{page}{1}
\section{Introduction}
During Run${-}1$ LHC has discovered a resonance which is a candidate for the Higgs boson of the
Standard Model (SM)~\cite{Chatrchyan:2012ufa,Aad:2012tfa}. The spin-$0$ nature of the resonance 
is well established~\cite{Bolognesi:2012mm} but there is no direct evidence for New Physics; 
furthermore, the available studies on the couplings of the resonance show compatibility with 
the Higgs boson of the SM. 
One possible scenario, in preparation for the results of Run${-}2$, requires a consistent theory 
of SM deviations. 
Ongoing and near future experiments can achieve an estimated per mille sensitivity on precision 
Higgs and electroweak (EW) observables. This level of precision provides a window to indirectly 
explore the theory space of Beyond-the-SM (BSM) physics and place constraints on specific UV 
models. For this purpose, a consistent procedure of constructing SM deviations is clearly 
desirable.

The first attempt to build a framework for SM-deviations is represented by the so-called
$\upkappa\,$-framework, introduced in 
\Brefs{LHCHiggsCrossSectionWorkingGroup:2012nn,Heinemeyer:2013tqa}. 
There is no need to repeat here the main argument, splitting and
shifting different loop contributions in the amplitudes for Higgs-mediated processes.
The $\upkappa\,$-framework is an intuitive language which misses internal consistency when
one moves beyond leading order (LO). As originally formulated, it violates gauge-invariance
and unitarity. In a Quantum Field Theory (QFT) approach to a spontaneously broken theory,
fermion masses and Yukawa couplings are deeply related and one cannot shift couplings while
keeping masses fixed.

To be more specific the original framework has the following limitations: kinematics is not
affected by $\upkappa\,$-parameters, therefore the framework works at the level of total 
cross-sections, not for differential distributions; it is LO, partially accomodating 
factorizable QCD but not EW corrections; it is not QFT-compatible (ad-hoc variation of the 
SM parameters, violates gauge symmetry and unitarity).

However, the original $\upkappa\,$-framework has one main virtue, to represent the first
attempt towards a fully consistent QFT of SM deviations. The question is: can we make it
fully consistent so that the experimental collaborations can simply upgrade their studies
of the Higgs boson couplings? The answer is evidently yes, although the construction of
a consistent theory of SM deviations (beyond LO) is far from trivial, especially from the 
technical point of view.

Recent years have witnessed an increasing interest in Higgs/SM EFT, see in particular
\Brefs{Contino:2013kra,Azatov:2014jga,Contino:2014aaa},
 \Brefs{Berthier:2015oma,Trott:2014dma,Alonso:2013hga,Jenkins:2013wua,Jenkins:2013sda,Jenkins:2013zja,Jenkins:2013fya}, 
 \Brefs{Artoisenet:2013puc,Alloul:2013naa},
 \Bref{Ellis:2014dva},
 \Bref{Falkowski:2014tna},
 \Bref{Low:2009di},
 \Brefs{Degrande:2012wf,Chen:2013kfa},
 \Bref{Grober:2015cwa},
 \Brefs{Englert:2015bwa,Englert:2015zra,Englert:2014uua} and
 \Brefs{Biekoetter:2014jwa,Gupta:2014rxa,Elias-Miro:2013mua,Elias-Miro:2013gya,Pomarol:2013zra,Masso:2014xra}.

In this work we will reestablish that Effective Field Theory (EFT) can provide an adequate answer
beyond LO. Furthermore, EFT represents the optimal approach towards Model Independence.
Of course, there is no formulation that is completely model independent and EFT, as any other 
approach, is based on a given set of (well defined) assumptions. Working within this set we 
will show how to use EFT for building a framework for SM deviations, generalizing the work
of \Bref{Passarino:2012cb}. A short version of our results, containing simple examples, was given 
in \Bref{Ghezzi:2014qpa} and presented in~\cite{POtalk,JS}. 

In full generality we can distinguish a top-down approach (model dependent) and a bottom-up
approach. The top-down approach is based on several steps. First one has to classify BSM models,
possibly respecting custodial symmetry and decoupling, then the corresponding EFT can be 
constructed, \eg via a covariant derivative expansion~\cite{Henning:2014wua}. Once the EFT is 
derived one can construct (model by model) the corresponding SM deviations.

The bottom-up approach starts with the inclusion of a basis of $\mrdim = 6$ operators
and proceeds directly to the classification of SM deviations, possibly respecting
the analytic structure of the SM amplitudes.

The Higgs EFT described and constructed in this work is based on several assumptions.
We consider one Higgs doublet with a linear representation; this is flexible.
We assume that there are no new ``light'' d.o.f. and decoupling of heavy d.o.f.; these are rigid 
assumptions. Absence of mass mixing of new heavy scalars with the SM Higgs doublet is 
also required.

We only work with $\mrdim = 6$ operators. Therefore the scale $\Lambda$ that characterizes the
EFT cannot be too small, otherwise neglecting $\mrdim = 8$ operators is not allowed.
Furthermore, $\Lambda$ cannot be too large, otherwise $\mrdim = 4$ higher-order loops are 
more important than $\mrdim = 6$ interference effects.
It is worth noting that these statements do not imply an inconsistency of EFT. It only means 
that higher dimensional operators and/or higher order EW effects (\eg \Bref{Actis:2008ts})
must be included as well.

To summarize the strategy that will be described in this work we identify
the following steps: start with EFT at a given order (here $\mrdim = 6$ and NLO)
and write any amplitude as a sum of $\upkappa\,$-deformed SM sub-amplitudes
(\eg $\PQt, \PQb$ and bosonic loops in $\PH \to \PGg\PGg$).
Another sum of $\upkappa\,$-deformed non-SM amplitudes is needed to complete the answer;
at this point we can show that the $\upkappa\,$-parameters are linear combinations of
Wilson coefficients.

The rationale for this course of action is better understood in terms of a comparison
between LEP and LHC. Physics is symmetry plus dynamics and symmetry is quintessential (gauge 
invariance \etc); however, symmetry without dynamics does not bring us this far.
At LEP dynamics was the SM, unknowns were $\mh\,(\alphas(\mz), \dots )$; at LHC (post the
discovery) unknowns are SM-deviations, dynamics? Specific BSM models are a choice but one
would like to try also a model-independent approach. Instead of inventing unknown form factors
we propose a decomposition where dynamics is controlled by $\mathrm{dim} = 4$ amplitudes
(with known analytical properties) and deviations (with a direct link to UV completion) are 
(constant) combinations of Wilson coefficients.
Only the comparison with experimental data will allow us to judge the goodness of a proposal
that, for us, is based on the belief that deviations need a SM basis.

On-shell studies at LHC will tell us a lot, off-shell ones will tell us (hopefully) much
more~\cite{Kauer:2012hd,Passarino:2012ri,Passarino:2013bha,Caola:2013yja,Campbell:2014gha}. 
If we run away from the $\PH$ peak with a SM-deformed 
theory, up to some reasonable value $s \muchless \Lambda^2$, we need to reproduce (deformed) SM
low-energy effects, \eg $\PV\PV$ and $\PQt\PQt$ thresholds. The BSM loops will remain unresolved
(as SM loops are unresolved in the Fermi theory). That is why we need to expand the
SM-deformations into a SM basis with the correct (low energy) behavior. If we stay in the 
neighbourhood of the peak any function will work, if we run away we have to know more of the 
analytical properties.

The outline of the paper is as follows: in \sect{Lag} we introduce the EFT Lagrangian.
In \sect{outl} we describe the various aspects of the calculation;
in \sect{UVren} we present details of the renormalization procedure, decays of the Higgs boson are
described in \sect{HBD}, EW precision data in \sect{EWPD}. Technical details, as well as the
complete list of counterterms and amplitudes are given in several appendices.

\section{The Lagrangian \label{Lag}}
In this Section we collect all definitions that are needed to write the Lagrangian
defined by
\bq
\Lag_{\EFT}= \Lag_4 + \sum_{n > 4}\,\sum_{i=1}^{N_n}\,\frac{a^n_i}{\Lambda^{n-4}}\,\Ope^{(d=n)}_i
\spc
\label{defEFT}
\eq
where $\Lag_4$ is the SM Lagrangian~\cite{Bardin:1999ak} and $a^n_i$ are arbitrary Wilson 
coefficients. Our EFT is defined by \eqn{defEFT} and it is based on a number of assumptions:
there is only one Higgs doublet (flexible),
a linear realization is used (flexible),
there are no new ``light'' d.o.f. and decoupling is assumed (rigid),
the UV completion is weakly-coupled and renormalizable (flexible).
Furthermore, neglecting $\mrdim = 8$ operators and NNLO EW corrections
implies the following range of applicability: $3\UTeV < \Lambda < 5\UTeV$.

We can anticipate the strategy by saying that we are at the border of two HEP phases.
A ``predictive'' phase: in any (strictly) renormalizable theory with $n$
parameters one needs to match $n$ data points, the $(n+1)$th calculation is
a prediction, \eg as doable in the SM.
A ``fitting'' (approximate predictive) phase: there are ($N_6{+}N_8{+}\,\dots = \infty)$ 
renormalized Wilson coefficients that have to be fitted, \eg measuring SM deformations due to 
a single $\mcO^{(6)}$ insertion ($N_6$ is enough for per mille accuracy).
\subsection{Conventions}
We begin by considering the field-content of the Lagrangian. The scalar field $\PK$ (with 
hypercharge $1/2$) is defined by
\bq
\PK = \frac{1}{\srt}\,\left(
\begin{array}{c}
\PH + 2\,\frac{M}{g} + i\,\Ppz \\
\srt\,i\,\Ppm
\end{array}
\right)
\eq
$\PH$ is the custodial singlet in $\lpar 2_{\ssL}\,\otimes\,2_{\ssR}\rpar = 1\,\oplus\,3$.
Charge conjugation gives $\upPhi^c_i = \ep_{ij}\,\upPhi^*_j$, or
\bq
\PK^c = -\,\frac{1}{\srt}\,\left(
\begin{array}{c}
\srt\,i\,\Ppp \\
\PH + 2\,\frac{M}{g} - i\,\Ppz 
\end{array}
\right)
\eq
The covariant derivative $D_{\mu}$ is 
\bq
D_{\mu}\,\PK = \Bigl( \pdmu - \frac{i}{2}\,g_{_0}\,B^a_{\mu}\,\uptau_a - 
\frac{i}{2}\,g\,g_{_1}\,B^0_{\mu}\Bigr)\,\PK
\spc
\eq
with $g_{_1} = -\stw/\ctw$ and where $\uptau^a$ are Pauli matrices while $\stw(\ctw)$ is the
sine(cosine) of the weak-mixing angle. Furthermore
\bq
\PWpmmu = \frac{1}{\srt}\,\lpar B^1_{\mu} \mp i\,B^2_{\mu}\rpar\spc
\qquad
\PZmu = \ctw\,B^3_{\mu} - \stw\,B^0_{\mu}\spc
\quad
\PA_{\mu} = \stw\,B^3_{\mu} + \ctw\,B^0_{\mu}\spc
\eq
\bq
F^a_{\mu\nu} = \pdmu\,B^a_{\nu} - \pdnu\,B^a_{\mu}
+ g_{_0}\,\epsilon^{a b c}\,B^b_{\mu}\,B^c_{\nu}\spc
\quad
F^0_{\mu\nu} = \pdmu\,B^0_{\nu} - \pdnu\,B^0_{\mu}\spp
\eq
Here $a,b,\dots = 1,\dots,3$. 
Furthermore, for the QCD part we introduce
\bq
\PG^a_{\mu\nu} = \pdmu\,\Pg^a_{\nu} - \pdnu\,\Pg^a_{\mu}
+ g_{\ssS}\,f^{a b c}\,\Pg^b_{\mu}\,\Pg^c_{\nu}\spp
\eq
Here $a,b,\dots = 1,\dots,8$ and the $f$ are the $SU(3)$ structure constants. 
Finally, we introduce fermions,
\bq
\PpsiL = \left(
\begin{array}{c}
\PQu \\ \PQd
\end{array}
\right)_{\ssL}
\qquad
\Pf_{\ssL\,,\,\ssR} = \frac{1}{2}\,\lpar 1 \pm \gamma^5\rpar \,\Pf
\eq
and their covariant derivatives
\bqa
D_{\mu}\,\PpsiL &=& \lpar \pdmu + g\,B^i_{\mu}\,T_i\rpar \,\PpsiL,
\quad i=0,\dots,3
\nl
T^a &=& -\frac{i}{2}\,\uptau^a\spc  \qquad
T^0 = -\frac{i}{2}\,g_{_2}\,I\spc
\eqa
\bq
D_{\mu}\,\PpsiR = \lpar \pdmu + g\,B^i_{\mu}\,t_i\rpar \,\PpsiR \spc
\qquad t^a = 0\;\; (a \not= 0) \spc
\eq
\bq
t^0 = -\frac{i}{2}\,\left(
\begin{array}{cc}
g_{_3} & 0 \\
0 & g_{_4}
\end{array}
\right)
\eq
with $g_i= -\stw/\ctw\,\uplambda_i$ and
\bq
\uplambda_2 = 1 - 2\,Q_{\PQu}\spc 
\quad
\uplambda_3 =  - 2\,Q_{\PQu}\spc 
\quad
\uplambda_4 =  - 2\,Q_{\PQd}\spp
\eq 
The Standard Model Lagrangian is the sum of several terms:
\bq
\Lag_{\mySM} = \Lag_{\myYM} + \Lag_{\PK} + \Lag{\gfix} +
\Lag_{\myFP} + \Lag_{\Pf}
\eq
\ie, Yang-Mills, scalar, gauge-fixing, Faddeev-Popov ghosts and fermions.
Furthermore, for a proper treatment of the neutral sector of the SM, we express $g_{_0}$
in terms of the coupling constant $g$, 
\bq
g_{_0} = g\,\lpar 1 + g^2\,\Gamma \rpar\spc
\label{Gammadef}
\eq
where $\Gamma$ is fixed by the request that the $\PZ-\PA$ transition is zero at $p^2= 0$,
see \Bref{Actis:2006ra}.
The scalar Lagrangian is given by
\bq
\Lag_{\PK} = - \,\lpar D_{\mu}\,\PK\rpar^{\dagger}\,D_{\mu}\,\PK -
\mu^2\,\PKdag\,\PK - \frac{1}{2}\,\uplambda\,\lpar \PKdag\,\PK\rpar^2\spp
\eq
We will work in the $\bh\,$-scheme of \Bref{Actis:2006ra}, where parameters are transformed
according to the following equations:
\bq
\mu^2 = \bh - 2\,\frac{\uplambda}{g^2}\,\Mbs\spc
\qquad
\uplambda = \frac{1}{4}\,g^2\,\frac{\mhs}{\Mbs}\spp
\eq
Furthermore, we introduce the Higgs VEV, $v= \srt\,M/g$, and fix $\bh$ order-by-order in 
perturbation theory by requiring $\langle\,0\,|\,\PH\,|\,0\,\rangle = 0$. Here we follow the 
approach described in \Brefs{Actis:2006ra,Passarino:1990xx}.
\subsection{$\mrdim = 6\;$ operators}
Our list of $d= 6$ operators is based on the work of 
\Brefs{Buchmuller:1985jz,Grzadkowski:2010es,AguilarSaavedra:2009mx,AguilarSaavedra:2010zi} 
and of \Brefs{Bonnet:2012nm,Bonnet:2011yx,Kanemura:2008ub,Horejsi:2004fs} (see also 
\Brefs{Hagiwara:1993qt,Hankele:2006ma}, 
\Bref{Anastasiou:2011pi},
\Brefs{Corbett:2012dm,Qi:2008ex,Hasegawa:2012mf,Degrande:2012gr,Azatov:2012rd},
\Brefs{GonzalezGarcia:1999fq,Eboli:1998vg,Barger:2003rs} and \Bref{delAguila:2010mx})
and is given in \refT{d6list}. We are not reporting the full set of $\mrdim = 6$
operators introduced in \Bref{Grzadkowski:2010es} but only those that are relevant for
our calculations, \eg CP-odd operators have not been considered in this work. It is worth noting
that we do not assume flavor universality. 

We need matching of UV models onto EFT, order-by-order in a loop expansion.
If $L = \{ \Ope^{(d)}_1,\,\dots\,\Ope^{(d)}_n \}$ is a list of operators in $V^{(d)}$
(the space of d-dimensional, gauge invariant operators), 
then these operators form a basis for $V^{(d)}$ iff every $\Ope^{(d)} \in V^{(d)}$ can be uniquely 
written as a linear combination of the elements in $L$.

While overcomplete sets (\eg those derived without using equations of motion) are useful for 
cross-checking, a set that is not a basis (discarding a priori subsets of operators) is 
questionable, \eg it is not closed under complete renormalization and may lead to violation of 
Ward-Slavnov-Taylor (WST) identities~\cite{Veltman:1970nh,Taylor:1971ff,Slavnov:1972fg}.
Finally, a basis is optimal insofar as it allows to write Feynman rules in arbitrary gauges.
Our choice is given by
\bq
\Lag_{\EFT_6} = \Lag_{\mySM} + \sum_i\,\frac{a_i}{\HSs}\,\Ope^{(6)}_i\spp
\label{FLag}
\eq
In \refT{d6list} we drop the superscript $(6)$ and write the explicit correspondence with the
operators of the so-called Warsaw basis, see \Bref{Grzadkowski:2010es}. We also introduce
\bq
\PQqL = \left(
\begin{array}{c}
\PQu \\ \PQd
\end{array}
\right)_{\ssL}\spc
\qquad
\PLL = \left(
\begin{array}{c}
\PGnl \\ \Pl
\end{array}
\right)_{\ssL}
\eq
where $\PQu$ stands for a generic up-quark ($\{\PQu\,,\,\PQc\,,\,\dots\}$), $\PQd$ stands for 
a generic down-quark ($\{\PQd\,,\,\PQs\,,\,\dots\}$) and $\Pl$ for $\{\Pe\,,\,\PGm\,,\,\dots\}$.
As usual, $\Pf_{\ssL\,,\,\ssR} = \frac{1}{2}\,\lpar 1 \pm \gamma^5\rpar \,\Pf$.
Furthermore,
\bq
\PKdag\,\Ddmu\,\PK =\PKdag D_{\mu} \PK - \lpar D_{\mu} \PK \rpar^{\dagger} \PK\spp
\eq
\begin{table}
\begin{center}
\caption[]{\label{d6list}{List of $\mrdim = 6$ operators, see \Bref{Grzadkowski:2010es}, 
entering the renormalization procedure and the phenomenological applications described in 
this paper}}
\vspace{0.2cm}
\begin{tabular}{ll}
\hline
&  \\
$\Ope_1 = g^3\,\Ope_{\Pphi} = g^3\,\KdK^3$ &
$\Ope_2 = g^2\,\Ope_{\Pphi\,\Box} = g^2\,\KdK\,\Box\,\KdK$ \\
$\Ope_3 = g^2\,\Ope_{\Pphi\,\ssD} = 
       g^2\,\lpar \PKdag \ssD_{\mu} \PK\rpar\,\lbra \lpar \ssD_{\mu} \PK\rpar^\dagger \PK\rbra $ &
$\Ope_4 = g^2\,\Ope_{\Pl\,\upphi} = g^2\,\KdK\,\PALL\,\PK^c\,\PlR$ \\
$\Ope_5 = g^2\,\Ope_{\PQu\,\upphi} = g^2\,\KdK\,\PAQqL\,\PK\,\PQuR$ &
$\Ope_6 = g^2\,\Ope_{\PQd\,\upphi} = g^2\,\KdK\,\PAQqL\,\PK^c\,\PQdR$ \\
$\Ope_7 = g^2\,\Ope^{(1)}_{\upphi\,\Pl} = g^2\,\PKdag \Ddmu \PK\,\PALL \gamma^{\mu} \PLL$ &
$\Ope_8 = g^2\,\Ope^{(1)}_{\upphi\,\PQq} = 
g^2\,\PKdag \Ddmu\,\PK\,\PAQqL \gamma^{\mu} \PQqL$ \\
$\Ope_9 = g^2\,\Ope_{\upphi\,\Pl} = 
g^2\,
\PKdag\,\Ddmu\,\PK 
\,\PAlR \gamma^{\mu} \PlR$ &
$\Ope_{10} = g^2\,\Ope_{\upphi\,\PQu} = 
g^2\,\PKdag \Ddmu\,\PK\,\PAQuR \gamma^{\mu} \PQuR$ \\
$\Ope_{11} = g^2\,\Ope_{\upphi\,\PQd} = 
g^2\,\PKdag \Ddmu\,\PK\,\PAQdR \gamma^{\mu} \PQdR$ &
$\Ope_{12} = g^2\,\Ope_{\upphi\,\PQu\PQd} = 
i\,g^2\,\lpar\,\PKdag\,D_{\mu}\,\PK \rpar\,\PAQuR\,\gamma^{\mu}\,\PQdR$ \\
$\Ope_{13} = g^2\,\Ope^{(3)}_{\upphi\,\Pl} = g^2\,
\PKdag\,\tau^a\,\Ddmu\,\PK \,\PALL\,\tau_a\,\gamma^{\mu}\,\PLL$ &
$\Ope_{14} = g^2\,\Ope^{(3)}_{\upphi\,\PQq} = g^2\,
\PKdag\,\tau^a\,\Ddmu\,\PK\,\PAQqL\,\tau_a\,\gamma^{\mu}\,\PQqL$ \\
$\Ope_{15} = g\,\Ope_{\upphi\,\PG} = g\,\KdK\,\PG^{a\,\mu\nu}\,\PG^a_{\mu\nu}$ &
$\Ope_{16} = g\,\Ope_{\upphi\,\PW} = g\,\KdK\,\ssF^{a\,\mu\nu}\,\ssF^a_{\mu\nu}$ \\
$\Ope_{17} = g\,\Ope_{\upphi\,\PB} = g\,\KdK\,\ssF^{0\,\mu\nu}\,\ssF^0_{\mu\nu}$ &
$\Ope_{18} = g\,\Ope_{\upphi\,\PW\PB} = g\,\PKdag\,\tau^a\,\PK\,
\ssF^{\mu\nu}_a\,\ssF^0_{\mu\nu}$ \\
$\Ope_{19} = g\,\Ope_{\Pl\,\PW} = g\,
\PALL\,\sigma^{\mu\nu}\,\PlR\,\tau_a\,\PK^c\,\ssF^a_{\mu\nu}$ &
$\Ope_{20} = g\,\Ope_{\PQu\,\PW} = g\,
\PAQqL\,\sigma^{\mu\nu}\,\PQuR\,\tau_a\,\PK\,\ssF^a_{\mu\nu}$ \\
$\Ope_{21} = g\,\Ope_{\PQd\,\PW} = g\,
\PAQqL\,\sigma^{\mu\nu}\,\PQdR\,\tau_a\,\PK^c\,\ssF^a_{\mu\nu}$ &
$\Ope_{22} = g\,\Ope_{\Pl\,\PB} = g\,
\PALL\,\sigma^{\mu\nu}\,\PlR\,\PK^c\,\ssF^0_{\mu\nu}$ \\
$\Ope_{23} = g\,\Ope_{\PQu\,\PB} = g\,
\PAQqL\,\sigma^{\mu\nu}\,\PQuR\,\PK\,\ssF^0_{\mu\nu}$ &
$\Ope_{24} = g\,\Ope_{\PQd\,\PB} = g\,
\PAQqL\,\sigma^{\mu\nu}\,\PQdR\,\PK^c\,\ssF^0_{\mu\nu}$ \\
$\Ope_{25} = g\,\Ope_{\PQu\,\PG} = g\,
\PAQqL\,\sigma^{\mu\nu}\,\PQuR\,\uplambda_a\,\PK\,\PG^a_{\mu\nu}$ &
$\Ope_{26} = g\,\Ope_{\PQd\,\PG} = g\,
\PAQqL\,\sigma^{\mu\nu}\,\PQdR\,\uplambda_a\,\PK^c\,\PG^a_{\mu\nu}$ \\
&  \\
\hline
\end{tabular}
\end{center}
\end{table}
We also transform Wilson coefficients according to \refT{nwc}.
\begin{table}
\begin{center}
\caption[]{\label{nwc}{Redefinition of Wilson coefficients}}
\vspace{0.2cm}
\begin{tabular}{llll}
\hline
& & & \\
$g\,a_1 = \ap             $ &  
$g^2\,a_2 = - \apBox   $ &
$g^2\,a_3 = - \apD   $ &  
$g\,\srt\,a_4 = - \frac{\mle}{M}\,\aLp  $ \\
$g\,\srt\,a_5 = - \frac{\mqu}{M}\,\aup $ &
$g\,\srt\,a_6 = - \frac{\mqd}{M}\,\adp $ &
$g^2\,a_7 = - a^{(1)}_{\upphi\,\Pl}                    $ & 
$g^2\,a_8 = - a^{(1)}_{\upphi\,\PQq}                   $ \\ 
$g^2\,a_9 = - a_{\upphi\,\Pl}                   $ & 
$g^2\,a_{10} = - a_{\upphi\,\PQu}               $ & 
$g^2\,a_{11} = - a_{\upphi\,\PQd}               $ & 
$g^2\,a_{12} = - a_{\upphi\,\PQu\PQd}           $ \\
$g^2\,a_{13} = - a^{(3)}_{\upphi\,\Pl}          $ & 
$g^2\,a_{14} = - a^{(3)}_{\upphi\,\PQq}         $ & 
$g^2\,a_{15} = g_{\ssS}\,a_{\upphi\,\PG}        $ &
$g\,a_{16} = a_{\upphi\,\PW}                    $ \\
$g\,a_{17} = a_{\upphi\,\PB}                         $ &
$g\,a_{18} = a_{\upphi\,\PW\PB}                      $ &
$g\,\srt\,a_{19} = \frac{\mle}{M}\,a_{\Pl\,\PW}   $ &
$g\,\srt\,a_{20} = \frac{\mqu}{M}\,a_{\PQu\,\PW} $ \\
$g\,\srt\,a_{21} = \frac{\mqd}{M}\,a_{\PQd\,\PW} $ &
$g\,\srt\,a_{22} = \frac{\mle}{M}\,a_{\Pl\,\PB}   $ &
$g\,\srt\,a_{23} = \frac{\mqu}{M}\,a_{\PQu\,\PB} $ &
$g\,\srt\,a_{24} = \frac{\mqd}{M}\,a_{\PQd\,\PB} $ \\
$g^2\,a_{25} = g_{\ssS}\,a_{\PQu\,\PG} $ &
$g^2\,a_{26} = g_{\ssS}\,a_{\PQd\,\PG} $ &
                                         &
                                         \\
& & & \\
\hline
\end{tabular}
\end{center}
\end{table}
As was pointed out in Tab. C.1 of \Bref{Einhorn:2013kja} the operators can be classified as
potentially-tree-generated (PTG) and loop-generated (LG).
If we assume that the high-energy theory is weakly-coupled and renormalizable
it follows that the PTG/LG classification of \Bref{Einhorn:2013kja} (used here) is correct. 
If we do not assume the above but work always in some EFT context (\ie also the next high-energy 
theory is EFT, possibly involving some strongly interacting theory) then classification changes, 
see Eqs.~(A1-A2) of \Bref{Jenkins:2013fya}.
\subsection{Four-fermion operators \label{ffO}}
For processes that involve external fermions and for the fermion self-energies we also 
need $\mrdim = 6$ four-fermion operators (see Tab.~3 of \Bref{Grzadkowski:2010es}). We show 
here one explicit example
\bqa
V_{\PQu\,\PQu\,\PQd\,\PQd} &=&
\frac{1}{4}\,\frac{g^2 \gds}{M^2}\,a^{(1)}_{\PQq\,\PQq}\,
\gamma^{\mu}\,\gamma_+\;\otimes\;\gamma_{\mu}\,\gamma_+
+ \frac{1}{8}\,\frac{g^2 \gds}{M^2}\,a^{(1)}_{\PQq\,\PQd}\,
\gamma^{\mu}\,\gamma_+\;\otimes\;\gamma_{\mu}\,\gamma_-
\nl
{}&+& \frac{1}{8}\,\frac{g^2 \gds}{M^2}\,a^{(1)}_{\PQq\,\PQu}\,
\gamma^{\mu}\,\gamma_-\;\otimes\;\gamma_{\mu}\,\gamma_+
+ \frac{1}{8}\,\frac{g^2 \gds}{M^2}\,a^{(1)}_{\PQu\,\PQd}\,
\gamma^{\mu}\,\gamma_-\;\otimes\;\gamma_{\mu}\,\gamma_-
\nl
{}&+& \frac{1}{16}\,\frac{g^2 \gds}{M^2}\,a^{(1)}_{\PQq\,\PQu\,\PQq\,\PQd}\,
\gamma_+\;\otimes\;\gamma_+
+ \frac{1}{16}\,\frac{g^2 \gds}{M^2}\,a^{(1)}_{\PQq\,\PQu\,\PQq\,\PQd}\,
\gamma_-\;\otimes\;\gamma_-\spc
\eqa
giving the $\PQu\PQu\PQd\PQd$ four-fermion vertex. Here $\gamma_{\pm} = 1/2\,(1 \pm \gamma^5)$
and $\gds$ is defined in \eqn{gsdef}.
\subsection{From the Lagrangian to the $S\,$-matrix\label{FL}}
There are several technical points that deserve a careful treatment when constructing $S\,$-matrix 
elements from the Lagrangian of \eqn{FLag}. 
We perform field and parameter redefinitions so that all kinetic and mass terms in the Lagrangian 
of \eqn{FLag} have the canonical normalization. First we define
\bq
\bh = 12\,\frac{M^4\,a_{\Pphi}}{g^2\,\HSs} + \bh'\spc
\qquad
\bh' = \lpar 1 + \ssdR_{\bh}\,\frac{\Mbs}{\HSs}\rpar\,\bhb\,
\label{bhbdef}
\eq
and $\bhb$ is fixed, order-by-order, to have zero vacuum expectation value for the 
(properly normalized) Higgs field.

Particular care should be devoted in selecting the starting gauge-fixing Lagrangian. 
In order to reproduce 
the free SM Lagrangian (after redefinitions) we fix an arbitrary gauge, described by four
$\xi$ parameters,
\bq
\Lag_{\mathrm{gf}} = - {\cal C}^+\,{\cal C}^-  - \frac{1}{2}\,{\cal C}^2_{\PZ}
- \frac{1}{2}\,{\cal C}^2_{\PA}\spc
\eq
\bq
{\cal C}^{\pm} = - \xi_{\PW}\,\pdmu\,\PWpmmu + \xi_{\pm}\,M\,\Pppm\spc
\quad
{\cal C}_{\PZ} = - \xi_{\PZ}\,\pdmu\,\PZmu + \xi_0\,\frac{M}{\ctw}\,\Ppz\spc
\quad
{\cal C}_{\PA} = \xi_{\PA}\,\pdmu\,\PAmu\spp
\eq
The full list of redefinitions is given in the following equations, where we have 
introduced $\RL = \Mbs/\HSs$. First the Lagrangian parameters,
\bq
\mhs = \OMHs\,\lpar 1 + \ssdR_{\PM_{\PH}}\,\RL\rpar\spc
\quad
\Mbs = \OMs\,\lpar 1 + \ssdR_{\PM}\,\RL\rpar\spc
\quad
M_{\Pf} = \OMf\,\lpar 1 + \ssdR_{\PM_{\Pf}}\,\RL\rpar\spc
\eq
\bq
\ctw = \lpar 1 + \ssdR_{\ctw}\,\RL\rpar\,\ctb\spc
\qquad
\stw = \lpar 1 + \ssdR_{\stw}\,\RL\rpar\,\stb\spc
\eq
secondly, the fields:
\bq
\begin{array}{lll}
\PH = \lpar 1 + \ssdR_{\PH}\,\RL\rpar\,\PHb \;\;&\;\;
\Ppz = \lpar 1 + \ssdR_{\Ppz}\,\RL\rpar\,\Ppzb \;\;&\;\;
\Pppm = \lpar 1 + \ssdR_{\Pppm}\,\RL\rpar\,\Pppmb \\
\PZmu = \lpar 1 + \ssdR_{\PZZ}\,\RL\rpar\,\PZbmud \;\;&\;\; 
\PAmu = \lpar 1 + \ssdR_{\PAA}\,\RL\rpar\,\PAbmud \;\;&\;\; 
\PWpmmu = \lpar 1 + \ssdR_{\PWpm}\,\RL\rpar\,\PWpmbmud \\
\PXpm = \lpar 1 + \ssdR_{\PXpm}\,\RL\rpar\,\PXpmb \;\;&\;\;
\PYz = \lpar 1 + \ssdR_{\PYz}\,\RL\rpar\,\PYzb \;\;&\;\;
\PYa = \lpar 1 + \ssdR_{\PYa}\,\RL\rpar\,\PYab \\
\end{array} 
\eq
where $\PXpm, \PYz$ and $\PYa$ are FP ghosts. Finally, the gauge parameters, normalized
to one:
\bq
\xi_i = 1 + \ssdR_{\xi_i}\,\RL
\qquad i= \PA, \PZ, \PW, \pm, 0 \spp
\eq
We introduce a new coupling constant
\bq
\gds = \frac{1}{\srt\,\myGF\,\HSs} = 0.0606\,\lpar \frac{\UTeV}{\Lambda}\rpar^2\spc
\label{gsdef}
\eq
where $\myGF$ is the Fermi coupling constant and derive the following solutions:
\bq
\ssdR_i\,\RL = \gds\,\Delta R_i\spc
\eq
where the $\Delta R_i$ are given in \refT{nc}. One could also write a more general relation
\bq
\PZmu = \ssR_{\PZZ}\,\PZbmu + \ssR_{\PZA}\,\PAbmu\spc \qquad
\PAmu = \ssR_{\PAZ}\,\PZbmu + \ssR_{\PAA}\,\PAbmu\spc
\eq
where non diagonal terms start at $\mcO(g^2)$. In this way we could also require cancellation 
of the $\PZ{-}\PA$ transition at $\mcO(\gds)$ but, in our experience, there is little to gain 
with this option.
\begin{table}
\begin{center}
\caption[]{\label{nc}{Normalization conditions}}
\vspace{0.2cm}
\begin{tabular}{lll}
\hline
& & \\
$\Delta R_{\PW} = a_{\upphi\,\sPW}   $ &
$\Delta R_{\PZZ} = \aZZ     $ &
$\Delta R_{\PAA} = \aAA     $ \\
$\Delta R_{\PH} = - \frac{1}{4}\,a_{\upphi\,\ssD} + \apBox  $ &
$\Delta R_{\Pppm} = 0                                   $ &
$\Delta R_{\Ppz} = - \frac{1}{4}\,\apD      $ \\
$\Delta R_{\PXpm} = \frac{1}{2}\,a_{\upphi\,\sPW}        $ &
$\Delta R_{\PYz} = \frac{1}{2}\,\aZZ                    $ &
$\Delta R_{\PYa} = \frac{1}{2}\,\aAA                    $ \\
$\Delta R_{\PQu} = - \frac{1}{2}\,\aup  $ &
$\Delta R_{\PQd} =  \frac{1}{2}\,\adp   $ &
$\Delta R_{\bh} = a_{\upphi\,\sPW} + \frac{1}{4}\,\apD - \apBox   $ \\
$\Delta R_{\xi_{\PW}} = - a_{\upphi\,\sPW}   $ &
$\Delta R_{\xi_{\PZ}} = - \aZZ        $ &
$\Delta R_{\xi_{\PA}} = - \aAA        $ \\
$\Delta R_{\xi_{\pm}} = a_{\upphi\,\sPW}          $ &
$\Delta R_{\xi_0} = \frac{1}{2}\,\apD + \aZZ     $ &
$\Delta R_{\mh} = \frac{1}{2}\,\apD - 2\,\apBox + 12\,\frac{\OMs}{\OMHs}\,\ap  $ \\
$\Delta R_{M} = - 2\,a_{\upphi\,\sPW} $ &
$\Delta R_{\ctw} = - \frac{1}{4}\,\apD + \stbs\,\lpar \aAA - \aZZ\rpar + \stb\,\ctb\,\aAZ   $ & \\
& &  \\
\hline
\end{tabular}
\end{center}
\end{table}
We have introduced the following combinations of Wilson coefficients:
\bqa
\aZZ &=& \stbs\,a_{\upphi\,\sPB} + \ctbs\,a_{\upphi\,\sPW} - \stb\,\ctb\,a_{\upphi\,\sPW\sPB} \spc
\nl
\aAA &=& \ctbs\,a_{\upphi\,\sPB} + \stbs\,a_{\upphi\,\sPW} + \stb\,\ctb\,a_{\upphi\,\sPW\sPB} \spc
\nl
\aAZ &=& 2\,\ctb\,\stb\,\lpar a_{\upphi\,\sPW} - a_{\upphi\,\sPB} \rpar + 
             \lpar 2\,\ctbs - 1 \rpar\,a_{\upphi\,\sPW\sPB} \spp
\eqa
With our choice of reparametrization the final result can be written as follows:
\bq
\Lag\lpar \{\upPhi\}\,,\,\{p\} \rpar =
\Lag_4\lpar \{{\overline{\upPhi}}\}\,,\,\{{\overline{p}}\} \rpar
+ \gds\,\aAZ\,\lpar
\partial_{\mu}\,{\overline{Z}}_{\nu}\,\partial^{\mu}\,{\overline{A}}^{\nu} -
\partial_{\mu}\,{\overline{Z}}_{\nu}\,\partial^{\nu}\,{\overline{A}}^{\mu} \rpar +
\Lag^{\intf}_6\lpar \{{\overline{\upPhi}}\}\,,\,\{{\overline{p}}\} \rpar\spc
\eq
where $\{\upPhi\}$ denotes the collection of fields and $\{p\}$ the collection of parameters.
In the following we will abandon the ${\overline{\upPhi}},\;{\overline{p}}$ notation since no 
confusion can arise.

\section{Overview of the calculation \label{outl}}
NLO EFT ($\mrdim = 6$) is constructed according to the following scheme: each amplitude, \eg 
$\PH \to \mid \mathrm{f} \rangle$, contains one-loop SM diagrams up to the relevant order in 
$g$, (tree) 
contact terms with one $\mrdim = 6$ operator and one-loop diagrams with one $\mrdim = 6$ 
operator insertion. Note that the latter contain also diagrams that do not have a counterpart 
in the SM (\eg bubbles with $3$ external lines). In full generality each amplitude is written 
as follows: 
\bq
\mcA = \sum_{n=\ssN}^{\infty}\,g^n\,\mcA^{(4)}_n +
       \sum_{n=\ssN_6}^{\infty}\,\sum_{l=0}^n\,\sum_{k=1}^{\infty}\,
       g^n\,g^l_{4+2\,k}\,\mcA^{(4+2\,k)}_{n\,l\,k}\spc
\eq
where $g$ is the $SU(2)$ coupling constant and $g_{4+2\,k} = 1/(\sqrt{2}\,G_{\ssF}\,\Lambda^2)^k$.
For each process the $\mathrm{dim} = 4$ LO defines the value of $N$ (\eg $N = 1$ for 
$\PH \to \PV\PV$, $N = 3$ for $\PH \to \PGg\PGg$ \etc). 
Furthermore, $N_6 = N$ for tree initiated processes and 
$N - 2$ for loop initiated ones.   
The full amplitude is obtained by inserting wave-function factors and finite renormalization
counterterms.
Renormalization makes UV finite all relevant, on-shell, $\mathrm{S}\,$-matrix elements. 
It is made in two steps: first we introduce counterterms
\bq
\Phi = \mrZ_{\Phi}\,\Phi_{\ren}\spc \qquad p = \mrZ_{p}\,p_{\ren} \spc
\eq
for fields and parameters. Counterterms are defined by
\bq
\mrZ_i = 1 + \frac{g^2}{16\,\pi^2}\,\lpar \dZ^{(4)}_i + \gds\,\dZ^{(6)}_i \rpar \spp
\eq
We construct self-energies, Dyson resum them and require that all propagators are UV finite.
In a second step we construct $3\,$-point (or higher) functions, check their 
$\Ope^{(4)}\,$-finiteness and remove the remaining $\Ope^{(6)}$ UV divergences
by mixing the Wilson coefficients $\WC{i}$:
\bq
\WC{i} = \sum_j\,Z^{\PW}_{ij}\,\PW^{\ren}_j \spp
\eq
Renormalized Wilson coefficients are scale dependent and the logarithm of the scale
can be resummed in terms of the LO coefficients of the anomalous dimension 
matrix~\cite{Alonso:2013hga}.

Our aim is to discuss Higgs couplings and their SM deviations which requires
precise definitions~\cite{Passarino:2010qk,Goria:2011wa,Gonzalez-Alonso:2015bha}: 
\begin{definition}
\begin{itshape}
The Higgs couplings can be extracted from Green's functions in well-defined 
kinematic limits, \eg residue of the poles after extracting the parts which are 
$1$P reducible. These are well-defined QFT objects, that we can probe both in production and in 
decays; from this perspective, $\PV\PH$ production or vector-boson-fusion are on equal footing 
with $\Pg\Pg$ fusion and Higgs decays. Therefore, the first step requires computing these residues
which is the main result of this paper.
\end{itshape}
\end{definition}

Every approach designed for studying SM deviations at LHC and beyond has to face a 
critical question: generally speaking, at LHC the EW core is embedded into a QCD environment, 
subject to large perturbative corrections and we expect considerable progress in the 
``evolution'' of these corrections; the same considerations apply to PDFs.
Therefore, does it make sense to `fit'' the EW core? Note that this is a general question 
which is not confined to our NLO approach.

In practice, our procedure is to write the answer in terms of SM deviations, \ie the dynamical 
parts are $\mathrm{dim} = 4$ and certain combinations of the deviation parameters will define the 
pseudo-observables (PO) to be fitted. Optimally, part of the factorizing QCD corrections could 
enter the PO definition.
The suggested procedure requires the parametrization to be as general as possible, \ie
no a priori dropping of terms in the basis of operators. This will allow us to ``reweight'' the
results when new (differential) $\mathrm{K}\,$-factors become available; new input will touch 
only the $\mathrm{dim} = 4$ components. PDFs changing is the most serious problem: at LEP the 
$\Pep\Pem$ structure functions were known to very high accuracy (the effect was tested by 
using different QED radiators, differing by higher orders treatment); a change of PDFs at 
LHC will change the convolution and make the reweighting less simple, but still possible. 
For recent progress on the impact of QCD corrections within the EFT approach we 
quote \Bref{Grober:2015cwa}.

\section{Renormalization \label{UVren}}
There are several steps in the renormalization procedure.
The orthodox approach to renormalization uses the language of ``counterterms''.
It is worth noting that this is not a mandatory step, since one could write directly 
renormalization equations that connect the bare parameters of the Lagrangian to a set of data, 
skipping the introduction of intermediate renormalized quantities and avoiding any 
unnecessary reference to a given renormalization scheme. 

In this approach, carried on at one loop in~\cite{Passarino:1989ta}, no special attention is 
paid to individual Green functions, and one is mainly concerned with UV finiteness of 
$\ssS$-matrix elements after the proper treatment of external legs in amputated Green functions, 
which greatly reduces the complexity of the calculation.

However, renormalization equations are usually organized through different building blocks, 
where gauge-boson self-energies embed process-independent (universal) higher-order corrections 
and play a privileged role. Therefore, their structure has to be carefully analyzed, and the 
language of counterterms allows to disentangle UV overlapping divergences which show up at two 
loops.

In a renormalizable gauge theory, in fact, the UV poles of any Green function can be removed 
order-by-order in perturbation theory. In addition, the imaginary part of a Green function 
at a given order is fixed, through unitarity constraints, by the previous orders.
Therefore, UV-subtraction terms have to be at most polynomials in the external momenta (in the 
following, ``local'' subtraction terms).
Therefore, we will express our results using the language of counterterms:
we promote bare quantities (parameters and fields) to renormalized ones and fix the 
counterterms at one loop in order to remove the UV poles.

Obviously, the absorption of UV divergences into local counterterms does not exhaust the 
renormalization procedure, because we have still to connect renormalized quantities to 
experimental data points, thus making the theory predictive. In the remainder of this section 
we discuss renormalization constants for all parameters and fields.
We introduce the following quantities
\bq
\DUV = \frac{2}{\ep} - \gamma_{\ssE} - \ln \pi - \ln\frac{\muRs}{\mu^2}\spc
\qquad
\DUV(x) = \frac{2}{\ep} - \gamma_{\ssE} - \ln \pi - \ln\frac{x}{\mu^2}\spc
\label{DUVfact}
\eq
where $\ep = 4 - \mathrm{d}$, $\mathrm{d}$ is the space-time dimension, $\gamma_{\ssE} = 0.5772$ is
the Euler - Mascheroni constant and $\muR$ is the renormalization scale. 
In \eqn{DUVfact} we have introduced an auxiliary mass $\mu$ which cancels in any UV-renormalized
quantity; $\muR$ cancels only after finite renormalization.
Furthermore, $x$ is positive definite. Only few functions are needed for renormalization
purposes,
\bq
A_0\lpar m \rpar = \frac{\mu^{\ep}}{i\,\pi^2}\,\int d^dq\,\frac{1}{q^2 + m^2} =
-\, m^2\,\lbra \DUVM + a^{\fin}_0\lpar m \rpar \rbra\spc
\quad
a^{\fin}_0\lpar m\rpar = 1 - \ln\frac{m^2}{\mws} \spc
\eq
\bq
B_0\lpar - s\,;\,m_1\,,\,m_2 \rpar = \frac{\mu^{\ep}}{i\,\pi^2}\,\int d^dq\,
\frac{1}{( q^2 + m^2_1)\,((q+p)^2 + m^2_2)} = \DUVM + \bfun{-s}{m_1}{m_2} \spc
\eq
where the finite part is
\bq
\bfun{-s}{m_1}{m_2} = 2 - \ln\frac{m_1 m_2}{\mws} - \ssR - \frac{1}{2}\,
\frac{m^2_1 - m^2_2}{s}\,\ln\frac{m^2_1}{m^2_2}\spc
\quad
\ssR = \frac{\Lambda}{s}\,\ln\frac{m^2_1 + m^2_2 - s - \Lambda - i\,0}{2\,m_2\,m_2}\spc
\label{defB0}
\eq
where $p^2 = - s$ and $\Lambda^2 = \lambda\lpar s\,,\,m^2_1\,,\,m^2_2\rpar$ is the 
K\"{a}llen lambda function. Furthermore we introduce
\bq
\LR = \ln \frac{\muRs}{\mws} \spc
\label{defLR}
\eq
with the choice of the EW scale, $x = \mws$, in \eqn{DUVfact}.

Technically speaking the renormalization program is complete only when UV poles are removed
from all, off-shell, Green functions, something that is beyond the scope of this paper. 
Furthermore, we introduce UV decompositions also for Green functions: given a one-loop Green 
function with $N$ external lines carrying Lorentz indices $\mu_j$, $j=1,\ldots,N$, we introduce 
form factors,
\bq
S_{\mu_1\,\,\dots\,\,\mu_{\ssN}}\, =\, \sum_{a=1}^{\ssA}
S_a \,K^a_{\mu_1\,\,\dots\,\,\mu_{\ssN}}\spp
\label{eq:one:ASDF}
\eq
Here the set $K^a$, with $a = 1,\dots,A$, contains independent tensor structures made up of 
external momenta, Kronecker-delta functions, elements of the Clifford algebra and Levi-Civita 
tensors. A large fraction of the form factors drops from the final answer when we make 
approximations, \eg vector bosons couple only to conserved currents \etc
Requiring that all (off-shell) form factors (including external unphysical lines) are
made UV finite by means of local counterterms implies working in the $R_{\xi\xi}\,$-gauge,
as shown (up to two loops in the SM) in \Bref{Actis:2006rb}.  

A full generality is beyond the scope of this paper, we will limit ourselves to the usual 
't Hooft-Feynman gauge and to those Green functions that are relevant for the phenomenological
applications considered in this paper.
\subsection{Tadpoles and transitions}
We begin by considering the treatment of tadpoles: we fix $\bhb$, \eqn{bhbdef}, such that 
$\langle\,0\,| \PHb |\,0\,\rangle = 0$~\cite{Actis:2006ra}. The solution is
\bq
\bhb = i\,g^2\,\mws\,\,\lpar \bhb^{(4)} + \gds\,\bhb^{(6)} \rpar\spc
\label{defbh}
\eq
where we split according to the following equation (see \eqn{DUVfact})
\bq
\bhb^{(n)} = \beta^{(n)}_{-1}\,\DUVM + \beta^{(n)}_0 + \beta^{(n)}_{\fin} \spp
\eq
The full result for the coefficients $\beta^{(n)}$ is given in \appendx{bhb}.
The parameter $\Gamma$, defined in \eqn{Gammadef}, is fixed by the request that the 
$\PZ-\PA$ transition is zero at $p^2= 0$; the corresponding expression is also reported in 
\appendx{bhb}.
\subsection{$\PH\;$ self-energy}
The one-loop $\PH$ self-energy is given by
\bq
\ssS_{\PHH} = \frac{g^2}{16\,\pi^2}\,\Sigma_{\PHH} =
\frac{g^2}{16\,\pi^2}\,\lpar \Sigma^{(4)}_{\PHH} +
                        \gds\,\Sigma^{(6)}_{\PHH} \rpar\spp
\eq
The bare $\PH$ self-energy is decomposed as follows:
\bq
\Sigma^{(n)}_{\PHH} = \Sigma^{(n)}_{\PHH\,;\,\mUV}\,\DUVM + \Sigma^{(n)}_{\PHH\,;\,\fin}\spp
\eq
Furthermore we introduce
\bq
\Sigma^{(n)}_{\PHH\,;\,\fin}(s) = \Delta^{(n)}_{\PHH\,;\,\fin}(s)\,\mws + 
                                  \Pi^{(n)}_{\PHH\,;\,\fin}(s)\,s\spp
\eq
The full result for the $\PH$ self-energy is given in \appendx{RSE}.
\subsection{$\PA\;$ self-energy}
The one-loop $\PA$ self-energy is given by
\bq
S^{\mu\nu}_{\PAA} = \frac{g^2}{16\,\pi^2}\,\Sigma^{\mu\nu}_{\PAA}\spc
\qquad
\Sigma^{\mu\nu}_{\PAA} = \Pi_{\PAA}\,\ssT^{\mu\nu}\spc
\label{AAself}
\eq
where the Lorentz structure is specified by the tensor
\bq
T^{\mu\nu} = - s\,\delta^{\mu\nu} - p^{\mu}\,p^{\nu} \spc
\eq
and $p^2 = - s$. Furthermore the bare $\Pi_{\PAA}$ is decomposed as follows:
\bq
\Pi_{\PAA} = \Pi^{(4)}_{\PAA} + \gds\,\Pi^{(6)}_{\PAA} \spc
\quad
\Pi^{(n)}_{\PAA} = \Pi^{(n)}_{\PAA\,;\,\mUV}\,\DUVM + \Pi^{(n)}_{\PAA\,;\,\fin}\spp
\eq
It is worth noting that the $\PA{-}\PA$ transition satisfies a doubly-contracted Ward identity
\bq
p_{\mu}\,S^{\mu\nu}_{\PAA}\,p_{\nu} = 0\spp
\eq
The full result for the $\PA$ self-energy is given in \appendx{RSE}.
\subsection{$\PW, \PZ\;$ self-energies}
The one-loop $\PW, \PZ$ self-energies are given by
\bq
S^{\mu\nu}_{\PVV} = \frac{g^2}{16\,\pi^2}\,\Sigma^{\mu\nu}_{\PVV}\spc
\qquad
\Sigma^{\mu\nu}_{\PVV} = \ssD_{\PVV}\,\delta^{\mu\nu} +
\ssP_{\PVV}\,p^{\mu}\,p^{\nu} \spc
\label{ZZself}
\eq
where the form factors are decomposed according to
\bq
\ssD_{\PVV} = \ssD^{(4)}_{\PVV} + \gds\,\ssD^{(6)}_{\PVV} \spc
\qquad
\ssP_{\PVV} = \ssP^{(4)}_{\PVV} + \gds\,\ssP^{(6)}_{\PVV} \spp
\eq
We also introduce the residue of the UV pole and the finite part:
\bq
\ssD^{(n)}_{\PVV} = \ssD^{(n)}_{\PVV\,;\,\mUV}\,\DUVM + \ssD^{(n)}_{\PVV\,;\,\fin}\spc
\eq
\etc The full result for these self-energies is given in \appendx{RSE}. We also introduce
\bqa
\ssD^{(n)}_{\PVV\,;\,\fin}(s) &=& 
\Delta^{(n)}_{\PVV\,;\,\fin}(s)\,\mws + \Pi^{(n)}_{\PVV\,;\,\fin}(s)\,s =
\Delta^{(n)}_{\PVV\,;\,\fin}(0)\,\mws 
+ \Bigl[ \Omega^{(n)}_{\PVV\,;\,\fin}(0) + \mrL^{\sPVV\,;\,\fin}_s \Bigr]\,s + {\mcO}(s^2)\spp
\nl
\mrL^{\sPZZ\,;\,\fin}_s &=& 
- \,\frac{1}{6}\,\ln\lpar - \frac{s}{M^2} \rpar \,\frac{1}{\ctws}\,\myNG
\spc \qquad L^{\sPWW}_s = 0 \spp
\label{DandO}
\eqa
\subsection{$\PZ{-}\PA\;$ transition}
The $\PZ{-}\PA$ transition (up to one loop) is given by
\bq
S^{\mu\nu}_{\PZA} = \frac{g^2}{16\,\pi^2}\,\Sigma^{\mu\nu}_{\PZA} +
\gds\,\ssT^{\mu\nu}\,\aAZ \spc
\qquad
\Sigma^{\mu\nu}_{\PZA} = \Pi_{\PZA}\,\ssT^{\mu\nu} + \ssP_{\PZA}\,p^{\mu}\,p^{\nu} \spc
\label{SZA}
\eq
\bq
\Pi_{\PZA} = \Pi^{(4)}_{\PZA} + \gds\,\Pi^{(6)}_{\PZA} \spc
\qquad
\ssP_{\PZA} = \ssP^{(4)}_{\PZA} + \gds\,\ssP^{(6)}_{\PZA} \spc
\eq
where we have included the term in the bare Lagrangian starting at $\mcO(\gds)$. The full 
result for the $\PZ{-}\PA$ transition is given in 
\appendx{RSE}.
\subsection{The fermion self-energy}
The fermion self-energy is given by
\bq
\ssS_{\Pf} = \frac{g^2}{16\,\pi^2}\,\lbra \Delta_{\Pf} +
\lpar \ssV_{\Pf} - \ssA_{\Pf}\,\gamma^5 \rpar\, i \sla{p} \rbra \spc
\label{fSE}
\eq
with a decomposition
\bq
\Delta_{\Pf} = \Delta^{(4)}_{\Pf} + \gds\,\Delta^{(6)}_{\Pf} \spc
\label{ktermf}
\eq
\etc The full result for the fermion self-energies ($\Pf = \PGn, \Pl, \PQu, \PQd$) is given 
in \appendx{RSE}.
\subsection{Ward{-}Slavnon{-}Taylor identities}
Let us consider doubly-contracted two-point WST 
identity~\cite{Veltman:1970nh,Taylor:1971ff,Slavnov:1972fg}, obtained by connecting two sources 
through vertices and propagators. Here we get, at every order in perturbation theory, the 
identities of \fig{fig:one:WIbare}.
\begin{figure}[ht]
\begin{eqnarray*}
\hbox{
\begin{picture}(80,20)(0,0)
\SetScale{1}
\SetWidth{1}
\Text(30,10)[b]{\scriptsize{$\PA$}}
\Text(70,10)[b]{\scriptsize{$\PA$}}
\Photon(20,0)(40,0){3}{4}
\Photon(60,0)(80,0){3}{4}
\CCirc(50,0){10}{Black}{Gray}
\CBoxc(20,0)(4,4){Black}{Black}
\CBoxc(80,0)(4,4){Black}{Black}
\Text(50,0)[]{\scriptsize{$\EFT$}}
\Text(90,0)[l]{\scriptsize{$=0$}}
\end{picture}
}
\qquad\qquad&&\\
\qquad&&\\
\qquad&&\\
\hspace{-1.5cm}
\hbox{
\begin{picture}(80,20)(0,0)
\SetScale{1}
\SetWidth{1}
\Text(30,10)[b]{\scriptsize{$\PZ$}}
\Text(70,10)[b]{\scriptsize{$\PZ$}}
\Photon(20,0)(40,0){3}{4}
\Photon(60,0)(80,0){3}{4}
\CCirc(50,0){10}{Black}{Gray}
\Text(50,0)[]{\scriptsize{$\EFT$}}
\CBoxc(20,0)(4,4){Black}{Black}
\CBoxc(80,0)(4,4){Black}{Black}
\Text(90,0)[l]{\scriptsize{$+$}}
\end{picture}
\begin{picture}(80,20)(0,0)
\SetScale{1}
\SetWidth{1}
\Text(30,10)[b]{\scriptsize{$\PZ$}}
\Text(70,10)[b]{\scriptsize{$\upphi^0$}}
\Photon(20,0)(40,0){3}{4}
\DashLine(60,0)(80,0){3}
\CCirc(50,0){10}{Black}{Gray}
\Text(50,0)[]{\scriptsize{$\EFT$}}
\CBoxc(20,0)(4,4){Black}{Black}
\CBoxc(80,0)(4,4){Black}{Black}
\Text(90,0)[l]{\scriptsize{$+$}}
\end{picture}
\begin{picture}(80,20)(0,0)
\SetScale{1}
\SetWidth{1}
\Text(30,10)[b]{\scriptsize{$\upphi^0$}}
\Text(70,10)[b]{\scriptsize{$\PZ$}}
\DashLine(20,0)(40,0){3}
\Photon(60,0)(80,0){3}{4}
\CCirc(50,0){10}{Black}{Gray}
\Text(50,0)[]{\scriptsize{$\EFT$}}
\CBoxc(20,0)(4,4){Black}{Black}
\CBoxc(80,0)(4,4){Black}{Black}
\Text(90,0)[l]{\scriptsize{$+$}}
\end{picture}}&&
\hspace{-0.7cm}
\hbox{
\begin{picture}(80,20)(0,0)
\SetScale{1}
\SetWidth{1}
\SetScale{1}
\Text(30,10)[b]{\scriptsize{$\upphi^0$}}
\Text(70,10)[b]{\scriptsize{$\upphi^0$}}
\DashLine(20,0)(40,0){3}
\DashLine(60,0)(80,0){3}
\CCirc(50,0){10}{Black}{Gray}
\Text(50,0)[]{\scriptsize{$\EFT$}}
\CBoxc(20,0)(4,4){Black}{Black}
\CBoxc(80,0)(4,4){Black}{Black}
\Text(90,0)[l]{\scriptsize{$=0$}}
\end{picture}
}\\
\qquad&&\\
\qquad&&\\
\hspace{-1.5cm}
\hbox{
\begin{picture}(80,20)(0,0)
\SetScale{1}
\SetWidth{1}
\Text(30,10)[b]{\scriptsize{$\PW$}}
\Text(70,10)[b]{\scriptsize{$\PW$}}
\Photon(20,0)(40,0){3}{4}
\Photon(60,0)(80,0){3}{4}
\CCirc(50,0){10}{Black}{Gray}
\Text(50,0)[]{\scriptsize{$\EFT$}}
\CBoxc(20,0)(4,4){Black}{Black}
\CBoxc(80,0)(4,4){Black}{Black}
\Text(90,0)[l]{\scriptsize{$+$}}
\end{picture}
\begin{picture}(80,20)(0,0)
\SetScale{1}
\SetWidth{1}
\Text(30,10)[b]{\scriptsize{$\PW$}}
\Text(70,10)[b]{\scriptsize{$\upphi$}}
\Photon(20,0)(40,0){3}{4}
\DashLine(60,0)(80,0){3}
\CCirc(50,0){10}{Black}{Gray}
\Text(50,0)[]{\scriptsize{$\EFT$}}
\CBoxc(20,0)(4,4){Black}{Black}
\CBoxc(80,0)(4,4){Black}{Black}
\Text(90,0)[l]{\scriptsize{$+$}}
\end{picture}
\begin{picture}(80,20)(0,0)
\SetScale{1}
\SetWidth{1}
\Text(30,10)[b]{\scriptsize{$\upphi$}}
\Text(70,10)[b]{\scriptsize{$\PW$}}
\DashLine(20,0)(40,0){3}
\Photon(60,0)(80,0){3}{4}
\CCirc(50,0){10}{Black}{Gray}
\Text(50,0)[]{\scriptsize{$\EFT$}}
\CBoxc(20,0)(4,4){Black}{Black}
\CBoxc(80,0)(4,4){Black}{Black}
\Text(90,0)[l]{\scriptsize{$+$}}
\end{picture}}
&&
\hspace{-0.7cm}
\hbox{
\begin{picture}(80,20)(0,0)
\SetScale{1}
\SetWidth{1}
\Text(30,10)[b]{\scriptsize{$\upphi$}}
\Text(70,10)[b]{\scriptsize{$\upphi$}}
\DashLine(20,0)(40,0){3}
\DashLine(60,0)(80,0){3}
\CCirc(50,0){10}{Black}{Gray}
\Text(50,0)[]{\scriptsize{$\EFT$}}
\CBoxc(20,0)(4,4){Black}{Black}
\CBoxc(80,0)(4,4){Black}{Black}
\Text(90,0)[l]{\scriptsize{$=0$}}
\end{picture}
}\\
\qquad&&\\
\qquad&&\\
\hspace{-1.5cm}
\hbox{
\begin{picture}(80,20)(0,0)
\SetScale{1}
\SetWidth{1}
\Text(30,10)[b]{\scriptsize{$\PA$}}
\Text(70,10)[b]{\scriptsize{$\PZ$}}
\Photon(20,0)(40,0){3}{4}
\Photon(60,0)(80,0){3}{4}
\CCirc(50,0){10}{Black}{Gray}
\Text(50,0)[]{\scriptsize{$\EFT$}}
\CBoxc(20,0)(4,4){Black}{Black}
\CBoxc(80,0)(4,4){Black}{Black}
\Text(90,0)[l]{\scriptsize{$+$}}
\end{picture}
\begin{picture}(80,20)(0,0)
\SetScale{1}
\SetWidth{1}
\Text(30,10)[b]{\scriptsize{$\PA$}}
\Text(70,10)[b]{\scriptsize{$\upphi^0$}}
\Photon(20,0)(40,0){3}{4}
\DashLine(60,0)(80,0){3}
\CCirc(50,0){10}{Black}{Gray}
\Text(50,0)[]{\scriptsize{$\EFT$}}
\CBoxc(20,0)(4,4){Black}{Black}
\CBoxc(80,0)(4,4){Black}{Black}
\Text(90,0)[l]{\scriptsize{$=0$}}
\end{picture}}&&
\end{eqnarray*}
\vspace{0.4cm}
\caption[]{Doubly-contracted WST identities with two external gauge bosons. Gray circles 
denote the sum of the needed Feynman diagrams at any given order in EFT.}
\label{fig:one:WIbare}
\end{figure}
WST identities~\cite{Veltman:1970nh,Taylor:1971ff,Slavnov:1972fg} require additional 
self-energies and transitions, \ie scalar-scalar and vector-scalar components
\bq
S_{\ssS\ssS} = \frac{g^2}{16\,\pi^2}\,\lbra \Sigma^{(4)}_{\ssS\ssS} +
\gds\,\Sigma^{(6)}_{\ssS\ssS} \rbra \spc
\qquad
S^{\mu}_{\ssV\ssS} = i\,\frac{g^2}{16\,\pi^2}\,\lbra \Sigma^{(4)}_{\ssV\ssS} +
\gds\,\Sigma^{(6)}_{\ssV\ssS} \rbra\,p^{\mu} \spp
\eq
\subsection{Dyson resummed propagators}
We will now present the Dyson resummed propagators for the electroweak gauge
bosons. The function ${\Pi}^{\ssI}_{ij}$ represents the sum of all $1$PI diagrams with two
external boson fields, $i$ and $j$, to all orders in perturbation theory (as
usual, the external Born propagators are not to be included in the
expression for ${\Pi}^{\ssI}_{ij}$). We write explicitly its Lorentz structure,
\bq 
{\Pi}^{\ssI}_{\mu \nu,\PVV} =  \ssD^{\ssI}_{\PVV}\,\delta_{\mu \nu} + 
    \ssP^{\ssI}_{\PVV}\,p_{\mu}\,p_{\nu}  \spc
\label{eq:PiLorentz}
\eq
where $\PV$ indicates SM vector fields, and $p_{\mu}$ is the incoming momentum of the vector 
boson.
The full propagator for a field $i$ which mixes with a field $j$ via the function 
${\Pi}^{\ssI}_{ij}$ is given by the perturbative series
\bqa
\bar{\Delta}_{ii} &=& 
    \Delta_{ii} \,+\, \Delta_{ii} \sum_{n=0}^{\infty} \,
      \prod_{l=1}^{n+1} \sum_{k_l} \Pi_{k_{l-1} k_{l}}^{\ssI} 
      \Delta_{k_{l} k_{l}} \spc
\label{eq:aa32}
    \\ &=& 
    \Delta_{ii} \,+\, \Delta_{ii}\,{\Pi}^{\ssI}_{ii}\,\Delta_{ii} 
    \,+\, \Delta_{ii} \!\sum_{k_1=i,j} {\Pi}^{\ssI}_{i k_1}
    \Delta_{k_1 k_1} {\Pi}^{\ssI}_{k_1 i} 
    \Delta_{ii} \,+\, \dots   
\nonumber
\eqa
where $k_0=k_{n+1}=i$, while for $l\neq n+1$, $k_l$ can be $i$ or
$j$. ${\Delta}_{ii}$ is the Born propagator of the field $i$. We write
\bq
\bar{\Delta}_{ii} = \Delta_{ii}\,\left[1 - \left({\Pi}\,\Delta\right)_{ii}\right]^{-1} \spc
\label{eq:dysonR}
\eq
and refer to $\bar{\Delta}_{ii}$ as the resummed propagator. The quantity $({\Pi}\,\Delta)_{ii}$ 
is the sum of all the possible products of Born propagators and self-energies, starting with 
a $1$PI self-energy ${\Pi}^{\ssI}_{ii}$, or transition ${\Pi}^{\ssI}_{ij}$, and
ending with a propagator $\Delta_{ii}$, such that each element of the sum cannot be obtained 
as a product of other elements in the sum. 

In practice it is useful to define, as an auxiliary quantity, the ``partially resummed'' 
propagator for the field $i$, $\hat{\Delta}_{ii}$, in which we resum only the proper $1$PI 
self-energy insertions ${\Pi}^{\ssI}_{ii}$, namely,
\bq 
\hat{\Delta}_{ii} = \Delta_{ii}\,\left[1 - {\Pi}^{\ssI}_{ii}\,\Delta_{ii} \right]^{-1} \spp
\label{eq:partialDyson}
\eq
If the particle $i$ were not mixing with $j$ through loops or two-leg vertex insertions, 
$\hat{\Delta}_{ii}$ would coincide with the resummed propagator
$\bar{\Delta}_{ii}$. 
Partially resummed propagators allow for a compact expression for $({\Pi}\,\Delta)_{ii}$,
\bq
({\Pi}\,\Delta)_{ii} = \Pi^{\ssI}_{ii} \Delta_{ii} + \Pi^{\ssI}_{ij} \hat{\Delta}_{jj} 
    \Pi^{\ssI}_{ji} \Delta_{ii} \spc
\eq
so that the resummed propagator of the field $i$ can be cast in the form
\bq
\bar{\Delta}_{ii} = \Delta_{ii}\,\left[1 - \left(\Pi^{\ssI}_{ii} +
     \Pi^{\ssI}_{ij} \hat{\Delta}_{jj} \Pi^{\ssI}_{ji} \right) \Delta_{ii} \right]^{-1}
\label{eq:dysonR2}
\eq
We can also define a resummed propagator for the $i${-}$j$ transition. In this case there is no 
corresponding Born propagator, and the resummed one is given by the sum of all possible products 
of 1PI $i$ and $j$ self-energies, transitions, and Born propagators starting with 
$\Delta_{ii}$ and ending with $\Delta_{jj}$.  This sum can be simply expressed in the following
compact form,
\bq
\bar{\Delta}_{ij} = \bar{\Delta}_{ii} \,{\Pi}^{\ssI}_{ij}\,\hat{\Delta}_{jj} \spp
\label{eq:dysonRmix}
\eq
\subsection{Renormalization of two-point functions}
Dyson resummed propagators are crucial for discussing several issues, from renormalization
to Ward-Slavnov-Taylor (WST) identities~\cite{Veltman:1970nh,Taylor:1971ff,Slavnov:1972fg}. 
Consider the $\PW$ or $\PZ$ self-energy; in general we have
\bq
\Sigma^{\PVV}_{\mu\nu}(s) = \frac{g^2}{16\,\pi^2}\,\lbra
\ssD^{\PVV}(s)\,\delta_{\mu\nu} + \ssP^{\PVV}(s)\,p_{\mu} p_{\nu} \rbra \spp
\eq
The corresponding partially resummed propagator is
\bq
\hD^{\PVV}_{\mu\nu} =
-\,\frac{\delta_{\mu\nu}}{s - M^2_{\PV} + \frac{g^2}{16\,\pi^2}\,\ssD^{\PVV}} +
\frac{g^2}{16\,\pi^2}\,\frac{\ssP^{\PVV}\,p_{\mu} p_{\nu}} 
{\lpar s - M^2_{\PV} + \frac{g^2}{16\,\pi^2}\,\ssD^{\PVV}\rpar\,
 \lpar s - M^2_{\PV} + \frac{g^2}{16\,\pi^2}\,\ssD^{\PVV} - 
\frac{g^2}{16\,\pi^2}\,\ssP^{\PVV}\,s\rpar} \spp
\eq
We only consider the case where $\PV$ couples to a conserved current; furthermore, we start
by including one-particle irreducible ($1$PI) self-energies. Therefore the inverse propagators 
are defined as follows:
\bei
\item $\PH$ partially resummed propagator is given by
\eei
\bq
g^{-2}\,\hat{\Delta}^{-1}_{\PHH}(s) =
   - g^{-2}\,\ssZ_{\PH}\,\lpar s - \mhs \rpar  - 
   \frac{1}{16\,\pi^2}\,\Sigma_{\PHH} \spp
\eq
\bei
\item $\PA$ partially resummed propagator is given by
\eei
\bq
g^{-2}\,\hat{\Delta}^{-1}_{\PAA}(s) =
   - g^{-2}\,s\,\lpar \ssZ_{\PA} - \frac{1}{16\,\pi^2}\,\Pi_{\PAA} \rpar \spp
\label{IAP}
\eq
\bei
\item $\PW$ partially resummed propagator is given by
\eei
\bq
g^{-2}\,\hat{\Delta}^{-1}_{\PWW}(s) =
   - g^{-2}\,\ssZ_{\PW}\,\lpar s - \Mbs\rpar   -  \frac{1}{16\,\pi^2}\,\ssD_{\PWW} \spp
\label{IWP}
\eq
\bei
\item $\PZ$ partially resummed propagator is given by
\eei
\bq
g^{-2}\,\hat{\Delta}^{-1}_{\PZZ}(s) =
   - g^{-2}\,\ssZ_{\PZ}\,\lpar s - \Mzbs\rpar   -  \frac{1}{16\,\pi^2}\,\ssD_{\PZZ} \spp
\label{IZP}
\eq
\bei
\item $\PZ{-}\PA$ transition is given by
\eei
\bq
S^{\PZA}_{\mu\nu} + S^{\PZA\,\ct}_{\mu\nu} 
\quad
S^{\PZA\,\ct}_{\mu\nu} = \frac{g^2}{16\,\pi^2}\,\Sigma^{\PZA\,\ct}_{\mu\nu}\,\DUV \spc
\label{ZAtr}
\eq
where $S^{\PZA}_{\mu\nu}$ is given in \eqn{SZA} and
\bq
\Sigma^{\PZA\,\ct}_{\mu\nu} = s\;\ssdZ^{(4)}_{\PAZ}\,\delta_{\mu\nu} +
\gds\,\lbra
\,s\;\ssdZ^{(6)}_{\PAZ}\,\delta_{\mu\nu} - 
\aAZ\,\lpar \ssdZ^{(4)}_{\PZ} + \ssdZ^{(4)}_{\PA} \rpar\,p_{\mu}\,p_{\nu} \rbra \spp
\eq
\bei
\item $\Pf$ resummed propagator is given by
\eei
\bq
\ssG^{-1}_{\Pf}(p) = {\overline{\ssZ}}_{\Pf}\,\lpar i\,\sla{p} 
                     + m_{\Pf} \rpar \ssZ_{\Pf} - \ssS_{\Pf} \spc
\label{IFP}
\eq
where the counterterms are
\bq 
\ssZ_{\Pf} = \ssZ_{\ssR\,\Pf}\,\gamma^{-} + \ssZ_{\ssL\,\Pf}\,\gamma^{+} \spc
\qquad 
{\overline{\ssZ}}_{\Pf} = \ssZ_{\ssL\,\Pf}\,\gamma^{+} + \ssZ_{\ssR\,\Pf}\,\gamma^{-}
\qquad
\gamma^{\pm} = \frac{1}{2}\,\lpar 1 \pm \gamma^5 \rpar \spc
\label{FSE}
\eq
\bq
\ssZ_{\ssI\,\Pf} = 1 - \frac{1}{2}\,\frac{g^2}{16\,\pi^2}\,\lbra
\ssdZ^{(4)}_{\ssI\,\Pf} + \gds\,\ssdZ^{(6)}_{\ssI\,\Pf}\,\DUV \rbra \spc
\qquad
m_{\Pf} = M_{\Pf}\,\lpar 1 + \frac{g^2}{16\,\pi^2}\,\ssdZ_{m_{\Pf}}\,\DUV \rpar \spc
\eq
where $M_{\Pf}$ denotes the renormalized fermion mass and $\ssI= \ssL, \ssR$. 
We have introduced counterterms for fields
\bq
\upPhi = \ssZ_{\upPhi}\,\upPhi_{\ren} \spc
\quad
\ssZ_{\upPhi} = 1 + \frac{g^2}{16\,\pi^2}\,\lpar \ssdZ^{(4)}_{\upPhi} +
\gds\,\ssdZ^{(6)}_{\upPhi}\rpar\,\DUV \spp
\eq
The bare photon field represents an exception, and here we use
\bq
\PA_{\mu} = \ssZ_{\PA}\,\PA^{\ren}_{\mu} + \ssZ_{\PAZ}\,\PZ^{\ren}_{\mu} \spc
\qquad
\ssZ_{\PAZ} = \frac{g^2}{16\,\pi^2}\,\lpar \ssdZ^{(4)}_{\PAZ} +
\gds\,\ssdZ^{(6)}_{\PAZ}\rpar\,\DUV \spp
\eq
In addition, bare fermion fields $\uppsi$ are written by means of bare left-handed and
right-handed chiral fields, $\uppsi_{\ssL}$ and $\uppsi_{\ssR}$. The latter are traded for 
renormalized fields.   

For masses we introduce 
\bq
\ssM^2 = \ssZ_{\ssM}\,\ssM^2_{\ren}
\quad
\ssZ_{\ssM} = 1 + \frac{g^2}{16\,\pi^2}\,\lpar \ssdZ^{(4)}_{\ssM} +
\gds\,\ssdZ^{(6)}_{\ssM}\rpar\,\DUV
\eq
and for parameters
\bq
\ssp = \ssZ_{\ssp}\,\ssp_{\ren}
\quad
\ssZ_{\ssp} = 1 + \frac{g^2}{16\,\pi^2}\,\lpar \ssdZ^{(4)}_{\ssp} +
\gds\,\ssdZ^{(6)}_{\ssp}\rpar\,\DUV \spp
\eq
The full list of counterterms is given in \appendx{LCT}.
It is worth noting that the insertion of $\mrdim = 6$ operators in the fermion self-energy
introduces UV divergences in $\Delta^{(6)}_{\Pf}$, \eqn{ktermf},  that are proportional to $s$ 
and cannot be absorbed by counterterms. They enter wave-function renormalization factors and will
be cancelled at the level of mixing among Wilson coefficients.
\subsection{One-particle reducible transitions}
Our procedure is such that there is a $\PZ{-}\PA$ vertex of $\mcO(\gds)$
\bq
\ssV^{\PZA}_{\mu\nu} = \gds\,\ssT_{\mu\nu}\,\aAZ \spc
\qquad
\ssT_{\mu\nu} = - s\,\delta_{\mu\nu} - p_{\mu}\,p_{\nu} \spc
\eq
inducing one-particle reducible ($1$PR) contributions to the self-energies. Since
$p^{\mu}\,\ssT_{\mu\nu} = 0$ we obtain
\bqa
\Pi^{\PAA}\bmid_{1\ssPR} &=& \frac{g^2 \gds}{16\,\pi^2}\,\frac{\stw}{\ctw}\,
\frac{s}{s - \Mzbs}\,\aAZ\,\Pi^{(4)}_{\PZA} \spc
\nl
\Pi^{\PZZ}\bmid_{1\ssPR} &=& \frac{g^2 \gds}{16\,\pi^2}\,\frac{\stw}{\ctw}\,
\aAZ\,\Pi^{(4)}_{\PZA} \spc
\nl
\Pi^{\PZA}\bmid_{1\ssPR} &=& \frac{g^2 \gds}{16\,\pi^2}\,\stws\,
\aAZ\,\Pi^{(4)}_{\PAA} \spp
\label{OPR}
\eqa
\subsection{$\PA{-}\PA$, $\PZ{-}\PA$, $\PZ{-}\PZ$ and $\PW{-}\PW$ transitions at $s = 0$}
The value $s = 0$ is particularly important since $\ssS,\ssT$ and $\ssU$ 
parameters~\cite{Peskin:1990zt} require self-energies and transitions at $s = 0$. We introduce 
the following functions:
\bq
\bfun{-s}{m_1}{m_2} = \bfun{0}{m_1}{m_2} - s\,\bfunp{0}{m_1}{m_2} + 
\frac{1}{2}\,s^2\,\bfuns{0}{m_1}{m_2} + {\mcO}(s^3)  \spc
\eq
where, with two different masses, we obtain
\bqa
\bfun{0}{m_1}{m_2} &=& \frac{m^2_2\,\afun{m_2} - m^2_1\,\afun{m_1}}{m^2_1 - m^2_2} \spc
\nl
\bfunp{0}{m_1}{m_2} &=& -\,\frac{1}{\lpar m^2_1 - m^2_2\rpar^3}\,
\Bigl[ \frac{1}{2}\,\lpar m^4_1 - m^4_2\rpar + m^4_2\,\afun{m_2} - m^4_1\,\afun{m_1} \Bigr]
\nl
{}&-& \frac{1}{\lpar m^2_1 - m^2_2\rpar^2}\,\Bigl[
m^2_1\,\afun{m_1} + m^2_2\,\afun{m_2} \Bigr] \spc
\nl
\bfuns{0}{m_1}{m_2} &=& 
          \frac{1}{\lpar m^2_1 - m^2_2\rpar^5}\,
\Bigl[ \frac{10}{3}\,\lpar m^6_1 - m^6_2\rpar + 4\,m^6_2\,\afun{m_2} - 4\,m^6_1\,\afun{m_1} \Bigr]
\nl
{}&+&          \frac{3}{\lpar m^2_1 - m^2_2\rpar^4}\,
\Bigl[ 2\,m^4_1\,\afun{m_1} + 2\,m^4_2\,\afun{m_2} - m^4_1 - m^4_2 \Bigr]
\nl
{}&+&          \frac{2}{\lpar m^2_1 - m^2_2\rpar^3}\,
\Bigl[ m^2_2\,\afun{m_2} - m^2_1\,\afun{m_1} \Bigr] \spp
\eqa
For equal masses we derive
\bq
\bfun{0}{m}{m} = 1 - \afun{m} \spc
\quad 
\bfunp{0}{m}{m} = -\,\frac{1}{6\,m^2} \spc
\quad
\bfuns{0}{m}{m} = \frac{1}{30\,m^4} \spp
\label{scaled}
\eq
After renormalization we obtain the results of \appendx{RSEZ}, with $\Pi$ defined in \eqn{AAself}
and $\Delta, \Omega$ defined in \eqn{DandO}. Furthermore $\myNG$ is the number of fermion 
generations and $\LR$ is defined in \eqn{defLR}. All functions defined in \eqn{scaled} are
successively scaled with $\mw$. In \appendx{RSEZ} we have used $s = \stw$, $c= \ctw$ and
$x_i$ are ratios of renormalized masses, \ie $\xph= \mh/M$, \etc

The expressions corresponding to $\mrdim = 6$ are rather long and we found convenient
to introduce linear combinations of Wilson coefficients, given in \eqn{LCall}.

\bq
\begin{array}{ll}
\apW = \stws\,\aAA + \ctw\stw\,\aAZ + \ctws\,\aZZ \quad & \quad
\apB = \ctws\,\aAA - \ctw\stw\,\aAZ + \stws\,\aZZ  \\
\apWB =  2\,\ctw\stw\,(\aAA - \aZZ) + ( 1 - 2\,\stws)\,\aAZ \quad & \quad
\apl= \frac{1}{2}\,\lpar \aplA - \aplV \rpar  \\
\apu= \frac{1}{2}\,\lpar \apuV - \apuA \rpar \quad & \quad
\apd= \frac{1}{2}\,\lpar \apdA - \apdV \rpar  \\
\aplo= \aplt - \frac{1}{2}\,\lpar \aplV + \aplA \rpar \quad & \quad
\aplt= \frac{1}{4}\,\lpar \aplV + \aplA + \apn \rpar  \\
\apqo= \frac{1}{4}\,\lpar \apuV + \apuA - \apdV - \apdA \rpar \quad & \quad
\apqt= \frac{1}{4}\,\lpar \apdV + \apdA + \apuV + \apuA \rpar  \\
\alW = \stw\,\alWB + \ctw\,\alBW \quad & \quad
\alB = \stw\,\alBW - \ctw\,\alWB  \nl
\end{array}
\eq
\bq
\begin{array}{ll}
\adW =  \stw\,\adWB + \ctw\,\adBW \quad & \quad
\adB = \stw\,\adBW - \ctw\,\adWB  \\
\auW = \stw\,\auWB + \ctw\,\auBW \quad & \quad
\auB = - \stw\,\auBW + \ctw\,\auWB  \\
\apWB = \ctw\,\apWA - \stw\,\apWZ \quad & \quad
\apW = \stw\,\apWA + \ctw\,\apWZ  \\
\apD = \apDB - 8\,\stws\,\apB  \quad & \quad
\apWDp= 4\,\apW + \apD  \\
\apWDm= 4\,\apW - \apD \quad & \quad
\aplWt= 4\,\aplt + 2\,\apW  \\
\apqWt= 4\,\apqt + 2\,\apW \quad & \quad
\apWDmB= \apWDp - 4\,\apBox  \\
\apWDpB= \apWDp + 4\,\apBox \quad & \quad
\apWBa = \apB - \apW  \\
\end{array}
\label{LCall}
\eq
The results for $\mrdim = 6$ simplify considerably if we neglect loop generated operators,
for instance one obtains full factorization for $\Pi_{\PAA}(0)$,
\bq
\Pi^{(6)}_{\PAA}(0) = - 8\,\frac{\ctws}{\stws}\,\apD\,\Pi^{(4)}_{\PAA}(0)
\eq
and partial factorization for the rest, \eg
\bqa
\Pi^{(6)}_{\PZA}(0) &=& 
- 4\,\frac{\ctws}{\stws}\,\apD\,\Pi^{(4)}_{\PZA}(0)
+ \frac{\stw}{\ctw}\,\apD\,\Pi^{(6)\,\nfact}_{\PZA}(0)
\nl
{}&-& \frac{2}{3}\,\lpar 1 - \LR \rpar\,\frac{\stw}{\ctw}\,\sum_{\gen}\,
\lpar \aplV + 2\,\apuV + \apdV \rpar
\nl
{}&-& \frac{2}{3}\,\frac{\stw}{\ctw}\,\sum_{\gen}\,
\Bigl[ 2\,\afun{\mqu}\,\apuV + \afun{\mqd}\,\apdV + \afun{\mle}\,\aplV \Bigr]
\eqa
\bqa
\Pi^{(6)\,\nfact}_{\PZA}(0) &=&
          - \frac{1}{24}\,\lpar 1 - 14\,\ctws \rpar
          - \frac{1}{24}\,\lpar 1 + 18\,\ctws \rpar\,\LR
          - \frac{1}{9}\,\lpar 5 + 8\,\ctws \rpar\,\lpar 1 - \LR \rpar\,\myNG
\nl
{}&-& \frac{1}{12}\,\lpar 3 + 4\,\ctws \rpar\,\sum_{\gen}\,\afun{\mle}
          - \frac{1}{18}\,\lpar 5 + 8\,\ctws \rpar\,\sum_{\gen}\,\afun{\mqu}
          - \frac{1}{36}\,\lpar 1 + 4\,\ctws \rpar\,\sum_{\gen}\,\afun{\mqd}
\nl
{}&+& \frac{1}{24}\,\lpar 1 + 18\,\ctws \rpar\,\afun{M}
\eqa
The rather long expressions with PTG and LG operator insertions are reported in \appendx{RSEZ}.
Results in this section and in \appendx{RSEZ} refer to the expansion of the $1$PI 
self-energies; inclusion of $1$PR components amounts to the following replacements
\bqa
\Sigma^{(\OPI + \OPR)}_{\PAA}(s) &=& - \Pi^{(\OPI)}_{\PAA}(0)\,s -
\frac{\bigl[ \Pi^{(\OPR)}_{\PZA}(0)\,s \bigr]^2}{s - M^2_0} + \mcO(s^2) =
- \Pi^{(\OPI)}_{\PAA}(0)\,s + \mcO(s^2) \spc
\nl
D^{(\OPI + \OPR)}_{\PZZ}(s) &=&
\Delta^{(\OPI)}_{\PZZ}(0) + \Bigl\{ \Omega^{(\OPI)}_{\PZZ}(0) -
\Bigl[ \Pi^{(\OPR)}_{\PZA}(0) \Bigr]^2 \Bigr\}\,s + \mcO(s^2) \spp
\eqa
\subsection{Finite renormalization \label{FiniteR}}
The last step in one-loop renormalization is the connection between renormalized quantities and 
POs. Since all quantities at this stage are UV-free, we term it ``finite renormalization''.
Note that the absorption of UV divergences into local counterterms is, to some extent, a trivial 
step; finite renormalization, instead, requires more attention. For example, beyond one loop one 
cannot use on-shell masses but only complex poles for all unstable 
particles~\cite{Actis:2006rc,Passarino:2010qk}. Let us show some examples where the concept of 
an on-shell mass can be employed. Suppose that we renormalize a physical (pseudo-)observable $F$,
\bq
F = F_{\ssB} + 
 \frac{g^2}{16\,\pi^2}\,\lbra F^{(4)}_{1\ssL}(m^2) + \gds\,F^{(6)}_{1\ssL}(m^2) \rbra
 + \mcO(g^4) \spc
\eq
where $m$ is some renormalized mass. Consider two cases: a) two-loop corrections are not included 
and b) $m$ appears at one and two loops in $F_{1\ssL}$ and $F_{2\ssL}$ but does not show up in 
the Born term $F_{\ssB}$.
In these cases we can use the concept of an on-shell mass performing a finite mass 
renormalization at one loop. If $m_0$ is the bare mass for the field $\PV$ we write
\bq
m^2_0 = M^2_{\ssOS} \, \left\{ 1\, +\, \frac{g^2}{16\,\pi^2}\,
\Re\,\Sigma_{\PVV\,;\,\fin}\bmid_{s= M^2_{\ssOS}}\, 
\right\} =
M^2_{\ssOS} + g^2\,\Delta M^2 \spc
\label{stillOMS}
\eq
where $M_{\ssOS}$ is the on-shell mass and $\Sigma$ is extracted from the required one-particle 
irreducible Green function; \eqn{stillOMS} is still meaningful (no dependence on gauge parameters) 
and will be used inside the result.

In the Complex Pole scheme we replace the conventional on-shell mass renormalization equation
with the associated expression for the complex pole
\bq
m^2_0 = M^2_{\ssOS}\,\lbra 1 + \frac{g^2}{16\,\pi^2}\,
\Re\,\Sigma_{\PVV\,;\,\fin}\lpar M^2_{\ssOS} \rpar \rbra
\Longrightarrow
m^2_0 = s_{\PV}\,\lbra 1 + \frac{g^2}{16\,\pi^2}\,
\Sigma_{\PVV\,;\,\fin}\lpar M^2_{\ssOS} \rpar \rbra \spc
\eq
where $s_{\PV}$ is the complex pole associated to $\PV$. In this Section we will discuss on-shell
finite renormalization; after removal of UV poles we have replaced $m_0 \to m_{\ren}$ \etc
and we introduce
\bq
M_{\PV\,\ren} = M_{\PV\,;\,\ssOS} + \frac{g^2_{\ren}}{16\,\pi^2}\,\lpar
\ssdCZ^{(4)}_{M_{\PV}} + \gds\,\ssdCZ^{(6)}_{M_{\PV}} \rpar
\eq
and require that $s = M_{\PV\,;\,\OS}$ is a zero of the real part of the inverse $\PV$ propagator,
up to$ \mcO(g^2 \gds)$. Therefore we introduce 
\bqa
M^2_{\ren} &=& M^2_{\PW\,;\,\ssOS}\,\Bigl[ 1 + \frac{g^2_{\ren}}{16\,\pi^2}\,\lpar
\ssdCZ^{(4)}_{\mw} + \gds\,\ssdCZ^{(6)}_{\mw} \Bigr] \rpar \spc
\nl
M^2_{\PH\,\ren} &=& M^2_{\PH\,;\,\ssOS}\,\Bigl[ 1 + \frac{g^2_{\ren}}{16\,\pi^2}\,\lpar
\ssdCZ^{(4)}_{M_{\PH}} + \gds\,\ssdCZ^{(6)}_{M_{\PH}} \rpar \Bigr] \spc
\nl
\ctwr &=& \ctW\,\Bigl[1 + \frac{g^2_{\ren}}{16\,\pi^2}\,\lpar
\ssdCZ^{(4)}_{\ctw} + \gds\,\ssdCZ^{(6)}_{\ctw} \rpar \Bigr] \spc
\label{IPSa}
\eqa
where $\ctWs = \mwsOS/\mzsOS$ and $s = \mzsOS$ will be a zero of the real part of the inverse
$\PZ$ propagator. 

Finite renormalization in the fermion sector requires the following steps:
if $M_{\Pf\,;\,\ssOS}$ denotes the on-shell fermion mass, using \eqn{FSE}, we write
\bq
M_{\Pf}\,\lbra \ssdCZ^{(4)}_{M_{\Pf}} 
+ \gds\,\ssdCZ^{(6)}_{M_{\Pf}}  \rbra =  
\Delta_{\Pf}\lpar M^2_{\Pf\,;\,\ssOS}\rpar + M_{\Pf\,;\,\ssOS}\,
V_{\Pf}\lpar M^2_{\Pf\,;\,\ssOS}\rpar
\eq
and determine the finite counterterms which are given in \appendx{FCT}.
\subsubsection{$\myGF\;$ renormalization scheme}
In the $\myGF\,$-scheme we write the following equation for the $g$ finite renormalization
\bq
g_{\ren} = g_{\exp} + \frac{g^2_{\exp}}{16\,\pi^2}\,\lpar
\ssdCZ^{(4)}_g + \gds\,\ssdCZ^{(6)}_g \rpar \spc
\label{gexp}
\eq
where $g_{\exp}$ will be expressed in terms of the Fermi coupling constant $\myGF$.
The $\mu\,$-lifetime can be written in the form
\bq
\frac{1}{\tau_{\mu}} = \frac{M^5_{\PGm}}{192\,\pi^3}\,\frac{g^4}{32\,\Mbq}\,
\lpar 1 + \delta_{\mu} \rpar \spp
\eq
The radiative corrections are $\delta_{\mu} = \delta^{\PW}_{\mu} + \delta_{\ssG}$
where $\delta_{\ssG}$ is the sum of vertices, boxes etc and $\delta^{\PW}_{\mu}$ is due
to the $\PW$ self-energy. The renormalization equation becomes
\bq
\frac{\myGF}{\srt} = \frac{g^2}{8\,\Mbs}\,\left\{
1 + \frac{g^2}{16\,\pi^2}\,\lbra \delta_{\ssG} + \frac{1}{\Mbs}\,\Sigma_{\PWW}(0)
\rbra
\right\} \spc
\label{GFscheme}
\eq
where we expand the solution for $g$
\bq
g^2_{\ren} = 4\,\srt\,\myGF\,M^2_{\PW\,;\,\ssOS}\,\left\{
 1 + \frac{\myGF M^2_{\PW\,;\,\ssOS}}{2\,\srt\,\pi^2}\,
\lbra \delta_{\ssG} + \frac{1}{\Mbs}\,\Sigma_{\PWW\,;\,\fin}(0)
\rbra
\right\}
\label{GFsol}
\eq
Note that the non universal part of the corrections is given by
\bq
\delta_{\ssG} = \delta^{(4)}_{\ssG} + \gds\,\delta^{(6)}_{\ssG} \spc
\quad
\delta^{(4)}_{\ssG} = 6 + \frac{7 - 4\,\stws}{2\,\stws}\,\ln\ctws \spc
\eq
but the contribution of $\mrdim = 6$ operators to muon decay is not available yet and
will not be included in the calculation.
It is worth noting that \eqns{IPSa}{GFscheme} define finite renormalization in
the $\{\myGF\,,\,\mw\,,\,\mz\}$ input parameter set.

We show few explicit examples of finite renormalization, \ie how to fix finite
counterterms. From the $\PH$ propagator and the definition of on-shell $\PH$ mass
one obtains
\bq
\ssdCZ^{(n)}_{\mh}= \frac{\mwsOS}{\mhsOS}\,\Re\,\Delta^{(n)}_{\PHH\,;\,\fin}\lpar \mhsOS \rpar +
                                           \Re\,\Pi^{(n)}_{\PHH\,;\,\fin}\lpar \mhsOS \rpar \spc
\label{fcta}
\eq
where $\mh$ is the renormalized $\PH$ mass and $\mhOS$ is the on-shell $\PH$ mass.
>From the $\PW$ propagator we have
\bq
\ssdCZ^{(n)}_{\mw}= \Re\,\Delta^{(n)}_{\PWW\,;\,\fin}\lpar \mwsOS \rpar +
                    \Re\,\Pi^{(n)}_{\PWW\,;\,\fin}\lpar \mwsOS \rpar \spp
\label{fctb}
\eq
>From the $\PZ$ propagator and the definition of on-shell $\PZ$ mass we have
\bq
\ssdCZ^{(n)}_{\ctw} = \frac{1}{2}\,\Re\,\Bigl[
\ssdCZ^{(n)}_{\mw} - \ctWs\,\Delta^{(n)}_{\PZZ\,;\,\fin}\lpar \mzsOS \rpar -
\Pi^{(n)}_{\PZZ\,;\,\fin}\lpar \mzsOS \rpar \Bigr] \spc
\label{fctc}
\eq
with $\ctWs = \mwsOS/\mzsOS$. All quantities in \eqns{fcta}{fctc} are the renormalized
ones.
\subsubsection{$\alpha\;$ renormalization scheme}
This scheme uses the fine structure constant $\alpha$. The new renormalization equation is
\bq
g^2\,\stws = 4\,\pi\alpha\,\Bigl[ 1 - \frac{\alpha}{4\,\pi}\,\frac{\Pi_{\PAA}(0)}{\stws} \Bigr]\spc
\eq
where $\alpha = \alpha_{\myQED}(0)$. Therefore, in this scheme, the finite counterterms are
\bq
g^2_{\ren} = g^2_{\sPA}\,\Bigl[ 1 + \frac{\alpha}{4\,\pi}\,\ssdCZ_{g} \Bigr] \spc
\quad
\ctwr = \cth\,\Bigl[ 1 + \frac{\alpha}{4\,\pi}\,\ssdCZ_{\ctw} \Bigr]  \spc
\quad
M_{\ren} = \mzOS\,\cths\,\Bigl[ 1 + \frac{\alpha}{8\,\pi}\,\ssdCZ_{\mw} \Bigr] \spc
\label{eqIPSb}
\eq
where the parameters $\cth$ and $g_{\sPA}$ are defined by
\bq
g^2_{\sPA} = \frac{4\,\pi\,\alpha}{\sths}
\qquad
\sths = \frac{1}{2}\,\Bigl[ 1 - \sqrt{1 - 4\,\frac{\pi\,\alpha}{\srt\,\myGF\,\mzsOS}}\Bigr] \spp
\label{defscipsb}
\eq
The reason for introducing this scheme is that the $\ssS, \ssT$ and $\ssU$ parameters 
(see \Bref{Peskin:1990zt}) have been originally given in the $\{\alpha\,,\,\myGF\,,\,\mz\}$ scheme
while, for the rest of the calculations  we have adopted the more convenient 
$\{\myGF\,,\,\mw\,,\,\mz\}$ scheme.
In this scheme, after requiring that $\mzsOS$ is a zero of the real part of the inverse
$\PZ$ propagator, we are left with one finite counterterm, $\ssdCZ_{g}$. The latter is fixed
by using $\myGF$ and requiring that
\bq
\frac{1}{\srt}\,\myGF = \frac{g^2}{8\,M^2}\,\Bigl\{ 1 + \frac{g^2}{16\,\pi^2}\,\Bigl[
                       \delta_{\ssG} + \frac{1}{M^2}\,\Delta_{\PWW}(0)
                     - \lpar \ssdZ_{\PW} + \ssdZ_{\mw} \rpar\,\DUV \Bigr] \Bigr\} \spc
\eq
where we use
\bq
g = g_{\ren}\,\lpar 1 + \frac{g^2_{\ren}}{16\,\pi^2}\,\ssdZ_{g}\,\DUV \rpar \spc
\qquad
g_{\ren} = g_{\sPA}\,\lpar 1 + \frac{\alpha}{8\,\pi}\,\ssdCZ_{g} \rpar \spc
\eq
for UV and finite renormalization.
\subsection{Wave function renormalization}
Let us summarize the various steps in renormalization. Consider the $\PV$ propagator, assuming 
that $\PV$ couples to conserved currents (in the following we will drop the label
$\PV$). We have
\bq
\oD_{\mu\nu} = 
-\,\frac{\delta_{\mu\nu}}{s - M^2 + \frac{g^2}{16\,\pi^2}\,\ssD} = 
-\,\delta_{\mu\nu}\,\Delta^{-1}(s) \spc
\eq
where $M$ is the $\PV$ bare mass. The procedure is as follows:
we introduce UV counterterms for the field and its mass,
\bq
\Delta(s)\,\bmid_{\ren} =
\ssZ_{\sPV}\,\lpar s - \ssZ_M\,M^2_{\ren} \rpar + \frac{g^2}{16\,\pi^2}\,\ssD(s) =
 s - M^2_{\ren} + \frac{g^2}{16\,\pi^2}\,\ssD(s)\,\bmid_{\ren}
\eq
and write the (finite) renormalization equation
\bq
M^2_{\ren} = M^2_{\ssOS} + \frac{g^2}{16\,\pi^2}\,\Re\,\ssD\lpar M^2_{\ssOS} \rpar\,\bmid_{\ren}
\spc
\eq
where $M_{\ssOS}$ is the (on-shell) physical mass. After UV and finite renormalization we can 
write the following Taylor expansion:
\bq
\Delta(s)\,\bmid_{\ren} = 
\lpar s - M^2_{\ssOS} \rpar\,\lpar 1 + \frac{g^2_{\exp}}{16\,\pi^2}\,\PW \rpar +
\mcO\lpar ( s - M^2_{\ssOS} )^2 \rpar \spc
\label{Texp}
\eq
where $g_{\exp}$ is defined in \eqn{gexp}.
The wave-function renormalization factor for the field $\upPhi$ will be denoted by
\bq
\ssZ^{-1/2}_{\ssWF\,;\,\upPhi} = \lpar 1 + \frac{g^2_{\exp}}{16\,\pi^2}\,\mrW_{\upPhi} 
\rpar^{-1/2} \spp
\label{WFdef}
\eq

For fermion fields we use \eqn{FSE} and introduce
\bq
V'_{\Pf}(s) = -\,\frac{d}{d s}\,V_{\Pf}(s) \spp
\eq
Next we multiply spinors by the appropriate factors, \ie
\bq
u_{\Pf}(p) \to \lpar 1 + \mrW_{\Pf\ssV} + \mrW_{\Pf\ssA}\,\gamma^5 \rpar\,u_{\Pf}(p)
\quad
{\overline{u}}_{\Pf}(p) \to {\overline{u}}_{\Pf}(p)\,
       \lpar 1 + \mrW_{\Pf\ssV} - \mrW_{\Pf\ssA}\,\gamma^5 \rpar \spc
\eq
where the wave-function renormalization factors are obtained from \eqn{fSE}
\bq
\mrW_{\Pf\ssV} = \frac{1}{2}\,\lbra V_{\Pf} + 2\,M_{\Pf}\,\Delta'_{\Pf}
                 - 2\,M^2_{\Pf}\,V'_{\Pf} \rbra\,\bmid_{s = M^2_{\Pf}}\spc
\quad
\mrW_{\Pf\ssA} = - \frac{1}{2}\,A_{\Pf}\,\bmid_{s = M^2_{\Pf}} \spp
\eq
For illustration we present the $\PH$ wave-function factor
\bqa
\mrW_{\PH} &=& \Re\,\lpar \mrW^{(4)}_{\PH} + \gds\,\mrW^{(6)}_{\PH} \rpar \spc
\nl 
\mrW^{(n)}_{\PH} &=& d\Pi^{(n)}_{\PHH}\lpar \mhsOS\rpar\,\mhsOS +
                     d\Delta^{(n)}_{\PHH}\lpar \mhsOS\rpar\,\mwsOS +
                     \Pi^{(n)}_{\PHH}\lpar \mhsOS\rpar -
                     2\,\ssdCZ^{(n)}_{g}
\eqa
and we expand any function of $s$ as follows:
\bq
f(s) = f\lpar M^2_{\ssOS}\rpar + \lpar s - M^2_{\ssOS} \rpar\,df\lpar M^2_{\ssOS}\rpar
+ {\mcO}\lpar ( s - M^2_{\ssOS})^2 \rpar \spc
\eq
with $f = \Pi^{(n)}, \Delta^{(n)}$.
For the $\PW, \PZ$ wave-function factor we obtain
\bqa
\mrW^{(n)}_{\PW} &=& d\Pi^{(n)}_{\PWW}\lpar \mwsOS\rpar\,\mwsOS +
                     d\Delta^{(n)}_{\PWW}\lpar \mwsOS\rpar\,\mwsOS +
                     \Pi^{(n)}_{\PWW}\lpar \mwsOS\rpar -
                     2\,\ssdCZ^{(n)}_{g} \spc
\nl
\mrW^{(n)}_{\PZ} &=& d\Pi^{(n)}_{\PZZ}\lpar \mzsOS\rpar\,\mzsOS +
                     d\Delta^{(n)}_{\PZZ}\lpar \mzsOS\rpar\,\mwsOS +
                     \Pi^{(n)}_{\PZZ}\lpar \mzsOS\rpar -
                     2\,\ssdCZ^{(n)}_{g} \spp
\eqa
Explicit expressions for the wave-function factors are given in \appendx{WFF}.
\subsection{Life and death of renormalization scale}
Consider the $\PA$ bare propagator
\bq
{\overline{\Delta}}^{-1}_{\PAA} = s + \frac{g^2}{16\,\pi^2}\,\Sigma_{\PAA}(s)
\quad
\Sigma_{\PAA}(s) =  \lpar \ssR^{(4)} + \gds\,\ssR^{(6)}\rpar\,\frac{1}{\ep} +
\sum_{x\,\in {\mcX}}\,\lpar \ssL^{(4)}_x + \gds\,\ssL^{(6)}_x\rpar\,\ln\frac{x}{\muRs} +
\Sigma^{\rest}_{\PAA} \spc
\eq
where $\ssR^{(n)}$ are the residues of the UV poles and $\ssL^{(n)}$ are arbitrary coefficients
of the scale-dependent logarithms. Furthermore,
\bq
\{ {\mcX} \} = \{ s\,,\,m^2\,,\,m^2_0\,,\,m^2_{\PH}\,,\,m^2_{\PQt}\,,\,m^2_{\PQb} \} \spp
\eq
The renormalized propagator is
\bq
{\overline{\Delta}}^{-1}_{\PAA}\bmid_{\ren} = 
\ssZ_{\PA}\,s  + \frac{g^2}{16\,\pi^2}\,\Sigma_{\PAA}(s) =
s + \frac{g^2}{16\,\pi^2}\,\Sigma^{\ren}_{\PAA}(s) \spp
\eq
Furthermore, we can write
\bq
\Sigma^{\ren}_{\PAA}(s) =  \sum_{x\,\in {\mcX}}\,\lpar \ssL^{(4)}_x + 
\gds\,\ssL^{(6)}_x\rpar\,\ln\frac{x}{\muRs} + \Sigma^{\rest}_{\PAA} \spp
\eq
Finite renormalization amounts to write $\Sigma^{\ren}_{\PAA}(s) = \Pi^{\ren}_{\PAA}(s)\,s$ and
to use $s = 0$ as subtraction point. Therefore, one can easily prove that
\bq
\frac{\partial}{\partial \muR}\,\Bigl[
\Pi^{\ren}_{\PAA}(s) - \Pi^{\ren}_{\PAA}(0) \Bigr] = 0 \spc
\eq
including $\Ope^{(6)}$ contribution. Therefore we may conclude that there is no $\muR$
problem when a subtraction point is available.
After discussing decays of the Higgs boson in \sect{HBD} we will see that an additional step is 
needed in the renormalization procedure, \ie mixing of the Wilson coefficients. At this point the
scale dependence problem will surface again and renormalized Wilson coefficients become scale 
dependent.
\section{Decays of the Higgs boson \label{HBD}}
In this Section we will present results for two-body decays of the Higgs boson while
four-body decays will be included in a forthcoming publication. 
Our approach is based on the fact that renormalizing a theory must be a fully
general procedure; only when this step is completed one may consider making approximations, 
\eg neglecting the lepton masses, keeping only PTG terms \etc
In particular, neglecting LG Wilson coefficients sensibly reduces the number of terms in any
amplitude. 

\begin{table}[t]
\begin{center}
\caption[]{\label{WClist}{Vector-like notation for Wilson coefficients.}}
\vspace{0.2cm}
\begin{tabular}{lll}
\hline
& &  \\
$\aAA = \WC{1}  $ & $\aZZ = \WC{2}  $ & $\aAZ = \WC{3}$   \\
$\apD = \WC{4}  $ & $\apBox = \WC{5}$ & $\ap = \WC{6}$    \\
$\alWB = \WC{7} $ & $\alBW = \WC{8} $ & $\adWB= \WC{9}$   \\
$\adBW = \WC{10}$ & $\auWB = \WC{11}$ & $\auBW = \WC{12}$ \\
$\alp = \WC{13} $ & $\adp = \WC{14} $ & $\aup= \WC{15}$  \\
$\aplA = \WC{16} $ & $\aplV = \WC{17} $ & $\apn= \WC{18}$  \\
$\apdA = \WC{19} $ & $\apdV = \WC{20} $ & $\apuA= \WC{21}$  \\
$\apuV = \WC{22}$ & $\apLldQ = \WC{23}$ & $ \aoQuQd= \WC{24}$ \\
& &  \\
\hline
\end{tabular}
\end{center}
\end{table}

It is useful to introduce a more compact notation for Wilson coefficients, given in \refT{WClist}
and to use the following definition:
\begin{definition} 
The PTG scenario: any amplitude computed at $\mcO(g^n\,\gds)$ has a SM component
of $\mcO(g^n)$ and two $\mrdim = 6$ components:
at $\mcO(g^{n-2}\,\gds)$ we allow both PTG and LG operator while
at $\mcO(g^n\,\gds)$ only PTG operators are included.
\end{definition} 
\subsection{Loop-induced processes: $\PH \to \PGg\PGg$}
The amplitude for the process $\PH(P) \to \PA_{\mu}(p_1)\PA_{\nu}(p_2)$ can be written as
\bq
\mrA^{\mu\nu}_{\sPHAA} = \mcT_{\sPHAA}\,T^{\mu\nu} \spc
\quad
\mhs\,T^{\mu\nu} = p^{\mu}_2\,p^{\nu}_1 - \spro{p_1}{p_2}\,\delta^{\mu\nu} \spp
\label{HAAamp}
\eq
The $S\,$-matrix element follows from \eqn{HAAamp} when we multiply the amplitude by the
photon polarizations $e_{\mu}(p_1)\,e_{\nu}(p_2)$; in writing \eqn{HAAamp} we have
used $p\,\cdot\,e(p)= 0$.

Next we introduce $\mrdim = 4$, LO, sub-amplitudes for $\PQt, \PQb$ loops and for the 
bosonic loops,
\bqa
\frac{3}{8}\,\frac{\mw}{\mqts}\,\mcT^{\PQt}_{\sPHAA\,;\,\myLO} &=&
2 + \lpar \mhs - 4\,\mqts \rpar\,\cfun{-\mhs}{0}{0}{\mqt}{\mqt}{\mqt} \spc
\nl
\frac{9}{2}\,\frac{\mw}{\mqbs}\,\mcT^{\PQb}_{\sPHAA\,;\,\myLO} &=&
2 + \lpar \mhs - 4\,\mqbs \rpar\,\cfun{-\mhs}{0}{0}{\mqb}{\mqb}{\mqb} \spc
\nl
\frac{1}{\mw}\,\mcT^{\PW}_{\sPHAA\,;\,\myLO} &=&
- 6 - 6\,\lpar \mhs - 2\,\mws \rpar\,\cfun{-\mhs}{0}{0}{\mw}{\mw}{\mw} \spc
\label{Ampfacts}
\eqa
where $\ssC_0$ is the scalar three-point function. The following result is obtained:
\bq
\mcT_{\sPHAA} = i\,\frac{g^3}{16\,\pi^2}\,\lpar 
                \mcT^{(4)}_{\sPHAA} + 
           \gds\,\mcT^{(6),b}_{\sPHAA} \rpar + 
           i\,g\,\gds\,\mcT^{(6),a}_{\sPHAA} \spc
\eq
where the $\mrdim = 4$ part of the amplitude
\bq
\mcT^{(4)}_{\sPHAA} = 2\,\stws\,\lpar 
\sumg\,\sum_{\Pf}\,\mcT^{\Pf}_{\sPHAA\,;\,\myLO} + \mcT^{\PW}_{\sPHAA\,;\,\myLO} \rpar 
\eq
is UV finite, as well as $\mcT^{(6),a}$ which is given by
\bq
\mcT^{(6),a}_{\sPHAA} = 2\,\frac{\mhs}{M}\,\lpar
                        \stws\,\apW + \ctws\,\apB + \stw\,\ctw\,\apWB \rpar  \spp
\eq
The $\mcT^{(6),b}$ component contains an UV-divergent part. UV renormalization requires
\bq
\mcT^{\ren}_{\sPHAA} = \mcT_{\sPHAA}\,
     \Bigl[ 1 + \frac{g^2}{16\,\pi^2}\,
     \lpar \ssdZ_{\PA} + \frac{1}{2}\,\ssdZ_{\PH} + 3\,\ssdZ_{g} \rpar\,\DUV \Bigr] \spc
\eq
\bq
\ctw = \ctwr\,\lpar 1 + \frac{g^2}{16\,\pi^2}\,\ssdZ_{\ctw}\,\DUV \rpar \spc
\qquad
g = g_{\ren}\,\lpar 1 + \frac{g^2_{\ren}}{16\,\pi^2}\,\ssdZ_{g}\,\DUV \rpar
\eq
and we obtain the renormalized version of the amplitude
\bqa
\mcT^{\ren}_{\sPHAA} &=& i\,\frac{g^3_{\ren}}{16\,\pi^2}\,\lpar \mcT^{(4)}_{\sPHAA} + 
           \gds\,\mcT^{(6),b}_{\sPHAA\,;\,\fin} \rpar 
    + i\,g_{\ren}\,\gds\,\mcT^{(6),a}_{\sPHAA\,;\,\ren} +   
      i\,\frac{g^3_{\ren}}{16\,\pi^2}\,\gds\,\mcT^{(6)}_{\sPHAA\,;\,div} \spc
\nl
\mcT^{(6)}_{\sPHAA\,;\,div} &=& \mcT^{(6),b}_{\sPHAA\,;\,\pole}\,\DUVM
    + \frac{M^2_{\PH\,\ren}}{M_{\ren}}\,\Bigl\{\Bigl[
      \ssdZ^{(4)}_{\PH} - \ssdZ^{(4)}_{\mw} + 2\,\ssdZ^{(4)}_{\PA} - 2\,\ssdZ^{(4)}_{g}
           \Bigr]\,\aAA - 2\,\frac{\ctwr}{\stwr}\,\ssdZ^{(4)}_{\ctw}\,\aAZ \Bigr\}\,\DUV \spc
\nl
\mcT^{(6),a}_{\sPHAA\,;\,\ren} &=& 2\,\frac{M^2_{\PH\,\ren}}{M_{\ren}}\,\aAA \spc
\label{HAAren}
\eqa
where $\aAA = \stw \ctw\,\apWB + \ctws\,\apB + \stws\,\apW$ and $\ctw = \ctwr$ \etc
The last step in the UV-renormalization procedure requires a mixing among Wilson coefficients
which cancels the remaining ($\mrdim = 6$) parts. To this purpose we define
\bq
\WC{i} = \sum_j\,Z^{\PW}_{ij}\,\PW^{\ren}_j \spc
\qquad
Z^{\PW}_{ij} = \delta_{ij} + \frac{g^2}{16\,\pi^2}\,\ssdZ^{\PW}_{ij}\,\DUV \spp
\label{MixW}
\eq
The matrix $\ssdZ^{\PW}$ is fixed by requiring cancellation of the residual UV poles and
we obtain
\bq
\mcT^{(6)}_{\sPHAA\,;\,\pole} \to \mcT^{(6),\ssR}_{\sPHAA}\,\ln\frac{\muRs}{\Mbs} \spp
\eq
Elements of the mixing matrix derived from $\PH \to \PAA$ are given in \appendx{MixWc}.
The result of \eqn{HAAren} becomes
\bqa
\mcT^{\ren}_{\sPHAA} &=& i\,\frac{g^3_{\ren}}{16\,\pi^2}\,\lpar \mcT^{(4)}_{\sPHAA} + 
           \gds\,\mcT^{(6),b}_{\sPHAA\,;\,\fin} +
           \gds\mcT^{(6),\ssR}_{\sPHAA}\,\ln\frac{\muRs}{\Mbs} \rpar 
           + i\,g_{\ren}\,\gds\,\mcT^{(6),a}_{\sPHAA}    \spp
\eqa
Inclusion of wave-function renormalization factors and of external leg factors
(due to field redefinition described in \sect{FL}) gives
\bqa
{}&{}& \mcT^{\ren}_{\sPHAA}\,
        \Bigl[ 1 - \frac{g^2_{\ren}}{16\,\pi^2}\,
        \lpar \mrW_{\PA} + \frac{1}{2}\,\mrW_{\PH} \rpar \Bigr]\,
        \Bigl[ 1 + \gds\,\lpar 2\,\aAA + \apBox - \frac{1}{4}\,\apD \rpar \Bigr] \spp
\eqa
Finite renormalization requires writing
\bqa
M^2_{\ren} &=& \mws\,\lpar 1 + \frac{g^2_{\ren}}{16\,\pi^2}\,\ssdCZ_{\mw} \rpar \spc
\nl
\ctwr &=& \ctW\,\lpar 1 + \frac{g^2_{\ren}}{16\,\pi^2}\,\ssdCZ_{\ctw} \rpar \spc
\nl
g_{\ren} &=& g_{\ssF}\,\lpar 1 + \frac{g^2_{\ssF}}{16\,\pi^2}\,\ssdCZ_{g} \rpar \spc
\label{frenp}
\eqa
where $g^2_{\ssF} = 4\,\srt\,\myGF\,\mws$ and $\ctW= \mw/\mz$.
Another convenient way for writing the answer is the following:
after renormalization we neglect all fermion masses but $\PQt, \PQb$ and write 
\bqa
\mcT_{\sPHAA} &=&
i\,\,\frac{g^3_{\ssF}\,\stWs}{8\,\pi^2}\,
\sum_{\ssI=\PW,\PQt,\PQb}\,\upkappa^{\sPHAA}_{\ssI}\,\mcT^{\ssI}_{\sPHAA\,;\,\myLO} + 
i\,g_{\ssF}\,\gds\,\frac{\mhs}{\mw}\,\WCr{1}
\nl
{}&+& i\,\frac{g^3_{\ssF}\,\gds}{\pi^2}\,\lbra
\sum_{i=1,3}\,\mcA^{\nfact}_{\PW,\,i}\,\WCr{i} +
\mcT^{\nfact}_{\sPHAA\,;\,\PQb}\,\WCr{9} +
\sum_{i=1,2}\,\mcT^{\nfact}_{\sPHAA\,;\,\PQt,\,i}\,\WCr{10 + i}
\rbra \spc
\label{linkappaHAA}
\eqa
where Wilson coefficients are those in \tabn{WClist}. The $\upkappa\,$-factors are given by
\bq
\upkappa^{\proc}_{\ssI} = 1 + \gds\,\Delta\upkappa^{\proc}_{\ssI}
\label{duk}
\eq
and there are additional, non-factorizable, contributions. The $\upkappa$ factors are
\bqa
\Delta \upkappa^{\sPHAA}_{\PQt} &=&
            \frac{3}{16}\,\frac{\mhs}{\stW\,\mws}\,\atWB          
            + ( 2 - \stWs )\,\frac{\ctW}{\stW}\,\aAZ
            + ( 6 - \stWs )\,\aAA 
\nl
{}&-& \frac{1}{2}\,\Bigl[
            \apD + 2\,\stWs\,( \ctWs\,\aZZ - \atp  - 2\,\apBox ) \Bigr]\,\frac{1}{\stWs} \spc
\nl
\Delta \upkappa^{\sPHAA}_{\PQb} &=&
          - \frac{3}{8}\,\frac{\mhs}{\stW\,\mws}\,\abWB
          + ( 2 - \stWs )\,\frac{\ctW}{\stW}\,\aAZ
          + ( 6 - \stWs )\,\aAA
\nl
{}&-& \frac{1}{2}\,\Bigl[
          \apD + 2\,\stWs\,( \ctWs\,\aZZ + \abp - 2\,\apBox) \Bigr]\,\frac{1}{\stWs} \spc
\nl
\Delta \upkappa^{\sPHAA}_{\PW} &=&
           ( 2 + \stWs )\,\frac{\ctW}{\stW}\,\aAZ
          + ( 6 + \stWs )\,\aAA
\nl
{}&-& \frac{1}{2}\,\Bigl[ 
          \apD - 2\,\stWs\,( 2\,\apBox + \ctWs\,\aZZ) \Bigr]\,\frac{1}{\stWs} \spc
\label{kappaHAA}
\eqa
where Wilson coefficients are the renormalized ones.
In the PTG scenario we only keep $\atp, \abp, \apD$ and $\apBox$ in \eqn{kappaHAA}.

The advantage of \eqn{linkappaHAA} is to establish a link between EFT and $\upkappa\,$-language
of \Bref{LHCHiggsCrossSectionWorkingGroup:2012nn}, which has a validity restricted to LO.
As a matter of fact \eqn{linkappaHAA} tells you that $\upkappa\,$-factors can be introduced
also at NLO level; they are combinations of Wilson coefficients but we have to
extend the scheme with the inclusion of process dependent, non-factorizable, contributions.

Returning to the original convention for Wilson coefficients we derive the following result
for the non-factorizable part of the amplitude:
\bq
\mcT^{\nfact}_{\sPHAA} = \mw\,\sum_{a\,\in\,\{A\}}\,\mcT^{\nfact}_{\sPHAA}(a)\,a \spc
\eq
where $\{ A \} = \{ \atWB, \abWB, \aAA, \aAZ, \aZZ \}$.
Finite counterterms, $\ssdCZ_{\mh}, \ssdCZ_g$ and $\ssdCZ_{\mw}$ are defined
in \eqn{IPSa} and in \eqn{gexp}. The results are reported on \appendx{Ampnf}.
In the PTG scenario all non-factorizable amplitudes for $\PH \to \PAA$ vanish.
\subsection{$\PH \to \PZ\PGg$}
The amplitude for $\PH(P) \to \PA_{\mu}(p_1)\PZ_{\nu}(p_2)$ can be written as
\bq
\mrA^{\mu\nu}_{\sPHAZ} = \mhs\,\mcD_{\sPHAZ}\,\delta^{\mu\nu}
+ \mcP^{11}_{\sPHAZ}\,p^{\mu}_1\,p^{\nu}_1 
+ \mcP^{12}_{\sPHAZ}\,p^{\mu}_1\,p^{\nu}_2 
+ \mcP^{21}_{\sPHAZ}\,p^{\mu}_2\,p^{\nu}_1 
+ \mcP^{22}_{\sPHAZ}\,p^{\mu}_2\,p^{\nu}_2 \spp
\label{HAZamp}
\eq
The result of \eqn{HAZamp} is fully general and can be used to prove WST identities. As far as the 
partial decay width is concerned only $\mcP^{21}_{\sPHAZ}$ will be relevant, due
to $p\,\cdot\,e(p) = 0$ where $e$ is the polarization vector.
We start by considering the $1$PI component of the amplitude and obtain
\bq
\mrA^{\mu\nu}_{\sPHAZ}\bmid_{\OPI} = 
\mhs\,\mcD^{(\OPI)}_{\sPHAZ}\,\delta^{\mu\nu} + \mcT^{(\OPI)}_{\sPHAZ}\,T^{\mu\nu} \spc
\eq
where $T$ is given by
\bq
\mhs\,T^{\mu\nu} = p^{\mu}_2\,p^{\nu}_1 - \spro{p_1}{p_2}\,\delta^{\mu\nu} \spp
\eq
Furthermore we can write the following decomposition:
\bqa
\mcT^{(\OPI)}_{\sPHAZ} &=& i\,\frac{g^3}{16\,\pi^2}\,\lpar
                              \mcT^{(4)}_{\sPHAZ} + \gds\,\mcT^{(6),b}_{\sPHAZ} \rpar
                            + i\,g\,\gds\,\mcT^{(6),a}_{\sPHAZ} \spc
\qquad
\mcD^{(\OPI)}_{\sPHAZ} = i\,\frac{g^3 \gds}{16\,\pi^2}\,\mcD^{(6)}_{\sPHAZ} \spc
\eqa
\bqa
\mcD^{(6)}_{\sPHAZ} &=& 3\,\ctw\,\stws\,\frac{M^3_0}{\mhs}\,
                             \cfun{-\mhs}{0}{- M^2_0}{M}{M}{M}\,\aAZ \spc
\nl
\mcT^{(6),a}_{\sPHAZ} &=& \frac{\mhs}{M}\,\Bigl[
2\,\stw \ctw\,\lpar \apW - \apB \rpar + \lpar 2\,\ctws  - 1 \rpar\,\apWB \Bigr] \spp
\eqa
Explicit expressions for $\mcT^{(4)}_{\sPHAZ}, \mcT^{(6),b}_{\sPHAZ}$ will not be reported here.
The $1$PR component of the amplitude is given by
\bq
\mrA^{\mu\nu}_{\sPHAZ}\bmid_{\OPR} = - \frac{1}{M^2_0}\,
\mrA^{\mu\alpha\,\off}_{\sPHAA}\lpar p_1\,,\,p_2 \rpar\,S^{\alpha\nu}_{\PAZ}\lpar p_2 \rpar =    
\mhs\,\mcD^{(\OPR)}_{\sPHAZ}\,\delta^{\mu\nu} + \mcT^{(\OPR)}_{\sPHAZ}\,T^{\mu\nu} \spc
\eq
where $\mrA^{\mu\alpha\,\off}_{\sPHAA}$ denotes the off-shell $\PH \to \PAA$ amplitude.
It is straightforward to derive
\bq
\mcD_{\sPHAZ} = \mcD^{(\OPI)}_{\sPHAZ} + \mcD^{(\OPR)}_{\sPHAZ} = 0
\eq
\ie the complete amplitude for $\PH \to \PAZ$ is proportional to $T^{\mu\nu}$ and, therefore, 
is transverse. UV renormalization requires the introduction of counterterms,
\bq
\mcT^{\ren}_{\sPHAZ} = \mcT_{\sPHAZ}\,
     \Bigl[ 1 + \frac{1}{2}\,\frac{g^2}{16\,\pi^2}\,
     \lpar \ssdZ_{\PH} + \ssdZ_{\PA} + \ssdZ_{\PZ} - 6\,\ssdZ_{g} \rpar\,\DUV \Bigr] \spc
\eq
\bq
\ctw = \ctwr\,\lpar 1 + \frac{g^2_{\ren}}{16\,\pi^2}\,\ssdZ_{\ctw}\,\DUV \rpar \spc
\qquad
g = g_{\ren}\,\lpar 1 + \frac{g^2_{\ren}}{16\,\pi^2}\,\ssdZ_{g}\,\DUV \rpar
\eq
and we obtain the following result for the renormalized amplitude:
\bqa
\mcT^{\ren}_{\sPHAZ} &=& i\,\frac{g^3_{\ren}}{16\,\pi^2}\,\lpar \mcT^{(4)}_{\sPHAZ} + 
           \gds\,\mcT^{(6),b}_{\sPHAZ\,;\,\fin} \rpar 
    + i\,g_{\ren}\,\gds\,\mcT^{(6),a}_{\sPHAZ\,;\,\ren} +   
      i\,\frac{g^3_{\ren}}{16\,\pi^2}\,\gds\,\mcT^{(6)}_{\sPHAZ\,;\,div} \spc
\nl
\mcT^{(6)}_{\sPHAZ\,;\,div} &=& \mcT^{(6),b}_{\sPHAZ\,;\,\pole}\,\DUVM +
      \frac{M^2_{\PH\,\ren}}{M_{\ren}}\,\Bigl\{\Bigl[
      \frac{1}{2}\,\ssdZ^{(4)}_{\PH} +
      \ssdZ^{(4)}_{\PA} -
      \frac{1}{2}\,\ssdZ^{(4)}_{\mw} -
      2\,\ssdZ^{(4)}_{g} \Bigr]\,\aAZ
\nl
{}&+& 2\,\frac{\ctwr}{\stwr}\,\ssdZ^{(4)}_{\ctw}\,\lpar \aAA - \aZZ \rpar \Bigr\}\,\DUV \spc
\nl
\mcT^{(6),a}_{\sPHAZ\,;\,\ren} &=& \frac{M^2_{\PH\,\ren}}{M_{\ren}}\,\aAZ \spc
\label{HAZren}
\eqa
where $\aAA = \stw \ctw\,\apWB + \ctws\,\apB + \stws\,\apW$ and $\ctw = \ctwr$ \etc
Once again, the last step in the UV-renormalization procedure requires a mixing among Wilson 
coefficients, performed according to \eqn{MixW}. We obtain
\bq
\mcT^{(6)}_{\sPHAZ\,;\,\pole} \to \mcT^{(6),\ssR}_{\sPHAZ}\,\ln\frac{\muRs}{\Mbs} \spp
\eq
Elements of the mixing matrix derived from the process $\PH \to \PAZ$ are given in \appendx{MixWc}.
After mixing the result of \eqn{HAZren} becomes
\bqa
\mcT^{\ren}_{\sPHAZ} &=& i\,\frac{g^3_{\ren}}{16\,\pi^2}\,\lpar \mcT^{(4)}_{\sPHAZ} + 
           \gds\,\mcT^{(6),b}_{\sPHAZ\,;\,\fin} +
           \gds\mcT^{(6),\ssR}_{\sPHAZ}\,\ln\frac{\muRs}{\Mbs} \rpar 
           + i\,g_{\ren}\,\gds\,\mcT^{(6),a}_{\sPHAZ}    \spp
\eqa
Inclusion of wave-function renormalization factors and of external leg factors
(due to field redefinition, defined in \sect{FL}) gives
\bqa
{}&{}& \mcT^{\ren}_{\sPHAZ}\,
        \Bigl[ 1 - \frac{1}{2}\,\frac{g^2_{\ren}}{16\,\pi^2}\,
        \lpar \mrW_{\PH} + \mrW_{\PA} + \mrW_{\PZ} \rpar \Bigr]\,
        \Bigl[ 1 + \gds\,\lpar \aAA + \aZZ + \apBox - \frac{1}{4}\,\apD \rpar \Bigr] \spp
\eqa
Finite renormalization is performed by using \eqn{frenp}. To write the final
answer it is convenient to define $\mrdim= 4$ sub-amplitudes $\mcT^{\ssI}_{\sPHAZ\,;\,\myLO}$ 
($I=\PW,\PQt,\PQb$): they are given in \appendx{LOsub}.
Another convenient way for writing $\mcT^{\ren}_{\sPHAZ}$ is the following:
\bqa
\mcT^{\ren}_{\sPHAZ} &=&
i\,\,\frac{g^3_{\ssF}}{\pi^2 \mz}\,
\sum_{\ssI=\PW,\PQt,\PQb}\,\upkappa^{\sPHAZ}_{\ssI}\,\mcT^{\ssI}_{\sPHAZ\,;\,\myLO} + 
i\,g_{\ssF}\,\gds\,\frac{\mhs}{M}\,\WCr{3}
\nl
{}&+& i\,\frac{g^3_{\ssF}\,\gds}{\pi^2}\,\lbra
\sum_{i=1,4}\,\mcT^{\nfact}_{\sPHAZ\,;\,\PW,\,i}\,\WCr{i} +
\sum_{i=11,12,22}\,\mcT^{\nfact}_{\sPHAZ\,;\,\PQt,\,i}\,\WCr{i} +
\sum_{i=9,10,20}\,\mcT^{\nfact}_{\sPHAZ\,;\,\PQb,\,i}\,\WCr{i}
\rbra \spp
\label{linkappaHAZ}
\eqa
The factorizable part is defined in terms of $\upkappa\,$-factors, see \eqn{duk}
\bqa
\Delta\upkappa^{\sPHAZ}_{\PQt} &=&
 \frac{1}{2}\,\lpar 2\,\atp + 4\,\apBox - \apD + 6\,\aAA + 2\,\aZZ \rpar \spc
\nl
\Delta\upkappa^{\sPHAZ}_{\PQb} &=&
 - \frac{1}{2}\,\lpar 2\,\abp - 4\,\apBox + \apD - 6\,\aAA - 2\,\aZZ \rpar \spc
\nl
\Delta\upkappa^{\sPHAZ}_{\PW} &=&
            \frac{1 + 6\,\ctWs}{\ctWs}\,\apBox
          - \frac{1}{4}\,\frac{1 + 4\,\ctWs}{\ctWs}\,\apD 
          - \frac{1}{2}\,\frac{1 + \ctWs - 24\,\ctWq}{\ctWs}\,\aAA \spc
\nl
{}&+& \frac{1}{4}\,\lpar 1 + 12\,\ctWs - 48\,\ctWq \rpar\,\frac{\stW}{\ctWc}\,\aAZ
          + \frac{1}{2}\,\frac{1 + 15\,\ctWs - 24\,\ctWq}{\ctWs}\,\aZZ \spp
\label{kappaHAZ}
\eqa
In the PTG scenario we only keep $\atp, \abp, \apD$ and $\apBox$ in \eqn{kappaHAZ}.

Returning to the original convention for Wilson coefficients we derive the following result
for the non-factorizable part of the amplitude:
\bq
\mcT^{\nfact}_{\sPHAZ} = \sum_{a\,\in\,\{A\}}\,\mcT^{\nfact}_{\sPHAZ}(a)\,a \spc
\eq
where $\{ A \} = \{ \aptV, \atBW, \atWB, \apbV, \abWB, \abBW, \apD, \aAZ, \aAA, \aZZ \}$.
In the PTG scenario there are only $3$ non-factorizable amplitudes for $\PH \to \PAZ$, those
proportional to $\aptV, \apbV$ and $\apD$. The full results are reported on \appendx{Ampnf}
where
\bq
\laz = \frac{\mhs}{\mws} - \frac{\mzs}{\mws} \spp
\label{deflaz}
\eq
\subsection{$\PH \to \PZZ$}
The amplitude for $\PH(P) \to \PZ_{\mu}(p_1)\PZ_{\nu}(p_2)$ can be written as
\bq
\mrA^{\mu\nu}_{\sPHZZ} = \mcD_{\sPHZZ}\,\delta^{\mu\nu} + 
\mcP^{11}_{\sPHZZ}\,p^{\mu}_1\,p^{\nu}_1 +
\mcP^{12}_{\sPHZZ}\,p^{\mu}_1\,p^{\nu}_2 +
\mcP^{21}_{\sPHZZ}\,p^{\mu}_2\,p^{\nu}_1 + 
\mcP^{22}_{\sPHZZ}\,p^{\mu}_2\,p^{\nu}_2 \spp
\label{HZZamp}
\eq
The result in \eqn{HZZamp} is fully general and can be used to prove WST identities. As far as the 
partial decay width is concerned only $\mcP^{21}_{\sPHZZ} \equiv \mcP_{\sPHZZ}$ will 
be relevant, due to $p\,\cdot\,e(p) = 0$ where $e$ is the polarization vector.
Note that computing WST identities requires additional amplitudes, \ie $\PH \to \Ppz\PGg$ 
and $\PH \to \Ppz\Ppz$.

We discuss first the $1$PI component of the process: as done before the form factors in 
\eqn{HZZamp} are decomposed as follows:
\bqa
\mcD^{\OPI}_{\sPHZZ} &=& - i\,g\,\frac{M}{\ctws} + i\,\frac{g^3}{16\,\pi^2}\,\lpar \,
                         \mcD^{(4)\,\OPI}_{\sPHZZ} + 
                         \gds\,\mcD^{(6)\,\OPI\,,b}_{\sPHZZ} \rpar
                         + i\,g\,\gds\,\mcD^{(6)\,\OPI\,,a}_{\sPHZZ} \spc
\nl
\mcP^{\OPI}_{\sPHZZ} &=& i\,\frac{g^3}{16\,\pi^2}\,\lpar \mcP^{(4)\,\OPI}_{\sPHZZ} + 
                         \gds\,\mcP^{(6)\,\OPI\,,b}_{\sPHZZ} \rpar
                         + i\,g\,\gds\,\mcP^{(6)\,\OPI\,,a}_{\sPHZZ} \spp
\eqa
It is easily seen that only $\mcD$ contains $\mrdim = 4$ UV divergences.
The $1$PR component of the process involves the $\PA{-}\PZ$ transition and it is given by
\bq
\mrA^{\mu\nu}_{\sPHZZ}\,\bmid_{\OPR} = \frac{i}{M^2_0}\,\Bigl[
\mcT^{\alpha\nu}_{\sPHAA}\lpar p_1\,,\,p_2 \rpar\,\Sigma^{\alpha\mu}_{\PAZ}\lpar p_1 \rpar +
\mcT^{\mu\alpha}_{\sPHAA}\lpar p_1\,,\,p_2 \rpar\,\Sigma^{\alpha\nu}_{\PAZ}\lpar p_2 \rpar 
\Bigr] \spc
\label{OPRHZZ}
\eq
where the $\PH \to \PA\PZ$ component is computed with off-shell $\PA$.
The r.h.s. of \eqn{OPRHZZ} is expanded up to $\mcO(g^3\,\gds)$ and we will use
\bq
\mcD_{\sPHZZ} = \mcD^{\OPI}_{\sPHZZ} + \mcD^{\OPR}_{\sPHZZ} \spc
\quad
\mcP_{\sPHZZ} = \mcP^{\OPI}_{\sPHZZ} + \mcP^{\OPR}_{\sPHZZ} \spp
\eq
Complete, bare, amplitudes are constructed
\bqa
\mcD_{\sPHZZ} &=& - i\,g\,\frac{M}{\ctws} + i\,\frac{g^3}{16\,\pi^2}\,\lpar\,
                  \mcD^{(4)}_{\sPHZZ} + \gds\,\mcD^{(6)\,,b}_{\sPHZZ} \rpar
                  + i\,g\,\gds\,\mcD^{(6)\,,a}_{\sPHZZ} \spc
\nl
\mcP_{\sPHZZ} &=& i\,\frac{g^3}{16\,\pi^2}\,\lpar\,\mcP^{(4)}_{\sPHZZ} + 
                  \gds\,\mcP^{(6)\,,b}_{\sPHZZ} \rpar
                  + i\,g\,\gds\,\mcP^{(6)\,,a}_{\sPHZZ} \spc
\eqa
where the $\mcO(g \gds)$ components are:
\bqa
\mcD^{(6)\,,a}_{\sPHZZ} &=&  - \frac{M}{\ctws}\,\lpar \apBox + \frac{1}{4}\,\apD \rpar
 + \lpar \frac{\mhs - 2\,\mzs}{M} - M \rpar\,\aZZ 
 - M\,\frac{\stws}{\ctws}\,\,\aAA
 - M\,\frac{\stw}{\ctw}\,\aAZ \spc
\nl
\mcP^{(6)\,,a}_{\sPHZZ} &=& 2\,\frac{1}{M}\,\aZZ \spp
\eqa
UV renormalization requires introduction of counterterms,
\bq
\mcF^{\ren}_{\sPHZZ} = \mcD_{\sPHZZ}\,
\Bigl[ 1 + \frac{g^2}{16\,\pi^2}\,\lpar
       \ssdZ_{\PZ} + \frac{1}{2}\,\ssdZ_{\PH} - 3\,\ssdZ_{g} \rpar \Bigr] \spc
\eq
where $\mcF= \mcD, \mcP$ and $\ssdZ_i = \ssdZ^{(4)}_i + \gds\,\ssdZ^{(6)}_i$.
We obtain
\bqa
\mcD_{\sPHZZ} &=& - i\,g_{\ren}\,\frac{M_{\ren}}{(\ctwr)^2} 
                  + i\,\frac{g^3_{\ren}}{16\,\pi^2}\,\lpar\,\mcD^{(4)}_{\sPHZZ}  
                  + \gds\,\mcD^{(6)}_{\sPHZZ\,;\,\fin} \rpar 
\nl
{}&+& i\,g_{\ren}\,\gds\,\mcD^{(6)\,,\ren\,,a}_{\sPHZZ}
                  + i\,\frac{g^3_{\ren}}{16\,\pi^2}\,\gds\,\mcD^{(6)}_{\sPHZZ\,;\,\pole} \spc
\nl
\mcP_{\sPHZZ} &=&  i\,\frac{g^3_{\ren}}{16\,\pi^2}\,\lpar\,\mcP^{(4)}_{\sPHZZ}  
                  + \gds\,\mcP^{(6)}_{\sPHZZ\,;\,\fin} \rpar
\nl
{}&+& i\,g_{\ren}\,\gds\,\mcP^{(6)\,,\ren\,,a}_{\sPHZZ}
                  + i\,\frac{g^3_{\ren}}{16\,\pi^2}\,\gds\,\mcP^{(6)}_{\sPHZZ\,;\,\pole} \spp
\eqa
The explicit expressions for $\mcF^{(6)}_{\sPHZZ\,;\,\pole}$ will not be reported here.
The last step in UV-renormalization requires a mixing among Wilson coefficients,
performed according to \eqn{MixW}. After the removal of the remaining ($\mrdim = 6$) UV 
parts we obtain
\bq
\mcF^{(6)}_{\sPHZZ\,;\,\pole} \to \mcF^{(6),\ssR}_{\sPHZZ}\,\ln\frac{\muRs}{\Mbs} \spp
\eq
Elements of the mixing matrix derived from $\PH \to \PZZ$ are given in \appendx{MixWc}.
Inclusion of wave-function renormalization factors and of external leg factors
(due to field redefinition, introduced in \sect{FL}) gives
\bqa
{}&{}& \mcF^{\ren}_{\sPHZZ}\,
        \Bigl[ 1 - \frac{1}{2}\,\frac{g^2_{\ren}}{16\,\pi^2}\,
        \lpar \mrW_{\PH} + 2\,\mrW_{\PZ} \rpar \Bigr]\,
        \Bigl[ 1 + \gds\,\lpar 2\,\aZZ + \apBox - \frac{1}{4}\,\apD \rpar \Bigr] \spp
\eqa
Finite renormalization is performed by using \eqn{frenp}. The process $\PH \to \PZZ$ 
starts at $\mcO(g)$, therefore, the full set of counterterms must be included, not only 
the $\mrdim = 4$ part, as we have done for the loop induced processes. 

It is convenient to define NLO sub-amplitudes; however, to respect a factorization into 
$\PQt, \PQb$ and bosonic components, we have to introduce the following quantities:
\bq
\begin{array}{ll}
\mrW_{\PH} = \mrW_{\PH\,\;\,\PW} + \mrW_{\PH\,\;\,\PQt}  + \mrW_{\PH\,\;\,\PQb} \quad & \quad
\mrW_{\PZ} = \mrW_{\PZ\,\;\,\PW} + \mrW_{\PZ\,\;\,\PQt}  + \mrW_{\PZ\,\;\,\PQb} +
               {\overline{\sum}}_{\gen}\,\mrW_{\PZ\,;\,\Pf} \\
\ssdCZ_{g} = \ssdCZ_{g\,;\,\PW} + \sum_{\gen}\,\ssdCZ_{g\,;\,\Pf} \quad & \quad
\ssdCZ_{\ctw} = \ssdCZ_{\ctw\,;\,\PW}  + \ssdCZ_{\ctw\,;\,\PQt}  + \ssdCZ_{\ctw\,;\,\PQb} 
               + {\overline{\sum}}_{\gen}\,\ssdCZ_{\ctw\,;\,\Pf} \\
\ssdCZ_{\mw} = \ssdCZ_{\mw\,;\,\PW} + \sum_{\gen}\,\ssdCZ_{\mw\,;\,\Pf} & \\
\end{array}
\eq
where $\mrW_{\upPhi\,;\,\upphi}$ denotes the $\upphi$ component of the $\upPhi$ wave-function 
factor \etc
Furthermore, $\sum_{\gen}$ implies summing over all fermions and all generations, while
${\overline{\sum}}_{\gen}$ excludes $\PQt$ and $\PQb$ from the sum.
We can now define $\upkappa\,$-factors for the process, see \eqn{duk}:
\bq
\Delta \upkappa^{\sPHZZ}_{\ssD\,;\,\myLO} = 
    \stWs\,\aAA + \Bigl[ 4 + \ctWs\,\lpar 1 - \frac{\mhs}{\mws} \rpar \Bigr]\,\aZZ
    + \ctW\,\stW\,\aAZ + 2\,\apBox \spc
\eq
\bqa
\Delta \upkappa^{\sPHZZ}_{\PQt\,;\,\ssD\,;\,\myNLO} &=&
 \atp + 2\,\apBox - \frac{1}{2}\,\apD + 2\,\aZZ + \stWs\,\aAA \spc
\nl
\Delta \upkappa^{\sPHZZ}_{\PQb\,;\,\ssD\,;\,\myNLO} &=&
 - \abp + 2\,\apBox - \frac{1}{2}\,\apD + 2\,\aZZ + \stWs\,\aAA \spc
\nl
\Delta \upkappa^{\sPHZZ}_{\PW\,;\,\ssD\,;\,\myNLO} &=&
          \frac{1}{12}\,\lpar 4 + \frac{1}{\ctWs} \rpar\,\apD
          + 2\,\apBox
          + \stWs\,\aAA
\nl
{}&+& \stWs\,\lpar 3\,\ctW + \frac{5}{3}\,\frac{1}{\ctW} \rpar\,\aAZ
          + \lpar 4 + \ctWs \rpar\,\aZZ \spc
\nl
\Delta \upkappa^{\sPHZZ}_{\PQt\,;\,\ssP\,;\,\myNLO} &=&
  \atp + 2\,\apBox - \frac{1}{2}\,\apD + 2\,\aZZ + \stWs\,\aAA \spc
\nl
\Delta \upkappa^{\sPHZZ}_{\PQb\,;\,\ssP\,;\,\myNLO} &=&
 - \abp + 2\,\apBox - \frac{1}{2}\,\apD + 2\,\aZZ + \stWs\,\aAA \spc
\nl
\Delta \upkappa^{\sPHZZ}_{\PW\,;\,\ssP\,;\,\myNLO} &=&
    4\,\apBox + \frac{5}{2}\,\apD + 12\,\aZZ + 3\,\stWs\,\aAA
\eqa
and obtain the final result for the amplitudes
\bqa
\mcD_{\sPHZZ} &=& - i\,g_{\ssF}\,\frac{\mw}{\ctWs}\,\upkappa^{\sPHZZ}_{\ssD\,;\,\myLO}
   + i\,\frac{g^3_{\ssF}}{\pi^2}\,\sum_{\PI{=} \PQt,\PQb,\PW}\,
   \upkappa^{\sPHZZ}_{\PI\,;\,\ssD\,;\,\myNLO}\,\mcD^{\PI}_{\sPHZZ\,;\,\myNLO}
\nl
{}&+& i\,\frac{g^3_{\ssF}}{\pi^2}\,\mcD^{(4)\,;\,\nfact}_{\sPHZZ}
 + i\,\frac{g^3_{\ssF}}{\pi^2}\,\gds\,\sum_{\{a\}}\,\mcD^{(6)\,;\,\nfact}_{\sPHZZ}(a) \spc
\nl\nl
\mcP_{\sPHZZ} &=& 2\,i\,g_{\ssF}\,\gds\,\frac{\aZZ}{\mw}
   + i\,\frac{g^3_{\ssF}}{\pi^2}\,\sum_{\PI{=} \PQt,\PQb,\PW}\,
   \upkappa^{\sPHZZ}_{\PI\,;\,\ssP\,;\,\myNLO}\,\mcP^{\PI}_{\sPHZZ\,;\,\myNLO}
\nl
{}&+& i\,\frac{g^3_{\ssF}}{\pi^2}\,\gds\,\sum_{\{a\}}\,\mcP^{(6)\,;\,\nfact}_{\sPHZZ}(a) \spp
\eqa
Here we have introduced
\bq
\mcD^{(4)\,;\,\nfact}_{\sPHZZ} = \frac{1}{32}\,\frac{\mw}{\ctWs}\,\lpar
   2\,{\overline{\sum}}_{\gen}\,\mrW_{\PZ\,;\,\Pf} - \sum_{\gen}\,\ssdZ_{\mw\,;\,\Pf}
   + 4\,{\overline{\sum}}_{\gen}\,\ssdZ_{\ctw\,;\,\Pf} 
- 2\,\sum_{\gen}\,\ssdZ_{g\,;\,\Pf} \rpar \spp
\eq
Dimension $4$ sub-amplitudes $\mcD^{\PI}_{\sPHZZ\,;\,\myNLO}$, $\mcP^{\PI}_{\sPHZZ\,;\,\myNLO}$ 
($I=\PW,\PQt,\PQb$) are defined by using
\bq
\lz = \mhs - 4\,\mzs \spc \qquad
\lzz = \frac{\mhs}{\mws} - 4\,\frac{\mzs}{\mws} \spc 
\label{deflzz}
\eq
and are given in \appendx{LOsub}. Non-factorizable $\mrdim = 6$ amplitudes are reported in 
\appendx{Ampnf}, using again \eqn{deflzz}.
\subsection{$\PH \to \PWW$}
The derivation of the amplitude for $\PH \to \PWW$ follows closely the one for
$\PH \to \PZZ$. There are two main differences, there are only $1$PI contributions 
for $\PH \to \PWW$ and the the process shows an infrared (IR) component.

The IR part originates from two different sources. Vertex diagrams generate an IR
$C_0$ function:
\bq
\cfun{-\mhs}{\mws}{\mws}{\mw}{0}{\mw} =
\frac{1}{\beta_{\sPW}\,\mhs}\,\ln\frac{\beta_{\sPW} - 1}{\beta_{\sPW} + 1}\,\DIR +
\cfunf{-\mhs}{\mws}{\mws}{\mw}{0}{\mw} \spc
\eq
where we have introduced
\bq
\beta^2_{\sPW} = 1 - 4\,\frac{\mws}{\mhs} \spc
\qquad 
\DIR = \frac{1}{\eph} + \ln\frac{\mws}{\mu^2} \spp
\eq
The second source of IR behavior is found in the $\PW$ wave-function factor:
\bq
\mrW_{\PW} =  - 2\,\stws\,\DIR + \mrW_{\PW}\,\bmid_{\fin} \spp
\eq
The lowest-order part of the amplitude is
\bq
\mcD^{\myLO}_{\sPHWW}= - g\,M \spp
\eq
The $\mcO(g \gds)$ components of $\PH \to \PWW$ are:
\bqa
\mcD^{(6)\,,a}_{\sPHWW} &=&  
\lpar \frac{\mhs - 2\,\mws}{M} + M \rpar\,\apW - M\,\apBox + \frac{1}{4}\,M\,\apD
    - M\,\frac{\stws}{\ctws}\,\,\aAA
    - M\,\frac{\stw}{\ctw}\,\aAZ \spc
\nl
\mcP^{(6)\,,a}_{\sPHWW} &=& 2\,\frac{1}{M}\,\apW \spp
\eqa
With their help we can isolate the IR part of the $\PH \to \PWW$ amplitude
\bq
\mcD_{\sPHWW}\,\bmid_{\IR} = \mrI^{\IR}_{\sPHWW}\,
\frac{1}{\beta_{\sPW}\,\mhs}\,\ln\frac{\beta_{\sPW} - 1}{\beta_{\sPW} + 1}\,\DIR \spc
\eq
\bqa
\mrI^{\IR}_{\sPHWW} &=&
\frac{1}{8}\,i\,\frac{g^2_{\ssF}}{\pi^2}\,\stWs\,\mhs\,\lpar 
\mcD^{\myLO}_{\sPHWW} + \gds\,\mcD^{(6)\,,a}_{\sPHWW} \rpar
\nl
{}&+& \frac{1}{16}\,i\,\frac{g^2_{\ssF}}{\pi^2}\,\gds\,\mcD^{\myLO}_{\sPHWW}\,
\Bigl[ 2\,\lpar \mhs + \mws \rpar\,\frac{\mhs}{\mws}\,\stws\,\apW
- 4\,\stws\,\mhs\,\apW - 2\,\stws\,\mhs \Bigr] \spp
\eqa
Having isolated the IR part of the amplitude we can repeat, step by step, the
procedure developed in the previous sections. There is a non trivial aspect in the 
mixing of Wilson coefficients: the $\mrdim = 6$ parts of $\PH \to \PAA,\PAZ$ and 
$\mcP^{(6)}_{\sPHZZ, \sPHWW}$ contain UV divercences proportional to $W_{1,2,3}$; once 
renormalization is completed for $\PH \to \PAA, \PAZ$ and $\PZZ$ there is no freedom 
left and UV finiteness of $\PH \to \PWW$ must follow, proving closure of the $\mrdim = 6$ 
basis with respect to renormalization.

We can now define $\upkappa\,$-factors for $\PH \to \PWW$, see \eqn{duk}. They are as follows:
\bq
\Delta \upkappa^{\sPHWW}_{\ssD\,;\,\myLO} = 
 \stWs\,\lpar \frac{\mhs}{\mw} - 5\,\mw \rpar\,\lpar \aAA + \aAZ + \aZZ \rpar
+ \frac{1}{2}\,\mw\,\apD - 2\,\mw\,\apBox \spc
\eq
\bqa
\Delta \upkappa^{\sPHWW}_{\PQq\,;\,\ssD\,;\,\myNLO} &=&
           \aptV + \aptA + \apbV + \apbA 
           - \adp + 2\,\apBox - \frac{1}{2}\,\apD
\nl
{}&+& \stW\,\abWB + \ctW\,\abBW + 5\,\stWs\,\aAA + 5\,\ctW\,\stW\,\aAZ + 5\,\ctWs\,\aZZ \spc
\nl
\Delta \upkappa^{\sPHWW}_{\PW\,;\,\ssD\,;\,\myNLO} &=&
           \frac{1}{96}\,\frac{\stWs}{\ctWs}\,\apD
          + \frac{23}{12}\,\apBox
          - \frac{35}{96}\,\apD
\nl
{}&+& 4\,\,\stWs\,\aAA
          + \frac{1}{12}\,\stw\,\lpar 3\,\frac{1}{\ctW} + 49\,\ctW \rpar\,\aAZ
          + \frac{1}{2}\,\lpar 9\,\ctWs + \stWs \rpar\,\aZZ \spc
\nl
\Delta \upkappa^{\sPHWW}_{\PQQ\,;\,\ssP\,;\,\myNLO} &=&
          \aptV + \aptA + \apbV + \apbA
          - \adp + 2\,\apBox - \frac{1}{2}\,\apD
\nl
{}&+& \stW\,\adWB + \ctW\,\adBW + 5\,\stWs\,\aAA + 5\,\ctW\,\stW\,\aAZ + 5\,\ctWs\,\aZZ \spc
\nl
\Delta \upkappa^{\sPHWW}_{\PW\,;\,\ssP\,;\,\myNLO} &=&
           7\,\apBox - 2\,\apD  + 5\,\stWs\,\aAA
          + 5\,\ctW\,\stW\,\aAZ + 5\,\ctWs\,\aZZ \spp
\eqa
Next we obtain the final result for the amplitudes
\bqa
\mcD_{\sPHWW} &=& - i\,g_{\ssF}\,\mw\,\upkappa^{\sPHWW}_{\ssD\,;\,\myLO}
   + i\,\frac{g^3_{\ssF}}{\pi^2}\,\sum_{\PI{=} \PQq,\PW}\,
   \upkappa^{\sPHWW}_{\PI\,;\,\ssD\,;\,\myNLO}\,\mcD^{\PI}_{\sPHWW\,;\,\myNLO}
\nl
{}&+& i\,\frac{g^3_{\ssF}}{\pi^2}\,\mcD^{(4)\,;\,\nfact}_{\sPHWW}
 + i\,\frac{g^3_{\ssF}}{\pi^2}\,\gds\,\sum_{\{a\}}\,\mcD^{(6)\,;\,\nfact}_{\sPHWW}(a) \spc
\nl\nl
\mcP_{\sPHWW} &=& 2\,i\,g_{\ssF}\,\gds\,\frac{1}{\mw}\,\lpar
\stWs\,\aAA + \ctWs\,\aZZ + \ctW\,\stW\,\aAZ \rpar
   + i\,\frac{g^3_{\ssF}}{\pi^2}\,\sum_{\PI{=} \PQq}\,
   \upkappa^{\sPHWW}_{\PI\,;\,\ssP\,;\,\myNLO}\,\mcP^{\PI}_{\sPHWW\,;\,\myNLO}
\nl
{}&+& i\,\frac{g^3_{\ssF}}{\pi^2}\,\gds\,\sum_{\{a\}}\,\mcP^{(6)\,;\,\nfact}_{\sPHWW}(a) \spc
\eqa
where we have introduced
\bq
\mcD^{(4)\,;\,\nfact}_{\sPHWW} = \frac{1}{32}\,\mw\,{\overline{\sum}}_{\gen}\,\lpar
   2\,\mrW_{\PW\,;\,\Pf}
   - \ssdCZ_{\mw\,;\,\Pf}
   - 2\,\ssdCZ_{g\,;\,\Pf} \rpar \spp
\eq
Dimension $4$ sub-amplitudes $\mcD^{\PI}_{\sPHWW\,;\,\myNLO}$, $\mcP^{\PI}_{\sPHWW\,;\,\myNLO}$ 
($I=\PW,\PQt,\PQb$) are defined by introducing
\bq
\lw = \mhs - 4\,\mws \spc \qquad
\lww = \frac{\mhs}{\mws} - 4 \spc
\label{deflww}
\eq
and are given in \appendx{LOsub}. Non-factorizable $\mrdim = 6$ amplitudes are reported in 
\appendx{Ampnf}, using again \eqn{deflww}.
\subsection{$\PH \to \PAQb\PQb(\PGtp\PGtm)$ and $\PH \to 4\,\Pl$}
These processes share the same level of complexity of $\PH \to \PZ\PZ(\PW\PW)$, including the
presence of IR singularities. They will be discussed in details in a forthcoming publication.
\section{ElectroWeak precision data \label{EWPD}}
EFT is not confined to describe Higgs couplings and their SM deviations. It can be used
to reformulate the constraints coming from electroweak precision data (EWPD), starting
from the $\ssS, \ssT$ and $\ssU$ parameters of \Bref{Peskin:1990zt} and including the
full list of LEP pseudo-observables (PO).

There are several ways for incorporating EWPD: the preferred option, so far, is
reducing (a priori) the number of $\mrdim = 6$ operators.
More generally, one could proceed by imposing penalty functions $\omega$ on the global LHC fit, 
that is functions defining an $\omega\,$-penalized LS estimator for a set of global penalty 
parameters (perhaps using {\sl{merit functions}} and the {\sl{homotopy method}}).
One could also consider using a Bayesian approach~\cite{deBlas:2014ula}, with a flat prior for 
the parameters.
Open questions are: one $\upkappa$ at the time? Fit first to the EWPD and then to 
$\PH$ observables? Combination of both? 

In the following we give a brief description of our procedure: from \eqn{AAself} and \eqn{OPR} 
we obtain
\bq
\Sigma^{\OPI + \OPR}_{\PAA}(s) = -\,\Bigl[ \Pi^{\OPI}_{\PAA}(0) + 
          \gds\,\frac{\stw}{\ctw}\,\frac{1}{\Mzbs}\,\aAZ\,\Pi^{\OPI}_{\PZA}(0) \Bigr]\,s + 
\mcO(s^2) = \Pi^c_{\PAA}\,s + \mcO(s^2) \spp
\eq
>From \eqn{SZA} and \eqn{OPR} we obtain
\bq
\Sigma^{\OPI + \OPR}_{\PZA}(s) = -\,\Bigl[ \Pi^{\OPI}_{\PZA}(0) 
+ \gds\,\stws\,\aAZ\,\Pi^{\OPI}_{\PAA}(0)\Bigr]\,s +
\mcO(s^2) = \Pi^c_{\PZA}\,s + \mcO(s^2)  \spp
\eq
>From \eqn{ZZself} and \eqn{OPR} we derive
\bq
\ssD^{\OPI + \OPR}_{\PZZ}(s) = \Delta_{\PZZ}(0) + \lbra \Omega_{\PZZ}(0)
- \gds\,\frac{\stw}{\ctw}\,\aAZ\,\Pi^{\OPI}_{\PZA}(0) \rbra\,s +
\mcO(s^2) = \Delta_{\PZZ}(0) + \Omega^c_{\PZZ}(0)\,s + \mcO(s^2) \spp
\eq
Similarly, we obtain
\bq
\ssD^{\OPI}_{\PWW}(s) = \Delta_{\PWW}(0) + \Omega_{\PWW}(0)\,s + \mcO(s^2) \spp
\eq
The $\ssS, \ssT$ and $\ssU$ parameters are defined in terms of (complete) self-energies
at $s = 0$ and of their (first) derivatives.
However, one has to be careful because the corresponding definition (see \Bref{Peskin:1990zt}) is 
given in the $\{\alpha\,,\,\myGF\,,\,\mz\}$ scheme, while we have adopted the more convenient 
$\{\myGF\,,\,\mw\,,\,\mz\}$ scheme.
Working in the $\alpha$-scheme has one advantage, the possibility of predicting the $\PW$ 
(on-shell) mass.
After UV renormalization and finite renormalization in the $\alpha\,$-scheme we define
$\mwsOS$ as the zero of the real part of the inverse $\PW$ propagator and derive the effect
of $\mrdim = 6$ operators.
\subsection{$\PW\;$ mass}
Working in the $\alpha\,$-scheme we can predict $\MW$. The solution is
\bqa
\frac{\mws}{\mzs} &=& \cths + \frac{\alpha}{\pi}\,\Re\,\Bigl\{
\lpar 1 - \frac{1}{2}\,\gds\,\apD\rpar\,\Delta^{(4)}_{\PB}\,\mw + 
\sum_{\gen}\,\Bigl[ 
\lpar 1 + 4\,\gds\,\aplt\rpar\,\Delta^{(4)}_{\Pl}\,\mw + 
\lpar 1 + 4\,\gds\apqt\rpar\,\Delta^{(4)}_{\PQq}\,\mw \Bigr]
\nl
{}&+& \gds\,\Bigl[
\Delta^{(6)}_{\PB}\,\mw + 
\sum_{\gen}\,\lpar 
\Delta^{(6)}_{\Pl}\,\mw + \Delta^{(6)}_{\PQq}\,\mw \rpar \Bigr]\Bigr\} \spp
\label{premw}
\eqa
where $\cths$ is defined in \eqn{defscipsb} and we drop the subscript OS (on-shell). 
Corrections are given in \appendx{DMW}. The expansion in \eqn{premw} can be improved when 
working within the SM ($\mrdim = 4$), see \Bref{Bardin:1999ak}: for instance, the expansion 
parameter is set to $\alpha(\mz)$ instead of $\alpha(0)$, \etc Any equation that gives 
$\mrdim = 6$ corrections to the SM result will always be understood as
\bq
{\mcO} = {\mcO}^{\mySM}\bmid_{{\mathrm{imp}}} + \frac{\alpha}{\pi}\,\gds\,{\mcO}^{(6)}
\eq
in order to match the TOPAZ0/Zfitter SM results when $\gds \to 0$, see
\Brefs{Montagna:1998kp,Montagna:1995ja,Montagna:1993ai} and
\Brefs{Akhundov:2014era,Arbuzov:2005ma}.
\subsection{$\ssS,\ssT$ and $\ssU$ parameters}
The $\ssS, \ssU$ and $\ssT$ (the original $\rho\,$-parameter of 
Veltman~\cite{Ross:1975fq}) are defined as follows:
\bqa
\alpha\,\ssT &=& \frac{1}{\mws}\,\ssD_{\PWW}(0) - \frac{1}{\mzs}\,\ssD_{\PZZ}(0) \spc
\nl
\frac{\alpha}{4\,\sths\cths}\,\ssS &=& 
\Omega_{\PZZ}(0) - \frac{\cths - \sths}{\sth\,\cth}\,\Pi_{\PZA}(0) - \Pi_{\PAA}(0) \spc
\nl
\frac{\alpha}{4\,\sths}\,\ssU &=& 
\Omega_{\PWW}(0) - \cths\,\Omega_{\PZZ}\,\Pi_{\PZA}(0) - \sths\,\Pi_{\PAA}(0) \spc
\label{PTpar}
\eqa
where all the self-energies are renormalized and $\sth$ is defined in \eqn{defscipsb}.
One of the interesting properties of these parameters is that, within the SM, they are
UV finite, \ie all UV divergences cancel in \eqn{PTpar} if they are written in terms
of bare parameters and bare self-energies.
When $\mrdim = 6$ operators are inserted we obtain the following results:
\bqa
\alpha\,\ssT &=& \alpha\,\mcT^{(4)} + \alpha\,\gds\,\mcT^{(6)} \spc
\nl 
\alpha\,\ssS &=& \alpha\,\mcS^{(4)} - 4\,\gds\,\frac{1 - 2\,\sths}{\sth\,\cth}\,\aAZ + 
              \alpha\,\gds\,\mcS^{(6)} \spc
\nl 
\alpha\,\ssU &=& \alpha\,\mcU^{(4)} - 4\,\gds\,\frac{1 - 2\,\sths}{\sth\,\cthc}\,\aAZ + 
              \alpha\,\gds\,\mcU^{(6)}
\label{PTparsix}
\eqa
and the introduction of counterterms is crucial to obtain an UV finite results.
Explicit results for the $\ssT$ parameter are given in \appendx{Tpar} where, for
simplicity, we only include PTG operators in loops.
\section{Conclusions}
In this paper we have developed a theory for Standard Model deviations based on the
Effective Field Theory approach. In particular, we have considered the introduction
of $\mrdim = 6$ operators and extended their application at the NLO level (for a very recent
development see \Bref{Hartmann:2015oia}).

The main result is represented by a consistent generalization of the LO 
$\upkappa\,$-framework, currently used by ATLAS and CMS.

This step forward is better understood when comparing the present situation with the one
at LEP; there the dynamics was fully described within the SM, with $\mh\,\alphas(\mz), \dots$
as unknowns. Today, post the LHC discovery of a $\PH\,$-candidate, unknowns are SM-deviations.
This fact poses precise questions on the next level of dynamics. A specific BSM model 
is certainly a choice but one would like to try a more model independent approach. 

The aim of this paper is to propose a decomposition where dynamics is controlled by $\mrdim = 4$ 
amplitudes (with known analytical properties) and deviations (with a direct link to UV 
completions) are (constant) combinations of Wilson coefficients for $\mrdim = 6$ operators.

Generalized $\upkappa\,$-parameters form hyperplanes in the space of Wilson coefficients;
each $\upkappa\,$-plane describes (tangent) flat directions while normal directions are blind.
Finally, $\upkappa\,$-planes intersect, providing correlations among different processes.
Our prescription allows a theoretically robust matching between theory and experiments.

Only the comparison with experimental data will allow us to judge the goodness of a proposal
that, for us, is based on the belief that SM deviations need a SM basis. 

\Acknowledgments
We gratefully acknowledge several important discussions with A.~David, M.~Duehrssen
and the participants in the ``Pseudo-observables: from LEP to LHC'' and
``ATLAS Higgs (N)NLO MC and Tools Workshop for LHC RUN-2'' events at Cern.

This work is supported by MIUR under contract $2001023713\_006$, by UniTo{-}Compagnia di 
San Paolo under contract ORTO$11$TPXK and by FP7-PEOPLE-$2012$-ITN HiggsTools 
PITN-GA-$2012{-}316704$. 

\clearpage
\appendix

\section{Appendix: $\bhb$ and $\Gamma$ \label{bhb}}

In this Appendix we present the full result for $\bhb$, defined in \eqn{defbh}. We have 
introduced ratios of masses
\bq
\xph = \frac{\mh}{\mw}, \quad x_{\Pf} = \frac{M_{\Pf}}{\mw}
\eq
\etc The various components are given by
\footnotesize
\bqa
\beta^{(4)}_{-1} &=& 
       - \sum_{\gen}\,(\xpls + 3\,\xpds + 3\,\xpus)
       + \frac{1}{8}\,(12 + 2\,\xph + 3\,\xphs)
       + \frac{1}{8}\,\frac{6 + \ctws\,\xph}{\ctwq}
\nl\nl
\beta^{(4)}_0 &=&  - \frac{1}{2}\,\frac{1 + 2\,\ctwq}{\ctwq}
\nl\nl
\beta^{(4)}_{\fin} &=&
       - \frac{1}{4}\,\afun{\mw}\,(6 + \xph)
       - \frac{3}{8}\,\afun{\mh}\,\xphs
       - \frac{1}{8}\,\frac{6 + \ctws\,\xph}{\ctwq}\,\afun{\mz}
\nl
{}&+& \sum_{\gen}\,\Bigl[ 3\,\afun{M_{\PQu}}\,\xpus + 
                     3\,\afun{M_{\PQd}}\,\xpds + 
                       \afun{M_{\Pl}}\,\xpls\Bigr]
\eqa
\bqa
\beta^{(6)}_{-1} &=& 
         \frac{1}{8}\,\frac{12 + \xph}{\ctws}\,\apW
       + \frac{1}{8}\,(36 + 2\,\xph + 3\,\xphs)\,\apW
       - \frac{1}{4}\,\sum_{\gen}\,\Bigl[
         3\,(4\,\aup + 4\,\apBox + 4\,\apW - \apD)\,\xpus
\nl
{}&-& 3\,(4\,\adp - 4\,\apBox - 4\,\apW + \apD)\,\xpds
   + (4\,\apBox + 4\,\apW - \apD - 4\,\aLp)\,\xpls
         \Bigr]
\nl
{}&+& \frac{3}{16}\,(4\,\apBox + 4\,\apW + \apD + 8\,\stws\,\apB - 8\,\ctw\,\stw\,\apWB)\,
       \frac{1}{\ctwq}
       + \frac{1}{32}\,(4\,\apBox - \apD)\,(12 - 2\,\xph + 7\,\xphs)
\nl
{}&-& \frac{1}{32}\,(4\,\apBox + \apD + 96\,\ctws\,\ap)\,\frac{\xph}{\ctws}
\nl\nl
\beta^{(6)}_0 &=& 
            - \frac{1}{8}\,\Bigl[
             8\,\stws\,\apB 
             + 4\,(1 + 2\,\ctwq)\,\apBox 
             + 4\,(1 + 2\,\ctws + 6\,\ctwq)\,\apW 
             + (1 - 2\,\ctwq)\,\apD 
             - 8\,\ctw\,\stw\,\apWB 
             \Bigr]\,\frac{1}{\ctwq}
\nl\nl
\beta^{(6)}_{\fin} &=&
         3\,\afun{\mh}\,\ap\,\xph
       - \frac{1}{4}\,\afun{\mw}\,(18 + \xph)\,\apW
       - \frac{1}{8}\,\afun{\mz}\,\frac{12 + \xph}{\ctws}\,\apW
\nl
{}&+& \frac{1}{4}\,\sum_{\gen}\,\Bigl[
        3\,(4\,\aup + 4\,\apBox + 4\,\apW - \apD)\,\afun{M_{\PQu}}\,\xpus
       - 3\,(4\,\adp - 4\,\apBox - 4\,\apW + \apD)\,\afun{M_{\PQd}}\,\xpds
\nl
{}&+& (4\,\apBox + 4\,\apW - \apD - 4\,\aLp)\,\afun{M_{\Pl}}\,\xpls
         \Bigr]
\nl
{}&-& \frac{3}{16}\,(4\,\apBox + 4\,\apW + \apD + 8\,\stws\,\apB - 8\,\ctw\,\stw\,\apWB)\,
        \frac{1}{\ctwq}\,\afun{\mz}
       - \frac{1}{16}\,(4\,\apBox - \apD)\,\afun{\mw}\,(6 - \xph)
\nl
{}&+& \frac{1}{32}\,(4\,\apBox + \apD)\,\afun{\mz}\,\frac{\xph}{\ctws}
    - \frac{1}{32}\,(28\,\apBox + 12\,\apW - 7\,\apD)\,\afun{\mh}\,\xphs
\eqa
\normalsize
We also present the full result for $\Gamma$, defined in \eqn{Gammadef}. We have 
\footnotesize
\bq
\Gamma^{(4)}_{-1} = - \frac{1}{8}
\quad
\Gamma^{(4)}_{0} = \frac{1}{8}
\quad
\Gamma^{(4)}_{\fin} = \frac{1}{8}\,a^{\fin}_0\lpar M \rpar
\eq
\bq
\Gamma^{(6)}_{-1} = - \frac{1}{4}\,\apW
\quad
\Gamma^{(6)}_{0} = \frac{1}{4}\,\apW
\quad
\Gamma^{(6)}_{\fin} = \frac{1}{4}\,a^{\fin}_0\lpar M \rpar\,\apW
\eq
\normalsize
\ie $\Gamma= \Gamma^{(4)}\,\lpar 1 + 2\,\gds\,\apW \rpar$.

\section{Appendix: Renormalized self-energies \label{RSE}}

In this Appendix we present the full set of renormalized self-energies. 
To keep the notation as compact as possible a number of auxiliary quantities has been introduced.
\subsection{Notations}
First we define the following set of polynomials:
\vspace{0.8cm}
\bei
\item[\fbox{$\mrF\,$}] where $s = \stw$ and $c = \ctw$ 
\eei

\scriptsize
\[
\begin{array}{lll}
\mrF^{a}_{0}= 1 - 6\,c \;\;&\;\;
\mrF^{a}_{1}= 4 - 9\,c \;\;&\;\;
\mrF^{a}_{2}= 1 - c \\
\mrF^{a}_{3}= 2 - 15\,c \;\;&\;\;
\mrF^{a}_{4}= 2 + 3\,c \;\;&\;\;
\mrF^{a}_{5}= 11 - 3\,s \\
\mrF^{a}_{6}= 8 - 3\,s & & \\
\end{array}
\]

\[
\begin{array}{lll}
\mrF^{b}_{0}= 1 + 2\,c \;\;&\;\;
\mrF^{b}_{1}= 1 + 3\,c \;\;&\;\;
\mrF^{b}_{2}= 1 + 18\,c \\
\mrF^{b}_{3}= 1 + c \;\;&\;\;
\mrF^{b}_{4}= 1 + 4\,c \;\;&\;\;
\mrF^{b}_{5}= 1 - 2\,s \\
\mrF^{b}_{6}= 1 + 24\,s^2\,c \;\;&\;\;
\mrF^{b}_{7}= 1 + \mrF^{a}_{0}\,c \;\;&\;\;
\mrF^{b}_{8}= 1 + 4\,\mrF^{a}_{1}\,c \\
\mrF^{b}_{9}= 3 + 4\,c \;\;&\;\;
\mrF^{b}_{10}= 5 + 8\,c \;\;&\;\;
\mrF^{b}_{11}= 1 - 40\,c + 36\,s\,c \\
\mrF^{b}_{12}= 1 + 20\,c - 12\,s\,c \;\;&\;\;
\mrF^{b}_{13}= 5 - 20\,c + 12\,s\,c \;\;&\;\;
\mrF^{b}_{14}= 5 - 3\,s \\
\mrF^{b}_{15}= 1 - 12\,\mrF^{a}_{2}\,c \;\;&\;\;
\mrF^{b}_{16}= 3 + 8\,s\,c \;\;&\;\;
\mrF^{b}_{17}= 13 + 4\,\mrF^{a}_{3}\,c \\
\mrF^{b}_{18}= 19 - 18\,s \;\;&\;\;
\mrF^{b}_{19}= 1 - 12\,c^2 \;\;&\;\;
\mrF^{b}_{20}= 1 - 24\,c^3 \\
\mrF^{b}_{21}= 1 + 4\,\mrF^{a}_{4}\,c \;\;&\;\;
\mrF^{b}_{22}= 1 + 8\,\mrF^{a}_{4}\,c^2 \;\;&\;\;
\mrF^{b}_{23}= 4 - 3\,s \\
\mrF^{b}_{24}= 5 + 12\,c^2 \;\;&\;\;
\mrF^{b}_{25}= 5 - 4\,\mrF^{a}_{4}\,c \;\;&\;\;
\mrF^{b}_{26}= 13 - \mrF^{a}_{5}\,s \\
\mrF^{b}_{27}= 21 - 4\,\mrF^{a}_{6}\,s \;\;&\;\;
\mrF^{b}_{28}= 39 - 40\,s \;\;&\;\;
\mrF^{b}_{29}= 1 - 4\,s\,c \\
\mrF^{b}_{30}= 3 - 10\,c \;\;&\;\;
\mrF^{b}_{31}= 3 - 2\,s \;\;&\;\;
\mrF^{b}_{32}= c + 2\,s \\~
\mrF^{b}_{33}= 4 - 7\,c \;\;&\;\;
\mrF^{b}_{34}= 2 + c & \\
\end{array}
\]

\normalsize

\vspace{0.8cm}
\bei
\item[\fbox{$\mrG\,$}] where we have introduced
\bqas
v_{\Pf} &=& 1 - 2\,\frac{Q_{\Pf}}{I^3_{\Pf}}\,\stws
\eqas
where $Q_{\Pl} = -1$, $Q_{\PQu} = \frac{2}{3}$ and $Q_{\PQd} = - \frac{1}{3}$; $I^3_{\Pf}$ is the
third component of isospin. Furthermore
\bqas
v^{(1)}_{\gen} = v^2_{\Pl} + 3\,\lpar v^2_{\PQu} + v_{\PQd} \rpar
& \qquad &
v^{(2)}_{\gen} = v^2_{\Pl} + 2\,v^2_{\PQu} + v_{\PQd} 
\eqas
\eei

\scriptsize
\[
\begin{array}{lll}
\mrG_{0}= 1 - 3\,\vqd \;\;&\;\;
\mrG_{1}= 3 - \vle \;\;&\;\;
\mrG_{2}= 5 - 3\,\vqu \\
\mrG_{3}= 20 - 3\,\vtg \;\;&\;\;
\mrG_{4}= 1 + \vqus \;\;&\;\;
\mrG_{5}= 1 + \vqds \\
\mrG_{6}= 1 + \vles \;\;&\;\;
\mrG_{7}= 9 + \vog \;\;&\;\;
\mrG_{8}= 2 - \vqus \\
\mrG_{9}= 2 - \vqds \;\;&\;\;
\mrG_{10}= 2 - \vles \;\;&\;\;
\mrG_{11}= 1 - \vle \\
\mrG_{12}= 1 - \vqu \;\;&\;\;
\mrG_{13}= 1 + \vqu \;\;&\;\;
\mrG_{14}= 1 - \vqd \\
\mrG_{15}= 1 + \vqd \;\;&\;\;
\mrG_{16}= 1 - 3\,\vle \;\;&\;\;
\mrG_{17}= 1 + \vle \\
\mrG_{18}= 2 + \vqu + \vqd \;\;&\;\;
\mrG_{19}= 3 - 5\,\vqu \;\;&\;\;
\mrG_{20}= 3 - \vqd \\
\mrG_{21}= 3 + \vle \;\;&\;\;
\mrG_{22}= 9 - 5\,\vqu - \vqd - 3\,\vle \;\;&\;\;
\mrG_{23}= \vqu - \vqd \\
\mrG_{24}= 2 - \vqu \;\;&\;\;
\mrG_{25}= 2 + \vqu \;\;&\;\;
\mrG_{26}= 2 - \vqd \\
\mrG_{27}= 2 + \vqd \;\;&\;\;
\mrG_{28}= 2 - \vle \;\;&\;\;
\mrG_{29}= 2 + \vle \\
\mrG_{30}= 2 + 3\,\vle \;\;&\;\;
\mrG_{31}= 6 + 5\,\vqu \;\;&\;\;
\mrG_{32}= 6 + \vqd \\
\mrG_{33}= 1 - \vles \;\;&\;\;
\mrG_{34}= 1 - 2\,\vle \;\;&\;\;
\mrG_{35}= 1 + 2\,\vle \\
\mrG_{36}= 1 + 7\,\vle \;\;&\;\;
\mrG_{37}= 3 - 4\,\vle \;\;&\;\;
\mrG_{38}= 3 + 4\,\vle \\
\mrG_{39}= 8 + 3\,\vle \;\;&\;\;
\mrG_{40}= 1 - 7\,\vle \;\;&\;\;
\mrG_{41}= 7 - \vle \\
\mrG_{42}= 1 - \vqus \;\;&\;\;
\mrG_{43}= 1 - 2\,\vqu \;\;&\;\;
\mrG_{44}= 1 + 2\,\vqu \\
\mrG_{45}= 3 - 4\,\vqu \;\;&\;\;
\mrG_{46}= 3 + 4\,\vqu \;\;&\;\;
\mrG_{47}= 3 + 13\,\vqu \\
\mrG_{48}= 4 + 3\,\vqu \;\;&\;\;
\mrG_{49}= 16 + 9\,\vqu \;\;&\;\;
\mrG_{50}= 3 - 13\,\vqu \\
\mrG_{51}= 13 - 3\,\vqu \;\;&\;\;
\mrG_{52}= 1 - \vqds \;\;&\;\;
\mrG_{53}= 1 - 2\,\vqd \\
\mrG_{54}= 1 + 2\,\vqd \;\;&\;\;
\mrG_{55}= 2 + 3\,\vqd \;\;&\;\;
\mrG_{56}= 3 - 4\,\vqd \\
\mrG_{57}= 3 + 4\,\vqd \;\;&\;\;
\mrG_{58}= 3 + 5\,\vqd \;\;&\;\;
\mrG_{59}= 8 + 9\,\vqd \\
\mrG_{60}= 3 - 5\,\vqd \;\;&\;\;
\mrG_{61}= 5 - 3\,\vqd & \\
\end{array}
\]

\normalsize

\vspace{0.8cm}
\bei
\item[\fbox{$\mrH\,$}] where we have introduced
\bqas
x_{\Pf} &=& \frac{M_{\Pf}}{M} \qquad \mbox{\etc}
\eqas
\eei

\scriptsize
\[
\begin{array}{lll}
\mrH_{0}= 3\,\xpus + 3\,\xpds + \xpls \;\;&\;\;
\mrH_{1}= 3\,\xpuq + 3\,\xpdq + \xplq \;\;&\;\;
\mrH_{2}= 4\,\xpus + \xpds + 3\,\xpls \\
\mrH_{3}=  - 2\,\xpus + \xpds \;\;&\;\;
\mrH_{4}= 2\,\xpus + \xpds \;\;&\;\;
\mrH_{5}= 10\,\xpus + \xpds + 9\,\xpls \\
\mrH_{6}=  - \xpus + \xpds \;\;&\;\;
\mrH_{7}= \xpus + \xpds \;\;&\;\;
\mrH_{8}= \xpuq - 2\,\xpds\,\xpus + \xpdq \\
\mrH_{9}= 1 + \xpds \;\;&\;\;
\mrH_{10}= 1 - \xpds \;\;&\;\;
\mrH_{11}= 2 - \xpds - \xpdq \\
\mrH_{12}= 2 + \xpds \;\;&\;\;
\mrH_{13}= 2 - \xpds \;\;&\;\;
\mrH_{14}= 1 + \xpus \\
\mrH_{15}= 1 - \xpus \;\;&\;\;
\mrH_{16}= 2 - \xpus - \xpuq \;\;&\;\;
\mrH_{17}= 2 + \xpus \\
\mrH_{18}= 2 - \xpus & & \\
\end{array}
\]

\normalsize

\vspace{0.8cm}
\bei
\item[\fbox{$\mrI\,$}] where we have introduced
\bqas
x_{\sPF} &=& \frac{M_{\sPF}}{M} \qquad \mbox{\etc}
\eqas
\eei

\scriptsize
\[
\begin{array}{lll}
\mrI_{0}= 2 + \xpLs \;\;&\;\;
\mrI_{1}= 2 - \xpLs \;\;&\;\;
\mrI_{2}= 1 + 2\,\xpLs \\
\mrI_{3}= 2 + 3\,\xpLs \;\;&\;\;
\mrI_{4}= 4 - \xpLs \;\;&\;\;
\mrI_{5}= 4 + \xpLs \\
\mrI_{6}= 1 + 2\,\xpUs \;\;&\;\;
\mrI_{7}= 1 + \xpUs \;\;&\;\;
\mrI_{8}= 2 + 3\,\xpUs \\
\mrI_{9}= 1 + 2\,\xpDs \;\;&\;\;
\mrI_{10}= 1 + \xpDs \;\;&\;\;
\mrI_{11}= 2 + 3\,\xpDs \\
\end{array}
\]

\normalsize

\vspace{0.8cm}
\bei
\item[\fbox{$\mrJ\,$}] where we have introduced
\bqas
\xph &=& \frac{\mh}{M} \qquad \xps = \frac{s}{M^2} 
\eqas
\eei

\scriptsize
\[
\begin{array}{lll}
\mrJ_{0}= 2 + \xphq \;\;&\;\;
\mrJ_{1}= 12 + \xphq \;\;&\;\;
\mrJ_{2}= 2 - \xphs \\
\mrJ_{3}= 4 - 7\,\xphs \;\;&\;\;
\mrJ_{4}= 4 - 5\,\xphs \;\;&\;\;
\mrJ_{5}= 12 - \xphq \\
\mrJ_{6}= 2 + 11\,\xphs \;\;&\;\;
\mrJ_{7}= 3 + 4\,\xphq \;\;&\;\;
\mrJ_{8}= 4 + 5\,\xphq \\
\mrJ_{9}= 10 + \xphq \;\;&\;\;
\mrJ_{10}= 36 + \xphq \;\;&\;\;
\mrJ_{11}= 4 + 3\,\xps \\
\mrJ_{12}= 2 + \xphs \;\;&\;\;
\mrJ_{13}= 3 + \xphs \;\;&\;\;
\mrJ_{14}= 4 + \xps \\
\mrJ_{15}= 6 - \xphs \;\;&\;\;
\mrJ_{16}= 8 - \xps \;\;&\;\;
\mrJ_{17}= 8 + \xps \\
\mrJ_{18}= 10 - \xphs \;\;&\;\;
\mrJ_{19}= 12 + \xps \;\;&\;\;
\mrJ_{20}= 2 - 3\,\xphs \\
\mrJ_{21}= 12 - \xps \;\;&\;\;
\mrJ_{22}= 12 + 5\,\xps \;\;&\;\;
\mrJ_{23}= 12 - \xphs \\
\mrJ_{24}= 32 - 3\,\xps \;\;&\;\;
\mrJ_{25}= 32 + \xps \;\;&\;\;
\mrJ_{26}= 48 + \xps \\
\mrJ_{27}= 2\,\xps - \xphs \;\;&\;\;
\mrJ_{28}= 2\,\xps + \xphs \;\;&\;\;
\mrJ_{29}= 5\,\xps - \xphs \\
\mrJ_{30}= 19 - 3\,\xphs \;\;&\;\;
\mrJ_{31}= 58 - 3\,\xphs \;\;&\;\;
\mrJ_{32}= 70 - 3\,\xphs \\
\mrJ_{33}= 106 - 3\,\xphs \;\;&\;\;
\mrJ_{34}= \xps - \xphs \;\;&\;\;
\mrJ_{35}= 1 + \xphs \\
\mrJ_{36}= 8 + 3\,\xphs \;\;&\;\;
\mrJ_{37}= 11\,\xps - \xphs \;\;&\;\;
\mrJ_{38}= 12\,\xps - 2\,\xphs\,\xps + \xphq \\
\end{array}
\]

\[
\begin{array}{lll}
\mrJ_{39}= 1 - 2\,\xps \;\;&\;\;
\mrJ_{40}= 1 + 2\,\xps \;\;&\;\;
\mrJ_{41}= 1 + 2\,\xps - \xphs \\
\mrJ_{42}= 1 + 10\,\xps \;\;&\;\;
\mrJ_{43}= 1 + 10\,\xps - 2\,\xphs - 2\,\xphs\,\xps + \xphq \;\;&\;\;
\mrJ_{44}= 1 + 18\,\xps \\
\mrJ_{45}= 2 - 68\,\xps - \xphs \;\;&\;\;
\mrJ_{46}= 3 + 5\,\xps \;\;&\;\;
\mrJ_{47}= 42 + \xphs \\
\mrJ_{48}= 9 - 5\,\xps \;\;&\;\;
\mrJ_{49}= 1 + 5\,\xps \;\;&\;\;
\mrJ_{50}= 3 + \xps \\
\mrJ_{51}= 7 - 5\,\xps \;\;&\;\;
\mrJ_{52}= 8 - 5\,\xps \;\;&\;\;
\mrJ_{53}= 13 + 5\,\xps \\
\mrJ_{54}= 20 - 3\,\xphs \;\;&\;\;
\mrJ_{55}= 23 - 15\,\xps \;\;&\;\;
\mrJ_{56}= 1 + 3\,\xps \\
\mrJ_{57}= 1 - 7\,\xps \;\;&\;\;
\mrJ_{58}= 1 - 9\,\xps \;\;&\;\;
\mrJ_{59}= 1 - 5\,\xps \\
\mrJ_{60}= 1 - 3\,\xps \;\;&\;\;
\mrJ_{61}= 1 - \xps \;\;&\;\;
\mrJ_{62}= 1 + \xps - \xphs \\
\mrJ_{63}= 1 + 4\,\xps \;\;&\;\;
\mrJ_{64}= 1 + 9\,\xps \;\;&\;\;
\mrJ_{65}= 1 + 12\,\xps \\
\mrJ_{66}= 1 + 14\,\xps \;\;&\;\;
\mrJ_{67}= 1 + 14\,\xps - \xphs \;\;&\;\;
\mrJ_{68}= 1 + 22\,\xps - 2\,\xphs - 14\,\xphs\,\xps + \xphq \\
\mrJ_{69}= 2 - 56\,\xps - \xphs \;\;&\;\;
\mrJ_{70}= 2 - \xps \;\;&\;\;
\mrJ_{71}= 2 + 3\,\xps \\
\mrJ_{72}= 2 + 5\,\xps \;\;&\;\;
\mrJ_{73}= 3 + 14\,\xps \;\;&\;\;
\mrJ_{74}= 8 + 24\,\xps - \xphs \\
\mrJ_{75}= 20 - \xphs \;\;&\;\;
\mrJ_{76}= 1 + \xps & \\
\end{array}
\]

\normalsize

\vspace{0.8cm}
\bei
\item[\fbox{$\mrK\,$}]
\eei

\scriptsize
\[
\begin{array}{lll}
\mrK_{0}= \xps + 2\,\xpus \;\;&\;\;
\mrK_{1}= \xps + 2\,\xpds \;\;&\;\;
\mrK_{2}= \xps + 2\,\xpls \\
\mrK_{3}= \xps - 6\,\xpls \;\;&\;\;
\mrK_{4}= \xps + \xpls \;\;&\;\;
\mrK_{5}= 2\,\xps - \xpls \\
\mrK_{6}= 2 - \xps \;\;&\;\;
\mrK_{7}= 2 + \xps & \\
\end{array}
\]

\normalsize

\vspace{0.8cm}
\bei
\item[\fbox{$\mrL\,$}]
\eei

\scriptsize
\[
\begin{array}{lll}
\mrL_{0}=  - \xpLs + \xps \;\;&\;\;
\mrL_{1}=  - 2\,\xpLs + \xps \;\;&\;\;
\mrL_{2}= \xpLs + \xps \\
\mrL_{3}=  - \xpLs + 2\,\xps \;\;&\;\;
\mrL_{4}= \xpLs + 2\,\xps \;\;&\;\;
\mrL_{5}= xL^4 - 2\,\xps\,\xpLs + \xpss \\
\mrL_{6}=  - \xpUs + \xps \;\;&\;\;
\mrL_{7}=  - 2\,\xpUs + \xps \;\;&\;\;
\mrL_{8}= \xpUs + \xps \\
\mrL_{9}=  - \xpUs + 2\,\xps \;\;&\;\;
\mrL_{10}= \xpUq - 2\,\xps\,\xpUs + \xpss \;\;&\;\;
\mrL_{11}=  - \xpDs + \xps \\
\mrL_{12}=  - 2\,\xpDs + \xps \;\;&\;\;
\mrL_{13}= \xpDs + \xps \;\;&\;\;
\mrL_{14}=  - \xpDs + 2\,\xps \\
\mrL_{15}= \xpDq - 2\,\xps\,\xpDs + \xpss & & \\
\end{array}
\]

\normalsize

\subsection{Renormalized self-energies}
The (renormalized) bosonic self-energies are decomposed according to
\bq
D_{ij}(s) = \frac{g^2}{16\,\pi^2}\,\Bigl[
\Delta^{(4)}_{ij}(s)\,\mws + \Pi^{(4)}_{ij}(s)\,s + \gds\,\lpar
\Delta^{(6)}_{ij}(s)\,\mws + \Pi^{(6)}_{ij}(s)\,s \rpar\Bigr]
\eq
while the (renormalized) fermionic self-energies are decomposed as
\bq
S_{\Pf} = \frac{g^2}{16\,\pi^2}\,\Bigl[
V^{(4)}_{\Pf\Pf}\,i\,\sla{p} + A^{(4)}_{\Pf\Pf}\,i\,\sla{p}\,\gamma^5 + \Sigma^{(4)}_{\Pf\Pf}\,\mw + 
\gds\,\lpar
V^{(6)}_{\Pf\Pf}\,i\,\sla{p} + A^{(6)}_{\Pf\Pf}\,i\,\sla{p}\,\gamma^5 + \Sigma^{(6)}_{\Pf\Pf}\,\mw \rpar
\Bigr]
\eq
In the following list we introduce a shorthand notation:
\bq
\sbfun{m_1}{m_2} = \bfun{-s}{m_1}{m_2}
\eq
and several linear combinations of Wilson coefficients
\bq
\begin{array}{ll}
\apW= \ctws\,\aZZ + \stws\,\aAA + \ctw\stw\,\aAZ \;\;&\;\;
\apB= \ctws\,\aAA + \sths\,\aZZ - \ctw\stw\,\aAZ \\
\apWB= \lpar 1 - 2\,\stws \rpar\,\aAZ + 2\,\ctw\stw\,\lpar \aAA - \aZZ \rpar  \;\;&\;\;
\apWB= \ctw\,\apWA - \stw\,\apWZ \\
\apW=  \stw\,\apWA + \ctw\,\apWZ \;\;&\;\;
\apWBa = \apB - \apW \\
\apd= \frac{1}{2}\,\lpar \apdA - \apdV \rpar \;\;&\;\;
\apu= \frac{1}{2}\,\lpar \apuV - \apuA \rpar \\
\apqo= \frac{1}{4}\,\lpar \apuV + \apuA - \apdV - \apdA \rpar \;\;&\;\;
\apqt= \frac{1}{4}\,\lpar \apdV + \apdA + \apuV + \apuA \rpar \\
\apuVA = 2\,\lpar \apqt + \apqo \rpar \;\;&\;\;
\apdVA = 2\,\lpar \apqt - \apqo \rpar \nl
\end{array}
\eq

\bq
\begin{array}{ll}
\apl= \frac{1}{2}\,\lpar \aplA - \aplV \rpar \;\;&\;\;
\aplo= \aplt - \frac{1}{2}\,\lpar \aplV + \aplA \rpar \\
\aplt= \frac{1}{4}\,\lpar \aplV + \aplA + \apn \rpar \;\;&\;\;
\aplVA = 2\,\lpar \aplt - \aplo \rpar \\
\alW = \stw\,\alWB + \ctw\,\alBW \;\;&\;\;
\alB = - \ctw\,\alWB + \stw\,\alBW \\
\auW = \stw\,\auWB + \ctw\,\auBW \;\;&\;\;
\auB = \ctw\,\auWB - \stw\,\auBW \\
\adW = \stw\,\adWB + \ctw\,\adBW \;\;&\;\;
\adB = - \ctw\,\adWB + \stw\,\adBW \\
\aplWt= 4\,\aplt + 2\,\apW \;\;&\;\;
\apqWt= 4\,\apqt + 2\,\apW \\
\apWDp + \apWDm = 8\,\apW \;\;&\;\;
\apWDp - \apWDm = 2\,\apD \\
\apWAB= \apWA - 2\,\stw\,\apB \;\;&\;\;
\apWAD= 4\,\stw\,\apWA - \ctws\,\apD \\
\alpBox= \alp - \apBox \;\;&\;\;
\aupBox= \aup + \apBox \\
\adpBox= \adp - \apBox & \\
\end{array}
\label{moreWC}
\eq

All functions in the following list are decomposed according to
\bq
F = \sum_{n=0}^{2}\,F_n 
\eq
where $F_0$ is the constant part (containing a dependence on $\muR$), $F_1$ contains
(finite) one-point functions and $F_2$ the (finite) two-point functions. Capital letters 
($\PQU$ \etc) denote a specific fermion, small letters ($\PQu$ \etc) are used when summing 
over fermions.

\footnotesize
\bei
\item \fbox{$\PH$ self-energy}
\eei
\bqas
\Pi^{(4)}_{\PH\PH\,;\,0} &=&
         \frac{1}{2}
         \,\mrF^{b}_{0}
         \,\frac{\LR}{c^2}
       -
         \frac{1}{2}
         \,\sumg 
         \,\mrH_{0}
         \,\LR
\eqas
\bqas
\Pi^{(4)}_{\PH\PH\,;\,1} &=& 0
\eqas
\bqas
\Pi^{(4)}_{\PH\PH\,;\,2} &=&
         \frac{1}{2}
         \,\frac{1}{c^2}
         \,\sbfun{\mzb}{\mzb}
       +
         \sbfun{M}{M}
       -
         \frac{1}{2}
         \,\sumg 
         \,\xpls
         \,\sbfun{\mle}{\mle}
\nl &-&
         \frac{3}{2}
         \,\sumg 
         \,\xpus
         \,\sbfun{\mqu}{\mqu}
       -
         \frac{3}{2}
         \,\sumg 
         \,\xpds
         \,\sbfun{\mqd}{\mqd}
\eqas
\bqas
\Delta^{(4)}_{\PH\PH\,;\,0} &=&
         2
       +
         \frac{1}{2}
         \,\frac{1}{c^4}
         \,( 2 - 3\,\LR )
       -
         \frac{3}{2}
         \,\mrJ_{0}
         \,\LR
       +
         2
         \,\sumg 
         \,\mrH_{1}
         \,\LR
\eqas
\bqas
\Delta^{(4)}_{\PH\PH\,;\,1} &=& 0
\eqas
\bqas
\Delta^{(4)}_{\PH\PH\,;\,2} &=& 
       -
         \frac{1}{4}
         \,\mrJ_{1}
         \,\sbfun{M}{M}
       -
         \frac{9}{8}
         \,\xphq
         \,\sbfun{\mh}{\mh}
       +
         2
         \,\sumg 
         \,\xplq
         \,\sbfun{\mle}{\mle}
\nl &+&
         6
         \,\sumg 
         \,\xpuq
         \,\sbfun{\mqu}{\mqu}
     +
         6
         \,\sumg 
         \,\xpdq
         \,\sbfun{\mqd}{\mqd}
     -
         \frac{1}{8}
         \,( 12 + c^4\,\xphq )
         \,\frac{1}{c^4}
         \,\sbfun{\mzb}{\mzb}
\eqas
\bqas
\Pi^{(6)}_{\PH\PH\,;\,0} &=&
       -
         4
         \,\apW
         \,( 1 - 2\,\LR )
       -
         \frac{1}{c^2}
         \,\aZZ
         \,( 2 - 3\,\LR )
\nl &+&
         \frac{1}{8}
         \,\Bigl[ ( 8\,\apW + \apD + 8\,\apBox ) 
         + ( 4\,\mrJ_{3}\,\apBox - \mrJ_{4}\,\apD )\,c^2 \Bigr]
         \,\frac{\LR}{c^2}
\nl &-&
         \frac{1}{4}
         \,\sumg \Bigl[ ( \apWDm + 4\,\apBox )\,\mrH_{0} + 4\,( 
         - 3\,\xpds\,\adp + 3\,\xpus\,\aup - \xpls\,\alp ) \Bigr]
         \,\LR
\eqas
\bqas
\Pi^{(6)}_{\PH\PH\,;\,1} &=&
       -
         \frac{1}{8}
         \,\frac{1}{c^2}
         \,\apD
         \,\afun{\mzb}
       -
         \frac{1}{8}
         \,( \apD - 4\,\apBox )
         \,\xphs
         \,\afun{\mh}
\eqas
\bqas
\Pi^{(6)}_{\PH\PH\,;\,2} &=&
         \frac{1}{4}
         \,\Bigl[ 32\,\apW - ( \apD - 4\,\apBox )\,\mrJ_{2} \Bigr]
         \,\sbfun{M}{M}
\nl &-&
         \frac{1}{8}
         \,\Bigl[  - 8\,( 3\,\aZZ + \apW + \apBox ) + ( \apD + 4\,
      \apBox )\,c^2\,\xphs \Bigr]
         \,\frac{1}{c^2}
         \,\sbfun{\mzb}{\mzb}
\nl &+&
         \frac{3}{8}
         \,( \apD - 4\,\apBox )
         \,\xphs
         \,\sbfun{\mh}{\mh}
\nl &-&
         \frac{1}{4}
         \,\sumg ( \apWDm - 4\,\alpBox )
         \,\xpls
         \,\sbfun{\mle}{\mle}
       -
         \frac{3}{4}
         \,\sumg ( \apWDm - 4\,\adpBox )
         \,\xpds
         \,\sbfun{\mqd}{\mqd}
\nl &-&
         \frac{3}{4}
         \,\sumg ( \apWDm + 4\,\aupBox )
         \,\xpus
         \,\sbfun{\mqu}{\mqu}
\eqas
\bqas
\Delta^{(6)}_{\PH\PH\,;\,0} &=&
         \frac{1}{2}
         \,\Bigl[ ( 6\,\ap + \apD\,\xphs ) + ( 6\,\mrJ_{6}\,\ap
         + \mrJ_{7}\,\apD - 3\,\mrJ_{8}\,\apBox - 6\,\mrJ_{9}\,\apW )\,c^2 \Bigr]
         \,\frac{\LR}{c^2}
\nl &+&
         \frac{1}{4}
         \,( 16\,\aZZ + 4\,\apW + 3\,\apD + 4\,\apBox )
         \,\frac{1}{c^4}
         \,( 2 - 3\,\LR )
       +
         ( 20\,\apW - \apD + 4\,\apBox )
\nl &+&
         \sumg \Bigl[ ( \apWDm + 4\,\apBox )\,\mrH_{1} 
         + 8\,(  - 3\,\xpdq\,\adp + 3\,\xpuq\,\aup - \xplq\,\alp ) \Bigr]
         \,\LR
\eqas
\bqas
\Delta^{(6)}_{\PH\PH\,;\,1} &=&
         4
         \,\sumg 
         \,\xplq\,\alp
         \,\afun{\mle}
     -
         12
         \,\sumg 
         \,\xpuq\,\aup
         \,\afun{\mqu}
     +
         12
         \,\sumg 
         \,\xpdq\,\adp
         \,\afun{\mqd}
\nl &-&
         \frac{3}{4}
         \,\Bigl[ 20\,\ap + ( \apD - 4\,\apBox )\,\xphs \Bigr]
         \,\xphs
         \,\afun{\mh}
\nl &-&
         \frac{1}{4}
         \,\Bigl[ c^2\,\apD\,\xphs - 6\,( 4\,\aZZ + \apD - 2\,c^2\,\ap ) \Bigr]
         \,\frac{1}{c^4}
         \,\afun{\mzb}
\nl &-&
         6
         \,( \ap - 2\,\apW )
         \,\afun{M}
\eqas
\bqas
\Delta^{(6)}_{\PH\PH\,;\,2} &=&
         \frac{3}{16}
         \,\Bigl[ 96\,\ap + (  - 12\,\apW + 7\,\apD - 28\,\apBox )\,\xphs \Bigr]
         \,\xphs
         \,\sbfun{\mh}{\mh}
\nl &+&
         \frac{1}{16}
         \,\Bigl[ 4\,c^2\,\apD\,\xphs - 12\,( \apWDp + 8\,\aZZ + 4\,\apBox )
       - ( \apWDm - 4\,\apBox )\,c^4\,\xphq \Bigr]
         \,\frac{1}{c^4}
         \,\sbfun{\mzb}{\mzb}
\nl &-&
         \frac{1}{8}
         \,\Bigl[ 4\,\mrJ_{10}\,\apW - ( \apD - 4\,\apBox )\,\mrJ_{5} \Bigr]
         \,\sbfun{M}{M}
\nl &+&
         \sumg ( \apWDm - 4\,\alpBox )
         \,\xplq
         \,\sbfun{\mle}{\mle}
\nl &+&
         3
         \,\sumg ( \apWDm - 4\,\adpBox )
         \,\xpdq
         \,\sbfun{\mqd}{\mqd}
\nl &+&
         3
         \,\sumg ( \apWDm + 4\,\aupBox )
         \,\xpuq
         \,\sbfun{\mqu}{\mqu}
\eqas
\vspace{0.5cm}
\bei
\item \fbox{$\PA$ self-energy}
\eei
\bqas
\Pi^{(4)}_{\PA\PA\,;\,0} &=&
         3
         \,s^2\,\LR
       +
         \frac{32}{27}
         \,s^2\,\myNG
         \,( 1 - 3\,\LR )
       +
         4
         \,\frac{s^2}{\xps}
       -
         \frac{8}{9}
         \,\sumg 
         \,\frac{s^2}{\xps}
         \,\mrH_{2}
\eqas
\bqas
\Pi^{(4)}_{\PA\PA\,;\,1} &=&
         4
         \,\frac{s^2}{\xps}
         \,\afun{M}
     -
         \frac{8}{3}
         \,\sumg 
         \,\frac{\xpls}{\xps}
         \,s^2
         \,\afun{\mle}
\nl &-&
         \frac{32}{9}
         \,\sumg 
         \,\frac{\xpus}{\xps}
         \,s^2
         \,\afun{\mqu}
     -
         \frac{8}{9}
         \,\sumg 
         \,\frac{\xpds}{\xps}
         \,s^2
         \,\afun{\mqd}
\eqas
\bqas
\Pi^{(4)}_{\PA\PA\,;\,2} &=&
         \frac{s^2}{\xps}
         \,\mrJ_{11}
         \,\sbfun{M}{M}
     -
         \frac{4}{3}
         \,\sumg 
         \,\frac{s^2}{\xps}
         \,\mrK_{2}
         \,\sbfun{\mle}{\mle}
\nl &-&
         \frac{16}{9}
         \,\sumg 
         \,\frac{s^2}{\xps}
         \,\mrK_{0}
         \,\sbfun{\mqu}{\mqu}
     -
         \frac{4}{9}
         \,\sumg 
         \,\frac{s^2}{\xps}
         \,\mrK_{1}
         \,\sbfun{\mqd}{\mqd}
\eqas
\bqas
\Pi^{(6)}_{\PA\PA\,;\,0} &=&
         \frac{16}{27}
         \,\myNG\,\apWAD
         \,( 1 - 3\,\LR )
     +
         2
         \,\frac{1}{\xps}
         \,\apWAD
     -
         \frac{4}{9}
         \,\sumg 
         \,\frac{1}{\xps}
         \,\mrH_{2}
         \,\apWAD
\nl &-&
         \frac{1}{2}
         \,\Bigl[ \mrJ_{13}\,c^2\,\apB - \mrJ_{15}\,s\,c^3\,\apWB
         + ( 3\,c^4\,\apD + s\,\apWA ) - ( \mrJ_{12}\,\apB + \mrJ_{18}\,\apW )\,s^2\,c^2 \Bigr]
         \,\frac{\LR}{c^2}
\nl &+&
         2
         \,\sumg (  - \xpds\,\adWB + 2\,\xpus\,\auWB - \xpls\,\alWB )
         \,s\,\LR
\eqas
\bqas
\Pi^{(6)}_{\PA\PA\,;\,1} &=&
         \frac{1}{2}
         \,\frac{1}{c^2}
         \,\aAA
         \,\afun{\mzb}
     +
         \frac{1}{2}
         \,\xphs\,\aAA
         \,\afun{\mh}
     -
         \frac{4}{3}
         \,\sumg 
         \,\frac{\xpls}{\xps}
         \,\apWAD
         \,\afun{\mle}
\nl &-&
         \frac{16}{9}
         \,\sumg 
         \,\frac{\xpus}{\xps}
         \,\apWAD
         \,\afun{\mqu}
     -
         \frac{4}{9}
         \,\sumg 
         \,\frac{\xpds}{\xps}
         \,\apWAD
         \,\afun{\mqd}
\nl &+&
         \Bigl[ \mrJ_{16}\,s\,c\,\apWB + \mrJ_{17}\,s^2\,\apW 
         + ( \xps\,\apB - 2\,\apD )\,c^2 \Bigr]
         \,\frac{1}{\xps}
         \,\afun{M}
\eqas
\bqas
\Pi^{(6)}_{\PA\PA\,;\,2} &=&
         \frac{1}{2}
         \,\Bigl[ 4\,\mrJ_{14}\,s\,c\,\apWB + (  - c^2\,\apD + 4\,s^2\,\apW )\,\mrJ_{11} \Bigr]
         \,\frac{1}{\xps}
         \,\sbfun{M}{M}
\nl &-&
         \frac{2}{9}
         \,\sumg ( 9\,s\,\xpds\,\xps\,\adWB + \mrK_{1}\,\apWAD )
         \,\frac{1}{\xps}
         \,\sbfun{\mqd}{\mqd}
\nl &+&
         \frac{4}{9}
         \,\sumg ( 9\,s\,\xpus\,\xps\,\auWB - 2\,\mrK_{0}\,\apWAD )
         \,\frac{1}{\xps}
         \,\sbfun{\mqu}{\mqu}
\nl &-&
         \frac{2}{3}
         \,\sumg ( 3\,s\,\xpls\,\xps\,\alWB + \mrK_{2}\,\apWAD )
         \,\frac{1}{\xps}
         \,\sbfun{\mle}{\mle}
\eqas
\bei
\item \fbox{$\PZ{-}\PA$ transition}
\eei
\bqas
\Pi^{(4)}_{\PZ\PA\,;\,0} &=&
       -
         \frac{1}{9}
         \,\frac{s}{c}
         \,\myNG\,\vtg
         \,( 1 - 3\,\LR )
       -
         \frac{1}{6}
         \,\frac{s}{c}
         \,\mrF^{b}_{2}
         \,\LR
\nl &+&
         \frac{2}{3}
         \,\sumg 
         \,( \xpl^2\,\vle + \vqd\,\xpd^2 + 2\,\vqu\,\xpu^2 )
         \,\frac{s}{c\,\xps}
     -
         \frac{1}{9}
         \,( 36\,c^2 + \mrJ_{19} )
         \,\frac{s}{c\,\xps}
\eqas
\bqas
\Pi^{(4)}_{\PZ\PA\,;\,1} &=&
     -
         \frac{4}{3}
         \,\frac{s}{c\,\xps}
         \,\mrF^{b}_{1}
         \,\afun{M}
     +
         \frac{2}{3}
         \,\sumg 
         \,\frac{s\,\xpds}{c\,\xps}
         \,\vqd
         \,\afun{\mqd}
\nl &+&
         \frac{4}{3}
         \,\sumg 
         \,\frac{s\,\xpus}{c\,\xps}
         \,\vqu
         \,\afun{\mqu}
     +
         \frac{2}{3}
         \,\sumg 
         \,\frac{s\,\xpls}{c\,\xps}
         \,\vle
         \,\afun{\mle}
\eqas
\bqas
\Pi^{(4)}_{\PZ\PA\,;\,2} &=&
         \frac{1}{3}
         \,\sumg 
         \,\frac{s}{c\,\xps}
         \,\mrK_{2}
         \,\vle
         \,\sbfun{\mle}{\mle}
     +
         \frac{2}{3}
         \,\sumg 
         \,\frac{s}{c\,\xps}
         \,\mrK_{0}
         \,\vqu
         \,\sbfun{\mqu}{\mqu}
\nl &+&
         \frac{1}{3}
         \,\sumg 
         \,\frac{s}{c\,\xps}
         \,\mrK_{1}
         \,\vqd
         \,\sbfun{\mqd}{\mqd}
     -
         \frac{1}{6}
         \,( 6\,\mrJ_{11}\,c^2 + \mrJ_{17} )
         \,\frac{s}{c\,\xps}
         \,\sbfun{M}{M}
\eqas
\bqas
\Pi^{(6)}_{\PZ\PA\,;\,0} &=&
         \frac{1}{36}
         \,\Bigl[ 144\,s^3\,c^2\,\apWAB - 4\,\mrJ_{26}\,s\,c\,\apWB 
         - 144\,\mrF^{b}_{3}\,s^2\,c\,\apWZ + 36\,\mrF^{b}_{5}\,c^2\,\apD 
\nl &+& ( \apD - 2\,s^2\,\apWDp )\,\mrJ_{19} \Bigr]\,\frac{1}{s\,c\,\xps}
\nl &-&
         \frac{1}{24}
         \,\Bigl[ \mrF^{b}_{8}\,c\,\apD + 4\,( 3\,\apB - \apW + 36\,s^2\,c^2\,\apWBa )\,s^2\,c 
         + 12\,( \mrJ_{12}\,\apB + \mrJ_{18}\,\apW )\,s^2\,c^3 
\nl &+& 2\,( \mrJ_{20}\,c^2 - 6\,\mrJ_{23}\,s^2\,c^2 + 3\,\mrF^{b}_{6} )\,s\,\apWB \Bigr]
         \,\frac{\LR}{s\,c^2}
\nl &+&
         \frac{1}{108}
         \,\Bigl\{ \Bigl[ 3\,c\,\apD\,\vtg - 4\,( 3\,\apWB\,\vtg 
         - 32\,(  - c\,\apWZ + s\,\apWAB )\,s\,c )\,s \Bigr]\,c 
\nl &+& 4\,( \mrG_{3}\,\apW - \mrF^{b}_{10}\,\apD )\,s^2 \Bigr\}
         \,\frac{\myNG}{s\,c}
         \,( 1 - 3\,\LR )
\nl &+&
         \frac{1}{18}
         \,\sumg \Bigl[ 4\,( \apD + 4\,c\,\apWZ - 4\,s\,\apWAB )\,\mrH_{2}\,s^2\,c^2 
         + (  - 24\,\xpds\,\apd + 48\,\xpus\,\apu - 4\,\mrG_{0}\,\xpds\,\apW 
\nl &-& 12\,\mrG_{1}\,\xpls\,\apW - 8\,\mrG_{2}\,\xpus\,\apW 
         - 24\,\mrH_{3}\,\apqo + 24\,\mrH_{4}\,\apqt
         + \mrH_{5}\,\apD - 4\,\mrK_{3}\,\aplV )\,s^2 
\nl &+& 3\,( - c\,\apD + 4\,s\,\apWB )\,( \xpl^2\,\vle + \vqd\,\xpd^2 + 2\,\vqu\,\xpu^2 )\,c \Bigr]
         \,\frac{1}{s\,c\,\xps}
\nl &+&
         \frac{1}{12}
         \,\sumg \Bigl\{ 8\,s\,\aplV - 3\,\Bigl[  - ( \alWB\,\vle + 4\,s\,c\,\alBW )\,\xpls 
         - ( 3\,\vqd\,\adWB + 4\,s\,c\,\adBW )\,\xpds 
\nl &+& ( 3\,\vqu\,\auWB + 8\,s\,c\,\auBW )\,\xpus \Bigr] \Bigr\}
         \,\frac{\LR}{c}
\nl &-&
         \frac{2}{9}
         \,\sumg ( \apdV + 2\,\apuV )
         \,( 1 - 3\,\LR )
         \,\frac{s}{c}
\eqas
\bqas
\Pi^{(6)}_{\PZ\PA\,;\,1} &=&
       -
         \frac{1}{4}
         \,\frac{1}{c^2}
         \,\aAZ
         \,\afun{\mzb}
       -
         \frac{1}{4}
         \,\xphs\,\aAZ
         \,\afun{\mh}
\nl &-&
         \frac{1}{6}
         \,\Bigl\{ 2\,\mrF^{b}_{7}\,\apD + 16\,\Bigl[ \apW + 3\,( c\,\apWBa
         + s\,\apWB )\,s^2\,c \Bigr]\,s^2 - 6\,( \xps\,\apB - \mrJ_{17}\,\apW )\,s^2\,c^2 
\nl &-& ( 6\,\mrJ_{21}\,s^2 - \mrJ_{24} )\,s\,c\,\apWB \Bigr\}
         \,\frac{1}{s\,c\,\xps}
         \,\afun{M}
\nl &-&
         \frac{1}{6}
         \,\sumg \Bigl\{ \Bigl[ c^2\,\apD\,\vle - 4\,( c\,\apWB\,\vle 
         + 2\,( \aplV + 2\,c^3\,\apWZ - 2\,s\,c^2\,\apWAB )\,s )\,s \Bigr] 
\nl &+& ( 4\,\mrG_{1}\,\apW - \mrF^{b}_{9}\,\apD )\,s^2 \Bigr\}
         \,\frac{\xpls}{s\,c\,\xps}
         \,\afun{\mle}
\nl &-&
         \frac{1}{18}
         \,\sumg \Bigl\{ \Bigl[ 3\,c^2\,\vqd\,\apD - 4\,( 3\,c\,\vqd\,\apWB 
         + 2\,( 3\,\apdV + 2\,c^3\,\apWZ - 2\,s\,c^2\,\apWAB )\,s )\,s \Bigr] 
\nl &+& ( 4\,\mrG_{0}\,\apW - \mrF^{b}_{4}\,\apD )\,s^2 \Bigr\}
         \,\frac{\xpds}{s\,c\,\xps}
         \,\afun{\mqd}
\nl &-&
         \frac{1}{9}
         \,\sumg \Bigl\{ \Bigl[ 3\,c^2\,\vqu\,\apD - 4\,( 3\,c\,\vqu\,\apWB 
         + 2\,( 3\,\apuV + 4\,c^3\,\apWZ - 4\,s\,c^2\,\apWAB )\,s )\,s \Bigr] 
\nl &+& ( 4\,\mrG_{2}\,\apW - \mrF^{b}_{10}\,\apD )\,s^2 \Bigr\}
         \,\frac{\xpus}{s\,c\,\xps}
         \,\afun{\mqu}
\eqas
\bqas
\Pi^{(6)}_{\PZ\PA\,;\,2} &=&
         \frac{1}{24}
         \,\Bigl\{  - 48\,\Bigl[ c^3\,\apW + ( c\,\apB + s\,\apWB )\,s^2 \Bigr]\,\mrJ_{11}\,s^2\,c 
         + ( \apD - 2\,s^2\,\apWDp )\,\mrJ_{17} + 6\,( \mrJ_{11}\,\mrF^{b}_{5} )\,c^2\,\apD 
\nl &+& 4\,( 6\,\mrJ_{22}\,s^2 - \mrJ_{25} )\,s\,c\,\apWB \Bigr\}
         \,\frac{1}{s\,c\,\xps}
         \,\sbfun{M}{M}
\nl &+&
         \frac{1}{12}
         \,\sumg \Bigl\{  - \Bigl[ c^2\,\apD\,\vle - 4\,( c\,\apWB\,\vle 
         + 2\,( \aplV + 2\,c^3\,\apWZ - 2\,s\,c^2\,\apWAB )\,s )\,s \Bigr]\,\mrK_{2} 
\nl &+& 3\,( \alWB\,\vle + 4\,s\,c\,\alBW )\,s\,\xpls\,\xps - ( 4\,\mrG_{1}\,\mrK_{2}\,\apW 
         - \mrK_{2}\,\mrF^{b}_{9}\,\apD )\,s^2 \Bigr\}
         \,\frac{1}{s\,c\,\xps}
         \,\sbfun{\mle}{\mle}
\nl &+&
         \frac{1}{36}
         \,\sumg \Bigl\{  - \Bigl[ 3\,c^2\,\vqd\,\apD - 4\,( 3\,c\,\vqd\,\apWB 
         + 2\,( 3\,\apdV + 2\,c^3\,\apWZ - 2\,s\,c^2\,\apWAB )\,s )\,s \Bigr]\,\mrK_{1} 
\nl &+& 9\,( 3\,\vqd\,\adWB + 4\,s\,c\,\adBW )\,s\,\xpds\,\xps 
         - ( 4\,\mrG_{0}\,\mrK_{1}\,\apW - \mrK_{1}\,\mrF^{b}_{4}\,\apD )\,s^2 \Bigr\}
         \,\frac{1}{s\,c\,\xps}
         \,\sbfun{\mqd}{\mqd}
\nl &-&
         \frac{1}{36}
         \,\sumg \Bigl\{ 2\,\Bigl[ 3\,c^2\,\vqu\,\apD - 4\,( 3\,c\,\vqu\,\apWB 
         + 2\,( 3\,\apuV + 4\,c^3\,\apWZ - 4\,s\,c^2\,\apWAB )\,s )\,s \Bigr]\,\mrK_{0} 
\nl &+& 9\,( 3\,\vqu\,\auWB + 8\,s\,c\,\auBW )\,s\,\xpus\,\xps 
         + 2\,( 4\,\mrG_{2}\,\mrK_{0}\,\apW - \mrK_{0}\,\mrF^{b}_{10}\,\apD )\,s^2 \Bigr\}
         \,\frac{1}{s\,c\,\xps}
         \,\sbfun{\mqu}{\mqu}
\eqas
\vspace{0.5cm}
\bei
\item \fbox{$\PZ$ self-energy}
\eei
\bqas
\Pi^{(4)}_{\PZ\PZ\,;\,0} &=&
       -
         3
         \,s^2\,\LR
       +
         \frac{1}{36}
         \,\frac{\myNG}{c^2}
         \,\mrG_{7}
         \,( 1 - 3\,\LR )
       +
         \frac{2}{9}
         \,( 1 + 15\,\LR )
       -
         \frac{1}{18}
         \,\frac{1}{c^2}
         \,( 2 + 3\,\LR )
\eqas
\bqas
\Pi^{(4)}_{\PZ\PZ\,;\,1} &=& 0
\eqas
\bqas
\Pi^{(4)}_{\PZ\PZ\,;\,2} &=& 
       -
         \frac{1}{12}
         \,\frac{1}{c^2}
         \,\mrF^{b}_{11}
         \,\sbfun{M}{M}
       -
         \frac{1}{12}
         \,\frac{1}{c^2}
         \,\sbfun{\mh}{\mzb}
       -
         \frac{1}{12}
         \,\sumg 
         \,\frac{1}{c^2}
         \,\mrG_{6}
         \,\sbfun{\mle}{\mle}
\nl &-&
         \frac{1}{4}
         \,\sumg 
         \,\frac{1}{c^2}
         \,\mrG_{4}
         \,\sbfun{\mqu}{\mqu}
     -
         \frac{1}{4}
         \,\sumg 
         \,\frac{1}{c^2}
         \,\mrG_{5}
         \,\sbfun{\mqd}{\mqd}
     -
         \frac{1}{6}
         \,\sumg 
         \,\frac{1}{c^2}
         \,\sbfun{0}{0}
\eqas
\bqas
\Delta^{(4)}_{\PZ\PZ\,;\,0} &=& 
         \frac{1}{6}
         \,\frac{1}{c^4}
         \,( 1 - 6\,\LR )
       -
         2
         \,\frac{\LR}{c^2}
       +
         \frac{1}{2}
         \,\sumg 
         \,\mrH_{0}
         \,\frac{\LR}{c^2}
\nl &+&
         \frac{1}{6}
         \,( \mrJ_{12} + 8\,\mrF^{b}_{14}\,c^2 )
         \,\frac{1}{c^2}
     -
         \frac{1}{6}
         \,\sumg ( 3\,\mrG_{4}\,\xpus + 3\,\mrG_{5}\,\xpds + \mrG_{6}\,\xpls )
         \,\frac{1}{c^2}
\eqas
\bqas
\Delta^{(4)}_{\PZ\PZ\,;\,1} &=& 
         \frac{1}{3}
         \,\frac{1}{c^2}
         \,\mrF^{b}_{12}
         \,\afun{M}
     -
         \frac{1}{6}
         \,\sumg 
         \,\frac{\xpls}{c^2}
         \,\mrG_{6}
         \,\afun{\mle}
     -
         \frac{1}{2}
         \,\sumg 
         \,\frac{\xpus}{c^2}
         \,\mrG_{4}
         \,\afun{\mqu}
\nl &-&
         \frac{1}{2}
         \,\sumg 
         \,\frac{\xpds}{c^2}
         \,\mrG_{5}
         \,\afun{\mqd}
     +
         \frac{1}{12}
         \,( 1 + \mrJ_{27}\,c^2 )
         \,\frac{\xphs}{c^4\,\xps}
         \,\afun{\mh}
     -
         \frac{1}{12}
         \,( 1 - \mrJ_{28}\,c^2 )
         \,\frac{1}{c^6\,\xps}
         \,\afun{\mzb}
\eqas
\bqas
\Delta^{(4)}_{\PZ\PZ\,;\,2} &=& 
       -
         \frac{1}{3}
         \,\frac{1}{c^2}
         \,\mrF^{b}_{13}
         \,\sbfun{M}{M}
       +
         \frac{1}{6}
         \,\sumg 
         \,\frac{\xpls}{c^2}
         \,\mrG_{10}
         \,\sbfun{\mle}{\mle}
       +
         \frac{1}{2}
         \,\sumg 
         \,\frac{\xpus}{c^2}
         \,\mrG_{8}
         \,\sbfun{\mqu}{\mqu}
\nl &+&
         \frac{1}{2}
         \,\sumg 
         \,\frac{\xpds}{c^2}
         \,\mrG_{9}
         \,\sbfun{\mqd}{\mqd}
     -
         \frac{1}{12}
         \,( 1 - \mrJ_{27}\,c^4\,\xphs + 2\,\mrJ_{29}\,c^2 )
         \,\frac{1}{c^6\,\xps}
         \,\sbfun{\mh}{\mzb}
\eqas
\bqas
\Pi^{(6)}_{\PZ\PZ\,;\,0} &=& 
         \frac{1}{9}
         \,\apD
         \,( 1 + 15\,\LR )
     -
         \frac{1}{18}
         \,\frac{1}{c^2}
         \,\apBox
         \,( 2 + 3\,\LR )
\nl &-&
         \frac{1}{18}
         \,\Bigl[ \apD - 4\,c^2\,\aZZ + 4\,( c\,\aAZ + s\,\aAA )\,s \Bigr]
         \,\frac{1}{c^2}
\nl &+&
         \frac{1}{72}
         \,\Bigl[ \mrG_{22}\,\apWDm + 4\,\vtg\,( \apD + 4\,c\,\apWZ - 4\,s\,\apWAB )\,c^2 \Bigr]
         \,\frac{\myNG}{c^2}
         \,( 1 - 3\,\LR )
\nl &+&
         \frac{1}{12}
         \,\Bigl\{ 2\,\mrJ_{30}\,c^2\,\apW - \Bigl[ \apWDp 
         + 6\,( 5\,\apB + 3\,c^2\,\apD + 24\,s^2\,c^2\,\apWBa )\,s^2 \Bigr] 
         - 2\,( \mrJ_{31}\,c^2 - 9\,\mrF^{b}_{16} )\,s\,c\,\apWB 
\nl &+& 2\,( \mrJ_{32}\,\apB - \mrJ_{33}\,\apW )\,s^2\,c^2 \Bigr\}
         \,\frac{\LR}{c^2}
\nl &-&
         \frac{1}{9}
         \,\sumg \Bigl[ 3\,\mrG_{12}\,\apu - 3\,\mrG_{14}\,\apd - 3\,\mrG_{18}\,\apqt 
         - \mrG_{21}\,\aplt - 3\,\mrG_{23}\,\apqo - ( \aplt - \aplV )\,\mrG_{11} \Bigr]
         \,\frac{1}{c^2}
         \,( 1 - 3\,\LR )
\nl &+&
         \frac{1}{2}
         \,\sumg (  - \xpls\,\vle\,\alBW - 3\,\vqd\,\xpds\,\adBW + 3\,\vqu\,\xpus\,\auBW )
         \,\frac{\LR}{c}
\eqas
\bqas
\Pi^{(6)}_{\PZ\PZ\,;\,1} &=& 
         \frac{1}{2}
         \,\frac{1}{c^2}
         \,\aZZ
         \,\afun{\mzb}
     +
         \frac{1}{2}
         \,\aZZ\,\xphs
         \,\afun{\mh}
     +
         \Bigl[ c^2\,\apW + ( c\,\apWB + s\,\apB )\,s \Bigr]
         \,\afun{M}
\eqas
\bqas
\Pi^{(6)}_{\PZ\PZ\,;\,2} &=& 
       -
         \frac{1}{24}
         \,( \apWDp + 48\,\aZZ + 4\,\apBox )
         \,\frac{1}{c^2}
         \,\sbfun{\mh}{\mzb}
\nl &-&
         \frac{1}{24}
         \,( \mrF^{b}_{11}\,\apD + 4\,\mrF^{b}_{15}\,s\,\apWA 
         + 4\,\mrF^{b}_{17}\,c\,\apWZ - 16\,\mrF^{b}_{18}\,s^2\,c^2\,\apB )
         \,\frac{1}{c^2}
         \,\sbfun{M}{M}
\nl &-&
         \frac{1}{24}
         \,\sumg \Bigl[ 12\,c\,\xpls\,\vle\,\alBW + 4\,( \apD + 4\,c\,\apWZ 
         - 4\,s\,\apWAB )\,c^2\,\vle + ( 8\,\mrG_{11}\,\apl + \mrG_{16}\,\apWDm 
\nl &+& 4\,\mrG_{17}\,\aplVA ) \Bigr]
         \,\frac{1}{c^2}
         \,\sbfun{\mle}{\mle}
\nl &-&
         \frac{1}{24}
         \,\sumg \Bigl[ 36\,c\,\vqd\,\xpds\,\adBW + 4\,( \apD + 4\,c\,\apWZ 
         - 4\,s\,\apWAB )\,c^2\,\vqd 
\nl &+& ( 24\,\mrG_{14}\,\apd + 12\,\mrG_{15}\,\apdVA + \mrG_{20}\,\apWDm ) \Bigr]
         \,\frac{1}{c^2}
         \,\sbfun{\mqd}{\mqd}
\nl &+&
         \frac{1}{24}
         \,\sumg \Bigl[ 36\,c\,\vqu\,\xpus\,\auBW 
         - 8\,( \apD + 4\,c\,\apWZ - 4\,s\,\apWAB )\,c^2\,\vqu 
\nl &+& ( 24\,\mrG_{12}\,\apu - 12\,\mrG_{13}\,\apuVA - \mrG_{19}\,\apWDm ) \Bigr]
         \,\frac{1}{c^2}
         \,\sbfun{\mqu}{\mqu}
\nl &-&
         \frac{1}{12}
         \,\sumg ( 4\,\apn + \apWDm )
         \,\frac{1}{c^2}
         \,\sbfun{0}{0}
\eqas
\bqas
\Delta^{(6)}_{\PZ\PZ\,;\,0} &=& 
         \frac{1}{6}
         \,\frac{1}{c^4}
         \,\apBox
         \,( 2 - 9\,\LR )
     +
         \frac{1}{24}
         \,\frac{1}{c^4}
         \,( 82 - 9\,\LR )
         \,\apD
\nl &+&
         \frac{1}{12}
         \,\Bigl[ 4\,\mrF^{b}_{20}\,s\,\apWA + 4\,\mrF^{b}_{22}\,c\,\apWZ 
         + 64\,\mrF^{b}_{23}\,s^2\,c^4\,\apB - 8\,\mrF^{b}_{26}\,s^2\,\apD 
         + ( 4\,\apBox\,\xphs + \mrJ_{12}\,\apWDp )\,c^2 \Bigr]
         \,\frac{1}{c^4}
\nl &-&
         \frac{1}{8}
         \,( \mrJ_{36}\,c^2\,\apD + 16\,\mrF^{b}_{0}\,\apW )
         \,\frac{\LR}{c^4}
\nl &-&
         \frac{1}{12}
         \,\sumg \Bigl[ 4\,( \apD + 4\,c\,\apWZ - 4\,s\,\apWAB )\,( \xpl^2\,\vle + \vqd\,\xpd^2 
         + 2\,\vqu\,\xpu^2 )\,c^2 
\nl &+& ( 8\,\mrG_{11}\,\xpls\,\apl - 24\,\mrG_{12}\,\xpus\,\apu + 12\,\mrG_{13}\,\xpus\,\apuVA 
         + 24\,\mrG_{14}\,\xpds\,\apd + 12\,\mrG_{15}\,\xpds\,\apdVA 
\nl &+& \mrG_{16}\,\xpls\,\apWDm + 4\,\mrG_{17}\,\xpls\,\aplVA 
         + \mrG_{19}\,\xpus\,\apWDm + \mrG_{20}\,\xpds\,\apWDm ) \Bigr]
         \,\frac{1}{c^2}
\nl &-&
         \frac{1}{4}
         \,\sumg (  - 24\,\xpds\,\apd + 24\,\xpus\,\apu - 8\,\xpls\,\aplA 
         - \mrH_{0}\,\apWDm + 24\,\mrH_{6}\,\apqo - 24\,\mrH_{7}\,\apqt )
         \,\frac{\LR}{c^2}
\eqas
\bqas
\Delta^{(6)}_{\PZ\PZ\,;\,1} &=& 
       -
         \frac{1}{24}
         \,\Bigl\{ \Bigl[ 48\,\aZZ\,\xps - 4\,\mrJ_{28}\,\apW 
         + ( \apD + 4\,\apBox )\,\mrJ_{34} \Bigr]\,c^2 + ( \apWDp + 4\,\apBox) \Bigr\}
         \,\frac{1}{c^6\,\xps}
         \,\afun{\mzb}
\nl &+&
         \frac{1}{24}
         \,\Bigl\{ \Bigl[ 48\,\aZZ\,\xps + \mrJ_{37}\,\apD 
         + 4\,( \apW + \apBox )\,\mrJ_{27} \Bigr]\,c^2 + ( \apWDp + 4\,\apBox ) \Bigr\}
         \,\frac{\xphs}{c^4\,\xps}
         \,\afun{\mh}
\nl &+&
         \frac{1}{6}
         \,( 4\,\mrF^{b}_{19}\,s\,\apWA + 4\,\mrF^{b}_{21}\,c\,\apWZ 
         + 32\,\mrF^{b}_{23}\,s^2\,c^2\,\apB + \mrF^{b}_{27}\,\apD )
         \,\frac{1}{c^2}
         \,\afun{M}
\nl &-&
         \frac{1}{12}
         \,\sumg \Bigl[ 4\,( \apD + 4\,c\,\apWZ - 4\,s\,\apWAB )\,c^2\,\vle 
\nl &+&  ( 8\,\mrG_{11}\,\apl + \mrG_{16}\,\apWDm + 4\,\mrG_{17}\,\aplVA ) \Bigr]
         \,\frac{\xpls}{c^2}
         \,\afun{\mle}
\nl &-&
         \frac{1}{12}
         \,\sumg \Bigl[ 4\,( \apD + 4\,c\,\apWZ - 4\,s\,\apWAB )\,c^2\,\vqd 
\nl &+& ( 24\,\mrG_{14}\,\apd + 12\,\mrG_{15}\,\apdVA + \mrG_{20}\,\apWDm ) \Bigr]
         \,\frac{\xpds}{c^2}
         \,\afun{\mqd}
\nl &-&
         \frac{1}{12}
         \,\sumg \Bigl[ 8\,( \apD + 4\,c\,\apWZ - 4\,s\,\apWAB )\,c^2\,\vqu 
\nl &-& ( 24\,\mrG_{12}\,\apu - 12\,\mrG_{13}\,\apuVA - \mrG_{19}\,\apWDm ) \Bigr]
         \,\frac{\xpus}{c^2}
         \,\afun{\mqu}
\eqas
\bqas
\Delta^{(6)}_{\PZ\PZ\,;\,2} &=& 
         \frac{1}{24}
         \,\Bigl\{ \Bigl[ 48\,\mrJ_{35}\,\xps\,\apB - 4\,\mrJ_{38}\,\apW 
         + ( \apD + 4\,\apBox )\,\mrJ_{27}\,\xphs \Bigr]\,c^4 
\nl &-& 2\,\Bigl[ 24\,( \apB - s\,c\,\apWB )\,\xps 
         + ( \apWDp + 4\,\apBox )\,\mrJ_{29} \Bigr]\,c^2 
\nl &+& 48\,( \apWZ - c\,\apB )\,c^5\,\xphs\,\xps - ( \apWDp + 4\,\apBox ) \Bigr\}
         \,\frac{1}{c^6\,\xps}
         \,\sbfun{\mh}{\mzb}
\nl &-&
         \frac{1}{6}
         \,( \mrF^{b}_{13}\,\apD - 32\,\mrF^{b}_{23}\,s^2\,c^2\,\apB + 4\,\mrF^{b}_{24}\,s\,\apWA 
         + 4\,\mrF^{b}_{25}\,c\,\apWZ )
         \,\frac{1}{c^2}
         \,\sbfun{M}{M}
\nl &-&
         \frac{1}{12}
         \,\sumg \Bigl[ 4\,( \apD + 4\,c\,\apWZ - 4\,s\,\apWAB )\,c^2\,\vle 
\nl &-& ( 4\,\mrG_{28}\,\aplVA + 8\,\mrG_{29}\,\apl + \mrG_{30}\,\apWDm ) \Bigr]
         \,\frac{\xpls}{c^2}
         \,\sbfun{\mle}{\mle}
\nl &-&
         \frac{1}{12}
         \,\sumg \Bigl[ 4\,( \apD + 4\,c\,\apWZ - 4\,s\,\apWAB )\,c^2\,\vqd 
\nl &-& ( 12\,\mrG_{26}\,\apdVA + 24\,\mrG_{27}\,\apd + \mrG_{32}\,\apWDm ) \Bigr]
         \,\frac{\xpds}{c^2}
         \,\sbfun{\mqd}{\mqd}
\nl &-&
         \frac{1}{12}
         \,\sumg \Bigl[ 8\,( \apD + 4\,c\,\apWZ - 4\,s\,\apWAB )\,c^2\,\vqu 
\nl &-& ( 12\,\mrG_{24}\,\apuVA - 24\,\mrG_{25}\,\apu + \mrG_{31}\,\apWDm ) \Bigr]
         \,\frac{\xpus}{c^2}
         \,\sbfun{\mqu}{\mqu}
\eqas
\vspace{0.5cm}
\bei
\item \fbox{$\PW$ self-energy}
\eei
\bqas
\Pi^{(4)}_{\PW\PW\,;\,0} &=& 
         \frac{4}{9}
         \,( 1 - 3\,\LR )
         \,\myNG
       +
         \frac{1}{18}
         \,( 2 + 57\,\LR )
\eqas
\bqas
\Pi^{(4)}_{\PW\PW\,;\,1} &=& 0
\eqas
\bqas
\Pi^{(4)}_{\PW\PW\,;\,2} &=& 
     -
         \frac{1}{12}
         \,\sbfun{M}{\mh}
     +
         \frac{1}{12}
         \,\sbfun{M}{\mzb}
         \,\mrF^{b}_{28}
\nl &+&
         \frac{10}{3}
         \,\sbfun{0}{M}
         \,s^2
     -
         \sumg 
         \,\sbfun{\mqu}{\mqd}
       -
         \frac{1}{3}
         \,\sumg 
         \,\sbfun{0}{\mle}
\eqas
\bqas
\Delta^{(4)}_{\PW\PW\,;\,0} &=& 
       -
         2
         \,\LR
       +
         \frac{1}{6}
         \,\frac{1}{c^2}
         \,( 1 - 6\,\LR )
       +
         \frac{1}{6}
         \,\mrJ_{47}
       -
         \frac{1}{6}
         \,\sumg 
         \,\mrH_{0}
         \,( 2 - 3\,\LR )
\eqas
\bqas
\Delta^{(4)}_{\PW\PW\,;\,1} &=& 
         \frac{1}{12}
         \,\frac{\xphs}{\xps}
         \,\mrJ_{41}
         \,\afun{\mh}
     -
         \frac{1}{6}
         \,\sumg 
         \,\frac{\xpls}{\xps}
         \,\mrK_{5}
         \,\afun{\mle}
\nl &-&
         \frac{1}{12}
         \,( 1 + 8\,\mrJ_{40}\,s^2\,c^2 - \mrJ_{44}\,c^2 )
         \,\frac{1}{c^4\,\xps}
         \,\afun{\mzb}
     +
         \frac{1}{12}
         \,( 1 - \mrJ_{45}\,c^2 )
         \,\frac{1}{c^2\,\xps}
         \,\afun{M}
\nl &-&
         \frac{1}{2}
         \,\sumg ( 2\,\xps - \mrH_{6} )
         \,\frac{\xpds}{\xps}
         \,\afun{\mqd}
     -
         \frac{1}{2}
         \,\sumg ( 2\,\xps + \mrH_{6} )
         \,\frac{\xpus}{\xps}
         \,\afun{\mqu}
\eqas
\bqas
\Delta^{(4)}_{\PW\PW\,;\,2} &=& 
       -
         \frac{2}{3}
         \,\frac{s^2}{\xps}
         \,\mrJ_{39}
         \,\sbfun{0}{M}
     -
         \frac{1}{12}
         \,\frac{1}{\xps}
         \,\mrJ_{43}
         \,\sbfun{M}{\mh}
\nl &+&
         \frac{1}{6}
         \,\sumg 
         \,\frac{\xpls}{\xps}
         \,\mrK_{4}
         \,\sbfun{0}{\mle}
     +
         \frac{1}{2}
         \,\sumg ( \mrH_{7}\,\xps + \mrH_{8} )
         \,\frac{1}{\xps}
         \,\sbfun{\mqu}{\mqd}
\nl &-&
         \frac{1}{12}
         \,( 1 - 8\,\mrJ_{39}\,s^2\,c^4 - 7\,\mrJ_{42}\,c^4 + 2\,\mrJ_{46}\,c^2 )
         \,\frac{1}{c^4\,\xps}
         \,\sbfun{M}{\mzb}
\eqas
\bqas
\Pi^{(6)}_{\PW\PW\,;\,0} &=& 
         \frac{2}{9}
         \,s^2\,\aAA
         \,( 1 - 9\,\LR )
     +
         \frac{8}{9}
         \,\myNG\,\apW
         \,( 1 - 3\,\LR )
     -
         \frac{1}{18}
         \,\apBox
         \,( 2 + 3\,\LR )
\nl &+&
         \frac{1}{6}
         \,\Bigl[ 6\,\mrF^{b}_{3}\,s\,c\,\apWB + ( \mrJ_{54}\,c^2 - 3\,\mrF^{b}_{29} )\,\apW \Bigr]
         \,\frac{\LR}{c^2}
     -
         \frac{2}{9}
         \,(  - c\,\aZZ + 2\,s\,\aAZ )
         \,c
\nl &+&
         \frac{4}{9}
         \,\sumg ( \aplt + 3\,\apqt )
         \,( 1 - 3\,\LR )
     +
         \frac{1}{2}
         \,\sumg (  - 3\,\xpds\,\adW + 3\,\xpus\,\auW - \xpls\,\alW )
         \,\LR
\eqas
\bqas
\Pi^{(6)}_{\PW\PW\,;\,1} &=& 
         \apW
         \,\afun{M}
       +
         \frac{1}{2}
         \,\apW\,\xphs
         \,\afun{\mh}
\nl &+&
         \frac{1}{6}
         \,\Bigl[ \mrF^{b}_{30}\,s\,c\,\aAZ + 3\,( c^2\,\aZZ + s^2\,\aAA ) \Bigr]
         \,\frac{1}{c^2}
         \,\afun{\mzb}
\eqas
\bqas
\Pi^{(6)}_{\PW\PW\,;\,2} &=& 
       -
         \frac{1}{3}
         \,\Bigl\{ 5\,c^2\,\apD - \mrJ_{53}\,s\,c\,\apWB 
         - \Bigl[ 2\,\mrJ_{49}\,\apW + ( 2\,\aAA + s\,c\,\aAZ )\,\mrJ_{48} \Bigr]\,s^2 \Bigr\}
         \,\sbfun{0}{M}
\nl &+&
         \frac{1}{24}
         \,\Bigl\{ 32\,\mrJ_{52}\,s^4\,c\,\apWBa 
         + \Bigl[ 39\,c\,\apWDp + 8\,( 3\,\apWB - 5\,s\,c\,\apD )\,s \Bigr] 
\nl &+& 16\,( \apWA - s\,\apB )\,\mrJ_{48}\,s^5\,c 
         - 16\,( \mrJ_{46}\,\apW + \mrJ_{51}\,\apB )\,s^2\,c 
\nl &-& 8\,( 5\,\mrJ_{50} + \mrJ_{55}\,s^2 )\,s\,c^2\,\apWB \Bigr\}
         \,\frac{1}{c}
         \,\sbfun{M}{\mzb}
\nl &+&
         \frac{1}{24}
         \,(  - 52\,\apW + \apD - 4\,\apBox )
         \,\sbfun{M}{\mh}
\nl &-&
         \frac{1}{2}
         \,\sumg ( 2\,\apqWt + 3\,\xpds\,\adW - 3\,\xpus\,\auW )
         \,\sbfun{\mqu}{\mqd}
\nl &-&
         \frac{1}{6}
         \,\sumg ( 2\,\aplWt + 3\,\xpls\,\alW )
         \,\sbfun{0}{\mle}
\eqas
\bqas
\Delta^{(6)}_{\PW\PW\,;\,0} &=& 
         \frac{1}{3}
         \,\frac{1}{c^2}
         \,\apW
         \,( 1 - 6\,\LR )
     -
         \frac{1}{12}
         \,\frac{1}{c^2}
         \,\apD
         \,( 11 - 15\,\LR )
\nl &+&
         \frac{1}{6}
         \,\Bigl[ 2\,c^3\,\apWB + ( 3\,\apD - 4\,c^2\,\aAA + 4\,c^2\,\apW )\,s \Bigr]
         \,\frac{s}{c^2}
         \,( 2 - 3\,\LR )
\nl &-&
         \frac{1}{12}
         \,\Bigl[ 40\,s\,\apWB - ( 4\,\mrJ_{35}\,\apBox 
         + 4\,\mrJ_{47}\,\apW + \mrJ_{75}\,\apD )\,c \Bigr]
         \,\frac{1}{c}
\nl &-&
         \frac{1}{2}
         \,( 8\,\apW + 3\,\apBox )
         \,\LR
     -
         \frac{1}{3}
         \,\sumg ( 2\,\xpls\,\aplt + \mrH_{0}\,\apW + 6\,\mrH_{7}\,\apqt )
         \,( 2 - 3\,\LR )
\eqas
\bqas
\Delta^{(6)}_{\PW\PW\,;\,1} &=& 
         \frac{1}{24}
         \,\Bigl[ \apWDp + 8\,\mrJ_{61}\,s\,c^3\,\aAZ - 8\,\mrJ_{71}\,s\,c\,\apWB 
\nl &-& ( 12\,\apD\,\xps + 4\,\mrJ_{62}\,\apBox 
         + 4\,\mrJ_{69}\,\apW - \mrJ_{74}\,\apD )\,c^2 \Bigr]
         \,\frac{1}{c^2\,\xps}
         \,\afun{M}
\nl &+&
         \frac{1}{24}
         \,\Bigl[ 16\,\mrJ_{63}\,s^5\,c^2\,\apWA - 16\,\mrJ_{64}\,s^4\,c^2\,\apW 
         - 16\,\mrF^{b}_{4}\,s^2\,c^4\,\xps\,\apB - ( \apWDp - 16\,s^4\,c^4\,\apB ) 
\nl &-& 8\,( \mrJ_{40}\,\apD + 2\,\mrJ_{70}\,\apW )\,s^2\,c^2 
         + ( 4\,\mrJ_{44}\,\apW + \mrJ_{65}\,\apD )\,c^2 
\nl &-& 8\,( \mrJ_{66}\,s^2\,c^2 - \mrJ_{71} + \mrJ_{72}\,c^2 )\,s\,c\,\apWB \Bigr]
         \,\frac{1}{c^4\,\xps}
         \,\afun{\mzb}
\nl &+&
         \frac{1}{24}
         \,\Bigl[ 4\,\mrJ_{67}\,\apW - ( \apD - 4\,\apBox )\,\mrJ_{41} \Bigr]
         \,\frac{\xphs}{\xps}
         \,\afun{\mh}
\nl &-&
         \frac{1}{2}
         \,\sumg ( 2\,\xps\,\apqWt - 3\,\xpds\,\xps\,\adW - 3\,\xpus\,\xps\,\auW 
         - \mrH_{6}\,\apqWt )
         \,\frac{\xpds}{\xps}
         \,\afun{\mqd}
\nl &-&
         \frac{1}{2}
         \,\sumg ( 2\,\xps\,\apqWt + 3\,\xpds\,\xps\,\adW + 3\,\xpus\,\xps\,\auW 
         + \mrH_{6}\,\apqWt )
         \,\frac{\xpus}{\xps}
         \,\afun{\mqu}
\nl &+&
         \frac{1}{6}
         \,\sumg ( 3\,\xpls\,\xps\,\alW - \mrK_{5}\,\aplWt )
         \,\frac{\xpls}{\xps}
         \,\afun{\mle}
\eqas
\bqas
\Delta^{(6)}_{\PW\PW\,;\,2} &=& 
         \frac{1}{3}
         \,\Bigl[ \mrJ_{39}\,c^2\,\apD - ( 2\,\aAA + s\,c\,\aAZ )\,\mrJ_{56}\,s^2 
         - ( c\,\apWB + 2\,s\,\apW )\,\mrJ_{57}\,s \Bigr]
         \,\frac{1}{\xps}
         \,\sbfun{0}{M}
\nl &-&
         \frac{1}{24}
         \,\Bigl[ 4\,\mrJ_{68}\,\apW - ( \apD - 4\,\apBox )\,\mrJ_{43} \Bigr]
         \,\frac{1}{\xps}
         \,\sbfun{M}{\mh}
\nl &-&
         \frac{1}{24}
         \,\Bigl\{  - 8\,\Bigl[ \mrJ_{39}\,\apD + 4\,\mrJ_{60}\,\apW 
         + 4\,( \apB - 3\,s^2\,\apWBa )\,\xps \Bigr]\,s^2\,c^4 
         + ( 1 - 7\,\mrJ_{42}\,c^4 + 2\,\mrJ_{46}\,c^2 )\,\apWDp 
\nl &+& 16\,( \apWA - s\,\apB )\,\mrJ_{56}\,s^5\,c^4 
         + 8\,( \mrJ_{58}\,s^2\,c^4 - \mrJ_{59}\,c^4 - \mrJ_{71} 
         + \mrJ_{73}\,c^2 )\,s\,c\,\apWB \Bigr\}
         \,\frac{1}{c^4\,\xps}
         \,\sbfun{M}{\mzb}
\nl &+&
         \frac{1}{6}
         \,\sumg ( 3\,\xpls\,\xps\,\alW + \mrK_{4}\,\aplWt )
         \,\frac{\xpls}{\xps}
         \,\sbfun{0}{\mle}
\nl &+&
         \frac{1}{2}
         \,\sumg ( 3\,\mrH_{6}\,\xpds\,\xps\,\adW + 3\,\mrH_{6}\,\xpus\,\xps\,\auW 
         + \mrH_{7}\,\xps\,\apqWt + \mrH_{8}\,\apqWt )
         \,\frac{1}{\xps}
         \,\sbfun{\mqu}{\mqd}
\eqas
\vspace{0.5cm}
\bei
\item \fbox{neutrino ($\PGN$) self-energy}
\eei
\bqas
V^{(4)}_{\PGN\PGN\,;\,0} &=&
       -
         \frac{1}{4}
       -
         \frac{1}{8}
         \,\frac{1}{c^2}
         \,( 1 - \LR )
       +
         \frac{1}{8}
         \,\mrI_{0}
         \,\LR
\eqas
\bqas
V^{(4)}_{\PGN\PGN\,;\,1} &=&
       -
         \frac{1}{8}
         \,\frac{1}{c^4\,\xps}
         \,\afun{\mzb}
       -
         \frac{1}{8}
         \,\frac{1}{\xps}
         \,\mrI_{0}
         \,\afun{M}
\eqas
\bqas
V^{(4)}_{\PGN\PGN\,;\,2} &=&
       -
         \frac{1}{8}
         \,\frac{1}{\xps}
         \,\mrI_{0}
         \,\mrJ_{61}
         \,\sbfun{0}{M}
       -
         \frac{1}{8}
         \,( 1 - c^2\,\xps )
         \,\frac{1}{c^4\,\xps}
         \,\sbfun{0}{\mzb}
\eqas
\bqas
V^{(6)}_{\PGN\PGN\,;\,0} &=&
       -
         \frac{1}{4}
         \,\apLWt
       +
         \frac{1}{8}
         \,\xpLs\,\aLW
         \,( 2 - 3\,\LR )
\nl &-&
         \frac{1}{16}
         \,( 4\,\apN + \apWDm )
         \,\frac{1}{c^2}
         \,( 1 - \LR )
     +
         \frac{1}{4}
         \,( \mrI_{0}\,\apW + 2\,\mrI_{1}\,\apLt )
         \,\LR
\eqas
\bqas
V^{(6)}_{\PGN\PGN\,;\,1} &=&
       -
         \frac{1}{16}
         \,( 4\,\apN + \apWDm )
         \,\frac{1}{c^4\,\xps}
         \,\afun{\mzb}
\nl &-&
         \frac{1}{8}
         \,( 3\,\xpLs\,\aLW + \mrI_{0}\,\apLWt )
         \,\frac{1}{\xps}
         \,\afun{M}
\eqas
\bqas
V^{(6)}_{\PGN\PGN\,;\,2} &=&
       -
         \frac{1}{16}
         \,( 4\,\apN + \apWDm )
         \,( 1 - c^2\,\xps )
         \,\frac{1}{c^4\,\xps}
         \,\sbfun{0}{\mzb}
\nl &-&
         \frac{1}{8}
         \,( 4\,\mrI_{0}\,\apLt + 2\,\mrI_{0}\,\mrJ_{61}\,\apW 
         - 4\,\mrI_{1}\,\xps\,\apLt + 3\,\mrJ_{76}\,\xpLs\,\aLW )
         \,\frac{1}{\xps}
         \,\sbfun{0}{M}
\eqas
\bqas
A^{(4)}_{\PGN\PGN\,;\,0} &=&
       -
         \frac{1}{4}
       -
         \frac{1}{8}
         \,\frac{1}{c^2}
         \,( 1 - \LR )
       +
         \frac{1}{8}
         \,\mrI_{0}
         \,\LR
\eqas
\bqas
A^{(4)}_{\PGN\PGN\,;\,1} &=&
       -
         \frac{1}{8}
         \,\frac{1}{c^4\,\xps}
         \,\afun{\mzb}
       -
         \frac{1}{8}
         \,\frac{1}{\xps}
         \,\mrI_{0}
         \,\afun{M}
\eqas
\bqas
A^{(4)}_{\PGN\PGN\,;\,2} &=&
       -
         \frac{1}{8}
         \,\frac{1}{\xps}
         \,\mrI_{0}
         \,\mrJ_{61}
         \,\sbfun{0}{M}
       -
         \frac{1}{8}
         \,( 1 - c^2\,\xps )
         \,\frac{1}{c^4\,\xps}
         \,\sbfun{0}{\mzb}
\eqas
\bqas
A^{(6)}_{\PGN\PGN\,;\,0} &=&
       -
         \frac{1}{4}
         \,\apLWt
       +
         \frac{1}{8}
         \,\xpLs\,\aLW
         \,( 2 - 3\,\LR )
\nl &-&
         \frac{1}{16}
         \,( 4\,\apN + \apWDm )
         \,\frac{1}{c^2}
         \,( 1 - \LR )
\nl &+&
         \frac{1}{4}
         \,( \mrI_{0}\,\apW + 2\,\mrI_{1}\,\apLt )
         \,\LR
\eqas
\bqas
A^{(6)}_{\PGN\PGN\,;\,1} &=&
       -
         \frac{1}{16}
         \,( 4\,\apN + \apWDm )
         \,\frac{1}{c^4\,\xps}
         \,\afun{\mzb}
\nl &-&
         \frac{1}{8}
         \,( 3\,\xpLs\,\aLW + \mrI_{0}\,\apLWt )
         \,\frac{1}{\xps}
         \,\afun{M}
\eqas
\bqas
A^{(6)}_{\PGN\PGN\,;\,2} &=&
       -
         \frac{1}{16}
         \,( 4\,\apN + \apWDm )
         \,( 1 - c^2\,\xps )
         \,\frac{1}{c^4\,\xps}
         \,\sbfun{0}{\mzb}
\nl &-&
         \frac{1}{8}
         \,( 4\,\mrI_{0}\,\apLt + 2\,\mrI_{0}\,\mrJ_{61}\,\apW 
         - 4\,\mrI_{1}\,\xps\,\apLt + 3\,\mrJ_{76}\,\xpLs\,\aLW )
         \,\frac{1}{\xps}
         \,\sbfun{0}{M}
\eqas
\vspace{0.5cm}
\bei
\item \fbox{lepton ($\PL$) self-energy}
\eei
\bqas
\Sigma^{(4)}_{\PL\PL\,;\,0} &=&
       -
         \frac{1}{8}
         \,( 16\,s^2\,c^2 - \mrG_{33} )
         \,\frac{1}{c^2}
         \,\xpL
         \,( 1 - 2\,\LR )
\eqas
\bqas
\Sigma^{(4)}_{\PL\PL\,;\,1} &=& 0
\eqas
\bqas
\Sigma^{(4)}_{\PL\PL\,;\,2} &=&
       -
         \frac{1}{4}
         \,\xpLc
         \,\sbfun{\mh}{\mLe}
       +
         4
         \,s^2\,\xpL
         \,\sbfun{0}{\mLe}
\nl &+&
         \frac{1}{4}
         \,( c^2\,\xpLs - \mrG_{33} )
         \,\frac{1}{c^2}
         \,\xpL
         \,\sbfun{\mzb}{\mLe}
\eqas
\bqas
\Sigma^{(6)}_{\PL\PL\,;\,0} &=&
       -
         \frac{1}{8}
         \,\Bigl[ c^2\,\vle\,\xpLs\,\aLBW + 4\,\mrF^{b}_{5}\,s\,c^2\,\apWB\,\vle
         + ( \vle\,\aLBW + 2\,c^3\,\aLW ) \Bigr]
         \,\frac{1}{c^3}
         \,\xpL
         \,( 2 - 3\,\LR )
\nl &+&
         \frac{1}{8}
         \,\Bigl\{ s\,c\,\xps\,\vle\,\xpL\,\aLB - 16\,\mrG_{29}\,s^2\,c^2\,\apW\,\xpL 
         + 4\,\mrL_{1}\,s\,c^2\,\xpL\,\aLWB 
\nl &+& \Bigl[ \mrG_{29}\,\aLW\,\xpL 
         + 128\,( M\,\DUV\,\UVsl ) \Bigr]\,c^2\,\xps 
         + 2\,( \mrG_{11}\,\apLVA + 2\,\mrG_{17}\,\apL )\,\xpL \Bigr\}
         \,\frac{1}{c^2}
\nl &-&
         \frac{1}{16}
         \,\Bigl\{ 32\,\mrG_{28}\,s\,c^3\,\apWB - 16\,\mrF^{b}_{31}\,s^2\,\apW\,\vle 
\nl &-& \Bigl[ 4\,\mrG_{11}\,\apW - \mrG_{36}\,\apD 
         + 4\,( 4\,c^4 + s^2\,\vle )\,\apD \Bigr] \Bigr\}
         \,\frac{1}{c^2}
         \,\xpL
         \,( 1 - 2\,\LR )
\nl &+&
         \frac{1}{4}
         \,\Bigl\{ 4\,\mrG_{39}\,s^2\,c^2\,\apW + \Bigl[ 3\,\xphs\,\aLp + 2\,\mrI_{0}\,\apLt 
         + \mrI_{2}\,\aLp - 2\,(  - \apL + \apLo + \apBox - 3\,s\,\aLWB )\,\xpLs \Bigr]\,c^2 
\nl &-& \Bigl[ \mrG_{34}\,\apLVA + 2\,\mrG_{35}\,\apL 
         - ( \aLp + 4\,s^2\,c^2\,\apB\,\vle ) \Bigr] \Bigr\}
         \,\frac{\LR}{c^2}
         \,\xpL
\nl &+&
         \frac{3}{2}
         \,\sumg (  - \xpds\,\xpd\,\aLldQ + \xpus\,\xpu\,\aoLlQu )
         \,\LR
\eqas
\bqas
\Sigma^{(6)}_{\PL\PL\,;\,1} &=&
       -
         \frac{3}{4}
         \,\xphs\,\xpL\,\aLp
         \,\afun{\mh}
       -
         \frac{3}{2}
         \,\sumg 
         \,\xpus\,\xpu\,\aoLlQu
         \,\afun{\mqu}
\nl &+&
         \frac{3}{2}
         \,\sumg 
         \,\xpds\,\xpd\,\aLldQ
         \,\afun{\mqd}
     -
         \frac{1}{8}
         \,( 2\,\aLp + 3\,\aLW + 4\,\apLt )
         \,\xpL
         \,\afun{M}
\nl &-&
         \frac{1}{16}
         \,( 3\,\vle\,\aLBW + 4\,c\,\apLA + 4\,c\,\aLp )
         \,\frac{1}{c^3}
         \,\xpL
         \,\afun{\mzb}
\nl &-&
         \frac{1}{16}
         \,( 3\,\vle\,\aLBW + 4\,c\,\apLA + 12\,s\,c\,\aLWB )
         \,\frac{1}{c}
         \,\xpLc
         \,\afun{\mLe}
\eqas
\bqas
\Sigma^{(6)}_{\PL\PL\,;\,2} &=&
         \frac{1}{16}
         \,\Bigl\{ 2\,c^3\,\apWDm\,\xpLs - 32\,\mrF^{b}_{3}\,s^2\,c\,\aAA\,\vle
         - 8\,\mrF^{b}_{4}\,s\,c^2\,\vle\,\aAZ 
\nl &+& \Bigl[ 3\,\aLBW - 8\,( \apD
         - 4\,c^2\,\aZZ )\,s^2\,c \Bigr]\,\vle - ( 3\,\vle\,\aLBW + 4\,c\,\apLA )\,\mrL_{0}\,c^2 
\nl &-& 2\,( 4\,\mrG_{11}\,\apW - \mrG_{36}\,\apD + \mrG_{37}\,\apLVA 
         + 2\,\mrG_{38}\,\apL )\,c \Bigr\}
         \,\frac{1}{c^3}
         \,\xpL
         \,\sbfun{\mzb}{\mLe}
\nl &+&
         \frac{1}{8}
         \,( 3\,\aLW + 4\,\apLt )
         \,\mrJ_{61}
         \,\xpL
         \,\sbfun{0}{M}
\nl &+&
         \frac{1}{4}
         \,( 8\,\apWAD - 3\,\mrL_{0}\,s\,\aLWB )
         \,\xpL
         \,\sbfun{0}{\mLe}
\nl &-&
         \frac{1}{8}
         \,( \apWDm - 4\,\alpBox )
         \,\xpLc
         \,\sbfun{\mh}{\mLe}
\eqas
\bqas
V^{(4)}_{\PL\PL\,;\,0} &=&
       -
         \frac{1}{4}
       +
         \frac{1}{8}
         \,\mrI_{3}
         \,\LR
       -
         \frac{1}{16}
         \,( 16\,s^2\,c^2 + \mrG_{6} )
         \,\frac{1}{c^2}
         \,( 1 - \LR )
\eqas
\bqas
V^{(4)}_{\PL\PL\,;\,1} &=&
       -
         \frac{1}{8}
         \,\frac{\xpLs}{\xps}
         \,\xphs
         \,\afun{\mh}
     -
         \frac{1}{8}
         \,\frac{1}{\xps}
         \,\mrI_{0}
         \,\afun{M}
\nl &-&
         \frac{1}{16}
         \,( 2\,c^2\,\xpLs + \mrG_{6} )
         \,\frac{1}{c^4\,\xps}
         \,\afun{\mzb}
     +
         \frac{1}{16}
         \,( 4\,c^2\,\xpLs + 16\,s^2\,c^2 + \mrG_{6} )
         \,\frac{\xpLs}{c^2\,\xps}
         \,\afun{\mLe}
\eqas
\bqas
V^{(4)}_{\PL\PL\,;\,2} &=&
         \frac{s^2}{\xps}
         \,\mrL_{2}
         \,\sbfun{0}{\mLe}
     -
         \frac{1}{8}
         \,\frac{1}{\xps}
         \,\mrI_{0}
         \,\mrJ_{61}
         \,\sbfun{0}{M}
\nl &-&
         \frac{1}{16}
         \,\Bigl[ ( \mrG_{6} - 2\,\mrL_{2}\,c^4\,\xpLs ) 
         - ( \mrG_{6}\,\xps - \mrG_{33}\,\xpLs )\,c^2 \Bigr]
         \,\frac{1}{c^4\,\xps}
         \,\sbfun{\mzb}{\mLe}
\nl &+&
         \frac{1}{8}
         \,( \xpLs + \mrJ_{34} )
         \,\frac{\xpLs}{\xps}
         \,\sbfun{\mh}{\mLe}
\eqas
\bqas
V^{(6)}_{\PL\PL\,;\,0} &=&
       -
         \frac{1}{32}
         \,\Bigl\{ 32\,\mrG_{28}\,s\,c^3\,\apWB - 16\,\mrF^{b}_{31}\,s^2\,\apW\,\vle 
\nl &+& \Bigl[ 8\,\mrG_{11}\,\apL - \mrG_{40}\,\apD 
         + 4\,( \apLVA + \apW )\,\mrG_{17} - 4\,( 4\,c^4+ s^2\,\vle )\,\apD \Bigr] \Bigr\}
         \,\frac{1}{c^2}
         \,( 1 - \LR )
\nl &+&
         \frac{1}{8}
         \,\Bigl\{ \Bigl[ 2\,\mrI_{3}\,\apW + 2\,\mrI_{4}\,\apLt 
         + (  - 2\,\apL + 2\,\apLo - \apD - 2\,\alpBox )\,\xpLs \Bigr] 
         + 8\,( \apB\,\vle + 2\,\apW )\,s^2 \Bigr\}
         \,\LR
\nl &+&
         \frac{1}{16}
         \,( \vle\,\aLBW + 4\,s\,c\,\aLWB )
         \,\frac{1}{c}
         \,\xpLs
         \,( 2 - 3\,\LR )
\nl &-&
         \frac{1}{4}
         \,( c\,\apLWt + 4\,\mrG_{29}\,s^2\,c\,\apW + 2\,\mrF^{b}_{5}\,s\,\apWB\,\vle )
         \,\frac{1}{c}
\eqas
\bqas
V^{(6)}_{\PL\PL\,;\,1} &=&
         \frac{1}{32}
         \,\Bigl\{ 4\,s^2\,\apWDp\,\vle - 16\,\mrF^{b}_{5}\,s\,c\,\apWB\,\vle 
\nl &-& 2\,\Bigl[ 3\,\vle\,\aLBW + 4\,c\,\apLA + c\,\apWDm 
         + 8\,( \apWZ - c\,\apB )\,s^2\,c^2\,\vle \Bigr]\,c\,\xpLs 
\nl &-& \Bigl[ 8\,\mrG_{11}\,\apL - \mrG_{40}\,\apD 
         + 4\,( \apLVA + \apW )\,\mrG_{17} \Bigr]
         + 8\,( \xps\,\aAZ + \mrI_{4}\,\apWB )\,s\,c^3\,\vle \Bigr\}
         \,\frac{1}{c^4\,\xps}
         \,\afun{\mzb}
\nl &+&
         \frac{1}{32}
         \,\Bigl\{ \Bigl[ 8\,\mrG_{11}\,\apL - \mrG_{40}\,\apD 
         + 4\,( \apLVA + \apW )\,\mrG_{17} - 4\,( 4\,c^4 + s^2\,\vle )\,\apD \Bigr] 
\nl &+& 2\,( 3\,\vle\,\aLBW + 4\,c\,\apLA + 2\,c\,\apWDm - 4\,c\,\alpBox 
         + 12\,s\,c\,\aLWB )\,c\,\xpLs + 16\,( 2\,\mrG_{28}\,c^2 
         + \mrF^{b}_{5}\,\vle )\,s\,c\,\apWB 
\nl &+& 16\,( 2\,\mrG_{29}\,c^2 - \mrF^{b}_{31}\,\vle )\,s^2\,\apW \Bigr\}
         \,\frac{\xpLs}{c^2\,\xps}
         \,\afun{\mLe}
\nl &-&
         \frac{1}{4}
         \,( 4\,\apLt + \mrI_{0}\,\apW )
         \,\frac{1}{\xps}
         \,\afun{M}
\nl &-&
         \frac{1}{16}
         \,( \apWDm - 4\,\alpBox )
         \,\frac{\xpLs}{\xps}
         \,\xphs
         \,\afun{\mh}
\eqas
\bqas
V^{(6)}_{\PL\PL\,;\,2} &=&
         \frac{1}{16}
         \,\Bigl[ ( \apWDm - 4\,\alpBox )\,\xpLs + ( \apWDm - 4\,\alpBox )\,\mrJ_{34} \Bigr]
         \,\frac{\xpLs}{\xps}
         \,\sbfun{\mh}{\mLe}
\nl &+&
         \frac{1}{32}
         \,\Bigl\{ 2\,\mrL_{2}\,c^4\,\apWDm\,\xpLs - \Bigl[ 8\,\mrG_{11}\,\apL 
         - \mrG_{40}\,\apD + 4\,( \apLVA + \apW )\,\mrG_{17} \Bigr] 
\nl &+& \Bigl[ 8\,\mrG_{11}\,\xps\,\apL - 4\,\mrG_{11}\,\apW\,\xpLs 
         + \mrG_{36}\,\apD\,\xpLs - \mrG_{40}\,\apD\,\xps 
         + 4\,( \apLVA + \apW )\,\mrG_{17}\,\xps \Bigr]\,c^2
\nl &-& 2\,( 3\,\aLBW - 4\,c\,\apLV - 8\,s^2\,c^3\,\apB )\,c\,\vle\,\xpLs 
         + 8\,( \aAZ\,\xpss - 2\,\mrI_{0}\,\xps\,\apWB - \mrI_{4}\,\apWB\,\xpLs )\,s\,c^5\,\vle 
\nl &-& 16\,( \apWZ - c\,\apB )\,\mrL_{3}\,s^2\,c^5\,\vle\,\xpLs 
\nl &-& 2\,( 3\,\vle\,\aLBW + 4\,c\,\apLA )\,\mrL_{0}\,c^3\,\xpLs 
         + 4\,( \apD + 4\,\mrF^{b}_{31}\,\apW )\,s^2\,\vle 
\nl &-& 8\,(  - c\,\apWB - 2\,s\,\aAA + 4\,s\,\apW )\,s\,c^2\,\xps\,\vle 
\nl &-& 4\,( 4\,\mrI_{3}\,\apW + \mrL_{2}\,\apD - 4\,\mrF^{b}_{32}\,\apW\,\xpLs )\,s^2\,c^2\,\vle 
\nl &+& 8\,( \mrI_{5}\,c^2 - 2\,\mrL_{4}\,s^2\,c^2 - 2\,\mrF^{b}_{5} )\,s\,c\,\apWB\,\vle \Bigr\}
         \,\frac{1}{c^4\,\xps}
         \,\sbfun{\mzb}{\mLe}
\nl &-&
         \frac{1}{4}
         \,( \mrI_{0}\,\mrJ_{61}\,\apW + 4\,\mrJ_{61}\,\apLt )
         \,\frac{1}{\xps}
         \,\sbfun{0}{M}
\nl &-&
         \frac{1}{4}
         \,( 3\,\mrL_{0}\,s\,\xpLs\,\aLWB - 2\,\mrL_{2}\,\apWAD + \mrL_{5}\,s\,c\,\vle\,\aAZ )
         \,\frac{1}{\xps}
         \,\sbfun{0}{\mLe}
\eqas
\bqas
A^{(4)}_{\PL\PL\,;\,0} &=&
       -
         \frac{1}{4}
       -
         \frac{1}{8}
         \,\frac{1}{c^2}
         \,\vle
         \,( 1 - \LR )
       +
         \frac{1}{8}
         \,\mrI_{1}
         \,\LR
\eqas
\bqas
A^{(4)}_{\PL\PL\,;\,1} &=&
         \frac{1}{8}
         \,\frac{\xpLs}{c^2\,\xps}
         \,\vle
         \,\afun{\mLe}
     -
         \frac{1}{8}
         \,\frac{1}{c^4\,\xps}
         \,\vle
         \,\afun{\mzb}
     -
         \frac{1}{8}
         \,\frac{1}{\xps}
         \,\mrI_{1}
         \,\afun{M}
\eqas
\bqas
A^{(4)}_{\PL\PL\,;\,2} &=&
       -
         \frac{1}{8}
         \,\frac{1}{\xps}
         \,\mrI_{1}
         \,\mrJ_{61}
         \,\sbfun{0}{M}
      -
         \frac{1}{8}
         \,( 1 - \mrL_{2}\,c^2 )
         \,\frac{1}{c^4\,\xps}
         \,\vle
         \,\sbfun{\mzb}{\mLe}
\eqas
\bqas
A^{(6)}_{\PL\PL\,;\,0} &=&
         \frac{1}{16}
         \,\frac{1}{c}
         \,\xpLs\,\aLBW
         \,( 2 - 3\,\LR )
\nl &+&
         \frac{1}{32}
         \,\Bigl\{ 4\,s^2\,\apD + \Bigl[ 8\,\mrG_{11}\,\apL - \mrG_{41}\,\apD 
         - 4\,( \apLVA + \apW )\,\mrG_{17} \Bigr] \Bigr\}
         \,\frac{1}{c^2}
         \,( 1 - \LR )
\nl &-&
         \frac{1}{4}
         \,\Bigl\{ 2\,\mrF^{b}_{31}\,s^2\,\apW - \Bigl[ \mrI_{1}\,\apW
         + \mrI_{4}\,\apLt + (  - \apLV + \apLt )\,\xpLs \Bigr]\,c^2
         - 4\,(  - c\,\apWB + s\,\apB )\,s\,c^2 \Bigr\}
         \,\frac{\LR}{c^2}
\nl &+&
         \frac{1}{2}
         \,(  - 2\,c^2\,\apLt - c^2\,\aZZ + s^2\,\aAA )
         \,\frac{1}{c^2}
\eqas
\bqas
A^{(6)}_{\PL\PL\,;\,1} &=&
         \frac{1}{32}
         \,\Bigl\{ 4\,s^2\,\apWDp - 16\,\mrF^{b}_{5}\,s\,c\,\apWB 
         - 2\,\Bigl[ 3\,\aLBW + 4\,c\,\apLV + 8\,( \apWZ - c\,\apB )\,s^2\,c^2 \Bigr]\,c\,\xpLs
\nl &+& \Bigl[ 8\,\mrG_{11}\,\apL - \mrG_{41}\,\apD 
         - 4\,( \apLVA + \apW )\,\mrG_{17} \Bigr] 
         + 8\,( \xps\,\aAZ + \mrI_{4}\,\apWB )\,s\,c^3 \Bigr\}
         \,\frac{1}{c^4\,\xps}
         \,\afun{\mzb}
\nl &-&
         \frac{1}{32}
         \,\Bigl\{ 16\,\mrF^{b}_{3}\,s^2\,\aAA + 4\,\Bigl[ 4\,c^3\,\aAZ 
         + ( \apD - 4\,c^2\,\aZZ )\,s \Bigr]\,s 
\nl &+& \Bigl[ 8\,\mrG_{11}\,\apL - \mrG_{41}\,\apD - 4\,( \apLVA + \apW )\,\mrG_{17} \Bigr] 
         - 2\,( 3\,\aLBW + 4\,c\,\apLV )\,c\,\xpLs \Bigr\}
         \,\frac{\xpLs}{c^2\,\xps}
         \,\afun{\mLe}
\nl &-&
         \frac{1}{4}
         \,( 4\,\apLt + \mrI_{1}\,\apW )
         \,\frac{1}{\xps}
         \,\afun{M}
\eqas
\bqas
A^{(6)}_{\PL\PL\,;\,2} &=&
       -
         \frac{1}{4}
         \,\frac{1}{\xps}
         \,\mrL_{5}
         \,s\,c\,\aAZ
         \,\sbfun{0}{\mLe}
\nl &-&
         \frac{1}{32}
         \,\Bigl\{  - \Bigl[ 8\,\mrG_{11}\,\apL - \mrG_{41}\,\apD
         - 4\,( \apLVA + \apW )\,\mrG_{17} \Bigr] + 2\,( 3\,\aLBW
         + 4\,c\,\apLV )\,\mrL_{0}\,c^3\,\xpLs 
\nl &+& 2\,( 3\,\aLBW - 4\,c\,\vle\,\apLA 
         - 8\,s^2\,c^3\,\apB )\,c\,\xpLs - 8\,( \aAZ\,\xpss - 2\,\mrI_{0}\,\xps\,\apWB 
         - \mrI_{4}\,\apWB\,\xpLs )\,s\,c^5 
\nl &+& 16\,( \apWZ - c\,\apB )\,\mrL_{3}\,s^2\,c^5\,\xpLs 
\nl &-& 4\,( \apD + 4\,\mrF^{b}_{31}\,\apW )\,s^2 + 8\,(  - c\,\apWB - 2\,s\,\aAA 
         + 4\,s\,\apW )\,s\,c^2\,\xps 
\nl &+& ( 8\,\mrG_{11}\,\xps\,\apL - 4\,\mrG_{17}\,\xps\,\apLVA 
         - 4\,\mrG_{17}\,\mrL_{2}\,\apW - \mrG_{41}\,\mrL_{2}\,\apD )\,c^2 
\nl &+& 4\,( 4\,\mrI_{3}\,\apW + \mrL_{2}\,\apD - 4\,\mrF^{b}_{32}\,\apW\,\xpLs )\,s^2\,c^2 
\nl &-& 8\,( \mrI_{5}\,c^2 - 2\,\mrL_{4}\,s^2\,c^2 - 2\,\mrF^{b}_{5} )\,s\,c\,\apWB \Bigr\}
         \,\frac{1}{c^4\,\xps}
         \,\sbfun{\mzb}{\mLe}
\nl &-&
         \frac{1}{4}
         \,( \mrI_{1}\,\mrJ_{61}\,\apW + 4\,\mrJ_{61}\,\apLt )
         \,\frac{1}{\xps}
         \,\sbfun{0}{M}
\eqas
We have introduced
\bqas
\UVsl &=& - \frac{1}{256}\,\frac{\mLes}{M}\,\Bigl[
          3\,( s\,\aLB + c\,\aLW )\,\vle 
\nl
{}&+& 4\,c\,( \apL - \apLo + 3\,\apLt + 3\,c\,s\,\aLB ) 
          + 6\,( 1 + 2\,s^2 )\,c\,\aLW \Bigr]\,\frac{1}{c}
\eqas
\vspace{0.5cm}
\bei
\item \fbox{$\PQU$ quark self-energy}
\eei
\bqas
\Sigma^{(4)}_{\PQU\PQU\,;\,0} &=&
         \frac{1}{2}
         \,\LR\,\xpDs\,\xpU
       -
         \frac{1}{72}
         \,( 64\,s^2\,c^2 - 9\,\mrG_{42} )
         \,\frac{1}{c^2}
         \,\xpU
         \,( 1 - 2\,\LR )
\eqas
\bqas
\Sigma^{(4)}_{\PQU\PQU\,;\,1} &=& 0
\eqas
\bqas
\Sigma^{(4)}_{\PQU\PQU\,;\,2} &=&
         \frac{1}{2}
         \,\xpDs\,\xpU
         \,\sbfun{M}{\mqD}
     -
         \frac{1}{4}
         \,\xpU\,\xpUs
         \,\sbfun{\mh}{\mqU}
\nl &+&
         \frac{16}{9}
         \,s^2\,\xpU
         \,\sbfun{0}{\mqU}
     +
         \frac{1}{4}
         \,( c^2\,\xpUs - \mrG_{42} )
         \,\frac{1}{c^2}
         \,\xpU
         \,\sbfun{\mzb}{\mqU}
\eqas
\bqas
\Sigma^{(6)}_{\PQU\PQU\,;\,0} &=&
         \frac{1}{3}
         \,\frac{s}{c}
         \,\vqu\,\apWB\,\xpU
         \,( 2 - 5\,\LR )
\nl &+&
         \frac{1}{8}
         \,\Bigl[ 2\,\mrH_{9}\,c^3\,\aUW + ( 1 + c^2\,\xpUs )\,\vqu\, \aUBW \Bigr]
         \,\frac{1}{c^3}
         \,\xpU
         \,( 2 - 3\,\LR )
\nl &+&
         \frac{1}{72}
         \,\Bigl\{ 9\,s\,c\,\vqu\,\xps\,\xpU\,\aUB - 32\,\mrG_{48}\,s^2\,c^2\,\apW\,\xpU 
         - 24\,\mrL_{7}\,s\,c^2\,\xpU\,\aUWB 
\nl &-& 9\,\Bigl[ \mrG_{25}\,\xpU\,\aUW 
         - 128\,( M\,\DUV\,\UVst ) \Bigr]\,c^2\,\xps 
         + 18\,( \mrG_{12}\,\apUVA - 2\,\mrG_{13}\,\apU )\,\xpU \Bigr\}
         \,\frac{1}{c^2}
\nl &+&
         \frac{1}{36}
         \,\Bigl\{ 8\,\mrG_{49}\,s^2\,c^2\,\apW - 9\,\Bigl[ 3\,\xphs\,\aUp 
         - 4\,\xpDs\,\apW - 4\,\mrH_{9}\,\apQt + \mrI_{6}\,\aUp 
         - 2\,( \apUA - \apBox - 2\,s\,\aUWB )\,\xpUs \Bigr]\,c^2 
\nl &-& 3\,\Bigl[ 3\,\mrG_{43}\,\apUVA - 6\,\mrG_{44}\,\apU + ( 3\,\aUp 
         - 8\,( c\,\apB + s\,\apWB )\,s^2\,c\,\vqu ) \Bigr] \Bigr\}
         \,\frac{\LR}{c^2}
         \,\xpU
\nl &+&
         \frac{1}{144}
         \,\Bigl\{ 96\,\mrF^{b}_{31}\,s^2\,\vqu\,\apW + \Bigl[ 36\,\mrG_{12}\,\apW 
         - 3\,\mrG_{47}\,\apD + 8\,( 8\,c^4\,\apD 
\nl &-& ( - 32\,c^3\,\apWB + 3\,s\,\vqu\,\apD )\,s ) \Bigr] \Bigr\}
         \,\frac{1}{c^2}
         \,\xpU
         \,( 1 - 2\,\LR )
\nl &+&
         \frac{1}{2}
         \,\sumg (  - 3\,\xpdc\,\aoQuQd + \xpls\,\xpl\,\aoLlQu )
         \,\LR
\eqas
\bqas
\Sigma^{(6)}_{\PQU\PQU\,;\,1} &=&
         \frac{3}{4}
         \,\xphs\,\xpU\,\aUp
         \,\afun{\mh}
     -
         \frac{1}{2}
         \,\sumg 
         \,\xplc\,\aoLlQu
         \,\afun{\mle}
     +
         \frac{3}{2}
         \,\sumg 
         \,\xpdc\,\aoQuQd
         \,\afun{\mqd}
\nl &+&
         \frac{1}{8}
         \,( 3\,\aUW - 4\,\apQt )
         \,\xpDs\,\xpU
         \,\afun{\mqD}
     +
         \frac{1}{8}
         \,( 3\,\aUW - 4\,\apQt + 2\,\aUp )
         \,\xpU
         \,\afun{M}
\nl &+&
         \frac{1}{16}
         \,( 3\,\vqu\,\aUBW - 4\,c\,\apUA + 4\,c\,\aUp )
         \,\frac{1}{c^3}
         \,\xpU
         \,\afun{\mzb}
\nl &+&
         \frac{1}{16}
         \,( 3\,\vqu\,\aUBW - 4\,c\,\apUA + 8\,s\,c\,\aUWB )
         \,\frac{1}{c}
         \,\xpU\,\xpUs
         \,\afun{\mqU}
\eqas
\bqas
\Sigma^{(6)}_{\PQU\PQU\,;\,2} &=&
         \frac{1}{8}
         \,\Bigl[ 8\,\xpDs\,\apW + ( 3\,\aUW - 4\,\apQt )\,\xps 
         - ( 3\,\aUW - 4\,\apQt )\,\mrH_{9} \Bigr]
         \,\xpU
         \,\sbfun{M}{\mqD}
\nl &+&
         \frac{1}{48}
         \,\Bigl\{ 6\,c^3\,\apWDm\,\xpUs - 64\,\mrF^{b}_{3}\,s^2\,c\,\vqu\,\aAA
         - 16\,\mrF^{b}_{4}\,s\,c^2\,\vqu\,\aAZ 
\nl &-& \Bigl[ 9\,\aUBW 
         + 16\,( \apD - 4\,c^2\,\aZZ )\,s^2\,c \Bigr]\,\vqu + 3\,( 3\,\vqu\,\aUBW 
         - 4\,c\,\apUA )\,\mrL_{6}\,c^2 
\nl &-& 2\,( 12\,\mrG_{12}\,\apW
         + 3\,\mrG_{45}\,\apUVA - 6\,\mrG_{46}\,\apU - \mrG_{47}\,\apD )\,c \Bigr\}
         \,\frac{1}{c^3}
         \,\xpU
         \,\sbfun{\mzb}{\mqU}
\nl &+&
         \frac{1}{18}
         \,( 16\,\apWAD + 9\,\mrL_{6}\,s\,\aUWB )
         \,\xpU
         \,\sbfun{0}{\mqU}
\nl &-&
         \frac{1}{8}
         \,( \apWDm + 4\,\aupBox )
         \,\xpU\,\xpUs
         \,\sbfun{\mh}{\mqU}
\eqas
\bqas
V^{(4)}_{\PQU\PQU\,;\,0} &=&
       -
         \frac{1}{4}
       +
         \frac{1}{8}
         \,( 3\,\xpUs + \mrH_{12} )
         \,\LR
       -
         \frac{1}{144}
         \,( 64\,s^2\,c^2 + 9\,\mrG_{4} )
         \,\frac{1}{c^2}
         \,( 1 - \LR )
\eqas
\bqas
V^{(4)}_{\PQU\PQU\,;\,1} &=&
       -
         \frac{1}{8}
         \,\frac{\xpUs}{\xps}
         \,\xphs
         \,\afun{\mh}
     +
         \frac{1}{8}
         \,( \xpUs + \mrH_{12} )
         \,\frac{\xpDs}{\xps}
         \,\afun{\mqD}
\nl &-&
         \frac{1}{8}
         \,( \xpUs + \mrH_{12} )
         \,\frac{1}{\xps}
         \,\afun{M}
     -
         \frac{1}{16}
         \,( 2\,c^2\,\xpUs + \mrG_{4} )
         \,\frac{1}{c^4\,\xps}
         \,\afun{\mzb}
\nl &+&
         \frac{1}{144}
         \,( 36\,c^2\,\xpUs + 64\,s^2\,c^2 + 9\,\mrG_{4} )
         \,\frac{\xpUs}{c^2\,\xps}
         \,\afun{\mqU}
\eqas
\bqas
V^{(4)}_{\PQU\PQU\,;\,2} &=&
         \frac{4}{9}
         \,\frac{s^2}{\xps}
         \,\mrL_{8}
         \,\sbfun{0}{\mqU}
\nl &-&
         \frac{1}{16}
         \,\Bigl[ ( \mrG_{4} - 2\,\mrL_{8}\,c^4\,\xpUs ) 
         - ( \mrG_{4}\,\xps - \mrG_{42}\,\xpUs )\,c^2 \Bigr]
         \,\frac{1}{c^4\,\xps}
         \,\sbfun{\mzb}{\mqU}
\nl &+&
         \frac{1}{8}
         \,( \xpUs + \mrJ_{34} )
         \,\frac{\xpUs}{\xps}
         \,\sbfun{\mh}{\mqU}
\nl &+&
         \frac{1}{8}
         \,( \xps\,\xpUs - \mrH_{10}\,\xpUs - \mrH_{11} + \mrH_{12}\,\xps )
         \,\frac{1}{\xps}
         \,\sbfun{M}{\mqD}
\eqas
\bqas
V^{(6)}_{\PQU\PQU\,;\,0} &=&
         \frac{1}{48}
         \,\Bigl[ 6\,c\,\xpDs\,\aDW - ( 3\,\vqu\,\aUBW + 8\,s\,c\,\aUWB )\,\xpUs \Bigr]
         \,\frac{1}{c}
         \,( 2 - 3\,\LR )
\nl &+&
         \frac{1}{96}
         \,\Bigl\{ 8\,s^2\,\vqu\,\apD + \Bigl[ 24\,\mrG_{12}\,\apU + \mrG_{50}\,\apD 
         - 12\,( \apUVA + \apW )\,\mrG_{13} \Bigr] \Bigr\}
         \,\frac{1}{c^2}
         \,( 1 - \LR )
\nl &-&
         \frac{1}{72}
         \,\Bigl\{  - \Bigl[ 18\,\mrH_{12}\,\apW + 36\,\mrH_{13}\,\apQt 
         - 9\,( 2\,\apUA - 6\,\apW + \apD - 2\,\aupBox )\,\xpUs 
\nl &+& 8\,( 2\,\apWAD + 3\,s^2\,\vqu\,\aZZ ) \Bigr]\,c^2 
         + 24\,( c\,\aAZ + s\,\aAA )\,\mrF^{b}_{3}\,s\,\vqu \Bigr\}
         \,\frac{\LR}{c^2}
\nl &-&
         \frac{1}{36}
         \,\Bigl\{ \Bigl[ ( 9\,\apQWt - 8\,c^2\,\apD )\,c 
         - 4\,( 3\,\vqu - 8\,c^2 )\,s\,\apWB \Bigr]\,c 
         + 4\,( 2\,\mrG_{48}\,c^2 - 3\,\mrF^{b}_{31}\,\vqu )\,s^2\,\apW \Bigr\}
         \,\frac{1}{c^2}
\eqas
\bqas
V^{(6)}_{\PQU\PQU\,;\,1} &=&
         \frac{1}{96}
         \,\Bigl\{ 16\,\mrK_{7}\,s\,c^3\,\vqu\,\aAZ 
         + \Bigl[ 24\,\mrG_{12}\,\apU + \mrG_{50}\,\apD - 12\,( \apUVA + \apW )\,\mrG_{13} \Bigr] 
\nl &+& 8\,\Bigl[ 4\,\mrF^{b}_{3}\,\aAA + ( \apD - 4\,c^2\,\aZZ ) \Bigr]\,s^2\,\vqu 
         + 2\,( 9\,\vqu\,\aUBW - 12\,c\,\apUA
\nl &-& 3\,c\,\apWDm - 8\,s\,c^2\,\vqu\,\aAZ )\,c\,\xpUs \Bigr\}
         \,\frac{1}{c^4\,\xps}
         \,\afun{\mzb}
\nl &-&
         \frac{1}{288}
         \,\Bigl\{ \Bigl[ 72\,\mrG_{12}\,\apU + 3\,\mrG_{50}\,\apD
         - 36\,( \apUVA + \apW )\,\mrG_{13} + 8\,( 8\,c^4\,\apD
\nl &+& ( 4\,c\,\apWB\,vta + 3\,s\,\vqu\,\apD )\,s ) \Bigr] 
         + 18\,( 3\,\vqu\,\aUBW - 4\,c\,\apUA - 2\,c\,\apWDm 
         - 4\,c\,\aupBox + 8\,s\,c\,\aUWB )\,c\,\xpUs 
\nl &-& 32\,( 2\,\mrG_{48}\,c^2 - 3\,\mrF^{b}_{31}\,\vqu )\,s^2\,\apW \Bigr\}
         \,\frac{\xpUs}{c^2\,\xps}
         \,\afun{\mqU}
\nl &-&
         \frac{1}{16}
         \,( \apWDm + 4\,\aupBox )
         \,\frac{\xpUs}{\xps}
         \,\xphs
         \,\afun{\mh}
\nl &+&
         \frac{1}{8}
         \,( 2\,\apW\,\xpUs + 3\,\xpDs\,\aDW + \mrH_{12}\,\apQWt )
         \,\frac{\xpDs}{\xps}
         \,\afun{\mqD}
\nl &-&
         \frac{1}{8}
         \,( 2\,\apW\,\xpUs + 3\,\xpDs\,\aDW + \mrH_{12}\,\apQWt )
         \,\frac{1}{\xps}
         \,\afun{M}
\eqas
\bqas
V^{(6)}_{\PQU\PQU\,;\,2} &=&
         \frac{1}{16}
         \,\Bigl[ ( \apWDm + 4\,\aupBox )\,\xpUs + ( \apWDm + 4\,\aupBox )\,\mrJ_{34} \Bigr]
         \,\frac{\xpUs}{\xps}
         \,\sbfun{\mh}{\mqU}
\nl &+&
         \frac{1}{96}
         \,\Bigl\{ 6\,\mrL_{8}\,c^4\,\apWDm\,\xpUs + 32\,\mrF^{b}_{31}\,s^2\,\vqu\,\apW 
         + 16\,\mrF^{b}_{33}\,s\,c^2\,\vqu\,\apWA\,\xpUs 
\nl &+& 8\,\Bigl[ 4\,c\,\apWB + ( \apD - 8\,c^2\,\apB )\,s \Bigr]\,s\,\vqu 
         + \Bigl[ 24\,\mrG_{12}\,\apU + \mrG_{50}\,\apD 
         - 12\,( \apUVA + \apW )\,\mrG_{13} \Bigr] 
\nl &-& \Bigl[ 24\,\mrG_{12}\,\xps\,\apU
         + 12\,\mrG_{12}\,\apW\,\xpUs - \mrG_{47}\,\apD\,\xpUs 
         + \mrG_{50}\,\apD\,\xps - 12\,( \apUVA + \apW )\,\mrG_{13}\,\xps \Bigr]\,c^2 
\nl &+& 2\,( 9\,\aUBW + 12\,c\,\apUV + 40\,s^2\,c^2\,\apWZ )\,c\,\vqu\,\xpUs 
         - 16\,( \aAZ - 2\,s\,c\,\aZZ )\,s\,c^3\,\vqu\,\xps 
\nl &+& 6\,( 3\,\vqu\,\aUBW - 4\,c\,\apUA )\,\mrL_{6}\,c^3\,\xpUs
         + 8\,( 8\,\mrI_{7}\,\apB - 4\,\mrI_{8}\,\apW - \mrL_{8}\,\apD - 4\,\mrF^{b}_{3}\,\xps\,\aAA 
         - 4\,\mrF^{b}_{32}\,\apB\,\xpUs )\,s^2\,c^2\,\vqu 
\nl &-& 16\,( \mrK_{6}\,\xps 
         + \mrL_{9}\,\xpUs )\,s\,c^5\,\vqu\,\aAZ \Bigr\}
         \,\frac{1}{c^4\,\xps}
         \,\sbfun{\mzb}{\mqU}
\nl &+&
         \frac{1}{8}
         \,( 2\,\apW\,\xps\,\xpUs - 3\,\xpDs\,\xps\,\aDW - 2\,\mrH_{10}\,\apW\,\xpUs 
         - 3\,\mrH_{10}\,\xpDs\,\aDW - \mrH_{11}\,\apQWt 
         + 2\,\mrH_{12}\,\apW\,\xps 
\nl &+& 4\,\mrH_{13}\,\xps\,\apQt )
         \,\frac{1}{\xps}
         \,\sbfun{M}{\mqD}
\nl &+&
         \frac{1}{18}
         \,( 9\,\mrL_{6}\,s\,\xpUs\,\aUWB + 4\,\mrL_{8}\,\apWAD - 3\,\mrL_{10}\,s\,c\,\vqu\,\aAZ )
         \,\frac{1}{\xps}
         \,\sbfun{0}{\mqU}
\eqas
\bqas
A^{(4)}_{\PQU\PQU\,;\,0} &=&
       -
         \frac{1}{4}
       -
         \frac{1}{8}
         \,\frac{1}{c^2}
         \,\vqu
         \,( 1 - \LR )
       -
         \frac{1}{8}
         \,( \xpUs - \mrH_{12} )
         \,\LR
\eqas
\bqas
A^{(4)}_{\PQU\PQU\,;\,1} &=&
         \frac{1}{8}
         \,\frac{\xpUs}{c^2\,\xps}
         \,\vqu
         \,\afun{\mqU}
     -
         \frac{1}{8}
         \,\frac{1}{c^4\,\xps}
         \,\vqu
         \,\afun{\mzb}
\nl &-&
         \frac{1}{8}
         \,( \xpUs - \mrH_{12} )
         \,\frac{\xpDs}{\xps}
         \,\afun{\mqD}
     +
         \frac{1}{8}
         \,( \xpUs - \mrH_{12} )
         \,\frac{1}{\xps}
         \,\afun{M}
\eqas
\bqas
A^{(4)}_{\PQU\PQU\,;\,2} &=&
       -
         \frac{1}{8}
         \,( 1 - \mrL_{8}\,c^2 )
         \,\frac{1}{c^4\,\xps}
         \,\vqu
         \,\sbfun{\mzb}{\mqU}
\nl &-&
         \frac{1}{8}
         \,( \xps\,\xpUs - \mrH_{10}\,\xpUs + \mrH_{11} - \mrH_{12}\,\xps )
         \,\frac{1}{\xps}
         \,\sbfun{M}{\mqD}
\eqas
\bqas
A^{(6)}_{\PQU\PQU\,;\,0} &=&
         \frac{1}{6}
         \,\frac{s^2}{c^2}
         \,\aAA
         \,( 1 - 2\,\LR )
     -
         \frac{1}{6}
         \,s\,c\,\aAZ
         \,( 1 + 2\,\LR )
\nl &-&
         \frac{1}{6}
         \,\Bigl[ \mrF^{b}_{34}\,c^2\,\aZZ - (  - 6\,c^2\,\apQt + s^4\,\aAA ) \Bigr]
         \,\frac{1}{c^2}
\nl &+&
         \frac{1}{96}
         \,\Bigl\{ 8\,s^2\,\apD - \Bigl[ 24\,\mrG_{12}\,\apU + \mrG_{51}\,\apD 
         + 12\,( \apUVA + \apW )\,\mrG_{13} \Bigr] \Bigr\}
         \,\frac{1}{c^2}
         \,( 1 - \LR )
\nl &-&
         \frac{1}{12}
         \,\Bigl\{  - 3\,\Bigl[ \mrH_{12}\,\apW + 2\,\mrH_{13}\,\apQt 
         - ( \apUV + \apW )\,\xpUs \Bigr]\,c + 4\,( \aAZ + s\,c\,\aAA - s\,c\,\aZZ )\,s \Bigr\}
         \,\frac{\LR}{c}
\nl &-&
         \frac{1}{16}
         \,( \xpUs\,\aUBW - 2\,c\,\xpDs\,\aDW )
         \,( 2 - 3\,\LR )
         \,\frac{1}{c}
\eqas
\bqas
A^{(6)}_{\PQU\PQU\,;\,1} &=&
         \frac{1}{96}
         \,\Bigl\{ 16\,\mrK_{7}\,s\,c^3\,\aAZ - \Bigl[ 24\,\mrG_{12}\,\apU 
         + \mrG_{51}\,\apD + 12\,( \apUVA + \apW )\,\mrG_{13} \Bigr] 
\nl &+& 8\,\Bigl[ 4\,\mrF^{b}_{3}\,\aAA + ( \apD - 4\,c^2\,\aZZ ) \Bigr]\,s^2 
         + 2\,( 9\,\aUBW - 12\,c\,\apUV - 8\,s\,c^2\,\aAZ )\,c\,\xpUs \Bigr\}
         \,\frac{1}{c^4\,\xps}
         \,\afun{\mzb}
\nl &-&
         \frac{1}{96}
         \,\Bigl\{ 32\,\mrF^{b}_{3}\,s^2\,\aAA + 8\,\Bigl[ 4\,c^3\,\aAZ 
         + ( \apD - 4\,c^2\,\aZZ )\,s \Bigr]\,s 
\nl &-& \Bigl[ 24\,\mrG_{12}\,\apU 
         + \mrG_{51}\,\apD + 12\,( \apUVA + \apW )\,\mrG_{13} \Bigr] 
         + 6\,( 3\,\aUBW - 4\,c\,\apUV )\,c\,\xpUs \Bigr\}
         \,\frac{\xpUs}{c^2\,\xps}
         \,\afun{\mqU}
\nl &-&
         \frac{1}{8}
         \,( 2\,\apW\,\xpUs - 3\,\xpDs\,\aDW - \mrH_{12}\,\apQWt )
         \,\frac{\xpDs}{\xps}
         \,\afun{\mqD}
\nl &+&
         \frac{1}{8}
         \,( 2\,\apW\,\xpUs - 3\,\xpDs\,\aDW - \mrH_{12}\,\apQWt )
         \,\frac{1}{\xps}
         \,\afun{M}
\eqas
\bqas
A^{(6)}_{\PQU\PQU\,;\,2} &=&
       -
         \frac{1}{6}
         \,\frac{1}{\xps}
         \,\mrL_{10}
         \,s\,c\,\aAZ
         \,\sbfun{0}{\mqU}
\nl &+&
         \frac{1}{96}
         \,\Bigl\{ 32\,\mrF^{b}_{31}\,s^2\,\apW + 16\,\mrF^{b}_{33}\,s\,c^2\,\apWA\,\xpUs 
         + 8\,\Bigl[ 4\,c\,\apWB + ( \apD - 8\,c^2\,\apB )\,s \Bigr]\,s 
\nl &-& \Bigl[ 24\,\mrG_{12}\,\apU + \mrG_{51}\,\apD 
         + 12\,( \apUVA + \apW )\,\mrG_{13} \Bigr] + 6\,( 3\,\aUBW
         - 4\,c\,\apUV )\,\mrL_{6}\,c^3\,\xpUs 
\nl &+& 2\,( 9\,\aUBW + 12\,c\,\vqu\,\apUA 
         + 40\,s^2\,c^2\,\apWZ )\,c\,\xpUs - 16\,( \aAZ - 2\,s\,c\,\aZZ )\,s\,c^3\,\xps 
\nl &+& ( 24\,\mrG_{12}\,\xps\,\apU + 12\,\mrG_{13}\,\xps\,\apUVA 
         + 12\,\mrG_{13}\,\mrL_{8}\,\apW + \mrG_{51}\,\mrL_{8}\,\apD )\,c^2 
\nl &+& 8\,( 8\,\mrI_{7}\,\apB - 4\,\mrI_{8}\,\apW - \mrL_{8}\,\apD 
         - 4\,\mrF^{b}_{3}\,\xps\,\aAA - 4\,\mrF^{b}_{32}\,\apB\,\xpUs )\,s^2\,c^2 
\nl &-& 16\,( \mrK_{6}\,\xps + \mrL_{9}\,\xpUs )\,s\,c^5\,\aAZ \Bigr\}
         \,\frac{1}{c^4\,\xps}
         \,\sbfun{\mzb}{\mqU}
\nl &-&
         \frac{1}{8}
         \,( 2\,\apW\,\xps\,\xpUs + 3\,\xpDs\,\xps\,\aDW - 2\,\mrH_{10}\,\apW\,\xpUs 
         + 3\,\mrH_{10}\,\xpDs\,\aDW + \mrH_{11}\,\apQWt 
         - 2\,\mrH_{12}\,\apW\,\xps 
\nl &-& 4\,\mrH_{13}\,\xps\,\apQt )
         \,\frac{1}{\xps}
         \,\sbfun{M}{\mqD}
\eqas
We have introduced
\bqas
\UVst &=& - \frac{1}{256}\,\frac{\mUs}{M}\,\Bigl[
          3\,( s\,\aUB - c\,\aUW )\,\vqu 
\nl
{}&+& 4\,c\,( \apQo - \apU + 3\,\apQt + 2\,c\,s\,\aUB ) 
          - 2\,( 3 + 4\,s^2 )\,c\,\aUW \Bigr]\,\frac{1}{c}
\eqas
\vspace{0.5cm}
\bei
\item \fbox{$\PQD$ quark self-energy}
\eei
\bqas
\Sigma^{(4)}_{\PQD\PQD\,;\,0} &=&
         \frac{1}{2}
         \,\LR\,\xpUs\,\xpD
       -
         \frac{1}{72}
         \,( 16\,s^2\,c^2 - 9\,\mrG_{52} )
         \,\frac{1}{c^2}
         \,\xpD
         \,( 1 - 2\,\LR )
\eqas
\bqas
\Sigma^{(4)}_{\PQD\PQD\,;\,1} &=& 0
\eqas
\bqas
\Sigma^{(4)}_{\PQD\PQD\,;\,2} &=&
         \frac{1}{2}
         \,\xpUs\,\xpD
         \,\sbfun{M}{\mqU}
     -
         \frac{1}{4}
         \,\xpD\,\xpDs
         \,\sbfun{\mh}{\mqD}
\nl &+&
         \frac{4}{9}
         \,s^2\,\xpD
         \,\sbfun{0}{\mqD}
     +
         \frac{1}{4}
         \,( c^2\,\xpDs - \mrG_{52} )
         \,\frac{1}{c^2}
         \,\xpD
         \,\sbfun{\mzb}{\mqD}
\eqas
\bqas
\Sigma^{(6)}_{\PQD\PQD\,;\,0} &=& 
         \frac{1}{6}
         \,\frac{s}{c}
         \,\vqd\,\apWB\,\xpD
         \,( 2 - 5\,\LR )
\nl &-&
         \frac{1}{8}
         \,\Bigl[ 2\,\mrH_{14}\,c^3\,\aDW + ( 1 + c^2\,\xpDs )\,\vqd\,\aDBW \Bigr]
         \,\frac{1}{c^3}
         \,\xpD
         \,( 2 - 3\,\LR )
\nl &+&
         \frac{1}{72}
         \,\Bigl\{ 9\,s\,c\,\vqd\,\xps\,\xpD\,\aDB - 16\,\mrG_{55}\,s^2\,c^2\,\apW\,\xpD 
         + 12\,\mrL_{12}\,s\,c^2\,\xpD\,\aDWB 
\nl &+& 9\,\Bigl[ \mrG_{27}\,\aDW\,\xpD + 128\,( M\,\DUV\,\UVsb ) \Bigr]\,c^2\,\xps 
         + 18\,( \mrG_{14}\,\apDVA + 2\,\mrG_{15}\,\apD )\,\xpD \Bigr\}
         \,\frac{1}{c^2}
\nl &+&
         \frac{1}{36}
         \,\Bigl\{ 4\,\mrG_{59}\,s^2\,c^2\,\apW + 9\,\Bigl[ 3\,\xphs\,\aDp 
         + 4\,\xpUs\,\apW + 4\,\mrH_{14}\,\apQt + \mrI_{9}\,\aDp 
\nl &+& 2\,( \apDA - \apBox + s\,\aDWB )\,\xpDs \Bigr]\,c^2 
         - 3\,\Bigl[ 3\,\mrG_{53}\,\apDVA + 6\,\mrG_{54}\,\apD 
\nl &-& ( 3\,\aDp + 4\,( c\,\apB + s\,\apWB )\,s^2\,c\,\vqd ) \Bigr] \Bigr\}
         \,\frac{\LR}{c^2}
         \,\xpD
\nl &+&
         \frac{1}{144}
         \,\Bigl\{ 48\,\mrF^{b}_{31}\,s^2\,\vqd\,\apW 
         + \Bigl[ 36\,\mrG_{14}\,\apW - 3\,\mrG_{58}\,\apD + 4\,( 4\,c^4\,\apD 
\nl &+& ( - 16\,c^3\,\apWB + 3\,s\,\vqd\,\apD )\,s ) \Bigr] \Bigr\}
         \,\frac{1}{c^2}
         \,\xpD
         \,( 1 - 2\,\LR )
\nl &-&
         \frac{1}{2}
         \,\sumg ( 3\,\xpuc\,\aoQuQd + \xpls\,\xpl\,\aLldQ )
         \,\LR
\eqas
\bqas
\Sigma^{(6)}_{\PQD\PQD\,;\,1} &=& 
       -
         \frac{3}{4}
         \,\xphs\,\xpD\,\aDp
         \,\afun{\mh}
     +
         \frac{1}{2}
         \,\sumg 
         \,\xplc\,\aLldQ
         \,\afun{\mle}
     +
         \frac{3}{2}
         \,\sumg 
         \,\xpuc\,\aoQuQd
         \,\afun{\mqu}
\nl &-&
         \frac{1}{8}
         \,( 2\,\aDp + 3\,\aDW + 4\,\apQt )
         \,\xpD
         \,\afun{M}
     -
         \frac{1}{8}
         \,( 3\,\aDW + 4\,\apQt )
         \,\xpUs\,\xpD
         \,\afun{\mqU}
\nl &-&
         \frac{1}{16}
         \,( 3\,\vqd\,\aDBW + 4\,c\,\apDA + 4\,c\,\aDp )
         \,\frac{1}{c^3}
         \,\xpD
         \,\afun{\mzb}
\nl &-&
         \frac{1}{16}
         \,( 3\,\vqd\,\aDBW + 4\,c\,\apDA + 4\,s\,c\,\aDWB )
         \,\frac{1}{c}
         \,\xpD\,\xpDs
         \,\afun{\mqD}
\eqas
\bqas
\Sigma^{(6)}_{\PQD\PQD\,;\,2} &=& 
         \frac{1}{8}
         \,\Bigl[ 8\,\xpUs\,\apW - ( 3\,\aDW + 4\,\apQt )\,\xps 
         + ( 3\,\aDW + 4\,\apQt )\,\mrH_{14} \Bigr]
         \,\xpD
         \,\sbfun{M}{\mqU}
\nl &+&
         \frac{1}{48}
         \,\Bigl\{ 6\,c^3\,\apWDm\,\xpDs - 32\,\mrF^{b}_{3}\,s^2\,c\,\vqd\,\aAA
         - 8\,\mrF^{b}_{4}\,s\,c^2\,\vqd\,\aAZ 
\nl &+& \Bigl[ 9\,\aDBW - 8\,( \apD
         - 4\,c^2\,\aZZ )\,s^2\,c \Bigr]\,\vqd 
         - 3\,( 3\,\vqd\,\aDBW + 4\,c\,\apDA )\,\mrL_{11}\,c^2 
\nl &-& 2\,( 12\,\mrG_{14}\,\apW + 3\,\mrG_{56}\,\apDVA + 6\,\mrG_{57}\,\apD 
         - \mrG_{58}\,\apD )\,c \Bigr\}
         \,\frac{1}{c^3}
         \,\xpD
         \,\sbfun{\mzb}{\mqD}
\nl &+&
         \frac{1}{36}
         \,( 8\,\apWAD - 9\,\mrL_{11}\,s\,\aDWB )
         \,\xpD
         \,\sbfun{0}{\mqD}
\nl &-&
         \frac{1}{8}
         \,( \apWDm - 4\,\adpBox )
         \,\xpD\,\xpDs
         \,\sbfun{\mh}{\mqD}
\eqas
\bqas
V^{(4)}_{\PQD\PQD\,;\,0} &=& 
       -
         \frac{1}{4}
       +
         \frac{1}{8}
         \,( 3\,\xpDs + \mrH_{17} )
         \,\LR
       -
         \frac{1}{144}
         \,( 16\,s^2\,c^2 + 9\,\mrG_{5} )
         \,\frac{1}{c^2}
         \,( 1 - \LR )
\eqas
\bqas
V^{(4)}_{\PQD\PQD\,;\,1} &=& 
       -
         \frac{1}{8}
         \,\frac{\xpDs}{\xps}
         \,\xphs
         \,\afun{\mh}
     +
         \frac{1}{8}
         \,( \xpDs + \mrH_{17} )
         \,\frac{\xpUs}{\xps}
         \,\afun{\mqU}
\nl &-&
         \frac{1}{8}
         \,( \xpDs + \mrH_{17} )
         \,\frac{1}{\xps}
         \,\afun{M}
     -
         \frac{1}{16}
         \,( 2\,c^2\,\xpDs + \mrG_{5} )
         \,\frac{1}{c^4\,\xps}
         \,\afun{\mzb}
\nl &+&
         \frac{1}{144}
         \,( 36\,c^2\,\xpDs + 16\,s^2\,c^2 + 9\,\mrG_{5} )
         \,\frac{\xpDs}{c^2\,\xps}
         \,\afun{\mqD}
\eqas
\bqas
V^{(4)}_{\PQD\PQD\,;\,2} &=& 
         \frac{1}{9}
         \,\frac{s^2}{\xps}
         \,\mrL_{13}
         \,\sbfun{0}{\mqD}
\nl &-&
         \frac{1}{16}
         \,\Bigl[ ( \mrG_{5} - 2\,\mrL_{13}\,c^4\,\xpDs )
       - ( \mrG_{5}\,\xps - \mrG_{52}\,\xpDs )\,c^2 \Bigr]
         \,\frac{1}{c^4\,\xps}
         \,\sbfun{\mzb}{\mqD}
\nl &+&
         \frac{1}{8}
         \,( \xpDs + \mrJ_{34} )
         \,\frac{\xpDs}{\xps}
         \,\sbfun{\mh}{\mqD}
\nl &+&
         \frac{1}{8}
         \,( \xps\,\xpDs - \mrH_{15}\,\xpDs - \mrH_{16} + \mrH_{17}\,\xps )
         \,\frac{1}{\xps}
         \,\sbfun{M}{\mqU}
\eqas
\bqas
V^{(6)}_{\PQD\PQD\,;\,0} &=& 
       -
         \frac{1}{48}
         \,\Bigl[ 6\,c\,\xpUs\,\aUW - ( 3\,\vqd\,\aDBW + 4\,s\,c\,\aDWB )\,\xpDs \Bigr]
         \,\frac{1}{c}
         \,( 2 - 3\,\LR )
\nl &+&
         \frac{1}{96}
         \,\Bigl\{ 4\,s^2\,\vqd\,\apD - \Bigl[ 24\,\mrG_{14}\,\apD - \mrG_{60}\,\apD 
         + 12\,( \apDVA + \apW )\,\mrG_{15} \Bigr] \Bigr\}
         \,\frac{1}{c^2}
         \,( 1 - \LR )
\nl &-&
         \frac{1}{72}
         \,\Bigl\{  - \Bigl[ 18\,\mrH_{17}\,\apW + 36\,\mrH_{18}\,\apQt 
         - 9\,( 2\,\apDA - 6\,\apW + \apD + 2\,\adpBox )\,\xpDs 
\nl &+& 4\,( \apWAD + 3\,s^2\,\vqd\,\aZZ ) \Bigr]\,c^2 
         + 12\,( c\,\aAZ + s\,\aAA )\,\mrF^{b}_{3}\,s\,\vqd \Bigr\}
         \,\frac{\LR}{c^2}
\nl &-&
         \frac{1}{36}
         \,\Bigl\{ \Bigl[ ( 9\,\apQWt - 2\,c^2\,\apD )\,c 
         - 2\,( 3\,\vqd - 4\,c^2 )\,s\,\apWB \Bigr]\,c + 2\,( 2\,\mrG_{55}\,c^2 
         - 3\,\mrF^{b}_{31}\,\vqd )\,s^2\,\apW \Bigr\}
         \,\frac{1}{c^2}
\eqas
\bqas
V^{(6)}_{\PQD\PQD\,;\,1} &=& 
         \frac{1}{96}
         \,\Bigl\{ 8\,\mrK_{7}\,s\,c^3\,\vqd\,\aAZ 
         - \Bigl[ 24\,\mrG_{14}\,\apD - \mrG_{60}\,\apD 
         + 12\,( \apDVA + \apW )\,\mrG_{15} \Bigr] 
\nl &+& 4\,\Bigl[ 4\,\mrF^{b}_{3}\,\aAA + ( \apD
         - 4\,c^2\,\aZZ ) \Bigr]\,s^2\,\vqd - 2\,( 9\,\vqd\,\aDBW + 12\,c\,\apDA
         + 3\,c\,\apWDm 
\nl &+& 4\,s\,c^2\,\vqd\,\aAZ )\,c\,\xpDs \Bigr\}
         \,\frac{1}{c^4\,\xps}
         \,\afun{\mzb}
\nl &+&
         \frac{1}{288}
         \,\Bigl\{ \Bigl[ 72\,\mrG_{14}\,\apD - 3\,\mrG_{60}\,\apD
         + 36\,( \apDVA + \apW )\,\mrG_{15} - 4\,( 4\,c^4\,\apD
\nl &+& ( 4\,c\,\apWB\,vba + 3\,s\,\vqd\,\apD )\,s ) \Bigr] 
         + 18\,( 3\,\vqd\,\aDBW + 4\,c\,\apDA + 2\,c\,\apWDm 
         - 4\,c\,\adpBox + 4\,s\,c\,\aDWB )\,c\,\xpDs 
\nl &+& 16\,( 2\,\mrG_{55}\,c^2 
         - 3\,\mrF^{b}_{31}\,\vqd )\,s^2\,\apW \Bigr\}
         \,\frac{\xpDs}{c^2\,\xps}
         \,\afun{\mqD}
\nl &-&
         \frac{1}{16}
         \,( \apWDm - 4\,\adpBox )
         \,\frac{\xpDs}{\xps}
         \,\xphs
         \,\afun{\mh}
\nl &+&
         \frac{1}{8}
         \,( 2\,\apW\,\xpDs - 3\,\xpUs\,\aUW + \mrH_{17}\,\apQWt )
         \,\frac{\xpUs}{\xps}
         \,\afun{\mqU}
\nl &-&
         \frac{1}{8}
         \,( 2\,\apW\,\xpDs - 3\,\xpUs\,\aUW + \mrH_{17}\,\apQWt )
         \,\frac{1}{\xps}
         \,\afun{M}
\eqas
\bqas
V^{(6)}_{\PQD\PQD\,;\,2} &=& 
         \frac{1}{16}
         \,\Bigl[ ( \apWDm - 4\,\adpBox )\,\xpDs + ( \apWDm - 4\,\adpBox )\,\mrJ_{34} \Bigr]
         \,\frac{\xpDs}{\xps}
         \,\sbfun{\mh}{\mqD}
\nl &+&
         \frac{1}{96}
         \,\Bigl\{ 6\,\mrL_{13}\,c^4\,\apWDm\,\xpDs + 16\,\mrF^{b}_{31}\,s^2\,\vqd\,\apW 
         + 8\,\mrF^{b}_{33}\,s\,c^2\,\vqd\,\apWA\,\xpDs 
\nl &+& 4\,\Bigl[ 4\,c\,\apWB + ( \apD - 8\,c^2\,\apB )\,s \Bigr]\,s\,\vqd 
         + \Bigl[ -  12\,\mrG_{14}\,\apW\,\xpDs + 24\,\mrG_{14}\,\apD\,\xps 
         + \mrG_{58}\,\apD\,\xpDs - \mrG_{60}\,\apD\,\xps 
\nl &+& 12\,( \apDVA + \apW )\,\mrG_{15}\,\xps \Bigr]\,c^2 
\nl &-& \Bigl[ 24\,\mrG_{14}\,\apD - \mrG_{60}\,\apD 
         + 12\,( \apDVA + \apW )\,\mrG_{15} \Bigr] - 2\,( 9\,\aDBW 
         - 12\,c\,\apDV - 20\,s^2\,c^2\,\apWZ )\,c\,\vqd\,\xpDs 
\nl &-& 8\,( \aAZ - 2\,s\,c\,\aZZ )\,s\,c^3\,\vqd\,\xps
         - 6\,( 3\,\vqd\,\aDBW + 4\,c\,\apDA )\,\mrL_{11}\,c^3\,\xpDs
\nl &+& 4\,( 8\,\mrI_{10}\,\apB - 4\,\mrI_{11}\,\apW 
         - \mrL_{13}\,\apD - 4\,\mrF^{b}_{3}\,\xps\,\aAA 
\nl &-& 4\,\mrF^{b}_{32}\,\apB\,\xpDs )\,s^2\,c^2\,\vqd - 8\,( \mrK_{6}\,\xps
         + \mrL_{14}\,\xpDs )\,s\,c^5\,\vqd\,\aAZ \Bigr\}
         \,\frac{1}{c^4\,\xps}
         \,\sbfun{\mzb}{\mqD}
\nl &+&
         \frac{1}{8}
         \,( 2\,\apW\,\xps\,\xpDs + 3\,\xpUs\,\xps\,\aUW - 2\,\mrH_{15}\,\apW\,\xpDs 
         + 3\,\mrH_{15}\,\xpUs\,\aUW - \mrH_{16}\,\apQWt 
\nl &+& 2\,\mrH_{17}\,\apW\,\xps 
         + 4\,\mrH_{18}\,\xps\,\apQt )
         \,\frac{1}{\xps}
         \,\sbfun{M}{\mqU}
\nl &-&
         \frac{1}{36}
         \,( 9\,\mrL_{11}\,s\,\xpDs\,\aDWB - 2\,\mrL_{13}\,\apWAD 
         + 3\,\mrL_{15}\,s\,c\,\vqd\,\aAZ )
         \,\frac{1}{\xps}
         \,\sbfun{0}{\mqD}
\eqas
\bqas
A^{(4)}_{\PQD\PQD\,;\,0} &=& 
       -
         \frac{1}{4}
       -
         \frac{1}{8}
         \,\frac{1}{c^2}
         \,\vqd
         \,( 1 - \LR )
       -
         \frac{1}{8}
         \,( \xpDs - \mrH_{17} )
         \,\LR
\eqas
\bqas
A^{(4)}_{\PQD\PQD\,;\,1} &=& 
         \frac{1}{8}
         \,\frac{\xpDs}{c^2\,\xps}
         \,\vqd
         \,\afun{\mqD}
     -
         \frac{1}{8}
         \,\frac{1}{c^4\,\xps}
         \,\vqd
         \,\afun{\mzb}
\nl &-&
         \frac{1}{8}
         \,( \xpDs - \mrH_{17} )
         \,\frac{\xpUs}{\xps}
         \,\afun{\mqU}
     +
         \frac{1}{8}
         \,( \xpDs - \mrH_{17} )
         \,\frac{1}{\xps}
         \,\afun{M}
\eqas
\bqas
A^{(4)}_{\PQD\PQD\,;\,2} &=& 
       -
         \frac{1}{8}
         \,( 1 - \mrL_{13}\,c^2 )
         \,\frac{1}{c^4\,\xps}
         \,\vqd
         \,\sbfun{\mzb}{\mqD}
\nl &-&
         \frac{1}{8}
         \,( \xps\,\xpDs - \mrH_{15}\,\xpDs + \mrH_{16} - \mrH_{17}\,\xps )
         \,\frac{1}{\xps}
         \,\sbfun{M}{\mqU}
\eqas
\bqas
A^{(6)}_{\PQD\PQD\,;\,0} &=& 
       -
         \frac{1}{6}
         \,\frac{s^2}{c^2}
         \,\aAA
         \,( 1 + \LR )
\nl &-&
         \frac{1}{6}
         \,s\,c\,\aAZ
         \,( 2 + \LR )
\nl &-&
         \frac{1}{6}
         \,\Bigl[ \mrF^{b}_{0}\,c^2\,\aZZ - 2\,(  - 3\,c^2\,\apQt + s^4\,\aAA ) \Bigr]
         \,\frac{1}{c^2}
\nl &+&
         \frac{1}{96}
         \,\Bigl\{ 4\,s^2\,\apD + \Bigl[ 24\,\mrG_{14}\,\apD - \mrG_{61}\,\apD 
         - 12\,( \apDVA + \apW )\,\mrG_{15} \Bigr] \Bigr\}
         \,\frac{1}{c^2}
         \,( 1 - \LR )
\nl &-&
         \frac{1}{12}
         \,\Bigl\{  - 3\,\Bigl[ \mrH_{17}\,\apW + 2\,\mrH_{18}\,\apQt 
         - ( \apDV + \apW )\,\xpDs \Bigr]\,c + 2\,( \aAZ + s\,c\,\aAA - s\,c\,\aZZ )\,s \Bigr\}
         \,\frac{\LR}{c}
\nl &+&
         \frac{1}{16}
         \,( \xpDs\,\aDBW - 2\,c\,\xpUs\,\aUW )
         \,( 2 - 3\,\LR )
         \,\frac{1}{c}
\eqas
\bqas
A^{(6)}_{\PQD\PQD\,;\,1} &=& 
         \frac{1}{96}
         \,\Bigl\{ 8\,\mrK_{7}\,s\,c^3\,\aAZ + \Bigl[ 24\,\mrG_{14}\,\apD 
         - \mrG_{61}\,\apD - 12\,( \apDVA + \apW )\,\mrG_{15} \Bigr] 
\nl &+& 4\,\Bigl[ 4\,\mrF^{b}_{3}\,\aAA + ( \apD - 4\,c^2\,\aZZ ) \Bigr]\,s^2 
         - 2\,( 9\,\aDBW + 12\,c\,\apDV + 4\,s\,c^2\,\aAZ )\,c\,\xpDs \Bigr\}
         \,\frac{1}{c^4\,\xps}
         \,\afun{\mzb}
\nl &-&
         \frac{1}{96}
         \,\Bigl\{ 16\,\mrF^{b}_{3}\,s^2\,\aAA + 4\,\Bigl[ 4\,c^3\,\aAZ 
         + ( \apD - 4\,c^2\,\aZZ )\,s \Bigr]\,s 
\nl &+& \Bigl[ 24\,\mrG_{14}\,\apD 
         - \mrG_{61}\,\apD - 12\,( \apDVA + \apW )\,\mrG_{15} \Bigr] 
         - 6\,( 3\,\aDBW + 4\,c\,\apDV )\,c\,\xpDs \Bigr\}
         \,\frac{\xpDs}{c^2\,\xps}
         \,\afun{\mqD}
\nl &-&
         \frac{1}{8}
         \,( 2\,\apW\,\xpDs + 3\,\xpUs\,\aUW - \mrH_{17}\,\apQWt )
         \,\frac{\xpUs}{\xps}
         \,\afun{\mqU}
\nl &+&
         \frac{1}{8}
         \,( 2\,\apW\,\xpDs + 3\,\xpUs\,\aUW - \mrH_{17}\,\apQWt )
         \,\frac{1}{\xps}
         \,\afun{M}
\eqas
\bqas
A^{(6)}_{\PQD\PQD\,;\,2} &=& 
       -
         \frac{1}{12}
         \,\frac{1}{\xps}
         \,\mrL_{15}
         \,s\,c\,\aAZ
         \,\sbfun{0}{\mqD}
\nl &+&
         \frac{1}{96}
         \,\Bigl\{ 16\,\mrF^{b}_{31}\,s^2\,\apW + 8\,\mrF^{b}_{33}\,s\,c^2\,\apWA\,\xpDs 
         + 4\,\Bigl[ 4\,c\,\apWB + ( \apD - 8\,c^2\,\apB )\,s \Bigr]\,s 
\nl &+& \Bigl[ 24\,\mrG_{14}\,\apD - \mrG_{61}\,\apD 
         - 12\,( \apDVA + \apW )\,\mrG_{15} \Bigr] - 6\,( 3\,\aDBW
         + 4\,c\,\apDV )\,\mrL_{11}\,c^3\,\xpDs 
\nl &-& 2\,( 9\,\aDBW - 12\,c\,\vqd\,\apDA 
         - 20\,s^2\,c^2\,\apWZ )\,c\,\xpDs - 8\,( \aAZ - 2\,s\,c\,\aZZ )\,s\,c^3\,\xps 
\nl &-& ( 24\,\mrG_{14}\,\apD\,\xps - 12\,\mrG_{15}\,\xps\,\apDVA 
         - 12\,\mrG_{15}\,\mrL_{13}\,\apW - \mrG_{61}\,\mrL_{13}\,\apD )\,c^2 
\nl &+& 4\,( 8\,\mrI_{10}\,\apB - 4\,\mrI_{11}\,\apW - \mrL_{13}\,\apD 
         - 4\,\mrF^{b}_{3}\,\xps\,\aAA - 4\,\mrF^{b}_{32}\,\apB\,\xpDs )\,s^2\,c^2 
\nl &-& 8\,( \mrK_{6}\,\xps + \mrL_{14}\,\xpDs )\,s\,c^5\,\aAZ \Bigr\}
         \,\frac{1}{c^4\,\xps}
         \,\sbfun{\mzb}{\mqD}
\nl &-&
         \frac{1}{8}
         \,( 2\,\apW\,\xps\,\xpDs - 3\,\xpUs\,\xps\,\aUW - 2\,\mrH_{15}\,\apW\,\xpDs 
         - 3\,\mrH_{15}\,\xpUs\,\aUW + \mrH_{16}\,\apQWt 
\nl &-& 2\,\mrH_{17}\,\apW\,\xps 
         - 4\,\mrH_{18}\,\xps\,\apQt )
         \,\frac{1}{\xps}
         \,\sbfun{M}{\mqU}
\eqas
We have introduced
\bqas
\UVsb &=& - \frac{1}{256}\,\frac{\mDs}{M}\,\Bigl[
           3\,( s\,\aDB + c\,\aDW )\,\vqd 
\nl
{}&+& 4\,c\,( \apD - \apQo + 3\,\apQt + c\,s\,\aDB )
           + 2\,( 3 + 2\,s^2 )\,c\,\aDW \Bigr]\,\frac{1}{c}
\eqas

\normalsize

\section{Appendix: Listing the counterterms \label{LCT}}

In this Appendix we give the full list of counterterms, dropping the $\ren\,$-subscript for the 
parameters, $\stw = s_{_{\theta\,\ren}}$ \etc
To keep the notation as compact as possible a number of auxiliary quantities are introduced.
First we define the following set of polynomials:

\vspace{0.8cm}
\bei
\item[\fbox{$\mrA(x)\,$}] where $x = \stws$:
\eei

\scriptsize
\[
\begin{array}{lll}
\mrA^{a}_{0}= 1 - x\;\;&\;\;
\mrA^{a}_{1}= 55 - 36\,x\;\;&\;\;
\mrA^{a}_{2}= 10 - 9\,x \\
\mrA^{a}_{3}= 5 - 2\,x\;\;&\;\;
\mrA^{a}_{4}= 7 - 4\,x\;\;&\;\;
\mrA^{a}_{5}= 3 - 2\,x \\
\end{array}
\]

\[
\begin{array}{lll}
\mrA^{b}_{0}= 19 - 72\,x \;\;&\;\;
\mrA^{b}_{1}= 19 + 9\,\mrA^{a}_{0}\,x \;\;&\;\;
\mrA^{b}_{2}= 29 - 2\,\mrA^{a}_{1}\,x \\
\mrA^{b}_{3}= 2 - x \;\;&\;\;
\mrA^{b}_{4}= 53 - 26\,x \;\;&\;\;
\mrA^{b}_{5}= 19 - 36\,x \\
\mrA^{b}_{6}= 61 + 36\,\mrA^{a}_{0}\,x \;\;&\;\;
\mrA^{b}_{7}= 56 - 29\,x \;\;&\;\;
\mrA^{b}_{8}= 13 + 12\,\mrA^{a}_{0}\,x \\
\mrA^{b}_{9}= 3 + 2\,\mrA^{a}_{0}\,x \;\;&\;\;
\mrA^{b}_{10}= 14 - 9\,x \;\;&\;\;
\mrA^{b}_{11}= 2 - 3\,x \\
\mrA^{b}_{12}= 1 - 4\,\mrA^{a}_{0}\,x \;\;&\;\;
\mrA^{b}_{13}= 29 + 4\,\mrA^{a}_{2}\,x \;\;&\;\;
\mrA^{b}_{14}= 50 - 23\,x \\
\mrA^{b}_{15}= 29 + 18\,\mrA^{a}_{0}\,x \;\;&\;\;
\mrA^{b}_{16}= 1 - x \;\;&\;\;
\mrA^{b}_{17}= 3 - \mrA^{a}_{3}\,x \\
\mrA^{b}_{18}= 3 - \mrA^{a}_{4}\,x \;\;&\;\;
\mrA^{b}_{19}= 1 - \mrA^{a}_{5}\,x & \\
\end{array}
\]

\[
\begin{array}{lll}
\mrA^{c}_{0}= 10 - 13\,x \;\;&\;\;
\mrA^{c}_{1}= 3 - 8\,x \;\;&\;\;
\mrA^{c}_{2}= 19 - 18\,x \\
\mrA^{c}_{3}= 1 - 7\,x \;\;&\;\;
\mrA^{c}_{4}= 1 - 4\,x \;\;&\;\;
\mrA^{c}_{5}= 1 - 2\,x \\
\mrA^{c}_{6}= 1 - \mrA^{b}_{0}\,x \;\;&\;\;
\mrA^{c}_{7}= 1 - 2\,\mrA^{b}_{1}\,x \;\;&\;\;
\mrA^{c}_{8}= 2 - \mrA^{b}_{2}\,x \\
\mrA^{c}_{9}= 3 - 2\,x \;\;&\;\;
\mrA^{c}_{10}= 4 + 3\,x \;\;&\;\;
\mrA^{c}_{11}= 19 - 24\,\mrA^{b}_{3}\,x \\
\mrA^{c}_{12}= 24 - \mrA^{b}_{4}\,x \;\;&\;\;
\mrA^{c}_{13}= 1 + 3\,x \;\;&\;\;
\mrA^{c}_{14}= 1 - \mrA^{b}_{5}\,x \\
\mrA^{c}_{15}= 2 - \mrA^{b}_{6}\,x \;\;&\;\;
\mrA^{c}_{16}= 3 - 4\,x \;\;&\;\;
\mrA^{c}_{17}= 3 + x \\
\mrA^{c}_{18}= 3 + 2\,x \;\;&\;\;
\mrA^{c}_{19}= 41 - 48\,\mrA^{b}_{3}\,x \;\;&\;\;
\mrA^{c}_{20}= 97 - 4\,\mrA^{b}_{7}\,x \\
\mrA^{c}_{21}= 1 - 2\,\mrA^{b}_{8}\,x \;\;&\;\;
\mrA^{c}_{22}= 3 - \mrA^{b}_{9}\,x \;\;&\;\;
\mrA^{c}_{23}= 97 - 4\,\mrA^{b}_{4}\,x \\
\mrA^{c}_{24}= 5 - 6\,x \;\;&\;\;
\mrA^{c}_{25}= 1 + 2\,x \;\;&\;\;
\mrA^{c}_{26}= 9 - 4\,x \\
\mrA^{c}_{27}= 11 - 6\,x \;\;&\;\;
\mrA^{c}_{28}= 13 - 8\,x \;\;&\;\;
\mrA^{c}_{29}= 19 - 4\,\mrA^{b}_{10}\,x \\
\mrA^{c}_{30}= 17 + 12\,\mrA^{b}_{11}\,x \;\;&\;\;
\mrA^{c}_{31}= 23 - 4\,x \;\;&\;\;
\mrA^{c}_{32}= 29 - 24\,x \\
\mrA^{c}_{33}= 73 - 72\,x \;\;&\;\;
\mrA^{c}_{34}= 1 - 2\,\mrA^{b}_{12}\,x \;\;&\;\;
\mrA^{c}_{35}= 2 - \mrA^{b}_{13}\,x \\
\mrA^{c}_{36}= 19 - 18\,\mrA^{b}_{12}\,x \;\;&\;\;
\mrA^{c}_{37}= 83 - 96\,\mrA^{b}_{3}\,x \;\;&\;\;
\mrA^{c}_{38}= 293 - 12\,\mrA^{b}_{14}\,x \\
\mrA^{c}_{39}= 1 - \mrA^{b}_{15}\,x \;\;&\;\;
\mrA^{c}_{40}= 7 - 6\,x \;\;&\;\;
\mrA^{c}_{41}= 1 + 2\,\mrA^{b}_{16}\,x^2 \\
\mrA^{c}_{42}= 1 - 4\,\mrA^{b}_{17}\,x \;\;&\;\;
\mrA^{c}_{43}= 1 - 8\,\mrA^{b}_{18}\,x \;\;&\;\;
\mrA^{c}_{44}= 1 - 4\,\mrA^{b}_{19}\,x \\
\end{array}
\]

\normalsize

\vspace{0.5cm}
\bei
\item[\fbox{$\mrB(x)\,$}] where $x = x_\sPH = \mh/\mw$:
\eei

\scriptsize
\[
\begin{array}{lll}
\mrB^{a}_{0}= 2 - 3\,x^2 \;\;&\;\;
\mrB^{a}_{1}= 11 + 15\,x^2 \;\;&\;\;
\mrB^{a}_{2}= 22 + 9\,x^2 \\
\mrB^{a}_{3}= 106 + 9\,x^2 \;\;&\;\;
\mrB^{a}_{4}= 74 + 9\,x^2 \;\;&\;\;
\mrB^{a}_{5}= 2 - 11\,x^2 \\
\mrB^{a}_{6}= 4 - 21\,x^2 \;\;&\;\;
\mrB^{a}_{7}= 11 - 3\,x^2 & \\
\end{array}
\]

\[
\begin{array}{lll}
\mrB^{b}_{0}= 6 - \mrB^{a}_{0}\,x^2 \;\;&\;\;
\mrB^{b}_{1}= 2 + x^2 \;\;&\;\;
\mrB^{b}_{2}= 7 - x^2 \\
\mrB^{b}_{3}= 10 - 3\,x^2 \;\;&\;\;
\mrB^{b}_{4}= 10 - x^2 \;\;&\;\;
\mrB^{b}_{5}= 11 + 6\,x^2 \\
\mrB^{b}_{6}= 31 + 15\,x^2 \;\;&\;\;
\mrB^{b}_{7}= 32 - 3\,x^2 \;\;&\;\;
\mrB^{b}_{8}= 96 - \mrB^{a}_{1}\,x^2 \\
\mrB^{b}_{9}= 132 - \mrB^{a}_{2}\,x^2 \;\;&\;\;
\mrB^{b}_{10}= 132 - \mrB^{a}_{3}\,x^2 \;\;&\;\;
\mrB^{b}_{11}= 196 - \mrB^{a}_{4}\,x^2 \\
\mrB^{b}_{12}= 2 + 11\,x^2 \;\;&\;\;
\mrB^{b}_{13}= 4 + x^2 \;\;&\;\;
\mrB^{b}_{14}= 6 - \mrB^{a}_{5}\,x^2 \\
\mrB^{b}_{15}= 12 - x^2 \;\;&\;\;
\mrB^{b}_{16}= 12 - \mrB^{a}_{6}\,x^2 \;\;&\;\;
\mrB^{b}_{17}= 30 - \mrB^{a}_{7}\,x^2 \\
\mrB^{b}_{18}= 1 - 9\,x^2 \;\;&\;\;
\mrB^{b}_{19}= 3 - 4\,x^2 \;\;&\;\;
\mrB^{b}_{20}= 7 + 8\,x^2 \\
\mrB^{b}_{21}= 15 + 4\,x^2 \;\;&\;\;
\mrB^{b}_{22}= 20 + 9\,x^2 \;\;&\;\;
\mrB^{b}_{23}= 55 - 3\,x^2 \\
\mrB^{b}_{24}= 80 + 27\,x^2 \;\;&\;\;
\mrB^{b}_{25}= 1 - 6\,x^2 \;\;&\;\;
\mrB^{b}_{26}= 2 - x^2 \\
\mrB^{b}_{27}= 10 - 11\,x^2 \;\;&\;\;
\mrB^{b}_{28}= 12 + 11\,x^2 \;\;&\;\;
\mrB^{b}_{29}= 23 - 3\,x^2 \\
\mrB^{b}_{30}= 34 + 15\,x^2 \;\;&\;\;
\mrB^{b}_{31}= 56 + 9\,x^2 \;\;&\;\;
\mrB^{b}_{32}= 144 - \mrB^{a}_{2}\,x^2 \\
\mrB^{b}_{33}= 4 + 3\,x^2 \;\;&\;\;
\mrB^{b}_{34}= 3 - x^2 \;\;&\;\;
\mrB^{b}_{35}= 6 - x^2 \\
\mrB^{b}_{36}= 9 - x^2 \;\;&\;\;
\mrB^{b}_{37}= 10 + 3\,x^2 \;\;&\;\;
\mrB^{b}_{38}= 22 - 3\,x^2 \\
\mrB^{b}_{39}= 82 - 9\,x^2 \;\;&\;\;
\mrB^{b}_{40}= 8 + 9\,x^2 \;\;&\;\;
\mrB^{b}_{41}= 22 - x^2 \\
\mrB^{b}_{42}= 34 - x^2 \;\;&\;\;
\mrB^{b}_{43}= 9 - 11\,x^2 \;\;&\;\;
\mrB^{b}_{44}= 31 - 6\,x^2 \\
\mrB^{b}_{45}= 32 + 3\,x^2 \;\;&\;\;
\mrB^{b}_{46}= 33 - 7\,x^2 \;\;&\;\;
\mrB^{b}_{47}= 40 + 33\,x^2 \\
\mrB^{b}_{48}= 51 - 7\,x^2 \;\;&\;\;
\mrB^{b}_{49}= 204 - \mrB^{a}_{2}\,x^2 \;\;&\;\;
\mrB^{b}_{50}= 204 - \mrB^{a}_{3}\,x^2 \\
\mrB^{b}_{51}= 5 + 4\,x^2 & & \\
\end{array}
\]

\normalsize

\vspace{0.5cm}
\bei
\item[\fbox{$\mrC\,$}] where we have introduced
$v_{\Pf} = 1 - 2\,\frac{Q_{\Pf}}{I^3_{\Pf}}\,\stws$
and
\bq
v^{(1)}_{\gen} = v^2_{\Pl} + 3\,\lpar v^2_{\PQu} + v_{\PQd} \rpar
\quad
v^{(2)}_{\gen} = v^2_{\Pl} + 2\,v^2_{\PQu} + v_{\PQd} 
\quad
v^{\pm}_{\Pf} = 1 \pm v_{\Pf} 
\eq
\eei

\scriptsize
\[
\begin{array}{lll}
\mrC^{a}_{0}= 9 + v^{(1)}_{\gen} \;\;&\;\;
\mrC^{a}_{1}= 17 + v^{(3)}_{\gen} \;\;&\;\;
\mrC^{a}_{2}= v^+_{\PQu} - v^+_{\PQd} \\
\mrC^{a}_{3}= v^+_{\PQu} + v^+_{\PQd} \;\;&\;\;
\mrC^{a}_{4}= 8 - v^{(2)}_{\gen} \;\;&\;\;
\mrC^{a}_{5}= 16 - 3\,v^{(2)}_{\gen} \\
\mrC^{a}_{6}= 41 - 4\,v^{(2)}_{\gen} + v^{(1)}_{\gen} \;\;&\;\;
\mrC^{a}_{7}= 53 - 12\,v^{(2)}_{\gen} - 3\,v^{(1)}_{\gen} \;\;&\;\;
\mrC^{a}_{8}= 3 + v_{\Pl} \\
\mrC^{a}_{9}= 4 - v^{(2)}_{\gen} \;\;&\;\;
\mrC^{a}_{10}= 5\,v^-_{\PQu} + v^-_{\PQd} + 3\,v^-_{\PL} \;\;&\;\;
\mrC^{a}_{11}= 2 + v^{(2)}_{\gen} \\
\mrC^{a}_{12}= 9 + 4\,v^{(2)}_{\gen} + v^{(1)}_{\gen} \;\;&\;\;
\mrC^{a}_{13}= 17 + v^{(3)}_{\gen} - 2\,v^{(2)}_{\gen} & \\
\end{array}
\]

\normalsize

\vspace{0.5cm}
\bei
\item[\fbox{$\mrD\,$}] where we have introduced
\bq
x_{\Pf} = \frac{M_{\Pf}}{\mw} \quad
x^{(1)}_{\gen} = x^2_{\Pl} + 3\,\lpar x^2_{\PQu} + x^2_{\PQd} \rpar \quad
x^{(2)}_{\gen} = x^4_{\Pl} + 3\,\lpar x^4_{\PQu} + x^4_{\PQd} \rpar 
\eq
\eei

\scriptsize
\[
\begin{array}{lll}
\mrD^{a}_{0}= 2 - 9\,x^2_{\Pl} \;\;&\;\;
\mrD^{a}_{1}= 2 - 9\,x^2_{\PQu} - 9\,x^2_{\PQd} \;\;&\;\;
\mrD^{a}_{2}=  - x^2_{\PQu} + x^2_{\PQd} \\
\mrD^{a}_{3}=  x^{(1)}_{\gen} - 2\,x^{(2)}_{\gen} \;\;&\;\;
\mrD^{a}_{4}= x^2_{\PQu} + x^2_{\PQd} \;\;&\;\;
\mrD^{a}_{5}= 2 - 3\,x^2_{\Pl} \\
\mrD^{a}_{6}= 2 - 3\,x^2_{\PQu} - 3\,x^2_{\PQd} & & \\
\end{array}
\]

\normalsize

\vspace{0.5cm}
\bei
\item[\fbox{$\mrE\,$}]
\eei

\scriptsize
\[
\begin{array}{lll}
\mrE^{a}_{0}= 2 - 9*x^2_{\PQU} + 9*x^2_{\PQD} \;\;&\;\;
\mrE^{a}_{1}= 2 - 9*x^2_{\PQD} \;\;&\;\;
\mrE^{a}_{2}= 26 + 9*x^2_{\PQU} + 9*x^2_{\PQD} \\
\mrE^{a}_{3}= 14 + 9*x^2_{\PQU} - 9*x^2_{\PQD} \;\;&\;\;
\mrE^{a}_{4}= 8 - 9*x^2_{\PQU} \;\;&\;\;
\mrE^{a}_{5}= 14 + 3*x^2_{\PL} \\
\mrE^{a}_{6}= 2 - x^2_{\PL} \;\;&\;\;
\mrE^{a}_{7}= 2 + x^2_{\PL} \;\;&\;\;
\mrE^{a}_{8}= 1 + 3*x^2_{\PQD} \\
\mrE^{a}_{9}= 4 + 4*x^2_{\PQU} + 3*x^2_{\PQD} \;\;&\;\;
\mrE^{a}_{10}= 4 + 6*x^2_{\PQU} + 3*x^2_{\PQD} \;\;&\;\;
\mrE^{a}_{11}= 4 - 3*x^2_{\PQD} \\
\mrE^{a}_{12}= 4 + 3*x^2_{\PQD} \;\;&\;\;
\mrE^{a}_{13}= 8 - 3*x^2_{\PQD} \;\;&\;\;
\mrE^{a}_{14}= 16 - 6*x^2_{\PQU} - 3*x^2_{\PQD} \\
\mrE^{a}_{15}= 1 + 3*x^2_{\PQU} \;\;&\;\;
\mrE^{a}_{16}= 4 + 3*x^2_{\PQU} + 4*x^2_{\PQD} \;\;&\;\;
\mrE^{a}_{17}= 8 - 3*x^2_{\PQU} \\
\mrE^{a}_{18}= 8 + 3*x^2_{\PQU} \;\;&\;\;
\mrE^{a}_{19}= 8 + 3*x^2_{\PQU} + 6*x^2_{\PQD} \;\;&\;\;
\mrE^{a}_{20}= 16 - 15*x^2_{\PQU} \\
\mrE^{a}_{21}= 20 - 3*x^2_{\PQU} - 6*x^2_{\PQD} \;\;&\;\;
\mrE^{a}_{22}= 1 + 3*x^2_{\PL} \;\;&\;\;
\mrE^{a}_{23}= 4 - x^2_{\PL} \\
\mrE^{a}_{24}= 4 + x^2_{\PL} \;\;&\;\;
\mrE^{a}_{25}= 4 + 3*x^2_{\PL} \;\;&\;\;
\mrE^{a}_{26}= 8 - 9*x^2_{\PL} \\
\mrE^{a}_{27}= 8 - x^2_{\PL} & & \\
\end{array}
\]

\normalsize

\subsection{$\mrdim = 4\;$ counterterms}
First we list the SM counterterms. In the following we use $s = \stw$ and $c= \ctw$.
\footnotesize
\bq
\dZ^{(4)}_{\PH} =
         \frac{4}{3}\,\myNG
       - \frac{1}{2}\,\sum_{\gen}\,x^{(1)}_{\gen}
       - \frac{1}{6}\,\frac{\mrA^{c}_0}{c^2}
\eq
\bq
\dZ^{(4)}_{\mh} =
       - \frac{1}{2}\,\Bigl\{ 
         \sum_{\gen} (4\,x^{(2)}_{\gen} - x^{(1)}_{\gen}\,x_{\PH}^2) - 
        \Bigl[3 - (x_{\PH}^2 - c^2\,\mrB^{b}_0)\,c^2\Bigr]\,
        \frac{1}{c^4}\Bigr\}
        \,\frac{1}{x_{\PH}^2}
\eq
\bq
\dZ^{(4)}_{\PA} =
       - \frac{1}{6}\,\mrA^{c}_2 + \frac{4}{9}\,\myNG\,\mrA^{c}_1
\qquad
\dZ^{(4)}_{\PAZ} =
         \frac{1}{6}\,\frac{s}{c}\,\mrA^{c}_2
       - \frac{1}{3}\,\frac{s}{c}\,\myNG\,v^{(2)}_{\gen}
\eq

\vspace{0.3cm}
\bq
\dZ^{(4)}_{\PW} = 0
\qquad
\dZ^{(4)}_{M_{\PW}} =
         \frac{4}{3}\,\myNG
       - \frac{1}{2}\,\sum_{\gen}\,x^{(1)}_{\gen}
       - \frac{1}{6}\,\frac{\mrA^{c}_3}{c^2}
\eq
\bq
\dZ^{(4)}_{\ctw} =
       - \frac{1}{12}\,\frac{s^2}{c^2}\,\mrA^{c}_2
       - \frac{1}{24}\,( \mrC^{a} - 16\,c^2 )\,\frac{\myNG}{c^2}
\qquad
\dZ^{(4)}_{\PZ} =
       - \frac{1}{6}\,\frac{s^2}{c^2}\,\mrA^{c}_2
       - \frac{1}{12}\,( \mrC^{a} - 16\,c^2 )\,\frac{\myNG}{c^2}
\eq
\bq
\dZ^{(4)}_{g} = - \frac{19}{12} + \frac{2}{3}\,\myNG
\eq

\vspace{0.3cm}
\bq
\dZ^{(4)}_{M_{\PQD}} =
         \frac{1}{48}\,\Bigl[ 3\,\mrC^{a}_1 
         + 2\,( 3\,\mrE^{a}_0 - 8\,s^2 )\,c^2 \Bigr]\,\frac{1}{c^2}
\eq
\bq
\dZ^{(4)}_{\ssR\,\PQD} =
         \frac{1}{144}\,\Bigl[ 9\,\mrC^{a}_2 
         + 8\,( 9\,x_{\PQd}^2 + 2\,s^2 )\,c^2 \Bigr]\,\frac{1}{c^2}
\qquad
\dZ^{(4)}_{\ssL\,\PQD} =
         \frac{1}{144}\,\Bigl[ 9\,\mrC^{a} 
         + 4\,( 9\,\mrE^{a}_1 + 4\,s^2 )\,c^2 \Bigr]\,\frac{1}{c^2}
\eq
\bq
\dZ^{(4)}_{M_{\PQU}} =
       - \frac{1}{24}\,( 5 - c^2\,\mrE^{a}_2 )\,\frac{1}{c^2}
\eq

\vspace{0.3cm}
\bq
\dZ^{(4)}_{\ssR\,\PQU} =
         \frac{1}{18}\,( 8 - c^2\,\mrE^{a}_3 )\,\frac{1}{c^2}
\qquad
\dZ^{(4)}_{\ssL\,\PQU} =
         \frac{1}{36}\,( 1 + c^2\,\mrE^{a}_4 )\,\frac{1}{c^2}
\eq
\bq
\dZ^{(4)}_{M_{\PL}} =
       - \frac{1}{8}\,( 11 - c^2\,\mrE^{a}_5 )\,\frac{1}{c^2}
\eq
\bq
\dZ^{(4)}_{\ssR\,\PL} =
         \frac{1}{2}\,( 2 - c^2\,\mrE^{a}_6 )\,\frac{1}{c^2}
\qquad
\dZ^{(4)}_{\ssL\,\Pl} =
         \frac{1}{4}\,( 1 + c^2\,\mrE^{a}_7 )\,\frac{1}{c^2}
\eq

\vspace{0.3cm}
\bq
\dZ^{(4)}_{\ssR\,\PGn} = 0
\qquad
\dZ^{(4)}_{\ssL\,\PGn} = 
         \frac{1}{4}\,( 1 + c^2\,\mrE^{a}_7 )\,\frac{1}{c^2}
\eq
\normalsize
\subsection{$\mrdim = 6\;$ counterterms}
To present $\mrdim = 6$ counterterms we define vectors; for counterterms

\bq
\begin{array}{lllll}
\CT_1 = \dZ_{\PH}^{(6)} \;\;&\;\;
\CT_2 = \dZ_{\mh}^{(6)} \;\;&\;\;
\CT_3 = \dZ_{\PA}^{(6)} \;\;&\;\;
\CT_4 = \dZ_{\PW}^{(6)} \;\;&\;\;
\CT_5 = \dZ_{\mw}^{(6)} \\
\CT_6 = \dZ_{\PAZ}^{(6)} \;\;&\;\;
\CT_7 = \dZ_{\cth}^{(6)} \;\;&\;\;
\CT_8 = \dZ_{\PZ}^{(6)} \;\;&\;\;
\CT_9 = \dZ_g^{(6)} \;\;&\;\;
\CT_{10} = \dZ_{M_{\PQD}}^{(6)} \\
\CT_{11} = \dZ_{\ssR\,\PQD}^{(6)} \;\;&\;\;
\CT_{12} = \dZ_{\ssL\,\PQD}^{(6)} \;\;&\;\;
\CT_{13} = \dZ_{M_{\PQU}}^{(6)} \;\;&\;\;
\CT_{14} = \dZ_{\ssR\,\PQU}^{(6)} \;\;&\;\;
\CT_{15} = \dZ_{\ssL\,\PQU}^{(6)} \\
\CT_{16} = \dZ_{M_{\PL}}^{(6)} \;\;&\;\;
\CT_{17} = \dZ_{\ssR\,\PL}^{(6)} \;\;&\;\;
\CT_{18} = \dZ_{\ssL\,\PL}^{(6)} \;\;&\;\;
\CT_{19} = \dZ_{\ssR\,\PGn}^{(6)} \;\;&\;\;
\CT_{20} = \dZ_{\ssL\,\PGn}^{(6)} \\
\end{array}
\eq

and for Wilson coefficients

\bq
\begin{array}{lllll}
\ap = \mrW_1 \;\;&\;\;
\apBox = \mrW_2 \;\;&\;\;
\apD = \mrW_3 \;\;&\;\;
\apW = \mrW_4 \\
\apB = \mrW_5 \;\;&\;\;
\apWB = \mrW_6 \;\;&\;\;
\aup = \mrW_7(\PQq\,,\,\PQu) \;\;&\;\;
\adp = \mrW_8(\PQq\,,\,\PQd) \\
\aLp = \mrW_9(\lambda\,,\,\Pl) \;\;&\;\;
\apqo = \mrW_{10}(\PQq) \;\;&\;\;
\aplo = \mrW_{11}(\Pl) \;\;&\;\;
\apu = \mrW_{12}(\PQu) \\
\apd = \mrW_{13}(\PQd) \;\;&\;\;
\apl = \mrW_{14}(\Pl) \;\;&\;\;
\apqt = \mrW_{15}(\PQq) \;\;&\;\:
\aplt = \mrW_{16}(\Pl) \\
\auW = \mrW_{17}(\PQq\,,\,\PQu) \;\;&\;\;
\adW = \mrW_{18}(\PQq\,,\,\PQd) \;\;&\;\;
\alW = \mrW_{19}(\lambda\,,\,\Pl) \;\;&\;\;
\auB = \mrW_{20}(\PQq\,,\,\PQu) \\
\adB = \mrW_{21}(\PQq\,,\,\PQd) \;\;&\;\;
\alB = \mrW_{22}(\Lambda\,,\,\Pl) \;\;&\;\;
\aLldQ = \mrW_{23}(\Lambda,\Pl,\PQd,\PQQ) \;\;&\;\;
\aoQuQd = \mrW_{24}(\PQQ,\PQu,\PQQ,\PQd) \\
\aoLlQu = \mrW_{25}(\Lambda,\Pl,\PQQ,\PQu) & & & \\
\end{array}
\eq
The result is given by introducing a matrix
\bq
\CT_i = \sum_{j=1,6}\,\mrM^{\ct}_{ij}\,\mrW_j +
        \sum_{j=7,25}\,\sum_{\gen}\,\mrM^{\ct}_{ij}(\mathrm{label})\,\mrW_j(\mathrm{label})
\eq
where, without assuming universality, we have
\bqa
\sum_{\gen}\,\mrM^{\ct}_{i,10}(\mathrm{label})\,\mrW_{10}(\mathrm{label}) &=&
\mrM^{\ct}_{i,10}(\PQu,\PQd)\,\mrW_{10}(\PQu,\PQd) +
\mrM^{\ct}_{i,10}(\PQc,\PQs)\,\mrW_{10}(\PQc,\PQs) +
\mrM^{\ct}_{i,10}(\PQt,\PQb)\,\mrW_{10}(\PQt,\PQb) 
\nl
\sum_{\gen}\,\mrM^{\ct}_{i,12}(\mathrm{label})\,\mrW_{12}(\mathrm{label}) &=&
\mrM^{\ct}_{i,12}(\PQu)\,\mrW_{12}(\PQu) +
\mrM^{\ct}_{i,12}(\PQc)\,\mrW_{12}(\PQc) +
\mrM^{\ct}_{i,12}(\PQt)\,\mrW_{12}(\PQt) 
\eqa
\etc In the following we introduce all non-zero entries of the matrix $\mrM^{\ct}$.
\footnotesize
\bei
\item {\underline{$\mrM^{\ct}_{1,i}$ entries}}
\eei
\bqas
   \mrM^{\ct}_{1,2} &=&
     -
         \frac{1}{6}
         \,\Bigl\{ 4\,\mrA^{c}_{11} - \mrB^{b}_{3}\,c^2 \Bigr\}
         \,\frac{1}{c^2}
\eqas
\bqas
   \mrM^{\ct}_{1,3} &=&
         \frac{1}{2}
         \,\sumg \,\xog
       -
         \frac{1}{6}
         \,\Bigl\{ 3\,\mrA^{c}_{12} - \mrB^{b}_{5}\,c^2 \Bigr\}
         \,\frac{1}{c^2}
\eqas
\bqas
   \mrM^{\ct}_{1,4} &=&
       -
         \sumg \Bigl\{ \xog 
         - \Bigl[ c^2\,\xog\,\xphs + 2\,( 2\,s^2\,\xtg + \mrD^{a}_{3} ) \Bigr]\,c^2 \Bigr\}
\nl &+&
         \frac{1}{6}
         \,\Bigl\{ 9\,\mrA^{c}_{9} - \Bigl[  - ( \mrA^{c}_{8}\,\xphs + 8\,\mrB^{b}_{2} ) 
         + (  - \mrB^{b}_{10}\,s^2 + \mrB^{b}_{11} )\,c^2 \Bigr]\,c^2 \Bigr\}
         \,\frac{1}{c^2}
\nl &+&
         \frac{1}{3}
         \,\Bigl\{ \mrC^{a}_{1} 
         + \Bigl[ 8\,c^2\,\xphs - ( 16 + \mrC^{a}_{0}\,\xphs ) \Bigr]\,c^2 \Bigr\}
         \,\myNG
\eqas
\bqas
   \mrM^{\ct}_{1,5} &=&
         \sumg \Bigl\{ 2\,\xog - \Bigl[ 4\,\xtg - \xog\,\xphs \Bigr]\,c^2 \Bigr\}
         \,s^2
\nl &+&
         \frac{1}{6}
         \,\Bigl\{ 6\,\mrA^{c}_{10} - \Bigl[ ( \mrA^{c}_{8}\,\xphs + 2\,\mrB^{b}_{7} ) 
         - ( 4\,\mrB^{b}_{4} - \mrB^{b}_{9}\,s^2 )\,c^2 \Bigr]\,c^2 \Bigr\}
         \,\frac{1}{c^2}
\nl &-&
         \frac{1}{6}
         \,\Bigl\{  - 16\,\Bigl[ 2 - c^2\,\xphs \Bigr]\,c^2 
         + \Bigl[ 2 - \mrA^{c}_{5}\,\xphs \Bigr]\,\mrC^{a}_{0} \Bigr\}
         \,\myNG
\eqas
\bqas
   \mrM^{\ct}_{1,6} &=&
       -
         \sumg \Bigl\{ 2\,\xog - \Bigl[ 4\,\xtg - \xog\,\xphs \Bigr]\,c^2 \Bigr\}
         \,s\,c
\nl &+&
         \frac{1}{12}
         \,\Bigl\{ 2\,\mrA^{c}_{5}\,\mrC^{a}_{0} + \Bigl[ 16\,\mrA^{c}_{5}\,c^2\,\xphs 
         - ( \mrA^{c}_{4}\,\mrC^{a}_{0}\,\xphs + 32\, \mrA^{c}_{5} ) \Bigr]\,c^2 \Bigr\}
         \,\frac{1}{s\,c}
         \,\myNG
\nl &+&
         \frac{1}{6}
         \,\Bigl\{ 2\,\mrA^{c}_{7} + \Bigl[ 2\,\mrB^{b}_{6}\,s^2 - (  - \mrA^{c}_{6}\,\xphs 
         + \mrB^{b}_{1} - \mrB^{b}_{8}\,s^2 )\,c^2 \Bigr]\,c^2 \Bigr\}
         \,\frac{1}{s\,c}
\eqas
\bqas
   \mrM^{\ct}_{1,7}\lpar \PQq\,,\,\PQu \rpar &=&
       -
         3
         \,\xpus
\qquad
   \mrM^{\ct}_{1,8}\lpar \PQq\,,\,\PQd \rpar =
         3
         \,\xpds
\qquad
   \mrM^{\ct}_{1,9}\lpar \lambda\,,\,\Pl \rpar =
         \xpls
\eqas
\bqas
   \mrM^{\ct}_{1,10}\lpar \PQq \rpar &=&
         2
         \,\Bigl\{ \mrC^{a}_{2} + 6\,\mrD^{a}_{2}\,c^2 \Bigr\}
         \,\frac{1}{c^2}
\qquad
   \mrM^{\ct}_{1,11}\lpar \Pl \rpar =
       -
         \frac{2}{3}
         \,\Bigl\{ \vple - 6\,c^2\,\xpls \Bigr\}
         \,\frac{1}{c^2}
\eqas
\bqas
   \mrM^{\ct}_{1,12}\lpar \PQu \rpar &=&
       -
         2
         \,\Bigl\{ \vmqu - 6\,c^2\,\xpus \Bigr\}
         \,\frac{1}{c^2}
\qquad
   \mrM^{\ct}_{1,13}\lpar \PQd \rpar =
         2
         \,\Bigl\{ \vmqd - 6\,c^2\,\xpds \Bigr\}
         \,\frac{1}{c^2}
\eqas
\bqas
   \mrM^{\ct}_{1,14}\lpar \Pl \rpar &=&
         \frac{2}{3}
         \,\Bigl\{ \vmle - 6\,c^2\,\xpls \Bigr\}
         \,\frac{1}{c^2}
\qquad
   \mrM^{\ct}_{1,15}\lpar \PQq \rpar =
         2
         \,\Bigl\{ \mrC^{a}_{3} + \mrD^{a}_{1}\,c^2 \Bigr\}
         \,\frac{1}{c^2}
\eqas
\bqas
   \mrM^{\ct}_{1,16}\lpar \Pl \rpar &=&
         \frac{2}{3}
         \,\Bigl\{ \vple + \mrD^{a}_{0}\,c^2 \Bigr\}
         \,\frac{1}{c^2}
\eqas
\bqas
   \mrM^{\ct}_{1,17}\lpar \PQq\,,\,\PQu \rpar &=&
       -
         \frac{3}{2}
         \,\xpus
\qquad
   \mrM^{\ct}_{1,18}\lpar \PQq\,,\,\PQd \rpar =
         \frac{3}{2}
         \,\xpds
\qquad
   \mrM^{\ct}_{1,19}\lpar \lambda\,,\,\Pl \rpar =
         \frac{1}{2}
         \,\xpls
\eqas
\bei
\item {\underline{$\mrM^{\ct}_{2,i}$ entries}}
\eei
\bqas
   \mrM^{\ct}_{2,1} &=&
       -
         3
         \,\Bigl\{ 1 + \mrB^{b}_{12}\,c^2 \Bigr\}
         \,\frac{1}{c^2\,\xphs}
\eqas
\bqas
   \mrM^{\ct}_{2,2} &=&
         \Bigl\{  - \sumg \Bigl[ 4\,\xtg - \xog\,\xphs \Bigr] 
         + \Bigl[ 3 - ( \xphs - \mrB^{b}_{14}\,c^2 )\,c^2 \Bigr]\,\frac{1}{c^4} \Bigr\}
         \,\frac{1}{\xphs}
\eqas
\bqas
   \mrM^{\ct}_{2,3} &=&
         \frac{1}{8}
         \,\Bigl\{ 2\,\sumg \Bigl[ 4\,\xtg - \xog\,\xphs \Bigr] + \Bigl[ 18 - ( 5\,\xphs
         + \mrB^{b}_{16}\,c^2 )\,c^2 \Bigr]\,\frac{1}{c^4} \Bigr\}
         \,\frac{1}{\xphs}
\eqas
\bqas
   \mrM^{\ct}_{2,4} &=&
         \Bigl\{  - \sumg \Bigl[ 4\,\xtg - \xog\,\xphs \Bigr] 
         + \Bigl[ 3 + ( \mrB^{b}_{15} + \mrB^{b}_{17}\,c^2 )\,c^2 \Bigr]\,\frac{1}{c^4} \Bigr\}
         \,\frac{1}{\xphs}
\eqas
\bqas
   \mrM^{\ct}_{2,5} &=&
         3
         \,\Bigl\{ 4 - \Bigl[  - c^2\,\xphs + \mrB^{b}_{13} \Bigr]\,c^2 \Bigr\}
         \,\frac{1}{c^4\,\xphs}
\eqas
\bqas
   \mrM^{\ct}_{2,6} &=&
       -
         3
         \,\Bigl\{ 4 - c^2\,\xphs \Bigr\}
         \,\frac{1}{\xphs}
         \,\frac{s}{c^3}
\qquad
   \mrM^{\ct}_{2,7}\lpar \PQq\,,\,\PQu \rpar =
       -
         3
         \,\Bigl\{ 8\,\xpus - \xphs \Bigr\}
         \,\frac{1}{\xphs}
         \,\xpus
\eqas
\bqas
   \mrM^{\ct}_{2,8}\lpar \PQq\,,\,\PQd \rpar &=&
         3
         \,\Bigl\{ 8\,\xpds - \xphs \Bigr\}
         \,\frac{1}{\xphs}
         \,\xpds
\qquad
   \mrM^{\ct}_{2,9}\lpar \lambda\,,\,\Pl \rpar =
         \Bigl\{ 8\,\xpls - \xphs \Bigr\}
         \,\frac{1}{\xphs}
         \,\xpls
\eqas
\bei
\item {\underline{$\mrM^{\ct}_{3,i}$ entries}}
\eei
\bqas
   \mrM^{\ct}_{3,2} &=&
         \sumg \,\xog
       -
         \frac{1}{3}
         \,\Bigl\{ \mrA^{c}_{19} + \mrB^{b}_{18}\,c^2 \Bigr\}
         \,\frac{1}{c^2}
\eqas
\bqas
   \mrM^{\ct}_{3,3} &=&
         \frac{16}{9}
         \,\myNG\,c^2
       +
         \frac{1}{4}
         \,\sumg \,\xog
       -
         \frac{1}{24}
         \,\Bigl\{ 3\,\mrA^{c}_{20} - \mrB^{b}_{22}\,c^2 \Bigr\}
         \,\frac{1}{c^2}
\eqas
\bqas
   \mrM^{\ct}_{3,4} &=&
         \sumg \Bigl\{ c^2\,\xog\,\xphs + 2\,\Bigl[ 2\,s^2\,\xtg + \mrD^{a}_{3} \Bigr] \Bigr\}
         \,c^2
\nl &+&
         \frac{1}{6}
         \,\Bigl\{ \Bigl[ 18 - \mrB^{b}_{20} + 2\,\mrB^{b}_{21}\,s^2 \Bigr]
         - \Bigl[  - 2\,\mrA^{c}_{14}\,\xphs - \mrB^{b}_{10}\,s^2 
         + \mrB^{b}_{11} \Bigr]\,c^2 \Bigr\}
\nl &+&
         \frac{1}{9}
         \,\Bigl\{ 3\,\Bigl[ 8\,c^2 - ( \mrC^{a}_{0} ) \Bigr]\,c^2\,\xphs + 
         \Bigl[  - 16\,\mrA^{c}_{17} + 3\,\mrC^{a}_{1} \Bigr] \Bigr\}
         \,\myNG
\eqas
\bqas
   \mrM^{\ct}_{3,5} &=&
         \sumg \Bigl\{ 2\,\xog - \Bigl[ 4\,\xtg - \xog\,\xphs \Bigr]\,c^2 \Bigr\}
         \,s^2
\nl &+&
         \frac{1}{6}
         \,\Bigl\{ 6\,\mrA^{c}_{13} - \Bigl[  - (  - 2\,\mrA^{c}_{14}\,\xphs + 4\,\mrB^{b}_{4} 
         - \mrB^{b}_{9}\,s^2 )\,c^2 + (  - 2\,\mrB^{b}_{19}\,s^2 
         + \mrB^{b}_{23} ) \Bigr]\,c^2 \Bigr\}
         \,\frac{1}{c^2}
\nl &-&
         \frac{1}{6}
         \,\Bigl\{  - 16\,\Bigl[ 2 - c^2\,\xphs \Bigr]\,c^2 
         + \Bigl[ 2 - \mrA^{c}_{5}\,\xphs \Bigr]\,\mrC^{a}_{0} \Bigr\}
         \,\myNG
\eqas
\bqas
   \mrM^{\ct}_{3,6} &=&
       -
         \sumg \Bigl\{ 2\,\xog - \Bigl[ 4\,\xtg - \xog\,\xphs \Bigr]\,c^2 \Bigr\}
         \,s\,c
\nl &+&
         \frac{1}{36}
         \,\Bigl\{ 6\,\mrA^{c}_{5}\,\mrC^{a}_{0} + \Bigl[ 48\,\mrA^{c}_{5}\,c^2\,\xphs 
         - ( 3\,\mrA^{c}_{4}\,\mrC^{a}_{0}\,\xphs + 32\,\mrA^{c}_{18} ) \Bigr]\,c^2 \Bigr\}
         \,\frac{1}{s\,c}
         \,\myNG
\nl &+&
         \frac{1}{6}
         \,\Bigl\{ \mrA^{c}_{15} + \Bigl[ \mrB^{b}_{24}\,s^2 
         - (  - \mrA^{c}_{6}\,\xphs + \mrB^{b}_{1} - \mrB^{b}_{8}\,s^2 )\,c^2 \Bigr]\,c^2 \Bigr\}
         \,\frac{1}{s\,c}
\eqas
\bqas
   \mrM^{\ct}_{3,10}\lpar \PQq \rpar &=&
         2
         \,\Bigl\{ \mrC^{a}_{2} + 6\,\mrD^{a}_{2}\,c^2 \Bigr\}
         \,\frac{1}{c^2}
\qquad
   \mrM^{\ct}_{3,11}\lpar \Pl \rpar =
       -
         \frac{2}{3}
         \,\Bigl\{ \vple - 6\,c^2\,\xpls \Bigr\}
         \,\frac{1}{c^2}
\eqas
\bqas
   \mrM^{\ct}_{3,12}\lpar \PQu \rpar &=&
       -
         2
         \,\Bigl\{ \vmqu - 6\,c^2\,\xpus \Bigr\}
         \,\frac{1}{c^2}
\qquad
   \mrM^{\ct}_{3,13}\lpar \PQd \rpar =
         2
         \,\Bigl\{ \vmqd - 6\,c^2\,\xpds \Bigr\}
         \,\frac{1}{c^2}
\eqas
\bqas
   \mrM^{\ct}_{3,14}\lpar \Pl \rpar &=&
         \frac{2}{3}
         \,\Bigl\{ \vmle - 6\,c^2\,\xpls \Bigr\}
         \,\frac{1}{c^2}
\qquad
   \mrM^{\ct}_{3,15}\lpar \PQq \rpar =
         2
         \,\Bigl\{ \mrC^{a}_{3} + \mrD^{a}_{1}\,c^2 \Bigr\}
         \,\frac{1}{c^2}
\qquad
   \mrM^{\ct}_{3,16}\lpar \Pl \rpar =
         \frac{2}{3}
         \,\Bigl\{ \vple + \mrD^{a}_{0}\,c^2 \Bigr\}
         \,\frac{1}{c^2}
\eqas
\bqas
   \mrM^{\ct}_{3,17}\lpar \PQq\,,\,\PQu \rpar &=&
       -
         \frac{1}{2}
         \,\mrA^{c}_{1}
         \,\xpus
\qquad
   \mrM^{\ct}_{3,18}\lpar \PQq\,,\,\PQd \rpar =
         \frac{1}{2}
         \,\mrA^{c}_{16}
         \,\xpds
\qquad
   \mrM^{\ct}_{3,19}\lpar \lambda\,,\,\Pl \rpar =
         \frac{1}{2}
         \,\mrA^{c}_{4}
         \,\xpls
\eqas
\bqas
   \mrM^{\ct}_{3,20}\lpar \PQq\,,\,\PQu \rpar &=&
         4
         \,s\,c\,\xpus
\qquad
   \mrM^{\ct}_{3,21}\lpar \PQq\,,\,\PQd \rpar =
         2
         \,s\,c\,\xpds
\qquad
   \mrM^{\ct}_{3,22}\lpar \lambda\,,\,\Pl \rpar =
         2
         \,s\,c\,\xpls
\eqas
\bei
\item {\underline{$\mrM^{\ct}_{4,i}$ entries}}
\eei
\bqas
   \mrM^{\ct}_{4,2} &=&
         \sumg \,\xog
       -
         \frac{1}{6}
         \,\Bigl\{ 2\,\mrA^{c}_{19} + 3\,\mrB^{b}_{25}\,c^2 \Bigr\}
         \,\frac{1}{c^2}
\eqas
\bqas
   \mrM^{\ct}_{4,3} &=&
         \frac{1}{4}
         \,\sumg \,\xog
       -
         \frac{1}{24}
         \,\Bigl\{ 3\,\mrA^{c}_{23} - \mrB^{b}_{31}\,c^2 \Bigr\}
         \,\frac{1}{c^2}
\eqas
\bqas
   \mrM^{\ct}_{4,4} &=&
         \sumg \Bigl\{ c^2\,\xog\,\xphs + 2\,\Bigl[ 2\,s^2\,\xtg + \mrD^{a}_{3} \Bigr] \Bigr\}
         \,c^2
\nl &+&
         \frac{1}{6}
         \,\Bigl\{ 6\,\mrA^{c}_{22} + \Bigl[  - (  - 2\,\mrA^{c}_{14}\,\xphs - \mrB^{b}_{10}\,s^2 
         + \mrB^{b}_{11} )\,c^2 + ( \mrB^{b}_{27} + \mrB^{b}_{28}\,s^2 ) \Bigr]\,c^2 \Bigr\}
         \,\frac{1}{c^2}
\nl &-&
         \frac{1}{6}
         \,\Bigl\{ 16\,\Bigl[ 2 - c^2\,\xphs \Bigr]\,c^2 
         - \Bigl[  - \mrA^{c}_{5}\,\xphs + \mrB^{b}_{26} \Bigr]\,\mrC^{a}_{0} \Bigr\}
         \,\myNG
\eqas
\bqas
   \mrM^{\ct}_{4,5} &=&
         \sumg \Bigl\{ 2\,\xog - \Bigl[ 4\,\xtg - \xog\,\xphs \Bigr]\,c^2 \Bigr\}
         \,s^2
\nl &+&
         \frac{1}{6}
         \,\Bigl\{ 6\,\mrA^{c}_{13} - \Bigl[ ( \mrA^{c}_{8}\,\xphs + 2\,\mrB^{b}_{29} ) 
         - ( 4\,\mrB^{b}_{4} - \mrB^{b}_{32}\,s^2 )\,c^2 \Bigr]\,c^2 \Bigr\}
         \,\frac{1}{c^2}
\nl &-&
         \frac{1}{6}
         \,\Bigl\{  - 16\,\Bigl[ 2 - c^2\,\xphs \Bigr]\,c^2 
         + \Bigl[ 2 - \mrA^{c}_{5}\,\xphs \Bigr]\,\mrC^{a}_{0} \Bigr\}
         \,\myNG
\eqas
\bqas
   \mrM^{\ct}_{4,6} &=&
       -
         \sumg \Bigl\{ 2\,\xog - \Bigl[ 4\,\xtg - \xog\,\xphs \Bigr]\,c^2 \Bigr\}
         \,s\,c
\nl &+&
         \frac{1}{12}
         \,\Bigl\{ 2\,\mrA^{c}_{5}\,\mrC^{a}_{0} + \Bigl[ 16\,\mrA^{c}_{5}\,c^2\,\xphs 
         - ( \mrA^{c}_{4}\,\mrC^{a}_{0}\,\xphs + 32\,\mrA^{c}_{5} ) \Bigr]\,c^2 \Bigr\}
         \,\frac{1}{s\,c}
         \,\myNG
\nl &+&
         \frac{1}{6}
         \,\Bigl\{ 2\,\mrA^{c}_{21} + \Bigl[ 2\,\mrB^{b}_{30}\,s^2 - ( 
         - \mrA^{c}_{6}\,\xphs + \mrB^{b}_{1} - \mrB^{b}_{8}\,s^2 )\,c^2 \Bigr]\,c^2 \Bigr\}
         \,\frac{1}{s\,c}
\eqas
\bqas
   \mrM^{\ct}_{4,10}\lpar \PQq \rpar &=&
         2
         \,\Bigl\{ \mrC^{a}_{2} + 6\,\mrD^{a}_{2}\,c^2 \Bigr\}
         \,\frac{1}{c^2}
\qquad
   \mrM^{\ct}_{4,11}\lpar \Pl \rpar =
       -
         \frac{2}{3}
         \,\Bigl\{ \vple - 6\,c^2\,\xpls \Bigr\}
         \,\frac{1}{c^2}
\eqas
\bqas
   \mrM^{\ct}_{4,12}\lpar \PQu \rpar &=&
       -
         2
         \,\Bigl\{ \vmqu - 6\,c^2\,\xpus \Bigr\}
         \,\frac{1}{c^2}
\qquad
   \mrM^{\ct}_{4,13}\lpar \PQd \rpar =
         2
         \,\Bigl\{ \vmqd - 6\,c^2\,\xpds \Bigr\}
         \,\frac{1}{c^2}
\eqas
\bqas
   \mrM^{\ct}_{4,14}\lpar \Pl \rpar &=&
         \frac{2}{3}
         \,\Bigl\{ \vmle - 6\,c^2\,\xpls \Bigr\}
         \,\frac{1}{c^2}
\qquad
   \mrM^{\ct}_{4,15}\lpar \PQq \rpar =
         2
         \,\Bigl\{ \mrC^{a}_{3} - 9\,\mrD^{a}_{4}\,c^2 \Bigr\}
         \,\frac{1}{c^2}
\qquad
   \mrM^{\ct}_{4,16}\lpar \Pl \rpar =
         \frac{2}{3}
         \,\Bigl\{ \vple - 9\,c^2\,\xpls \Bigr\}
         \,\frac{1}{c^2}
\eqas
\bei
\item {\underline{$\mrM^{\ct}_{5,i}$ entries}}
\eei
\bqas
   \mrM^{\ct}_{5,2} &=&
         \frac{5}{3}
\qquad
   \mrM^{\ct}_{5,3} =
       -
         \frac{1}{4}
         \,\mrA^{c}_{24}
         \,\frac{1}{c^2}
\eqas
\bqas
   \mrM^{\ct}_{5,4} &=&
         \frac{8}{3}
         \,\myNG
       -
         \sumg \,\xog
       +
         \frac{1}{6}
         \,\Bigl\{ 15 + \mrB^{b}_{33}\,c^2 \Bigr\}
         \,\frac{1}{c^2}
\eqas
\bqas
   \mrM^{\ct}_{5,6} &=&
       -
         \frac{s}{c}
\qquad
   \mrM^{\ct}_{5,15}\lpar \PQq \rpar =
         2
         \,\mrD^{a}_{6}
\qquad
   \mrM^{\ct}_{5,16}\lpar \Pl \rpar =
         \frac{2}{3}
         \,\mrD^{a}_{5}
\eqas
\bqas
   \mrM^{\ct}_{5,17}\lpar \PQq\,,\,\PQu \rpar &=&
       -
         \frac{3}{2}
         \,\xpus
\qquad
   \mrM^{\ct}_{5,18}\lpar \PQq\,,\,\PQd \rpar =
         \frac{3}{2}
         \,\xpds
\qquad
   \mrM^{\ct}_{5,19}\lpar \lambda\,,\,\Pl \rpar =
         \frac{1}{2}
         \,\xpls
\eqas
\bei
\item {\underline{$\mrM^{\ct}_{6,i}$ entries}}
\eei
\bqas
   \mrM^{\ct}_{6,3} &=&
       -
         \frac{1}{24}
         \,\mrA^{c}_{29}
         \,\frac{1}{s\,c}
       -
         \frac{1}{36}
         \,\Bigl\{  - 3\,c^2\,\vtg + 4\,\mrA^{c}_{28}\,s^2 \Bigr\}
         \,\frac{1}{s\,c}
         \,\myNG
\eqas
\bqas
   \mrM^{\ct}_{6,4} &=&
         \frac{1}{36}
         \,\Bigl\{ 64\,\mrA^{c}_{25}\,c^2 + \Bigl[ 24\,s^2\,\vtg + \mrC^{a}_{7} \Bigr] \Bigr\}
         \,\frac{s}{c}
         \,\myNG
\nl &+&
         \frac{1}{6}
         \,\Bigl\{ \mrA^{c}_{27} - 6\,\Bigl[ \mrA^{c}_{25}\,\mrB^{b}_{15} 
         - 2\,\mrB^{b}_{35}\,s^2 \Bigr]\,c^2 \Bigr\}
         \,\frac{s}{c}
\eqas
\bqas
   \mrM^{\ct}_{6,5} &=&
         \frac{1}{6}
         \,\Bigl\{ 5\,\mrA^{c}_{24} - 6\,\Bigl[ \mrB^{b}_{1}\,c^2 - ( 
       - 2\,\mrB^{b}_{35} + \mrB^{b}_{38} )\,s^2 \Bigr]\,c^2 \Bigr\}
         \,\frac{s}{c}
\nl &+&
         \frac{1}{36}
         \,\Bigl\{ 3\,\mrC^{a}_{0} - 8\,\Bigl[ 16\,s^2 + 3\,\mrC^{a}_{4} \Bigr]\,c^2 \Bigr\}
         \,\frac{s}{c}
         \,\myNG
\eqas
\bqas
   \mrM^{\ct}_{6,6} &=&
       -
         \frac{1}{72}
         \,\Bigl\{ 3\,\mrA^{c}_{16}\,\mrC^{a}_{0} - 2\,\Bigl[ 128\,s^2\,c^2
         + (  - 8\,\mrC^{a}_{5}\,s^2 + 3\,\mrC^{a}_{6} ) \Bigr]\,c^2 \Bigr\}
         \,\frac{1}{c^2}
         \,\myNG
\nl &-&
         \frac{1}{12}
         \,\Bigl\{ \mrA^{c}_{26} - 2\,\Bigl[  - ( 3\,\mrB^{b}_{13} 
         + 12\,\mrB^{b}_{34}\,s^4 - \mrB^{b}_{39}\,s^2 ) 
         + (  - 12\,\mrB^{b}_{36}\,s^2 + \mrB^{b}_{37} )\,c^2 \Bigr]\,c^2 \Bigr\}
         \,\frac{1}{c^2}
\eqas
\bqas
   \mrM^{\ct}_{6,10}\lpar \PQq \rpar &=&
       -
         \frac{2}{3}
         \,\frac{s}{c}
\qquad
   \mrM^{\ct}_{6,11}\lpar \Pl \rpar =
         \frac{2}{3}
         \,\frac{s}{c}
\qquad
   \mrM^{\ct}_{6,12}\lpar \PQu \rpar =
       -
         \frac{4}{3}
         \,\frac{s}{c}
\qquad
   \mrM^{\ct}_{6,13}\lpar \PQd \rpar =
         \frac{2}{3}
         \,\frac{s}{c}
\eqas
\bqas
   \mrM^{\ct}_{6,14}\lpar \Pl \rpar &=&
         \frac{2}{3}
         \,\frac{s}{c}
\qquad
   \mrM^{\ct}_{6,15}\lpar \PQq \rpar =
       -
         2
         \,\frac{s}{c}
\qquad
   \mrM^{\ct}_{6,16}\lpar \Pl \rpar =
       -
         \frac{2}{3}
         \,\frac{s}{c}
\eqas
\bqas
   \mrM^{\ct}_{6,17}\lpar \PQq\,,\,\PQu \rpar &=&
         \frac{1}{4}
         \,\Bigl\{ 8\,c^2 + 3\,v_{\PQu} \Bigr\}
         \,\frac{s}{c}
         \,\xpus
\qquad
   \mrM^{\ct}_{6,18}\lpar \PQq\,,\,\PQd \rpar =
       -
         \frac{1}{4}
         \,\Bigl\{ 4\,c^2 + 3\,v_{\PQd} \Bigr\}
         \,\frac{s}{c}
         \,\xpds
\eqas
\bqas
   \mrM^{\ct}_{6,19}\lpar \lambda\,,\,\Pl \rpar &=&
       -
         \frac{1}{4}
         \,\Bigl\{ 4\,c^2 + v_{\Pl} \Bigr\}
         \,\frac{s}{c}
         \,\xpls
\qquad
   \mrM^{\ct}_{6,20}\lpar \PQq\,,\,\PQu \rpar =
         \frac{1}{4}
         \,\Bigl\{ 3\,v_{\PQu} - 8\,s^2 \Bigr\}
         \,\xpus
\eqas
\bqas
   \mrM^{\ct}_{6,21}\lpar \PQq\,,\,\PQd \rpar &=&
         \frac{1}{4}
         \,\Bigl\{ 3\,v_{\PQd} - 4\,s^2 \Bigr\}
         \,\xpds
\qquad
   \mrM^{\ct}_{6,22}\lpar \lambda\,,\,\Pl \rpar =
         \frac{1}{4}
         \,\Bigl\{ v_{\Pl} - 4\,s^2 \Bigr\}
         \,\xpls
\eqas
\bei
\item {\underline{$\mrM^{\ct}_{7,i}$ entries}}
\eei
\bqas
   \mrM^{\ct}_{7,2} &=&
       -
         \frac{5}{6}
         \,\frac{s^2}{c^2}
\eqas
\bqas
   \mrM^{\ct}_{7,3} &=&
       -
         \frac{1}{8}
         \,\sumg \,\xog
       +
         \frac{1}{48}
         \,\Bigl\{  - 4\,c^2\,\vtg + \mrC^{a}_{10} \Bigr\}
         \,\frac{1}{c^2}
         \,\myNG
       -
         \frac{1}{48}
         \,\Bigl\{ \mrA^{c}_{30} + \mrB^{b}_{40}\,c^2 \Bigr\}
         \,\frac{1}{c^2}
\eqas
\bqas
   \mrM^{\ct}_{7,4} &=&
         \frac{1}{12}
         \,\Bigl\{ \mrA^{c}_{33} - 3\,\mrB^{b}_{42}\,c^2 \Bigr\}
         \,\frac{s^2}{c^2}
       -
         \frac{1}{12}
         \,\Bigl\{  - 4\,\mrC^{a}_{9}\,c^2 + \mrC^{a}_{10} \Bigr\}
         \,\frac{1}{c^2}
         \,\myNG
\eqas
\bqas
   \mrM^{\ct}_{7,5} &=&
       -
         \frac{1}{4}
         \,\Bigl\{ \mrA^{c}_{32} - \mrB^{b}_{41}\,c^2 \Bigr\}
         \,\frac{s^2}{c^2}
\qquad
   \mrM^{\ct}_{7,6} =
         \frac{1}{3}
         \,\frac{s}{c}
         \,\myNG\,\vtg
       +
         \frac{1}{12}
         \,\Bigl\{ \mrA^{c}_{31} - \mrB^{b}_{38}\,c^2 \Bigr\}
         \,\frac{s}{c}
\eqas
\bqas
   \mrM^{\ct}_{7,10}\lpar \PQq \rpar &=&
       -
         \frac{1}{2}
         \,\Bigl\{ \mrC^{a}_{2} + 6\,\mrD^{a}_{2}\,c^2 \Bigr\}
         \,\frac{1}{c^2}
\qquad
   \mrM^{\ct}_{7,11}\lpar \Pl \rpar =
       -
         \frac{1}{6}
         \,\Bigl\{ \vmle + 6\,c^2\,\xpls \Bigr\}
         \,\frac{1}{c^2}
\eqas
\bqas
   \mrM^{\ct}_{7,12}\lpar \PQu \rpar &=&
         \frac{1}{2}
         \,\Bigl\{ \vmqu - 6\,c^2\,\xpus \Bigr\}
         \,\frac{1}{c^2}
\qquad
   \mrM^{\ct}_{7,13}\lpar \PQd \rpar =
       -
         \frac{1}{2}
         \,\Bigl\{ \vmqd - 6\,c^2\,\xpds \Bigr\}
         \,\frac{1}{c^2}
\eqas
\bqas
   \mrM^{\ct}_{7,14}\lpar \Pl \rpar &=&
       -
         \frac{1}{6}
         \,\Bigl\{ \vmle - 6\,c^2\,\xpls \Bigr\}
         \,\frac{1}{c^2}
\qquad
   \mrM^{\ct}_{7,15}\lpar \PQq \rpar =
       -
         \frac{1}{2}
         \,\Bigl\{  - 4\,c^2 + \mrC^{a}_{3} \Bigr\}
         \,\frac{1}{c^2}
\eqas
\bqas
   \mrM^{\ct}_{7,16}\lpar \Pl \rpar &=&
       -
         \frac{1}{6}
         \,\Bigl\{  - 4\,c^2 + \mrC^{a}_{8} \Bigr\}
         \,\frac{1}{c^2}
\qquad
   \mrM^{\ct}_{7,17}\lpar \PQq\,,\,\PQu \rpar =
       -
         \frac{3}{4}
         \,\xpus\,\vmqu
\eqas
\bqas
   \mrM^{\ct}_{7,18}\lpar \PQq\,,\,\PQd \rpar &=&
         \frac{3}{4}
         \,\xpds\,\vmqd
\qquad
   \mrM^{\ct}_{7,19}\lpar \lambda\,,\,\Pl \rpar =
         \frac{1}{4}
         \,\xpls\,\vmle
\eqas
\bqas
   \mrM^{\ct}_{7,20}\lpar \PQq\,,\,\PQu \rpar &=&
       -
         \frac{3}{4}
         \,\frac{s}{c}
         \,v_{\PQu}\,\xpus
\qquad
   \mrM^{\ct}_{7,21}\lpar \PQq\,,\,\PQd \rpar =
       -
         \frac{3}{4}
         \,\frac{s}{c}
         \,v_{\PQd}\,\xpds
\eqas
\bqas
   \mrM^{\ct}_{7,22}\lpar \lambda\,,\,\Pl \rpar &=&
       -
         \frac{1}{4}
         \,\frac{s}{c}
         \,v_{\Pl}\,\xpls
\eqas
\bei
\item {\underline{$\mrM^{\ct}_{8,i}$ entries}}
\eei
\bqas
   \mrM^{\ct}_{8,2} &=&
         \sumg \,\xog
       -
         \frac{1}{6}
         \,\Bigl\{ \mrA^{c}_{37} + 2\,\mrB^{b}_{18}\,c^2 \Bigr\}
         \,\frac{1}{c^2}
\eqas
\bqas
   \mrM^{\ct}_{8,3} &=&
         \frac{1}{4}
         \,\sumg \,\xog
       +
         \frac{1}{24}
         \,\Bigl\{  - 4\,c^2\,\vtg + \mrC^{a}_{10} \Bigr\}
         \,\frac{1}{c^2}
         \,\myNG
       -
         \frac{1}{24}
         \,\Bigl\{ \mrA^{c}_{38} - 3\,\mrB^{b}_{45}\,c^2 \Bigr\}
         \,\frac{1}{c^2}
\eqas
\bqas
   \mrM^{\ct}_{8,4} &=&
         \sumg \Bigl\{ c^2\,\xog\,\xphs + 2\,\Bigl[ 2\,s^2\,\xtg + \mrD^{a}_{3} \Bigr] \Bigr\}
         \,c^2
\nl &+&
         \frac{1}{6}
         \,\Bigl\{ \mrA^{c}_{36} + \Bigl[  - (  
         - 2\,\mrA^{c}_{14}\,\xphs + \mrB^{b}_{11} - \mrB^{b}_{50}\,s^2 )\,c^2 
         + ( \mrB^{b}_{43} - 2\,\mrB^{b}_{48}\,s^2 ) \Bigr]\,c^2 \Bigr\}
         \,\frac{1}{c^2}
\nl &-&
         \frac{1}{6}
         \,\Bigl\{ \mrC^{a}_{10} + 2\,\Bigl[ ( 16 - 8\,c^2\,\xphs 
         + \mrC^{a}_{0}\,\xphs )\,c^2 - ( \mrC^{a}_{13} ) \Bigr]\,c^2 \Bigr\}
         \,\frac{1}{c^2}
         \,\myNG
\eqas
\bqas
   \mrM^{\ct}_{8,5} &=&
         \sumg \Bigl\{ 2\,\xog - \Bigl[ 4\,\xtg - \xog\,\xphs \Bigr]\,c^2 \Bigr\}
         \,s^2
\nl &-&
         \frac{1}{6}
         \,\Bigl\{ 9\,\mrA^{c}_{34} + \Bigl[  - (  - 2\,\mrA^{c}_{14}\,\xphs + 4\,\mrB^{b}_{4} 
         - \mrB^{b}_{49}\,s^2 )\,c^2 + ( \mrB^{b}_{44} 
         - 2\,\mrB^{b}_{46}\,s^2 ) \Bigr]\,c^2 \Bigr\}
         \,\frac{1}{c^2}
\nl &-&
         \frac{1}{6}
         \,\Bigl\{  - 16\,\Bigl[ 2 - c^2\,\xphs \Bigr]\,c^2 
         + \Bigl[ 2 - \mrA^{c}_{5}\,\xphs \Bigr]\,\mrC^{a}_{0} \Bigr\}
         \,\myNG
\eqas
\bqas
   \mrM^{\ct}_{8,6} &=&
       -
         \sumg \Bigl\{ 2\,\xog - \Bigl[ 4\,\xtg - \xog\,\xphs \Bigr]\,c^2 \Bigr\}
         \,s\,c
\nl &+&
         \frac{1}{6}
         \,\Bigl\{ \mrA^{c}_{35} + \Bigl[ \mrB^{b}_{47}\,s^2 
         - (  - \mrA^{c}_{6}\,\xphs + \mrB^{b}_{1} - \mrB^{b}_{8}\,s^2 )\,c^2 \Bigr]\,c^2 \Bigr\}
         \,\frac{1}{s\,c}
\nl &-&
         \frac{1}{12}
         \,\Bigl\{ \Bigl[ 16\,( 2 - \mrA^{c}_{5}\,\xphs )\,c^2 
         - (  - \mrA^{c}_{4}\,\mrC^{a}_{0}\,\xphs 
         + 2\,\mrC^{a}_{0} + 16\,\mrC^{a}_{11}\,s^2 ) \Bigr]\,c^2 
\nl &+&
         2\,\Bigl[ (  - 8\,s^2\,\vtg + \mrC^{a}_{12} ) \Bigr]\,s^2 \Bigr\}
         \,\frac{1}{s\,c}
         \,\myNG
\eqas
\bqas
   \mrM^{\ct}_{8,10}\lpar \PQq \rpar &=&
         \Bigl\{ \mrC^{a}_{2} + 12\,\mrD^{a}_{2}\,c^2 \Bigr\}
         \,\frac{1}{c^2}
\qquad
   \mrM^{\ct}_{8,11}\lpar \Pl \rpar =
       -
         \frac{1}{3}
         \,\Bigl\{  - 12\,c^2\,\xpls + \mrC^{a}_{8} \Bigr\}
         \,\frac{1}{c^2}
\eqas
\bqas
   \mrM^{\ct}_{8,12}\lpar \PQu \rpar &=&
       -
         \Bigl\{ \vmqu - 12\,c^2\,\xpus \Bigr\}
         \,\frac{1}{c^2}
\qquad
   \mrM^{\ct}_{8,13}\lpar \PQd \rpar =
         \Bigl\{ \vmqd - 12\,c^2\,\xpds \Bigr\}
         \,\frac{1}{c^2}
\eqas
\bqas
   \mrM^{\ct}_{8,14}\lpar \Pl \rpar &=&
         \frac{1}{3}
         \,\Bigl\{ \vmle - 12\,c^2\,\xpls \Bigr\}
         \,\frac{1}{c^2}
\qquad
   \mrM^{\ct}_{8,15}\lpar \PQq \rpar =
         \Bigl\{ \mrC^{a}_{3} + 2\,\mrD^{a}_{1}\,c^2 \Bigr\}
         \,\frac{1}{c^2}
\eqas
\bqas
   \mrM^{\ct}_{8,16}\lpar \Pl \rpar &=&
       -
         \frac{1}{3}
         \,\Bigl\{ \vmle - 2\,\mrD^{a}_{0}\,c^2 \Bigr\}
         \,\frac{1}{c^2}
\qquad
   \mrM^{\ct}_{8,17}\lpar \PQq\,,\,\PQu \rpar =
       -
         \frac{3}{2}
         \,\xpus\,\vmqu
\eqas
\bqas
   \mrM^{\ct}_{8,18}\lpar \PQq\,,\,\PQd \rpar &=&
         \frac{3}{2}
         \,\xpds\,\vmqd
\qquad
   \mrM^{\ct}_{8,19}\lpar \lambda\,,\,\Pl \rpar =
         \frac{1}{2}
         \,\xpls\,\vmle
\eqas
\bqas
   \mrM^{\ct}_{8,20}\lpar \PQq\,,\,\PQu \rpar &=&
       -
         \frac{3}{2}
         \,\frac{s}{c}
         \,v_{\PQu}\,\xpus
\qquad
   \mrM^{\ct}_{8,21}\lpar \PQq\,,\,\PQd \rpar =
       -
         \frac{3}{2}
         \,\frac{s}{c}
         \,v_{\PQd}\,\xpds
\eqas
\bqas
   \mrM^{\ct}_{8,22}\lpar \lambda\,,\,\Pl \rpar &=&
       -
         \frac{1}{2}
         \,\frac{s}{c}
         \,v_{\Pl}\,\xpls
\eqas
\bei
\item {\underline{$\mrM^{\ct}_{9,i}$ entries}}
\eei
\bqas
   \mrM^{\ct}_{9,2} &=&
         \frac{1}{2}
         \,\sumg \,\xog
       -
         \frac{1}{6}
         \,\Bigl\{ \mrA^{c}_{19} + \mrB^{b}_{18}\,c^2 \Bigr\}
         \,\frac{1}{c^2}
\eqas
\bqas
   \mrM^{\ct}_{9,3} &=&
         \frac{1}{8}
         \,\sumg \,\xog
       -
         \frac{1}{48}
         \,\Bigl\{ 3\,\mrA^{c}_{23} - \mrB^{b}_{31}\,c^2 \Bigr\}
         \,\frac{1}{c^2}
\eqas
\bqas
   \mrM^{\ct}_{9,4} &=&
         \frac{1}{2}
         \,\sumg \Bigl\{ c^2\,\xog\,\xphs + 2\,\Bigl[ 2\,s^2\,\xtg + \mrD^{a}_{3} \Bigr] \Bigr\}
         \,c^2
\nl &+&
         \frac{1}{12}
         \,\Bigl\{ 3\,\mrA^{c}_{40} - \Bigl[ (  - \mrA^{c}_{8}\,\xphs
         + 2\,\mrB^{b}_{51} ) + (  - \mrB^{b}_{10}\,s^2 + \mrB^{b}_{11} )\,c^2 \Bigr]\,c^2 \Bigr\}
         \,\frac{1}{c^2}
\nl &+&
         \frac{1}{6}
         \,\Bigl\{ \mrC^{a}_{1} + \Bigl[ 8\,c^2\,\xphs 
         - ( 16 + \mrC^{a}_{0}\,\xphs ) \Bigr]\,c^2 \Bigr\}
         \,\myNG
\eqas
\bqas
   \mrM^{\ct}_{9,5} &=&
         \frac{1}{2}
         \,\sumg \Bigl\{ 2\,\xog - \Bigl[ 4\,\xtg - \xog\,\xphs \Bigr]\,c^2 \Bigr\}
         \,s^2
\nl &+&
         \frac{1}{12}
         \,\Bigl\{ 6\,\mrA^{c}_{13} - \Bigl[ ( \mrA^{c}_{8}\,\xphs + 2\,\mrB^{b}_{29} ) 
         - ( 4\,\mrB^{b}_{4} - \mrB^{b}_{9}\,s^2 )\,c^2 \Bigr]\,c^2 \Bigr\}
         \,\frac{1}{c^2}
\nl &-&
         \frac{1}{12}
         \,\Bigl\{  - 16\,\Bigl[ 2 - c^2\,\xphs \Bigr]\,c^2 
         + \Bigl[ 2 - \mrA^{c}_{5}\,\xphs \Bigr]\,\mrC^{a}_{0} \Bigr\}
         \,\myNG
\eqas
\bqas
   \mrM^{\ct}_{9,6} &=&
       -
         \frac{1}{2}
         \,\sumg \Bigl\{ 2\,\xog - \Bigl[ 4\,\xtg - \xog\,\xphs \Bigr]\,c^2 \Bigr\}
         \,s\,c
\nl &+&
         \frac{1}{24}
         \,\Bigl\{ 2\,\mrA^{c}_{5}\,\mrC^{a}_{0} + \Bigl[ 16\,\mrA^{c}_{5}\,c^2\,\xphs 
         - ( \mrA^{c}_{4}\,\mrC^{a}_{0}\,\xphs + 32\,\mrA^{c}_{5} ) \Bigr]\,c^2 \Bigr\}
         \,\frac{1}{s\,c}
         \,\myNG
\nl &+&
         \frac{1}{12}
         \,\Bigl\{ 2\,\mrA^{c}_{39} + \Bigl[ 2\,\mrB^{b}_{6}\,s^2 - ( 
         - \mrA^{c}_{6}\,\xphs + \mrB^{b}_{1} - \mrB^{b}_{8}\,s^2 )\,c^2 \Bigr]\,c^2 \Bigr\}
         \,\frac{1}{s\,c}
\eqas
\bqas
   \mrM^{\ct}_{9,10}\lpar \PQq \rpar &=&
         \Bigl\{ \mrC^{a}_{2} + 6\,\mrD^{a}_{2}\,c^2 \Bigr\}
         \,\frac{1}{c^2}
\qquad
   \mrM^{\ct}_{9,11}\lpar \Pl \rpar =
       -
         \frac{1}{3}
         \,\Bigl\{ \vple - 6\,c^2\,\xpls \Bigr\}
         \,\frac{1}{c^2}
\eqas
\bqas
   \mrM^{\ct}_{9,12}\lpar \PQu \rpar &=&
       -
         \Bigl\{ \vmqu - 6\,c^2\,\xpus \Bigr\}
         \,\frac{1}{c^2}
\qquad
   \mrM^{\ct}_{9,13}\lpar \PQd \rpar =
         \Bigl\{ \vmqd - 6\,c^2\,\xpds \Bigr\}
         \,\frac{1}{c^2}
\eqas
\bqas
   \mrM^{\ct}_{9,14}\lpar \Pl \rpar &=&
         \frac{1}{3}
         \,\Bigl\{ \vmle - 6\,c^2\,\xpls \Bigr\}
         \,\frac{1}{c^2}
\qquad
   \mrM^{\ct}_{9,15}\lpar \PQq \rpar =
         \Bigl\{ \mrC^{a}_{3} + \mrD^{a}_{1}\,c^2 \Bigr\}
         \,\frac{1}{c^2}
\eqas
\bqas
   \mrM^{\ct}_{9,16}\lpar \Pl \rpar &=&
         \frac{1}{3}
         \,\Bigl\{ \vple + \mrD^{a}_{0}\,c^2 \Bigr\}
         \,\frac{1}{c^2}
\qquad
   \mrM^{\ct}_{9,17}\lpar \PQq\,,\,\PQu \rpar =
       -
         \frac{3}{4}
         \,\xpus
\eqas
\bqas
   \mrM^{\ct}_{9,18}\lpar \PQq\,,\,\PQd \rpar &=&
         \frac{3}{4}
         \,\xpds
\qquad
   \mrM^{\ct}_{9,19}\lpar \lambda\,,\,\Pl \rpar =
         \frac{1}{4}
         \,\xpls
\eqas
\bei
\item {\underline{$\mrM^{\ct}_{10,i}$ entries}}
\eei
\bqas
   \mrM^{\ct}_{10,2} &=&
         \frac{3}{4}
         \,\xpDs
\qquad
   \mrM^{\ct}_{10,3} =
       -
         \frac{1}{48}
         \,\Bigl\{ 7 + 6\,c^2\,\xpDs \Bigr\}
         \,\frac{1}{c^2}
\eqas
\bqas
   \mrM^{\ct}_{10,4} &=&
         \frac{1}{12}
         \,\Bigl\{ 7 + \mrE^{a}_{0}\,c^2 \Bigr\}
         \,\frac{1}{c^2}
\qquad
   \mrM^{\ct}_{10,6} =
       -
         \frac{1}{6}
         \,\frac{s}{c}
\eqas
\bqas
   \mrM^{\ct}_{10,8}\lpar \PQQ\,,\,\PQD \rpar &=&
       -
         \frac{1}{4}
         \,\Bigl\{ 1 + \Bigl[ 3\,\xphs + \mrE^{a}_{8} \Bigr]\,c^2 \Bigr\}
         \,\frac{1}{c^2}
\qquad
   \mrM^{\ct}_{10,10}\lpar \PQQ \rpar =
       -
         \frac{1}{4}
         \,\Bigl\{ 4 - \mrE^{a}_{12}\,c^2 \Bigr\}
         \,\frac{1}{c^2}
\eqas
\bqas
   \mrM^{\ct}_{10,13}\lpar \PQD \rpar &=&
         \frac{1}{4}
         \,\Bigl\{ 2 + \mrE^{a}_{11}\,c^2 \Bigr\}
         \,\frac{1}{c^2}
\qquad
   \mrM^{\ct}_{10,15}\lpar \PQQ \rpar =
         \frac{1}{4}
         \,\Bigl\{ 4 - \mrE^{a}_{10}\,c^2 \Bigr\}
         \,\frac{1}{c^2}
\eqas
\bqas
   \mrM^{\ct}_{10,17}\lpar \PQQ\,,\,\PQU \rpar &=&
         \frac{3}{8}
         \,\xpUs
\qquad
   \mrM^{\ct}_{10,18}\lpar \PQQ\,,\,\PQD \rpar =
       -
         \frac{1}{16}
         \,\Bigl\{ 2\,\mrA^{c}_{16} + 3\,\mrE^{a}_{9}\,c^2 \Bigr\}
         \,\frac{1}{c^2}
\eqas
\bqas
   \mrM^{\ct}_{10,21}\lpar \PQQ\,,\,\PQD \rpar &=&
         \frac{1}{16}
         \,\Bigl\{ 2 - \mrE^{a}_{13}\,c^2 \Bigr\}
         \,\frac{s}{c^3}
\qquad
   \mrM^{\ct}_{10,23}\lpar \lambda\,,\,\Pl\,,\,\PQD\,,\,\PQQ \rpar =
         \frac{1}{2}
         \,\sumg \frac{\xplc}{\xpD}
\eqas
\bqas
   \mrM^{\ct}_{10,24}\lpar \PQq\,,\,\PQu\,,\,\PQQ\,,\,\PQD \rpar &=&
         \frac{3}{2}
         \,\sumg \frac{\xpuc}{\xpD}
\eqas
\bei
\item {\underline{$\mrM^{\ct}_{11,i}$ entries}}
\eei
\bqas
   \mrM^{\ct}_{11,2} &=&
         \frac{1}{4}
         \,\xpDs
\qquad
   \mrM^{\ct}_{11,3} =
       -
         \frac{1}{72}
         \,\Bigl\{ 4 + 9\,c^2\,\xpDs \Bigr\}
         \,\frac{1}{c^2}
\eqas
\bqas
   \mrM^{\ct}_{11,4} &=&
         \frac{1}{9}
         \,\Bigl\{ 2\,\mrA^{c}_{41} - \mrE^{a}_{1}\,c^2 \Bigr\}
         \,\frac{1}{c^2}
\qquad
   \mrM^{\ct}_{11,5} =
       -
         \frac{4}{9}
         \,s^4
\eqas
\bqas
   \mrM^{\ct}_{11,6} &=&
         \frac{2}{9}
         \,\mrA^{c}_{25}
         \,s\,c
\qquad
   \mrM^{\ct}_{11,8}\lpar \PQQ\,,\,\PQD \rpar =
       -
         \frac{1}{4}
         \,\xpDs
\eqas
\bqas
   \mrM^{\ct}_{11,13}\lpar \PQD \rpar &=&
         \frac{1}{6}
         \,\Bigl\{ 4 - \mrE^{a}_{12}\,c^2 \Bigr\}
         \,\frac{1}{c^2}
\qquad
   \mrM^{\ct}_{11,21}\lpar \PQQ\,,\,\PQD \rpar =
         \frac{1}{4}
         \,\frac{s}{c}
         \,\xpDs
\eqas
\bei
\item {\underline{$\mrM^{\ct}_{12,i}$ entries}}
\eei
\bqas
   \mrM^{\ct}_{12,2} &=&
         \frac{1}{4}
         \,\xpDs
\qquad
   \mrM^{\ct}_{12,3} =
       -
         \frac{1}{72}
         \,\Bigl\{ 1 + 9\,c^2\,\xpDs \Bigr\}
         \,\frac{1}{c^2}
\eqas
\bqas
   \mrM^{\ct}_{12,4} &=&
         \frac{1}{18}
         \,\Bigl\{ \mrA^{c}_{42} + \mrE^{a}_{2}\,c^2 \Bigr\}
         \,\frac{1}{c^2}
\qquad
   \mrM^{\ct}_{12,5} =
         \frac{2}{9}
         \,\mrA^{c}_{9}
         \,s^2
\eqas
\bqas
   \mrM^{\ct}_{12,6} &=&
       -
         \frac{4}{9}
         \,s\,c^3
\qquad
   \mrM^{\ct}_{12,8}\lpar \PQQ\,,\,\PQD \rpar =
       -
         \frac{1}{4}
         \,\xpDs
\eqas
\bqas
   \mrM^{\ct}_{12,10}\lpar \PQQ \rpar &=&
       -
         \frac{1}{6}
         \,\Bigl\{ 2 + \mrE^{a}_{11}\,c^2 \Bigr\}
         \,\frac{1}{c^2}
\qquad
   \mrM^{\ct}_{12,15}\lpar \PQQ \rpar =
         \frac{1}{6}
         \,\Bigl\{ 2 + \mrE^{a}_{14}\,c^2 \Bigr\}
         \,\frac{1}{c^2}
\eqas
\bqas
   \mrM^{\ct}_{12,17}\lpar \PQQ\,,\,\PQU \rpar &=&
         \frac{3}{4}
         \,\xpUs
\qquad
   \mrM^{\ct}_{12,18}\lpar \PQQ\,,\,\PQD \rpar =
       -
         \frac{3}{8}
         \,\xpDs
\eqas
\bqas
   \mrM^{\ct}_{12,21}\lpar \PQQ\,,\,\PQD \rpar &=&
       -
         \frac{1}{8}
         \,\frac{s}{c}
         \,\xpDs
\eqas
\bei
\item {\underline{$\mrM^{\ct}_{13,i}$ entries}}
\eei
\bqas
   \mrM^{\ct}_{13,2} &=&
         \frac{3}{4}
         \,\xpUs
\qquad
   \mrM^{\ct}_{13,3} =
         \frac{1}{48}
         \,\Bigl\{ 5 - 6\,c^2\,\xpUs \Bigr\}
         \,\frac{1}{c^2}
\eqas
\bqas
   \mrM^{\ct}_{13,4} &=&
       -
         \frac{1}{12}
         \,\Bigl\{ 5 - \mrE^{a}_{3}\,c^2 \Bigr\}
         \,\frac{1}{c^2}
\qquad
   \mrM^{\ct}_{13,6} =
       -
         \frac{5}{3}
         \,\frac{s}{c}
\eqas
\bqas
   \mrM^{\ct}_{13,7}\lpar \PQQ\,,\,\PQU \rpar &=&
         \frac{1}{4}
         \,\Bigl\{ 1 + \Bigl[ 3\,\xphs + \mrE^{a}_{15} \Bigr]\,c^2 \Bigr\}
         \,\frac{1}{c^2}
\qquad
   \mrM^{\ct}_{13,10}\lpar \PQQ \rpar =
         \frac{1}{4}
         \,\Bigl\{ 8 - \mrE^{a}_{18}\,c^2 \Bigr\}
         \,\frac{1}{c^2}
\eqas
\bqas
   \mrM^{\ct}_{13,12}\lpar \PQU \rpar &=&
         \frac{1}{4}
         \,\Bigl\{ 2 - \mrE^{a}_{17}\,c^2 \Bigr\}
         \,\frac{1}{c^2}
\qquad
   \mrM^{\ct}_{13,15}\lpar \PQQ \rpar =
         \frac{1}{4}
         \,\Bigl\{ 8 - \mrE^{a}_{19}\,c^2 \Bigr\}
         \,\frac{1}{c^2}
\eqas
\bqas
   \mrM^{\ct}_{13,17}\lpar \PQQ\,,\,\PQU \rpar &=&
         \frac{1}{16}
         \,\Bigl\{ 2\,\mrA^{c}_{1} + 3\,\mrE^{a}_{16}\,c^2 \Bigr\}
         \,\frac{1}{c^2}
\qquad
   \mrM^{\ct}_{13,18}\lpar \PQQ\,,\,\PQD \rpar =
       -
         \frac{3}{8}
         \,\xpDs
\eqas
\bqas
   \mrM^{\ct}_{13,20}\lpar \PQQ\,,\,\PQU \rpar &=&
         \frac{1}{16}
         \,\Bigl\{ 10 - \mrE^{a}_{20}\,c^2 \Bigr\}
         \,\frac{s}{c^3}
\qquad
   \mrM^{\ct}_{13,24}\lpar \PQQ\,,\,\PQU\,,\,\PQq\,,\,\PQd \rpar =
         \frac{3}{2}
         \,\sumg \frac{\xpdc}{\xpU}
\eqas
\bqas
   \mrM^{\ct}_{13,25}\lpar\lambda\,,\,\Pl\,,\,\PQQ\,,\,\PQU \rpar &=&
       -
         \frac{1}{2}
         \,\sumg \frac{\xplc}{\xpU}
\eqas
\bei
\item {\underline{$\mrM^{\ct}_{14,i}$ entries}}
\eei
\bqas
   \mrM^{\ct}_{14,2} &=&
         \frac{1}{4}
         \,\xpUs
\qquad
   \mrM^{\ct}_{14,3} =
       -
         \frac{1}{72}
         \,\Bigl\{ 16 + 9\,c^2\,\xpUs \Bigr\}
         \,\frac{1}{c^2}
\eqas
\bqas
   \mrM^{\ct}_{14,4} &=&
         \frac{1}{9}
         \,\Bigl\{ 8\,\mrA^{c}_{41} - \mrE^{a}_{4}\,c^2 \Bigr\}
         \,\frac{1}{c^2}
\qquad
   \mrM^{\ct}_{14,5} =
       -
         \frac{16}{9}
         \,s^4
\eqas
\bqas
   \mrM^{\ct}_{14,6} &=&
         \frac{8}{9}
         \,\mrA^{c}_{25}
         \,s\,c
\qquad
   \mrM^{\ct}_{14,7}\lpar \PQQ\,,\,\PQU \rpar =
         \frac{1}{4}
         \,\xpUs
\eqas
\bqas
   \mrM^{\ct}_{14,12}\lpar \PQU \rpar &=&
       -
         \frac{1}{6}
         \,\Bigl\{ 8 - \mrE^{a}_{18}\,c^2 \Bigr\}
         \,\frac{1}{c^2}
\qquad
   \mrM^{\ct}_{14,20}\lpar \PQQ\,,\,\PQU \rpar =
         \frac{1}{2}
         \,\frac{s}{c}
         \,\xpUs
\eqas
\bei
\item {\underline{$\mrM^{\ct}_{15,i}$ entries}}
\eei
\bqas
   \mrM^{\ct}_{15,2} &=&
         \frac{1}{4}
         \,\xpUs
\qquad
   \mrM^{\ct}_{15,3} =
       -
         \frac{1}{72}
         \,\Bigl\{ 1 + 9\,c^2\,\xpUs \Bigr\}
         \,\frac{1}{c^2}
\eqas
\bqas
   \mrM^{\ct}_{15,4} &=&
         \frac{1}{18}
         \,\Bigl\{ \mrA^{c}_{43} + \mrE^{a}_{2}\,c^2 \Bigr\}
         \,\frac{1}{c^2}
\qquad
   \mrM^{\ct}_{15,5} =
         \frac{4}{9}
         \,\mrA^{c}_{16}
         \,s^2
\eqas
\bqas
   \mrM^{\ct}_{15,6} &=&
       -
         \frac{4}{9}
         \,\mrA^{c}_{4}
         \,s\,c
\qquad
   \mrM^{\ct}_{15,7}\lpar \PQQ\,,\,\PQU \rpar =
         \frac{1}{4}
         \,\xpUs
\eqas
\bqas
   \mrM^{\ct}_{15,10}\lpar \PQQ \rpar &=&
       -
         \frac{1}{6}
         \,\Bigl\{ 2 - \mrE^{a}_{17}\,c^2 \Bigr\}
         \,\frac{1}{c^2}
\qquad
   \mrM^{\ct}_{15,15}\lpar \PQQ \rpar =
       -
         \frac{1}{6}
         \,\Bigl\{ 2 - \mrE^{a}_{21}\,c^2 \Bigr\}
         \,\frac{1}{c^2}
\eqas
\bqas
   \mrM^{\ct}_{15,17}\lpar \PQQ\,,\,\PQU \rpar &=&
         \frac{3}{8}
         \,\xpUs
\qquad
   \mrM^{\ct}_{15,18}\lpar \PQQ\,,\,\PQD \rpar =
       -
         \frac{3}{4}
         \,\xpDs
\eqas
\bqas
   \mrM^{\ct}_{15,20}\lpar \PQQ\,,\,\PQU \rpar &=&
         \frac{1}{8}
         \,\frac{s}{c}
         \,\xpUs
\eqas
\bei
\item {\underline{$\mrM^{\ct}_{16,i}$ entries}}
\eei
\bqas
   \mrM^{\ct}_{16,2} &=&
         \frac{3}{4}
         \,\xpLs
\qquad
   \mrM^{\ct}_{16,3} =
         \frac{1}{16}
         \,\Bigl\{ 11 - 2\,c^2\,\xpLs \Bigr\}
         \,\frac{1}{c^2}
\eqas
\bqas
   \mrM^{\ct}_{16,4} &=&
       -
         \frac{1}{4}
         \,\Bigl\{ 11 - \mrE^{a}_{5}\,c^2 \Bigr\}
         \,\frac{1}{c^2}
\qquad
   \mrM^{\ct}_{16,6} =
       -
         \frac{9}{2}
         \,\frac{s}{c}
\eqas
\bqas
   \mrM^{\ct}_{16,9}\lpar \Lambda\,,\,\PL \rpar &=&
       -
         \frac{1}{4}
         \,\Bigl\{ 1 + \Bigl[ 3\,\xphs + \mrE^{a}_{22} \Bigr]\,c^2 \Bigr\}
         \,\frac{1}{c^2}
\qquad
   \mrM^{\ct}_{16,11}\lpar \PL \rpar =
       -
         \frac{3}{4}
         \,\Bigl\{ 4 - \mrE^{a}_{24}\,c^2 \Bigr\}
         \,\frac{1}{c^2}
\eqas
\bqas
   \mrM^{\ct}_{16,14}\lpar \PL \rpar &=&
       -
         \frac{3}{4}
         \,\Bigl\{ 2 - \mrE^{a}_{23}\,c^2 \Bigr\}
         \,\frac{1}{c^2}
\qquad
   \mrM^{\ct}_{16,16}\lpar \PL \rpar =
         \frac{3}{4}
         \,\Bigl\{ 4 - \mrE^{a}_{24}\,c^2 \Bigr\}
         \,\frac{1}{c^2}
\eqas
\bqas
   \mrM^{\ct}_{16,19}\lpar \Lambda\,,\,\PL \rpar &=&
       -
         \frac{3}{16}
         \,\Bigl\{ 2\,\mrA^{c}_{4} + \mrE^{a}_{25}\,c^2 \Bigr\}
         \,\frac{1}{c^2}
\qquad
   \mrM^{\ct}_{16,22}\lpar \Lambda\,,\,\PL \rpar =
         \frac{3}{16}
         \,\Bigl\{ 6 - \mrE^{a}_{26}\,c^2 \Bigr\}
         \,\frac{s}{c^3}
\eqas
\bqas
   \mrM^{\ct}_{16,23}\lpar \Lambda\,,\,\PL\,,\,\PQd\,,\,\PQq \rpar &=&
         \frac{3}{2}
         \,\sumg \frac{\xpdc}{\xpL}
\qquad
   \mrM^{\ct}_{16,25}\lpar \Lambda\,,\,\PL\,,\,\PQq\,,\,\PQu \rpar =
       -
         \frac{3}{2}
         \,\sumg \frac{\xpuc}{\xpL}
\eqas
\bei
\item {\underline{$\mrM^{\ct}_{17,i}$ entries}}
\eei
\bqas
   \mrM^{\ct}_{17,2} &=&
         \frac{1}{4}
         \,\xpLs
\qquad
   \mrM^{\ct}_{17,3} =
       -
         \frac{1}{8}
         \,\Bigl\{ 4 + c^2\,\xpLs \Bigr\}
         \,\frac{1}{c^2}
\eqas
\bqas
   \mrM^{\ct}_{17,4} &=&
         \Bigl\{ 2\,\mrA^{c}_{41} - \mrE^{a}_{6}\,c^2 \Bigr\}
         \,\frac{1}{c^2}
\qquad
   \mrM^{\ct}_{17,5} =
       -
         4
         \,s^4
\eqas
\bqas
   \mrM^{\ct}_{17,6} &=&
         2
         \,\mrA^{c}_{25}
         \,s\,c
\qquad
   \mrM^{\ct}_{17,9}\lpar \Lambda\,,\,\PL \rpar =
       -
         \frac{1}{4}
         \,\xpLs
\eqas
\bqas
   \mrM^{\ct}_{17,14}\lpar \PL \rpar &=&
         \frac{1}{2}
         \,\Bigl\{ 4 - \mrE^{a}_{24}\,c^2 \Bigr\}
         \,\frac{1}{c^2}
\qquad
   \mrM^{\ct}_{17,22}\lpar \Lambda\,,\,\PL \rpar =
         \frac{3}{4}
         \,\frac{s}{c}
         \,\xpLs
\eqas
\bei
\item {\underline{$\mrM^{\ct}_{18,i}$ entries}}
\eei
\bqas
   \mrM^{\ct}_{18,2} &=&
         \frac{1}{4}
         \,\xpLs
\qquad
   \mrM^{\ct}_{18,3} =
       -
         \frac{1}{8}
         \,\Bigl\{ 1 + c^2\,\xpLs \Bigr\}
         \,\frac{1}{c^2}
\eqas
\bqas
   \mrM^{\ct}_{18,4} &=&
         \frac{1}{2}
         \,\Bigl\{ \mrA^{c}_{44} + \mrE^{a}_{7}\,c^2 \Bigr\}
         \,\frac{1}{c^2}
\qquad
   \mrM^{\ct}_{18,5} =
         2
         \,\mrA^{c}_{5}
         \,s^2
\eqas
\bqas
   \mrM^{\ct}_{18,6} &=&
         4
         \,s^3\,c
\qquad
   \mrM^{\ct}_{18,9}\lpar \Lambda\,,\,\PL \rpar =
       -
         \frac{1}{4}
         \,\xpLs
\eqas
\bqas
   \mrM^{\ct}_{18,11}\lpar \PL \rpar &=&
         \frac{1}{2}
         \,\Bigl\{ 2 - \mrE^{a}_{23}\,c^2 \Bigr\}
         \,\frac{1}{c^2}
\qquad
   \mrM^{\ct}_{18,16}\lpar \PL \rpar =
       -
         \frac{1}{2}
         \,\Bigl\{ 2 - \mrE^{a}_{27}\,c^2 \Bigr\}
         \,\frac{1}{c^2}
\eqas
\bqas
   \mrM^{\ct}_{18,19}\lpar \Lambda\,,\,\PL \rpar &=&
       -
         \frac{3}{8}
         \,\xpLs
\qquad
   \mrM^{\ct}_{18,22}\lpar \Lambda\,,\,\PL \rpar =
         \frac{3}{8}
         \,\frac{s}{c}
         \,\xpLs
\eqas
\bei
\item {\underline{$\mrM^{\ct}_{20,i}$ entries}}
\eei
\bqas
   \mrM^{\ct}_{20,3} &=&
       -
         \frac{1}{8}
         \,\frac{1}{c^2}
\qquad
   \mrM^{\ct}_{20,4} =
         \frac{1}{2}
         \,\Bigl\{ 1 + \mrE^{a}_{7}\,c^2 \Bigr\}
         \,\frac{1}{c^2}
\eqas
\bqas
   \mrM^{\ct}_{20,11}\lpar \PL \rpar &=&
         \frac{1}{c^2}
\qquad
   \mrM^{\ct}_{20,16}\lpar \PL \rpar =
         \Bigl\{ 1 + \mrE^{a}_{6}\,c^2 \Bigr\}
         \,\frac{1}{c^2}
\eqas
\bqa
   \mrM^{\ct}_{20,19}\lpar \Lambda\,,\,\PL \rpar &=&
       -
         \frac{3}{4}
         \,\xpLs
\eqa

\normalsize

\section{Appendix: Self-energies at $s = 0$ \label{RSEZ}}

In this Appendix we present the full list of bosonic self energies evaluated at $s = 0$.
We have introduced a simplified notation,
\bq
\zbfunp{m_1}{m_1} = \bfunp{0}{m_1}{m_2}
\eq
\etc We introduce the following polynomials:

\vspace{0.5cm}
\bei
\item[\fbox{$\mrM\,$}] where $s = \stw$, $c = \ctw$ and 
\eei

\scriptsize
\[
\begin{array}{lll}
\mrM^{a}_{0}= 5 - 3\,c & & \\
\mrM^{b}_{0}= 1 - 2\,c \;\;&\;\;
\mrM^{b}_{1}= 4 - 9\,c \;\;&\;\;
\mrM^{b}_{2}= 7 - 10\,c \\
\mrM^{b}_{3}= 5 + 4\,c \;\;&\;\;
\mrM^{b}_{4}= 1 - 3\,s \;\;&\;\;
\mrM^{b}_{5}= 4 - c \\
\mrM^{b}_{6}= 17 + 16\,c \;\;&\;\;
\mrM^{b}_{7}= 31 - 11\,c \;\;&\;\;
\mrM^{b}_{8}= 4 - s \\
\mrM^{b}_{9}= 1 + s \;\;&\;\;
\mrM^{b}_{10}= 3 + c \;\;&\;\;
\mrM^{b}_{11}= 15 + \mrM^{a}_{0}\,c \\
\mrM^{b}_{12}= 17 - 8\,c \;\;&\;\;
\mrM^{b}_{13}= 8 - 5\,c \;\;&\;\;
\mrM^{b}_{14}= 41 - 24\,s \\
\end{array}
\]

\[
\begin{array}{lll}
\mrM^{c}_{0}= 1 + 18\,c \;\;&\;\;
\mrM^{c}_{1}= 1 + 4\,c \;\;&\;\;
\mrM^{c}_{2}= 1 + 6\,c^2 \\
\mrM^{c}_{3}= 1 + 24\,s^2\,c \;\;&\;\;
\mrM^{c}_{4}= 1 + 3\,\mrM^{b}_{0}\,c \;\;&\;\;
\mrM^{c}_{5}= 1 + 4\,\mrM^{b}_{1}\,c \\
\mrM^{c}_{6}= 2 - \mrM^{b}_{2}\,c \;\;&\;\;
\mrM^{c}_{7}= 3 - c \;\;&\;\;
\mrM^{c}_{8}= 3 + 4\,c \\
\mrM^{c}_{9}= 5 + 8\,c \;\;&\;\;
\mrM^{c}_{10}= 7 - 38\,c \;\;&\;\;
\mrM^{c}_{11}= c-2 \\
\mrM^{c}_{12}= 1 - 40\,c + 36\,s\,c \;\;&\;\;
\mrM^{c}_{13}= 3 - 2\,\mrM^{b}_{3}\,c \;\;&\;\;
\mrM^{c}_{14}= 9 - 8\,s \\
\mrM^{c}_{15}= 11 + 4\,c \;\;&\;\;
\mrM^{c}_{16}= 1 + 10\,s \;\;&\;\;
\mrM^{c}_{17}= 1 + 2\,\mrM^{b}_{4}\,s \\
\mrM^{c}_{18}= 1 - 2\,\mrM^{b}_{5}\,c \;\;&\;\;
\mrM^{c}_{19}= 2 - c \;\;&\;\;
\mrM^{c}_{20}= 3 - \mrM^{b}_{6}\,c \\
\end{array}
\]

\[
\begin{array}{lll}
\mrM^{c}_{21}= 4 + c \;\;&\;\;
\mrM^{c}_{22}= 5 - 2\,c \;\;&\;\;
\mrM^{c}_{23}= 6 - \mrM^{b}_{7}\,c \\
\mrM^{c}_{24}= 7 - 2\,\mrM^{b}_{8}\,s \;\;&\;\;
\mrM^{c}_{25}= 9 - 2\,\mrM^{b}_{9}\,s \;\;&\;\;
\mrM^{c}_{26}= 18 - 11\,c \\
\mrM^{c}_{27}= 5 - 35\,c + 8\,s\,c \;\;&\;\;
\mrM^{c}_{28}= 39 - 40\,s \;\;&\;\;
\mrM^{c}_{29}= 1 + c \\
\mrM^{c}_{30}= 1 - 4\,s\,c \;\;&\;\;
\mrM^{c}_{31}= 2 - 3\,\mrM^{b}_{10}\,c \;\;&\;\;
\mrM^{c}_{32}= 3 - \mrM^{b}_{11}\,c \\
\mrM^{c}_{33}= 4 + 3\,c \;\;&\;\;
\mrM^{c}_{34}= 5 - 2\,\mrM^{b}_{12}\,c \;\;&\;\;
\mrM^{c}_{35}= 9 - 8\,\mrM^{b}_{13}\,c \\
\mrM^{c}_{36}= 29 - 16\,c \;\;&\;\;
\mrM^{c}_{37}= 37 - 48\,s \;\;&\;\;
\mrM^{c}_{38}= 37 - 2\,\mrM^{b}_{14}\,s \\
\mrM^{c}_{39}= 49 - 34\,c \;\;&\;\;
\mrM^{c}_{40}= 79 - 40\,c & \\
\end{array}
\]

\normalsize

\vspace{0.5cm}
\bei
\item[\fbox{$\mrN\,$}] 
\eei

\scriptsize
\[
\begin{array}{lll}
\mrN_{0}= 1 - 3*\vqd \;\;&\;\;
\mrN_{1}= 3 - \vle \;\;&\;\;
\mrN_{2}= 5 - 3*\vqu \\
\mrN_{3}= 20 - 3*\vtg \;\;&\;\;
\mrN_{4}= 1 + \vqu^2 \;\;&\;\;
\mrN_{5}= 1 + \vqd^2 \\
\mrN_{6}= 1 + \vle^2 \;\;&\;\;
\mrN_{7}= 9 + \vog \;\;&\;\;
\mrN_{8}= 38 + 3\,\vog \\
\mrN_{9}= 1 - \vle \;\;&\;\;
\mrN_{10}= 1 - \vqu \;\;&\;\;
\mrN_{11}= 1 + \vqu \\
\mrN_{12}= 1 - \vqd \;\;&\;\;
\mrN_{13}= 1 + \vqd \;\;&\;\;
\mrN_{14}= 1 - 3\,\vle \\
\mrN_{15}= 1 + \vle \;\;&\;\;
\mrN_{16}= 2 - \vqu \;\;&\;\;
\mrN_{17}= 2 + \vqu + \vqd \\
\mrN_{18}= 2 - \vqd \;\;&\;\;
\mrN_{19}= 2 - \vle \;\;&\;\;
\mrN_{20}= 2 + 3\,\vle \\
\mrN_{21}= 3 - 5\,\vqu \;\;&\;\;
\mrN_{22}= 3 - \vqd \;\;&\;\;
\mrN_{23}= 3 + \vle \\
\mrN_{24}= 4 + \vqu + \vqd \;\;&\;\;
\mrN_{25}= 9 - 5\,\vqu - \vqd - 3\,\vle \;\;&\;\;
\mrN_{26}= 10 + 3\,\vle \\
\mrN_{27}= 38 - 15\,\vqu - 3\,\vqd - 9\,\vle \;\;&\;\;
\mrN_{28}= \vqu - \vqd & \\
\end{array}
\]

\normalsize

\vspace{0.5cm}
\bei
\item[\fbox{$\mrO\,$}] 
\eei

\scriptsize
\[
\begin{array}{lll}
\mrO_{0}= \xpls + 3\,\xpds + 3\,\xpus \;\;&\;\;
\mrO_{1}= \xpds + \xpus \;\;&\;\;
\mrO_{2}= \xpds - 3\,\xpus \\
\mrO_{3}= 3\,\xpds - \xpus \;\;&\;\;
\mrO_{4}= ( \xpus - \xpds )^2 & \\
\end{array}
\]

\normalsize

\vspace{0.5cm}
\bei
\item[\fbox{$\mrP\,$}] 
\eei

\scriptsize
\[
\begin{array}{lll}
\mrP_{0}= 1 - \xphs \;\;&\;\;
\mrP_{1}= 2 - \xphs - \xphq \;\;&\;\;
\mrP_{2}= 3 - 2\,\xphs - \xphq \\
\mrP_{3}= 6 - 7\,\xphs + \xphq \;\;&\;\;
\mrP_{4}= 10 - 11\,\xphs + \xphq \;\;&\;\;
\mrP_{5}= 2 - 5\,\xphs + 3\,\xphq \\
\mrP_{6}= 12 - 13\,\xphs + \xphq \;\;&\;\;
\mrP_{7}= 12 - 11\,\xphs - \xphq \;\;&\;\;
\mrP_{8}= 8 - 5\,\xphs - 3\,\xphq \\
\mrP_{9}= 1 - 3\,\xphs + 3\,\xphq - \xphvi \;\;&\;\;
\mrP_{10}= 42 - 41\,\xphs - \xphq \;\;&\;\;
\mrP_{11}= 1 - \xphq \\
\mrP_{12}= 7 - 6\,\xphs \;\;&\;\;
\mrP_{13}= 9 - \xphs \;\;&\;\;
\mrP_{14}= 20 - 21\,\xphs + \xphq \\
\mrP_{15}= 5 - 6\,\xphs + \xphq \;\;&\;\;
\mrP_{16}= 78 - 79\,\xphs \;\;&\;\;
\mrP_{17}= 10 + 3\,\xphs \\
\mrP_{18}= 11 - 18\,\xphs + 7\,\xphq \;\;&\;\;
\mrP_{19}= 20 - 23\,\xphs + 3\,\xphq \;\;&\;\;
\mrP_{20}= 32 - 45\,\xphs \\
\mrP_{21}= 80 - 79\,\xphs & & \\
\end{array}
\]

\normalsize

With their help we derive the vector-vector transitions at $s = 0$

\footnotesize
\vspace{0.5cm}
\bei
\item \fbox{$\PA$ self-energy}
\eei
\bqas
\Pi^{(4)}_{\PA\PA\,;\,0}(0) &=&
       -
         \frac{32}{9}
         \,s^2\,\myNG
         \,( 1 - \LR )
       +
         \frac{1}{3}
         \,s^2
         \,( 7 - 9\,\LR )
\eqas
\bqas
\Pi^{(4)}_{\PA\PA\,;\,1}(0) &=&
         3
         \,s^2
         \,\afun{M}
       -
         \frac{4}{3}
         \,\sumg 
         \,s^2
         \,\afun{\mle}
\nl &-&
         \frac{16}{9}
         \,\sumg 
         \,s^2
         \,\afun{\mqu}
       -
         \frac{4}{9}
         \,\sumg 
         \,s^2
         \,\afun{\mqd}
\eqas
\bqas
\Pi^{(4)}_{\PA\PA\,;\,2}(0) &=& 0
\eqas
\bqas
\Pi^{(6)}_{\PA\PA\,;\,0}(0) &=& 
       -
         \frac{16}{9}
         \,\myNG\,\apWAD
         \,( 1 - \LR )
       -
         \frac{1}{6}
         \,c^2\,\apD
         \,( 7 - 9\,\LR )
\nl &+&
         \frac{1}{2}
         \,\Bigl[ s\,\apWA - \frac{c^2}{\xphs - 1}\,\mrP_{2}\,\apB 
         + \frac{s\,c^3}{\xphs - 1}\,\mrP_{3}\,\apWB 
         + ( \mrP_{1}\,\apB + \mrP_{4}\,\apW )\,\frac{s^2\,c^2}{\xphs - 1} \Bigr]
         \,\frac{\LR}{c^2}
\nl &+&
         \frac{2}{3}
         \,( c\,\apWB + 7\,s\,\apW )
         \,s
       -
         2
         \,\sumg ( \adWB\,\xpds - 2\,\auWB\,\xpus + \alWB\,\xpls )
         \,s
         \,( 1 - \LR )
\eqas
\bqas
\Pi^{(6)}_{\PA\PA\,;\,1}(0) &=& 
         \frac{1}{2}
         \,\frac{\xphs}{\xphs - 1}
         \,\mrP_{0}
         \,\aAA
         \,\afun{\mh}
       -
         \frac{1}{2}
         \,\frac{1}{c^2}
         \,\aAA
         \,\afun{\mzb}
\nl &-&
         \frac{1}{2}
         \,\Bigl[ 2\,c^2\,\apB + 3\,c^2\,\apD - 2\,( 3\,c\,\apWB + 5\,s\,\apW )\,s \Bigr]
         \,\afun{M}
       -
         \frac{2}{9}
         \,\sumg ( \apWAD + 9\,s\,\adWB\,\xpds )
         \,\afun{\mqd}
\nl &-&
         \frac{2}{3}
         \,\sumg ( \apWAD + 3\,s\,\alWB\,\xpls )
         \,\afun{\mle}
       -
         \frac{4}{9}
         \,\sumg ( 2\,\apWAD - 9\,s\,\auWB\,\xpus )
         \,\afun{\mqu}
\eqas
\bqas
\Pi^{(6)}_{\PA\PA\,;\,2}(0) &=& 0
\eqas
\vspace{0.5cm}
\bei
\item \fbox{$\PZ{-}\PA$ transition}
\eei

\bqas
\Pi^{(4)}_{\PZ\PA\,;\,0}(0) &=& 
       -
         \frac{1}{3}
         \,\frac{s}{c}
         \,\myNG\,\vtg
         \,( 1 - \LR )
       -
         \frac{1}{6}
         \,\frac{s}{c}
         \,( 1 + \LR )
       +
         \frac{1}{3}
         \,s\,c
         \,( 7 - 9\,\LR )
\eqas
\bqas
\Pi^{(4)}_{\PZ\PA\,;\,1}(0) &=& 
         \frac{1}{6}
         \,\frac{s}{c}
         \,\mrM^{c}_{0}
         \,\afun{M}
       -
         \frac{1}{3}
         \,\sumg 
         \,\frac{s}{c}
         \,\vle
         \,\afun{\mle}
\nl &-&
         \frac{2}{3}
         \,\sumg 
         \,\frac{s}{c}
         \,\vqu
         \,\afun{\mqu}
       -
         \frac{1}{3}
         \,\sumg 
         \,\frac{s}{c}
         \,\vqd
         \,\afun{\mqd}
\eqas
\bqas
\Pi^{(4)}_{\PZ\PA\,;\,2}(0) &=& 0 
\eqas
\bqas
\Pi^{(6)}_{\PZ\PA\,;\,0}(0) &=& 
         \frac{2}{3}
         \,\frac{1}{s}
         \,c\,\apD
         \,( 1 - \LR )
       -
         \frac{1}{24}
         \,\frac{1}{s\,c}
         \,\apD
         \,( 1 + \LR )
       -
         \frac{1}{6}
         \,\frac{1}{s}
         \,c^3\,\apD
         \,( 7 - 9\,\LR )
\nl &-&
         \frac{1}{12}
         \,\Bigl[ 3\,\mrM^{c}_{3}\,\apWB - 2\,(- 3\,\apB + \apW - 36\,s^2\,c^2\,\apWBa )\,s\,c 
         - 6\,( \mrP_{1}\,\apB + \mrP_{4}\,\apW )\,\frac{s\,c^3}{\xphs - 1} 
\nl &-& ( \mrP_{5} - 6\,\mrP_{6}\,s^2 )\,\frac{c^2}{\xphs - 1}\,\apWB \Bigr]
         \,\frac{\LR}{c^2}
\nl &-&
         \frac{1}{6}
         \,\Bigl[ \mrM^{c}_{6}\,\apWA + \mrM^{c}_{10}\,s\,c\,\apWZ 
         - 28\,s^3\,c\,\apB \,c \Bigr]
         \,\frac{1}{c}
\nl &+&
         \frac{1}{36}
         \,\Bigl\{ \Bigl[ 3\,c\,\apD\,\vtg - 4\,( 3\,\apWB\,\vtg 
         - 32\,(- c\,\apWZ + s\,\apWAB )\,s\,c )\,s \Bigr]\,c 
\nl &+& 4\,( \mrN_{3}\,\apW - \mrM^{c}_{9}\,\apD )\,s^2 \Bigr\}
         \,\frac{\myNG}{s\,c}
         \,( 1 - \LR )
\nl &-&
         \frac{1}{12}
         \,\sumg \Bigl\{  - 3\,\Bigl[  - ( \alWB\,\vle + 4\,s\,c\,\alBW )\,\xpls
         - ( 3\,\vqd\,\adWB + 4\,s\,c\,\adBW )\,\xpds + ( 3\,\vqu\,\auWB 
\nl &+& 8\,s\,c\,\auBW )\,\xpus \Bigr] + 8\,( \aplV + \apdV + 2\,\apuV )\,s \Bigr\}
         \,\frac{1}{c}
         \,( 1 - \LR )
\eqas
\bqas
\Pi^{(6)}_{\PZ\PA\,;\,1}(0) &=& 
         \frac{1}{4}
         \,\frac{\xphs}{\xphs - 1}
         \,\mrP_{0}
         \,\aAZ
         \,\afun{\mh}
       -
         \frac{1}{4}
         \,\frac{1}{c^2}
         \,\aAZ
         \,\afun{\mzb}
\nl &+&
         \frac{1}{24}
         \,( 8\,\mrM^{c}_{2}\,s^2\,\aAA - 8\,\mrM^{c}_{4}\,s\,c\,\aAZ + \mrM^{c}_{5}\,\apD 
         + 48\,\mrM^{c}_{7}\,s^2\,c^2\,\aZZ )
         \,\frac{1}{s\,c}
         \,\afun{M}
\nl &-&
         \frac{1}{12}
         \,\sumg \Bigl\{  - \Bigl[ c^2\,\apD\,\vle - 4\,( c\,\apWB\,\vle 
         + 2\,( \aplV + 2\,c^3\,\apWZ - 2\,s\,c^2\,\apWAB )\,s )\,s \Bigr] 
\nl &+& 3\,( \alWB\,\vle + 4\,s\,c\,\alBW )\,s\,\xpls - ( 4\,\mrN_{1}\,\apW
         - \mrM^{c}_{8}\,\apD )\,s^2 \Bigr\}
         \,\frac{1}{s\,c}
         \,\afun{\mle}
\nl &-&
         \frac{1}{36}
         \,\sumg \Bigl\{  - \Bigl[ 3\,c^2\,\vqd\,\apD - 4\,( 3\,c\,\vqd\,\apWB 
         + 2\,( 3\,\apdV + 2\,c^3\,\apWZ - 2\,s\,c^2\,\apWAB )\,s )\,s \Bigr] 
\nl &+& 9\,( 3\,\vqd\,\adWB + 4\,s\,c\,\adBW )\,s\,\xpds 
         - ( 4\,\mrN_{0}\,\apW - \mrM^{c}_{1}\,\apD )\,s^2 \Bigr\}
         \,\frac{1}{s\,c}
         \,\afun{\mqd}
\nl &+&
         \frac{1}{36}
         \,\sumg \Bigl\{ 2\,\Bigl[ 3\,c^2\,\vqu\,\apD - 4\,( 3\,c\,\vqu\,\apWB 
         + 2\,( 3\,\apuV + 4\,c^3\,\apWZ - 4\,s\,c^2\,\apWAB )\,s )\,s \Bigr] 
\nl &+& 9\,( 3\,\vqu\,\auWB + 8\,s\,c\,\auBW )\,s\,\xpus 
         + 2\,( 4\,\mrN_{2}\,\apW - \mrM^{c}_{9}\,\apD )\,s^2 \Bigr\}
         \,\frac{1}{s\,c}
         \,\afun{\mqu}
\eqas
\bqas
\Pi^{(6)}_{\PZ\PA\,;\,2}(0) &=& 0 
\eqas
\vspace{0.5cm}
\bei
\item \fbox{$\PZ$ self-energy}
\eei
\bqas
\Delta^{(4)}_{\PZ\PZ\,;\,0}(0) &=&
       -
         \frac{1}{6}
         \,\frac{1}{c^4}
         \,( 1 - 6\,\LR )
       +
         2
         \,\frac{\LR}{c^2}
\nl &+&
         \frac{1}{6}
         \,\frac{1}{c^2}
         \,\frac{1}{\xphs - 1}
         \,\mrP_{7}
       +
         \frac{1}{2}
         \,\sumg 
         \,\frac{1}{c^2}
         \,\mrO_{0}
         \,( 1 - \LR )
\eqas
\bqas
\Delta^{(4)}_{\PZ\PZ\,;\,1}(0) &=&
       -
         2
         \,\frac{1}{c^2}
         \,\afun{M}
       +
         \frac{1}{2}
         \,\sumg 
         \,\frac{\xpls}{c^2}
         \,\afun{\mle}
       +
         \frac{3}{2}
         \,\sumg 
         \,\frac{\xpus}{c^2}
         \,\afun{\mqu}
\nl &+&
         \frac{3}{2}
         \,\sumg 
         \,\frac{\xpds}{c^2}
         \,\afun{\mqd}
\nl &-&
         \frac{2}{3}
         \,( c^2 - \frac{1}{1 - c^2\,\xphs}\,\mrM^{c}_{11} )
         \,\frac{1}{c^6}
         \,\afun{\mh}
     -
         \frac{1}{3}
         \,( c^2 + 2\,\frac{1}{1 - c^2\,\xphs}\,\mrM^{c}_{11} )
         \,\frac{1}{c^6}
         \,\afun{\mzb}
\eqas
\bqas
\Delta^{(4)}_{\PZ\PZ\,;\,2}(0) &=&
       -
         \frac{1}{12}
         \,\Bigl[ 1 + ( 2\,\mrP_{0} - \mrP_{0}\,c^2\,\xphs )\,\frac{c^2}{\xphs - 1}\,\xphs \Bigr]
         \,\frac{1}{c^6}
         \,\zbfunp{\mh}{\mzb}
\eqas
\bqas
\Delta^{(6)}_{\PZ\PZ\,;\,0}(0) &=&
       -
         \frac{1}{8}
         \,\frac{1}{\xphs - 1}
         \,\mrP_{8}
         \,\apD
         \,\frac{\LR}{c^2}
       -
         \frac{1}{24}
         \,( 4\,\apBox + \apD )
         \,\frac{1}{c^4}
         \,( 2 - 9\,\LR )
\nl &+&
         \frac{1}{12}
         \,( 4\,\mrP_{0}\,\xphs\,\apBox + \mrP_{7}\,\apD )
         \,\frac{1}{c^2}
         \,\frac{1}{\xphs - 1}
\nl &+&
         \frac{1}{4}
         \,\sumg ( 24\,\xpds\,\apd - 24\,\xpus\,\apu + 8\,\xpls\,\aplA 
         - \mrO_{0}\,\apD + 12\,\mrO_{1}\,\apuVA )
         \,\frac{1}{c^2}
         \,( 1 - \LR )
\eqas
\bqas
\Delta^{(6)}_{\PZ\PZ\,;\,1}(0) &=&
       -
         \frac{1}{c^2}
         \,\apD
         \,\afun{M}
\nl &+&
         \frac{1}{24}
         \,\Bigl[ 9\,\frac{c^4}{\xphs - 1}\,\mrP_{0}\,\apD\,\xphs
         - 8\,( 4\,\apBox + \apD )\,c^2 
         + 8\,( 4\,\apBox + \apD )\,\frac{1}{1 - c^2\,\xphs}\,\mrM^{c}_{11} \Bigr]
         \,\frac{1}{c^6}
         \,\afun{\mh}
\nl &-&
         \frac{1}{24}
         \,\Bigl[ ( 4\,\apBox + \apD )\,c^2 
         + 8\,( 4\,\apBox + \apD )\,\frac{1}{1 - c^2\,\xphs}\,\mrM^{c}_{11} \Bigr]
         \,\frac{1}{c^6}
         \,\afun{\mzb}
\nl &+&
         \frac{1}{4}
         \,\sumg ( 8\,\aplA - \apD )
         \,\frac{\xpls}{c^2}
         \,\afun{\mle}
       +
         \frac{3}{4}
         \,\sumg ( 8\,\apdA - \apD )
         \,\frac{\xpds}{c^2}
         \,\afun{\mqd}
\nl &+&
         \frac{3}{4}
         \,\sumg ( 8\,\apuA - \apD )
         \,\frac{\xpus}{c^2}
         \,\afun{\mqu}
\eqas
\bqas
\Delta^{(6)}_{\PZ\PZ\,;\,2}(0) &=&
       -
         \frac{1}{24}
         \,\Bigl[ ( 4\,\apBox + \apD ) 
       + ( 4\,\apBox + \apD )\,( 2\,\mrP_{0} - \mrP_{0}\,c^2\,\xphs )\,
         \frac{c^2}{\xphs - 1}\,\xphs \Bigr]
         \,\frac{1}{c^6}
         \,\zbfunp{\mh}{\mzb}
\eqas
\bqas
\Omega^{(4)}_{\PZ\PZ\,;\,0}(0) &=&
         \frac{2}{3}
         \,( 3 - 5\,\LR )
       -
         \frac{1}{3}
         \,s^2
         \,( 7 - 9\,\LR )
       +
         \frac{1}{36}
         \,\frac{1}{c^2}
         \,( 11 + 6\,\LR )
\nl &+&
         \frac{1}{12}
         \,\frac{\myNG}{c^2}
         \,\mrN_{7}
         \,\LR
       -
         \frac{1}{36}
         \,( \mrN_{8} + 6\,\Lir )
         \,\frac{\myNG}{c^2}
\eqas
\bqas
\Omega^{(4)}_{\PZ\PZ\,;\,1}(0) &=&
       -
         \frac{1}{12}
         \,\frac{1}{c^2}
         \,\mrM^{c}_{12}
         \,\afun{M}
       -
         \frac{1}{12}
         \,\frac{1}{c^6}
         \,\frac{1}{1 - c^2\,\xphs}
         \,\afun{\mzb}
\nl &-&
         \frac{1}{12}
         \,\sumg 
         \,\frac{1}{c^2}
         \,\mrN_{6}
         \,\afun{\mle}
       -
         \frac{1}{4}
         \,\sumg 
         \,\frac{1}{c^2}
         \,\mrN_{4}
         \,\afun{\mqu}
\nl &-&
         \frac{1}{4}
         \,\sumg 
         \,\frac{1}{c^2}
         \,\mrN_{5}
         \,\afun{\mqd}
       +
         \frac{1}{12}
         \,(- c^4 + \frac{1}{1 - c^2\,\xphs} )
         \,\frac{1}{c^6}
         \,\afun{\mh}
\eqas
\bqas
\Omega^{(4)}_{\PZ\PZ\,;\,2}(0) &=&
         \frac{1}{24}
         \,\Bigl[ 1 + ( 2\,\mrP_{0} - \mrP_{0}\,c^2\,\xphs )\,\frac{c^2}{\xphs - 1}\,\xphs \Bigr]
         \,\frac{1}{c^6}
         \,\zbfuns{\mh}{\mzb}
\nl &-&
         \frac{1}{6}
         \,( 5 + \frac{c^2}{\xphs - 1}\,\mrP_{0}\,\xphs )
         \,\frac{1}{c^4}
         \,\zbfunp{\mh}{\mzb}
\eqas
\bqas
\Omega^{(6)}_{\PZ\PZ\,;\,0}(0) &=&
       -
         \frac{1}{6}
         \,\myNG\,\apD\,\vtg
         \,( 1 - \LR )
       +
         \frac{1}{18}
         \,\frac{1}{c^2}
         \,\apBox
         \,( 2 + 3\,\LR )
\nl &+&
         \frac{1}{3}
         \,\apD
         \,( 3 - 5\,\LR )
       -
         \frac{1}{6}
         \,s^2\,\apD
         \,( 7 - 9\,\LR )
\nl &+&
         \frac{1}{72}
         \,\frac{1}{c^2}
         \,\apD
         \,( 11 + 6\,\LR )
       -
         \frac{1}{24}
         \,\frac{\myNG}{c^2}
         \,\mrN_{25}
         \,\LR\,\apD
\nl &-&
         \sumg 
         \,( 1 - \LR )
         \,\frac{1}{c^2}
         \,\mrN_{28}
         \,\apqo
       +
         \frac{1}{72}
         \,( \mrN_{27} + 6\,\Lir )
         \,\frac{\myNG}{c^2}
         \,\apD
\nl &-&
         \frac{1}{3}
         \,\sumg \Bigl[ 3\,\mrN_{10}\,\apu - 3\,\mrN_{12}\,\apd 
         - 3\,\mrN_{17}\,\apqt - \mrN_{23}\,\aplt - ( \aplt - \aplV )\,\mrN_{9} \Bigr]
         \,\frac{\LR}{c^2}
\nl &+&
         \frac{1}{9}
         \,\sumg \Bigl[ 9\,\mrN_{16}\,\apu 
         - 9\,\mrN_{18}\,\apd - 3\,\mrN_{19}\,\apl - \mrN_{20}\,\aplt 
         - 9\,\mrN_{24}\,\apqt + \mrN_{26}\,\aplo - 3\,\Lir\,\apn \Bigr]
         \,\frac{1}{c^2}
\eqas
\bqas
\Omega^{(6)}_{\PZ\PZ\,;\,1}(0) &=&
       -
         \frac{1}{24}
         \,\frac{1}{c^2}
         \,\mrM^{c}_{12}
         \,\apD
         \,\afun{M}
\nl &+&
         \frac{1}{24}
         \,\Bigl[  - ( 4\,\apBox + \apD )\,c^4 
         + ( 4\,\apBox + \apD )\,\frac{1}{1 - c^2\,\xphs} \Bigr]
         \,\frac{1}{c^6}
         \,\afun{\mh}
\nl &-&
         \frac{1}{24}
         \,( 4\,\apBox + \apD )
         \,\frac{1}{c^6}
         \,\frac{1}{1 - c^2\,\xphs}
         \,\afun{\mzb}
\nl &-&
         \frac{1}{24}
         \,\sumg \Bigl[ 4\,c^2\,\apD\,\vle 
         + ( 8\,\mrN_{9}\,\apl - \mrN_{14}\,\apD + 4\,\mrN_{15}\,\aplVA ) \Bigr]
         \,\frac{1}{c^2}
         \,\afun{\mle}
\nl &-&
         \frac{1}{24}
         \,\sumg \Bigl[ 4\,c^2\,\vqd\,\apD 
         + ( 24\,\mrN_{12}\,\apd + 12\,\mrN_{13}\,\apdVA - \mrN_{22}\,\apD ) \Bigr]
         \,\frac{1}{c^2}
         \,\afun{\mqd}
\nl &-&
         \frac{1}{24}
         \,\sumg \Bigl[ 8\,c^2\,\vqu\,\apD 
         - ( 24\,\mrN_{10}\,\apu - 12\,\mrN_{11}\,\apuVA + \mrN_{21}\,\apD ) \Bigr]
         \,\frac{1}{c^2}
         \,\afun{\mqu}
\eqas
\bqas
\Omega^{(6)}_{\PZ\PZ\,;\,2}(0) &=&
         \frac{1}{48}
         \,\Bigl[ ( 4\,\apBox + \apD ) 
         + ( 4\,\apBox + \apD )\,( 2\,\mrP_{0} - \mrP_{0}\,c^2\,\xphs )\,
         \frac{c^2}{\xphs - 1}\,\xphs \Bigr]
         \,\frac{1}{c^6}
         \,\zbfuns{\mh}{\mzb}
\nl &-&
         \frac{1}{12}
         \,\Bigl[ 5\,( 4\,\apBox + \apD ) + ( 4\,\apBox + \apD )\,
         \frac{c^2}{\xphs - 1}\,\mrP_{0}\,\xphs \Bigr]
         \,\frac{1}{c^4}
         \,\zbfunp{\mh}{\mzb}
\eqas
\vspace{0.5cm}
\bei
\item \fbox{$\PW$ self-energy}
\eei
\bqas
\Delta^{(4)}_{\PW\PW\,;\,0}(0) &=&
         2
         \,\LR
       -
         \frac{1}{6}
         \,\frac{1}{c^2}
         \,( 1 - 6\,\LR )
       +
         \frac{1}{6}
         \,\frac{1}{\xphs - 1}
         \,\mrP_{10}
       +
         \frac{1}{6}
         \,\sumg 
         \,\mrO_{0}
         \,( 2 - 3\,\LR )
\eqas
\bqas
\Delta^{(4)}_{\PW\PW\,;\,1}(0) &=&
       -
         \frac{2}{3}
         \,\frac{\xphs}{\xphs - 1}
         \,\afun{\mh}
       -
         \frac{1}{3}
         \,\frac{1}{s^2\,c^2}
         \,\mrM^{c}_{13}
         \,\afun{\mzb}
\nl &-&
         \frac{1}{2}
         \,\sumg 
         \,\frac{\xpus}{ \xpus - \xpds}
         \,\mrO_{2}
         \,\afun{\mqu}
       -
         \frac{1}{2}
         \,\sumg 
         \,\frac{\xpds}{ \xpus - \xpds}
         \,\mrO_{3}
         \,\afun{\mqd}
\nl &+&
         \frac{1}{2}
         \,\sumg 
         \,\xpls
         \,\afun{\mle}
       +
         \frac{1}{3}
         \,( 2\,\frac{s^2}{\xphs - 1} - \mrM^{c}_{15} )
         \,\frac{1}{s^2}
         \,\afun{M}
\eqas
\bqas
\Delta^{(4)}_{\PW\PW\,;\,2}(0) &=&
       -
         \frac{1}{12}
         \,\frac{s^4}{c^4}
         \,\mrM^{c}_{14}
         \,\zbfunp{M}{\mzb}
       +
         \frac{1}{12}
         \,\frac{1}{\xphs - 1}
         \,\mrP_{9}
         \,\zbfunp{M}{\mh}
\nl &-&
         \frac{2}{3}
         \,s^2
         \,\zbfunp{0}{M}
       +
         \frac{1}{2}
         \,\sumg 
         \,\mrO_{4}
         \,\zbfunp{\mqu}{\mqd}
       +
         \frac{1}{6}
         \,\sumg 
         \,\xplq
         \,\zbfunp{0}{\mle}
\eqas
\bqas
\Delta^{(6)}_{\PW\PW\,;\,0}(0) &=&
       -
         \frac{1}{3}
         \,\frac{1}{c^2}
         \,\apW
         \,( 1 - 6\,\LR )
       +
         \frac{1}{12}
         \,\frac{1}{c^2}
         \,\apD
         \,( 11 - 15\,\LR )
\nl &-&
         \frac{1}{6}
         \,\Bigl[ 2\,c^3\,\apWB + ( 3\,\apD - 4\,c^2\,\aAA + 4\,c^2\,\apW )\,s \Bigr]
         \,\frac{s}{c^2}
         \,( 2 - 3\,\LR )
\nl &+&
         \frac{1}{12}
         \,\Bigl[ 40\,s\,\apWB 
         + ( 4\,\mrP_{10}\,\apW + 4\,\mrP_{11}\,\apBox 
         + \mrP_{14}\,\apD )\,\frac{c}{\xphs - 1} \Bigr]
         \,\frac{1}{c}
\nl &+&
         \frac{1}{2}
         \,( 3\,\apBox + 8\,\apW )
         \,\LR
       +
         \frac{1}{3}
         \,\sumg ( 2\,\xpls\,\aplt + \mrO_{0}\,\apW + 6\,\mrO_{1}\,\apqt )
         \,( 2 - 3\,\LR )
\eqas
\bqas
\Delta^{(6)}_{\PW\PW\,;\,1}(0) &=&
       -
         \frac{1}{2}
         \,\sumg 
         \,\frac{\xpus}{ \xpus - \xpds}
         \,\mrO_{2}
         \,\apqWt
         \,\afun{\mqu}
\nl &-&
         \frac{1}{2}
         \,\sumg 
         \,\frac{\xpds}{ \xpus - \xpds}
         \,\mrO_{3}
         \,\apqWt
         \,\afun{\mqd}
       +
         \frac{1}{2}
         \,\sumg 
         \,\xpls\,\aplWt
         \,\afun{\mle}
\nl &-&
         \frac{1}{6}
         \,\Bigl[ 4\,\mrM^{c}_{16}\,s^2\,\aAA + 3\,\mrM^{c}_{21}\,\apD
         + 20\,\mrM^{c}_{22}\,c^2\,\aZZ + 4\,\mrM^{c}_{26}\,s\,c\,\aAZ
\nl &-& ( 8\,\apW - 2\,\mrP_{12}\,\apD + \mrP_{13}\,\apBox )\,\frac{s^2}{\xphs - 1} \Bigr]
         \,\frac{1}{s^2}
         \,\afun{M}
\nl &-&
         \frac{1}{3}
         \,( \apWDm + 4\,\apBox )
         \,\frac{\xphs}{\xphs - 1}
         \,\afun{\mh}
\nl &+&
         \frac{1}{12}
         \,( 8\,\mrM^{c}_{17}\,s^2\,\aAA - 24\,\mrM^{c}_{18}\,c^2\,\aZZ - \mrM^{c}_{20}\,\apD 
         - 4\,\mrM^{c}_{23}\,s\,c\,\aAZ )
         \,\frac{1}{s^2\,c^2}
         \,\afun{\mzb}
\eqas
\bqas
\Delta^{(6)}_{\PW\PW\,;\,2}(0) &=&
         \frac{1}{2}
         \,\sumg 
         \,\mrO_{4}
         \,\apqWt
         \,\zbfunp{\mqu}{\mqd}
       +
         \frac{1}{6}
         \,\sumg 
         \,\xplq\,\aplWt
         \,\zbfunp{0}{\mle}
\nl &-&
         \frac{1}{3}
         \,\Bigl[ \mrM^{c}_{19}\,s\,c\,\aAZ + (- c^2\,\apD + 4\,s^2\,\aAA ) \Bigr]
         \,\zbfunp{0}{M}
\nl &+&
         \frac{1}{24}
         \,( \apWDm + 4\,\apBox )
         \,\frac{1}{\xphs - 1}
         \,\mrP_{9}
         \,\zbfunp{M}{\mh}
\nl &+&
         \frac{1}{24}
         \,( 16\,s^2\,c^4\,\apB - \mrM^{c}_{14}\,\apD - 4\,\mrM^{c}_{24}\,s\,\apWA 
         - 4\,\mrM^{c}_{25}\,c\,\apWZ )
         \,\frac{s^4}{c^4}
         \,\zbfunp{M}{\mzb}
\eqas
\bqas
\Omega^{(4)}_{\PW\PW\,;\,0}(0) &=&
       -
         \frac{4}{9}
         \,\myNG
         \,( 1 - 3\,\LR )
       -
         \frac{1}{18}
         \,( 2 + 57\,\LR )
\eqas
\bqas
\Omega^{(4)}_{\PW\PW\,;\,1}(0) &=&
       -
         \frac{1}{12}
         \,\frac{\xphs}{\xphs - 1}
         \,\afun{\mh}
       +
         \frac{1}{12}
         \,\frac{1}{s^2}
         \,\mrM^{c}_{28}
         \,\afun{\mzb}
\nl &-&
         \sumg 
         \,\frac{\xpus}{ \xpus - \xpds}
         \,\afun{\mqu}
       +
         \sumg 
         \,\frac{\xpds}{ \xpus - \xpds}
         \,\afun{\mqd}
\nl &-&
         \frac{1}{3}
         \,\sumg 
         \,\afun{\mle}
       -
         \frac{1}{12}
         \,( 39 + \frac{s^2}{\xphs - 1}\,\mrP_{16} )
         \,\frac{1}{s^2}
         \,\afun{M}
\eqas
\bqas
\Omega^{(4)}_{\PW\PW\,;\,2}(0) &=&
         \frac{1}{24}
         \,\frac{s^4}{c^4}
         \,\mrM^{c}_{14}
         \,\zbfuns{M}{\mzb}
       +
         \frac{1}{6}
         \,\frac{1}{\xphs - 1}
         \,\mrP_{15}
         \,\zbfunp{M}{\mh}
\nl &-&
         \frac{1}{24}
         \,\frac{1}{\xphs - 1}
         \,\mrP_{9}
         \,\zbfuns{M}{\mh}
       -
         \frac{1}{6}
         \,\frac{1}{c^2}
         \,\mrM^{c}_{27}
         \,\zbfunp{M}{\mzb}
\nl &+&
         \frac{4}{3}
         \,s^2
         \,\zbfunp{0}{M}
       +
         \frac{1}{3}
         \,s^2
         \,\zbfuns{0}{M}
\nl &+&
         \frac{1}{2}
         \,\sumg 
         \,\mrO_{1}
         \,\zbfunp{\mqu}{\mqd}
       +
         \frac{1}{6}
         \,\sumg 
         \,\xpls
         \,\zbfunp{0}{\mle}
\nl &-&
         \frac{1}{4}
         \,\sumg 
         \,\mrO_{4}
         \,\zbfuns{\mqu}{\mqd}
       -
         \frac{1}{12}
         \,\sumg 
         \,\xplq
         \,\zbfuns{0}{\mle}
\eqas
\bqas
\Omega^{(6)}_{\PW\PW\,;\,0}(0) &=&
       -
         \frac{2}{9}
         \,s^2\,\aAA
         \,( 1 - 9\,\LR )
       -
         \frac{8}{9}
         \,\myNG\,\apW
         \,( 1 - 3\,\LR )
\nl &+&
         \frac{1}{18}
         \,\apBox
         \,( 2 + 3\,\LR )
       +
         \frac{2}{9}
         \,(- c\,\aZZ + 2\,s\,\aAZ )
         \,c
\nl &+&
         \frac{1}{6}
         \,( \frac{c^2}{\xphs - 1}\,\mrP_{19}\,\apW 
         - 6\,\mrM^{c}_{29}\,s\,c\,\apWB + 3\,\mrM^{c}_{30}\,\apW )
         \,\frac{\LR}{c^2}
\nl &-&
         \frac{4}{9}
         \,\sumg ( \aplt + 3\,\apqt )
         \,( 1 - 3\,\LR )
       +
         \frac{1}{2}
         \,\sumg ( 3\,\xpds\,\adW - 3\,\xpus\,\auW + \xpls\,\alW )
         \,\LR
\eqas
\bqas
\Omega^{(6)}_{\PW\PW\,;\,1}(0) &=&
       -
         \frac{1}{24}
         \,\Bigl[ 4\,\mrP_{17}\,\apW + ( 4\,\apBox - \apD ) \Bigr]
         \,\frac{\xphs}{\xphs - 1}
         \,\afun{\mh}
\nl &-&
         \frac{1}{24}
         \,\Bigl[ 4\,\mrM^{c}_{35}\,s\,c\,\aAZ + 4\,\mrM^{c}_{38}\,s^2\,\aAA 
         - ( \mrM^{c}_{28}\,\apD - 12\,\mrM^{c}_{34}\,\aZZ )\,c^2 \Bigr]
         \,\frac{1}{s^2\,c^2}
         \,\afun{\mzb}
\nl &-&
         \frac{1}{24}
         \,\Bigl[ 12\,\mrM^{c}_{36}\,c^2\,\aZZ - 4\,\mrM^{c}_{37}\,s^2\,\aAA 
         + 4\,\mrM^{c}_{39}\,s\,c\,\aAZ + \mrM^{c}_{40}\,\apD 
\nl &-& ( 4\,\apBox - 4\,\mrP_{20}\,\apW - \mrP_{21}\,\apD )\,\frac{s^2}{\xphs - 1} \Bigr]
         \,\frac{1}{s^2}
         \,\afun{M}
\nl &-&
         \frac{1}{2}
         \,\sumg ( 2\,\apqWt + 3\,\xpds\,\adW - 3\,\xpus\,\auW )
         \,\frac{\xpus}{ \xpus - \xpds}
         \,\afun{\mqu}
\nl &+&
         \frac{1}{2}
         \,\sumg ( 2\,\apqWt + 3\,\xpds\,\adW - 3\,\xpus\,\auW )
         \,\frac{\xpds}{ \xpus - \xpds}
         \,\afun{\mqd}
\nl &-&
         \frac{1}{6}
         \,\sumg ( 2\,\aplWt + 3\,\xpls\,\alW )
         \,\afun{\mle}
\eqas
\bqas
\Omega^{(6)}_{\PW\PW\,;\,2}(0) &=&
       -
         \frac{1}{4}
         \,\sumg 
         \,\mrO_{4}
         \,\apqWt
         \,\zbfuns{\mqu}{\mqd}
       -
         \frac{1}{12}
         \,\sumg 
         \,\xplq\,\aplWt
         \,\zbfuns{0}{\mle}
\nl &+&
         \frac{1}{12}
         \,\Bigl[ 4\,s^2\,c\,\aAA - 4\,\mrM^{c}_{32}\,s\,\aAZ 
         - ( \mrM^{c}_{27}\,\apD + 12\,\mrM^{c}_{31}\,\aZZ )\,c \Bigr]
         \,\frac{1}{c^3}
         \,\zbfunp{M}{\mzb}
\nl &+&
         \frac{1}{12}
         \,\Bigl[ 4\,\mrP_{18}\,\apW + ( 4\,\apBox - \apD )\,\mrP_{15} \Bigr]
         \,\frac{1}{\xphs - 1}
         \,\zbfunp{M}{\mh}
\nl &+&
         \frac{1}{6}
         \,\Bigl[ \mrM^{c}_{19}\,s\,c\,\aAZ + (- c^2\,\apD + 4\,s^2\,\aAA ) \Bigr]
         \,\zbfuns{0}{M}
\nl &+&
         \frac{1}{3}
         \,\Bigl[ \mrM^{c}_{33}\,s\,c\,\aAZ + 2\,(- c^2\,\apD + 4\,s^2\,\aAA ) \Bigr]
         \,\zbfunp{0}{M}
\nl &-&
         \frac{1}{48}
         \,( \apWDm + 4\,\apBox )
         \,\frac{1}{\xphs - 1}
         \,\mrP_{9}
         \,\zbfuns{M}{\mh}
\nl &-&
         \frac{1}{48}
         \,( 16\,s^2\,c^4\,\apB - \mrM^{c}_{14}\,\apD 
         - 4\,\mrM^{c}_{24}\,s\,\apWA - 4\,\mrM^{c}_{25}\,c\,\apWZ )
         \,\frac{s^4}{c^4}
         \,\zbfuns{M}{\mzb}
\nl &+&
         \frac{1}{6}
         \,\sumg ( \aplWt + 3\,\xpls\,\alW )
         \,\xpls
         \,\zbfunp{0}{\mle}
\nl &+&
         \frac{1}{2}
         \,\sumg ( \mrO_{1}\,\apqWt - 3\,\mrO_{1}\,\xpds\,\adW
         - 3\,\mrO_{1}\,\xpus\,\auW )
         \,\zbfunp{\mqu}{\mqd}
\eqas

\normalsize

\section{Appendix: Finite counterterms \label{FCT}}

In this Appendix we present the list of finite counterterms for fields and parameters,
as defined in \sect{FiniteR}. It should be understood that only the real part of the loop
functions has to be included, \ie $\ssB_0 \equiv \Re\,\ssB_0$ \etc

\footnotesize
\bqas
\ssdCZ^{(4)}_{\mw} &=&
       -
         \frac{1}{18}
         \,\Bigl[ 3\,c^2\,\xphs + ( 3 + 128\,c^2 ) \Bigr]\,\frac{1}{c^2}
     -
         \frac{4}{9}
         \,( 1 - 3\,\LR )
         \,\myNG
\nl &+&
         \frac{1}{6}
         \,( 6 - 7\,c^2 )\,\frac{1}{c^2}
         \,\LR
     +
         \frac{1}{6}
         \,\sumg( 3\,\xpds + 3\,\xpus + \xpls )
         \,( 2 - 3\,\LR )
\nl &+&
         \frac{1}{2}
         \,\sumg( 2 - \xpds + \xpus )\,\xpds
         \,\afun{\mb}
     +
         \frac{1}{2}
         \,\sumg( 2 + \xpds - \xpus )\,\xpus
         \,\afun{\mt}
\nl &+&
         \frac{1}{6}
         \,\sumg( 2 - \xpls )\,\xpls
         \,\afun{\ml}
     -
         \frac{1}{12}
         \,\Bigl[ c^2\,\xphs + ( 1 + 66\,c^2 ) \Bigr]\,\frac{1}{c^2}
         \,\afun{\mw}
\nl &-&
         \frac{1}{12}
         \,( 3 - \xphs )\,\xphs
         \,\afun{\mh}
     +
         \frac{1}{12}
         \,( 1 - 19\,c^2 + 24\,s^2\,c^2 )\,\frac{1}{c^4}
         \,\afun{\mz}
\nl &-&
         4
         \,s^2
         \,\bfun{ - \mws}{0}{\mw}
\nl &+&
         \frac{1}{6}
         \,\sumg\Bigl[ 2 - ( 1 + \xpls )\,\xpls \Bigr]
         \,\bfun{ - \mws}{0}{\ml}
\nl &+&
         \frac{1}{2}
         \,\sumg\Bigl[ 2 - \xpds - \xpus - ( \xpus - \xpds )^2 \Bigr]
         \,\bfun{ - \mws}{\mt}{\mb}
\nl &+&
         \frac{1}{12}
         \,\Bigl[ 1 + 48\,s^2\,c^4 + 4\,( 4 - 29\,c^2 )\,c^2 \Bigr]\,\frac{1}{c^4}
         \,\bfun{ - \mws}{\mw}{\mz}
\nl &+&
         \frac{1}{12}
         \,\Bigl[ 12 - ( 4 - \xphs )\,\xphs \Bigr]
         \,\bfun{ - \mws}{\mw}{\mh}
\eqas
\bqas
\ssdCZ^{(4)}_{\ctw} &=&
       -
         \frac{1}{36}
         \,\Bigl\{  - 3\,\sumg( 1 - \vles )\,c^2\,\xpls - 9\,\sumg( 1 - \vqds )\,c^2\,\xpds 
         - 9\,\sumg( 1 - \vqus )\,c^2\,\xpus 
\nl &+& 2\,\Bigl[ 97 - 12\,( 11 - 3\,s^2 )\,s^2 \Bigr]\,s^2 \Bigr\}\,\frac{1}{c^2}
\nl &-&
         \frac{1}{12}
         \,( 19 - 18\,s^2 )\,\frac{s^2}{c^2}
         \,\LR
     +
         \frac{1}{72}
         \,\Bigl[ ( 9 - 16\,c^2 ) + ( \vles + 3\,\vqds + 3\,\vqus ) \Bigr]\,\frac{1}{c^2}
         \,( 1 - 3\,\LR )
         \,\myNG
\nl &-&
         \frac{1}{12}
         \,\sumg\Bigl[ \xpls - ( 1 - \vles ) \Bigr]\,\xpls
         \,\afun{\ml}
\nl &+&
         \frac{1}{4}
         \,\sumg\Bigl[ ( 1 - \vqds ) - ( \xpds - \xpus ) \Bigr]\,\xpds
         \,\afun{\mb}
\nl &+&
         \frac{1}{4}
         \,\sumg\Bigl[ ( 1 - \vqus ) + ( \xpds - \xpus ) \Bigr]\,\xpus
         \,\afun{\mt}
\nl &+&
         \frac{1}{24}
         \,\Bigl[ c^4\,\xphs + ( 1 - 18\,c^2 + 24\,s^2\,c^2 ) \Bigr]\,\frac{1}{c^4}
         \,\afun{\mz}
\nl &-&
         \frac{1}{24}
         \,\Bigl\{ c^2\,\xphs + \Bigl[ 1 + 48\,s^2\,c^4 
         + 2\,( 31 - 40\,c^2 )\,c^2 \Bigr] \Bigr\}\,\frac{1}{c^2}
         \,\afun{\mw}
\nl &-&
         \frac{1}{24}
         \,( 3 - \xphs )\,\xphs
         \,\afun{\mh}
     +
         \frac{1}{24}
         \,( 3 - c^2\,\xphs )\,\xphs
         \,\afun{\mh}
\nl &-&
         2
         \,s^2
         \,\bfun{ - \mws}{0}{\mw}
     -
         \frac{1}{12}
         \,\frac{1}{c^2}\,\sumg
         \,\bfun{ - \mzs}{0}{0}
\nl &+&
         \frac{1}{12}
         \,\sumg\Bigl[ 2 - ( 1 + \xpls )\,\xpls \Bigr]
         \,\bfun{ - \mws}{0}{\ml}
\nl &-&
         \frac{1}{24}
         \,\sumg\Bigl[ ( 1 + \vles ) - 2\,( 2 - \vles )\,c^2\,\xpls \Bigr]\,\frac{1}{c^2}
         \,\bfun{ - \mzs}{\ml}{\ml}
\nl &-&
         \frac{1}{8}
         \,\sumg\Bigl[ ( 1 + \vqds ) - 2\,( 2 - \vqds )\,c^2\,\xpds \Bigr]\,\frac{1}{c^2}
         \,\bfun{ - \mzs}{\mb}{\mb}
\nl &-&
         \frac{1}{8}
         \,\sumg\Bigl[ ( 1 + \vqus ) - 2\,( 2 - \vqus )\,c^2\,\xpus \Bigr]\,\frac{1}{c^2}
         \,\bfun{ - \mzs}{\mt}{\mt}
\nl &+&
         \frac{1}{4}
         \,\sumg\Bigl[ 2 - \xpds - \xpus - ( \xpus - \xpds)^2 \Bigr]
         \,\bfun{ - \mws}{\mt}{\mb}
\nl &+&
         \frac{1}{24}
         \,\Bigl[ 1 + 48\,s^2\,c^4 + 4\,( 4 - 29\,c^2 )\,c^2 \Bigr]\,\frac{1}{c^4}
         \,\bfun{ - \mws}{\mw}{\mz}
\nl &+&
         \frac{1}{24}
         \,\Bigl[ 12 - ( 4 - \xphs )\,\xphs \Bigr]
         \,\bfun{ - \mws}{\mw}{\mh}
\nl &-&
         \frac{1}{24}
         \,\Bigl\{ 1 + 4\,\Bigl[ 4 - ( 17 + 12\,c^2 )\,c^2 \Bigr]\,c^2 \Bigr\}\,\frac{1}{c^2}
         \,\bfun{ - \mzs}{\mw}{\mw}
\nl &-&
         \frac{1}{24}
         \,( 12 - 4\,c^2\,\xphs + c^4\,\xphq )\,\frac{1}{c^2}
         \,\bfun{ - \mzs}{\mh}{\mz}
\eqas
\bqas
\ssdCZ^{(4)}_{g} &=&
       -
         \frac{1}{2}
         \,\delta^{(4)}_{\ssG} 
     -
         \frac{2}{9}
         \,( 1 - 3\,\LR )
         \,\myNG
     -
         \frac{1}{36}
         \,( 2 + 57\,\LR )
\nl &+&
         \frac{1}{4}
         \,\sumg\Bigl[ 2\,\frac{\xpds}{\xpus-\xpds} + ( 1 - \xpds + \xpus ) \Bigr]\,\xpds
         \,\afun{\mb}
\nl &-&
         \frac{1}{4}
         \,\sumg\Bigl\{ 2\,\frac{\xpdq}{\xpus-\xpds} 
         + \Bigl[ 2\,\xpds + ( 1 - \xpds + \xpus )\,\xpus \Bigr] \Bigr\}
         \,\afun{\mt}
\nl &-&
         \frac{1}{12}
         \,\sumg( 1 + \xpls )\,\xpls
         \,\afun{\ml}
\nl &+&
         \frac{1}{24}
         \,\Bigl[ 8\,\frac{\xphs}{\xphs-1} - ( 11 - \xphs ) \Bigr]\,\xphs
         \,\afun{\mh}
\nl &+&
         \frac{1}{24}
         \,\Bigl\{ 1 + \Bigl[ 16 - ( 69 + 8\,c^2 )\,c^2 \Bigr]\,c^2 \Bigr\}\,\frac{1}{s^2\,c^4}
         \,\afun{\mz}
\nl &-&
         \frac{1}{24}
         \,\Bigl\{  - 7\,s^2\,c^2\,\xphs + 8\,\frac{\xphq}{\xphs-1}\,s^2\,c^2 
         + \Bigl[ 1 + ( 13 - 74\,c^2 )\,c^2 \Bigr] \Bigr\}\,\frac{1}{s^2\,c^2}
         \,\afun{\mw}
\nl &-&
         2
         \,s^2
         \,\bfun{ - \mws}{0}{\mw}
\nl &+&
         \frac{1}{12}
         \,\sumg\Bigl[ 2 - ( 1 + \xpls )\,\xpls \Bigr]
         \,\bfun{ - \mws}{0}{\ml}
\nl &+&
         \frac{1}{4}
         \,\sumg\Bigl[ 2 - \xpds - \xpus - ( \xpus - \xpds)^2 \Bigr]
         \,\bfun{ - \mws}{\mt}{\mb}
\nl &+&
         \frac{1}{24}
         \,\Bigl[ 1 + 48\,s^2\,c^4 + 4\,( 4 - 29\,c^2 )\,c^2 \Bigr]\,\frac{1}{c^4}
         \,\bfun{ - \mws}{\mw}{\mz}
\nl &+&
         \frac{1}{24}
         \,\Bigl[ 12 - ( 4 - \xphs )\,\xphs \Bigr]
         \,\bfun{ - \mws}{\mw}{\mh}
\nl &-&
         \frac{1}{12}
         \,\sumg\,\xplq
         \,\bfunp{0}{0}{\ml}
     +
         \frac{1}{3}
         \,s^2
         \,\bfunp{0}{0}{\mw}
\nl &-&
         \frac{1}{4}
         \,\sumg( - \xpds + \xpus )^2
         \,\bfunp{0}{\mt}{\mb}
\nl &+&
         \frac{1}{24}
         \,( - 1 + \xphs )^2
         \,\bfunp{0}{\mw}{\mh}
\nl &+&
         \frac{1}{24}
         \,( 9 - 8\,s^2 )\,\frac{s^4}{c^4}
         \,\bfunp{0}{\mw}{\mz}
\eqas
\bqas
\ssdCZ^{(4)}_{\mh} &=&
         \frac{1}{2}
         \,\Bigl\{ 3\,c^2\,\xphs - \sumg\Bigl[ 3\,( - \xphs + 4\,\xpds )\,\xpds 
         + 3\,( - \xphs + 4\,\xpus )\,\xpus 
\nl &+& ( - \xphs + 4\,\xpls )\,\xpls \Bigr]\,\frac{1}{\xphs}\,c^2 
         - ( 1 + 2\,c^2 ) \Bigr\}\,\frac{1}{c^2}
         \,\LR
\nl &-&
         \frac{1}{2}
         \,( 1 + 2\,c^4 )\,\frac{1}{c^4}\,\frac{1}{\xphs}
         \,( 2 - 3\,\LR )
\nl &+&
         \frac{9}{8}
         \,\xphs
         \,\bfun{ - \mhs}{\mh}{\mh}
\nl &-&
         \frac{3}{2}
         \,\sumg( - \xphs + 4\,\xpds )\,\frac{\xpds}{\xphs}
         \,\bfun{ - \mhs}{\mb}{\mb}
\nl &-&
         \frac{3}{2}
         \,\sumg( - \xphs + 4\,\xpus )\,\frac{\xpus}{\xphs}
         \,\bfun{ - \mhs}{\mt}{\mt}
\nl &-&
         \frac{1}{2}
         \,\sumg( - \xphs + 4\,\xpls )\,\frac{\xpls}{\xphs}
         \,\bfun{ - \mhs}{\ml}{\ml}
\nl &+&
         \frac{1}{4}
         \,\Bigl[ 12 - ( 4 - \xphs )\,\xphs \Bigr]\,\frac{1}{\xphs}
         \,\bfun{ - \mhs}{\mw}{\mw}
\nl &+&
         \frac{1}{8}
         \,( - 4\,c^2 + c^4\,\xphs + 12\,\frac{1}{\xphs} )\,\frac{1}{c^4}
         \,\bfun{ - \mhs}{\mz}{\mz}
\eqas
\bqas
\ssdCZ^{(6)}_{\mw} &=&
         \frac{8}{9}
         \,\apW
         \,( 1 - 3\,\LR )
         \,\myNG
\nl &-&
         \frac{1}{36}
         \,\Bigl[ 3\,c^2\,\apWDm\,\xphs + 4\,( 2 + 3\,\xphs )\,c^2\,\apBox
         + 4\,( 3 + 128\,c^2 )\,\apW + 3\,( 9 - 8\,s^2 )\,\apD \Bigr]\,\frac{1}{c^2}
\nl &+&
         \frac{1}{12}
         \,\Bigl\{ 20\,c^2\,\apBox + 6\,\sumg\,c^2\,\alW\,\xpls 
         + 18\,\sumg\,c^2\,\adW\,\xpds - 18\,\sumg\,c^2\,\auW\,\xpus 
\nl &+& 2\,\Bigl[ 3\,c^2\,\xphs + ( 15 + 4\,c^2 ) \Bigr]\,\apW 
         - 3\,( 5 - 6\,s^2 )\,\apD \Bigr\}\,\frac{1}{c^2}
         \,\LR
     +
         \frac{1}{3}
         \,\frac{s}{c}\,\apWB
         \,( 10 - 3\,\LR )
\nl &+&
         \frac{1}{6}
         \,\sumg\Bigl[ \aplWt\,\xpls + 3\,( \xpds + \xpus )\,\apqWt \Bigr]
         \,( 2 - 3\,\LR )
     -
         \frac{4}{9}
         \,\sumg( \aplt + 3\,\apqt )
         \,( 1 - 3\,\LR )
\nl &-&
         \frac{1}{6}
         \,\sumg\Bigl[ 3\,\alW\,\xpls - ( 2 - \xpls )\,\aplWt \Bigr]\,\xpls
         \,\afun{\ml}
\nl &-&
         \frac{1}{2}
         \,\sumg\Bigl[ 3\,\adW\,\xpds + 3\,\auW\,\xpus 
         - ( 2 - \xpds + \xpus)\,\apqWt \Bigr]\,\xpds
         \,\afun{\mb}
\nl &+&
         \frac{1}{2}
         \,\sumg\Bigl[ 3\,\adW\,\xpds + 3\,\auW\,\xpus 
         + ( 2 + \xpds - \xpus )\,\apqWt \Bigr]\,\xpus
         \,\afun{\mt}
\nl &+&
         \frac{1}{24}
         \,\Bigl[ \apWDm\,\xphs - 72\,\apW + 3\,\apD - 4\,( 3 - \xphs )\,\apBox \Bigr]\,\xphs
         \,\afun{\mh}
\nl &+&
         \frac{1}{24}
         \,\Bigl[  - c^2\,\apWDm\,\xphs + 40\,s\,c\,\apWB 
         - 4\,( 1 + 60\,c^2 )\,\apW + 4\,( 2 - \xphs )\,c^2\,\apBox 
         - ( 21 - 20\,s^2 )\,\apD \Bigr]\,\frac{1}{c^2}
         \,\afun{\mw}
\nl &+&
         \frac{1}{24}
         \,\Bigl[ ( 1 - 13\,c^2 + 24\,s^2\,c^2 )\,\apD + 4\,( 1 - 8\,c^2)\,s\,\apWA 
         + 4\,( 11 - 32\,c^2 )\,c\,\apWZ \Bigr]\,\frac{1}{c^4}
         \,\afun{\mz}
\nl &+&
         \frac{1}{2}
         \,\sumg\Bigl[ 3\,( 1 - \xpds + \xpus )\,\adW\,\xpds 
         - 3\,( 1 + \xpds - \xpus )\,\auW\,\xpus 
         + ( 2 - \xpds - \xpus - ( \xpus - \xpds)^2)\,\apqWt \Bigr]
\nl &\times& \bfun{ - \mws}{\mt}{\mb}
\nl &+&
         \frac{1}{6}
         \,\sumg\Bigl\{ \Bigl[ 2 - ( 1 + \xpls )\,\xpls \Bigr]\,\aplWt 
         + 3\,( 1 - \xpls )\,\alW\,\xpls \Bigr\}
         \,\bfun{ - \mws}{0}{\ml}
\nl &+&
         \frac{1}{24}
         \,\Bigl\{ \apWDm\,\xphq + 4\,\Bigl[ 12 - ( 4 - \xphs )\,\xphs \Bigr]\,\apBox 
         - 4\,( 3 - \xphs )\,\apD + 16\,( 9 - 4\,\xphs )\,\apW \Bigr\}
         \,\bfun{ - \mws}{\mw}{\mh}
\nl &+&
         \frac{1}{24}
         \,\Bigl\{ \Bigl[ 1 + 48\,s^2\,c^4 + 4\,( 4 - 29\,c^2 )\,c^2 \Bigr]\,\apD
         + 44\,\Bigl[ 1 - 2\,( 1 + 4\,c^2 )\,c^2 \Bigr]\,c\,\apWZ 
\nl &+& 4\,\Bigl[ 1 + 2\,( 3 - 20\,c^2 )\,c^2 \Bigr]\,s\,\apWA \Bigr\}\,\frac{1}{c^4}
         \,\bfun{ - \mws}{\mw}{\mz}
\nl &-&
         2
         \,\apWAD
         \,\bfun{ - \mws}{0}{\mw}
\eqas
\bqas
\ssdCZ^{(6)}_{\ctw} &=&
         \frac{1}{18}
         \,\frac{s^2}{c^2}\,\apBox
         \,( 2 - 15\,\LR )
\nl &+&
         \frac{1}{144}
         \,\Bigl\{ 9\,\apWDm - 64\,c^2\,\apW + \Bigl[ ( 1 + 4\,c^2 )\,\vqd 
         + ( 3 + 4\,c^2 )\,\vle + ( 5 + 8\,c^2 )\,\vqu \Bigr]\,\apD 
\nl &-& 4\,\Bigl[ ( 1 + 8\,c^2 )\,\vqd + ( 3 + 8\,c^2 )\,\vle 
         + ( 5 + 16\,c^2 )\,\vqu \Bigr]\,s\,\apWA 
         - 4\,\Bigl[ ( 5 - 8\,c^2 )\,\vqd + ( 7 - 8\,c^2 )\,\vle 
\nl &+& ( 13 - 16\,c^2 )\,\vqu \Bigr]\,c\,\apWZ 
         + 64\,\Bigl[ ( \vle + \vqd + 2\,\vqu ) \Bigr]\,s^2\,c^2\,\apB \Bigr\}\,\frac{1}{c^2}
         \,( 1 - 3\,\LR )
         \,\myNG
\nl &+&
         \frac{1}{72}
         \,\Bigl\{ 3\,c^2\,\apWDp\,\xphs - 3\,c^2\,\apWDm\,\xphs 
         + 6\,\sumg\,c^2\,\aplWt\,\xpls - 24\,\sumg\,c^2\,\aplV\,\xpls\,\vle 
         - 72\,\sumg\,c^2\,\vqd\,\apdV\,\xpds 
\nl &-& 72\,\sumg\,c^2\,\vqu\,\apuV\,\xpus 
         + 12\,\sumg\Bigl[ ( 1 + 8\,c^2 )\,\vqd\,\xpds + ( 3 + 8\,c^2 )\,\xpls\,\vle 
         + ( 5 + 16\,c^2 )\,\vqu\,\xpus \Bigr]\,s\,c^2\,\apWA 
\nl &+& 12\,\sumg\Bigl[ ( 5 - 8\,c^2 )\,\vqd\,\xpds + ( 7 - 8\,c^2)\,\xpls\,\vle 
         + ( 13 - 16\,c^2 )\,\vqu\,\xpus \Bigr]\,c^3\,\apWZ 
         + 18\,\sumg( \xpds + \xpus )\,c^2\,\apqWt 
\nl &-& 192\,\sumg( \xpls\,\vle + \vqd\,\xpds + 2\,\vqu\,\xpus )\,s^2\,c^4\,\apB 
         - 8\,\Bigl[ 1 + ( 31 + 36\,c^4 )\,c^2 \Bigr]\,s^2\,\aAA 
\nl &-& 8\,\Bigl[ 17 + ( 43 + 36\,( 1 + c^2)\,c^2 )\,c^2 \Bigr]\,s\,c\,\aAZ 
         - 8\,\Bigl[ 29 + 36\,( 1 - s^2 )\,s^2 \Bigr]\,s^2\,c^2\,\aZZ 
\nl &-& \Bigl[ 3\,\sumg( 1 + 4\,c^2 )\,c^2\,\vqd\,\xpds 
         + 3\,\sumg( 3 + 4\,c^2 )\,c^2\,\xpls\,\vle 
\nl &+& 3\,\sumg( 5 + 8\,c^2 )\,c^2\,\vqu\,\xpus 
         - 2\,( 52 - ( 149 - 12\,( 11 - 3\,s^2)\,s^2)\,s^2 ) \Bigr]\,\apD 
         \Bigr\}\,\frac{1}{c^2}
\nl &+&
         \frac{1}{72}
         \,\sumg\Bigl[ 4\,\apn + 4\,\aplA + 4\,\aplV\,\vle + 12\,\apdA + 12\,\apuA
         + 12\,\vqd\,\apdV + 12\,\vqu\,\apuV 
\nl &-& 16\,c^2\,\aplt - 48\,c^2\,\apqt 
         - 24\,c^2\,\apl\,\xpls - 72\,c^2\,\apd\,\xpds 
         + 72\,c^2\,\apu\,\xpus + 24\,c^2\,\aplo\,\xpls
\nl &+& 72\,( \xpds - \xpus )\,c^2\,\apqo 
         + 3\,( 3\,\xpds + 3\,\xpus + \xpls )\,c^2\,\apD \Bigr]\,\frac{1}{c^2}
         \,( 1 - 3\,\LR )
\nl &+&
         \frac{1}{48}
         \,\Bigl\{ 12\,c^2\,\apW\,\xphs - 12\,\sumg\,c\,\alBW\,\xpls\,\vle 
         - 36\,\sumg\,c\,\vqd\,\adBW\,\xpds + 36\,\sumg\,c\,\vqu\,\auBW\,\xpus 
\nl &+& 12\,\sumg\,c^2\,\alW\,\xpls + 36\,\sumg\,c^2\,\adW\,\xpds 
         - 36\,\sumg\,c^2\,\auW\,\xpus 
         - 4\,\Bigl[ 7 + 4\,( 14 + 3\,c^2 )\,c^2 \Bigr]\,s\,c\,\aAZ 
\nl &-& 4\,\Bigl[ 19 - 4\,( 8 - 3\,s^2 )\,s^2 \Bigr]\,s^2\,\aAA 
         - 4\,\Bigl[ 3\,c^2\,\xphs + ( 31 - 4\,( 1 + 3\,s^2)\,s^2 )\,s^2\Bigr]\,\aZZ 
\nl &-& \Bigl[ 9\,c^2\,\xphs + ( 25 + 4\,( 4 - 9\,s^2)\,s^2 ) \Bigr]\,\apD
         \Bigr\}\,\frac{1}{c^2}
         \,\LR
\nl &+&
         \frac{1}{24}
         \,\sumg\Bigl\{  - 8\,\aplV\,\vle - 6\,\alW\,\xpls - 8\,\apl + 8\,\aplo 
         - 64\,s^2\,c^2\,\apB\,\vle + \Bigl[ 1 - ( 3 + 4\,c^2 )\,\vle \Bigr]\,\apD 
\nl &+& 2\,( 1 - \xpls )\,\aplWt + 4\,( 3 + 8\,c^2 )\,s\,\apWA\,\vle 
         + 4\,( 7 - 8\,c^2 )\,c\,\apWZ\,\vle \Bigr\}\,\xpls
         \,\afun{\ml}
\nl &+&
         \frac{1}{24}
         \,\sumg\Bigl\{  - 18\,\adW\,\xpds - 18\,\auW\,\xpus - 24\,\apd 
         + 24\,\apqo - 24\,\vqd\,\apdV - 64\,s^2\,c^2\,\vqd\,\apB 
\nl &+& \Bigl[ 3 - ( 1 + 4\,c^2 )\,\vqd \Bigr]\,\apD 
         + 6\,( 1 - \xpds + \xpus )\,\apqWt + 4\,( 1 + 8\,c^2 )\,s\,\vqd\,\apWA 
\nl &+& 4\,( 5 - 8\,c^2 )\,c\,\vqd\,\apWZ \Bigr\}\,\xpds
         \,\afun{\mb}
\nl &-&
         \frac{1}{24}
         \,\sumg\Bigl\{  - 18\,\adW\,\xpds - 18\,\auW\,\xpus - 24\,\apu 
         + 24\,\apqo + 24\,\vqu\,\apuV + 128\,s^2\,c^2\,\vqu\,\apB 
\nl &-& \Bigl[ 3 - ( 5 + 8\,c^2 )\,\vqu \Bigr]\,\apD 
         - 6\,( 1 + \xpds - \xpus )\,\apqWt - 4\,( 5 + 16\,c^2 )\,s\,\vqu\,\apWA 
\nl &-& 4\,( 13 - 16\,c^2 )\,c\,\vqu\,\apWZ \Bigr\}\,\xpus
         \,\afun{\mt}
\nl &+&
         \frac{1}{48}
         \,\Bigl[ \apWDm\,\xphs - 72\,\apW + 3\,\apD - 4\,( 3 - \xphs )\,\apBox \Bigr]\,\xphs
         \,\afun{\mh}
\nl &+&
         \frac{1}{48}
         \,\Bigl[ 12\,\apD - c^2\,\apWDp\,\xphs + 12\,s\,c\,\aAZ + 12\,s^2\,\aAA
         + 4\,( 3 - c^2\,\xphs )\,\apBox + 12\,( 5 + c^2 )\,\aZZ \Bigr]\,\xphs
         \,\afun{\mh}
\nl &-&
         \frac{1}{48}
         \,\Bigl\{ c^2\,\apWDm\,\xphs 
         + 4\,\Bigl[ 11 + 8\,( 8 + 3\,( 1 + 2\,c^2)\,c^2 )\,c^2 \Bigr]\,s\,c\,\aAZ 
\nl &+& 4\,\Bigl[ 15 + 4\,( 7 - 6\,( 5 - 2\,c^2)\,c^2 )\,c^2 \Bigr]\,c^2\,\aZZ 
         + 4\,\Bigl[ 61 - 12\,( 11 - 2\,( 5 - 2\,s^2)\,s^2 )\,s^2\Bigr]\,s^2\,\aAA 
\nl &-& \Bigl[ 63 - 16\,( 12 - ( 11 - 3\,s^2)\,s^2 )\,s^2 \Bigr]\,\apD 
         - 4\,( 2 - \xphs )\,c^2\,\apBox \Bigr\}\,\frac{1}{c^2}
         \,\afun{\mw}
\nl &+&
         \frac{1}{48}
         \,\Bigl\{ c^4\,\apWDp\,\xphs + 4\,\Bigl[ 8 - ( 31 - 24\,s^2 )\,s^2 \Bigr]\,s^2\,\aAA 
         + 4\,\Bigl[ 11 - ( 41 - 24\,c^2 )\,c^2 \Bigr]\,s\,c\,\aAZ 
\nl &+& 4\,\Bigl[ 12 - ( 65 - 24\,c^2 )\,c^2 \Bigr]\,c^2\,\aZZ 
         + ( 1 - 15\,c^2 + 24\,s^2\,c^2 )\,\apD 
         - 4\,( 2 - c^2\,\xphs )\,c^2\,\apBox \Bigr\}\,\frac{1}{c^4}
         \,\afun{\mz}
\nl &+&
         \frac{1}{4}
         \,\sumg\Bigl[ 3\,( 1 - \xpds + \xpus )\,\adW\,\xpds 
         - 3\,( 1 + \xpds - \xpus )\,\auW\,\xpus 
         + ( 2 - \xpds - \xpus - ( \xpus - \xpds)^2)\,\apqWt \Bigr]
\nl &\times& \bfun{ - \mws}{\mt}{\mb}
\nl &-&
         \frac{1}{48}
         \,\sumg\Bigl\{ 12\,c\,\alBW\,\xpls\,\vle + \Bigl[ ( 3 + 4\,c^2 ) 
         + 2\,( 3 + 4\,c^2 )\,c^2\,\xpls \Bigr]\,\apD\,\vle 
\nl &-& 4\,\Bigl[ ( 3 + 8\,c^2 ) + 2\,( 3 + 8\,c^2 )\,c^2\,\xpls \Bigr]\,s\,\apWA\,\vle 
         - 4\,\Bigl[ ( 7 - 8\,c^2 ) + 2\,( 7 - 8\,c^2 )\,c^2\,\xpls \Bigr]\,c\,\apWZ\,\vle 
\nl &+& ( 1 - 4\,c^2\,\xpls )\,\apWDm + 8\,( 1 - 4\,c^2\,\xpls )\,\aplA 
         + 8\,( 1 + 2\,c^2\,\xpls )\,\aplV\,\vle 
\nl &+& 64\,( 1 + 2\,c^2\,\xpls )\,s^2\,c^2\,\apB\,\vle \Bigr\}\,\frac{1}{c^2}
         \,\bfun{ - \mzs}{\ml}{\ml}
\nl &-&
         \frac{1}{48}
         \,\sumg\Bigl\{ 36\,c\,\vqd\,\adBW\,\xpds + \Bigl[ ( 1 + 4\,c^2 ) 
         + 2\,( 1 + 4\,c^2 )\,c^2\,\xpds \Bigr]\,\vqd\,\apD - 4\,\Bigl[ ( 1 + 8\,c^2 ) 
\nl &+& 2\,( 1 + 8\,c^2 )\,c^2\,\xpds \Bigr]\,s\,\vqd\,\apWA 
         - 4\,\Bigl[ ( 5 - 8\,c^2 ) + 2\,( 5 - 8\,c^2 )\,c^2\,\xpds \Bigr]\,c\,\vqd\,\apWZ 
         + 3\,( 1 - 4\,c^2\,\xpds )\,\apWDm 
\nl &+&  24\,( 1 - 4\,c^2\,\xpds )\,\apdA 
         + 24\,( 1 + 2\,c^2\,\xpds )\,\vqd\,\apdV 
         + 64\,( 1 + 2\,c^2\,\xpds )\,s^2\,c^2\,\vqd\,\apB \Bigr\}\,
         \frac{1}{c^2}
\nl &\times& \bfun{ - \mzs}{\mb}{\mb}
\nl &+&
         \frac{1}{48}
         \,\sumg\Bigl\{ 36\,c\,\vqu\,\auBW\,\xpus - \Bigl[ ( 5 + 8\,c^2 ) 
         + 2\,( 5 + 8\,c^2 )\,c^2\,\xpus \Bigr]\,\vqu\,\apD + 4\,\Bigl[ ( 5 + 16\,c^2 )
\nl &+& 2\,( 5 + 16\,c^2 )\,c^2\,\xpus \Bigr]\,s\,\vqu\,\apWA 
         + 4\,\Bigl[ ( 13 - 16\,c^2 ) + 2\,( 13 - 16\,c^2 )\,c^2\,\xpus \Bigr]\,c\,\vqu\,\apWZ 
\nl &-& 3\,( 1 - 4\,c^2\,\xpus )\,\apWDm - 24\,( 1 - 4\,c^2\,\xpus )\,\apuA 
        - 24\,( 1 + 2\,c^2\,\xpus )\,\vqu\,\apuV 
         - 128\,( 1 + 2\,c^2\,\xpus )\,s^2\,c^2\,\vqu\,\apB \Bigr\}\,\frac{1}{c^2}
\nl &\times&         \,\bfun{ - \mzs}{\mt}{\mt}
\nl &+&
         \frac{1}{12}
         \,\sumg\Bigl\{ \Bigl[ 2 - ( 1 + \xpls )\,\xpls \Bigr]\,\aplWt 
         + 3\,( 1 - \xpls )\,\alW\,\xpls \Bigr\}
         \,\bfun{ - \mws}{0}{\ml}
\nl &-&
         \frac{1}{24}
         \,\sumg( \apWDm + 4\,\apn )\,\frac{1}{c^2}
         \,\bfun{ - \mzs}{0}{0}
\nl &-&
         \frac{1}{48}
         \,\Bigl[ 48\,( 2 - c^2\,\xphs )\,\aZZ + ( 12 - 4\,c^2\,\xphs
         + c^4\,\xphq )\,\apWDp + 4\,( 12 - 4\,c^2\,\xphs + c^4\,\xphq )\,\apBox
         \Bigr]\,\frac{1}{c^2}
\nl &\times& \bfun{ - \mzs}{\mh}{\mz}
\nl &+&
         \frac{1}{48}
         \,\Bigl\{ \apWDm\,\xphq + 4\,\Bigl[ 12 
         - ( 4 - \xphs )\,\xphs \Bigr]\,\apBox - 4\,( 3 - \xphs )\,\apD 
         + 16\,( 9 - 4\,\xphs )\,\apW \Bigr\}
         \,\bfun{ - \mws}{\mw}{\mh}
\nl &+&
         \frac{1}{48}
         \,\Bigl\{ \Bigl[ 1 + 48\,s^2\,c^4 + 4\,( 4 - 29\,c^2 )\,c^2 \Bigr]\,\apD
         + 44\,\Bigl[ 1 - 2\,( 1 + 4\,c^2 )\,c^2 \Bigr]\,c\,\apWZ 
\nl &+& 4\,\Bigl[ 1 + 2\,( 3 - 20\,c^2 )\,c^2 \Bigr]\,s\,\apWA \Bigr\}\,\frac{1}{c^4}
         \,\bfun{ - \mws}{\mw}{\mz}
\nl &-&
         \frac{1}{48}
         \,\Bigl\{ \Bigl[ 1 + 4\,( 4 - ( 17 + 12\,c^2)\,c^2 )\,c^2 \Bigr]\,\apD
         + 12\,\Bigl[ 5 + 4\,( 6 + ( 3 + 4\,c^2)\,c^2 )\,c^2 \Bigr]\,s\,c\,\aAZ 
\nl &+& 12\,\Bigl[ 7 - 4\,( 5 + ( 9 - 4\,c^2)\,c^2 )\,c^2 \Bigr]\,c^2\,\aZZ 
         + 4\,\Bigl[ 33 - 4\,( 29 - 3\,( 11 - 4\,s^2)\,s^2 )\,s^2 \Bigr]\,s^2\,\aAA 
         \Bigr\}\,\frac{1}{c^2}
\nl &\times& \bfun{ - \mzs}{\mw}{\mw}
\nl &-&
         \apWAD
         \,\bfun{ - \mws}{0}{\mw}
\eqas
\bqas
\ssdCZ^{(6)}_{g} &=&
       -
         \frac{1}{9}
         \,( c^2\,\aZZ - 2\,s\,c\,\aAZ + s^2\,\aAA )
     +
         \frac{1}{36}
         \,\apBox
         \,( 2 + 3\,\LR )
\nl &-&
         \frac{4}{9}
         \,\apW
         \,( 1 - 3\,\LR )
         \,\myNG
     -
         \frac{2}{9}
         \,\sumg( \aplt + 3\,\apqt )
         \,( 1 - 3\,\LR )
\nl &+&
         \frac{1}{12}
         \,\Bigl[ 3\,c^2\,\apW\,\xphs + 3\,\sumg\,c^2\,\alW\,\xpls 
         + 9\,\sumg\,c^2\,\adW\,\xpds - 9\,\sumg\,c^2\,\auW\,\xpus 
         + ( 9 - 38\,c^2 )\,s\,c\,\aAZ 
\nl &+& ( 15 - 32\,c^2 )\,c^2\,\aZZ 
         - ( 29 - 32\,s^2 )\,s^2\,\aAA \Bigr]\,\frac{1}{c^2}
         \,\LR
\nl &-&
         \frac{1}{12}
         \,\sumg\Bigl[ 3\,\alW\,\xpls + ( 1 + \xpls )\,\aplWt \Bigr]\,\xpls
         \,\afun{\ml}
\nl &+&
         \frac{1}{4}
         \,\sumg\Bigl[ 3\,( 1 - \xpds + \xpus )\,\adW\,\xpds 
         - 3\,( 1 + \xpds - \xpus )\,\auW\,\xpus 
         + ( 2 - \xpds - \xpus - ( \xpus - \xpds)^2)\,\apqWt \Bigr]
\nl &\times& \bfun{ - \mws}{\mt}{\mb}
\nl &-&
         \frac{1}{4}
         \,\sumg\Bigl\{ 3\,\adW\,\xpds + 3\,\auW\,\xpus 
         - \Bigl[ 2\,\frac{\xpds}{\xpus-\xpds} 
         + ( 1 - \xpds + \xpus ) \Bigr]\,\apqWt \Bigr\}\,\xpds
         \,\afun{\mb}
\nl &+&
         \frac{1}{4}
         \,\sumg\Bigl\{ 3\,\adW\,\xpus\,\xpds + 3\,\auW\,\xpuq 
         - \Bigl[ 2\,\frac{\xpdq}{\xpus-\xpds} + ( 2\,\xpds 
         + ( 1 - \xpds + \xpus)\,\xpus ) \Bigr]\,\apqWt \Bigr\}
         \,\afun{\mt}
\nl &+&
         \frac{1}{48}
         \,\Bigl\{  - 104\,\apW + 11\,\apD 
         + 4\,\Bigl[ 8\,\frac{\xphs}{\xphs-1} - ( 11 - \xphs ) \Bigr]\,\apBox 
         + ( 1 + 8\,\frac{1}{\xphs-1} )\,\apWDm\,\xphs \Bigr\}\,\xphs
         \,\afun{\mh}
\nl &-&
         \frac{1}{48}
         \,\Bigl\{ \Bigl[ 1 + ( 27 - 88\,c^2 )\,c^2 \Bigr]\,\apD 
         - 4\,\Bigl[ 4 + ( 3 - 8\,s^2 )\,s^2 \Bigr]\,s^2\,\aAA 
         + 4\,\Bigl[ 11 - 3\,( 17 - 4\,c^2 )\,c^2 \Bigr]\,s\,c\,\aAZ 
\nl &+& 4\,\Bigl[ 21 - ( 89 - 8\,c^2 )\,c^2 \Bigr]\,c^2\,\aZZ 
         + 4\,\Bigl[ 8\,\frac{\xphq}{\xphs-1} 
         - ( 11 + 7\,\xphs ) \Bigr]\,s^2\,c^2\,\apBox 
\nl &-& ( 7 - 8\,\frac{\xphs}{\xphs-1} )\,s^2\,c^2\,\apWDm\,\xphs 
         \Bigr\}\,\frac{1}{s^2\,c^2}
         \,\afun{\mw}
\nl &+&
         \frac{1}{48}
         \,\Bigl\{ \Bigl[ 1 + ( 16 - ( 69 + 8\,c^2)\,c^2 )\,c^2 \Bigr]\,\apD 
         - 4\,\Bigl[ 4 - ( 3 + 2\,s^2 )\,s^2 \Bigr]\,s^2\,\aAA 
         + 12\,\Bigl[ 7 - ( 25 + 2\,c^2 )\,c^2 \Bigr]\,c^2\,\aZZ 
\nl &+& 4\,\Bigl[ 11 - ( 41 - 2\,( 2 - c^2)\,c^2 )\,c^2 \Bigr]\,s\,c\,\aAZ 
         \Bigr\}\,\frac{1}{s^2\,c^4}
         \,\afun{\mz}
\nl &+&
         \frac{1}{12}
         \,\sumg\Bigl\{ \Bigl[ 2 - ( 1 + \xpls )\,\xpls \Bigr]\,\aplWt 
         + 3\,( 1 - \xpls )\,\alW\,\xpls \Bigr\}
         \,\bfun{ - \mws}{0}{\ml}
\nl &+&
         \frac{1}{48}
         \,\Bigl\{ \apWDm\,\xphq + 4\,\Bigl[ 12 
         - ( 4 - \xphs )\,\xphs \Bigr]\,\apBox - 4\,( 3 - \xphs )\,\apD 
         + 16\,( 9 - 4\,\xphs )\,\apW \Bigr\}
         \,\bfun{ - \mws}{\mw}{\mh}
\nl &+&
         \frac{1}{48}
         \,\Bigl\{ \Bigl[ 1 + 48\,s^2\,c^4 + 4\,( 4 - 29\,c^2 )\,c^2 \Bigr]\,\apD
         + 44\,\Bigl[ 1 - 2\,( 1 + 4\,c^2 )\,c^2 \Bigr]\,c\,\apWZ 
\nl &+& 4\,\Bigl[ 1 + 2\,( 3 - 20\,c^2 )\,c^2 \Bigr]\,s\,\apWA \Bigr\}\,\frac{1}{c^4}
         \,\bfun{ - \mws}{\mw}{\mz}
\nl &-&
         \apWAD
         \,\bfun{ - \mws}{0}{\mw}
\nl &-&
         \frac{1}{4}
         \,\sumg( - \xpds + \xpus )^2\,\apqWt
         \,\bfunp{0}{\mt}{\mb}
\nl &-&
         \frac{1}{12}
         \,\sumg\,\aplWt\,\xplq
         \,\bfunp{0}{0}{\ml}
\nl &+&
         \frac{1}{6}
         \,\Bigl[  - c^2\,\apD + 4\,s^2\,\aAA + ( 2 - c^2 )\,s\,c\,\aAZ \Bigr]
         \,\bfunp{0}{0}{\mw}
\nl &+&
         \frac{1}{48}
         \,( \xphs - 1)^2\,( \apWDm + 4\,\apBox )
         \,\bfunp{0}{\mw}{\mh}
\nl &-&
         \frac{1}{48}
         \,\Bigl\{ 16\,s^2\,c^4\,\apB - 4\,\Bigl[ 7 - 2\,( 4 - s^2 )\,s^2 \Bigr]\,s\,\apWA 
         - 4\,\Bigl[ 9 - 2\,( 1 + s^2 )\,s^2 \Bigr]\,c\,\apWZ 
\nl &-& ( 9 - 8\,s^2 )\,\apD \Bigr\}\,\frac{s^4}{c^4}
         \,\bfunp{0}{\mw}{\mz}
\eqas
\bqas
\ssdCZ^{(6)}_{\mh} &=&
         4
         \,\apW
\nl &-&
         \frac{1}{4}
         \,\Bigl[ 4\,( 1 + 2\,c^4 )\,\frac{1}{\xphs}\,\apBox 
         + 4\,( 1 + 10\,c^4 )\,\frac{1}{\xphs}\,\apW 
         + ( 3 - 2\,c^4 )\,\frac{1}{\xphs}\,\apD 
         + 4\,( - c^2 + 4\,\frac{1}{\xphs} )\,\aZZ \Bigr]\,\frac{1}{c^4}
         \,( 2 - 3\,\LR )
\nl &+&
         \frac{1}{8}
         \,\Bigl\{ 192\,\frac{\xpds}{\xphs}\,\sumg\,c^2\,\adp\,\xpds 
         - 192\,\frac{\xpus}{\xphs}\,\sumg\,c^2\,\aup\,\xpus 
         + 64\,\frac{\xpls}{\xphs}\,\sumg\,c^2\,\aLp\,\xpls 
\nl &-& 2\,\sumg\Bigl[ 3\,( - \xphs + 4\,\xpds )\,\xpds 
         + 3\,( - \xphs + 4\,\xpus )\,\xpus 
         + ( - \xphs + 4\,\xpls )\,\xpls \Bigr]\,\frac{1}{\xphs}\,c^2\,\apWDm 
\nl &-& 8\,\sumg( \alpBox\,\xpls + 3\,\adpBox\,\xpds - 3\,\aupBox\,\xpus )\,c^2 
         - 24\,\Bigl[ 11\,c^2 + ( 1 + 2\,c^2 )\,\frac{1}{\xphs} \Bigr]\,\ap 
         + 8\,\Bigl[ 3\,c^2\,\xphs - ( 1 + 8\,c^2 ) \Bigr]\,\apW 
\nl &+& 8\,\Bigl[ 11\,c^2\,\xphs 
         - 4\,\sumg( 3\,\xpdq + 3\,\xpuq + \xplq )\,\frac{1}{\xphs}\,c^2 
         - ( 1 + 2\,c^2 ) \Bigr]\,\apBox 
\nl &-& \Bigl[ 21\,c^2\,\xphs + ( 5 - 4\,c^2 ) \Bigr]\,\apD \Bigr\}\,\frac{1}{c^2}
         \,\LR
\nl &-&
         12
         \,\frac{\xpds}{\xphs}\,\sumg\,\adp\,\xpds
         \,\afun{\mb}
     +
         12
         \,\frac{\xpus}{\xphs}\,\sumg\,\aup\,\xpus
         \,\afun{\mt}
     -
         4
         \,\frac{\xpls}{\xphs}\,\sumg\,\aLp\,\xpls
         \,\afun{\ml}
\nl &-&
         \frac{3}{8}
         \,\Bigl[ 16\,\frac{1}{\xphs}\,\aZZ - 8\,\frac{1}{\xphs}\,c^2\,\ap 
         + ( - c^2 + 4\,\frac{1}{\xphs} )\,\apD \Bigr]\,\frac{1}{c^4}
         \,\afun{\mz}
\nl &+&
         6
         \,( - 2\,\apW + \ap )\,\frac{1}{\xphs}
         \,\afun{\mw}
     +
         \frac{1}{8}
         \,( 7\,\apD\,\xphs - 28\,\apBox\,\xphs + 120\,\ap )
         \,\afun{\mh}
\nl &-&
         \frac{3}{4}
         \,\sumg\Bigl\{  - 4\,\Bigl[ ( - \xphs + 4\,\xpds )\,\adpBox \Bigr] + 
         ( - \xphs + 4\,\xpds )\,\apWDm \Bigr\}\,\frac{\xpds}{\xphs}
         \,\bfun{ - \mhs}{\mb}{\mb}
\nl &-&
         \frac{3}{4}
         \,\sumg\Bigl\{ 4\,\Bigl[ ( - \xphs + 4\,\xpus )\,\aupBox \Bigr] 
         + ( - \xphs + 4\,\xpus )\,\apWDm \Bigr\}\,\frac{\xpus}{\xphs}
         \,\bfun{ - \mhs}{\mt}{\mt}
\nl &-&
         \frac{1}{4}
         \,\sumg\Bigl\{  - 4\,\Bigl[ ( - \xphs + 4\,\xpls )\,\alpBox \Bigr] + 
         ( - \xphs + 4\,\xpls )\,\apWDm \Bigr\}\,\frac{\xpls}{\xphs}
         \,\bfun{ - \mhs}{\ml}{\ml}
\nl &+&
         \frac{1}{16}
         \,\Bigl[ ( - 4\,c^2 + c^4\,\xphs + 12\,\frac{1}{\xphs} )\,\apWDp 
         + 4\,( - 4\,c^2 + c^4\,\xphs + 12\,\frac{1}{\xphs} )\,\apBox
         + 48\,( - c^2 + 2\,\frac{1}{\xphs} )\,\aZZ \Bigr]\,\frac{1}{c^4}
\nl &\times& \bfun{ - \mhs}{\mz}{\mz}
\nl &+&
         \frac{1}{8}
         \,\Bigl\{ \apWDm\,\xphs + 4\,\Bigl[ 12 - ( 4 - \xphs )\,\xphs \Bigr]\,
         \frac{1}{\xphs}\,\apBox - 4\,( 3 - \xphs )\,\frac{1}{\xphs}\,\apD
         + 16\,( 9 - 4\,\xphs )\,\frac{1}{\xphs}\,\apW \Bigr\}
\nl &\times& \bfun{ - \mhs}{\mw}{\mw}
\nl &-&
         \frac{9}{16}
         \,( - 4\,\apW\,\xphs + 3\,\apD\,\xphs - 12\,\apBox\,\xphs + 32\,\ap )
         \,\bfun{ - \mhs}{\mh}{\mh}
\eqas

\normalsize

\section{Appendix: Wave-function factors \label{WFF}}

In this Appendix we present the full list of wave-function renormalization factors
for $\PH, \PZ$ and $\PW$ fields. For the $\PW$ factor we present only the IR finite part.

\footnotesize
\bqas
\mrW^{(4)}_{\PH} &=&
       -
         \frac{1}{2}
         \,\Bigl[  - \sumg( 3\,\xpds + 3\,\xpus + \xpls )\,c^2 
         + ( 1 + 2\,c^2 ) \Bigr]\,\frac{1}{c^2}
         \,\LR
\nl &+&
         \frac{3}{2}
         \,\sumg\,\xpds
         \,\bfun{ - \mhs}{\mb}{\mb}
     +
         \frac{3}{2}
         \,\sumg\,\xpus
         \,\bfun{ - \mhs}{\mt}{\mt}
\nl &+&
         \frac{1}{2}
         \,\sumg\,\xpls
         \,\bfun{ - \mhs}{\ml}{\ml}
     -
         \frac{1}{2}
         \,\frac{1}{c^2}
         \,\bfun{ - \mhs}{\mz}{\mz}
\nl &-&
         \bfun{ - \mhs}{\mw}{\mw}
     +
         \frac{9}{8}
         \,\xphq
         \,\bfunp{ - \mhs}{\mh}{\mh}
\nl &-&
         \frac{3}{2}
         \,\sumg(  - \xphs + 4\,\xpds )\,\xpds
         \,\bfunp{ - \mhs}{\mb}{\mb}
\nl &-&
         \frac{3}{2}
         \,\sumg(  - \xphs + 4\,\xpus )\,\xpus
         \,\bfunp{ - \mhs}{\mt}{\mt}
\nl &-&
         \frac{1}{2}
         \,\sumg(  - \xphs + 4\,\xpls )\,\xpls
         \,\bfunp{ - \mhs}{\ml}{\ml}
\nl &+&
         \frac{1}{4}
         \,\Bigl[ 12 - ( 4 - \xphs )\,\xphs \Bigr]
         \,\bfunp{ - \mhs}{\mw}{\mw}
\nl &+&
         \frac{1}{8}
         \,( 12 - 4\,c^2\,\xphs + c^4\,\xphq )\,\frac{1}{c^4}
         \,\bfunp{ - \mhs}{\mz}{\mz}
\eqas
\bqas
\mrW^{(4)}_{\PZ} &=&
         \frac{1}{9}
         \,( 1 - 2\,c^2 )\,\frac{1}{c^2}
     -
         \frac{1}{36}
         \,\Bigl[ ( 9 + \vles + 3\,\vqds + 3\,\vqus ) \Bigr]\,\frac{1}{c^2}
         \,( 1 - 3\,\LR )
         \,\myNG
\nl &+&
         \frac{1}{6}
         \,( 1 - 20\,c^2 + 18\,s^2\,c^2 )\,\frac{1}{c^2}
         \,\LR
\nl &+&
         \frac{1}{12}
         \,( 1 - c^2\,\xphs )\,\xphs
         \,\afun{\mh}
     -
         \frac{1}{12}
         \,( 1 - c^2\,\xphs )\,\frac{1}{c^2}
         \,\afun{\mz}
\nl &+&
         \frac{1}{12}
         \,\sumg\Bigl[ ( 1 + \vles ) \Bigr]\,\frac{1}{c^2}
         \,\bfun{ - \mzs}{\ml}{\ml}
\nl &+&
         \frac{1}{6}
         \,\frac{1}{c^2}\,\sumg
         \,\bfun{ - \mzs}{0}{0}
\nl &+&
         \frac{1}{4}
         \,\sumg\Bigl[ ( 1 + \vqds ) \Bigr]\,\frac{1}{c^2}
         \,\bfun{ - \mzs}{\mb}{\mb}
\nl &+&
         \frac{1}{4}
         \,\sumg\Bigl[ ( 1 + \vqus ) \Bigr]\,\frac{1}{c^2}
         \,\bfun{ - \mzs}{\mt}{\mt}
\nl &+&
         \frac{1}{12}
         \,( 1 - 40\,c^2 + 36\,s^2\,c^2 )\,\frac{1}{c^2}
         \,\bfun{ - \mzs}{\mw}{\mw}
\nl &+&
         \frac{1}{6}
         \,\frac{1}{c^4}\,\sumg
         \,\bfunp{ - \mzs}{0}{0}
\nl &+&
         \frac{1}{12}
         \,\sumg\Bigl[ ( 1 + \vles ) - 2\,( 2 - \vles )\,c^2\,\xpls \Bigr]\,\frac{1}{c^4}
         \,\bfunp{ - \mzs}{\ml}{\ml}
\nl &+&
         \frac{1}{4}
         \,\sumg\Bigl[ ( 1 + \vqds ) - 2\,( 2 - \vqds )\,c^2\,\xpds \Bigr]\,\frac{1}{c^4}
         \,\bfunp{ - \mzs}{\mb}{\mb}
\nl &+&
         \frac{1}{4}
         \,\sumg\Bigl[ ( 1 + \vqus ) - 2\,( 2 - \vqus )\,c^2\,\xpus \Bigr]\,\frac{1}{c^4}
         \,\bfunp{ - \mzs}{\mt}{\mt}
\nl &+&
         \frac{1}{12}
         \,\Bigl\{ 1 + 4\,\Bigl[ 4 - ( 17 + 12\,c^2 )\,c^2 \Bigr]\,c^2 \Bigr\}\,\frac{1}{c^4}
         \,\bfunp{ - \mzs}{\mw}{\mw}
\nl &+&
         \frac{1}{12}
         \,( 2 - c^2\,\xphs )\,\xphs
         \,\bfun{ - \mzs}{\mh}{\mz}
\nl &+&
         \frac{1}{12}
         \,( 12 - 4\,c^2\,\xphs + c^4\,\xphq )\,\frac{1}{c^4}
         \,\bfunp{ - \mzs}{\mh}{\mz}
\eqas
\bqas
\mrW^{(4)}_{\PW} &=&
       -
         \frac{1}{9}
         \,( 1 - 36\,s^2 )
     -
         \frac{19}{6}
         \,\LR
     -
         \frac{4}{9}
         \,( 1 - 3\,\LR )
         \,\myNG
\nl &+&
         \frac{1}{6}
         \,\sumg\,\xplq
         \,\afun{\ml}
     +
         \frac{1}{2}
         \,\sumg( \xpds - \xpus )\,\xpds
         \,\afun{\mb}
\nl &-&
         \frac{1}{2}
         \,\sumg( \xpds - \xpus )\,\xpus
         \,\afun{\mt}
     +
         \frac{1}{12}
         \,\Bigl[ c^2\,\xphs + ( 1 - 2\,c^2 ) \Bigr]\,\frac{1}{c^2}
         \,\afun{\mw}
\nl &+&
         \frac{1}{12}
         \,( 1 - \xphs )\,\xphs
         \,\afun{\mh}
     -
         \frac{1}{12}
         \,( 9 - 8\,s^2 )\,\frac{s^2}{c^4}
         \,\afun{\mz}
\nl &-&
         4
         \,s^2
         \,\bfun{ - \mws}{0}{\mw}
\nl &+&
         \frac{1}{6}
         \,\sumg( 2 + \xplq )
         \,\bfun{ - \mws}{0}{\ml}
\nl &+&
         \frac{1}{2}
         \,\sumg( 2 + ( \xpus-\xpds )^2 )
         \,\bfun{ - \mws}{\mt}{\mb}
\nl &-&
         \frac{1}{12}
         \,\Bigl[ 1 - 48\,s^2\,c^4 + 2\,( 3 + 16\,c^2 )\,c^2 \Bigr]\,\frac{1}{c^4}
         \,\bfun{ - \mws}{\mw}{\mz}
\nl &+&
         \frac{1}{12}
         \,( 2 - \xphs )\,\xphs
         \,\bfun{ - \mws}{\mw}{\mh}
\nl &+&
         \frac{1}{6}
         \,\sumg\Bigl[ 2 - ( 1 + \xpls )\,\xpls \Bigr]
         \,\bfunp{ - \mws}{0}{\ml}
\nl &+&
         \frac{1}{2}
         \,\sumg( 2 - \xpds - \xpus - ( \xpus-\xpds )^2 )
         \,\bfunp{ - \mws}{\mt}{\mb}
\nl &+&
         \frac{1}{12}
         \,\Bigl[ 1 + 48\,s^2\,c^4 + 4\,( 4 - 29\,c^2 )\,c^2 \Bigr]\,\frac{1}{c^4}
         \,\bfunp{ - \mws}{\mw}{\mz}
\nl &+&
         \frac{1}{12}
         \,\Bigl[ 12 - ( 4 - \xphs )\,\xphs \Bigr]
         \,\bfunp{ - \mws}{\mw}{\mh}
\eqas
\bqas
\mrW^{(6)}_{\PH} &=&
         4
         \,\apW
     +
         \frac{1}{c^2}\,\aZZ
         \,( 2 - 3\,\LR )
\nl &-&
         \frac{1}{8}
         \,\Bigl\{  - 2\,\sumg( 3\,\xpds + 3\,\xpus + \xpls )\,c^2\,\apWDm
         + 8\,\sumg( \alpBox\,\xpls + 3\,\adpBox\,\xpds - 3\,\aupBox\,\xpus )\,c^2 
\nl &+& \Bigl[ 5\,c^2\,\xphs + ( 1 - 4\,c^2 ) \Bigr]\,\apD 
         - 4\,\Bigl[ 7\,c^2\,\xphs - 2\,( 1 + 2\,c^2 ) \Bigr]\,\apBox 
         + 8\,( 1 + 8\,c^2 )\,\apW \Bigr\}\,\frac{1}{c^2}
         \,\LR
\nl &+&
         \frac{1}{8}
         \,\frac{1}{c^2}\,\apD
         \,\afun{\mz}
     +
         \frac{1}{8}
         \,( \apD - 4\,\apBox )\,\xphs
         \,\afun{\mh}
\nl &+&
         \frac{3}{4}
         \,\sumg\Bigl[ \apWDm + 4\,\aupBox \Bigr]\,\xpus
         \,\bfun{ - \mhs}{\mt}{\mt}
     +
         \frac{3}{4}
         \,\sumg\Bigl[ \apWDm - 4\,\adpBox \Bigr]\,\xpds
         \,\bfun{ - \mhs}{\mb}{\mb}
\nl &+&
         \frac{1}{4}
         \,\sumg\Bigl[ \apWDm - 4\,\alpBox \Bigr]\,\xpls
         \,\bfun{ - \mhs}{\ml}{\ml}
\nl &-&
         \frac{1}{8}
         \,\Bigl[ 24\,\aZZ + 8\,\apW - c^2\,\apD\,\xphs 
         + 4\,( 2 - c^2\,\xphs )\,\apBox \Bigr]\,\frac{1}{c^2}
         \,\bfun{ - \mhs}{\mz}{\mz}
\nl &-&
         \frac{1}{4}
         \,\Bigl[ 32\,\apW - ( 2 - \xphs )\,\apD + 4\,( 2 - \xphs )\,\apBox \Bigr]
         \,\bfun{ - \mhs}{\mw}{\mw}
\nl &-&
         \frac{3}{8}
         \,( \apD - 4\,\apBox )\,\xphs
         \,\bfun{ - \mhs}{\mh}{\mh}
\nl &-&
         \frac{3}{4}
         \,\sumg\Bigl\{  - 4\,\Bigl[ (  - \xphs + 4\,\xpds )\,\adpBox \Bigr] 
         + (  - \xphs + 4\,\xpds )\,\apWDm \Bigr\}\,\xpds
         \,\bfunp{ - \mhs}{\mb}{\mb}
\nl &-&
         \frac{3}{4}
         \,\sumg\Bigl\{ 4\,\Bigl[ (  - \xphs + 4\,\xpus )\,\aupBox \Bigr] 
         + (  - \xphs + 4\,\xpus )\,\apWDm \Bigr\}\,\xpus
         \,\bfunp{ - \mhs}{\mt}{\mt}
\nl &-&
         \frac{1}{4}
         \,\sumg\Bigl\{  - 4\,\Bigl[ (  - \xphs + 4\,\xpls )\,\alpBox \Bigr] 
         + (  - \xphs + 4\,\xpls )\,\apWDm \Bigr\}\,\xpls
         \,\bfunp{ - \mhs}{\ml}{\ml}
\nl &+&
         \frac{1}{16}
         \,\Bigl[ 48\,( 2 - c^2\,\xphs )\,\aZZ + ( 12 - 4\,c^2\,\xphs
         + c^4\,\xphq )\,\apWDp 
         + 4\,( 12 - 4\,c^2\,\xphs + c^4\,\xphq )\,\apBox \Bigr]\,\frac{1}{c^4}
\nl &\times& \bfunp{ - \mhs}{\mz}{\mz}
\nl &+&
         \frac{1}{8}
         \,\Bigl\{ \apWDm\,\xphq + 4\,\Bigl[ 12 - ( 4 - \xphs )\,\xphs \Bigr]\,\apBox 
         - 4\,( 3 - \xphs )\,\apD + 16\,( 9 - 4\,\xphs )\,\apW \Bigr\}
         \,\bfunp{ - \mhs}{\mw}{\mw}
\nl &-&
         \frac{9}{16}
         \,(  - 4\,\apW\,\xphs + 3\,\apD\,\xphs - 12\,\apBox\,\xphs + 32\,\ap )\,\xphs
         \,\bfunp{ - \mhs}{\mh}{\mh}
\eqas
\bqas
\mrW^{(6)}_{\PZ} &=&
         \frac{1}{18}
         \,\frac{1}{c^2}\,\apBox
         \,( 2 + 3\,\LR )
\nl &-&
         \frac{1}{9}
         \,\sumg( \apn + \aplA + \aplV\,\vle + 3\,\apdA + 3\,\apuA + 3\,\vqd\,\apdV 
         + 3\,\vqu\,\apuV )\,\frac{1}{c^2}
         \,( 1 - 3\,\LR )
\nl &-&
         \frac{1}{72}
         \,\Bigl\{ 9\,\apWDm + \Bigl[ ( 1 + 4\,c^2 )\,\vqd + ( 3 + 4\,c^2 )\,\vle 
         + ( 5 + 8\,c^2 )\,\vqu \Bigr]\,\apD 
\nl &-& 4\,\Bigl[ ( 1 + 8\,c^2 )\,\vqd 
         + ( 3 + 8\,c^2 )\,\vle + ( 5 + 16\,c^2 )\,\vqu \Bigr]\,s\,\apWA 
\nl &-& 4\,\Bigl[ ( 5 - 8\,c^2 )\,\vqd + ( 7 - 8\,c^2 )\,\vle 
         + ( 13 - 16\,c^2 )\,\vqu \Bigr]\,c\,\apWZ 
\nl &+& 64\,\Bigl[ ( \vle + \vqd + 2\,\vqu ) \Bigr]\,s^2\,c^2\,\apB \Bigr\}\,\frac{1}{c^2}
         \,( 1 - 3\,\LR )
         \,\myNG
\nl &-&
         \frac{1}{12}
         \,\Bigl\{  - 6\,\sumg\,c\,\alBW\,\xpls\,\vle - 18\,\sumg\,c\,\vqd\,\adBW\,\xpds
         + 18\,\sumg\,c\,\vqu\,\auBW\,\xpus 
\nl &-& 8\,\Bigl[ 4 + 3\,( 2 + c^2 )\,c^2 \Bigr]\,s\,c\,\aAZ 
         - 2\,\Bigl[ 3\,c^2\,\xphs + ( 15 + 4\,(7 - 3\,(6 - c^2)\,c^2)\,c^2 ) \Bigr]\,\aZZ 
\nl &-& ( 1 - 20\,c^2 + 18\,s^2\,c^2 )\,\apD 
         + 4\,( 5 - 6\,s^4 )\,s^2\,\aAA \Bigr\}\,\frac{1}{c^2}
         \,\LR
\nl &+&
         \frac{1}{18}
         \,\Bigl[  - 4\,c^2\,\aZZ + 8\,s\,c\,\aAZ + 4\,s^2\,\aAA 
         + ( 1 - 2\,c^2 )\,\apD \Bigr]\,\frac{1}{c^2}
\nl &+&
         \frac{1}{24}
         \,\Bigl[ \apD - 8\,c\,\apWZ - c^2\,\apWDp\,\xphs + 4\,s\,\apWA - 12\,s^2\,\apB 
         + 4\,( 1 - c^2\,\xphs )\,\apBox \Bigr]\,\xphs
         \,\afun{\mh}
\nl &-&
         \frac{1}{24}
         \,\Bigl[ \apD + 16\,c\,\apWZ - c^2\,\apWDp\,\xphs + 4\,s\,\apWA 
         + 12\,s^2\,\apB + 4\,( 1 - c^2\,\xphs )\,\apBox \Bigr]\,\frac{1}{c^2}
         \,\afun{\mz}
\nl &-&
         ( c^2\,\apW + s\,c\,\apWB + s^2\,\apB )
         \,\afun{\mw}
\nl &+&
         \frac{1}{24}
         \,\sumg\Bigl[ \apWDm + 8\,\aplA + 8\,\aplV\,\vle + 12\,c\,\alBW\,\xpls\,\vle 
         + 64\,s^2\,c^2\,\apB\,\vle + ( 3 + 4\,c^2 )\,\apD\,\vle 
\nl &-& 4\,( 3 + 8\,c^2 )\,s\,\apWA\,\vle 
         - 4\,( 7 - 8\,c^2 )\,c\,\apWZ\,\vle \Bigr]\,\frac{1}{c^2}
         \,\bfun{ - \mzs}{\ml}{\ml}
\nl &+&
         \frac{1}{24}
         \,\sumg\Bigl[ 3\,\apWDm + 24\,\apdA + 24\,\vqd\,\apdV 
         + 36\,c\,\vqd\,\adBW\,\xpds + 64\,s^2\,c^2\,\vqd\,\apB 
         + ( 1 + 4\,c^2 )\,\vqd\,\apD 
\nl &-& 4\,( 1 + 8\,c^2 )\,s\,\vqd\,\apWA 
         - 4\,( 5 - 8\,c^2 )\,c\,\vqd\,\apWZ \Bigr]\,\frac{1}{c^2}
         \,\bfun{ - \mzs}{\mb}{\mb}
\nl &+&
         \frac{1}{24}
         \,\sumg\Bigl[ 3\,\apWDm + 24\,\apuA + 24\,\vqu\,\apuV 
         - 36\,c\,\vqu\,\auBW\,\xpus + 128\,s^2\,c^2\,\vqu\,\apB 
         + ( 5 + 8\,c^2 )\,\vqu\,\apD 
\nl &-& 4\,( 5 + 16\,c^2 )\,s\,\vqu\,\apWA 
         - 4\,( 13 - 16\,c^2 )\,c\,\vqu\,\apWZ \Bigr]\,\frac{1}{c^2}
         \,\bfun{ - \mzs}{\mt}{\mt}
\nl &+&
         \frac{1}{12}
         \,\sumg( \apWDm + 4\,\apn )\,\frac{1}{c^2}
         \,\bfun{ - \mzs}{0}{0}
\nl &+&
         \frac{1}{24}
         \,\Bigl[ 48\,\aZZ + ( 2 - c^2\,\xphs )\,c^2\,\apWDp\,\xphs 
         + 4\,( 2 - c^2\,\xphs )\,c^2\,\apBox\,\xphs \Bigr]\,\frac{1}{c^2}
         \,\bfun{ - \mzs}{\mh}{\mz}
\nl &+&
         \frac{1}{24}
         \,\Bigl\{ 12\,\Bigl[ 5 + 4\,( 3 - c^2 )\,c^2 \Bigr]\,s\,c\,\aAZ 
         + 12\,( 7 - 4\,( 4 + c^2 )\,c^2 \Bigr]\,c^2\,\aZZ 
\nl &-& 4\,\Bigl[ 35 - 12\,( 4 - s^2 )\,s^2 \Bigr]\,s^2\,\aAA 
         + ( 1 - 40\,c^2 + 36\,s^2\,c^2 )\,\apD \Bigr\}\,\frac{1}{c^2}
         \,\bfun{ - \mzs}{\mw}{\mw}
\nl &+&
         \frac{1}{24}
         \,\sumg\Bigl\{ 12\,c\,\alBW\,\xpls\,\vle + \Bigl[ ( 3 + 4\,c^2 ) 
         + 2\,( 3 + 4\,c^2 )\,c^2\,\xpls \Bigr]\,\apD\,\vle 
\nl &-& 4\,\Bigl[ ( 3 + 8\,c^2 ) 
         + 2\,( 3 + 8\,c^2 )\,c^2\,\xpls \Bigr]\,s\,\apWA\,\vle 
         - 4\,\Bigl[ ( 7 - 8\,c^2,rp) + 2\,( 7 - 8\,c^2 )\,c^2\,\xpls \Bigr]\,c\,\apWZ\,\vle 
\nl &+& ( 1 - 4\,c^2\,\xpls )\,\apWDm + 8\,( 1 - 4\,c^2\,\xpls )\,\aplA 
         + 8\,( 1 + 2\,c^2\,\xpls )\,\aplV\,\vle 
\nl &+& 64\,( 1 + 2\,c^2\,\xpls )\,s^2\,c^2\,\apB\,\vle \Bigr\}\,\frac{1}{c^4}
         \,\bfunp{ - \mzs}{\ml}{\ml}
\nl &+&
         \frac{1}{24}
         \,\sumg\Bigl\{ 36\,c\,\vqd\,\adBW\,\xpds + \Bigl[ ( 1 + 4\,c^2 ) 
         + 2\,( 1 + 4\,c^2 )\,c^2\,\xpds \Bigr]\,\vqd\,\apD 
\nl &-& 4\,\Bigl[ ( 1 + 8\,c^2 ) + 2\,( 1 + 8\,c^2 )\,c^2\,\xpds \Bigr]\,s\,\vqd\,\apWA 
         - 4\,\Bigl[ ( 5 - 8\,c^2,rp) + 2\,( 5 - 8\,c^2 )\,c^2\,\xpds \Bigr]\,c\,\vqd\,\apWZ 
\nl &+& 3\,( 1 - 4\,c^2\,\xpds )\,\apWDm + 24\,( 1 - 4\,c^2\,\xpds )\,\apdA 
         + 24\,( 1 + 2\,c^2\,\xpds )\,\vqd\,\apdV 
\nl &+& 64\,( 1 + 2\,c^2\,\xpds )\,s^2\,c^2\,\vqd\,\apB \Bigr\}\,\frac{1}{c^4}
         \,\bfunp{ - \mzs}{\mb}{\mb}
\nl &-&
         \frac{1}{24}
         \,\sumg\Bigl\{ 36\,c\,\vqu\,\auBW\,\xpus - \Bigl[ ( 5 + 8\,c^2 ) 
         + 2\,( 5 + 8\,c^2 )\,c^2\,\xpus \Bigr]\,\vqu\,\apD 
\nl &+& 4\,\Bigl[ ( 5 + 16\,c^2 ) + 2\,( 5 + 16\,c^2 )\,c^2\,\xpus \Bigr]\,s\,\vqu\,\apWA 
         + 4\,\Bigl[ ( 13 - 16\,c^2 ) + 2\,( 13 - 16\,c^2 )\,c^2\,\xpus \Bigr]\,c\,\vqu\,\apWZ 
\nl &-& 3\,( 1 - 4\,c^2\,\xpus )\,\apWDm - 24\,( 1 - 4\,c^2\,\xpus )\,\apuA 
         - 24\,( 1 + 2\,c^2\,\xpus )\,\vqu\,\apuV 
\nl &-& 128\,( 1 + 2\,c^2\,\xpus )\,s^2\,c^2\,\vqu\,\apB \Bigr\}\,\frac{1}{c^4}
         \,\bfunp{ - \mzs}{\mt}{\mt}
\nl &+&
         \frac{1}{12}
         \,\sumg( \apWDm + 4\,\apn )\,\frac{1}{c^4}
         \,\bfunp{ - \mzs}{0}{0}
\nl &+&
         \frac{1}{24}
         \,\Bigl[ 48\,( 2 - c^2\,\xphs )\,\aZZ 
         + ( 12 - 4\,c^2\,\xphs + c^4\,\xphq )\,\apWDp 
         + 4\,( 12 - 4\,c^2\,\xphs + c^4\,\xphq )\,\apBox \Bigr]\,\frac{1}{c^4}
\nl &\times& \bfunp{ - \mzs}{\mh}{\mz}
\nl &+&
         \frac{1}{24}
         \,\Bigl\{ \Bigl[ 1 + 4\,( 4 - (17 + 12\,c^2)\,c^2 )\,c^2 \Bigr]\,\apD
         + 12\,\Bigl[ 5 + 4\,( 6 + (3 + 4\,c^2)\,c^2 )\,c^2 \Bigr]\,s\,c\,\aAZ 
\nl &+& 12\,\Bigl[ 7 - 4\,( 5 + (9 - 4\,c^2)\,c^2 )\,c^2 \Bigr]\,c^2\,\aZZ 
         + 4\,\Bigl[ 33 
         - 4\,( 29 - 3\,(11 - 4\,s^2)\,s^2 )\,s^2 \Bigr]\,s^2\,\aAA \Bigr\}\,\frac{1}{c^4}
\nl &\times& \bfunp{ - \mzs}{\mw}{\mw}
\eqas
\bqas
\mrW^{(6)}_{\PW} &=&
       -
         \frac{1}{9}
         \,( 3\,\apqWt + \aplWt )
         \,( 1 - 3\,\LR )
         \,\myNG
     +
         \frac{1}{18}
         \,\apBox
         \,( 2 + 3\,\LR )
\nl &+&
         \frac{1}{6}
         \,\Bigl[ 3\,c^2\,\apW\,\xphs + 3\,\sumg\,c^2\,\alW\,\xpls + 9\,\sumg\,c^2\,\adW\,\xpds 
         - 9\,\sumg\,c^2\,\auW\,\xpus + ( 9 - 38\,c^2 )\,s\,c\,\aAZ 
\nl &+& ( 15 - 32\,c^2 )\,c^2\,\aZZ - ( 29 - 32\,s^2 )\,s^2\,\aAA \Bigr]\,\frac{1}{c^2}
         \,\LR
\nl &-&
         \frac{2}{9}
         \,\Bigl[ 9\,c^2\,\apD - 39\,s\,c\,\apWB + 6\,s^2\,\aAA + ( 1 - 42\,s^2 )\,\apW \Bigr]
\nl &+&
         \frac{1}{6}
         \,\sumg\,\aplWt\,\xplq
         \,\afun{\ml}
\nl &+&
         \frac{1}{2}
         \,\sumg( \xpds - \xpus )\,\apqWt\,\xpds
         \,\afun{\mb}
\nl &-&
         \frac{1}{2}
         \,\sumg( \xpds - \xpus )\,\apqWt\,\xpus
         \,\afun{\mt}
\nl &-&
         \frac{1}{24}
         \,\Bigl[ \apWDm\,\xphs + 8\,\apW + \apD - 4\,( 1 - \xphs )\,\apBox \Bigr]\,\xphs
         \,\afun{\mh}
\nl &+&
         \frac{1}{24}
         \,\Bigl[ c^2\,\apWDm\,\xphs - 4\,( 1 - \xphs )\,c^2\,\apBox 
         + ( 1 + 8\,c^2 )\,\apD + 4\,( 5 - 14\,c^2 )\,s\,c\,\aAZ 
\nl &+& 4\,( 9 - 16\,c^2 )\,c^2\,\aZZ - 4\,( 15 - 16\,s^2 )\,s^2\,\aAA \Bigr]\,\frac{1}{c^2}
         \,\afun{\mw}
\nl &+&
         \frac{1}{6}
         \,\sumg\Bigl[ 3\,\alW\,\xpls + ( 2 + \xplq )\,\aplWt \Bigr]
         \,\bfun{ - \mws}{0}{\ml}
\nl &+&
         \frac{1}{2}
         \,\sumg\Bigl[ 3\,\adW\,\xpds - 3\,\auW\,\xpus + ( 2 + ( \xpus-\xpds )^2 )\,\apqWt \Bigr]
         \,\bfun{ - \mws}{\mt}{\mb}
\nl &+&
         \frac{1}{24}
         \,\Bigl[ 48\,\apW + ( 2 - \xphs )\,\apWDm\,\xphs + 4\,( 2 - \xphs )\,\apBox\,\xphs \Bigr]
         \,\bfun{ - \mws}{\mw}{\mh}
\nl &-&
         \frac{1}{24}
         \,\Bigl\{ \Bigl[ 1 - 48\,s^2\,c^4 + 2\,( 3 + 16\,c^2 )\,c^2 \Bigr]\,\apD
         + 4\,\Bigl[ 1 + 2\,( 1 - 4\,c^4 )\,c^2 \Bigr]\,s\,\apWA 
\nl &+& 4\,\Bigl[ 5 - 2\,( 5 - 2\,( 9 + 2\,c^2)\,c^2 )\,c^2 \Bigr]\,c\,\apWZ 
         + 16\,( 3 - 4\,s^2 )\,s^2\,c^4\,\apB \Bigr\}\,\frac{1}{c^4}
         \,\bfun{ - \mws}{\mw}{\mz}
\nl &-&
         2
         \,\apWAD
         \,\bfun{ - \mws}{0}{\mw}
\nl &+&
         \frac{1}{2}
         \,\sumg\Bigl[ 3\,( 1 - \xpds + \xpus )\,\adW\,\xpds - 3\,( 1
         + \xpds - \xpus )\,\auW\,\xpus + ( 2 - \xpds - \xpus - ( \xpus-\xpds )^2 )\,\apqWt \Bigr]
\nl &\times& \bfunp{ - \mws}{\mt}{\mb}
\nl &+&
         \frac{1}{6}
         \,\sumg\Bigl\{ \Bigl[ 2 - ( 1 + \xpls )\,\xpls \Bigr]\,\aplWt 
         + 3\,( 1 - \xpls )\,\alW\,\xpls \Bigr\}
         \,\bfunp{ - \mws}{0}{\ml}
\nl &-&
         \frac{1}{24}
         \,\Bigl[ 12\,( 3 - 2\,c^2 )\,c^2\,\aZZ + 4\,( 3 - 2\,s^2 )\,s^2\,\aAA 
         + 4\,( 5 - 12\,c^4 )\,s\,c\,\aAZ + ( 9 - 8\,s^2 )\,s^2\,\apD \Bigr]\,\frac{1}{c^4}
         \,\afun{\mz}
\nl &+&
         \frac{1}{24}
         \,\Bigl\{ \apWDm\,\xphq + 4\,\Bigl[ 12 - ( 4 - \xphs )\,\xphs \Bigr]\,\apBox 
         - 4\,( 3 - \xphs )\,\apD + 16\,( 9 - 4\,\xphs )\,\apW \Bigr\}
         \,\bfunp{ - \mws}{\mw}{\mh}
\nl &+&
         \frac{1}{24}
         \,\Bigl\{ \Bigl[ 1 + 48\,s^2\,c^4 + 4\,( 4 - 29\,c^2 )\,c^2 \Bigr]\,\apD
         + 44\,\Bigl[ 1 - 2\,( 1 + 4\,c^2 )\,c^2 \Bigr]\,c\,\apWZ 
\nl &+& 4\,\Bigl[ 1 + 2\,( 3 - 20\,c^2 )\,c^2 \Bigr]\,s\,\apWA \Bigr\}\,\frac{1}{c^4}
         \,\bfunp{ - \mws}{\mw}{\mz}
\eqas

\normalsize

\section{Appendix: Mixing of Wilson coefficients \label{MixWc}}

In this Appendix we present the entries of the mixing matrix, \eqn{MixW}, that can be derived from
the renormalization of $\PH \to \PVV$. 
\subsection{Notations}
First we define

\vspace{0.5cm}
\bei
\item[\fbox{$\mrR\,$}] 
\eei

\scriptsize
\[
\begin{array}{lll}
\mrR^{a}_{0}= 1 + 2\,c^2 \;\;&\;\;
\mrR^{a}_{1}= 7 - 12\,c^2 \;\;&\;\;
\mrR^{a}_{2}= 23 - 12\,c^2 \\
\mrR^{a}_{3}= 13 - 9\,c^2 \;\;&\;\;
\mrR^{a}_{4}= 1 - 12\,c^2 \;\;&\;\;
\mrR^{a}_{5}= 7 + 6\,c^2 \\
\mrR^{a}_{6}= 77 + 48\,c^2 \;\;&\;\;
\mrR^{a}_{7}= 5 - 12\,c^2 \;\;&\;\;
\mrR^{a}_{8}= 11 - 4\,c^2 \\
\mrR^{a}_{9}= 5 + 48\,c^2 & & \\
\end{array}
\]

\[
\begin{array}{lll}
\mrR^{b}_{0}= 1 + 6\,s^2 \;\;&\;\;
\mrR^{b}_{1}= 1 - s^2 \;\;&\;\;
\mrR^{b}_{2}= 11 - 12\,s^2 \\
\mrR^{b}_{3}= 1 - c^2 \;\;&\;\;
\mrR^{b}_{4}= 13 - 18\,\mrR^{a}_{0}\,c^2 \;\;&\;\;
\mrR^{b}_{5}= 41 + 6\,\mrR^{a}_{1}\,c^2 \\
\mrR^{b}_{6}= 65 - 6\,\mrR^{a}_{2}\,c^2 \;\;&\;\;
\mrR^{b}_{7}= 9 - 4\,s^2 \;\;&\;\;
\mrR^{b}_{8}= 7 - 6\,s^2 \\
\mrR^{b}_{9}= 11 - 9\,s^2 \;\;&\;\;
\mrR^{b}_{10}= 2 - c^2 \;\;&\;\;
\mrR^{b}_{11}= 43 - 18\,s^2 \\
\mrR^{b}_{12}= 31 - 8\,\mrR^{a}_{3}\,c^2 \;\;&\;\;
\mrR^{b}_{13}= 7 - 8\,c^4 \;\;&\;\;
\mrR^{b}_{14}= 179 + 16\,\mrR^{a}_{4}\,c^2 \\
\mrR^{b}_{15}= 77 - 12\,\mrR^{a}_{5}\,c^2 \;\;&\;\;
\mrR^{b}_{16}= 79 - 4\,\mrR^{a}_{6}\,c^2 \;\;&\;\;
\mrR^{b}_{17}= 7\,c^2 + 3\,s^2 \\
\mrR^{b}_{18}= 35 + 3\,\mrR^{a}_{7}\,c^2 \;\;&\;\;
\mrR^{b}_{19}= 107 - 32\,\mrR^{a}_{8}\,c^2 \;\;&\;\;
\mrR^{b}_{20}= 267 - 4\,\mrR^{a}_{9}\,c^2 \\
\mrR^{b}_{21}= 7 - 78\,c^2 & & \\
\end{array}
\]

\[
\begin{array}{lll}
\mrR^{c}_{0}= 1 + 2\,c^2 - 4\,\mrR^{b}_{0}\,s^2\,c^2 \;\;&\;\;
\mrR^{c}_{1}= 3 - 4\,\mrR^{b}_{1}\,s^2 \;\;&\;\;
\mrR^{c}_{2}= 1 - 2\,s^2 \\
\mrR^{c}_{3}= 7 + 6\,\mrR^{b}_{2}\,s^2 \;\;&\;\;
\mrR^{c}_{4}= 9 - 16\,c^2 \;\;&\;\;
\mrR^{c}_{5}= 4 - 3\,s^2 \\
\mrR^{c}_{6}= 3 - 4\,\mrR^{b}_{3}\,c^2 \;\;&\;\;
\mrR^{c}_{7}= 17 + 2\,\mrR^{b}_{4}\,c^2 \;\;&\;\;
\mrR^{c}_{8}= 1 - 2\,c^2 \\
\mrR^{c}_{9}= 11 - \mrR^{b}_{5}\,c^2 \;\;&\;\;
\mrR^{c}_{10}= 1 + \mrR^{b}_{6}\,c^2 \;\;&\;\;
\mrR^{c}_{11}= 91 - 18\,\mrR^{b}_{7}\,s^2 \\
\mrR^{c}_{12}= 1 + 8\,\mrR^{b}_{1}\,s^2 \;\;&\;\;
\mrR^{c}_{13}= 5 + 5\,c^2 - 12\,\mrR^{b}_{8}\,s^2\,c^2 \;\;&\;\;
\mrR^{c}_{14}= 27 + 128\,s^2\,c^2 \\
\mrR^{c}_{15}= 1 - 4\,s^2 \;\;&\;\;
\mrR^{c}_{16}= 19 - 8\,\mrR^{b}_{9}\,s^2 \;\;&\;\;
\mrR^{c}_{17}= 1 - \mrR^{b}_{10}\,c^2 \\
\mrR^{c}_{18}= 117 - 4\,\mrR^{b}_{11}\,s^2 \;\;&\;\;
\mrR^{c}_{19}= 1 + 4\,s^2 \;\;&\;\;
\mrR^{c}_{20}= 1 + \mrR^{b}_{12}\,c^2 \\
\mrR^{c}_{21}= 1 + 8\,c^2 \;\;&\;\;
\mrR^{c}_{22}= 1 + c^2 \;\;&\;\;
\mrR^{c}_{23}= 3 - 5\,c^2 \\
\mrR^{c}_{24}= 1 - 7\,c^2 \;\;&\;\;
\mrR^{c}_{25}= 5 - 11\,c^2 \;\;&\;\;
\mrR^{c}_{26}= 1 + \mrR^{b}_{13}\,c^2 \\
\mrR^{c}_{27}= 3 - \mrR^{b}_{14}\,c^2 \;\;&\;\;
\mrR^{c}_{28}= 27 - 48\,c^2 + 128\,s^4 \;\;&\;\;
\mrR^{c}_{29}= 9 + 8\,c^4 \\
\mrR^{c}_{30}= 13 - \mrR^{b}_{15}\,c^2 \;\;&\;\;
\mrR^{c}_{31}= 58 + \mrR^{b}_{16}\,c^2 \;\;&\;\;
\mrR^{c}_{32}= 3\,c^2 + \mrR^{b}_{17}\,s^2 \\
\mrR^{c}_{33}= 2 + c^2 \;\;&\;\;
\mrR^{c}_{34}= 14 - \mrR^{b}_{18}\,c^2 \;\;&\;\;
\mrR^{c}_{35}= 54 + \mrR^{b}_{19}\,c^2 \\
\mrR^{c}_{36}= 55 - \mrR^{b}_{20}\,c^2 \;\;&\;\;
\mrR^{c}_{37}= 7 - 2\,\mrR^{b}_{21}\,c^2 \\
\end{array}
\]

\normalsize

\vspace{0.5cm}
\bei
\item[\fbox{$\mrS\,$}] 
\eei

\scriptsize
\[
\begin{array}{lll}
\mrS_{0}= 3\,\xpds + 3\,\xpus + \xpls \;\;&\;\;
\mrS_{1}= 5\,\xpds + \xpus \;\;&\;\;
\mrS_{2}= \xpds + \xpus \\
\mrS_{3}= \xpds + 5\,\xpus \;\;&\;\;
\mrS_{4}= 3\,\xpdq + 3\,\xpuq + \xplq & \\
\end{array}
\]

\normalsize

With their help we derive the relevant elements of the mixing matrix.
\subsection{Mixing matrix}
\footnotesize
\bqas
d\mrZ^{\PW}_{1,1} &=&
         \frac{32}{9}
         \,s^2\,\myNG
       +
         \frac{1}{4}
         \,( \mrR^{c}_{0} + \mrR^{c}_{1}\,c^2\,\xphs )
         \,\frac{1}{c^2}
\eqas
\bqas
d\mrZ^{\PW}_{1,2} &=&
       -
         ( 6 - \xphs )
         \,s^2\,c^2
\eqas
\bqas
d\mrZ^{\PW}_{1,3} &=&
       -
         \frac{1}{24}
         \,( \vog + \mrR^{c}_{4} )
         \,\frac{1}{s\,c}
         \,\myNG
       -
         \frac{1}{12}
         \,( 6\,\mrR^{c}_{2}\,c^2\,\xphs + \mrR^{c}_{3} )
         \,\frac{s}{c}
\eqas
\bqas
d\mrZ^{\PW}_{2,1} &=&
         ( c^2\,\xphs - 2\,\mrR^{c}_{5} )
         \,s^2
\eqas
\bqas
d\mrZ^{\PW}_{2,2} &=&
         \frac{1}{12}
         \,( 9 + \vog )
         \,\frac{1}{c^2}
         \,\myNG
       +
         \frac{1}{12}
         \,( 3\,\mrR^{c}_{6}\,c^2\,\xphs + \mrR^{c}_{7} )
         \,\frac{1}{c^2}
\eqas
\bqas
d\mrZ^{\PW}_{2,3} &=&
         \frac{1}{24}
         \,( \vog + 8\,s^2\,\vtg + \mrR^{c}_{4} )
         \,\frac{1}{s\,c}
         \,\myNG
       -
         \frac{1}{12}
         \,( 6\,\mrR^{c}_{8}\,s^2\,c^2\,\xphs - \mrR^{c}_{9} )
         \,\frac{1}{s\,c}
\eqas
\bqas
d\mrZ^{\PW}_{3,1} &=&
         \frac{1}{12}
         \,( \vog + 8\,s^2\,\vtg + \mrR^{c}_{4} )
         \,\frac{1}{s\,c}
         \,\myNG
       -
         \frac{1}{6}
         \,( 6\,\mrR^{c}_{2}\,s^2\,c^2\,\xphs + \mrR^{c}_{10} )
         \,\frac{1}{s\,c}
\eqas
\bqas
d\mrZ^{\PW}_{3,2} &=&
       -
         \frac{1}{12}
         \,( \vog + \mrR^{c}_{4} )
         \,\frac{1}{s\,c}
         \,\myNG
       +
         \frac{1}{6}
         \,( 6\,\mrR^{c}_{2}\,c^2\,\xphs - \mrR^{c}_{11} )
         \,\frac{s}{c}
\eqas
\bqas
d\mrZ^{\PW}_{3,3} &=&
         \frac{1}{72}
         \,( 3\,\vog + \mrR^{c}_{14} )
         \,\frac{1}{c^2}
         \,\myNG
       +
         \frac{1}{12}
         \,( 3\,\mrR^{c}_{12}\,c^2\,\xphs + 2\,\mrR^{c}_{13} )
         \,\frac{1}{c^2}
\eqas
\bqas
d\mrZ^{\PW}_{4,1} &=&
         \frac{16}{3}
         \,s^2\,\myNG
       -
         \frac{1}{3}
         \,( 3\,\mrR^{c}_{15}\,c^2\,\xphs - \mrR^{c}_{16} )
         \,\frac{s^2}{c^2}
\eqas
\bqas
d\mrZ^{\PW}_{4,2} &=&
       -
         \frac{80}{9}
         \,\frac{1}{c^2}
         \,\mrR^{c}_{17}
         \,\myNG
       +
         \frac{1}{3}
         \,( 3\,\mrR^{c}_{15}\,c^2\,\xphs - \mrR^{c}_{18} )
         \,\frac{s^2}{c^2}
\eqas
\bqas
d\mrZ^{\PW}_{4,3} &=&
         \frac{80}{9}
         \,\frac{s^3}{c}
         \,\myNG
       +
         \frac{1}{3}
         \,( 3\,\mrR^{c}_{19}\,s^2\,c^2\,\xphs - \mrR^{c}_{20} )
         \,\frac{1}{s\,c}
\eqas
\bqas
d\mrZ^{\PW}_{4,4} &=&
         \frac{1}{2}
         \,\sumg
         \,\mrS_{0}
       +
         \frac{1}{12}
         \,( 9\,c^2\,\xphs + 7\,\mrR^{c}_{21} )
         \,\frac{1}{c^2}
\eqas
\bqas
d\mrZ^{\PW}_{4,5} &=&
         \frac{10}{3}
         \,\frac{s^2}{c^2}
\eqas
\bqas
d\mrZ^{\PW}_{4,16} &=&
       -
         5
         \,\sumg\,\xpls
       +
         \frac{2}{3}
         \,\frac{1}{c^2}
         \,\mrR^{c}_{22}
         \,\myNG
\eqas
\bqas
d\mrZ^{\PW}_{4,17} &=&
       -
         \sumg\,\xpls
       -
         \frac{2}{3}
         \,\frac{1}{c^2}
         \,\mrR^{c}_{23}
         \,\myNG
\eqas
\bqas
d\mrZ^{\PW}_{4,18} &=&
         \frac{2}{3}
         \,\myNG
       -
         \sumg\,\xpls
\eqas
\bqas
d\mrZ^{\PW}_{4,19} &=&
         2
         \,\frac{1}{c^2}
         \,\mrR^{c}_{22}
         \,\myNG
       -
         3
         \,\sumg
         \,\mrS_{1}
\eqas
\bqas
d\mrZ^{\PW}_{4,20} &=&
       -
         \frac{2}{3}
         \,\frac{1}{c^2}
         \,\mrR^{c}_{24}
         \,\myNG
       -
         3
         \,\sumg
         \,\mrS_{2}
\eqas
\bqas
d\mrZ^{\PW}_{4,21} &=&
         2
         \,\frac{1}{c^2}
         \,\mrR^{c}_{22}
         \,\myNG
       -
         3
         \,\sumg
         \,\mrS_{3}
\eqas
\bqas
d\mrZ^{\PW}_{4,22} &=&
       -
         \frac{2}{3}
         \,\frac{1}{c^2}
         \,\mrR^{c}_{25}
         \,\myNG
       -
         3
         \,\sumg
         \,\mrS_{2}
\eqas
\bqas
d\mrZ^{\PW}_{5,1} &=&
       -
         \frac{1}{72}
         \,( 3\,\vog + 24\,s^2\,\vtg + \mrR^{c}_{28} )
         \,\myNG
       +
         \frac{1}{24}
         \,( 12\,s^2\,c^6\,\xphq - 9\,\mrR^{c}_{26}\,c^2\,\xphs - \mrR^{c}_{27} )
         \,\frac{1}{c^2}
\eqas
\bqas
d\mrZ^{\PW}_{5,2} &=&
       -
         \frac{1}{24}
         \,( 2\,\vog - 9\,c^2\,\xphs - c^2\,\vog\,\xphs + 2\,\mrR^{c}_{29} )
         \,\frac{1}{c^2}
         \,\myNG
\nl &-&
         \frac{1}{24}
         \,( \mrR^{c}_{30}\,c^2\,\xphs + \mrR^{c}_{31} + 3\,\mrR^{c}_{32}\,c^4\,\xphq )
         \,\frac{1}{c^2}
\nl &-&
         \frac{1}{4}
         \,\sumg
         \,( 2\,\mrS_{0} + \mrS_{0}\,c^2\,\xphs - 4\,\mrS_{4}\,c^2 )
\eqas
\bqas
d\mrZ^{\PW}_{5,3} &=&
         \frac{1}{144}
         \,( 3\,c^2\,\vog\,\xphs + 3\,\mrR^{c}_{4}\,c^2\,\xphs - 3\,\mrR^{c}_{33}\,\vog 
         - 24\,\mrR^{c}_{33}\,s^2\,\vtg - \mrR^{c}_{35} )
         \,\frac{1}{s\,c}
         \,\myNG
\nl &+&
         \frac{1}{24}
         \,( 6\,\mrR^{c}_{2}\,s^2\,c^4\,\xphq + 2\,\mrR^{c}_{34}\,c^2\,\xphs - \mrR^{c}_{36} )
         \,\frac{1}{s\,c}
\eqas
\bqas
d\mrZ^{\PW}_{5,4} &=&
         \frac{1}{48}
         \,\frac{1}{c^2}
         \,\mrR^{c}_{37}
\eqas
\bqas
d\mrZ^{\PW}_{5,5} &=&
         4
         \,c^2
\eqas

\normalsize

\section{Appendix: Non-factorizable amplitudes \label{Ampnf}}

In this Appendix we present the explicit expressions for the non-factorizable part
of the $\PH \to \PAA, \PAZ, \PZZ$ and $\PWW$ amplitudes.
\subsection{Notations}
It is useful to introduce the following sets of polynomials:
\vspace{0.8cm}
\bei
\item[\fbox{$\mrT\,$}] where $s = \stw$ and $c = \ctw$ 
\eei

\scriptsize
\[
\begin{array}{llll}
\mrT^{a}_{0}= 81 - 56\,c^2 \;\;&\;\;
\mrT^{a}_{1}= 53 - 39\,c^2 \;\;&\;\;
\mrT^{a}_{2}= 35 - 26\,c^2 \;\;&\;\;
\mrT^{a}_{3}= 35 - 18\,c^2 \\
\mrT^{a}_{4}= 19 - 10\,c^2 \;\;&\;\;
\mrT^{a}_{5}= 21 - 11\,c^2 \;\;&\;\;
\mrT^{a}_{6}= 5 + 4\,c^2 \;\;&\;\;
\mrT^{a}_{7}= 1 - 6\,c^2 \\
\mrT^{a}_{8}= 7 - 12\,c^2 \;\;&\;\;
\mrT^{a}_{9}= 47 + 12\,c^2 \;\;&\;\;
\mrT^{a}_{10}= 7 - c^2 \;\;&\;\;
\mrT^{a}_{11}= 37 + 4\,c^2 \\
\mrT^{a}_{12}= 213 - 68\,c^2 \;\;&\;\;
\mrT^{a}_{13}= 29 - 4\,c^2 \;\;&\;\;
\mrT^{a}_{14}= 49 - 12\,c^2 \;\;&\;\;
\mrT^{a}_{15}= 173 - 192\,c^2 \\
\mrT^{a}_{16}= 331 - 200\,c^2 \;\;&\;\;
\mrT^{a}_{17}= 33 - 16\,c^2 \;\;&\;\;
\mrT^{a}_{18}= 53 - 16\,c^2 \;\;&\;\;
\mrT^{a}_{19}= 8 - 3\,c^2 \\
\mrT^{a}_{20}= 16 - 9\,c^2 \;\;&\;\;
\mrT^{a}_{21}= 3 - c^2 & & \\
\end{array}
\]


\[
\begin{array}{llll}
\mrT^{b}_{0}= 7 - 18\,c^2 \;\;&\;\;
\mrT^{b}_{1}= 2 - 27\,c^2 \;\;&\;\;
\mrT^{b}_{2}= 49 - 2\,\mrT^{a}_{0}\,c^2 \;\;&\;\;
\mrT^{b}_{3}= 11 - \mrT^{a}_{1}\,c^2 \\
\mrT^{b}_{4}= 41 - 6\,\mrT^{a}_{2}\,c^2 \;\;&\;\;
\mrT^{b}_{5}= 37 - 2\,\mrT^{a}_{3}\,c^2 \;\;&\;\;
\mrT^{b}_{6}= 39 - 4\,\mrT^{a}_{4}\,c^2 \;\;&\;\;
\mrT^{b}_{7}= 11 - \mrT^{a}_{5}\,c^2 \\
\mrT^{b}_{8}= 33 - 56\,c^2 \;\;&\;\;
\mrT^{b}_{9}= 37 - 78\,c^2 \;\;&\;\;
\mrT^{b}_{10}= 6 - 13\,c^2 \;\;&\;\;
\mrT^{b}_{11}= 4 - 5\,c^2 \\
\mrT^{b}_{12}= 7 - 9\,c^2 \;\;&\;\;
\mrT^{b}_{13}= 19 - 22\,c^2 \;\;&\;\;
\mrT^{b}_{14}= 1 - c^2 \;\;&\;\;
\mrT^{b}_{15}= 23 - 22\,c^2 \\
\mrT^{b}_{16}= 1 + \mrT^{a}_{6}\,c^2 \;\;&\;\;
\mrT^{b}_{17}= 1 + 4\,c^2 \;\;&\;\;
\mrT^{b}_{18}= 11 + 12\,c^2 \;\;&\;\;
\mrT^{b}_{19}= 1 + 2\,\mrT^{a}_{7}\,c^2 \\
\mrT^{b}_{20}= 7 - 12\,c^2 \;\;&\;\;
\mrT^{b}_{21}= 1 - 6\,c^2 \;\;&\;\;
\mrT^{b}_{22}= 3 - 2\,c^2 \;\;&\;\;
\mrT^{b}_{23}= 25 + 6\,\mrT^{a}_{8}\,c^2 \\
\mrT^{b}_{24}= 1 + 2\,c^2 \;\;&\;\;
\mrT^{b}_{25}= 7 - 6\,c^2 \;\;&\;\;
\mrT^{b}_{26}= 1 - 2\,c^2 \;\;&\;\;
\mrT^{b}_{27}= 1 + 3\,c^2 \\
\mrT^{b}_{28}= 1 + 12\,c^2 \;\;&\;\;
\mrT^{b}_{29}= 7 + 6\,c^2 \;\;&\;\;
\mrT^{b}_{30}= 1 + c^2 \;\;&\;\;
\mrT^{b}_{31}= 3 - 4\,c^2 \\
\end{array}
\]
\[
\begin{array}{llll}
\mrT^{b}_{32}= 13 - 34\,c^2 \;\;&\;\;
\mrT^{b}_{33}= 1 - 5\,c^2 \;\;&\;\;
\mrT^{b}_{34}= 3 - 5\,c^2 \;\;&\;\;
\mrT^{b}_{35}= 1 + 20\,c^2 \\
\mrT^{b}_{36}= 3 + 4\,c^2 \;\;&\;\;
\mrT^{b}_{37}= 27 + 68\,c^2 \;\;&\;\;
\mrT^{b}_{38}= 11 + 3\,c^2 \;\;&\;\;
\mrT^{b}_{39}= 13 - 3\,c^2 \\
\mrT^{b}_{40}= 5 + 4\,c^2 \;\;&\;\;
\mrT^{b}_{41}= 12 + \mrT^{a}_{9}\,c^2 \;\;&\;\;
\mrT^{b}_{42}= 1 - 4\,c^2 \;\;&\;\;
\mrT^{b}_{43}= 7 - 3\,c^2 \\
\mrT^{b}_{44}= 13 - 12\,c^2 \;\;&\;\;
\mrT^{b}_{45}= 37 - 12\,\mrT^{a}_{10}\,c^2 \;\;&\;\;
\mrT^{b}_{46}= 11 - 23\,c^2 \;\;&\;\;
\mrT^{b}_{47}= 7 + c^2 \\
\mrT^{b}_{48}= 7 - c^2 \;\;&\;\;
\mrT^{b}_{49}= 44 - \mrT^{a}_{11}\,c^2 \;\;&\;\;
\mrT^{b}_{50}= 11 + c^2 \;\;&\;\;
\mrT^{b}_{51}= 23 - 7\,c^2 \\
\mrT^{b}_{52}= 78 - \mrT^{a}_{12}\,c^2 \;\;&\;\;
\mrT^{b}_{53}= 58 - 3\,\mrT^{a}_{13}\,c^2 \;\;&\;\;
\mrT^{b}_{54}= 37 - 3\,c^2 \;\;&\;\;
\mrT^{b}_{55}= 8 - \mrT^{a}_{14}\,c^2 \\
\mrT^{b}_{56}= 55 - 56\,c^2 \;\;&\;\;
\mrT^{b}_{57}= 7 + 4\,c^2 \;\;&\;\;
\mrT^{b}_{58}= 1 - 36\,c^2 \;\;&\;\;
\mrT^{b}_{59}= 115 - 4\,\mrT^{a}_{15}\,c^2 \\
\mrT^{b}_{60}= 77 - 192\,c^2 \;\;&\;\;
\mrT^{b}_{61}= 3 - 8\,c^2 \;\;&\;\;
\mrT^{b}_{62}= 119 + 360\,c^2 \;\;&\;\;
\mrT^{b}_{63}= 31 + 18\,c^2 \\
\end{array}
\]
\[
\begin{array}{llll}
\mrT^{b}_{64}= 37\,c^2 - s^2 \;\;&\;\;
\mrT^{b}_{65}= 14 - 11\,c^2 \;\;&\;\;
\mrT^{b}_{66}= 231 - 200\,c^2 \;\;&\;\;
\mrT^{b}_{67}= 184 - \mrT^{a}_{16}\,c^2 \\
\mrT^{b}_{68}= 809 - 552\,c^2 \;\;&\;\;
\mrT^{b}_{69}= 365 - 496\,c^2 \;\;&\;\;
\mrT^{b}_{70}= 103 - 96\,c^2 \;\;&\;\;
\mrT^{b}_{71}= 185 - 52\,c^2 \\
\mrT^{b}_{72}= 51 + 11\,c^2 \;\;&\;\;
\mrT^{b}_{73}= 22 - \mrT^{a}_{17}\,c^2 \;\;&\;\;
\mrT^{b}_{74}= 23 - 4\,c^2 \;\;&\;\;
\mrT^{b}_{75}= 45 - 16\,c^2 \\
\mrT^{b}_{76}= 79 - 2\,\mrT^{a}_{18}\,c^2 \;\;&\;\;
\mrT^{b}_{77}= 121 - 40\,c^2 \;\;&\;\;
\mrT^{b}_{78}= 6 + c^2 \;\;&\;\;
\mrT^{b}_{79}= 11 - 8\,c^2 \\
\mrT^{b}_{80}= 87 - 40\,c^2 \;\;&\;\;
\mrT^{b}_{81}= 3\,c^2 - s^2 \;\;&\;\;
\mrT^{b}_{82}= 31 - 36\,c^2 \;\;&\;\;
\mrT^{b}_{83}= 13 + 30\,c^2 \\
\mrT^{b}_{84}= 47 - 40\,c^2 \;\;&\;\;
\mrT^{b}_{85}= 37 - 42\,c^2 \;\;&\;\;
\mrT^{b}_{86}= 1 + 45\,c^2 \;\;&\;\;
\mrT^{b}_{87}= 37 - 32\,c^2 \\
\mrT^{b}_{88}= 1 + 16\,c^2 \;\;&\;\;
\mrT^{b}_{89}= 9 + 16\,c^2 \;\;&\;\;
\mrT^{b}_{90}= 2 - c^2 \;\;&\;\;
\mrT^{b}_{91}= 11 - 2\,\mrT^{a}_{19}\,c^2 \\
\mrT^{b}_{92}= 5 - 2\,\mrT^{a}_{20}\,c^2 \;\;&\;\;
\mrT^{b}_{93}= 9 - 4\,c^2 \;\;&\;\;
\mrT^{b}_{94}= 13 - 6\,c^2 \;\;&\;\;
\mrT^{b}_{95}= 49 - 16\,c^2 \\
\mrT^{b}_{96}= 69 - 32\,c^2 \;\;&\;\;
\mrT^{b}_{97}= 97 - 32\,\mrT^{a}_{21}\,c^2 \;\;&\;\;
\mrT^{b}_{98}= 1 - 2\,s^2 & \\
\end{array}
\]


\[
\begin{array}{llll}
\mrT^{c}_{0}= 1 - s^2 \;\;&\;\;
\mrT^{c}_{1}= 3 + 2\,\mrT^{b}_{0}\,c^2 \;\;&\;\;
\mrT^{c}_{2}= 3 + 2\,\mrT^{b}_{1}\,c^2 \;\;&\;\;
\mrT^{c}_{3}= 3 - \mrT^{b}_{2}\,c^2 \\
\mrT^{c}_{4}= 5 - 6\,c^2 \;\;&\;\;
\mrT^{c}_{5}= 5 + 8\,\mrT^{b}_{3}\,c^2 \;\;&\;\;
\mrT^{c}_{6}= 6 - 7\,s^2 \;\;&\;\;
\mrT^{c}_{7}= 7 + 2\,\mrT^{b}_{4}\,c^2 \\
\mrT^{c}_{8}= 9 - 2\,\mrT^{b}_{5}\,c^2 \;\;&\;\;
\mrT^{c}_{9}= 9 - 2\,\mrT^{b}_{6}\,c^2 \;\;&\;\;
\mrT^{c}_{10}= 11 - 8\,\mrT^{b}_{7}\,c^2 \;\;&\;\;
\mrT^{c}_{11}= 2 - c^2 \\
\mrT^{c}_{12}= 4 - 3\,s^2 \;\;&\;\;
\mrT^{c}_{13}= 49 - 78\,s^2 \;\;&\;\;
\mrT^{c}_{14}= 15 - 8\,c^2 \;\;&\;\;
\mrT^{c}_{15}= 2 - \mrT^{b}_{8}\,c^2 \\
\mrT^{c}_{16}= 3 + 2\,\mrT^{b}_{9}\,c^2 \;\;&\;\;
\mrT^{c}_{17}= 5 + 12\,\mrT^{b}_{10}\,c^2 \;\;&\;\;
\mrT^{c}_{18}= 7 - 8\,\mrT^{b}_{11}\,c^2 \;\;&\;\;
\mrT^{c}_{19}= 7 - 4\,\mrT^{b}_{12}\,c^2 \\
\mrT^{c}_{20}= 9 - 2\,\mrT^{b}_{13}\,c^2 \;\;&\;\;
\mrT^{c}_{21}= 35 + 6\,c^2 \;\;&\;\;
\mrT^{c}_{22}= 25 - 66\,c^2 \;\;&\;\;
\mrT^{c}_{23}= 5 - 6\,s^2 \\
\mrT^{c}_{24}= 4 - \mrT^{b}_{8}\,c^2 \;\;&\;\;
\mrT^{c}_{25}= 11 - 40\,\mrT^{b}_{14}\,c^2 \;\;&\;\;
\mrT^{c}_{26}= 11 - 36\,\mrT^{b}_{14}\,c^2 \;\;&\;\;
\mrT^{c}_{27}= 13 - 2\,\mrT^{b}_{15}\,c^2 \\
\mrT^{c}_{28}= 1 - 2\,c^2 \;\;&\;\;
\mrT^{c}_{29}= 1 - 12\,c^2 \;\;&\;\;
\mrT^{c}_{30}= 5 - 4\,\mrT^{b}_{16}\,c^2 \;\;&\;\;
\mrT^{c}_{31}= 1 + 4\,c^2 \\
\end{array}
\]
\[
\begin{array}{llll}
\mrT^{c}_{32}= 2 - 9\,\mrT^{b}_{17}\,c^2 \;\;&\;\;
\mrT^{c}_{33}= 11 + 16\,c^2 \;\;&\;\;
\mrT^{c}_{34}= 12 - \mrT^{b}_{18}\,c^2 \;\;&\;\;
\mrT^{c}_{35}= 1 + c^2 \\
\mrT^{c}_{36}= 3 - 4\,c^2 \;\;&\;\;
\mrT^{c}_{37}= 5 - 2\,\mrT^{b}_{19}\,c^2 \;\;&\;\;
\mrT^{c}_{38}= 11 - 36\,c^2 \;\;&\;\;
\mrT^{c}_{39}= 11 + 12\,c^2 \\
\mrT^{c}_{40}= 2 + \mrT^{b}_{20}\,c^2 \;\;&\;\;
\mrT^{c}_{41}= 5 + \mrT^{b}_{21}\,c^2 \;\;&\;\;
\mrT^{c}_{42}= 29 - 18\,\mrT^{b}_{22}\,c^2 \;\;&\;\;
\mrT^{c}_{43}= 47 - 36\,c^4 \\
\mrT^{c}_{44}= 5 - 2\,\mrT^{b}_{23}\,c^2 \;\;&\;\;
\mrT^{c}_{45}= 19 - 36\,\mrT^{b}_{24}\,c^2 \;\;&\;\;
\mrT^{c}_{46}= 1 + 6\,c^2 \;\;&\;\;
\mrT^{c}_{47}= 1 - 2\,\mrT^{b}_{14}\,c^2 \\
\mrT^{c}_{48}= 1 - 3\,c^2 \;\;&\;\;
\mrT^{c}_{49}= 1 - 4\,\mrT^{b}_{25}\,c^2 \;\;&\;\;
\mrT^{c}_{50}= 5 + 6\,c^2 \;\;&\;\;
\mrT^{c}_{51}= 5 - 2\,\mrT^{b}_{22}\,c^2 \\
\mrT^{c}_{52}= 4 + c^2 \;\;&\;\;
\mrT^{c}_{53}= 1 - 4\,c^2 \;\;&\;\;
\mrT^{c}_{54}= 1 - c^2 \;\;&\;\;
\mrT^{c}_{55}= 1 + 4\,\mrT^{b}_{14}\,c^2 \\
\mrT^{c}_{56}= 1 + \mrT^{b}_{26}\,c^2 \;\;&\;\;
\mrT^{c}_{57}= 2 - \mrT^{b}_{27}\,c^2 \;\;&\;\;
\mrT^{c}_{58}= 3 - \mrT^{b}_{28}\,c^2 \;\;&\;\;
\mrT^{c}_{59}= 5 - 2\,\mrT^{b}_{29}\,c^2 \\
\mrT^{c}_{60}= 3 + 4\,c^2 \;\;&\;\;
\mrT^{c}_{61}= 5 + 12\,\mrT^{b}_{26}\,c^2 \;\;&\;\;
\mrT^{c}_{62}= 1 + 16\,c^2 \;\;&\;\;
\mrT^{c}_{63}= 2 - \mrT^{b}_{20}\,c^2 \\
\end{array}
\]
\[
\begin{array}{llll}
\mrT^{c}_{64}= 3 - 4\,\mrT^{b}_{14}\,c^2 \;\;&\;\;
\mrT^{c}_{65}= 31 - 12\,\mrT^{b}_{30}\,c^2 \;\;&\;\;
\mrT^{c}_{66}= 41 - 12\,\mrT^{b}_{31}\,c^4 \;\;&\;\;
\mrT^{c}_{67}= 5 + 2\,\mrT^{b}_{32}\,c^2 \\
\mrT^{c}_{68}= 1 + 8\,c^2 \;\;&\;\;
\mrT^{c}_{69}= 1 + 4\,\mrT^{b}_{33}\,c^2 \;\;&\;\;
\mrT^{c}_{70}= 59 + 12\,\mrT^{b}_{34}\,c^2 \;\;&\;\;
\mrT^{c}_{71}= 61 + 12\,\mrT^{b}_{35}\,c^4 \\
\mrT^{c}_{72}= 2 - 3\,\mrT^{b}_{36}\,c^2 \;\;&\;\;
\mrT^{c}_{73}= 22 - \mrT^{b}_{37}\,c^2 \;\;&\;\;
\mrT^{c}_{74}= 1 + 12\,\mrT^{b}_{30}\,c^2 \;\;&\;\;
\mrT^{c}_{75}= 23 - 4\,\mrT^{b}_{38}\,c^2 \\
\mrT^{c}_{76}= 3 + c^2 \;\;&\;\;
\mrT^{c}_{77}= 11 - 7\,c^2 \;\;&\;\;
\mrT^{c}_{78}= 15 + 4\,\mrT^{b}_{39}\,c^2 \;\;&\;\;
\mrT^{c}_{79}= 12 - \mrT^{b}_{40}\,c^2 \\
\mrT^{c}_{80}= 55 + 117\,c^2 \;\;&\;\;
\mrT^{c}_{81}= 15 - 4\,\mrT^{b}_{41}\,c^2 \;\;&\;\;
\mrT^{c}_{82}= 8 - 3\,s^2 \;\;&\;\;
\mrT^{c}_{83}= 23 + 3\,c^2 \\
\mrT^{c}_{84}= 1 + 12\,\mrT^{b}_{14}\,c^2 \;\;&\;\;
\mrT^{c}_{85}= 1 + 3\,\mrT^{b}_{42}\,c^2 \;\;&\;\;
\mrT^{c}_{86}= 8 + 3\,c^2 \;\;&\;\;
\mrT^{c}_{87}= 4 - 3\,c^2 \\
\mrT^{c}_{88}= 5 - 4\,c^2 \;\;&\;\;
\mrT^{c}_{89}= 19 - 4\,\mrT^{b}_{43}\,c^2 \;\;&\;\;
\mrT^{c}_{90}= 13 + \mrT^{b}_{44}\,c^2 \;\;&\;\;
\mrT^{c}_{91}= 1 - 3\,c^4 \\
\mrT^{c}_{92}= 3 + 4\,\mrT^{b}_{39}\,c^2 \;\;&\;\;
\mrT^{c}_{93}= 17 - 2\,\mrT^{b}_{45}\,c^2 \;\;&\;\;
\mrT^{c}_{94}= 19 - 4\,\mrT^{b}_{46}\,c^2 \;\;&\;\;
\mrT^{c}_{95}= 35 - 4\,\mrT^{b}_{47}\,c^2 \\
\end{array}
\]
\[
\begin{array}{llll}
\mrT^{c}_{96}= 37 - 12\,\mrT^{b}_{48}\,c^2 \;\;&\;\;
\mrT^{c}_{97}= 37 - 4\,\mrT^{b}_{49}\,c^2 \;\;&\;\;
\mrT^{c}_{98}= 51 - 4\,\mrT^{b}_{50}\,c^2 \;\;&\;\;
\mrT^{c}_{99}= 67 - 4\,\mrT^{b}_{51}\,c^2 \\
\mrT^{c}_{100}= 75 + 4\,\mrT^{b}_{52}\,c^2 \;\;&\;\;
\mrT^{c}_{101}= 77 - 4\,\mrT^{b}_{53}\,c^2 \;\;&\;\;
\mrT^{c}_{102}= 5 - 7\,c^2 \;\;&\;\;
\mrT^{c}_{103}= 41 - 4\,\mrT^{b}_{47}\,c^2 \\
\mrT^{c}_{104}= 53 - 4\,\mrT^{b}_{54}\,c^2 \;\;&\;\;
\mrT^{c}_{105}= 67 - 4\,\mrT^{b}_{55}\,c^2 \;\;&\;\;
\mrT^{c}_{106}= 1 - 24\,c^2 \;\;&\;\;
\mrT^{c}_{107}= 13 - \mrT^{b}_{56}\,c^2 \\
\mrT^{c}_{108}= 27 - 56\,c^2 \;\;&\;\;
\mrT^{c}_{109}= 41 - 64\,c^2 \;\;&\;\;
\mrT^{c}_{110}= 1 - 2\,\mrT^{b}_{57}\,c^2 \;\;&\;\;
\mrT^{c}_{111}= 2\,c^2 - s^2 \\
\mrT^{c}_{112}= 7 + 8\,c^2 \;\;&\;\;
\mrT^{c}_{113}= 9 - 2\,\mrT^{b}_{58}\,c^2 \;\;&\;\;
\mrT^{c}_{114}= 11 + 36\,c^2 \;\;&\;\;
\mrT^{c}_{115}= 17 - 72\,c^2 \\
\mrT^{c}_{116}= 20 + \mrT^{b}_{59}\,c^2 \;\;&\;\;
\mrT^{c}_{117}= 39 + 4\,\mrT^{b}_{60}\,c^2 \;\;&\;\;
\mrT^{c}_{118}= 41 - 156\,c^2 \;\;&\;\;
\mrT^{c}_{119}= 49 - 204\,c^2 \\
\mrT^{c}_{120}= 79 + 228\,\mrT^{b}_{61}\,c^2 \;\;&\;\;
\mrT^{c}_{121}= 151 - 4\,\mrT^{b}_{62}\,c^2 \;\;&\;\;
\mrT^{c}_{122}= 5 - 12\,c^2 \;\;&\;\;
\mrT^{c}_{123}= 25 - 8\,\mrT^{b}_{63}\,c^2 \\
\mrT^{c}_{124}= 35\,c^2 - s^2 \;\;&\;\;
\mrT^{c}_{125}= 36\,c^2 - s^2 \;\;&\;\;
\mrT^{c}_{126}= 71\,c^4 - \mrT^{b}_{64}\,s^2 \;\;&\;\;
\mrT^{c}_{127}= 8 - 7\,\mrT^{b}_{22}\,c^2 \\
\end{array}
\]
\[
\begin{array}{llll}
\mrT^{c}_{128}= 12 - 11\,c^2 \;\;&\;\;
\mrT^{c}_{129}= 11 - 6\,c^2 \;\;&\;\;
\mrT^{c}_{130}= 13 - 24\,c^2 \;\;&\;\;
\mrT^{c}_{131}= 17 - 22\,c^2 \\
\mrT^{c}_{132}= 19 - 286\,c^2 \;\;&\;\;
\mrT^{c}_{133}= 23 - 4\,\mrT^{b}_{65}\,c^2 \;\;&\;\;
\mrT^{c}_{134}= 137 - 2\,\mrT^{b}_{66}\,c^2 \;\;&\;\;
\mrT^{c}_{135}= 143 - 4\,\mrT^{b}_{67}\,c^2 \\
\mrT^{c}_{136}= 164 - \mrT^{b}_{68}\,c^2 \;\;&\;\;
\mrT^{c}_{137}= 251 - 152\,c^2 \;\;&\;\;
\mrT^{c}_{138}= 19 - \mrT^{b}_{69}\,c^2 \;\;&\;\;
\mrT^{c}_{139}= 18 + \mrT^{b}_{70}\,c^2 \\
\mrT^{c}_{140}= 55 + 2\,\mrT^{b}_{71}\,c^2 \;\;&\;\;
\mrT^{c}_{141}= 85 - 96\,c^2 \;\;&\;\;
\mrT^{c}_{142}= 85 - 8\,\mrT^{b}_{72}\,c^2 \;\;&\;\;
\mrT^{c}_{143}= 3 - 11\,c^2 \\
\mrT^{c}_{144}= 3 + 13\,c^2 \;\;&\;\;
\mrT^{c}_{145}= 9 - 16\,c^2 \;\;&\;\;
\mrT^{c}_{146}= 21 - 40\,c^2 \;\;&\;\;
\mrT^{c}_{147}= 21 - 4\,\mrT^{b}_{73}\,c^2 \\
\mrT^{c}_{148}= 3 - 8\,c^2 \;\;&\;\;
\mrT^{c}_{149}= 5 + 2\,c^2 \;\;&\;\;
\mrT^{c}_{150}= 9 - 4\,c^2 \;\;&\;\;
\mrT^{c}_{151}= 17 - 14\,c^2 \\
\mrT^{c}_{152}= 7 + \mrT^{b}_{74}\,c^2 \;\;&\;\;
\mrT^{c}_{153}= 31 - 8\,c^2 \;\;&\;\;
\mrT^{c}_{154}= 17 - \mrT^{b}_{75}\,c^2 \;\;&\;\;
\mrT^{c}_{155}= 31 - 2\,\mrT^{b}_{76}\,c^2 \\
\mrT^{c}_{156}= 69 - 2\,\mrT^{b}_{77}\,c^2 \;\;&\;\;
\mrT^{c}_{157}= 11 - 8\,\mrT^{b}_{78}\,c^2 \;\;&\;\;
\mrT^{c}_{158}= 19 + 2\,c^2 \;\;&\;\;
\mrT^{c}_{159}= 4 - 7\,c^2 \\
\end{array}
\]
\[
\begin{array}{llll}
\mrT^{c}_{160}= 13 + 2\,\mrT^{b}_{79}\,c^2 \;\;&\;\;
\mrT^{c}_{161}= 21 + 2\,\mrT^{b}_{80}\,c^2 \;\;&\;\;
\mrT^{c}_{162}= 11 - 8\,c^2 \;\;&\;\;
\mrT^{c}_{163}= 3\,c^4 - \mrT^{b}_{81}\,s^2 \\
\mrT^{c}_{164}= 1 - 8\,c^2 \;\;&\;\;
\mrT^{c}_{165}= 5 - 24\,c^2 \;\;&\;\;
\mrT^{c}_{166}= 4 - \mrT^{b}_{82}\,c^2 \;\;&\;\;
\mrT^{c}_{167}= 7 - 3\,c^2 \\
\mrT^{c}_{168}= 13 - 36\,c^2 \;\;&\;\;
\mrT^{c}_{169}= 19 - 2\,\mrT^{b}_{83}\,c^2 \;\;&\;\;
\mrT^{c}_{170}= 25 - 112\,c^2 \;\;&\;\;
\mrT^{c}_{171}= 105 - 8\,\mrT^{b}_{84}\,c^2 \\
\mrT^{c}_{172}= 49 - 120\,c^2 \;\;&\;\;
\mrT^{c}_{173}= 2 - 47\,c^2 \;\;&\;\;
\mrT^{c}_{174}= 2 - 7\,c^2 \;\;&\;\;
\mrT^{c}_{175}= 2 + 313\,c^2 \\
\mrT^{c}_{176}= 19 - 48\,c^2 \;\;&\;\;
\mrT^{c}_{177}= 8 - 21\,c^2 \;\;&\;\;
\mrT^{c}_{178}= 9 - 8\,c^2 \;\;&\;\;
\mrT^{c}_{179}= 11 - 2\,\mrT^{b}_{85}\,c^2 \\
\mrT^{c}_{180}= 13 - 56\,c^2 \;\;&\;\;
\mrT^{c}_{181}= 13 - 2\,\mrT^{b}_{86}\,c^2 \;\;&\;\;
\mrT^{c}_{182}= 15 - 58\,c^2 \;\;&\;\;
\mrT^{c}_{183}= 49 - 4\,\mrT^{b}_{87}\,c^2 \\
\mrT^{c}_{184}= 2 - 77\,c^2 \;\;&\;\;
\mrT^{c}_{185}= 2 - 251\,c^2 \;\;&\;\;
\mrT^{c}_{186}= 1 + 2\,c^2 \;\;&\;\;
\mrT^{c}_{187}= 3 - 2\,\mrT^{b}_{22}\,c^2 \\
\mrT^{c}_{188}= 21 + 4\,\mrT^{b}_{88}\,c^2 \;\;&\;\;
\mrT^{c}_{189}= 7 - 4\,c^2 \;\;&\;\;
\mrT^{c}_{190}= 17 - 18\,c^2 \;\;&\;\;
\mrT^{c}_{191}= 15 + \mrT^{b}_{89}\,c^2 \\
\end{array}
\]
\[
\begin{array}{llll}
\mrT^{c}_{192}= 41 - 16\,c^2 \;\;&\;\;
\mrT^{c}_{193}= 23 - 16\,c^2 \;\;&\;\;
\mrT^{c}_{194}= 39 - 34\,c^2 \;\;&\;\;
\mrT^{c}_{195}= 23 - 10\,c^2 \\
\mrT^{c}_{196}= 53 - 4\,\mrT^{b}_{79}\,c^2 \;\;&\;\;
\mrT^{c}_{197}= 3 + 16\,c^2 \;\;&\;\;
\mrT^{c}_{198}= 5 - 4\,\mrT^{b}_{90}\,c^2 \;\;&\;\;
\mrT^{c}_{199}= 1 - 2\,s^2 \\
\mrT^{c}_{200}= 1 + 3\,\mrT^{b}_{14}\,c^2 \;\;&\;\;
\mrT^{c}_{201}= 2 - \mrT^{b}_{91}\,c^2 \;\;&\;\;
\mrT^{c}_{202}= 3 + 2\,\mrT^{b}_{92}\,c^2 \;\;&\;\;
\mrT^{c}_{203}= 3 - 2\,\mrT^{b}_{93}\,c^2 \\
\mrT^{c}_{204}= 8 - 9\,c^2 \;\;&\;\;
\mrT^{c}_{205}= 8 + c^2 \;\;&\;\;
\mrT^{c}_{206}= 9 - 2\,\mrT^{b}_{94}\,c^2 \;\;&\;\;
\mrT^{c}_{207}= 10 - 13\,c^2 \\
\mrT^{c}_{208}= 10 + c^2 \;\;&\;\;
\mrT^{c}_{209}= 13 - 5\,c^2 \;\;&\;\;
\mrT^{c}_{210}= 14 - \mrT^{b}_{95}\,c^2 \;\;&\;\;
\mrT^{c}_{211}= 18 - \mrT^{b}_{96}\,c^2 \\
\mrT^{c}_{212}= 21 - 8\,c^2 \;\;&\;\;
\mrT^{c}_{213}= 46 - \mrT^{b}_{97}\,c^2 \;\;&\;\;
\mrT^{c}_{214}= 37 - 16\,c^2 \;\;&\;\;
\mrT^{c}_{215}= c^2 - \mrT^{b}_{98}\,s^2 \\
\end{array}
\]


\[
\begin{array}{llll}
\mrT^{d}_{0}= 1 - 6\,s^2 \;\;&\;\;
\mrT^{d}_{1}= 1 - 3\,s^2 \;\;&\;\;
\mrT^{d}_{2}= 1 - 2\,s^2 \;\;&\;\;
\mrT^{d}_{3}= 1 - 4\,\mrT^{c}_{0}\,s^2 \\
\mrT^{d}_{4}= 2 - 3\,s^2 \;\;&\;\;
\mrT^{d}_{5}= 3 - 4\,\mrT^{c}_{0}\,s^2 \;\;&\;\;
\mrT^{d}_{6}= 1 - 6\,c^2 - 72\,s^2\,c^2 \;\;&\;\;
\mrT^{d}_{7}= 1 - 2\,c^2 \\
\mrT^{d}_{8}= 1 + 4\,c^2 \;\;&\;\;
\mrT^{d}_{9}= 5 + 8\,c^2 \;\;&\;\;
\mrT^{d}_{10}= c^2 - \mrT^{c}_{1}\,s^2 \;\;&\;\;
\mrT^{d}_{11}= 3\,c^4 - \mrT^{c}_{2}\,s^2 \\
\mrT^{d}_{12}= 1 - 4\,s^2 \;\;&\;\;
\mrT^{d}_{13}= 1 - 8\,\mrT^{c}_{0}\,s^2 \;\;&\;\;
\mrT^{d}_{14}= 1 + 8\,\mrT^{c}_{0}\,s^2 \;\;&\;\;
\mrT^{d}_{15}= 1 + 2\,\mrT^{c}_{3}\,c^2 \\
\mrT^{d}_{16}= 1 - 4\,\mrT^{c}_{4}\,c^4 \;\;&\;\;
\mrT^{d}_{17}= 1 - \mrT^{c}_{5}\,c^2 \;\;&\;\;
\mrT^{d}_{18}= 1 + 8\,\mrT^{c}_{6}\,s^2\,c^2 \;\;&\;\;
\mrT^{d}_{19}= 1 - \mrT^{c}_{7}\,c^2 \\
\mrT^{d}_{20}= 1 + \mrT^{c}_{8}\,c^2 \;\;&\;\;
\mrT^{d}_{21}= 1 + \mrT^{c}_{9}\,c^2 \;\;&\;\;
\mrT^{d}_{22}= 1 + \mrT^{c}_{10}\,c^2 \;\;&\;\;
\mrT^{d}_{23}= 3 - 4\,\mrT^{c}_{11}\,c^2 \\
\mrT^{d}_{24}= 3 - 4\,\mrT^{c}_{12}\,s^2 \;\;&\;\;
\mrT^{d}_{25}= 5 - 8\,s^2 \;\;&\;\;
\mrT^{d}_{26}= 5 - 6\,s^2 \;\;&\;\;
\mrT^{d}_{27}= 5 + \mrT^{c}_{13}\,s^2 \\
\mrT^{d}_{28}= 11 - 2\,\mrT^{c}_{14}\,c^2 \;\;&\;\;
\mrT^{d}_{29}= 1 - c^2 \;\;&\;\;
\mrT^{d}_{30}= 1 + 2\,\mrT^{c}_{15}\,c^2 \;\;&\;\;
\mrT^{d}_{31}= 1 - \mrT^{c}_{16}\,c^2 \\
\end{array}
\]
\[
\begin{array}{llll}
\mrT^{d}_{32}= 1 - \mrT^{c}_{17}\,c^2 \;\;&\;\;
\mrT^{d}_{33}= 1 + \mrT^{c}_{18}\,c^2 \;\;&\;\;
\mrT^{d}_{34}= 1 + \mrT^{c}_{19}\,c^2 \;\;&\;\;
\mrT^{d}_{35}= 1 + \mrT^{c}_{20}\,c^2 \\
\mrT^{d}_{36}= 1 + \mrT^{c}_{21}\,c^2 \;\;&\;\;
\mrT^{d}_{37}= 2 - 3\,\mrT^{c}_{22}\,c^2 \;\;&\;\;
\mrT^{d}_{38}= 3 - 2\,c^2 \;\;&\;\;
\mrT^{d}_{39}= 5 - 21\,c^2 \\
\mrT^{d}_{40}= 5 - 4\,s^2 \;\;&\;\;
\mrT^{d}_{41}= 7 + 6\,\mrT^{c}_{23}\,s^2 \;\;&\;\;
\mrT^{d}_{42}= 11 - 8\,c^2 \;\;&\;\;
\mrT^{d}_{43}= 13 - 8\,s^2 \\
\mrT^{d}_{44}= 20 - 39\,s^2 \;\;&\;\;
\mrT^{d}_{45}= 1 + 2\,\mrT^{c}_{24}\,c^2 \;\;&\;\;
\mrT^{d}_{46}= 1 + \mrT^{c}_{25}\,c^2 \;\;&\;\;
\mrT^{d}_{47}= 1 + \mrT^{c}_{26}\,c^2 \\
\mrT^{d}_{48}= 1 + \mrT^{c}_{27}\,c^2 \;\;&\;\;
\mrT^{d}_{49}= 3 + 4\,s^2 \;\;&\;\;
\mrT^{d}_{50}= 3 + 8\,s^2 \;\;&\;\;
\mrT^{d}_{51}= 7 - 6\,s^2 \\
\mrT^{d}_{52}= 1 - 4\,s^2\,c^2 \;\;&\;\;
\mrT^{d}_{53}= 1 - 14\,c^2 \;\;&\;\;
\mrT^{d}_{54}= 1 - 11\,c^2 \;\;&\;\;
\mrT^{d}_{55}= 1 - 8\,c^2 \\
\mrT^{d}_{56}= 1 - 2\,c^2 - 4\,\mrT^{c}_{28}\,s^2\,c^2 \;\;&\;\;
\mrT^{d}_{57}= 1 + 6\,c^2 \;\;&\;\;
\mrT^{d}_{58}= 1 + 8\,c^2 \;\;&\;\;
\mrT^{d}_{59}= 1 + 10\,c^2 \\
\mrT^{d}_{60}= 1 + 18\,c^2 \;\;&\;\;
\mrT^{d}_{61}= 1 - 4\,\mrT^{c}_{29}\,c^4 \;\;&\;\;
\mrT^{d}_{62}= 1 - \mrT^{c}_{30}\,c^2 \;\;&\;\;
\mrT^{d}_{63}= 2 - 7\,c^2 - 6\,s^2\,c^2 \\
\end{array}
\]
\[
\begin{array}{llll}
\mrT^{d}_{64}= 2 - c^2 \;\;&\;\;
\mrT^{d}_{65}= 3 + 10\,c^2 - 4\,\mrT^{c}_{31}\,s^2\,c^2 \;\;&\;\;
\mrT^{d}_{66}= 3 + 192\,c^6 + 8\,\mrT^{c}_{32}\,s^2\,c^2 \;\;&\;\;
\mrT^{d}_{67}= 3 - 4\,\mrT^{c}_{33}\,c^2 \\
\mrT^{d}_{68}= 7 + 4\,c^2 - 24\,s^2\,c^2 \;\;&\;\;
\mrT^{d}_{69}= 13 - 4\,\mrT^{c}_{34}\,c^2 \;\;&\;\;
\mrT^{d}_{70}= 1 + 3\,c^2 \;\;&\;\;
\mrT^{d}_{71}= 1 - 4\,c^4 \\
\mrT^{d}_{72}= 1 + 12\,\mrT^{c}_{35}\,c^2 \;\;&\;\;
\mrT^{d}_{73}= 1 - 6\,\mrT^{c}_{36}\,c^2 \;\;&\;\;
\mrT^{d}_{74}= 1 - \mrT^{c}_{37}\,c^2 \;\;&\;\;
\mrT^{d}_{75}= 1 - 2\,\mrT^{c}_{38}\,c^2 \\
\mrT^{d}_{76}= 2 - \mrT^{c}_{39}\,c^2 \;\;&\;\;
\mrT^{d}_{77}= 3 - 8\,c^2 \;\;&\;\;
\mrT^{d}_{78}= 3 - 4\,c^2 \;\;&\;\;
\mrT^{d}_{79}= 5 - 4\,\mrT^{c}_{40}\,c^2 \\
\mrT^{d}_{80}= 5 - 24\,\mrT^{c}_{41}\,c^4 \;\;&\;\;
\mrT^{d}_{81}= 7 - 2\,\mrT^{c}_{42}\,c^2 \;\;&\;\;
\mrT^{d}_{82}= 11 - \mrT^{c}_{43}\,c^2 \;\;&\;\;
\mrT^{d}_{83}= 15 - \mrT^{c}_{44}\,c^2 \\
\mrT^{d}_{84}= 17 - 12\,c^2 \;\;&\;\;
\mrT^{d}_{85}= 25 - 24\,c^2 \;\;&\;\;
\mrT^{d}_{86}= 45 - 2\,\mrT^{c}_{45}\,c^2 \;\;&\;\;
\mrT^{d}_{87}= 1 - 6\,c^2 \\
\mrT^{d}_{88}= 1 - 4\,c^2 \;\;&\;\;
\mrT^{d}_{89}= 1 - 2\,\mrT^{c}_{46}\,c^2 \;\;&\;\;
\mrT^{d}_{90}= 1 + 4\,\mrT^{c}_{47}\,c^2 \;\;&\;\;
\mrT^{d}_{91}= 2 + \mrT^{c}_{48}\,c^2 \\
\mrT^{d}_{92}= 3 + 4\,c^2 \;\;&\;\;
\mrT^{d}_{93}= 3 - 2\,s^2 \;\;&\;\;
\mrT^{d}_{94}= 3 + \mrT^{c}_{49}\,c^2 \;\;&\;\;
\mrT^{d}_{95}= 3 - 2\,\mrT^{c}_{50}\,c^2 \\
\end{array}
\]
\[
\begin{array}{llll}
\mrT^{d}_{96}= 3 - \mrT^{c}_{51}\,c^2 \;\;&\;\;
\mrT^{d}_{97}= 1 - 3\,c^6 - \mrT^{c}_{52}\,s^2\,c^2 \;\;&\;\;
\mrT^{d}_{98}= 1 - 2\,\mrT^{c}_{36}\,c^2 \;\;&\;\;
\mrT^{d}_{99}= 1 + \mrT^{c}_{53}\,c^2 \\
\mrT^{d}_{100}= 1 - 4\,\mrT^{c}_{54}\,c^2 \;\;&\;\;
\mrT^{d}_{101}= 1 - \mrT^{c}_{55}\,c^2 \;\;&\;\;
\mrT^{d}_{102}= 1 - 2\,\mrT^{c}_{56}\,c^2 \;\;&\;\;
\mrT^{d}_{103}= 1 - 4\,\mrT^{c}_{57}\,c^2 \\
\mrT^{d}_{104}= 1 - 2\,\mrT^{c}_{58}\,c^2 \;\;&\;\;
\mrT^{d}_{105}= 1 - \mrT^{c}_{59}\,c^2 \;\;&\;\;
\mrT^{d}_{106}= 2 + c^2 \;\;&\;\;
\mrT^{d}_{107}= 3 + c^2 \\
\mrT^{d}_{108}= 3 - 4\,\mrT^{c}_{54}\,c^2 \;\;&\;\;
\mrT^{d}_{109}= 3 - \mrT^{c}_{60}\,c^2 \;\;&\;\;
\mrT^{d}_{110}= 4 + c^2 \;\;&\;\;
\mrT^{d}_{111}= 4 + \mrT^{c}_{61}\,c^2 \\
\mrT^{d}_{112}= 7 - 4\,c^2 \;\;&\;\;
\mrT^{d}_{113}= 1 + 8\,c^2 - 12\,s^2\,c^2 \;\;&\;\;
\mrT^{d}_{114}= 1 + 32\,c^4 - 4\,\mrT^{c}_{62}\,s^2\,c^2 \;\;&\;\;
\mrT^{d}_{115}= 1 - 4\,\mrT^{c}_{36}\,c^4 \\
\mrT^{d}_{116}= 1 - 4\,\mrT^{c}_{63}\,c^2 \;\;&\;\;
\mrT^{d}_{117}= 2 - 3\,\mrT^{c}_{64}\,c^2 \;\;&\;\;
\mrT^{d}_{118}= 3 - 16\,s^2\,c^2 \;\;&\;\;
\mrT^{d}_{119}= 4 - 3\,\mrT^{c}_{64}\,c^2 \\
\mrT^{d}_{120}= 4 - \mrT^{c}_{65}\,c^2 \;\;&\;\;
\mrT^{d}_{121}= 8 - \mrT^{c}_{66}\,c^2 \;\;&\;\;
\mrT^{d}_{122}= 1 - 2\,\mrT^{c}_{67}\,c^2 \;\;&\;\;
\mrT^{d}_{123}= 2 - 60\,s^2\,c^4 + 5\,\mrT^{c}_{68}\,c^2 \\
\mrT^{d}_{124}= 2 - 3\,\mrT^{c}_{69}\,c^2 \;\;&\;\;
\mrT^{d}_{125}= 4 - 3\,\mrT^{c}_{69}\,c^2 \;\;&\;\;
\mrT^{d}_{126}= 4 + \mrT^{c}_{70}\,c^2 \;\;&\;\;
\mrT^{d}_{127}= 8 + \mrT^{c}_{71}\,c^2 \\
\end{array}
\]
\[
\begin{array}{llll}
\mrT^{d}_{128}= 15 - 16\,s^2\,c^2 \;\;&\;\;
\mrT^{d}_{129}= 17 - 24\,s^2\,c^2 \;\;&\;\;
\mrT^{d}_{130}= 1 + 4\,\mrT^{c}_{72}\,c^2 \;\;&\;\;
\mrT^{d}_{131}= 1 + 4\,\mrT^{c}_{73}\,c^2 \\
\mrT^{d}_{132}= 3 + 2\,\mrT^{c}_{74}\,c^2 \;\;&\;\;
\mrT^{d}_{133}= 4 - \mrT^{c}_{75}\,c^2 \;\;&\;\;
\mrT^{d}_{134}= 5 + 4\,\mrT^{c}_{54}\,c^2 \;\;&\;\;
\mrT^{d}_{135}= 7 - 4\,\mrT^{c}_{76}\,c^2 \\
\mrT^{d}_{136}= 11 - 2\,\mrT^{c}_{75}\,c^2 \;\;&\;\;
\mrT^{d}_{137}= 15 - 4\,\mrT^{c}_{77}\,c^2 \;\;&\;\;
\mrT^{d}_{138}= 16 + \mrT^{c}_{78}\,c^2 \;\;&\;\;
\mrT^{d}_{139}= 17 - 4\,\mrT^{c}_{79}\,c^2 \\
\mrT^{d}_{140}= 19 - 4\,\mrT^{c}_{80}\,c^2 \;\;&\;\;
\mrT^{d}_{141}= 24 - \mrT^{c}_{81}\,c^2 \;\;&\;\;
\mrT^{d}_{142}= 55 - 12\,\mrT^{c}_{82}\,s^2 \;\;&\;\;
\mrT^{d}_{143}= 73 - 12\,\mrT^{c}_{83}\,c^2 \\
\mrT^{d}_{144}= 1 + 2\,c^2 \;\;&\;\;
\mrT^{d}_{145}= 1 - 12\,c^4 \;\;&\;\;
\mrT^{d}_{146}= 1 - \mrT^{c}_{84}\,c^2 \;\;&\;\;
\mrT^{d}_{147}= 1 - 4\,\mrT^{c}_{85}\,c^2 \\
\mrT^{d}_{148}= 1 - 4\,\mrT^{c}_{86}\,c^2 \;\;&\;\;
\mrT^{d}_{149}= 3 + 10\,c^2 \;\;&\;\;
\mrT^{d}_{150}= 3 - 4\,\mrT^{c}_{87}\,c^2 \;\;&\;\;
\mrT^{d}_{151}= 3 - 2\,\mrT^{c}_{88}\,c^2 \\
\mrT^{d}_{152}= 3 - \mrT^{c}_{89}\,c^2 \;\;&\;\;
\mrT^{d}_{153}= 4 - 3\,s^2 \;\;&\;\;
\mrT^{d}_{154}= 5 - 2\,\mrT^{c}_{31}\,c^2 \;\;&\;\;
\mrT^{d}_{155}= 7 - 4\,\mrT^{c}_{90}\,c^2 \\
\mrT^{d}_{156}= 1 + 2\,\mrT^{c}_{91}\,c^2 \;\;&\;\;
\mrT^{d}_{157}= 1 + 2\,\mrT^{c}_{92}\,c^2 \;\;&\;\;
\mrT^{d}_{158}= 2 + \mrT^{c}_{92}\,c^2 \;\;&\;\;
\mrT^{d}_{159}= 4 - 3\,c^2 \\
\end{array}
\]
\[
\begin{array}{llll}
\mrT^{d}_{160}= 4 + 3\,c^2 \;\;&\;\;
\mrT^{d}_{161}= 5 - 2\,c^2 \;\;&\;\;
\mrT^{d}_{162}= 5 + 6\,c^2 \;\;&\;\;
\mrT^{d}_{163}= 6 - c^2 \\
\mrT^{d}_{164}= 8 - 3\,c^2 \;\;&\;\;
\mrT^{d}_{165}= 8 + \mrT^{c}_{93}\,c^2 \;\;&\;\;
\mrT^{d}_{166}= 9 - 2\,c^2 \;\;&\;\;
\mrT^{d}_{167}= 10 - 3\,c^2 \\
\mrT^{d}_{168}= 10 - \mrT^{c}_{94}\,c^2 \;\;&\;\;
\mrT^{d}_{169}= 10 - \mrT^{c}_{95}\,c^2 \;\;&\;\;
\mrT^{d}_{170}= 10 - \mrT^{c}_{96}\,c^2 \;\;&\;\;
\mrT^{d}_{171}= 10 + \mrT^{c}_{97}\,c^2 \\
\mrT^{d}_{172}= 10 - \mrT^{c}_{98}\,c^2 \;\;&\;\;
\mrT^{d}_{173}= 10 - \mrT^{c}_{99}\,c^2 \;\;&\;\;
\mrT^{d}_{174}= 10 - \mrT^{c}_{100}\,c^2 \;\;&\;\;
\mrT^{d}_{175}= 10 - \mrT^{c}_{101}\,c^2 \\
\mrT^{d}_{176}= 13 - 24\,s^2\,c^2 + 2\,\mrT^{c}_{102}\,c^4 \;\;&\;\;
\mrT^{d}_{177}= 14 - 3\,c^2 \;\;&\;\;
\mrT^{d}_{178}= 16 - 15\,c^2 \;\;&\;\;
\mrT^{d}_{179}= 30 - \mrT^{c}_{103}\,c^2 \\
\mrT^{d}_{180}= 32 - 21\,c^2 \;\;&\;\;
\mrT^{d}_{181}= 118 + \mrT^{c}_{104}\,c^2 \;\;&\;\;
\mrT^{d}_{182}= 122 + \mrT^{c}_{105}\,c^2 \;\;&\;\;
\mrT^{d}_{183}= 1 + 9\,c^2 \\
\mrT^{d}_{184}= 1 + 8\,c^4 \;\;&\;\;
\mrT^{d}_{185}= 1 + 5\,\mrT^{c}_{106}\,c^2 \;\;&\;\;
\mrT^{d}_{186}= 1 + 4\,\mrT^{c}_{107}\,c^2 \;\;&\;\;
\mrT^{d}_{187}= 1 + 2\,\mrT^{c}_{108}\,c^2 \\
\mrT^{d}_{188}= 3 + 158\,c^2 \;\;&\;\;
\mrT^{d}_{189}= 10 - \mrT^{c}_{109}\,c^2 \;\;&\;\;
\mrT^{d}_{190}= 12 + 13\,c^2 \;\;&\;\;
\mrT^{d}_{191}= 12 - \mrT^{c}_{110}\,c^2 \\
\end{array}
\]
\[
\begin{array}{llll}
\mrT^{d}_{192}= 23 - 146\,c^2 \;\;&\;\;
\mrT^{d}_{193}= 23 - 48\,c^2 \;\;&\;\;
\mrT^{d}_{194}= 49 + 8\,c^2 \;\;&\;\;
\mrT^{d}_{195}= c^2 - s^2 \\
\mrT^{d}_{196}= c^4 - \mrT^{c}_{111}\,s^2 \;\;&\;\;
\mrT^{d}_{197}= 1 - 84\,c^2 \;\;&\;\;
\mrT^{d}_{198}= 1 - 12\,\mrT^{c}_{112}\,c^2 \;\;&\;\;
\mrT^{d}_{199}= 1 - 4\,\mrT^{c}_{113}\,c^2 \\
\mrT^{d}_{200}= 1 + \mrT^{c}_{114}\,c^2 \;\;&\;\;
\mrT^{d}_{201}= 1 + 4\,\mrT^{c}_{115}\,c^2 \;\;&\;\;
\mrT^{d}_{202}= 1 - 4\,\mrT^{c}_{116}\,c^2 \;\;&\;\;
\mrT^{d}_{203}= 1 - 2\,\mrT^{c}_{117}\,c^2 \\
\mrT^{d}_{204}= 1 - 2\,\mrT^{c}_{118}\,c^2 \;\;&\;\;
\mrT^{d}_{205}= 1 - 2\,\mrT^{c}_{119}\,c^2 \;\;&\;\;
\mrT^{d}_{206}= 1 - \mrT^{c}_{120}\,c^2 \;\;&\;\;
\mrT^{d}_{207}= 2 - \mrT^{c}_{121}\,c^2 \\
\mrT^{d}_{208}= 3 + 19\,\mrT^{c}_{122}\,c^2 \;\;&\;\;
\mrT^{d}_{209}= 23 + 2\,\mrT^{c}_{123}\,c^2 \;\;&\;\;
\mrT^{d}_{210}= 25 + 108\,c^2 \;\;&\;\;
\mrT^{d}_{211}= 4\,c^2 + s^2 \\
\mrT^{d}_{212}= 13\,c^2 + s^2 \;\;&\;\;
\mrT^{d}_{213}= 29\,c^2 + s^2 \;\;&\;\;
\mrT^{d}_{214}= 48\,c^2 - \mrT^{c}_{124}\,s^2 \;\;&\;\;
\mrT^{d}_{215}= 59\,c^2 + 23\,s^2 \\
\mrT^{d}_{216}= 61\,c^2 + s^2 \;\;&\;\;
\mrT^{d}_{217}= 77\,c^2 + s^2 \;\;&\;\;
\mrT^{d}_{218}= 83\,c^2 - s^2 \;\;&\;\;
\mrT^{d}_{219}= 85\,c^2 + s^2 \\
\mrT^{d}_{220}= 93\,c^2 + s^2 \;\;&\;\;
\mrT^{d}_{221}= 109\,c^2 + s^2 \;\;&\;\;
\mrT^{d}_{222}= 115\,c^2 - s^2 \;\;&\;\;
\mrT^{d}_{223}= 133\,c^2 + s^2 \\
\end{array}
\]
\[
\begin{array}{llll}
\mrT^{d}_{224}= 157\,c^2 + s^2 \;\;&\;\;
\mrT^{d}_{225}= 167\,c^2 + 3\,s^2 \;\;&\;\;
\mrT^{d}_{226}= 173\,c^2 + s^2 \;\;&\;\;
\mrT^{d}_{227}= 35\,c^4 - \mrT^{c}_{125}\,s^2 \\
\mrT^{d}_{228}= 35\,c^6 - \mrT^{c}_{126}\,s^2 \;\;&\;\;
\mrT^{d}_{229}= 1 - 3\,c^2 \;\;&\;\;
\mrT^{d}_{230}= 1 + c^2 \;\;&\;\;
\mrT^{d}_{231}= 1 - 8\,\mrT^{c}_{28}\,c^2 \\
\mrT^{d}_{232}= 1 - 2\,\mrT^{c}_{28}\,c^2 \;\;&\;\;
\mrT^{d}_{233}= 1 - 2\,\mrT^{c}_{31}\,c^2 \;\;&\;\;
\mrT^{d}_{234}= 1 - \mrT^{c}_{127}\,c^2 \;\;&\;\;
\mrT^{d}_{235}= 1 - \mrT^{c}_{128}\,c^2 \\
\mrT^{d}_{236}= 3 - 131\,c^2 \;\;&\;\;
\mrT^{d}_{237}= 3 - 107\,c^2 \;\;&\;\;
\mrT^{d}_{238}= 3 - 83\,c^2 \;\;&\;\;
\mrT^{d}_{239}= 3 - 41\,c^2 \\
\mrT^{d}_{240}= 3 - 11\,c^2 \;\;&\;\;
\mrT^{d}_{241}= 3 - \mrT^{c}_{129}\,c^2 \;\;&\;\;
\mrT^{d}_{242}= 3 + \mrT^{c}_{130}\,c^2 \;\;&\;\;
\mrT^{d}_{243}= 3 - \mrT^{c}_{131}\,c^2 \\
\mrT^{d}_{244}= 3 + \mrT^{c}_{132}\,c^2 \;\;&\;\;
\mrT^{d}_{245}= 3 - \mrT^{c}_{133}\,c^2 \;\;&\;\;
\mrT^{d}_{246}= 3 - \mrT^{c}_{134}\,c^2 \;\;&\;\;
\mrT^{d}_{247}= 3 - \mrT^{c}_{135}\,c^2 \\
\mrT^{d}_{248}= 3 - \mrT^{c}_{136}\,c^2 \;\;&\;\;
\mrT^{d}_{249}= 3 - \mrT^{c}_{137}\,c^2 \;\;&\;\;
\mrT^{d}_{250}= 6 + \mrT^{c}_{138}\,c^2 \;\;&\;\;
\mrT^{d}_{251}= 9 + 71\,c^2 \\
\mrT^{d}_{252}= 15 - 23\,c^2 \;\;&\;\;
\mrT^{d}_{253}= 15 - \mrT^{c}_{139}\,c^2 \;\;&\;\;
\mrT^{d}_{254}= 15 + \mrT^{c}_{140}\,c^2 \;\;&\;\;
\mrT^{d}_{255}= 21 - \mrT^{c}_{141}\,c^2 \\
\end{array}
\]
\[
\begin{array}{llll}
\mrT^{d}_{256}= 21 - \mrT^{c}_{142}\,c^2 \;\;&\;\;
\mrT^{d}_{257}= 31 - 47\,c^2 \;\;&\;\;
\mrT^{d}_{258}= 24\,c^2 - \mrT^{c}_{143}\,s^2 \;\;&\;\;
\mrT^{d}_{259}= 24\,c^2 - \mrT^{c}_{144}\,s^2 \\
\mrT^{d}_{260}= 1 + 22\,c^2 \;\;&\;\;
\mrT^{d}_{261}= 1 - 2\,\mrT^{c}_{145}\,c^2 \;\;&\;\;
\mrT^{d}_{262}= 1 - \mrT^{c}_{146}\,c^2 \;\;&\;\;
\mrT^{d}_{263}= 1 - \mrT^{c}_{147}\,c^2 \\
\mrT^{d}_{264}= 2 + \mrT^{c}_{148}\,c^2 \;\;&\;\;
\mrT^{d}_{265}= 3 - 4\,\mrT^{c}_{149}\,c^2 \;\;&\;\;
\mrT^{d}_{266}= 3 - 2\,\mrT^{c}_{150}\,c^2 \;\;&\;\;
\mrT^{d}_{267}= 3 - \mrT^{c}_{151}\,c^2 \\
\mrT^{d}_{268}= 7 - 40\,c^2 \;\;&\;\;
\mrT^{d}_{269}= 151 - 150\,s^2 \;\;&\;\;
\mrT^{d}_{270}= 4\,c^2 + 5\,s^2 \;\;&\;\;
\mrT^{d}_{271}= 1 - 44\,c^4 \\
\mrT^{d}_{272}= 1 - \mrT^{c}_{68}\,c^2 \;\;&\;\;
\mrT^{d}_{273}= 1 + 4\,\mrT^{c}_{152}\,c^2 \;\;&\;\;
\mrT^{d}_{274}= 1 - 2\,\mrT^{c}_{153}\,c^2 \;\;&\;\;
\mrT^{d}_{275}= 2 - 11\,c^2 \\
\mrT^{d}_{276}= 3 + 130\,c^2 \;\;&\;\;
\mrT^{d}_{277}= 3 + 4\,\mrT^{c}_{154}\,c^2 \;\;&\;\;
\mrT^{d}_{278}= 3 + 2\,\mrT^{c}_{155}\,c^2 \;\;&\;\;
\mrT^{d}_{279}= 3 + \mrT^{c}_{156}\,c^2 \\
\mrT^{d}_{280}= 7 - 2\,\mrT^{c}_{157}\,c^2 \;\;&\;\;
\mrT^{d}_{281}= 8 - 5\,\mrT^{c}_{158}\,c^2 \;\;&\;\;
\mrT^{d}_{282}= 9 + 4\,\mrT^{c}_{159}\,c^2 \;\;&\;\;
\mrT^{d}_{283}= 9 - \mrT^{c}_{160}\,c^2 \\
\mrT^{d}_{284}= 10 - \mrT^{c}_{161}\,c^2 \;\;&\;\;
\mrT^{d}_{285}= 11 - 2\,\mrT^{c}_{162}\,c^2 \;\;&\;\;
\mrT^{d}_{286}= c^2 - 2\,s^2 \;\;&\;\;
\mrT^{d}_{287}= 3\,c^2 + s^2 \\
\end{array}
\]
\[
\begin{array}{llll}
\mrT^{d}_{288}= 6\,c^2 - s^2 \;\;&\;\;
\mrT^{d}_{289}= 13\,c^2 - s^2 \;\;&\;\;
\mrT^{d}_{290}= 17\,c^2 - s^2 \;\;&\;\;
\mrT^{d}_{291}= 21\,c^2 - s^2 \\
\mrT^{d}_{292}= c^6 - \mrT^{c}_{163}\,s^2 \;\;&\;\;
\mrT^{d}_{293}= 1 + 40\,c^2 \;\;&\;\;
\mrT^{d}_{294}= 1 + 4\,\mrT^{c}_{164}\,c^2 \;\;&\;\;
\mrT^{d}_{295}= 1 + \mrT^{c}_{165}\,c^2 \\
\mrT^{d}_{296}= 5 - 56\,c^2 \;\;&\;\;
\mrT^{d}_{297}= 5 - 8\,c^2 \;\;&\;\;
\mrT^{d}_{298}= 5 + 72\,c^2 \;\;&\;\;
\mrT^{d}_{299}= 5 + 88\,c^2 \\
\mrT^{d}_{300}= 5 + 4\,\mrT^{c}_{166}\,c^2 \;\;&\;\;
\mrT^{d}_{301}= 5 - 8\,\mrT^{c}_{167}\,c^2 \;\;&\;\;
\mrT^{d}_{302}= 5 + 2\,\mrT^{c}_{168}\,c^2 \;\;&\;\;
\mrT^{d}_{303}= 5 - 4\,\mrT^{c}_{169}\,c^2 \\
\mrT^{d}_{304}= 5 + \mrT^{c}_{170}\,c^2 \;\;&\;\;
\mrT^{d}_{305}= 5 + \mrT^{c}_{171}\,c^2 \;\;&\;\;
\mrT^{d}_{306}= 7 - 50\,c^2 \;\;&\;\;
\mrT^{d}_{307}= 7 - 18\,c^2 \\
\mrT^{d}_{308}= 7 - 2\,c^2 \;\;&\;\;
\mrT^{d}_{309}= 9 - 8\,s^2 \;\;&\;\;
\mrT^{d}_{310}= 11 + 4\,\mrT^{c}_{172}\,c^2 \;\;&\;\;
\mrT^{d}_{311}= s^4 - \mrT^{c}_{173}\,c^2 \\
\mrT^{d}_{312}= s^4 - \mrT^{c}_{174}\,c^2 \;\;&\;\;
\mrT^{d}_{313}= 7\,s^4 + \mrT^{c}_{175}\,c^2 \;\;&\;\;
\mrT^{d}_{314}= 1 - 7\,c^2 \;\;&\;\;
\mrT^{d}_{315}= 1 + 14\,c^2 \\
\mrT^{d}_{316}= 1 + 16\,c^2 \;\;&\;\;
\mrT^{d}_{317}= 1 + 2\,\mrT^{c}_{176}\,c^2 \;\;&\;\;
\mrT^{d}_{318}= 5 - 28\,c^2 \;\;&\;\;
\mrT^{d}_{319}= 5 - 12\,c^2 \\
\end{array}
\]
\[
\begin{array}{llll}
\mrT^{d}_{320}= 5 + 14\,c^2 \;\;&\;\;
\mrT^{d}_{321}= 5 + 66\,c^2 \;\;&\;\;
\mrT^{d}_{322}= 5 - 4\,\mrT^{c}_{86}\,c^2 \;\;&\;\;
\mrT^{d}_{323}= 5 + 8\,\mrT^{c}_{88}\,c^2 \\
\mrT^{d}_{324}= 5 + 4\,\mrT^{c}_{177}\,c^2 \;\;&\;\;
\mrT^{d}_{325}= 5 + 4\,\mrT^{c}_{178}\,c^2 \;\;&\;\;
\mrT^{d}_{326}= 5 + 2\,\mrT^{c}_{179}\,c^2 \;\;&\;\;
\mrT^{d}_{327}= 5 + 2\,\mrT^{c}_{180}\,c^2 \\
\mrT^{d}_{328}= 5 - 4\,\mrT^{c}_{181}\,c^2 \;\;&\;\;
\mrT^{d}_{329}= 5 + 2\,\mrT^{c}_{182}\,c^2 \;\;&\;\;
\mrT^{d}_{330}= 5 + 2\,\mrT^{c}_{183}\,c^2 \;\;&\;\;
\mrT^{d}_{331}= 9\,c^2 + s^2 \\
\mrT^{d}_{332}= s^4 - \mrT^{c}_{184}\,c^2 \;\;&\;\;
\mrT^{d}_{333}= 5\,s^4 - \mrT^{c}_{185}\,c^2 \;\;&\;\;
\mrT^{d}_{334}= 1 - 20\,c^2 \;\;&\;\;
\mrT^{d}_{335}= 1 + 5\,c^2 \\
\mrT^{d}_{336}= 1 - \mrT^{c}_{4}\,c^2 \;\;&\;\;
\mrT^{d}_{337}= 1 - \mrT^{c}_{178}\,c^2 \;\;&\;\;
\mrT^{d}_{338}= 1 - 4\,\mrT^{c}_{186}\,c^2 \;\;&\;\;
\mrT^{d}_{339}= 1 - 2\,\mrT^{c}_{187}\,c^2 \\
\mrT^{d}_{340}= 1 - \mrT^{c}_{188}\,c^2 \;\;&\;\;
\mrT^{d}_{341}= 2 - 7\,\mrT^{c}_{54}\,c^2 \;\;&\;\;
\mrT^{d}_{342}= 3 + 2\,s^2 \;\;&\;\;
\mrT^{d}_{343}= 3 - \mrT^{c}_{189}\,c^2 \\
\mrT^{d}_{344}= 3 - 4\,\mrT^{c}_{190}\,c^2 \;\;&\;\;
\mrT^{d}_{345}= 4 - 9\,c^2 \;\;&\;\;
\mrT^{d}_{346}= 4 - \mrT^{c}_{191}\,c^2 \;\;&\;\;
\mrT^{d}_{347}= 5 - 13\,c^2 \\
\mrT^{d}_{348}= 5 + c^2 \;\;&\;\;
\mrT^{d}_{349}= 6 - 13\,c^2 \;\;&\;\;
\mrT^{d}_{350}= 6 - \mrT^{c}_{192}\,c^2 \;\;&\;\;
\mrT^{d}_{351}= 7 - \mrT^{c}_{193}\,c^2 \\
\end{array}
\]
\[
\begin{array}{llll}
\mrT^{d}_{352}= 7 - \mrT^{c}_{194}\,c^2 \;\;&\;\;
\mrT^{d}_{353}= 8 - \mrT^{c}_{195}\,c^2 \;\;&\;\;
\mrT^{d}_{354}= 9 - 35\,c^2 \;\;&\;\;
\mrT^{d}_{355}= 11 - 10\,c^2 \\
\mrT^{d}_{356}= 14 - \mrT^{c}_{196}\,c^2 \;\;&\;\;
\mrT^{d}_{357}= 1 + 68\,c^2 \;\;&\;\;
\mrT^{d}_{358}= 1 - \mrT^{c}_{87}\,c^2 \;\;&\;\;
\mrT^{d}_{359}= 1 - \mrT^{c}_{88}\,c^2 \\
\mrT^{d}_{360}= 1 - 4\,\mrT^{c}_{197}\,c^2 \;\;&\;\;
\mrT^{d}_{361}= 1 - \mrT^{c}_{198}\,c^2 \;\;&\;\;
\mrT^{d}_{362}= 4 - c^2 \;\;&\;\;
\mrT^{d}_{363}= 9 - 16\,c^2 \\
\mrT^{d}_{364}= 13 - 32\,c^2 \;\;&\;\;
\mrT^{d}_{365}= 13 - 8\,c^2 \;\;&\;\;
\mrT^{d}_{366}= 25 + 32\,c^4 \;\;&\;\;
\mrT^{d}_{367}= c^2 - \mrT^{c}_{199}\,s^2 \\
\mrT^{d}_{368}= 1 + 17\,c^2 \;\;&\;\;
\mrT^{d}_{369}= 1 + 8\,\mrT^{c}_{11}\,c^2 \;\;&\;\;
\mrT^{d}_{370}= 1 - 8\,\mrT^{c}_{200}\,c^2 \;\;&\;\;
\mrT^{d}_{371}= 1 + 8\,\mrT^{c}_{201}\,c^2 \\
\mrT^{d}_{372}= 1 - 4\,\mrT^{c}_{202}\,c^2 \;\;&\;\;
\mrT^{d}_{373}= 1 + 4\,\mrT^{c}_{203}\,c^2 \;\;&\;\;
\mrT^{d}_{374}= 1 + 2\,\mrT^{c}_{204}\,c^2 \;\;&\;\;
\mrT^{d}_{375}= 1 + 2\,\mrT^{c}_{205}\,c^2 \\
\mrT^{d}_{376}= 1 + 2\,\mrT^{c}_{206}\,c^2 \;\;&\;\;
\mrT^{d}_{377}= 1 + 2\,\mrT^{c}_{207}\,c^2 \;\;&\;\;
\mrT^{d}_{378}= 1 + 2\,\mrT^{c}_{208}\,c^2 \;\;&\;\;
\mrT^{d}_{379}= 1 - \mrT^{c}_{209}\,c^2 \\
\mrT^{d}_{380}= 1 + \mrT^{c}_{210}\,c^2 \;\;&\;\;
\mrT^{d}_{381}= 1 + \mrT^{c}_{211}\,c^2 \;\;&\;\;
\mrT^{d}_{382}= 1 + \mrT^{c}_{212}\,c^2 \;\;&\;\;
\mrT^{d}_{383}= 1 + \mrT^{c}_{213}\,c^2 \\
\end{array}
\]
\[
\begin{array}{llll}
\mrT^{d}_{384}= 3 + 4\,c^4 \;\;&\;\;
\mrT^{d}_{385}= 9 - 20\,c^2 \;\;&\;\;
\mrT^{d}_{386}= 17 - 8\,c^2 \;\;&\;\;
\mrT^{d}_{387}= 23 - 68\,c^2 \\
\mrT^{d}_{388}= 23 - 4\,\mrT^{c}_{214}\,c^2 \;\;&\;\;
\mrT^{d}_{389}= 9\,c^2 - 8\,s^2 \;\;&\;\;
\mrT^{d}_{390}= 8\,c^4 - \mrT^{c}_{215}\,s^2 & \\
\end{array}
\]

\normalsize

\vspace{0.8cm}
\bei
\item[\fbox{$\mrU\,$}] 
\eei

\scriptsize
\[
\begin{array}{llll}
\mrU_{0}= 2\,\vtq + \vbq + \vle \;\;&\;\;
\mrU_{1}= 3 - \vtqs \;\;&\;\;
\mrU_{2}= 3 + \vtqs \;\;&\;\;
\mrU_{3}= 3 - \vbqs \\
\mrU_{4}= 3 + \vbqs \;\;&\;\;
\mrU_{5}= 27 - 16\,\vtq \;\;&\;\;
\mrU_{6}= 27 - 8\,\vbq \;\;&\;\;
\mrU_{7}= 21 - \vtqs \\
\mrU_{8}= 21 - \vbqs \;\;&\;\;
\mrU_{9}= 27 + \vtqs \;\;&\;\;
\mrU_{10}= 27 + \vbqs & \\
\end{array}
\]

\normalsize

\vspace{0.8cm}
\bei
\item[\fbox{$\mrV\,$}] 
\eei

\scriptsize
\bqas
\mrV_{0} &=& 2 - \xphs 
\qquad
\mrV_{1} = 3 - \xphs 
\qquad
\mrV_{2} = 6 - \xphs 
\nl      
\mrV_{3} &=& 4\,\xpts - \xphs 
\qquad
\mrV_{4} = 4\,\xpbs - \xphs 
\qquad
\mrV_{5} = 8 + \xphs 
\nl      
\mrV_{6} &=& \xpbs + \xpts 
\qquad
\mrV_{7} = 6\,\xpbs + 6\,\xpts + 7\,\xphs 
\nl      
\mrV_{8} &=& - ( 2\,\xpts - \xphs )\,\frac{1}{\xphs}
\qquad
\mrV_{9} = \frac{1}{\xphs}
\nl      
\mrV_{10} &=& ( 2\,\xpts + \xphs )\,\frac{1}{\xphs}
\qquad
\mrV_{11} = - ( 2\,\xpbs - \xphs )\,\frac{1}{\xphs}
\nl      
\mrV_{12} &=& ( 2\,\xpbs + \xphs )\,\frac{1}{\xphs}
\qquad
\mrV_{13} = ( \xpbs + \xpts )\,\frac{1}{\xphs}
\nl      
\mrV_{14} &=& 2 - \xpbs - \xpts 
\qquad
\mrV_{15} = 2 - \xpbs + \xpts 
\qquad
\mrV_{16} = 2 + \xpbs - \xpts 
\nl      
\mrV_{17} &=& \Bigl[ 6 - ( 1 - 2\,\xpbs + \xpts )\,\xpts - ( 1 + \xpbs )\,\xpbs \Bigr]
\qquad      
\mrV_{18} = 1 + \xpbs - \xpts 
\nl      
\mrV_{19} &=& \Bigl[ 1 + ( 2 + \xpbs )\,\xpbs - ( 2 + 2\,\xpbs - \xpts )\,\xpts \Bigr]
\qquad      
\mrV_{20} = \Bigl[ 1 - \xptq + ( 2 + \xpbs )\,\xpbs \Bigr]
\nl      
\mrV_{21} &=& \Bigl[ 2 - 2\,\xpbs + 2\,\xpts - ( 1 + \xpbs - \xpts )\,\xphs \Bigr]
\qquad      
\mrV_{22} = 2 + 2\,\xpbs - 2\,\xpts - \xphs 
\nl
\mrV_{23} &=& 2 + 2\,\xpbs + 2\,\xpts - \xphs 
\qquad      
\mrV_{24} = \Bigl[ 1 - \xptq - ( 2 - \xpbs )\,\xpbs \Bigr]
\nl      
\mrV_{25} &=& \Bigl[ 2 + 2\,\xpbq + \xphs\,\xpbs - 2\,( 1 + \xpbs )\,\xpts \Bigr]
\qquad      
\mrV_{26} = \Bigl\{ \Bigl[ 2 - 2\,\xpbs + ( 1 - \xpbs )\,\xphs \Bigr]\,\xpbs 
              + ( 4 + 2\,\xpbs + \xphs\,\xpbs )\,\xpts \Bigr\}
\nl      
\mrV_{27} &=& \Bigl[ 4\,\xpts + ( 4 - \xphs )\,\xpbs \Bigr]
\qquad      
\mrV_{28} = \Bigl[ 4 - ( 4 - \xphs )\,\xphs \Bigr]
\nl      
\mrV_{29} &=& \Bigl[ 32 - ( 81 - 8\,\xphs )\,\xphs \Bigr]
\qquad      
\mrV_{30} = \Bigl[ 61 - 2\,( 4 - \xphs )\,\xphs \Bigr]
\nl      
\mrV_{31} &=& \Bigl[ 64 + ( 50 - 41\,\xphs )\,\xphs \Bigr]
\qquad      
\mrV_{32} = \Bigl[ 96 + ( 148 - 61\,\xphs )\,\xphs \Bigr]
\nl      
\mrV_{33} &=& \Bigl[ 100 - ( 50 + 49\,\xphs )\,\xphs \Bigr]
\qquad      
\mrV_{34} = \Bigl\{ 128 + \Bigl[ 36 - ( 132 - 41\,\xphs )\,\xphs \Bigr]\,\xphs \Bigr\}
\nl      
\mrV_{35} &=& 146 + \xphs 
\qquad
\mrV_{36} = 4 - \xphs 
\qquad
\mrV_{37} = 4 + \xphq 
\nl      
\mrV_{38} &=& 2 + \xphs 
\qquad
\mrV_{39} = 5 - \xphs 
\qquad
\mrV_{40} = 10 - \xphs 
\nl      
\mrV_{41} &=& 9 - 4\,\xphs 
\qquad
\mrV_{42} = 14 - \xphs 
\nl      
\mrV_{43} &=& \Bigl[ 14 + ( 8 + \xphs )\,\xphs \Bigr]
\qquad      
\mrV_{44} = \Bigl[ 18 - ( 10 - \xphs )\,\xphs \Bigr]
\nl      
\mrV_{45} &=& 22 + \xphs 
\qquad      
\mrV_{46} = \Bigl[ 28 - 9\,( 2 + \xphs )\,\xphs \Bigr]
\nl      
\mrV_{47} &=& 3 - \xpbs + \xpts 
\qquad
\mrV_{48} = 3 + \xpbs - \xpts 
\nl      
\mrV_{49} &=& \Bigl[ 5 - ( 6 + 2\,\xpbs - \xpts )\,\xpts + ( 10 + \xpbs )\,\xpbs \Bigr]
\qquad      
\mrV_{50} = \Bigl[ 1 - ( 2 - \xpbs )\,\xpbs 
              - ( 2 + 2\,\xpbs - \xpts )\,\xpts \Bigr]\,\frac{1}{\xphs}
\nl      
\mrV_{51} &=& ( 1 - \xpbs + \xpts )\,\frac{1}{\xphs}
\qquad      
\mrV_{52} = ( 1 + \xpbs - \xpts )\,\frac{1}{\xphs}
\nl      
\mrV_{53} &=& 1 - \xpbs + \xpts 
\qquad      
\mrV_{54} = \Bigl[ 1 - \xptq + ( 4 + \xpbs )\,\xpbs \Bigr]
\nl      
\mrV_{55} &=& \xpbs - \xpts 
\qquad      
\mrV_{56} = ( 1 + \xpbs - \xpts + \xphs )\,\frac{1}{\xphs}
\nl      
\mrV_{57} &=& \Bigl[ 3 + ( 2 - 2\,\xpbs - \xpts )\,\xpts - 3\,( 2 - \xpbs )\,\xpbs \Bigr]
\qquad
\mrV_{58} =  3 - 3\,\xpbs - \xpts 
\nl
\mrV_{59} &=&  3 + 5\,\xpbs - \xpts 
\qquad      
\mrV_{60} = \Bigl[ 5 - ( 6 + 6\,\xpbs - \xpts )\,\xpts + ( 14 + 5\,\xpbs )\,\xpbs \Bigr]
\nl      
\mrV_{61} &=& \Bigl[ 1 - 2\,\xphs\,\xpbs - ( 2 - \xpbs )\,\xpbs 
               - ( 2 + 2\,\xpbs - \xpts )\,\xpts \Bigr]\,\frac{1}{\xphs}
\qquad      
\mrV_{62} = \Bigl\{ \Bigl[ 1 - \xpbs + ( 1 - \xpbs )\,\xphs \Bigr]\,\xpbs 
              + ( 1 + 2\,\xpbs - \xpts + \xphs\,\xpbs )\,\xpts \Bigr\}\,\frac{1}{\xphs}
\nl      
\mrV_{63} &=& - \Bigl[ 2\,\xpts - ( 2 + \xphs )\,\xpbs \Bigr]\,\frac{1}{\xphs}
\qquad      
\mrV_{64} = \Bigl[ 2\,\xpts + ( 2 - \xphs )\,\xpbs \Bigr]\,\frac{1}{\xphs}
\nl      
\mrV_{65} &=& \Bigl[ 3\,( 1 - \xpbs )\,\xpbs + ( 3 + 2\,\xpbs + \xpts )\,\xpts \Bigr]
\qquad      
\mrV_{66} = - ( 2 - \xphs )\,\frac{1}{\xphs}
\nl      
\mrV_{67} &=& \Bigl[ 8 - ( 4 - \xphs )\,\xphs \Bigr]
\qquad      
\mrV_{68} = 1 - \xphs 
\nl
\mrV_{69} &=& 2 - 5\,\xphs 
\qquad        
\mrV_{70} = \Bigl[ 16 - 5\,( 4 - \xphs )\,\xphs \Bigr]
\nl        
\mrV_{71} &=& \Bigl[ 16 - ( 18 - 5\,\xphs )\,\xphs \Bigr]
\qquad        
\mrV_{72} = \Bigl[ 24 - ( 2 + 5\,\xphs )\,\xphs \Bigr]
\nl        
\mrV_{73} &=& \Bigl\{ 32 - \Bigl[ 52 - ( 28 - 5\,\xphs )\,\xphs \Bigr]\,\xphs \Bigr\}
\qquad        
\mrV_{74} = \Bigl[ 48 - ( 32 - 5\,\xphs )\,\xphs \Bigr]
\nl        
\mrV_{75} &=& \Bigl[ 88 - ( 6 - 5\,\xphs )\,\xphs \Bigr]
\qquad        
\mrV_{76} = \Bigl\{ 160 - \Bigl[ 168 - ( 52 - 5\,\xphs )\,\xphs \Bigr]\,\xphs \Bigr\}
\nl        
\mrV_{77} &=& \Bigl[ 2 - ( 4 - \xphs )\,\xphs \Bigr]
\qquad        
\mrV_{78} = \Bigl[ 2 - ( 3 - \xphs )\,\xphs \Bigr]
\nl        
\mrV_{79} &=& \Bigl\{ 4 - \Bigl[ 8 - ( 5 - \xphs )\,\xphs \Bigr]\,\xphs \Bigr\}
\qquad        
\mrV_{80} = \Bigl[ 6 - ( 4 - \xphs )\,\xphs \Bigr]
\nl        
\mrV_{81} &=& \Bigl[ 8 - ( 12 - \xphs )\,\xphs \Bigr]
\qquad        
\mrV_{82} = 12 - \xphs 
\nl
\mrV_{83} &=& 16 - 3\,\xphq 
\qquad        
\mrV_{84} = \Bigl\{ 20 - \Bigl[ 24 - ( 8 - \xphs )\,\xphs \Bigr]\,\xphs \Bigr\}
\nl        
\mrV_{85} &=& \Bigl[ 48 - ( 12 - \xphs )\,\xphs \Bigr]
\qquad        
\mrV_{86} = 64 + 3\,\xphq 
\qquad
\mrV_{87} = 1 + 2\,\xphs 
\qquad
\mrV_{88} = 7 - 4\,\xphs 
\nl        
\mrV_{89} &=& 12 + 7\,\xphs 
\qquad
\mrV_{90} = 48 - 25\,\xphs 
\qquad
\mrV_{91} = 68 + 9\,\xphs 
\nl        
\mrV_{92} &=& \frac{1}{\xphq}
\qquad
\mrV_{93} = ( 9 + 2\,\xphs )\,\frac{1}{\xphs}
\nl        
\mrV_{94} &=& 3 + 2\,\xphs 
\qquad
\mrV_{95} = 4 + \xphs 
\eqas

\normalsize

\vspace{0.8cm}
\bei
\item[\fbox{$\mrW\,$}] 
\eei

\scriptsize
\[
\begin{array}{llll}
\mrW_{0}= 1 - 4\,\frac{\xpts}{\laz} \;\;&\;\;
\mrW_{1}= 1 - 4\,\frac{\xpbs}{\laz} \;\;&\;\;
\mrW_{2}= 1 + \frac{\xphs}{\laz} \;\;&\;\;
\mrW_{3}= 1 + 2\,\frac{\xphs}{\laz} \\
\mrW_{4}= 2 + \frac{\xphs}{\laz} \;\;&\;\;
\mrW_{5}= 3 + 2\,\frac{\xphs}{\laz} \;\;&\;\;
\mrW_{6}= 1 - 6\,\frac{\xpts}{\lzz} \;\;&\;\;
\mrW_{7}= 1 - 6\,\frac{\xpbs}{\lzz} \\
\mrW_{8}= 1 + 4\,\frac{\xphs}{\lzz} \;\;&\;\;
\mrW_{9}= 1 - \frac{\xphs}{\lzz} \;\;&\;\;
\mrW_{10}= 1 + \frac{\xphs}{\lzz} \;\;&\;\;
\mrW_{11}= 1 + 3\,\frac{\xphs}{\lzz} \\
\mrW_{12}= 1 + 5\,\frac{\xphs}{\lzz} \;\;&\;\;
\mrW_{13}= 1 + 6\,\frac{\xphs}{\lzz} \;\;&\;\;
\mrW_{14}= 4 + 15\,\frac{\xphs}{\lzz} \;\;&\;\;
\mrW_{15}= 5 + 6\,\frac{\xphs}{\lzz} \\
\mrW_{16}= 5 + 8\,\frac{\xphs}{\lzz} \;\;&\;\;
\mrW_{17}= 1 + 2\,\frac{\xphs}{\lzz} \;\;&\;\;
\mrW_{18}= \frac{1}{\lzz} - \mcX_{9} \;\;&\;\;
\mrW_{19}= \frac{1}{\lzz} + \mcX_{9} \\
\mrW_{20}= 2\,\frac{\xpts}{\lzz} - \mcX_{8} \;\;&\;\;
\mrW_{21}= 7\,\frac{1}{\lzz} + \mcX_{9} \;\;&\;\;
\mrW_{22}= 9\,\frac{1}{\lzz} - \mcX_{9} \;\;&\;\;
\mrW_{23}= 2\,\frac{\xpts}{\lzz} + \mcX_{10} \\
\mrW_{24}= 2\,\frac{\xpbs}{\lzz} - \mcX_{11} \;\;&\;\;
\mrW_{25}= 2\,\frac{\xpbs}{\lzz} + \mcX_{12} \;\;&\;\;
\mrW_{26}= 1 + 3\,\frac{\xphq}{\lzzs} - \frac{\xphs}{\lzz} \;\;&\;\;
\mrW_{27}= 1 - 6\,\frac{\xphs}{\lzz} \\
\mrW_{28}= 6\,\frac{\xphs}{\lzzs} + \frac{1}{\lzz} - \mcX_{9} \;\;&\;\;
\mrW_{29}= 1 - 9\,\frac{\xphq}{\lzzs} - \frac{\xphs}{\lzz} \;\;&\;\;
\mrW_{30}= 1 - 6\,\frac{\xphq}{\lzzs} - 6\,\frac{\xphs}{\lzz} \;\;&\;\;
\mrW_{31}= 1 - 3\,\frac{\xphq}{\lzzs} + 3\,\frac{\xphs}{\lzz} \\
\end{array}
\]


\[
\begin{array}{llll}
\mrW_{32}= 1 + 3\,\frac{\xphq}{\lzzs} + \frac{\xphs}{\lzz} \;\;&\;\;
\mrW_{33}= 1 + 6\,\frac{\xphq}{\lzzs} - 9\,\frac{\xphs}{\lzz} \;\;&\;\;
\mrW_{34}= 2 + 3\,\frac{\xphs}{\lzz} \;\;&\;\;
\mrW_{35}= 5 - 6\,\frac{\xphs}{\lzz} \\
\mrW_{36}= 1 - 96\,\frac{\xphq}{\lzzs} + 9\,\frac{\xphs}{\lzz} \;\;&\;\;
\mrW_{37}= 1 + 15\,\frac{\xphq}{\lzzs} + 3\,\frac{\xphs}{\lzz} \;\;&\;\;
\mrW_{38}= 1 + 30\,\frac{\xphq}{\lzzs} - 9\,\frac{\xphs}{\lzz} \;\;&\;\;
\mrW_{39}= 1 + 15\,\frac{\xphs}{\lzz} \\
\mrW_{40}= 2 + 3\,\frac{\xphq}{\lzzs} + 9\,\frac{\xphs}{\lzz} \;\;&\;\;
\mrW_{41}= 4 + 51\,\frac{\xphs}{\lzz} \;\;&\;\;
\mrW_{42}= 5 + 39\,\frac{\xphs}{\lzz} \;\;&\;\;
\mrW_{43}= 6 + 17\,\frac{\xphs}{\lzz} \\
\mrW_{44}= 17 + 51\,\frac{\xphq}{\lzzs} - 13\,\frac{\xphs}{\lzz} \;\;&\;\;
\mrW_{45}= 21 - 72\,\frac{\xphq}{\lzzs} - 29\,\frac{\xphs}{\lzz} \;\;&\;\;
\mrW_{46}= 6\,\frac{\xphs}{\lzzs} + 15\,\frac{1}{\lzz} + 2\,\mcX_{9} \;\;&\;\;
\mrW_{47}= \mcX_{6}\,\frac{1}{\lzz} + \mcX_{13} \\
\mrW_{48}= 1 - 3\,\frac{\xphq}{\lzzs} - \frac{\xphs}{\lzz} \;\;&\;\;
\mrW_{49}= 1 + 3\,\frac{\xphq}{\lzzs} + 5\,\frac{\xphs}{\lzz} \;\;&\;\;
\mrW_{50}= 1 + 6\,\frac{\xphq}{\lzzs} + 3\,\frac{\xphs}{\lzz} \;\;&\;\;
\mrW_{51}= 6\,\frac{\xphs}{\lzzs} + 7\,\frac{1}{\lzz} + \mcX_{9} \\
\mrW_{52}= 1 + 6\,\frac{\xphs}{\lzzs} + 3\,\frac{\xphs}{\lzz} \;\;&\;\;
\mrW_{53}= 1 + 3\,\frac{\xphq}{\lzzs} \;\;&\;\;
\mrW_{54}= 1 + 2\,\mcX_{18}\,\frac{1}{\lww} \;\;&\;\;
\mrW_{55}= \mcX_{18} + 2\,\mcX_{19}\,\frac{1}{\lww} \\
\mrW_{56}= 8\,\mcX_{18}\,\frac{1}{\lww} + \mcX_{21} \;\;&\;\;
\mrW_{57}= 8\,\mcX_{18}\,\frac{1}{\lww} + \mcX_{23} \;\;&\;\;
\mrW_{58}= 4\,\mcX_{19}\,\frac{1}{\lww} + \mcX_{20} \;\;&\;\;
\mrW_{59}= \xpbs + 2\,\mcX_{18}\,\frac{1}{\lww} \\
\mrW_{60}= 8\,\mcX_{18}\,\frac{\xpts}{\lww} + \mcX_{26} \;\;&\;\;
\mrW_{61}= 4\,\mcX_{19}\,\frac{1}{\lww} + \mcX_{25} \;\;&\;\;
\mrW_{62}= 1 - \mcX_{28}\,\frac{1}{\lww} \;\;&\;\;
\mrW_{63}= 36 + \frac{\xphs}{\lww} \\
\end{array}
\]


\[
\begin{array}{lll}
\mrW_{64}= 8\,\mcX_{0} + \mcX_{31}\,\frac{1}{\lww} \;\;&\;\;
\mrW_{65}= \mcX_{29} + \mcX_{34}\,\frac{1}{\lww} \;\;&\;\;
\mrW_{66}= 2\,\mcX_{30} + \mcX_{32}\,\frac{1}{\lww} \\

\mrW_{67}= \mcX_{33} - \mcX_{35}\,\frac{\xphq}{\lww} \;\;&\;\;
\mrW_{68}= 2 - \frac{\xphq}{\lww} \;\;&\;\;
\mrW_{69}= 8\,\frac{1}{\lww} + \mcX_{0} \\

\mrW_{70}= 16\,\frac{1}{\lww} + \mcX_{37} \;\;&\;\;
\mrW_{71}= 24\,\frac{1}{\lww} - \mcX_{36}\,\xphs \;\;&\;\;
\mrW_{72}= \frac{\xphq}{\lww} + \mcX_{43} \\ 

\mrW_{73}= 2\,\mcX_{0}\,\frac{1}{\lww} - \mcX_{40} \;\;&\;\;
\mrW_{74}= \mcX_{28}\,\frac{1}{\lww} - \mcX_{41} \;\;&\;\;
\mrW_{75}= \mcX_{42}\,\frac{\xphs}{\lww} + \mcX_{44} \\
\mrW_{76}= \mcX_{45}\,\frac{\xphq}{\lww} - \mcX_{46} \;\;&\;\;
\mrW_{77}= 8 + \frac{\xphs}{\lww} \;\;&\;\;
\mrW_{78}= \frac{1}{\lww} - \mcX_{9} \\
\mrW_{79}= \frac{1}{\lww} + \mcX_{9} \;\;&\;\;
\mrW_{80}= 12\,\mcX_{18}\,\frac{1}{\lwws} + \mcX_{47}\,\frac{1}{\lww} + \mcX_{52} \;\;&\;\;
\mrW_{81}= 12\,\mcX_{18}\,\frac{1}{\lwws} + \mcX_{48}\,\frac{1}{\lww} + \mcX_{51} \\
\mrW_{82}= 12\,\mcX_{19}\,\frac{1}{\lwws} + \mcX_{49}\,\frac{1}{\lww} - \mcX_{50} \;\;&\;\;
\mrW_{83}= 1 + 2\,\frac{1}{\lww} \;\;&\;\;
\mrW_{84}= 1 - 2\,\mcX_{6}\,\frac{1}{\lww} - 12\,\mcX_{18}\,\frac{1}{\lwws} \\
\mrW_{85}= 1 + 2\,\mcX_{53}\,\frac{1}{\lww} \;\;&\;\;
\mrW_{86}= \mcX_{18} - 12\,\mcX_{18}\,\frac{1}{\lwws} + 4\,\mcX_{55}\,\frac{1}{\lww} \;\;&\;\;
\mrW_{87}= 6\,\mcX_{19}\,\frac{1}{\lww} + \mcX_{54} \\
\mrW_{88}= 2\,\frac{\xpts}{\lww} + \mcX_{63} \;\;&\;\;
\mrW_{89}= 2\,\frac{\xpts}{\lww} + \mcX_{64} \;\;&\;\;
\mrW_{90}= 12\,\mcX_{18}\,\frac{1}{\lwws} + \mcX_{51} + \mcX_{59}\,\frac{1}{\lww} \\
\mrW_{91}= 12\,\mcX_{18}\,\frac{\xpts}{\lwws} + \mcX_{62} + \mcX_{65}\,\frac{1}{\lww} \;\;&\;\;
\mrW_{92}= 12\,\mcX_{19}\,\frac{1}{\lwws} + \mcX_{60}\,\frac{1}{\lww} - \mcX_{61} \;\;&\;\;
\mrW_{93}= \mcX_{50} - \mcX_{57}\,\frac{1}{\lww} \\
\mrW_{94}= \mcX_{56} + \mcX_{58}\,\frac{1}{\lww} \;\;&\;\;
\mrW_{95}= 1 - \frac{\xphs}{\lww} & \\
\end{array}
\]


\[
\begin{array}{lll}
\mrW_{96}= 1 - \frac{\xphs}{\lww} + 12\,\mcX_{0}\,\frac{1}{\lwws} \;\;&\;\;
\mrW_{97}= 12\,\mcX_{0}\,\frac{1}{\lwws} - \mcX_{0}\,\frac{1}{\lww} - \mcX_{66} \;\;&\;\;
\mrW_{98}= 12\,\mcX_{28}\,\frac{1}{\lwws} + \mcX_{36} + \mcX_{67}\,\frac{1}{\lww} \\
\mrW_{99}= 1 - 12\,\frac{\xphs}{\lwws} - \frac{\xphs}{\lww} \;\;&\;\;
\mrW_{100}= 1 + 12\,\frac{\xphs}{\lwws} - \frac{\xphs}{\lww} \;\;&\;\;
\mrW_{101}= 1 + 3\,\frac{\xphs}{\lww} \\
\mrW_{102}= 1 + \mcX_{36}\,\frac{1}{\lww} - 12\,\mcX_{68}\,\frac{1}{\lwws} \;\;&\;\;
\mrW_{103}= 5\,\xphs - 12\,\mcX_{71}\,\frac{1}{\lwws} - \mcX_{74}\,\frac{1}{\lww} \;\;&\;\;
\mrW_{104}= 12\,\mcX_{69}\,\frac{\xphq}{\lwws} - \mcX_{72} - \mcX_{75}\,\frac{\xphs}{\lww}\\
\mrW_{105}= \mcX_{70} + 12\,\mcX_{73}\,\frac{1}{\lwws} + \mcX_{76}\,\frac{1}{\lww} \;\;&\;\;
\mrW_{106}= 3 + 4\,\frac{\xphs}{\lww} \;\;&\;\;
\mrW_{107}= 36\,\frac{\xphvi}{\lwws} + \mcX_{83} + \mcX_{86}\,\frac{\xphs}{\lww} \\
\mrW_{108}= \xphs - 12\,\mcX_{78}\,\frac{1}{\lwws} - \mcX_{80}\,\frac{1}{\lww} \;\;&\;\;
\mrW_{109}= \mcX_{36} - \mcX_{36}\,\frac{\xphs}{\lww} \;\;&\;\;
\mrW_{110}= \mcX_{77} + 12\,\mcX_{79}\,\frac{1}{\lwws} + \mcX_{84}\,\frac{1}{\lww} \\
\mrW_{111}= 12\,\mcX_{81}\,\frac{1}{\lwws} - \mcX_{82} - \mcX_{85}\,\frac{1}{\lww} \;\;&\;\;
\mrW_{112}= 1 - 8\,\frac{1}{\lww} \;\;&\;\;
\mrW_{113}= 60\,\frac{\xphs}{\lwws} - \mcX_{88} + \mcX_{90}\,\frac{1}{\lww} \\
\mrW_{114}= 60\,\frac{\xphq}{\lwws} + \mcX_{89} + \mcX_{91}\,\frac{\xphs}{\lww} \;\;&\;\;
\mrW_{115}= 5\,\frac{\xphs}{\lww} - \mcX_{87} \;\;&\;\;
\mrW_{116}= 60\,\frac{\xphs}{\lwws} + 17\,\frac{\xphs}{\lww} + \mcX_{94} \\
\mrW_{117}= 13\,\frac{1}{\lww} - \mcX_{93} \;\;&\;\;
\mrW_{118}= 24\,\frac{1}{\lww} + \mcX_{36} \;\;&\;\;
\mrW_{119}= 1 - \mcX_{95}\,\frac{1}{\lww} \\
\mrW_{120}= 60\,\frac{\xphs}{\lwws} + \mcX_{94} & & \\
\end{array}
\]

\normalsize

\vspace{0.8cm}
\bei
\item[\fbox{$\mrX\,$}] 
\eei

Here $\mrW_{\upPhi\,|\,\upphi}$ denotes the $\upphi$ component of the $\upPhi$ wave-function 
factor \etc
Furthermore, $\sum_{\gen}$ implies summing over all fermions and all generations, while
${\overline{\sum}}_{\gen}$ excludes $\PQt$ and $\PQb$ from the sum.

\scriptsize
\bqa
\mrX_{0} &=& \WFH + 2\,\WFA - 2\,\degf + \dMf
\quad
\mrX_{1} = 3 - \WFZ - \WFH - \WFA + 2\,\degf - \dMf
\nl
\mrX_{2} &=& 2\,\WFZt + \WFHt + 4\,\dcthft
\quad
\mrX_{3} = 2\,\WFZb + \WFHb + 4\,\dcthfb
\nl
\mrX_{4} &=& 2\,\degff - 4\,\dcthfrf + \dMff - 2\,\WFZrf
\quad
\mrX_{5} = 4 - 2\,\degfW + 2\,\WFZW + \WFHW - \dMfW + 4\,\dcthfW
\nl
\mrX_{6} &=& 6 + 4\,\degfW - 4\,\WFZW - 2\,\WFHW + 6\,\WFZb + 3\,\WFHb + 6\,\WFZt + 3\,
      \WFHt + 2\,\dMfW - 8\,\dcthfW + 12\,\dcthfb + 12\,\dcthft
\nl
\mrX_{7} &=& 11 + 30\,\degfW - 30\,\WFZW - 15\,\WFHW + 15\,\dMfW - 60\,\dcthfW
\nl
\mrX_{8} &=& 46 + 12\,\degfW - 12\,\WFZW - 6\,\WFHW - 6\,\degff + 6\,\dcthfrf - 3\,
      \dMff + 6\,\WFZrf + 6\,\WFZb + 3\,\WFHb + 6\,\WFZt 
\nl &+& 3\,\WFHt + 6\,\dMfW - 30\,\dcthfW + 6\,\dcthfb + 6\,\dcthft
\nl
\mrX_{9} &=& \dcthfrf + \dcthfW + \dcthfb + \dcthft
\quad
\mrX_{10} = 1 - 2\,\degff + 4\,\dcthfrf - \dMff + 2\,\WFZrf
\nl
\mrX_{11} &=& 5 + 4\,\dcthfrf + 4\,\dcthfW + 4\,\dcthfb + 4\,\dcthft
\nl
\mrX_{12} &=& 1 - 8\,\degff + 16\,\dcthfrf + 8\,\WFZrf + 4\,\WFZb + 2\,\WFHb + 4\,
      \WFZt + 2\,\WFHt + 4\,\dMfW + 8\,\dcthfb + 8\,\dcthft
\nl
\mrX_{13} &=& 3 + 2\,\degfW - 2\,\WFZW - \WFHW + 2\,\degff - \dMff - 2\,\WFZrf - 2\,
      \WFZb - \WFHb - 2\,\WFZt - \WFHt - \dMfW
\nl
\mrX_{14} &=& 44 - 2\,\degff - \dMff + 2\,\WFZrf + 2\,\WFZb + \WFHb + 2\,\WFZt + 
      \WFHt - 4\,\dcthfW
\nl
\mrX_{15} &=& \WFH6 + 2\,\WFZ6 + 4\,\dcthf6 - \dMf6 - 2\,\degf6
\nl
\mrX_{16} &=& 1 - 8\,\degfW + 8\,\WFZW + 4\,\WFHW - 8\,\degff + 4\,\dMff + 8\,\WFZrf
       + 8\,\WFZb + 4\,\WFHb + 8\,\WFZt + 4\,\WFHt + 4\,\dMfW
\nl
\mrX_{17} &=& 2 + 2\,\degftb + \dMftb - 2\,\WFWtb - \WFHtb
\quad
\mrX_{18} = 2\,\degftb + \dMftb - 2\,\WFWtb - \WFHtb
\nl
\mrX_{19} &=& 103 + 144\,\degfrf + 72\,\dMfrf - 6\,\WFWW - 144\,\WFWrf + 6\,\degfW
       - 3\,\WFHW + 3\,\dMfW
\nl
\mrX_{20} &=& 17 - 2\,\WFWW + 2\,\degfW - \WFHW + \dMfW
\nl
\mrX_{21} &=& 163 - 96\,\degfrf - 48\,\dMfrf + 26\,\WFWW + 96\,\WFWrf - 26\,\degfW
       + 13\,\WFHW - 13\,\dMfW
\nl
\mrX_{22} &=& 11 - 16\,\degfrf + 8\,\dMfrf + 16\,\WFWW + 16\,\WFWrf - 16\,\degftb + 
      8\,\dMftb + 16\,\WFWtb + 8\,\WFHtb - 16\,\degfW 
\nl &+& 8\,\WFHW + 8\,\dMfW - 16\,\dcthf
\nl
\mrX_{23} &=& 35 + 120\,\degfrf + 12\,\dMfrf - 22\,\WFWW - 120\,\WFWrf - 48\,\dMftb
       + 22\,\degfW - 11\,\WFHW - 37\,\dMfW + 120\,\dcthf
\nl
\mrX_{24} &=& 293 + 12\,\WFWW - 12\,\degfW + 6\,\WFHW - 6\,\dMfW
\quad
\mrX_{25} = \dcthf
\nl
\mrX_{26} &=& 3 - 4\,\degfrf + 2\,\dMfrf + 4\,\WFWW + 4\,\WFWrf - 4\,\degftb + 2\,
      \dMftb + 4\,\WFWtb + 2\,\WFHtb - 4\,\degfW + 2\,\WFHW + 2\,\dMfW
\nl
\mrX_{27} &=& 72 - 10\,\degfrf - \dMfrf + 2\,\WFWW + 10\,\WFWrf + 4\,\dMftb - 2\,
      \degfW + \WFHW + 3\,\dMfW
\nl
\mrX_{28} &=& 3 - 8\,\degfrf + 4\,\dMfrf + 8\,\WFWW + 8\,\WFWrf - 8\,\degftb + 4\,
      \dMftb + 8\,\WFWtb + 4\,\WFHtb 
\nl &-& 8\,\degfW + 4\,\WFHW + 4\,\dMfW - 16\,\dcthf
\nl
\mrX_{29} &=& 11 - 20\,\degfrf - 2\,\dMfrf + 2\,\WFWW + 20\,\WFWrf + 8\,\dMftb - 2\,
      \degfW + \WFHW + 7\,\dMfW - 40\,\dcthf
\nl
\mrX_{30} &=& 2\,\WFWs + \WFHs - \dMfs - 2\,\degfs
\quad
\mrX_{31} = 2\,\WFW + \WFH - 2\,\degf - 2\,\dcthf + \dMf
\nl
\mrX_{32} &=& 2\,\WFW + \WFH - 2\,\degf + \dMf
\quad
\mrX_{33} = 2\,\WFW + \WFH - 2\,\degf - 4\,\dcthf + \dMf
\eqa

\normalsize

\subsection{Amplitudes}
We use the following notazion: $\mcT^{\nfact}_{\PA\PA}(\auWB)$ is the non-factorizable
part of the $\mcT_{\PA\PA}$ amplitude that is proportional to the Wilson coefficient $\auWB$ \etc
\bq
\mcD^{\nfact}_{\sPHVV} = \mw\,\mcT^{\nfact}_{\ssD\,;\,\PV\PV} \spc
\qquad
\mcP^{\nfact}_{\sPHVV} = \frac{1}{\mw}\,\mcT^{\nfact}_{\ssP\,;\,\PV\PV} \spc
\eq
with $\PV = \PZ, \PW$, while $\mcT^{\nfact}_{\sPHAA, \sPHAZ}$ should be multiplied by $\mw$
to restore its dimensionality. Furthermore, $\laz$ is defined in \eqn{deflaz}, $\lzz$ in 
\eqn{deflzz} and $\lww$ in \eqn{deflww}. The function $\ssC_0$ in this Appendix is the scalar 
three-point function, scaled with $\mw$. The amplitudes are listed in the following equations:

\footnotesize
\bei
\item {\underline{$\PH\PA\PA$}} Amplitudes
\eei
\bqas
\mcT^{\nfact}_{\PA\PA}(\auWB) &=&
       -
         \frac{1}{8}
         \,s\,\xphs\,\xpts
       +
         \frac{1}{8}
         \,s\,\xphs\,\xpts
         \,\afun{\mt}
\nl &+&
         \frac{1}{8}
         \,s\,\xphs\,\xpts
         \,\bfun{ - \mhs}{\mt}{\mt}
\nl &-&
         \frac{1}{16}
         \,s\,\xphq\,\xpts
         \,\cfun{ - \mhs}{0}{0}{\mt}{\mt}{\mt}
\eqas
\bqas
\mcT^{\nfact}_{\PA\PA}(\adWB) &=&
         \frac{1}{16}
         \,s\,\xphs\,\xpbs
     -
         \frac{1}{16}
         \,s\,\xphs\,\xpbs
         \,\afun{\mb}
\nl &-&
         \frac{1}{16}
         \,s\,\xphs\,\xpbs
         \,\bfun{ - \mhs}{\mb}{\mb}
\nl &+&
         \frac{1}{32}
         \,s\,\xphq\,\xpbs
         \,\cfun{ - \mhs}{0}{0}{\mb}{\mb}{\mb}
\eqas
\bqas
\mcT^{\nfact}_{\PA\PA}(\aAZ) &=&
       -
         \frac{1}{2}
         \,s^3\,c\,\xphs
         \,\afun{\mw}
     +
         \frac{1}{4}
         \,\mrT^{d}_{1}\,s\,c\,\xphs
\nl &-&
         \frac{1}{16}
         \,\mrT^{d}_{2}\,\mrV_{0}\,s\,c\,\xphs
         \,\bfun{ - \mhs}{\mw}{\mw}
\nl &+&
         \frac{1}{2}
         \,( s^2\,\xphs + \mrT^{d}_{1} )\,s\,c\,\xphs
         \,\cfun{ - \mhs}{0}{0}{\mw}{\mw}{\mw}
\nl &-&
         \frac{1}{16}
         \,( 2\,\mrT^{d}_{0} - \mrT^{d}_{2}\,\xphs )\,s\,c\,\xphs
         \,\LR
\eqas
\bqas
\mcT^{\nfact}_{\PA\PA}(\aAA) &=&
       -
         \frac{1}{16}
         \,\xphs
         \,\mrX_{0}
     +
         \frac{1}{4}
         \,\mrT^{d}_{4}\,s^2\,\xphs
     -
         \frac{1}{2}
         \,s^4\,\xphs
         \,\afun{\mw}
\nl &-&
         \frac{1}{64}
         \,\xphq
         \,\bfun{ - \mhs}{\mz}{\mz}
\nl &-&
         \frac{3}{64}
         \,\xphq
         \,\bfun{ - \mhs}{\mh}{\mh}
\nl &+&
         \frac{1}{2}
         \,( s^2\,\xphs + \mrT^{d}_{4} )\,s^2\,\xphs
         \,\cfun{ - \mhs}{0}{0}{\mw}{\mw}{\mw}
\nl &-&
         \frac{1}{32}
         \,( 8\,s^2\,c^2 + \mrT^{d}_{3}\,\xphs )\,\xphs
         \,\bfun{ - \mhs}{\mw}{\mw}
\nl &-&
         \frac{1}{32}
         \,( 8\,\mrT^{d}_{1}\,s^2 + \mrT^{d}_{5}\,\xphs )
      \,\xphs
         \,\LR
\eqas
\bqas
\mcT^{\nfact}_{\PA\PA}(\aZZ) &=&
       -
         \frac{3}{4}
         \,s^2\,c^2\,\xphs
     -
         \frac{1}{2}
         \,s^2\,c^2\,\xphs
         \,\afun{\mw}
\nl &+&
         \frac{1}{8}
         \,\mrV_{0}\,s^2\,c^2\,\xphs
         \,\bfun{ - \mhs}{\mw}{\mw}
\nl &-&
         \frac{1}{2}
         \,\mrV_{1}\,s^2\,c^2\,\xphs
         \,\cfun{ - \mhs}{0}{0}{\mw}{\mw}{\mw}
\nl &+&
         \frac{1}{8}
         \,\mrV_{2}\,s^2\,c^2\,\xphs
         \,\LR
\eqas
\vspace{0.5cm}
\bei
\item {\underline{$\PH\PA\PZ$}} Amplitudes
\eei
\bqas
\mcT^{\nfact}_{\PA\PZ}(\aptV) &=&
         \frac{1}{2}
         \,\frac{s}{c^3}\,\frac{\xphs}{\lazs}\,\xpts
         \,\bfun{ - \mhs}{\mt}{\mt}
\nl &-&
         \frac{1}{2}
         \,\frac{s}{c^3}\,\frac{\xphs}{\lazs}\,\xpts
         \,\bfun{ - \mzs}{\mt}{\mt}
\nl &+&
         \frac{1}{2}
         \,\frac{s}{c}\,\frac{\xphs}{\laz}\,\xpts
\nl &+&
         \frac{1}{4}
         \,\frac{s}{c}\,\mrW_{0}\,\xphs\,\xpts
         \,\cfun{ - \mhs}{0}{ - \mzs}{\mt}{\mt}{\mt}
\eqas
\bqas
\mcT^{\nfact}_{\PA\PZ}(\apbV) &=&
         \frac{1}{4}
         \,\frac{s}{c^3}\,\frac{\xphs}{\lazs}\,\xpbs
         \,\bfun{ - \mhs}{\mb}{\mb}
\nl &-&
         \frac{1}{4}
         \,\frac{s}{c^3}\,\frac{\xphs}{\lazs}\,\xpbs
         \,\bfun{ - \mzs}{\mb}{\mb}
\nl &+&
         \frac{1}{4}
         \,\frac{s}{c}\,\frac{\xphs}{\laz}\,\xpbs
\nl &+&
         \frac{1}{8}
         \,\frac{s}{c}\,\mrW_{1}\,\xphs\,\xpbs
         \,\cfun{ - \mhs}{0}{ - \mzs}{\mb}{\mb}{\mb}
\eqas
\bqas
\mcT^{\nfact}_{\PA\PZ}(\atBW) &=&
         \frac{1}{16}
         \,s\,\xphs\,\xpts
         \,\afun{\mt}
     -
         \frac{1}{8}
         \,\frac{s}{c^2}\,\frac{\xphs}{\laz}\,\xpts
\nl &-&
         \frac{1}{16}
         \,\Bigl[ 2\,\frac{1}{\lazs} 
         + (  - c^2 + \frac{1}{\laz} )\,c^2 \Bigr]\,\frac{s}{c^4}\,\xphs\,\xpts
         \,\bfun{ - \mhs}{\mt}{\mt}
\nl &+&
         \frac{1}{8}
         \,(  - c^2 + 2\,\frac{1}{\laz} )\,\frac{s}{c^2}\,\xphs\,\xptq
         \,\cfun{ - \mhs}{0}{ - \mzs}{\mt}{\mt}{\mt}
\nl &+&
         \frac{1}{16}
         \,( c^2 + 2\,\frac{1}{\laz} )\,\frac{s}{c^4}\,\frac{\xphs}{\laz}\,\xpts
         \,\bfun{ - \mzs}{\mt}{\mt}
\eqas
\bqas
\mcT^{\nfact}_{\PA\PZ}(\atWB) &=&
       -
         \frac{3}{128}
         \,\frac{\vtq}{c}\,\xphs\,\xpts
\nl &-&
         \frac{3}{64}
         \,\frac{\vtq}{c}\,\xphs\,\xptq
         \,\cfun{ - \mhs}{0}{ - \mzs}{\mt}{\mt}{\mt}
\nl &+&
         \frac{3}{128}
         \,( c^2 + \frac{1}{\laz} )\,\frac{\vtq}{c^3}\,\xphs\,\xpts
         \,\bfun{ - \mhs}{\mt}{\mt}
\nl &-&
         \frac{3}{128}
         \,( c^2 + \frac{1}{\laz} )\,\frac{\vtq}{c^3}\,\xphs\,\xpts
         \,\bfun{ - \mzs}{\mt}{\mt}
\eqas
\bqas
\mcT^{\nfact}_{\PA\PZ}(\abBW) &=&
       -
         \frac{1}{32}
         \,s\,\xphs\,\xpbs
         \,\afun{\mb}
     +
         \frac{1}{16}
         \,\frac{s}{c^2}\,\frac{\xphs}{\laz}\,\xpbs
\nl &+&
         \frac{1}{32}
         \,\Bigl[ 2\,\frac{1}{\lazs} 
          + (  - c^2 + \frac{1}{\laz} )\,c^2 \Bigr]\,\frac{s}{c^4}\,\xphs\,\xpbs
         \,\bfun{ - \mhs}{\mb}{\mb}
\nl &-&
         \frac{1}{16}
         \,(  - c^2 + 2\,\frac{1}{\laz} )\,\frac{s}{c^2}\,\xphs\,\xpbq
         \,\cfun{ - \mhs}{0}{ - \mzs}{\mb}{\mb}{\mb}
\nl &-&
         \frac{1}{32}
         \,( c^2 + 2\,\frac{1}{\laz} )\,\frac{s}{c^4}\,\frac{\xphs}{\laz}\,\xpbs
         \,\bfun{ - \mzs}{\mb}{\mb}
\eqas
\bqas
\mcT^{\nfact}_{\PA\PZ}(\abWB) &=&
         \frac{3}{128}
         \,\frac{\vbq}{c}\,\xphs\,\xpbs
\nl &+&
         \frac{3}{64}
         \,\frac{\vbq}{c}\,\xphs\,\xpbq
         \,\cfun{ - \mhs}{0}{ - \mzs}{\mb}{\mb}{\mb}
\nl &-&
         \frac{3}{128}
         \,( c^2 + \frac{1}{\laz} )\,\frac{\vbq}{c^3}\,\xphs\,\xpbs
         \,\bfun{ - \mhs}{\mb}{\mb}
\nl &+&
         \frac{3}{128}
         \,( c^2 + \frac{1}{\laz} )\,\frac{\vbq}{c^3}\,\xphs\,\xpbs
         \,\bfun{ - \mzs}{\mb}{\mb}
\eqas
\bqas
\mcT^{\nfact}_{\PA\PZ}(\apD) &=&
         \frac{1}{32}
         \,\Bigl[ 2\,\frac{1}{\laz}\,\mrT^{d}_{6}\,c^2 
         - ( \frac{\xphs}{\laz}\,\mrT^{d}_{10} - 2\,\mrT^{d}_{11} ) \Bigr]\,\frac{1}{s\,c^3}\,\xphs
         \,\cfun{ - \mhs}{0}{ - \mzs}{\mw}{\mw}{\mw}
\nl &+&
         \frac{1}{192}
         \,\Bigl\{ 3\,\frac{\xphs}{\laz}\,\mrT^{d}_{10} 
         - 6\,\Bigl[ \mrT^{d}_{6} + ( \xpbs\,\vbq + 2\,\xpts\,\vtq )\,c^2 \Bigr]\,
         \frac{1}{\laz}\,c^2 
\nl &+& 2\,\Bigl[ ( \frac{\xpbs}{\laz}\,\mrT^{d}_{8} 
         + 2\,\frac{\xpts}{\laz}\,\mrT^{d}_{9} + 9\,\mrT^{d}_{7} ) \Bigr]\,s^2\,c^2 \Bigr\}\,
         \frac{1}{s\,c^3}\,\xphs
\nl &-&
         \frac{1}{192}
         \,( 3\,c^2\,\vbq - \mrT^{d}_{8}\,s^2 )\,\frac{1}{s\,c}\,\mrW_{1}\,\xphs\,\xpbs
         \,\cfun{ - \mhs}{0}{ - \mzs}{\mb}{\mb}{\mb}
\nl &-&
         \frac{1}{96}
         \,( 3\,c^2\,\vbq - \mrT^{d}_{8}\,s^2 )\,\frac{1}{s\,c^3}\,\frac{\xphs}{\lazs}\,\xpbs
         \,\bfun{ - \mhs}{\mb}{\mb}
\nl &+&
         \frac{1}{96}
         \,( 3\,c^2\,\vbq - \mrT^{d}_{8}\,s^2 )\,\frac{1}{s\,c^3}\,\frac{\xphs}{\lazs}\,\xpbs
         \,\bfun{ - \mzs}{\mb}{\mb}
\nl &-&
         \frac{1}{96}
         \,( 3\,c^2\,\vtq - \mrT^{d}_{9}\,s^2 )\,\frac{1}{s\,c}\,\mrW_{0}\,\xphs\,\xpts
         \,\cfun{ - \mhs}{0}{ - \mzs}{\mt}{\mt}{\mt}
\nl &-&
         \frac{1}{48}
         \,( 3\,c^2\,\vtq - \mrT^{d}_{9}\,s^2 )\,\frac{1}{s\,c^3}\,\frac{\xphs}{\lazs}\,\xpts
         \,\bfun{ - \mhs}{\mt}{\mt}
\nl &+&
         \frac{1}{48}
         \,( 3\,c^2\,\vtq - \mrT^{d}_{9}\,s^2 )\,\frac{1}{s\,c^3}\,\frac{\xphs}{\lazs}\,\xpts
         \,\bfun{ - \mzs}{\mt}{\mt}
\nl &-&
         \frac{1}{64}
         \,(  - \frac{\xphs}{\laz}\,\mrT^{d}_{10} + 2\,\frac{1}{\laz}\,\mrT^{d}_{6}\,c^2 
         - 6\,\mrT^{d}_{7}\,s^2\,c^2 )\,\frac{1}{s\,c^5}\,\frac{\xphs}{\laz}
         \,\bfun{ - \mhs}{\mw}{\mw}
\nl &+&
         \frac{1}{64}
         \,(  - \frac{\xphs}{\laz}\,\mrT^{d}_{10} + 2\,\frac{1}{\laz}\,\mrT^{d}_{6}\,c^2 
         - 6\,\mrT^{d}_{7}\,s^2\,c^2 )\,\frac{1}{s\,c^5}\,\frac{\xphs}{\laz}
         \,\bfun{ - \mzs}{\mw}{\mw}
\eqas
\bqas
\mcT^{\nfact}_{\PA\PZ}(\aAZ) &=&
         \frac{1}{32}
         \,\xphs
         \,\mrX_{1}
     -
         \frac{1}{64}
         \,\xphq
         \,\afun{\mh}
     +
         \frac{1}{64}
         \,\frac{1}{c^2}\,\xphs
         \,\afun{\mz}
\nl &-&
         \frac{1}{64}
         \,\Bigl[ 2\,\frac{1}{\laz} 
         - ( c^2\,\xphs + 2\,\mrW_{4} )\,c^2 \Bigr]\,\frac{1}{c^6}\,\frac{\xphs}{\laz}
         \,\bfun{ - \mzs}{\mh}{\mz}
\nl &+&
         \frac{1}{16}
         \,\Bigl[ \frac{1}{\laz}\,\mrT^{d}_{7} 
         + ( \frac{\xphs}{\laz}\,\mrT^{d}_{2} - \mrT^{d}_{26} )\,c^2 \Bigr]\,
         \frac{s^2}{c^2}\,\xphs
         \,\afun{\mw}
\nl &+&
         \frac{1}{32}
         \,\Bigl\{ 16\,s^2\,c^6\,\xphs + 8\,\frac{1}{\laz}\,\mrT^{d}_{27}\,c^4 
         - \Bigl[ (  - \frac{\xphs}{\laz}\,\mrT^{d}_{22} 
         + 2\,\mrT^{d}_{15} ) \Bigr] \Bigr\}\,\frac{1}{c^4}\,\xphs
\nl &\times& \cfun{ - \mhs}{0}{ - \mzs}{\mw}{\mw}{\mw}
\nl &-&
         \frac{1}{64}
         \,\Bigl\{ 4\,\frac{\xphs}{\laz}\,\mrT^{d}_{2}\,s^2\,c^6 
         - \Bigl[ ( 2\,\mrT^{d}_{19} + \mrT^{d}_{20}\,\xphs ) \Bigr]\,
         \frac{1}{\lazs} + 2\,( 3\,\frac{1}{\laz}\,\mrT^{d}_{23}
         + \mrT^{d}_{24}\,c^2 )\,c^4 \Bigr\}\,\frac{1}{c^6}\,\xphs
\nl &\times& \bfun{ - \mzs}{\mw}{\mw}
\nl &+&
         \frac{1}{64}
         \,\Bigl\{ 8\,\frac{\xphs}{\laz}\,\mrT^{d}_{2}\,s^2\,c^6 
         + \Bigl[ 2\,\frac{1}{\laz}\,\mrT^{d}_{28} - ( 2\,\mrT^{d}_{2}
         - \mrT^{d}_{13}\,\xphs )\,c^2 \Bigr]\,c^4 - \Bigl[ ( 2\,\mrT^{d}_{19} 
         + \mrT^{d}_{20}\,\xphs ) \Bigr]\,\frac{1}{\lazs} \Bigr\}\,
         \frac{1}{c^6}\,\xphs
\nl &\times& \bfun{ - \mhs}{\mw}{\mw}
\nl &+&
         \frac{1}{32}
         \,\Bigl\{ \frac{1}{\lazs} - \Bigl[ 2\,\frac{1}{\laz}\,\mrW_{2} 
         - ( \frac{1}{\laz}\,\mrW_{2} 
         + \mrW_{2}\,c^2 )\,c^2\,\xphs \Bigr]\,c^2 \Bigr\}\,\frac{1}{c^8}\,\xphs
         \,\cfun{ - \mhs}{0}{ - \mzs}{\mz}{\mh}{\mz}
\nl &+&
         \frac{1}{128}
         \,\Bigl\{ 4\,\frac{1}{\lazs} - \Bigl[ 2\,\frac{1}{\laz}\,\mrW_{5} 
         + ( c^2\,\xphs + 2\,\mrW_{3} )\,c^2 \Bigr]\,c^2 \Bigr\}\,\frac{1}{c^6}\,\xphs
         \,\bfun{ - \mhs}{\mz}{\mz}
\nl &-&
         \frac{1}{192}
         \,\Bigl\{ 3\,\Bigl[ \mrT^{d}_{18} 
         - 8\,( \xpbs\,\vbq + 2\,\xpts\,\vtq )\,\frac{1}{\laz}\,c^2 \Bigr]\,c^2 
         + 3\,\Bigl[ ( 2\,\mrT^{d}_{17} + \mrT^{d}_{21}\,\xphs ) \Bigr]\,\frac{1}{\laz} 
\nl &+& 8\,( \mrT^{d}_{12}\,\xpbs 
         + 2\,\mrT^{d}_{25}\,\xpts )\,\frac{1}{\laz}\,s^2\,c^4 \Bigr\}\,\frac{1}{c^4}\,\xphs
\nl &-&
         \frac{1}{64}
         \,( 1 - c^2\,\xphs )\,\frac{1}{c^4}\,\frac{\xphs}{\laz}
         \,\bfun{0}{\mz}{\mh}
\nl &-&
         \frac{3}{32}
         \,( 1 - c^2\,\xphs )\,\frac{1}{c^4}\,\frac{\xphs}{\laz}\,\xphs
         \,\cfun{ - \mhs}{0}{ - \mzs}{\mh}{\mz}{\mh}
\nl &+&
         \frac{1}{24}
         \,( 3\,\vbq - \mrT^{d}_{12}\,s^2 )\,\frac{1}{c^2}\,\frac{\xphs}{\lazs}\,\xpbs
         \,\bfun{ - \mhs}{\mb}{\mb}
\nl &-&
         \frac{1}{24}
         \,( 3\,\vbq - \mrT^{d}_{12}\,s^2 )\,\frac{1}{c^2}\,\frac{\xphs}{\lazs}\,\xpbs
         \,\bfun{ - \mzs}{\mb}{\mb}
\nl &+&
         \frac{1}{48}
         \,( 3\,\vbq - \mrT^{d}_{12}\,s^2 )\,\mrW_{1}\,\xphs\,\xpbs
         \,\cfun{ - \mhs}{0}{ - \mzs}{\mb}{\mb}{\mb}
\nl &+&
         \frac{1}{12}
         \,( 3\,\vtq - \mrT^{d}_{25}\,s^2 )\,\frac{1}{c^2}\,\frac{\xphs}{\lazs}\,\xpts
         \,\bfun{ - \mhs}{\mt}{\mt}
\nl &-&
         \frac{1}{12}
         \,( 3\,\vtq - \mrT^{d}_{25}\,s^2 )\,\frac{1}{c^2}\,\frac{\xphs}{\lazs}\,\xpts
         \,\bfun{ - \mzs}{\mt}{\mt}
\nl &+&
         \frac{1}{24}
         \,( 3\,\vtq - \mrT^{d}_{25}\,s^2 )\,\mrW_{0}\,\xphs\,\xpts
         \,\cfun{ - \mhs}{0}{ - \mzs}{\mt}{\mt}{\mt}
\nl &+&
         \frac{3}{128}
         \,(  - c^2 + 4\,\frac{1}{\laz} )\,\frac{1}{c^2}\,\xphq
         \,\bfun{ - \mhs}{\mh}{\mh}
\nl &-&
         \frac{1}{64}
         \,( c^2 + 6\,\frac{1}{\laz} )\,\frac{1}{c^2}\,\xphq
         \,\bfun{ - \mzs}{\mz}{\mh}
\nl &-&
         \frac{1}{64}
         \,( \mrT^{d}_{14}\,c^2\,\xphs + 2\,\mrT^{d}_{16} )\,\frac{1}{c^2}\,\xphs
         \,\LR
\eqas
\bqas
\mcT^{\nfact}_{\PA\PZ}(\aAA) &=&
       -
         \frac{1}{12}
         \,s\,c\,\xphs\,\xpbs\,\vbq
         \,\afun{\mb}
     -
         \frac{1}{6}
         \,s\,c\,\xphs\,\xpts\,\vtq
         \,\afun{\mt}
\nl &-&
         \frac{1}{24}
         \,\frac{s^3}{c^3}\,\frac{\xphs}{\lazs}\,\mrT^{d}_{40}\,\xpbs
         \,\bfun{ - \mhs}{\mb}{\mb}
\nl &-&
         \frac{1}{12}
         \,\frac{s^3}{c^3}\,\frac{\xphs}{\lazs}\,\mrT^{d}_{43}\,\xpts
         \,\bfun{ - \mhs}{\mt}{\mt}
\nl &-&
         \frac{1}{48}
         \,\frac{s^3}{c}\,\mrT^{d}_{40}\,\mrW_{1}\,\xphs\,\xpbs
         \,\cfun{ - \mhs}{0}{ - \mzs}{\mb}{\mb}{\mb}
\nl &-&
         \frac{1}{24}
         \,\frac{s^3}{c}\,\mrT^{d}_{43}\,\mrW_{0}\,\xphs\,\xpts
         \,\cfun{ - \mhs}{0}{ - \mzs}{\mt}{\mt}{\mt}
\nl &-&
         \frac{1}{24}
         \,\frac{s}{c}\,\xphs\,\vle
         \,\bfun{ - \mzs}{0}{0}
\nl &+&
         \frac{1}{24}
         \,\frac{s}{c}\,\mrU_{0}\,\xphs
         \,( 1 - 3\,\LR )
\nl &+&
         \frac{1}{24}
         \,\Bigl[ \frac{\xpbs}{\lazs}\,\mrT^{d}_{40}\,s^2 
         - ( 1 + 2\,c^2\,\xpbs )\,c^2\,\vbq \Bigr]\,\frac{s}{c^3}\,\xphs
         \,\bfun{ - \mzs}{\mb}{\mb}
\nl &+&
         \frac{1}{12}
         \,\Bigl[ \frac{\xpts}{\lazs}\,\mrT^{d}_{43}\,s^2 
         - ( 1 + 2\,c^2\,\xpts )\,c^2\,\vtq \Bigr]\,\frac{s}{c^3}\,\xphs
         \,\bfun{ - \mzs}{\mt}{\mt}
\nl &+&
         \frac{1}{16}
         \,\Bigl\{ 8\,s^2\,c^4\,\xphs - 8\,\frac{1}{\laz}\,\mrT^{d}_{44}\,c^4 
         + \Bigl[ (  - \frac{\xphs}{\laz}\,\mrT^{d}_{35} 
         + 2\,\mrT^{d}_{30} ) \Bigr] \Bigr\}\,\frac{s}{c^3}\,\xphs
\nl &\times& \cfun{ - \mhs}{0}{ - \mzs}{\mw}{\mw}{\mw}
\nl &+&
         \frac{1}{32}
         \,\Bigl\{ 8\,\frac{\xphs}{\laz}\,s^2\,c^6 - 2\,\Bigl[ \frac{1}{\laz}\,\mrT^{d}_{42} 
         + ( 1 - \mrT^{d}_{2}\,\xphs )\,c^2 \Bigr]\,c^4 
         + \Bigl[ ( 2\,\mrT^{d}_{32} 
         + \mrT^{d}_{34}\,\xphs ) \Bigr]\,\frac{1}{\lazs} \Bigr\}\,\frac{s}{c^5}\,\xphs
\nl &\times& \bfun{ - \mhs}{\mw}{\mw}
\nl &-&
         \frac{1}{96}
         \,\Bigl\{ 12\,\frac{\xphs}{\laz}\,s^2\,c^6 
         + 3\,\Bigl[ ( 2\,\mrT^{d}_{32} + \mrT^{d}_{34}\,\xphs ) \Bigr]\,\frac{1}{\lazs} 
         - 2\,( 9\,\frac{1}{\laz}\,\mrT^{d}_{38} + \mrT^{d}_{36} )\,c^4 \Bigr\}\,\frac{s}{c^5}\,\xphs
\nl &\times& \bfun{ - \mzs}{\mw}{\mw}
\nl &+&
         \frac{1}{288}
         \,\Bigl\{ 2\,\Bigl[ \mrT^{d}_{37} 
         - 12\,( \xpbs\,\vbq + 2\,\xpts\,\vtq )\,c^2 \Bigr]\,c^2 
         + 9\,\Bigl[ ( 2\,\mrT^{d}_{31} + \mrT^{d}_{33}\,\xphs ) \Bigr]\,\frac{1}{\laz} 
\nl &-& 12\,( \mrT^{d}_{40}\,\xpbs 
         + 2\,\mrT^{d}_{43}\,\xpts )\,\frac{1}{\laz}\,s^2\,c^2 \Bigr\}\,\frac{s}{c^3}\,\xphs
\nl &+&
         \frac{1}{24}
         \,( 3\,\frac{\xphs}{\laz}\,s^2\,c^2 - 3\,\frac{1}{\laz}\,\mrT^{d}_{29} 
         - \mrT^{d}_{39}\,c^2 )\,\frac{s}{c}\,\xphs
         \,\afun{\mw}
\nl &+&
         \frac{1}{48}
         \,( 3\,\mrT^{d}_{2}\,c^2\,\xphs + \mrT^{d}_{41} )\,\frac{s}{c}\,\xphs
         \,\LR
\eqas
\bqas
\mcT^{\nfact}_{\PA\PZ}(\aZZ) &=&
         \frac{1}{24}
         \,\frac{s}{c}\,\frac{\xphs}{\lazs}\,\mrT^{d}_{49}\,\xpbs
         \,\bfun{ - \mhs}{\mb}{\mb}
\nl &-&
         \frac{1}{24}
         \,\frac{s}{c}\,\frac{\xphs}{\lazs}\,\mrT^{d}_{49}\,\xpbs
         \,\bfun{ - \mzs}{\mb}{\mb}
\nl &+&
         \frac{1}{12}
         \,\frac{s}{c}\,\frac{\xphs}{\lazs}\,\mrT^{d}_{50}\,\xpts
         \,\bfun{ - \mhs}{\mt}{\mt}
\nl &-&
         \frac{1}{12}
         \,\frac{s}{c}\,\frac{\xphs}{\lazs}\,\mrT^{d}_{50}\,\xpts
         \,\bfun{ - \mzs}{\mt}{\mt}
\nl &+&
         \frac{1}{48}
         \,\mrT^{d}_{49}\,\mrW_{1}\,s\,c\,\xphs\,\xpbs
         \,\cfun{ - \mhs}{0}{ - \mzs}{\mb}{\mb}{\mb}
\nl &+&
         \frac{1}{24}
         \,\mrT^{d}_{50}\,\mrW_{0}\,s\,c\,\xphs\,\xpts
         \,\cfun{ - \mhs}{0}{ - \mzs}{\mt}{\mt}{\mt}
\nl &+&
         \frac{1}{32}
         \,\Bigl\{ 4\,\frac{\xphs}{\laz}\,s^2\,c^6 + \Bigl[ ( 2\,\mrT^{d}_{32} 
         + \mrT^{d}_{47}\,\xphs ) \Bigr]\,\frac{1}{\lazs}
         - 2\,( 3\,\frac{1}{\laz}\,\mrT^{d}_{38} 
         - \mrT^{d}_{51}\,c^2 )\,c^4 \Bigr\}\,\frac{s}{c^5}\,\xphs
\nl &\times& \bfun{ - \mzs}{\mw}{\mw}
\nl &-&
         \frac{1}{32}
         \,\Bigl\{ 8\,\frac{\xphs}{\laz}\,s^2\,c^6 - 2\,\Bigl[ \frac{1}{\laz}\,\mrT^{d}_{42} 
         + ( 1 - \mrT^{d}_{2}\,\xphs )\,c^2\Bigr]\,c^4 
         + \Bigl[ ( 2\,\mrT^{d}_{32} 
         + \mrT^{d}_{47}\,\xphs ) \Bigr]\,\frac{1}{\lazs} \Bigr\}\,\frac{s}{c^5}\,\xphs
\nl &\times& \bfun{ - \mhs}{\mw}{\mw}
\nl &-&
         \frac{1}{96}
         \,\Bigl\{ 2\,\Bigl[ (  - 2\,\frac{\xpbs}{\laz}\,\mrT^{d}_{49} 
         - 4\,\frac{\xpts}{\laz}\,\mrT^{d}_{50} + 21\,\mrT^{d}_{2} ) \Bigr]\,c^4 
         + 3\,\Bigl[ ( 2\,\mrT^{d}_{31} 
         + \mrT^{d}_{46}\,\xphs ) \Bigr]\,\frac{1}{\laz} \Bigr\}\,\frac{s}{c^3}\,\xphs
\nl &+&
         \frac{1}{16}
         \,\Bigl\{  - \Bigl[ (  - \frac{\xphs}{\laz}\,\mrT^{d}_{48} 
         + 2\,\mrT^{d}_{45} ) \Bigr] 
         + 8\,( c^2\,\xphs + \frac{1}{\laz}\,\mrT^{d}_{44} )\,c^4 \Bigr\}\,\frac{s}{c^3}\,\xphs
\nl &\times& \cfun{ - \mhs}{0}{ - \mzs}{\mw}{\mw}{\mw}
\nl &+&
         \frac{1}{16}
         \,( 12\,c^2 - \mrT^{d}_{2}\,\xphs )\,s\,c\,\xphs
         \,\LR
\nl &-&
         \frac{1}{8}
         \,( \frac{\xphs}{\laz}\,s^2\,c^2 - \frac{1}{\laz}\,\mrT^{d}_{29} 
         + \mrT^{d}_{4}\,c^2 )\,\frac{s}{c}\,\xphs
         \,\afun{\mw}
\eqas
\vspace{0.5cm}
\bei
\item {\underline{$\PH\PZ\PZ$}} Amplitudes
\eei
\bqas
\mcT^{\nfact}_{\ssD\,;\,\PZ\PZ}(\aptV) &=&
         \frac{5}{16}
         \,\frac{\vtq}{c^2}\,\xpts
     +
         \frac{1}{8}
         \,\frac{\vtq}{c^2}\,\xpts
         \,\afun{\mt}
     -
         \frac{1}{48}
         \,\frac{\vtq}{c^4}
         \,( 1 - 3\,\LR )
\nl &+&
         \frac{3}{8}
         \,\frac{\vtq}{c^4}\,\frac{\xpts}{\lzz}
         \,\bfun{ - \mhs}{\mt}{\mt}
\nl &-&
         \frac{3}{32}
         \,\Bigl[ 4\,\frac{1}{\lzz} + ( 2 + \mrV_{3}\,c^2 )\,c^2 \Bigr]\,\frac{\vtq}{c^6}\,\xpts
         \,\cfun{ - \mhs}{ - \mzs}{ - \mzs}{\mt}{\mt}{\mt}
\nl &+&
         \frac{1}{16}
         \,( s\mrW_{6} + 2\,c^2\,\xpts )\,\frac{\vtq}{c^4}
         \,\bfun{ - \mzs}{\mt}{\mt}
\eqas
\bqas
\mcT^{\nfact}_{\ssD\,;\,\PZ\PZ}(\aptA) &=&
         \frac{1}{8}
         \,\frac{1}{c^2}\,\xpts
         \,\afun{\mt}
     +
         \frac{1}{16}
         \,\frac{1}{c^2}\,\xpts
         \,( 5 - 12\,\LR )
     -
         \frac{1}{48}
         \,\frac{1}{c^4}
         \,( 1 - 3\,\LR )
\nl &+&
         \frac{3}{8}
         \,\frac{1}{c^4}\,\frac{\xpts}{\lzz}
         \,\bfun{ - \mhs}{\mt}{\mt}
\nl &-&
         \frac{3}{32}
         \,\Bigl[ 4\,\frac{1}{\lzz} + ( 2 - \mrV_{3}\,c^2 )\,c^2 \Bigr]\,\frac{1}{c^6}\,\xpts
         \,\cfun{ - \mhs}{ - \mzs}{ - \mzs}{\mt}{\mt}{\mt}
\nl &+&
         \frac{1}{16}
         \,( s\mrW_{6} - 10\,c^2\,\xpts )\,\frac{1}{c^4}
         \,\bfun{ - \mzs}{\mt}{\mt}
\eqas
\bqas
\mcT^{\nfact}_{\ssD\,;\,\PZ\PZ}(\apbV) &=&
         \frac{5}{16}
         \,\frac{\vbq}{c^2}\,\xpbs
     +
         \frac{1}{8}
         \,\frac{\vbq}{c^2}\,\xpbs
         \,\afun{\mb}
     -
         \frac{1}{48}
         \,\frac{\vbq}{c^4}
         \,( 1 - 3\,\LR )
\nl &+&
         \frac{3}{8}
         \,\frac{\vbq}{c^4}\,\frac{\xpbs}{\lzz}
         \,\bfun{ - \mhs}{\mb}{\mb}
\nl &-&
         \frac{3}{32}
         \,\Bigl[ 4\,\frac{1}{\lzz} + ( 2 + \mrV_{4}\,c^2 )\,c^2 \Bigr]\,\frac{\vbq}{c^6}\,\xpbs
         \,\cfun{ - \mhs}{ - \mzs}{ - \mzs}{\mb}{\mb}{\mb}
\nl &+&
         \frac{1}{16}
         \,( s\mrW_{7} + 2\,c^2\,\xpbs )\,\frac{\vbq}{c^4}
         \,\bfun{ - \mzs}{\mb}{\mb}
\eqas
\bqas
\mcT^{\nfact}_{\ssD\,;\,\PZ\PZ}(\apbA) &=&
         \frac{1}{8}
         \,\frac{1}{c^2}\,\xpbs
         \,\afun{\mb}
     +
         \frac{1}{16}
         \,\frac{1}{c^2}\,\xpbs
         \,( 5 - 12\,\LR )
     -
         \frac{1}{48}
         \,\frac{1}{c^4}
         \,( 1 - 3\,\LR )
\nl &+&
         \frac{3}{8}
         \,\frac{1}{c^4}\,\frac{\xpbs}{\lzz}
         \,\bfun{ - \mhs}{\mb}{\mb}
\nl &-&
         \frac{3}{32}
         \,\Bigl[ 4\,\frac{1}{\lzz} + ( 2 - \mrV_{4}\,c^2 )\,c^2 \Bigr]\,\frac{1}{c^6}\,\xpbs
         \,\cfun{ - \mhs}{ - \mzs}{ - \mzs}{\mb}{\mb}{\mb}
\nl &+&
         \frac{1}{16}
         \,( s\mrW_{7} - 10\,c^2\,\xpbs )\,\frac{1}{c^4}
         \,\bfun{ - \mzs}{\mb}{\mb}
\eqas
\bqas
\mcT^{\nfact}_{\ssD\,;\,\PZ\PZ}(\aplV) &=&
         \frac{1}{48}
         \,\frac{1}{c^4}\,\vle
         \,\bfun{ - \mzs}{0}{0}
     -
         \frac{1}{144}
         \,\frac{1}{c^4}\,\vle
         \,( 1 - 3\,\LR )
\eqas
\bqas
\mcT^{\nfact}_{\ssD\,;\,\PZ\PZ}(\aplA) &=&
         \frac{1}{48}
         \,\frac{1}{c^4}
         \,\bfun{ - \mzs}{0}{0}
     -
         \frac{1}{144}
         \,\frac{1}{c^4}
         \,( 1 - 3\,\LR )
\eqas
\bqas
\mcT^{\nfact}_{\ssD\,;\,\PZ\PZ}(\atp) &=&
       -
         \frac{1}{32}
         \,\frac{1}{c^2}
         \,\mrX_{2}
\eqas
\bqas
\mcT^{\nfact}_{\ssD\,;\,\PZ\PZ}(\abp) &=&
         \frac{1}{32}
         \,\frac{1}{c^2}
         \,\mrX_{3}
\eqas
\bqas
\mcT^{\nfact}_{\ssD\,;\,\PZ\PZ}(\atBW) &=&
       -
         \frac{3}{128}
         \,\frac{\vtq}{c}\,\xphs\,\xpts
\nl &-&
         \frac{3}{128}
         \,\Bigl[ 8\,\frac{1}{\lzz} + ( 2 - c^2\,\xphs )\,c^2 \Bigr]\,\frac{\vtq}{c^5}\,\xpts
         \,\bfun{ - \mhs}{\mt}{\mt}
\nl &+&
         \frac{3}{128}
         \,\Bigl[ 8\,\frac{1}{\lzz} + ( 2 - c^2\,\xphs )\,c^2 \Bigr]\,\frac{\vtq}{c^5}\,\xpts
         \,\bfun{ - \mzs}{\mt}{\mt}
\nl &+&
         \frac{3}{64}
         \,\Bigl\{ 4\,\frac{1}{\lzz} 
         + \Bigl[ 1 
         + ( 4 - c^2\,\xphs )\,c^2\,\xpts \Bigr]\,c^2 \Bigr\}\,\frac{1}{c^7}\,\xpts\,\vtq
         \,\cfun{ - \mhs}{ - \mzs}{ - \mzs}{\mt}{\mt}{\mt}
\eqas
\bqas
\mcT^{\nfact}_{\ssD\,;\,\PZ\PZ}(\abBW) &=&
         \frac{3}{128}
         \,\frac{\vbq}{c}\,\xphs\,\xpbs
\nl &+&
         \frac{3}{128}
         \,\Bigl[ 8\,\frac{1}{\lzz} + ( 2 - c^2\,\xphs )\,c^2 \Bigr]\,\frac{\vbq}{c^5}\,\xpbs
         \,\bfun{ - \mhs}{\mb}{\mb}
\nl &-&
         \frac{3}{128}
         \,\Bigl[ 8\,\frac{1}{\lzz} + ( 2 - c^2\,\xphs )\,c^2 \Bigr]\,\frac{\vbq}{c^5}\,\xpbs
         \,\bfun{ - \mzs}{\mb}{\mb}
\nl &-&
         \frac{3}{64}
         \,\Bigl\{ 4\,\frac{1}{\lzz} 
         + \Bigl[ 1 
         + ( 4 - c^2\,\xphs )\,c^2\,\xpbs \Bigr]\,c^2 \Bigr\}\,\frac{1}{c^7}\,\xpbs\,\vbq
         \,\cfun{ - \mhs}{ - \mzs}{ - \mzs}{\mb}{\mb}{\mb}
\eqas
\bqas
\mcT^{\nfact}_{\ssD\,;\,\PZ\PZ}(\ap) &=&
         \frac{3}{16}
         \,\frac{1}{c^2}
\nl &-&
         \frac{3}{8}
         \,\Bigl[ 4\,\frac{1}{\lzz} 
         + ( \frac{\xphs}{\lzz}\,c^2\,\xphs - \mrW_{8} )\,c^2 \Bigr]\,\frac{1}{c^6}
         \,\cfun{ - \mhs}{ - \mzs}{ - \mzs}{\mh}{\mz}{\mh}
\nl &+&
         \frac{3}{8}
         \,( 2 - c^2\,\xphs )\,\frac{1}{c^4\,\lzz}
         \,\bfun{ - \mhs}{\mh}{\mh}
\nl &-&
         \frac{3}{8}
         \,( 2 - c^2\,\xphs )\,\frac{1}{c^4\,\lzz}
         \,\bfun{ - \mzs}{\mh}{\mz}
\eqas
\bqas
\mcT^{\nfact}_{\ssD\,;\,\PZ\PZ}(\apBox) &=&
       -
         \frac{1}{16}
         \,\frac{1}{c^2}
         \,\mrX_{4}
     -
         \frac{1}{192}
         \,\frac{1}{c^2}\,\xphs
         \,( 25 + 3\,\LR )
     -
         \frac{1}{288}
         \,\frac{1}{c^4}
         \,( 13 - 57\,\LR )
\nl &-&
         \frac{1}{16}
         \,\Bigl[ 2 - ( c^2\,\xphs + \mrW_{11} )\,c^2\,\xphs \Bigr]\,\frac{1}{c^6}
         \,\cfun{ - \mhs}{ - \mzs}{ - \mzs}{\mz}{\mh}{\mz}
\nl &-&
         \frac{1}{32}
         \,\Bigl[ 8\,\frac{1}{\lzz} 
         + ( 2\,\mrW_{11} - \mrW_{12}\,c^2\,\xphs )\,c^2 \Bigr]\,\frac{1}{c^6}
         \,\bfun{ - \mhs}{\mh}{\mh}
\nl &+&
         \frac{1}{32}
         \,\Bigl\{ 16\,\frac{1}{\lzz} + \Bigl[ 4\,\mrW_{10} 
         + ( \mrW_{12}\,c^2\,\xphs - 2\,\mrW_{16} )\,c^2\,\xphs \Bigr]\,c^2 \Bigr\}\,\frac{1}{c^8}
\nl &\times& \cfun{ - \mhs}{ - \mzs}{ - \mzs}{\mh}{\mz}{\mh}
\nl &+&
         \frac{1}{96}
         \,\Bigl\{ 24\,\frac{1}{\lzz} + \Bigl[ 6\,\mrW_{15} 
         + ( c^2\,\xphs - \mrW_{14} )\,c^2\,\xphs \Bigr]\,c^2 \Bigr\}\,\frac{1}{c^6}
         \,\bfun{ - \mzs}{\mh}{\mz}
\nl &+&
         \frac{1}{96}
         \,( 2 - c^2\,\xphs )\,\frac{1}{c^4}
         \,\afun{\mz}
\nl &-&
         \frac{1}{96}
         \,( 3 - c^2\,\xphs )\,\frac{1}{c^2}\,\xphs
         \,\afun{\mh}
\nl &-&
         \frac{1}{64}
         \,( 2\,s\mrW_{13} + 3\,c^2\,\xphs )\,\frac{1}{c^4}
         \,\bfun{ - \mhs}{\mz}{\mz}
\nl &-&
         \frac{1}{16}
         \,( 4\,\frac{1}{\lzz}\,\mrT^{d}_{52} + \mrT^{d}_{52}\,c^2\,s\mrW_{9} )\,\frac{1}{c^6}
         \,\bfun{ - \mhs}{\mw}{\mw}
\nl &+&
         \frac{1}{16}
         \,( 4\,\frac{1}{\lzz}\,\mrT^{d}_{52} + \mrT^{d}_{52}\,c^2\,s\mrW_{9} )\,\frac{1}{c^6}
         \,\bfun{ - \mzs}{\mw}{\mw}
\nl &+&
         \frac{1}{16}
         \,( 4\,\frac{1}{\lzz}\,\mrT^{d}_{52} + \mrT^{d}_{52}\,c^2\,s\mrW_{9} )\,\frac{1}{c^8}
         \,\cfun{ - \mhs}{ - \mzs}{ - \mzs}{\mw}{\mw}{mw}
\eqas
\bqas
\mcT^{\nfact}_{\ssD\,;\,\PZ\PZ}(\apD) &=&
         \frac{1}{192}
         \,\frac{1}{c^2}
         \,\mrX_{6}
     -
         \frac{1}{384}
         \,\frac{1}{c^4}
         \,\mrX_{5}
\nl &-&
         \frac{1}{384}
         \,\Bigl[ 4\,\mrT^{d}_{63}\,c^2 
         + ( \mrT^{d}_{56}\,\xphs + 2\,\mrT^{d}_{69} )\,\frac{1}{\lzz} \Bigr]\,\frac{1}{c^6}
         \,\bfun{ - \mzs}{\mw}{\mw}
\nl &+&
         \frac{1}{384}
         \,\Bigl[ \mrT^{d}_{68}\,c^2 
         + ( \mrT^{d}_{56}\,\xphs + 2\,\mrT^{d}_{69} )\,\frac{1}{\lzz} \Bigr]\,\frac{1}{c^6}
         \,\bfun{ - \mhs}{\mw}{\mw}
\nl &-&
         \frac{1}{384}
         \,\Bigl[ ( \mrT^{d}_{56}\,\xphs + 2\,\mrT^{d}_{69} )\,\frac{1}{\lzz} 
         - ( \mrT^{d}_{61}\,c^2\,\xphs - 6\,\mrT^{d}_{62} )\,c^2 \Bigr]\,\frac{1}{c^8}
\nl &\times& \cfun{ - \mhs}{ - \mzs}{ - \mzs}{\mw}{\mw}{\mw}
\nl &+&
         \frac{1}{64}
         \,\Bigl\{ 3\,c^2 + \Bigl[ ( 3 - \mrT^{d}_{8}\,\vbq ) \Bigr]
         \,\frac{1}{\lzz} \Bigr\}\,\frac{1}{c^4}\,\xpbs
         \,\bfun{ - \mzs}{\mb}{\mb}
\nl &+&
         \frac{1}{64}
         \,\Bigl\{ 3\,c^2 + \Bigl[ ( 3 - \mrT^{d}_{9}\,\vtq ) \Bigr]
         \,\frac{1}{\lzz} \Bigr\}\,\frac{1}{c^4}\,\xpts
         \,\bfun{ - \mzs}{\mt}{\mt}
\nl &-&
         \frac{1}{256}
         \,\Bigl\{ 4\,\frac{1}{\lzz}\,\mrV_{5} 
         + \Bigl[ ( 8 - 4\,\frac{\xphs}{\lzz}\,\mrT^{d}_{57}\,\xphs - \mrT^{d}_{60}\,\xphs ) 
         + ( 2\,c^2 
         + \frac{\xphs}{\lzz}\,\mrT^{d}_{58} )\,c^2\,\xphq \Bigr]\,c^2 \Bigr\}\,\frac{1}{c^8}
\nl &\times& \cfun{ - \mhs}{ - \mzs}{ - \mzs}{\mh}{\mz}{\mh}
\nl &+&
         \frac{1}{768}
         \,\Bigl\{ 4\,\mrT^{d}_{7} 
         - \Bigl[ ( \frac{\xphs}{\lzz})\,\mrT^{d}_{8} + \mrT^{d}_{55} )\,c^2\,\xphs 
         + 2\,( \frac{ \xphs}{\lzz}\,\mrT^{d}_{53} 
         + \mrT^{d}_{7} ) \Bigr]\,c^2\,\xphs \Bigr\}\,\frac{1}{c^8}
\nl &\times& \cfun{ - \mhs}{ - \mzs}{ - \mzs}{\mz}{\mh}{\mz}
\nl &+&
         \frac{1}{2304}
         \,\Bigl\{ \mrT^{d}_{66} + 3\,\Bigl[ \mrT^{d}_{65}\,\xphs
         + 6\,( \mrT^{d}_{8}\,\xpbs\,\vbq + \mrT^{d}_{9}\,\xpts\,\vtq
         - 3\,\mrV_{6} )\,c^2 \Bigr]\,c^2 \Bigr\}\,\frac{1}{c^6}
\nl &+&
         \frac{1}{256}
         \,\Bigl\{ \Bigl[ 7\,c^2\,\xphs 
         + ( 4 - \frac{\xphs}{\lzz}\,\mrT^{d}_{58}\,\xphs ) \Bigr]\,c^2 
         + 2\,\Bigl[ ( 8 + \mrT^{d}_{8}\,\xphs ) \Bigr]\,\frac{1}{\lzz} \Bigr\}\,\frac{1}{c^6}
         \,\bfun{ - \mhs}{\mh}{\mh}
\nl &+&
         \frac{1}{384}
         \,\Bigl\{ \Bigl[ \frac{\xphs}{\lzz}\,\mrT^{d}_{59} 
         - 2\,( 1 - c^2\,\xphs )\,c^2 \Bigr]\,c^2\,\xphs 
         - 2\,\Bigl[ ( \frac{\xphs}{\lzz}\,\mrT^{d}_{64} 
         + 12\,\frac{1}{\lzz} + \mrT^{d}_{54} ) \Bigr] \Bigr\}\,\frac{1}{c^6}
\nl &\times& \bfun{ - \mzs}{\mh}{\mz}
\nl &+&
         \frac{1}{256}
         \,\Bigl\{ 4\,\Bigl[ ( 3 - \mrT^{d}_{8}\,\vbq ) \Bigr]\,\frac{1}{\lzz} 
         + \Bigl[ 2\,( 3 - \mrT^{d}_{8}\,\vbq ) - ( \mrT^{d}_{8}\,\mrV_{4}\,\vbq 
         + 3\,\mrV_{4} )\,c^2 \Bigr]\,c^2 \Bigr\}\,\frac{1}{c^6}\,\xpbs
\nl &\times& \cfun{ - \mhs}{ - \mzs}{ - \mzs}{\mb}{\mb}{\mb}
\nl &+&
         \frac{1}{256}
         \,\Bigl\{ 4\,\Bigl[ ( 3 - \mrT^{d}_{9}\,\vtq ) \Bigr]\,\frac{1}{\lzz} 
         + \Bigl[ 2\,( 3 - \mrT^{d}_{9}\,\vtq ) - ( \mrT^{d}_{9}\,\mrV_{3}\,\vtq 
         + 3\,\mrV_{3} )\,c^2 \Bigr]\,c^2 \Bigr\}\,\frac{1}{c^6}\,\xpts
\nl &\times& \cfun{ - \mhs}{ - \mzs}{ - \mzs}{\mt}{\mt}{\mt}
\nl &-&
         \frac{1}{768}
         \,\Bigl\{  - \Bigl[ ( 2\,\frac{\xphs}{\lzz}\,\mrT^{d}_{53} 
         + \mrT^{d}_{55} ) \Bigr] 
         + ( 9\,c^2 - \frac{\xphs}{\lzz})\,\mrT^{d}_{8} )\,c^2\,\xphs \Bigr\}\,\frac{1}{c^6}
         \,\bfun{ - \mhs}{\mz}{\mz}
\nl &+&
         \frac{1}{384}
         \,( 1 - 2\,c^2\,\xphs )\,\frac{1}{c^4}
         \,\afun{\mz}
\nl &-&
         \frac{1}{64}
         \,( 3 - \mrT^{d}_{8}\,\vbq )\,\frac{1}{c^4}\,\frac{\xpbs}{\lzz}
         \,\bfun{ - \mhs}{\mb}{\mb}
\nl &-&
         \frac{1}{64}
         \,( 3 - \mrT^{d}_{9}\,\vtq )\,\frac{1}{c^4}\,\frac{\xpts}{\lzz}
         \,\bfun{ - \mhs}{\mt}{\mt}
\nl &-&
         \frac{1}{384}
         \,( 15 - 2\,c^2\,\xphs )\,\frac{1}{c^2}\,\xphs
         \,\afun{\mh}
\nl &-&
         \frac{1}{768}
         \,( \mrT^{d}_{67} - 6\,\mrV_{7}\,c^4 )\,\frac{1}{c^6}
         \,\LR
\eqas
\bqas
\mcT^{\nfact}_{\ssD\,;\,\PZ\PZ}(\aAZ) &=&
         \frac{1}{288}
         \,\frac{s}{c^3}
         \,\mrX_{7}
     +
         \frac{1}{96}
         \,\frac{s}{c}
         \,\mrX_{8}
     +
         \frac{1}{16}
         \,\frac{c}{s}
         \,\mrX_{9}
     +
         \frac{1}{12}
         \,\mrT^{d}_{70}\,s\,c\,\xphs
         \,\afun{\mw}
\nl &+&
         \frac{5}{192}
         \,\Bigl[ 4 - ( 2\,\mrW_{10} + \mrW_{10}\,c^2\,\xphs )\,c^2\,\xphs \Bigr]\,\frac{s}{c^7}
         \,\cfun{ - \mhs}{ - \mzs}{ - \mzs}{\mz}{\mh}{\mz}
\nl &-&
         \frac{5}{64}
         \,\Bigl[ 4\,\frac{1}{\lzz} 
         + ( \frac{\xphs}{\lzz}\,c^2\,\xphs - \mrW_{8} )\,c^2 \Bigr]\,\frac{s}{c^7}\,\xphs
         \,\cfun{ - \mhs}{ - \mzs}{ - \mzs}{\mh}{\mz}{\mh}
\nl &-&
         \frac{1}{192}
         \,\Bigl[ 2\,\mrT^{d}_{82}\,c^4\,\xphs + \mrT^{d}_{83} 
         - 6\,( \mrT^{d}_{7}\,c^2\,\xphq - 9\,\mrV_{6} )\,s^2\,c^4 \Bigr]\,\frac{1}{s\,c^5}
         \,\LR
\nl &-&
         \frac{1}{96}
         \,\Bigl[ 4\,( \mrT^{d}_{72} + \mrT^{d}_{76}\,c^2\,\xphs )\,c^2 
         + ( \mrT^{d}_{79}\,\xphs - 2\,\mrT^{d}_{81} )\,\frac{1}{\lzz} \Bigr]\,\frac{s}{c^5}
         \,\bfun{ - \mzs}{\mw}{\mw}
\nl &+&
         \frac{1}{576}
         \,\Bigl\{ 3\,\mrT^{d}_{80} + \Bigl[ \mrT^{d}_{86}\,\xphs
         - 3\,( 8\,c^2\,\xphs\,\xpbs\,\vbq + 16\,c^2\,\xphs\,\xpts\,\vtq 
         - 2\,\mrT^{d}_{84}\,\xpbs\,\vbq - 2\,\mrT^{d}_{85}\,\xpts\,\vtq 
\nl &-& 9\,\mrU_{2}\,\xpts - 9\,\mrU_{4}\,\xpbs )\,c^2 \Bigr]\,c^2\Bigr\}\,\frac{s}{c^5}
\nl &-&
         \frac{1}{96}
         \,\Bigl\{ \Bigl[ 4\,c^2\,\xphs\,\xpbs\,\vbq 
         + ( 2\,\xphs\,\vbq + \mrU_{6}\,\xpbs ) \Bigr]\,c^2 
         - \Bigl[ ( 4\,\vbq - 6\,\frac{\xpbs}{\lzz}\,\mrT^{d}_{78}\,\vbq 
         - 9\,\frac{\xpbs}{\lzz}\,\mrU_{4} ) \Bigr] \Bigr\}\,\frac{s}{c^3}
\nl &\times& \bfun{ - \mzs}{\mb}{\mb}
\nl &-&
         \frac{1}{96}
         \,\Bigl\{ \Bigl[ 8\,c^2\,\xphs\,\xpts\,\vtq 
         + ( 4\,\xphs\,\vtq + \mrU_{5}\,\xpts ) \Bigr]\,c^2 
         - \Bigl[ ( 8\,\vtq - 6\,\frac{\xpts}{\lzz}\,\mrT^{d}_{77}\,\vtq 
         - 9\,\frac{\xpts}{\lzz}\,\mrU_{2} ) \Bigr] \Bigr\}\,\frac{s}{c^3}
\nl &\times& \bfun{ - \mzs}{\mt}{\mt}
\nl &-&
         \frac{1}{96}
         \,\Bigl\{ \Bigl[ \mrT^{d}_{73} 
         - 3\,( \mrT^{d}_{7}\,c^2\,\xphs - \mrT^{d}_{71} )\,c^2\,\xphs \Bigr]\,c^2 
         - ( \mrT^{d}_{79}\,\xphs - 2\,\mrT^{d}_{81} )\,\frac{1}{\lzz} \Bigr\}\,
         \frac{s}{c^5}
\nl &\times& \bfun{ - \mhs}{\mw}{\mw}
\nl &+&
         \frac{1}{96}
         \,\Bigl\{ \Bigl[ 6\,\mrT^{d}_{74} 
         + ( 24\,c^6\,\xphs - \mrT^{d}_{75} )\,c^2\,\xphs \Bigr]\,c^2 
         - ( \mrT^{d}_{79}\,\xphs - 2\,\mrT^{d}_{81} )\,\frac{1}{\lzz} \Bigr\}\,\frac{s}{c^7}
\nl &\times& \cfun{ - \mhs}{ - \mzs}{ - \mzs}{\mw}{\mw}{\mw}
\nl &-&
         \frac{1}{128}
         \,\Bigl\{ 4\,\Bigl[ ( 2\,\mrT^{d}_{77}\,\vtq + 3\,\mrU_{2} ) \Bigr]\,\frac{1}{\lzz} 
         + \Bigl[ 2\,( 2\,\mrT^{d}_{77}\,\vtq + 3\,\mrU_{2} ) 
         + ( 2\,\mrT^{d}_{77}\,\mrV_{3}\,\vtq - 3\,\mrV_{3}\,\mrU_{1} )\,c^2 \Bigr]\,c^2 \Bigr\}\,
         \frac{s}{c^5}\,\xpts
\nl &\times& \cfun{ - \mhs}{ - \mzs}{ - \mzs}{\mt}{\mt}{\mt}
\nl &-&
         \frac{1}{128}
         \,\Bigl\{ 4\,\Bigl[ ( 2\,\mrT^{d}_{78}\,\vbq + 3\,\mrU_{4} ) \Bigr]\,\frac{1}{\lzz} 
         + \Bigl[ 2\,( 2\,\mrT^{d}_{78}\,\vbq + 3\,\mrU_{4} ) 
         + ( 2\,\mrT^{d}_{78}\,\mrV_{4}\,\vbq - 3\,\mrV_{4}\,\mrU_{3} )\,c^2 \Bigr]\,c^2 \Bigr\}\,
         \frac{s}{c^5}\,\xpbs
\nl &\times& \cfun{ - \mhs}{ - \mzs}{ - \mzs}{\mb}{\mb}{\mb}
\nl &+&
         \frac{1}{48}
         \,( 2 - c^2\,\xphs )\,\frac{s}{c^3}\,\vle
         \,\bfun{ - \mzs}{0}{0}
\nl &+&
         \frac{5}{64}
         \,( 2 - c^2\,\xphs )\,\frac{s}{c^5}\,\frac{\xphs}{\lzz}
         \,\bfun{ - \mhs}{\mh}{\mh}
\nl &+&
         \frac{1}{24}
         \,( 2 - c^2\,\xphs )\,\frac{s}{c}\,\xpbs\,\vbq
         \,\afun{\mb}
\nl &+&
         \frac{1}{12}
         \,( 2 - c^2\,\xphs )\,\frac{s}{c}\,\xpts\,\vtq
         \,\afun{\mt}
\nl &+&
         \frac{5}{192}
        \,( s\mrW_{17} + \frac{\xphs}{\lzz}\,c^2\,\xphs )\,\frac{s}{c^5}
         \,\bfun{ - \mhs}{\mz}{\mz}
\nl &-&
         \frac{5}{96}
         \,( 2\,s\mrW_{17} - \frac{\xphs}{\lzz}\,c^2\,\xphs )\,\frac{s}{c^5}
         \,\bfun{ - \mzs}{\mh}{\mz}
\nl &+&
         \frac{1}{32}
         \,( 2\,\mrT^{d}_{77}\,\vtq + 3\,\mrU_{2} )\,\frac{s}{c^3}\,\frac{\xpts}{\lzz}
         \,\bfun{ - \mhs}{\mt}{\mt}
\nl &+&
         \frac{1}{32}
         \,( 2\,\mrT^{d}_{78}\,\vbq + 3\,\mrU_{4} )\,\frac{s}{c^3}\,\frac{\xpbs}{\lzz}
         \,\bfun{ - \mhs}{\mb}{\mb}
\nl &-&
         \frac{1}{48}
         \,( 2\,\mrU_{0} - \mrU_{0}\,c^2\,\xphs )\,\frac{s}{c^3}
         \,( 1 - 3\,\LR )
\eqas
\bqas
\mcT^{\nfact}_{\ssD\,;\,\PZ\PZ}(\aAA) &=&
         \frac{1}{32}
         \,\mrX_{11}
     +
         \frac{1}{32}
         \,\frac{s^2}{c^2}
         \,\mrX_{10}
\nl &+&
         \frac{3}{32}
         \,\Bigl[ 4\,\frac{1}{\lzz} 
         + ( \frac{\xphs}{\lzz}\,c^2\,\xphs - \mrW_{8} )\,c^2 \Bigr]\,\frac{s^2}{c^6}\,\xphs
         \,\cfun{ - \mhs}{ - \mzs}{ - \mzs}{\mh}{\mz}{\mh}
\nl &-&
         \frac{1}{32}
         \,\Bigl[ 4\,\mrT^{d}_{29} 
         - ( 2\,\mrW_{10} + \mrW_{10}\,c^2\,\xphs )\,s^2\,c^2\,\xphs \Bigr]\,\frac{1}{c^6}
         \,\cfun{ - \mhs}{ - \mzs}{ - \mzs}{\mz}{\mh}{\mz}
\nl &-&
         \frac{1}{32}
         \,\Bigl[ \mrT^{d}_{90} + \mrT^{d}_{96}\,c^2\,\xphs 
         + ( \mrT^{d}_{8}\,\xpbs\,\vbq + \mrT^{d}_{9}\,\xpts\,\vtq )\,s^2\,c^2 \Bigr]\,\frac{1}{c^4}
\nl &+&
         \frac{1}{16}
         \,\Bigl[ ( \mrT^{d}_{71}\,\xphs - 4\,\mrT^{d}_{87} )\,\frac{1}{\lzz} 
         + ( 2\,\mrT^{d}_{88} + \mrT^{d}_{92}\,c^2\,\xphs )\,c^2 \Bigr]\,\frac{s^2}{c^4}
         \,\bfun{ - \mzs}{\mw}{\mw}
\nl &-&
         \frac{1}{16}
         \,\Bigl\{ 3\,c^2 + \Bigl[ ( 3 - \mrT^{d}_{8}\,\vbq ) \Bigr]
         \,\frac{1}{\lzz} \Bigr\}\,\frac{s^2}{c^4}\,\xpbs
         \,\bfun{ - \mzs}{\mb}{\mb}
\nl &-&
         \frac{1}{16}
         \,\Bigl\{ 3\,c^2 + \Bigl[ ( 3 - \mrT^{d}_{9}\,\vtq ) \Bigr]
         \,\frac{1}{\lzz} \Bigr\}\,\frac{s^2}{c^4}\,\xpts
         \,\bfun{ - \mzs}{\mt}{\mt}
\nl &-&
         \frac{1}{16}
         \,\Bigl\{ \Bigl[ c^2\,\xphq + ( 4 - \mrT^{d}_{93}\,\xphs ) \Bigr]\,c^4 
         + ( \mrT^{d}_{71}\,\xphs - 4\,\mrT^{d}_{87} )\,\frac{1}{\lzz} \Bigr\}\,\frac{s^2}{c^4}
         \,\bfun{ - \mhs}{\mw}{\mw}
\nl &+&
         \frac{1}{16}
         \,\Bigl\{  - \Bigl[ 2\,\mrT^{d}_{89} 
         - ( 4\,c^4\,\xphs + \mrT^{d}_{95} )\,c^2\,\xphs \Bigr]\,c^2 
         + ( \mrT^{d}_{71}\,\xphs - 4\,\mrT^{d}_{87} )\,\frac{1}{\lzz} \Bigr\}\,\frac{s^2}{c^6}
\nl &\times& \cfun{ - \mhs}{ - \mzs}{ - \mzs}{\mw}{\mw}{\mw}
\nl &-&
         \frac{1}{64}
         \,\Bigl\{ 4\,\Bigl[ ( 3 - \mrT^{d}_{8}\,\vbq ) \Bigr]\,\frac{1}{\lzz} 
         + \Bigl[ 2\,( 3 - \mrT^{d}_{8}\,\vbq ) - ( \mrT^{d}_{8}\,\mrV_{4}\,\vbq 
         + 3\,\mrV_{4} )\,c^2 \Bigr]\,c^2 \Bigr\}\,\frac{s^2}{c^6}\,\xpbs
\nl &\times& \cfun{ - \mhs}{ - \mzs}{ - \mzs}{\mb}{\mb}{\mb}
\nl &-&
         \frac{1}{64}
         \,\Bigl\{ 4\,\Bigl[ ( 3 - \mrT^{d}_{9}\,\vtq ) \Bigr]\,\frac{1}{\lzz} 
         + \Bigl[ 2\,( 3 - \mrT^{d}_{9}\,\vtq ) - ( \mrT^{d}_{9}\,\mrV_{3}\,\vtq 
         + 3\,\mrV_{3} )\,c^2 \Bigr]\,c^2 \Bigr\}\,\frac{s^2}{c^6}\,\xpts
\nl &\times& \cfun{ - \mhs}{ - \mzs}{ - \mzs}{\mt}{\mt}{\mt}
\nl &-&
         \frac{3}{32}
         \,( 2 - c^2\,\xphs )\,\frac{s^2}{c^4}\,\frac{\xphs}{\lzz}
         \,\bfun{ - \mhs}{\mh}{\mh}
\nl &+&
         \frac{1}{16}
         \,( 3 - \mrT^{d}_{8}\,\vbq )\,\frac{s^2}{c^4}\,\frac{\xpbs}{\lzz}
         \,\bfun{ - \mhs}{\mb}{\mb}
\nl &+&
         \frac{1}{16}
         \,( 3 - \mrT^{d}_{9}\,\vtq )\,\frac{s^2}{c^4}\,\frac{\xpts}{\lzz}
         \,\bfun{ - \mhs}{\mt}{\mt}
\nl &-&
         \frac{1}{32}
         \,( s\mrW_{17} + \frac{\xphs}{\lzz}\,c^2\,\xphs )\,\frac{s^2}{c^4}
         \,\bfun{ - \mhs}{\mz}{\mz}
\nl &+&
         \frac{1}{16}
         \,( 2\,s\mrW_{17} - \frac{\xphs}{\lzz}\,c^2\,\xphs )\,\frac{s^2}{c^4}
         \,\bfun{ - \mzs}{\mh}{\mz}
\nl &-&
         \frac{1}{32}
         \,( 2\,s^2\,c^6\,\xphq - 4\,\mrT^{d}_{91}\,c^4\,\xphs - \mrT^{d}_{94} )\,\frac{1}{c^4}
         \,\LR
\eqas
\bqas
\mcT^{\nfact}_{\ssD\,;\,\PZ\PZ}(\aZZ) &=&
         \frac{1}{32}
         \,\mrX_{14}
     +
         \frac{1}{32}
         \,\xphs
         \,\mrX_{13}
     +
         \frac{1}{32}
         \,\frac{1}{c^2}
         \,\mrX_{12}
\nl &+&
         \frac{1}{32}
         \,\Bigl[ 2\,\frac{\xphs}{\lzz}\,s^2\,c^2\,\xphs 
         - ( 10 - c^2\,\xphs )\,c^2\,\xphs + 2\,( \frac{\xphs}{\lzz}\,\mrT^{d}_{8} 
         + 2\,\mrT^{d}_{107} ) \Bigr]\,\frac{1}{c^4}
         \,\bfun{ - \mzs}{\mh}{\mz}
\nl &+&
         \frac{1}{128}
         \,\Bigl[ 4\,\frac{\xphs}{\lzz}\,s^2\,c^2\,\xphs - 4\,\mrT^{d}_{110}\,s\mrW_{17} 
         - ( 10 + c^2\,\xphs )\,c^2\,\xphs \Bigr]\,\frac{1}{c^4}
         \,\bfun{ - \mhs}{\mz}{\mz}
\nl &-&
         \frac{3}{32}
         \,\Bigl[ 4\,\frac{1}{\lzz}\,\mrT^{d}_{29} 
         - ( 4 - c^2\,\xphs )\,\frac{\xphs}{\lzz}\,s^2\,c^2 
         - ( 2\,c^2\,\xphs - \mrT^{d}_{107} )\,c^2 \Bigr]\,\frac{1}{c^6}\,\xphs
\nl &\times& \cfun{ - \mhs}{ - \mzs}{ - \mzs}{\mh}{\mz}{\mh}
\nl &+&
         \frac{3}{128}
         \,\Bigl[ 4\,( 2 - c^2\,\xphs )\,\frac{1}{\lzz}\,s^2 
         + ( 10 - c^2\,\xphs )\,c^2 \Bigr]\,\frac{1}{c^4}\,\xphs
         \,\bfun{ - \mhs}{\mh}{\mh}
\nl &-&
         \frac{1}{32}
         \,\Bigl\{ 9\,c^2 + \Bigl[ ( 2\,\mrT^{d}_{42}\,\vtq 
         + 3\,\mrU_{2} ) \Bigr]\,\frac{1}{\lzz} \Bigr\}\,\frac{1}{c^2}\,\xpts
         \,\bfun{ - \mzs}{\mt}{\mt}
\nl &-&
         \frac{1}{32}
         \,\Bigl\{ 9\,c^2 + \Bigl[ ( 2\,\mrT^{d}_{112}\,\vbq 
         + 3\,\mrU_{4} ) \Bigr]\,\frac{1}{\lzz} \Bigr\}\,\frac{1}{c^2}\,\xpbs
         \,\bfun{ - \mzs}{\mb}{\mb}
\nl &-&
         \frac{1}{32}
         \,\Bigl\{ \frac{\xphs}{\lzz}\,s^2\,c^4\,\xphq + 4\,\mrT^{d}_{106} 
         - \Bigl[ \mrT^{d}_{107}\,c^2\,\xphs + 2\,( \frac{\xphs}{lzz}\,\mrT^{d}_{110} 
         + \mrT^{d}_{107} ) \Bigr]\,c^2\,\xphs \Bigr\}\,\frac{1}{c^6}
\nl &\times& \cfun{ - \mhs}{ - \mzs}{ - \mzs}{\mz}{\mh}{\mz}
\nl &+&
         \frac{1}{64}
         \,\Bigl\{ 4\,\mrT^{d}_{97}\,\xphs + \Bigl[ ( 2\,\mrT^{d}_{42}\,\xpts\,\vtq 
         - 32\,\mrT^{d}_{107} + 2\,\mrT^{d}_{112}\,\xpbs\,\vbq 
         + 3\,\mrU_{2}\,\xpts + 3\,\mrU_{4}\,\xpbs ) \Bigr]\,c^2 \Bigr\}\,\frac{1}{c^2}
\nl &+&
         \frac{1}{64}
         \,\Bigl\{ 2\,\mrT^{d}_{111} + \Bigl[ 2\,\mrT^{d}_{103}\,\xphs 
         - ( \mrT^{d}_{108}\,\xphq + 18\,\mrV_{6} )\,c^2 \Bigr]\,c^2 \Bigr\}\,\frac{1}{c^4}
         \,\LR
\nl &-&
         \frac{1}{64}
         \,\Bigl\{ \Bigl[ \mrT^{d}_{100}\,c^2\,\xphq + 2\,( 2\,\mrT^{d}_{100} 
         - \mrT^{d}_{102}\,\xphs ) \Bigr]\,c^2 + 4\,\Bigl[ ( 6\,\mrT^{d}_{98} 
         - \mrT^{d}_{101}\,\xphs ) \Bigr]\,\frac{1}{\lzz} \Bigr\}\,\frac{1}{c^4}
\nl &\times& \bfun{ - \mhs}{\mw}{\mw}
\nl &+&
         \frac{1}{16}
         \,\Bigl\{ \Bigl[ 2\,\mrT^{d}_{104} 
         + ( 4\,c^6\,\xphs - \mrT^{d}_{105} )\,c^2\,\xphs \Bigr]\,c^2 
         + \Bigl[ ( 6\,\mrT^{d}_{98} - \mrT^{d}_{101}\,\xphs ) \Bigr]\,\frac{1}{\lzz} \Bigr\}\,
         \frac{1}{c^6}
\nl &\times& \cfun{ - \mhs}{ - \mzs}{ - \mzs}{\mw}{\mw}{\mw}
\nl &-&
         \frac{1}{128}
         \,\Bigl\{ 4\,\Bigl[ ( 2\,\mrT^{d}_{42}\,\vtq + 3\,\mrU_{2} ) \Bigr]\,\frac{1}{\lzz} 
         + \Bigl[ 2\,( 2\,\mrT^{d}_{42}\,\vtq + 3\,\mrU_{2} ) 
         + ( 2\,\mrT^{d}_{42}\,\mrV_{3}\,\vtq - 3\,\mrV_{3}\,\mrU_{1} )\,c^2 \Bigr]\,c^2 \Bigr\}\,
         \frac{1}{c^4}\,\xpts
\nl &\times& \cfun{ - \mhs}{ - \mzs}{ - \mzs}{\mt}{\mt}{\mt}
\nl &+&
         \frac{1}{16}
         \,\Bigl\{ \Bigl[ ( 6\,\mrT^{d}_{98} - \mrT^{d}_{101}\,\xphs ) \Bigr]\,\frac{1}{\lzz} 
         + ( 2\,\mrT^{d}_{99} - \mrT^{d}_{109}\,c^2\,\xphs )\,c^2 \Bigr\}\,\frac{1}{c^4}
         \,\bfun{ - \mzs}{\mw}{\mw}
\nl &-&
         \frac{1}{128}
         \,\Bigl\{ 4\,\Bigl[ ( 2\,\mrT^{d}_{112}\,\vbq + 3\,\mrU_{4} ) \Bigr]\,\frac{1}{\lzz} 
         + \Bigl[ 2\,( 2\,\mrT^{d}_{112}\,\vbq + 3\,\mrU_{4} ) 
         + ( 2\,\mrT^{d}_{112}\,\mrV_{4}\,\vbq - 3\,\mrV_{4}\,\mrU_{3} )\,c^2 \Bigr]\,c^2
         \Bigl\}
         \,\frac{1}{c^4}\,\xpbs
\nl &\times& \cfun{ - \mhs}{ - \mzs}{ - \mzs}{\mb}{\mb}{\mb}
\nl &-&
         \frac{1}{32}
         \,( 4 - c^2\,\xphs )\,\frac{1}{c^2}\,\xphs
         \,\afun{\mh}
\nl &+&
         \frac{1}{32}
         \,( 4 - c^2\,\xphs )\,\frac{1}{c^4}
         \,\afun{\mz}
\nl &+&
         \frac{1}{32}
         \,( 2\,\mrT^{d}_{42}\,\vtq + 3\,\mrU_{2} )\,\frac{1}{c^2}\,\frac{\xpts}{\lzz}
         \,\bfun{ - \mhs}{\mt}{\mt}
\nl &+&
         \frac{1}{32}
         \,( 2\,\mrT^{d}_{112}\,\vbq + 3\,\mrU_{4} )\,\frac{1}{c^2}\,\frac{\xpbs}{\lzz}
         \,\bfun{ - \mhs}{\mb}{\mb}
\eqas
\bqas
\mcT^{\nfact}_{\ssD\,;\,\PZ\PZ}(\ren) &=&
         \frac{1}{32}
         \,\frac{1}{c^2}
         \,\mrX_{15}
\eqas
\bqas
\mcT^{\nfact}_{\ssP\,;\,\PZ\PZ}(\aptV) &=&
         \frac{3}{16}
         \,\frac{\vtq}{c^2}\,\mrW_{19}\,\xpts
\nl &-&
         \frac{3}{32}
         \,\Bigl[ 12\,\frac{1}{\lzzs} 
         + ( \mrW_{18} + 2\,\mrW_{20}\,c^2 )\,c^2 \Bigr]\,\frac{\vtq}{c^6}\,\xpts
         \,\cfun{ - \mhs}{ - \mzs}{ - \mzs}{\mt}{\mt}{\mt}
\nl &+&
         \frac{3}{32}
         \,(  - c^2\,s\mrW_{18} + 12\,\frac{1}{\lzzs} )\,\frac{\vtq}{c^4}\,\xpts
         \,\bfun{ - \mhs}{\mt}{\mt}
\nl &-&
         \frac{3}{32}
         \,(  - c^2\,s\mrW_{18} + 12\,\frac{1}{\lzzs} )\,\frac{\vtq}{c^4}\,\xpts
         \,\bfun{ - \mzs}{\mt}{\mt}
\eqas
\bqas
\mcT^{\nfact}_{\ssP\,;\,\PZ\PZ}(\aptA) &=&
         \frac{3}{16}
         \,\frac{1}{c^2}\,\mrW_{19}\,\xpts
\nl &-&
         \frac{3}{32}
         \,\Bigl[ 12\,\frac{1}{\lzzs} 
         + ( \mrW_{22} + 2\,\mrW_{23}\,c^2 )\,c^2 \Bigr]\,\frac{1}{c^6}\,\xpts
         \,\cfun{ - \mhs}{ - \mzs}{ - \mzs}{\mt}{\mt}{\mt}
\nl &+&
         \frac{3}{32}
         \,( c^2\,s\mrW_{21} + 12\,\frac{1}{\lzzs} )\,\frac{1}{c^4}\,\xpts
         \,\bfun{ - \mhs}{\mt}{\mt}
\nl &-&
         \frac{3}{32}
        \,( c^2\,s\mrW_{21} + 12\,\frac{1}{\lzzs} )\,\frac{1}{c^4}\,\xpts
         \,\bfun{ - \mzs}{\mt}{\mt}
\eqas
\bqas
\mcT^{\nfact}_{\ssP\,;\,\PZ\PZ}(\apbV) &=&
         \frac{3}{16}
         \,\frac{\vbq}{c^2}\,\mrW_{19}\,\xpbs
\nl &-&
         \frac{3}{32}
         \,\Bigl[ 12\,\frac{1}{\lzzs} 
         + ( \mrW_{18} + 2\,\mrW_{24}\,c^2 )\,c^2 \Bigr]\,\frac{\vbq}{c^6}\,\xpbs
         \,\cfun{ - \mhs}{ - \mzs}{ - \mzs}{\mb}{\mb}{\mb}
\nl &+&
         \frac{3}{32}
         \,(  - c^2\,s\mrW_{18} + 12\,\frac{1}{\lzzs} )\,\frac{\vbq}{c^4}\,\xpbs
         \,\bfun{ - \mhs}{\mb}{\mb}
\nl &-&
         \frac{3}{32}
         \,(  - c^2\,s\mrW_{18} + 12\,\frac{1}{\lzzs} )\,\frac{\vbq}{c^4}\,\xpbs
         \,\bfun{ - \mzs}{\mb}{\mb}
\eqas
\bqas
\mcT^{\nfact}_{\ssP\,;\,\PZ\PZ}(\apbA) &=&
         \frac{3}{16}
         \,\frac{1}{c^2}\,\mrW_{19}\,\xpbs
\nl &-&
         \frac{3}{32}
         \,\Bigl[ 12\,\frac{1}{\lzzs} + ( \mrW_{22} 
         + 2\,\mrW_{25}\,c^2 )\,c^2 \Bigr]\,\frac{1}{c^6}\,\xpbs
         \,\cfun{ - \mhs}{ - \mzs}{ - \mzs}{\mb}{\mb}{\mb}
\nl &+&
         \frac{3}{32}
         \,( 12\,\frac{1}{\lzzs} + \mrW_{21}\,c^2 )\,\frac{1}{c^4}\,\xpbs
         \,\bfun{ - \mhs}{\mb}{\mb}
\nl &-&
         \frac{3}{32}
         \,( 12\,\frac{1}{\lzzs} + \mrW_{21}\,c^2 )\,\frac{1}{c^4}\,\xpbs
         \,\bfun{ - \mzs}{\mb}{\mb}
\eqas
\bqas
\mcT^{\nfact}_{\ssP\,;\,\PZ\PZ}(\atBW) &=&
         \frac{3}{32}
         \,\Bigl\{ 6\,\frac{1}{\lzzs} + \Bigl[ \frac{1}{\lzz} 
         + ( - c^2 
         + 2\,\frac{1}{\lzz} )\,c^2\,\xpts \Bigr]\,c^2 \Bigr\}\,\frac{1}{c^7}\,\xpts\,\vtq
\nl &\times& \cfun{ - \mhs}{ - \mzs}{ - \mzs}{\mt}{\mt}{\mt}
\nl &-&
         \frac{3}{64}
         \,( c^2 + 2\,\frac{1}{\lzz} )\,\frac{\vtq}{c^3}\,\xpts
\nl &-&
         \frac{3}{64}
         \,(  - c^4 + 12\,\frac{1}{\lzzs} )\,\frac{\vtq}{c^5}\,\xpts
         \,\bfun{ - \mhs}{\mt}{\mt}
\nl &+&
         \frac{3}{64}
         \,(  - c^4 + 12\,\frac{1}{\lzzs} )\,\frac{\vtq}{c^5}\,\xpts
         \,\bfun{ - \mzs}{\mt}{\mt}
\eqas
\bqas
\mcT^{\nfact}_{\ssP\,;\,\PZ\PZ}(\abBW) &=&
       -
         \frac{3}{32}
         \,\Bigl\{ 6\,\frac{1}{\lzzs} + \Bigl[ \frac{1}{\lzz} 
         + ( - c^2 
         + 2\,\frac{1}{\lzz} )\,c^2\,\xpbs \Bigr]\,c^2 \Bigr\}\,\frac{1}{c^7}\,\xpbs\,\vbq
\nl &\times& \cfun{ - \mhs}{ - \mzs}{ - \mzs}{\mb}{\mb}{\mb}
\nl &+&
         \frac{3}{64}
         \,( c^2 + 2\,\frac{1}{\lzz} )\,\frac{\vbq}{c^3}\,\xpbs
\nl &+&
         \frac{3}{64}
         \,(  - c^4 + 12\,\frac{1}{\lzzs} )\,\frac{\vbq}{c^5}\,\xpbs
         \,\bfun{ - \mhs}{\mb}{\mb}
\nl &-&
         \frac{3}{64}
         \,(  - c^4 + 12\,\frac{1}{\lzzs} )\,\frac{\vbq}{c^5}\,\xpbs
         \,\bfun{ - \mzs}{\mb}{\mb}
\eqas
\bqas
\mcT^{\nfact}_{\ssP\,;\,\PZ\PZ}(\ap) &=&
         \frac{3}{16}
         \,\frac{1}{c^2}\,\mrW_{18}
         \,\afun{\mz}
     +
         \frac{3}{16}
         \,\frac{1}{c^2}\,\mrW_{19}
     +
         \frac{3}{16}
         \,\mrW_{9}
         \,\afun{\mh}
\nl &+&
         \frac{3}{32}
         \,\Bigl[ 24\,\frac{1}{\lzzs} - ( 12\,\frac{\xphs}{\lzzs} 
         - \mrW_{9}\,c^2 )\,c^2 \Bigr]\,\frac{1}{c^4}
         \,\bfun{ - \mhs}{\mh}{\mh}
\nl &-&
         \frac{3}{32}
         \,\Bigl[ 24\,\frac{1}{\lzzs} - ( \mrW_{9}\,c^2
         + 2\,\mrW_{27} )\,c^2 \Bigr]\,\frac{1}{c^4}
         \,\bfun{ - \mzs}{\mh}{\mz}
\nl &-&
         \frac{3}{32}
         \,\Bigl\{ 48\,\frac{1}{\lzzs} + \Bigl[ 8\,\frac{1}{\lzz}\,\mrW_{26} 
         - ( \mrW_{9}\,c^2\,\xphs - 4\,\mrW_{25} )\,c^2 \Bigr]\,c^2 \Bigr\}\,\frac{1}{c^6}
         \,\cfun{ - \mhs}{ - \mzs}{ - \mzs}{\mh}{\mz}{\mh}
\eqas
\bqas
\mcT^{\nfact}_{\ssP\,;\,\PZ\PZ}(\apBox) &=&
         \frac{3}{64}
         \,\frac{1}{c^2}\,\mrW_{9}
         \,\afun{\mz}
     -
         \frac{3}{64}
         \,\mrW_{9}\,\xphs
         \,\afun{\mh}
\nl &+&
         \frac{1}{128}
         \,\Bigl[ 8\,\frac{1}{\lzz}\,s\mrW_{26} 
         - ( \mrW_{9}\,c^2\,\xphs - 2\,\mrW_{32} )\,c^2 \Bigr]\,\frac{1}{c^4}
         \,\bfun{ - \mhs}{\mz}{\mz}
\nl &+&
         \frac{1}{128}
         \,\Bigl[ 16\,\frac{1}{\lzz}\,s\mrW_{11} 
         + ( \mrW_{9}\,c^2\,\xphs + 4\,\mrW_{30} )\,c^2 \Bigr]\,\frac{1}{c^4}\,\xphs
         \,\cfun{ - \mhs}{ - \mzs}{ - \mzs}{\mz}{\mh}{\mz}
\nl &-&
         \frac{1}{64}
         \,\Bigl\{ 48\,\frac{1}{\lzzs} + \Bigl[ 12\,\frac{1}{\lzz}
         + (  - 4\,\frac{1}{\lzz}\,\mrW_{33} 
         + \mrW_{9}\,c^2 )\,c^2\,\xphs \Bigr]\,c^2 \Bigr\}\,\frac{1}{c^6}
         \,\bfun{ - \mhs}{\mh}{\mh}
\nl &+&
         \frac{1}{128}
         \,\Bigl\{ 96\,\frac{1}{\lzzs} + \Bigl[ 16\,\frac{1}{\lzz}\,\mrW_{11} 
         - ( 3\,\mrW_{9}\,c^2\,\xphs - 4\,\mrW_{28} )\,c^2 \Bigr]\,c^2 \Bigr\}\,\frac{1}{c^6}
         \,\bfun{ - \mzs}{\mh}{\mz}
\nl &+&
         \frac{1}{64}
         \,\Bigl\{ 96\,\frac{1}{\lzzs} + \Bigl[ 8\,\frac{1}{\lzz}\,\mrW_{34} 
         - ( \mrW_{9}\,c^4\,\xphq - 4\,\mrW_{29} 
         - 4\,\mrW_{31}\,c^2\,\xphs )\,c^2 \Bigr]\,c^2 \Bigr\}\,\frac{1}{c^8}
\nl &\times& \cfun{ - \mhs}{ - \mzs}{ - \mzs}{\mh}{\mz}{\mh}
\nl &-&
         \frac{1}{32}
         \,\Bigl\{ 12\,\frac{1}{\lzzs}\,\mrT^{d}_{117} 
         - \Bigl[ ( 12\,\frac{\xphs}{\lzz}\,\mrT^{d}_{52} - \mrT^{d}_{120} )\,\frac{1}{\lzz} 
         + ( \mrT^{d}_{52}\,\mrW_{9} 
         + \mrT^{d}_{113}\,\mrV_{9} )\,c^2 \Bigr]\,c^2 \Bigr\}\,\frac{1}{c^6}
\nl &\times& \bfun{ - \mhs}{\mw}{\mw}
\nl &+&
         \frac{1}{32}
         \,\Bigl\{ 12\,\frac{1}{\lzzs}\,\mrT^{d}_{117} 
         - \Bigl[ ( 12\,\frac{\xphs}{\lzz}\,\mrT^{d}_{52} - \mrT^{d}_{120} )\,\frac{1}{\lzz} 
         + ( \mrT^{d}_{52}\,\mrW_{9} 
         + \mrT^{d}_{113}\,\mrV_{9} )\,c^2 \Bigr]\,c^2 \Bigr\}\,\frac{1}{c^6}
\nl &\times& \bfun{ - \mzs}{\mw}{\mw}
\nl &+&
         \frac{1}{32}
         \,\Bigl\{ 12\,\frac{1}{\lzzs}\,\mrT^{d}_{117} 
         - \Bigl[ ( 12\,\frac{\xphs}{\lzz}\,\mrT^{d}_{52} - \mrT^{d}_{121} )\,\frac{1}{\lzz} 
         - (  - \frac{\xphs}{\lzz}\,\mrT^{d}_{115} 
         + \mrT^{d}_{114} + \mrT^{d}_{116}\,\mrV_{9} )\,c^2 \Bigr]\,c^2 \Bigr\}\,\frac{1}{c^8}
\nl &\times& \cfun{ - \mhs}{ - \mzs}{ - \mzs}{\mw}{\mw}{\mw}
\nl &-&
         \frac{1}{64}
         \,\Bigl\{ 4\,\frac{1}{\lzz}\,\mrT^{d}_{119} 
         + \Bigl[ ( 5 - \frac{\xphs}{\lzz}\,\mrT^{d}_{118} 
         - 4\,\mrT^{d}_{113}\,\mrV_{9} ) \Bigr]\,c^2 \Bigr\}\,\frac{1}{c^4}
\eqas
\bqas
\mcT^{\nfact}_{\ssP\,;\,\PZ\PZ}(\apD) &=&
         \frac{1}{512}
         \,\Bigl[ 8\,s\mrW_{45} 
         - ( 5\,\mrW_{9}\,c^2\,\xphs - 2\,\mrW_{37} )\,c^2 \Bigr]\,\frac{1}{c^4}
         \,\bfun{ - \mhs}{\mz}{\mz}
\nl &-&
         \frac{1}{512}
         \,\Bigl[ 16\,s\mrW_{39} 
         - ( 5\,\mrW_{9}\,c^2\,\xphs - 4\,\mrW_{36} )\,c^2\,\xphs \Bigr]\,\frac{1}{c^4}
         \,\cfun{ - \mhs}{ - \mzs}{ - \mzs}{\mz}{\mh}{\mz}
\nl &+&
         \frac{1}{128}
         \,\Bigl[ ( \frac{\xphs}{\lzz}\,\mrT^{d}_{128} 
         + \mrT^{d}_{8}\,\mrW_{19}\,\xpbs\,\vbq + \mrT^{d}_{9}\,\mrW_{19}\,\xpts\,\vtq 
         + \mrT^{d}_{129} - 3\,\mrW_{47} )\,c^2 
\nl &2&\,( \frac{1}{\lzz}\,\mrT^{d}_{125} 
         + \mrT^{d}_{123}\,\mrV_{9} ) \Bigr]\,\frac{1}{c^4}
\nl &-&
         \frac{1}{256}
         \,\Bigl\{ 48\,\frac{1}{\lzzs} + \Bigl[ 12\,\frac{1}{\lzz}\,\mrW_{42} 
         - ( 6\,\mrW_{9}\,c^2\,\xphs - \mrW_{44} )\,c^2 \Bigr]\,c^2 \Bigr\}\,\frac{1}{c^6}
         \,\bfun{ - \mzs}{\mh}{\mz}
\nl &+&
         \frac{1}{512}
         \,\Bigl\{ 96\,\frac{1}{\lzzs} + \Bigl[ 24\,\frac{1}{\lzz}\,\mrW_{38} 
         + (  - 4\,\frac{1}{\lzz}\,\mrW_{40} 
         + 17\,\mrW_{9}\,c^2 )\,c^2\,\xphs \Bigr]\,c^2 \Bigr\}\,\frac{1}{c^6}
         \,\bfun{ - \mhs}{\mh}{\mh}
\nl &-&
         \frac{1}{512}
         \,\Bigl\{ 192\,\frac{1}{\lzzs} + \Bigl[ 16\,\frac{1}{\lzz}\,\mrW_{41} 
         - ( 17\,\mrW_{9}\,c^4\,\xphq - 8\,\mrW_{35} 
         - 4\,\mrW_{43}\,c^2\,\xphs )\,c^2 \Bigr]\,c^2 \Bigr\}\,\frac{1}{c^8}
\nl &\times& \cfun{ - \mhs}{ - \mzs}{ - \mzs}{\mh}{\mz}{\mh}
\nl &+&
         \frac{1}{128}
         \,\Bigl\{ 12\,\frac{1}{\lzzs}\,\mrT^{d}_{124} 
         + \Bigl[ ( 24\,\frac{\xphs}{\lzz}\,\mrT^{d}_{52} + \mrT^{d}_{126} )\,\frac{1}{\lzz} 
         + ( 2\,\mrT^{d}_{52}\,\mrW_{9} 
         + 5\,\mrT^{d}_{113}\,\mrV_{9} )\,c^2 \Bigr]\,c^2 \Bigr\}\,\frac{1}{c^6}
\nl &\times& \bfun{ - \mhs}{\mw}{\mw}
\nl &-&
         \frac{1}{128}
         \,\Bigl\{ 12\,\frac{1}{\lzzs}\,\mrT^{d}_{124} 
         + \Bigl[ ( 24\,\frac{\xphs}{\lzz}\,\mrT^{d}_{52} + \mrT^{d}_{126} )\,\frac{1}{\lzz} 
         + ( 2\,\mrT^{d}_{52}\,\mrW_{9} 
         + 5\,\mrT^{d}_{113}\,\mrV_{9} )\,c^2 \Bigr]\,c^2 \Bigr\}\,\frac{1}{c^6}
\nl &\times& \bfun{ - \mzs}{\mw}{\mw}
\nl &-&
         \frac{1}{128}
         \,\Bigl\{ 12\,\frac{1}{\lzzs}\,\mrT^{d}_{124} 
         + \Bigl[ ( 24\,\frac{\xphs}{\lzz}\,\mrT^{d}_{52} + \mrT^{d}_{127} )\,\frac{1}{\lzz} 
         + ( 2\,\frac{\xphs}{\lzz}\,\mrT^{d}_{115} - 5\,\mrT^{d}_{116}\,\mrV_{9} 
         - 2\,\mrT^{d}_{122} )\,c^2 \Bigr]\,c^2 \Bigr\}\,\frac{1}{c^8}
\nl &\times& \cfun{ - \mhs}{ - \mzs}{ - \mzs}{\mw}{\mw}{\mw}
\nl &+&
         \frac{1}{256}
         \,\Bigl\{ 12\,\Bigl[ ( 3 - \mrT^{d}_{8}\,\vbq ) \Bigr]\,\frac{1}{\lzzs} 
         - \Bigl[ ( \mrT^{d}_{8}\,\mrW_{18}\,\vbq - 3\,\mrW_{22} ) 
         + 2\,( \mrT^{d}_{8}\,\mrW_{24}\,\vbq 
         - 3\,\mrW_{46} )\,c^2 \Bigr]\,c^2 \Bigr\}\,\frac{1}{c^6}\,\xpbs
\nl &\times& \cfun{ - \mhs}{ - \mzs}{ - \mzs}{\mb}{\mb}{\mb}
\nl &-&
         \frac{1}{256}
         \,\Bigl\{ 12\,\Bigl[ ( 3 - \mrT^{d}_{8}\,\vbq ) \Bigr]\,\frac{1}{\lzzs} 
         + ( \mrT^{d}_{8}\,\mrW_{18}\,\vbq + 3\,\mrW_{21} )\,c^2 \Bigr\}\,\frac{1}{c^4}\,\xpbs
         \,\bfun{ - \mhs}{\mb}{\mb}
\nl &+&
         \frac{1}{256}
         \,\Bigl\{ 12\,\Bigl[ ( 3 - \mrT^{d}_{8}\,\vbq ) \Bigr]\,\frac{1}{\lzzs} 
         + ( \mrT^{d}_{8}\,\mrW_{18}\,\vbq + 3\,\mrW_{21} )\,c^2 \Bigr\}\,\frac{1}{c^4}\,\xpbs
         \,\bfun{ - \mzs}{\mb}{\mb}
\nl &+&
         \frac{1}{256}
         \,\Bigl\{ 12\,\Bigl[ ( 3 - \mrT^{d}_{9}\,\vtq ) \Bigr]\,\frac{1}{\lzzs} 
         - \Bigl[ ( \mrT^{d}_{9}\,\mrW_{18}\,\vtq - 3\,\mrW_{22} ) 
         + 2\,( \mrT^{d}_{9}\,\mrW_{20}\,\vtq 
         - 3\,\mrW_{23} )\,c^2 \Bigr]\,c^2 \Bigr\}\,\frac{1}{c^6}\,\xpts
\nl &\times& \cfun{ - \mhs}{ - \mzs}{ - \mzs}{\mt}{\mt}{\mt}
\nl &-&
         \frac{1}{256}
         \,\Bigl\{ 12\,\Bigl[ ( 3 - \mrT^{d}_{9}\,\vtq ) \Bigr]\,\frac{1}{\lzzs} 
         + ( \mrT^{d}_{9}\,\mrW_{18}\,\vtq + 3\,\mrW_{21} )\,c^2 \Bigr\}\,\frac{1}{c^4}\,\xpts
         \,\bfun{ - \mhs}{\mt}{\mt}
\nl &+&
         \frac{1}{256}
         \,\Bigl\{ 12\,\Bigl[ ( 3 - \mrT^{d}_{9}\,\vtq ) \Bigr]\,\frac{1}{\lzzs} 
         + ( \mrT^{d}_{9}\,\mrW_{18}\,\vtq + 3\,\mrW_{21} )\,c^2 \Bigr\}\,\frac{1}{c^4}\,\xpts
         \,\bfun{ - \mzs}{\mt}{\mt}
\nl &-&
         \frac{1}{64}
         \,( 2 - 3\,c^2\,s\mrW_{9}\,\xphs )\,\frac{1}{c^2}
         \,\afun{\mh}
\nl &-&
         \frac{1}{64}
         \,( 3\,c^2\,s\mrW_{9} - 2\,\mrV_{9} )\,\frac{1}{c^4}
         \,\afun{\mz}
\eqas
\bqas
\mcT^{\nfact}_{\ssP\,;\,\PZ\PZ}(\aAZ) &=&
       -
         \frac{1}{12}
         \,s\,c\,\xpbs\,\vbq
         \,\afun{\mb}
     -
         \frac{1}{6}
         \,s\,c\,\xpts\,\vtq
         \,\afun{\mt}
\nl &-&
         \frac{1}{24}
         \,\frac{s}{c}\,\vle
         \,\bfun{ - \mzs}{0}{0}
     +
         \frac{1}{24}
         \,\frac{s}{c}\,\mrU_{0}
         \,( 1 - 3\,\LR )
     +
         \frac{1}{6}
         \,\mrT^{d}_{70}\,s\,c
         \,\afun{\mw}
\nl &-&
         \frac{3}{256}
         \,\Bigl[ 4\,s\mrW_{51} + ( 12\,\frac{\xphs}{\lzzs} 
         - \mrW_{9}\,c^2 )\,c^2\,\xphs \Bigr]\,\frac{s}{c^3}
         \,\bfun{ - \mhs}{\mz}{\mz}
\nl &+&
         \frac{3}{256}
         \,\Bigl[ 8\,s\mrW_{49} 
         - ( \mrW_{9}\,c^2\,\xphs - 2\,\mrW_{50} )\,c^2\,\xphs \Bigr]\,\frac{s}{c^3}
         \,\cfun{ - \mhs}{ - \mzs}{ - \mzs}{\mz}{\mh}{\mz}
\nl &-&
         \frac{9}{256}
         \,\Bigl[ 24\,\frac{1}{\lzzs} - ( 12\,\frac{\xphs}{\lzzs} 
         - \mrW_{9}\,c^2 )\,c^2 \Bigr]\,\frac{s}{c^3}\,\xphs
         \,\bfun{ - \mhs}{\mh}{\mh}
\nl &+&
         \frac{3}{128}
         \,\Bigl[ 16\,\frac{1}{\lzz}\,s\mrW_{11} - ( \mrW_{9}\,c^2\,\xphs 
         - 4\,\mrW_{48} )\,c^2 \Bigr]\,\frac{s}{c^3}
         \,\bfun{ - \mzs}{\mh}{\mz}
\nl &+&
         \frac{9}{256}
         \,\Bigl\{ 48\,\frac{1}{\lzzs} + \Bigl[ 8\,\frac{1}{\lzz}\,\mrW_{26} 
         - ( \mrW_{9}\,c^2\,\xphs - 4\,\mrW_{25} )\,c^2 \Bigr]\,c^2 \Bigr\}\,\frac{s}{c^5}\,\xphs
\nl &\times& \cfun{ - \mhs}{ - \mzs}{ - \mzs}{\mh}{\mz}{\mh}
\nl &+&
         \frac{1}{128}
         \,\Bigl\{ 24\,\frac{1}{\lzzs}\,\mrT^{d}_{133} 
         + \Bigl[ ( 64\,c^6\,\xphs - \frac{\xphs}{\lzz}\,\mrT^{d}_{139} 
         - 2\,\mrT^{d}_{130}\,\mrV_{9} + \mrT^{d}_{131} )\,c^2 
\nl &-& 2\,( 6\,\frac{\xphs}{\lzz}\,\mrT^{d}_{100} 
         - \mrT^{d}_{141} )\,\frac{1}{\lzz} \Bigr]\,c^2 \Bigr\}\,\frac{s}{c^7}
         \,\cfun{ - \mhs}{ - \mzs}{ - \mzs}{\mw}{\mw}{\mw}
\nl &-&
         \frac{1}{128}
         \,\Bigl\{ 24\,\frac{1}{\lzzs}\,\mrT^{d}_{133} 
         - \Bigl[ 2\,( 6\,\frac{\xphs}{\lzz}\,\mrT^{d}_{100} 
         - \mrT^{d}_{138} )\,\frac{1}{\lzz} + ( \frac{\xphs}{\lzz}\,\mrT^{d}_{137} 
         + 8\,\mrT^{d}_{7}\,c^2\,\xphs 
\nl &-& 2\,\mrT^{d}_{72}\,\mrV_{9} 
         - \mrT^{d}_{135} )\,c^2 \Bigr]\,c^2 \Bigr\}\,\frac{s}{c^5}
         \,\bfun{ - \mhs}{\mw}{\mw}
\nl &+&
         \frac{1}{384}
         \,\Bigl\{ 72\,\frac{1}{\lzzs}\,\mrT^{d}_{133} 
         + \Bigl[ - 6\,( 6\,\frac{\xphs}{\lzz}\,\mrT^{d}_{100} 
         - \mrT^{d}_{138} )\,\frac{1}{\lzz} 
\nl &+& (  - 3\,\frac{\xphs}{\lzz}\,\mrT^{d}_{137} 
         + 6\,\mrT^{d}_{72}\,\mrV_{9} - \mrT^{d}_{140} )\,c^2 \Bigr]\,c^2 \Bigr\}\,\frac{s}{c^5}
         \,\bfun{ - \mzs}{\mw}{\mw}
\nl &-&
         \frac{1}{384}
         \,\Bigl\{ \Bigl[ 32\,c^2\,\xpbs\,\vbq 
         + ( 16\,\vbq + 9\,\frac{\xpbs}{\lzz}\,\mrU_{8} 
         - 6\,\mrT^{d}_{78}\,\mrW_{18}\,\xpbs\,\vbq + 9\,\mrV_{9}\,\mrU_{4}\,\xpbs ) \Bigr]\,c^2 
\nl &+& 36\,\Bigl[ ( 2\,\mrT^{d}_{78}\,\vbq 
         + 3\,\mrU_{4} ) \Bigr]\,\frac{\xpbs}{\lzzs} \Bigr\}\,\frac{s}{c^3}
         \,\bfun{ - \mzs}{\mb}{\mb}
\nl &-&
         \frac{1}{384}
         \,\Bigl\{ \Bigl[ 64\,c^2\,\xpts\,\vtq + ( 32\,\vtq + 9\,\frac{\xpts}{\lzz}\,\mrU_{7} 
         - 6\,\mrT^{d}_{77}\,\mrW_{18}\,\xpts\,\vtq + 9\,\mrV_{9}\,\mrU_{2}\,\xpts ) \Bigr]\,c^2 
\nl &+& 36\,\Bigl[ ( 2\,\mrT^{d}_{77}\,\vtq 
         + 3\,\mrU_{2}) \Bigr]\,\frac{\xpts}{\lzzs} \Bigr\}\,\frac{s}{c^3}
         \,\bfun{ - \mzs}{\mt}{\mt}
\nl &-&
         \frac{1}{576}
         \,\Bigl\{ \Bigl[ 48\,( \xpbs\,\vbq + 2\,\xpts\,\vtq )\,c^2 
         - (  - 9\,\frac{\xphs}{\lzz}\,\mrT^{d}_{134} + 18\,\mrT^{d}_{77}\,\mrW_{19}\,\xpts\,\vtq 
\nl &+& 18\,\mrT^{d}_{78}\,\mrW_{19}\,\xpbs\,\vbq - \mrT^{d}_{143} 
         + 27\,\mrU_{2}\,\mrW_{19}\,\xpts + 27\,\mrU_{4}\,\mrW_{19}\,\xpbs ) \Bigr]\,c^2 
         + 9\,( \frac{1}{\lzz}\,\mrT^{d}_{136} + \mrT^{d}_{132}\,\mrV_{9} ) \Bigr\}\,\frac{s}{c^3}
\nl &+&
         \frac{1}{128}
         \,\Bigl\{  - \Bigl[ (  - 3\,\frac{1}{\lzz}\,\mrU_{7} 
         + 2\,\mrT^{d}_{77}\,\mrW_{18}\,\vtq - 3\,\mrV_{9}\,\mrU_{2} ) \Bigr]\,c^2 
         + 12\,\Bigl[ ( 2\,\mrT^{d}_{77}\,\vtq
         + 3\,\mrU_{2} ) \Bigr]\,\frac{1}{\lzzs} \Bigr\}\,\frac{s}{c^3}\,\xpts
\nl &\times& \bfun{ - \mhs}{\mt}{\mt}
\nl &+&
         \frac{1}{128}
         \,\Bigl\{  - \Bigl[ (  - 3\,\frac{1}{\lzz}\,\mrU_{8} 
         + 2\,\mrT^{d}_{78}\,\mrW_{18}\,\vbq - 3\,\mrV_{9}\,\mrU_{4} ) \Bigr]\,c^2 
         + 12\,\Bigl[ ( 2\,\mrT^{d}_{78}\,\vbq 
         + 3\,\mrU_{4} ) \Bigr]\,\frac{1}{\lzzs} \Bigr\}\,\frac{s}{c^3}\,\xpbs
\nl &\times& \bfun{ - \mhs}{\mb}{\mb}
\nl &-&
         \frac{1}{128}
         \,\Bigl\{ \Bigl[ ( 3\,\frac{1}{\lzz}\,\mrU_{9} + 2\,\mrT^{d}_{77}\,\mrW_{18}\,\vtq 
         - 3\,\mrV_{9}\,\mrU_{2} ) + 2\,( 2\,\mrT^{d}_{77}\,\mrW_{20}\,\vtq
         + 3\,\mrU_{1} + 6\,\mrU_{2}\,\mrW_{19}\,\xpts )\,c^2 \Bigr]\,c^2 
\nl &+& 12\,\Bigl[ ( 2\,\mrT^{d}_{77}\,\vtq 
         + 3\,\mrU_{2} ) \Bigr]\,\frac{1}{\lzzs} \Bigr\}\,\frac{s}{c^5}\,\xpts
         \,\cfun{ - \mhs}{ - \mzs}{ - \mzs}{\mt}{\mt}{\mt}
\nl &-&
         \frac{1}{128}
         \,\Bigl\{ \Bigl[ ( 3\,\frac{1}{\lzz}\,\mrU_{10} 
         + 2\,\mrT^{d}_{78}\,\mrW_{18}\,\vbq - 3\,\mrV_{9}\,\mrU_{4} ) 
         + 2\,( 2\,\mrT^{d}_{78}\,\mrW_{24}\,\vbq
         + 3\,\mrU_{3} + 6\,\mrU_{4}\,\mrW_{19}\,\xpbs )\,c^2 \Bigr]\,c^2 
\nl &+& 12\,\Bigl[ ( 2\,\mrT^{d}_{78}\,\vbq 
         + 3\,\mrU_{4} ) \Bigr]\,\frac{1}{\lzzs} \Bigr\}\,\frac{s}{c^5}\,\xpbs
         \,\cfun{ - \mhs}{ - \mzs}{ - \mzs}{\mb}{\mb}{\mb}
\nl &+&
         \frac{3}{64}
         \,( 1 - c^2\,\xphs )\,\frac{s}{c}\,\mrW_{9}
         \,\afun{\mh}
\nl &+&
         \frac{3}{64}
         \,( s\mrW_{18} + c^2\,s\mrW_{9} )\,\frac{s}{c^3}
         \,\afun{\mz}
\nl &+&
         \frac{1}{48}
         \,( 3\,\mrT^{d}_{7}\,c^2\,\xphs + \mrT^{d}_{142} )\,\frac{s}{c}
         \,\LR
\eqas
\bqas
\mcT^{\nfact}_{\ssP\,;\,\PZ\PZ}(\aAA) &=&
         \frac{1}{32}
         \,\Bigl\{ 12\,\frac{1}{\lzzs}\,\mrT^{d}_{150} 
         - \Bigl[ 2\,( 2\,c^2\,\xphs + \frac{\xphs}{\lzz}\,\mrT^{d}_{38} - \mrT^{d}_{144} )\,c^2 
         + ( 24\,\frac{\xphs}{\lzzs}\,\mrT^{d}_{7} 
\nl &-& \frac{1}{\lzz}\,\mrT^{d}_{148} 
         + \mrT^{d}_{145}\,\mrV_{9} ) \Bigr]\,c^2 \Bigr\}\,\frac{s^2}{c^4}
         \,\bfun{ - \mhs}{\mw}{\mw}
\nl &-&
         \frac{1}{32}
         \,\Bigl\{ 12\,\frac{1}{\lzzs}\,\mrT^{d}_{150} 
         - \Bigl[ 2\,( 8\,c^4\,\xphs + \frac{\xphs}{\lzz}\,\mrT^{d}_{154} + \mrT^{d}_{151} )\,c^2 
         + ( 24\,\frac{\xphs}{\lzzs}\,\mrT^{d}_{7} 
\nl &-& \frac{1}{\lzz}\,\mrT^{d}_{155} 
         - \mrT^{d}_{147}\,\mrV_{9} ) \Bigr]\,c^2 \Bigr\}\,\frac{s^2}{c^6}
         \,\cfun{ - \mhs}{ - \mzs}{ - \mzs}{\mw}{\mw}{\mw}
\nl &-&
         \frac{1}{32}
         \,\Bigl\{ 12\,\frac{1}{\lzzs}\,\mrT^{d}_{150} 
         - \Bigl[ ( 24\,\frac{\xphs}{\lzzs}\,\mrT^{d}_{7} - \frac{1}{\lzz}\,\mrT^{d}_{148} 
         + \mrT^{d}_{145}\,\mrV_{9} ) + 2\,( \frac{\xphs}{\lzz}\,\mrT^{d}_{38} 
         + \mrT^{d}_{149} )\,c^2 \Bigr]\,c^2 \Bigr\}\,\frac{s^2}{c^4}
\nl &\times& \bfun{ - \mzs}{\mw}{\mw}
\nl &-&
         \frac{1}{32}
         \,\Bigl\{ 8\,\mrT^{d}_{29}\,c^4 + \Bigl[ 4\,\frac{\xphs}{\lzz}\,\mrT^{d}_{7}\,c^2 
         + ( \mrT^{d}_{8}\,\mrW_{19}\,\xpbs\,\vbq + \mrT^{d}_{9}\,\mrW_{19}\,\xpts\,\vtq 
         - 3\,\mrW_{47} ) \Bigr]\,s^2 
\nl &+& 2\,(  - \frac{1}{\lzz}\,\mrT^{d}_{152} 
         + \mrT^{d}_{146}\,\mrV_{9} ) \Bigr\}\,\frac{1}{c^2}
\nl &-&
         \frac{1}{64}
         \,\Bigl\{ 12\,\Bigl[ ( 3 - \mrT^{d}_{8}\,\vbq ) \Bigr]\,\frac{1}{\lzzs} 
         - \Bigl[ ( \mrT^{d}_{8}\,\mrW_{18}\,\vbq - 3\,\mrW_{22} ) 
         + 2\,( \mrT^{d}_{8}\,\mrW_{24}\,\vbq 
         - 3\,\mrW_{46} )\,c^2 \Bigr]\,c^2 \Bigr\}\,\frac{s^2}{c^6}\,\xpbs
\nl &\times& \cfun{ - \mhs}{ - \mzs}{ - \mzs}{\mb}{\mb}{\mb}
\nl &+&
         \frac{1}{64}
         \,\Bigl\{ 12\,\Bigl[ ( 3 - \mrT^{d}_{8}\,\vbq ) \Bigr]\,\frac{1}{\lzzs} 
         + ( \mrT^{d}_{8}\,\mrW_{18}\,\vbq + 3\,\mrW_{21} )\,c^2 \Bigr\}\,\frac{s^2}{c^4}\,\xpbs
         \,\bfun{ - \mhs}{\mb}{\mb}
\nl &-&
         \frac{1}{64}
         \,\Bigl\{ 12\,\Bigl[ ( 3 - \mrT^{d}_{8}\,\vbq ) \Bigr]\,\frac{1}{\lzzs} 
         + ( \mrT^{d}_{8}\,\mrW_{18}\,\vbq + 3\,\mrW_{21} )\,c^2 \Bigr\}\,\frac{s^2}{c^4}\,\xpbs
         \,\bfun{ - \mzs}{\mb}{\mb}
\nl &-&
         \frac{1}{64}
         \,\Bigl\{ 12\,\Bigl[ ( 3 - \mrT^{d}_{9}\,\vtq ) \Bigr]\,\frac{1}{\lzzs} 
         - \Bigl[ ( \mrT^{d}_{9}\,\mrW_{18}\,\vtq - 3\,\mrW_{22} ) 
         + 2\,( \mrT^{d}_{9}\,\mrW_{20}\,\vtq 
         - 3\,\mrW_{23} )\,c^2 \Bigr]\,c^2 \Bigr\}\,\frac{s^2}{c^6}\,\xpts
\nl &\times& \cfun{ - \mhs}{ - \mzs}{ - \mzs}{\mt}{\mt}{\mt}
\nl &+&
         \frac{1}{64}
         \,\Bigl\{ 12\,\Bigl[ ( 3 - \mrT^{d}_{9}\,\vtq ) \Bigr]\,\frac{1}{\lzzs} 
         + ( \mrT^{d}_{9}\,\mrW_{18}\,\vtq + 3\,\mrW_{21} )\,c^2 \Bigr\}\,\frac{s^2}{c^4}\,\xpts
         \,\bfun{ - \mhs}{\mt}{\mt}
\nl &-&
         \frac{1}{64}
         \,\Bigl\{ 12\,\Bigl[ ( 3 - \mrT^{d}_{9}\,\vtq ) \Bigr]\,\frac{1}{\lzzs} 
         + ( \mrT^{d}_{9}\,\mrW_{18}\,\vtq + 3\,\mrW_{21} )\,c^2 \Bigr\}\,\frac{s^2}{c^4}\,\xpts
         \,\bfun{ - \mzs}{\mt}{\mt}
\nl &-&
         \frac{1}{8}
         \,( c^2\,\xphs - 2\,\mrT^{d}_{153} )\,s^2
         \,\LR
\eqas
\bqas
\mcT^{\nfact}_{\ssP\,;\,\PZ\PZ}(\aZZ) &=&
       -
         \frac{1}{64}
         \,\mrX_{16}
\nl &+&
         \frac{3}{256}
         \,\Bigl\{ 24\,\frac{1}{\lzzs}\,\mrT^{d}_{167} 
         + \Bigl[ 4\,( 8 - 3\,\frac{\xphs}{\lzz}\,\mrT^{d}_{167} )\,\frac{1}{\lzz} 
         + (  - \frac{\xphs}{\lzz}\,\mrT^{d}_{167} 
         + 3\,\mrT^{d}_{64} )\,c^2 \Bigr]\,c^2 \Bigr\}\,\frac{1}{c^4}\,\xphs
\nl &\times& \bfun{ - \mhs}{\mh}{\mh}
\nl &-&
         \frac{3}{256}
         \,\Bigl\{ 48\,\frac{1}{\lzzs}\,\mrT^{d}_{167} 
         + \Bigl[ - ( 12\,\frac{\xphs}{\lzz}\,\mrT^{d}_{163} 
         + \mrT^{d}_{167}\,\mrW_{9}\,c^2\,\xphs - 4\,\mrT^{d}_{167}\,\mrW_{53} )\,c^2 
\nl &+& 24\,(  - 2\,\frac{\xphs}{\lzz}\,\mrT^{d}_{167} 
         + \mrT^{d}_{163} )\,\frac{1}{\lzz} \Bigr]\,c^2 \Bigr\}\,\frac{1}{c^6}\,\xphs
         \,\cfun{ - \mhs}{ - \mzs}{ - \mzs}{\mh}{\mz}{\mh}
\nl &-&
         \frac{1}{64}
         \,\Bigl\{ \mrT^{d}_{164} + \Bigl[ ( \frac{\xphs}{\lzz}\,\mrT^{d}_{167} 
         + 3\,\frac{1}{\lzz} - \mrT^{d}_{177} ) \Bigr]\,c^2\,\xphs \Bigr\}\,\frac{1}{c^2}
         \,\afun{\mh}
\nl &+&
         \frac{1}{64}
         \,\Bigl\{ \mrT^{d}_{164}\,\mrV_{9} 
         + \Bigl[ ( \frac{\xphs}{\lzz}\,\mrT^{d}_{167} + 3\,\frac{1}{\lzz} 
         - \mrT^{d}_{177} ) \Bigr]\,c^2 \Bigr\}\,\frac{1}{c^4}
         \,\afun{\mz}
\nl &+&
         \frac{1}{64}
         \,\Bigl\{ \mrT^{d}_{165}\,\mrV_{9} 
         + \Bigl[ ( \frac{\xphs}{\lzz}\,\mrT^{d}_{179} + \frac{1}{\lzz}\,\mrT^{d}_{157} 
         + 2\,\mrT^{d}_{176} ) + ( 2\,\mrT^{d}_{42}\,\mrW_{19}\,\xpts\,\vtq 
\nl &+& 2\,\mrT^{d}_{112}\,\mrW_{19}\,\xpbs\,\vbq + 3\,\mrU_{2}\,\mrW_{19}\,\xpts 
         + 3\,\mrU_{4}\,\mrW_{19}\,\xpbs )\,c^2 \Bigr]\,c^2 \Bigr\}\,\frac{1}{c^4}
\nl &+&
         \frac{1}{128}
         \,\Bigl\{ \Bigl[ 64\,c^8\,\xphs - ( \frac{\xphs}{\lzz}\,\mrT^{d}_{171} 
         + 2\,\frac{1}{\lzz}\,\mrT^{d}_{182} - \mrT^{d}_{174} - 2\,\mrT^{d}_{175}\,\mrV_{9} ) \Bigr]\,c^2
\nl &-& 12\,\Bigl[ ( 2\,\mrT^{d}_{158} + \mrT^{d}_{169}\,\xphs )
         \Bigr]\,\frac{1}{\lzzs} \Bigr\}\,\frac{1}{c^6}
         \,\cfun{ - \mhs}{ - \mzs}{ - \mzs}{\mw}{\mw}{\mw}
\nl &-&
         \frac{1}{128}
         \,\Bigl\{ \Bigl[ 4\,\mrT^{d}_{100}\,c^2\,\xphs 
         - (  - \frac{\xphs}{\lzz}\,\mrT^{d}_{173} + 2\,\frac{1}{\lzz}\,\mrT^{d}_{181} 
         + 2\,\mrT^{d}_{170}\,\mrV_{9} + \mrT^{d}_{172} ) \Bigr]\,c^2 
\nl &-& 12\,\Bigl[ ( 2\,\mrT^{d}_{158} 
         + \mrT^{d}_{169}\,\xphs ) \Bigr]\,\frac{1}{\lzzs} \Bigr\}\,\frac{1}{c^4}
         \,\bfun{ - \mhs}{\mw}{\mw}
\nl &-&
         \frac{1}{256}
         \,\Bigl\{  - \Bigl[ \mrT^{d}_{167}\,\mrW_{9}\,c^2\,\xphs 
         + 2\,( 36\,\frac{\xphs}{\lzzs} - 6\,\frac{\xphs}{\lzzs}\,\mrT^{d}_{167}\,\xphs 
         - 3\,\frac{\xphs}{\lzz}\,\mrT^{d}_{159} 
         + \mrT^{d}_{160} ) \Bigr]\,c^2\,\xphs 
\nl &+& 8\,( \frac{\xphs}{\lzz}\,\mrT^{d}_{178} 
         + \mrT^{d}_{159} ) \Bigr\}\,\frac{1}{c^4}
         \,\cfun{ - \mhs}{ - \mzs}{ - \mzs}{\mz}{\mh}{\mz}
\nl &-&
         \frac{1}{128}
         \,\Bigl\{ \Bigl[ 2\,(  - 6\,\frac{\xphs}{\lzzs}\,\mrT^{d}_{167}\,\xphs 
         + \frac{\xphs}{\lzz}\,\mrT^{d}_{162} + 3\,\mrT^{d}_{166} ) 
         - (  - \frac{\xphs}{\lzz}\,\mrT^{d}_{167} 
         + 3\,\mrT^{d}_{163} )\,c^2\,\xphs \Bigr]\,c^2 
\nl &+& 8\,\Bigl[ ( 9\,\frac{\xphs}{\lzz}\,\mrT^{d}_{161} 
         + 2\,\mrT^{d}_{159} ) \Bigr]\,\frac{1}{\lzz} \Bigr\}\,\frac{1}{c^4}
         \,\bfun{ - \mzs}{\mh}{\mz}
\nl &+&
         \frac{1}{256}
         \,\Bigl\{ \Bigl[ 12\,( \frac{\xphs}{\lzzs}\,\mrT^{d}_{167}\,\xphs - \mrW_{52} ) 
         + ( \frac{\xphs}{\lzz}\,\mrT^{d}_{167} - \mrT^{d}_{177} )\,c^2\,\xphs \Bigr]\,c^2 
         + 4\,( \frac{1}{\lzz}\,\mrT^{d}_{180} + \mrT^{d}_{164}\,\mrV_{9} ) \Bigr\}\,\frac{1}{c^4}
\nl &\times& \bfun{ - \mhs}{\mz}{\mz}
\nl &+&
         \frac{1}{128}
         \,\Bigl\{  - \Bigl[ (  - 3\,\frac{1}{\lzz}\,\mrU_{7} 
         + 2\,\mrT^{d}_{42}\,\mrW_{18}\,\vtq - 3\,\mrV_{9}\,\mrU_{2} ) \Bigr]\,c^2 
         + 12\,\Bigl[ ( 2\,\mrT^{d}_{42}\,\vtq
        + 3\,\mrU_{2} ) \Bigr]\,\frac{1}{\lzzs} \Bigr\}\,\frac{1}{c^2}\,\xpts
\nl &\times& \bfun{ - \mhs}{\mt}{\mt}
\nl &-&
         \frac{1}{128}
         \,\Bigl\{  - \Bigl[ (  - 3\,\frac{1}{\lzz}\,\mrU_{7} 
         + 2\,\mrT^{d}_{42}\,\mrW_{18}\,\vtq - 3\,\mrV_{9}\,\mrU_{2} ) \Bigr]\,c^2 
         + 12\,\Bigl[ ( 2\,\mrT^{d}_{42}\,\vtq 
         + 3\,\mrU_{2} ) \Bigr]\,\frac{1}{\lzzs} \Bigr\}\,\frac{1}{c^2}\,\xpts
\nl &\times& \bfun{ - \mzs}{\mt}{\mt}
\nl &+&
         \frac{1}{128}
         \,\Bigl\{  - \Bigl[ (  - 3\,\frac{1}{\lzz}\,\mrU_{8} 
         + 2\,\mrT^{d}_{112}\,\mrW_{18}\,\vbq - 3\,\mrV_{9}\,\mrU_{4} ) \Bigr]\,c^2 
         + 12\,\Bigl[ ( 2\,\mrT^{d}_{112}\,\vbq 
         + 3\,\mrU_{4} ) \Bigr]\,\frac{1}{\lzzs} \Bigr\}\,\frac{1}{c^2}\,\xpbs
\nl &\times& \bfun{ - \mhs}{\mb}{\mb}
\nl &-&
         \frac{1}{128}
         \,\Bigl\{  - \Bigl[ (  - 3\,\frac{1}{\lzz}\,\mrU_{8} 
         + 2\,\mrT^{d}_{112}\,\mrW_{18}\,\vbq - 3\,\mrV_{9}\,\mrU_{4} ) \Bigr]\,c^2 
         + 12\,\Bigl[ ( 2\,\mrT^{d}_{112}\,\vbq 
         + 3\,\mrU_{4} ) \Bigr]\,\frac{1}{\lzzs} \Bigr\}\,\frac{1}{c^2}\,\xpbs
\nl &\times& \bfun{ - \mzs}{\mb}{\mb}
\nl &-&
         \frac{1}{128}
         \,\Bigl\{ \Bigl[ ( 3\,\frac{1}{\lzz}\,\mrU_{9} + 2\,\mrT^{d}_{42}\,\mrW_{18}\,\vtq 
         - 3\,\mrV_{9}\,\mrU_{2} ) + 2\,( 2\,\mrT^{d}_{42}\,\mrW_{20}\,\vtq
         + 3\,\mrU_{1} + 6\,\mrU_{2}\,\mrW_{19}\,\xpts )\,c^2 \Bigr]\,c^2 
\nl &+& 12\,\Bigl[ ( 2\,\mrT^{d}_{42}\,\vtq 
         + 3\,\mrU_{2} ) \Bigr]\,\frac{1}{\lzzs} \Bigr\}\,\frac{1}{c^4}\,\xpts
         \,\cfun{ - \mhs}{ - \mzs}{ - \mzs}{\mt}{\mt}{\mt}
\nl &-&
         \frac{1}{128}
         \,\Bigl\{ \Bigl[ ( 3\,\frac{1}{\lzz}\,\mrU_{10} + 2\,\mrT^{d}_{112}\,\mrW_{18}\,\vbq 
         - 3\,\mrV_{9}\,\mrU_{4} ) + 2\,( 2\,\mrT^{d}_{112}\,\mrW_{24}\,\vbq
         + 3\,\mrU_{3} + 6\,\mrU_{4}\,\mrW_{19}\,\xpbs )\,c^2 \Bigr]\,c^2 
\nl&+& 12\,\Bigl[ ( 2\,\mrT^{d}_{112}\,\vbq 
         + 3\,\mrU_{4} ) \Bigr]\,\frac{1}{\lzzs} \Bigr\}\,\frac{1}{c^4}\,\xpbs
         \,\cfun{ - \mhs}{ - \mzs}{ - \mzs}{\mb}{\mb}{\mb}
\nl &-&
         \frac{1}{128}
         \,\Bigl\{ 12\,\Bigl[ ( 2\,\mrT^{d}_{158} + \mrT^{d}_{169}\,\xphs ) \Bigr]\,\frac{1}{\lzzs} 
         + (  - \frac{\xphs}{\lzz}\,\mrT^{d}_{173} + 2\,\frac{1}{\lzz}\,\mrT^{d}_{181} 
         + \mrT^{d}_{168} + 2\,\mrT^{d}_{170}\,\mrV_{9} )\,c^2,rc)\,\frac{1}{c^4}
\nl &\times& \bfun{ - \mzs}{\mw}{\mw}
\nl &-&
         \frac{1}{32}
         \,( \mrT^{d}_{108}\,c^2\,\xphs + 4\,\mrT^{d}_{156})\,\frac{1}{c^2}
         \,\LR
\eqas
\vspace{0.5cm}
\bei
\item {\underline{$\PH\PW\PW$}} Amplitudes
\eei
\bqas
\mcT^{\nfact}_{\ssD\,;\,\PW\PW}(\aptV) &=&
         \frac{1}{32}
         \,\mrX_{17}
     -
         \frac{1}{16}
         \,\mrV_{6}
     -
         \frac{3}{32}
         \,\mrV_{14}
         \,\LR
\nl &-&
         \frac{1}{32}
         \,\mrV_{15}\,\xpbs
         \,\afun{\mb}
     -
         \frac{1}{32}
         \,\mrV_{16}\,\xpts
         \,\afun{\mt}
     -
         \frac{1}{32}
         \,\mrV_{17}
         \,\bfun{ - \mws}{\mt}{\mb}
\eqas
\bqas
\mcT^{\nfact}_{\ssD\,;\,\PW\PW}(\aptA) &=&
         \frac{1}{32}
         \,\mrX_{17}
     -
         \frac{1}{16}
         \,\mrV_{6}
     -
         \frac{3}{32}
         \,\mrV_{14}
         \,\LR
\nl &-&
         \frac{1}{32}
         \,\mrV_{15}\,\xpbs
         \,\afun{\mb}
     -
         \frac{1}{32}
         \,\mrV_{16}\,\xpts
         \,\afun{\mt}
     -
         \frac{1}{32}
         \,\mrV_{17}
         \,\bfun{ - \mws}{\mt}{\mb}
\eqas
\bqas
\mcT^{\nfact}_{\ssD\,;\,\PW\PW}(\apbV) &=&
         \frac{1}{32}
         \,\mrX_{17}
     -
         \frac{1}{16}
         \,\mrV_{6}
     -
         \frac{3}{32}
         \,\mrV_{14}
         \,\LR
\nl &-&
         \frac{1}{32}
         \,\mrV_{15}\,\xpbs
         \,\afun{\mb}
     -
         \frac{1}{32}
         \,\mrV_{16}\,\xpts
         \,\afun{\mt}
     -
         \frac{1}{32}
         \,\mrV_{17}
         \,\bfun{ - \mws}{\mt}{\mb}
\eqas
\bqas
\mcT^{\nfact}_{\ssD\,;\,\PW\PW}(\apbA) &=&
         \frac{1}{32}
         \,\mrX_{17}
     -
         \frac{1}{16}
         \,\mrV_{6}
     -
         \frac{3}{32}
         \,\mrV_{14}
         \,\LR
\nl &-&
         \frac{1}{32}
         \,\mrV_{15}\,\xpbs
         \,\afun{\mb}
     -
         \frac{1}{32}
         \,\mrV_{16}\,\xpts
         \,\afun{\mt}
     -
         \frac{1}{32}
         \,\mrV_{17}
         \,\bfun{ - \mws}{\mt}{\mb}
\eqas
\bqas
\mcT^{\nfact}_{\ssD\,;\,\PW\PW}(\apn) &=&
       -
         \frac{1}{16}
         \,\bfun{ - \mws}{0}{0}
       +
         \frac{1}{48}
         \,( 1 - 3\,\LR )
\eqas
\bqas
\mcT^{\nfact}_{\ssD\,;\,\PW\PW}(\aplV) &=&
       -
         \frac{1}{16}
         \,\bfun{ - \mws}{0}{0}
       +
         \frac{1}{48}
         \,( 1 - 3\,\LR )
\eqas
\bqas
\mcT^{\nfact}_{\ssD\,;\,\PW\PW}(\aplA) &=&
       -
         \frac{1}{16}
         \,\bfun{ - \mws}{0}{0}
       +
         \frac{1}{48}
         \,( 1 - 3\,\LR )
\eqas
\bqas
\mcT^{\nfact}_{\ssD\,;\,\PW\PW}(\atp) &=&
         \frac{3}{32}
         \,\xpts
         \,( 1 - \LR )
\nl &+&
         \frac{3}{16}
         \,\frac{\xpts}{\lww}\,\mrV_{18}
         \,\bfun{ - \mhs}{\mt}{\mt}
\nl &-&
         \frac{3}{32}
         \,\mrW_{54}\,\xpts
         \,\bfun{ - \mws}{\mt}{\mb}
\nl &-&
         \frac{3}{32}
         \,\mrW_{55}\,\xpts
         \,\cfun{ - \mhs}{ - \mws}{ - \mws}{\mt}{\mb}{\mt}
\eqas
\bqas
\mcT^{\nfact}_{\ssD\,;\,\PW\PW}(\abp) &=&
       -
         \frac{1}{32}
         \,\mrX_{18}
     +
         \frac{3}{32}
         \,\xpts
         \,( 1 - \LR )
\nl &+&
         \frac{3}{16}
         \,\frac{\xpts}{\lww}\,\mrV_{18}
         \,\bfun{ - \mhs}{\mt}{\mt}
\nl &-&
         \frac{3}{32}
         \,\mrW_{54}\,\xpts
         \,\bfun{ - \mws}{\mt}{\mb}
\nl &-&
         \frac{3}{32}
         \,\mrW_{55}\,\xpts
         \,\cfun{ - \mhs}{ - \mws}{ - \mws}{\mt}{\mb}{\mt}
\eqas
\bqas
\mcT^{\nfact}_{\ssD\,;\,\PW\PW}(\atBW) &=&
     -
         \frac{3}{128}
         \,c\,\xphs\,\xpts
\nl &-&
         \frac{3}{128}
         \,c\,\xphs\,\xpts\,\xpbs
         \,\afun{\mb}
     +
         \frac{3}{128}
         \,c\,\xphs\,\xptq
         \,\afun{\mt}
\nl &+& \frac{3}{32}
         \,c\,\xpts\,\xpbs
         \,\bfun{ - \mhs}{\mb}{\mb}
\nl &+&
         \frac{3}{64}
         \,\mrV_{22}\,c\,\xpts\,\xpbs
         \,\cfun{ - \mhs}{ - \mws}{ - \mws}{\mb}{\mt}{\mb}
\nl &+&
         \frac{3}{128}
         \,\mrW_{56}\,c\,\xpts
         \,\bfun{ - \mws}{\mt}{\mb}
\nl &-&
         \frac{3}{128}
         \,\mrW_{57}\,c\,\xpts
         \,\bfun{ - \mhs}{\mt}{\mt}
\nl &+&
         \frac{3}{64}
         \,\mrW_{58}\,c\,\xpts
         \,\cfun{ - \mhs}{ - \mws}{ - \mws}{\mt}{\mb}{\mt}
\eqas
\bqas
\mcT^{\nfact}_{\ssD\,;\,\PW\PW}(\atWB) &=&
     -
         \frac{3}{128}
         \,s\,\xphs\,\xpts
\nl &-&
         \frac{3}{128}
         \,s\,\xphs\,\xpts\,\xpbs
         \,\afun{\mb}
     +
         \frac{3}{128}
         \,s\,\xphs\,\xptq
         \,\afun{\mt}
\nl &+&
         \frac{3}{32}
         \,s\,\xpts\,\xpbs
         \,\bfun{ - \mhs}{\mb}{\mb}
\nl &+&
         \frac{3}{64}
         \,\mrV_{22}\,s\,\xpts\,\xpbs
         \,\cfun{ - \mhs}{ - \mws}{ - \mws}{\mb}{\mt}{\mb}
\nl &+&
         \frac{3}{128}
         \,\mrW_{56}\,s\,\xpts
         \,\bfun{ - \mws}{\mt}{\mb}
\nl &-&
         \frac{3}{128}
         \,\mrW_{57}\,s\,\xpts
         \,\bfun{ - \mhs}{\mt}{\mt}
\nl &+&
         \frac{3}{64}
         \,\mrW_{58}\,s\,\xpts
         \,\cfun{ - \mhs}{ - \mws}{ - \mws}{\mt}{\mb}{\mt}
\eqas
\bqas
\mcT^{\nfact}_{\ssD\,;\,\PW\PW}(\abBW) &=&
         \frac{1}{32}
         \,c
         \,\mrX_{18}
     +
         \frac{3}{32}
         \,\mrV_{6}\,c
         \,\LR
\nl &-&
         \frac{3}{128}
         \,c\,\xphs\,\xpbq
         \,\afun{\mb}
     +
         \frac{3}{128}
         \,c\,\xphs\,\xpts\,\xpbs
         \,\afun{\mt}
\nl &+&
         \frac{3}{128}
         \,\mrV_{23}\,c\,\xpbs
         \,\bfun{ - \mhs}{\mb}{\mb}
\nl &+&
         \frac{3}{64}
         \,\mrV_{24}\,c\,\xpbs
         \,\cfun{ - \mhs}{ - \mws}{ - \mws}{\mb}{\mt}{\mb}
\nl &-&
         \frac{3}{128}
         \,\mrV_{27}\,c
\nl &-&
         \frac{3}{32}
         \,\mrW_{59}\,c\,\xpts
         \,\bfun{ - \mhs}{\mt}{\mt}
\nl &+&
         \frac{3}{128}
         \,\mrW_{60}\,c
         \,\bfun{ - \mws}{\mt}{\mb}
\nl &+&
         \frac{3}{64}
         \,\mrW_{61}\,c\,\xpts
         \,\cfun{ - \mhs}{ - \mws}{ - \mws}{\mt}{\mb}{\mt}
\eqas
\bqas
\mcT^{\nfact}_{\ssD\,;\,\PW\PW}(\abWB) &=&
        \frac{1}{32}
         \,s
         \,\mrX_{18}
     +
         \frac{3}{32}
         \,\mrV_{6}\,s
         \,\LR
\nl &-&
         \frac{3}{128}
         \,s\,\xphs\,\xpbq
         \,\afun{\mb}
     +
         \frac{3}{128}
         \,s\,\xphs\,\xpts\,\xpbs
         \,\afun{\mt}
\nl &+&
         \frac{3}{128}
         \,\mrV_{23}\,s\,\xpbs
         \,\bfun{ - \mhs}{\mb}{\mb}
\nl &+&
         \frac{3}{64}
         \,\mrV_{24}\,s\,\xpbs
         \,\cfun{ - \mhs}{ - \mws}{ - \mws}{\mb}{\mt}{\mb}
\nl &-&
         \frac{3}{128}
         \,\mrV_{27}\,s
\nl &-&
         \frac{3}{32}
         \,\mrW_{59}\,s\,\xpts
         \,\bfun{ - \mhs}{\mt}{\mt}
\nl &+&
         \frac{3}{128}
         \,\mrW_{60}\,s
         \,\bfun{ - \mws}{\mt}{\mb}
\nl &+&
         \frac{3}{64}
         \,\mrW_{61}\,s\,\xpts
         \,\cfun{ - \mhs}{ - \mws}{ - \mws}{\mt}{\mb}{\mt}
\eqas
\bqas
\mcT^{\nfact}_{\ssD\,;\,\PW\PW}(\apB) &=&
         \frac{3}{16}
\nl &+&
         \frac{3}{8}
         \,\frac{1}{\lww}\,\mrV_{0}
         \,\bfun{ - \mhs}{\mh}{\mh}
\nl &-&
         \frac{3}{8}
         \,\frac{1}{\lww}\,\mrV_{0}
         \,\bfun{ - \mws}{\mw}{\mh}
\nl &+&
         \frac{3}{8}
         \,\mrW_{62}
         \,\cfun{ - \mhs}{ - \mws}{ - \mws}{\mh}{\mw}{\mh}
\eqas
\bqas
\mcT^{\nfact}_{\ssD\,;\,\PW\PW}(\apBox) &=&
       -
         \frac{1}{1152}
         \,\mrX_{19}
     -
         \frac{1}{96}
         \,\xphs
         \,\afun{\mw}
\nl &-&
         \frac{1}{768}
         \,\xphs
         \,( 103 + 12\,\LR )
     -
         \frac{1}{768}
         \,\frac{1}{c^2}
     +
         \frac{1}{768}
         \,\frac{1}{c^2}\,\mrT^{d}_{188}
         \,\LR
\nl &+&
         \frac{1}{96}
         \,\mrV_{0}\,s^2
         \,\cfunf{ - \mhs}{ - \mws}{ - \mws}{\mw}{0}{\mw}
\nl &-&
         \frac{1}{96}
         \,\mrV_{1}\,\xphs
         \,\afun{\mh}
\nl &-&
         \frac{1}{256}
         \,\mrW_{64}
         \,\bfun{ - \mhs}{\mh}{\mh}
\nl &+&
         \frac{1}{256}
         \,\mrW_{65}
         \,\cfun{ - \mhs}{ - \mws}{ - \mws}{\mh}{\mw}{\mh}
\nl &+&
         \frac{1}{384}
         \,\mrW_{66}
         \,\bfun{ - \mws}{\mw}{\mh}
\nl &-&
         \frac{1}{768}
         \,\mrW_{67}
         \,\cfun{ - \mhs}{ - \mws}{ - \mws}{\mw}{\mh}{\mw}
\nl &-&
         \frac{1}{384}
         \,\Bigl[ \frac{1}{\lww}\,\mrT^{d}_{185} 
         + ( 23\,\frac{\xphs}{\lww}\,c^2 - 2\,\mrT^{d}_{183} )\,c^2 \Bigr]\,\frac{1}{c^4}
         \,\bfun{ - \mws}{\mw}{\mz}
\nl &+&
         \frac{1}{768}
         \,\Bigl[ 2\,\frac{1}{\lww}\,\mrT^{d}_{187} 
         + ( 23\,\frac{\xphs}{\lww}\,\mrT^{d}_{195} + \mrT^{d}_{193} )\,c^2 \Bigr]\,\frac{1}{c^4}
         \,\bfun{ - \mhs}{\mz}{\mz}
\nl &-&
         \frac{1}{768}
         \,\Bigl\{ c^2\,s\mrW_{63}\,\xphs + \Bigl[ (  - \frac{\xphs}{\lww}\,\mrT^{d}_{192} 
         + 2\,\frac{1}{\lww}\,\mrT^{d}_{194} + 2\,\mrT^{d}_{190} ) \Bigr] \Bigr\}\,\frac{1}{c^2}
         \,\bfun{ - \mhs}{\mw}{\mw}
\nl &+&
         \frac{1}{768}
         \,\Bigl\{ 2\,\frac{1}{\lww}\,\mrT^{d}_{186} - \Bigl[ 9\,c^4\,\xphs 
         - (  - 23\,\frac{\xphs}{\lww}\,\mrT^{d}_{196} 
         + 2\,\mrT^{d}_{189} ) \Bigr]\,c^2 \Bigr\}\,\frac{1}{c^6}
\nl &\times& \cfun{ - \mhs}{ - \mws}{ - \mws}{\mz}{\mw}{\mz}
\nl &-&
         \frac{1}{768}
         \,\Bigl\{ \mrT^{d}_{184}\,c^2\,\xphs 
         + \Bigl[ ( 23\,\frac{\xphs}{\lww} - 2\,\frac{1}{\lww}\,\mrT^{d}_{194} 
         - 2\,\mrT^{d}_{191} ) \Bigr] \Bigr\}\,\frac{1}{c^4}
\nl &\times& \cfun{ - \mhs}{ - \mws}{ - \mws}{\mw}{\mz}{\mw}
\eqas
\bqas
\mcT^{\nfact}_{\ssD\,;\,\PW\PW}(\apD) &=&
       -
         \frac{1}{3072}
         \,\mrX_{21}
     +
         \frac{1}{3072}
         \,\frac{s^2}{c^2}
         \,\mrX_{20}
\nl &+&
         \frac{1}{384}
         \,\xphs
         \,\afun{\mw}
     +
         \frac{1}{384}
         \,\frac{1}{c^4}\,\mrT^{d}_{211}
         \,\afun{\mz}
     +
         \frac{1}{384}
         \,\mrV_{1}\,\xphs
         \,\afun{\mh}
\nl &+&
         \frac{1}{3072}
         \,\Bigl[ \frac{1}{\lww}\,\mrT^{d}_{206} 
         + ( \frac{\xphs}{\lww}\,\mrT^{d}_{215}\,c^2 + 2\,\mrT^{d}_{208} )\,c^2 \Bigr]\,\frac{1}{c^6}
         \,\bfun{ - \mws}{\mw}{\mz}
\nl &-&
         \frac{1}{6144}
         \,\Bigl[ ( \mrT^{d}_{197}\,\xphs + 2\,\mrT^{d}_{201} )\,\frac{1}{\lww} 
         - ( \mrT^{d}_{199}\,\xphs - 2\,\mrT^{d}_{209} )\,c^2 \Bigr]\,\frac{1}{c^6}
\nl &\times& \cfun{ - \mhs}{ - \mws}{ - \mws}{\mw}{\mz}{\mw}
\nl &-&
         \frac{1}{3072}
         \,\Bigl\{ 8\,c^2\,s\mrW_{71} + \Bigl[ ( 4\,\frac{\xphs}{\lww}\,\mrT^{d}_{219} 
         + \mrT^{d}_{223}\,\mrW_{68} ) \Bigr] \Bigr\}\,\frac{1}{c^2}
         \,\bfun{ - \mws}{\mw}{\mh}
\nl &-&
         \frac{1}{2048}
         \,\Bigl\{ 16\,c^2\,s\mrW_{70} + \Bigl[ (  - 4\,\frac{\xphs}{\lww}\,\mrT^{d}_{217} 
         + \frac{\xphs}{\lww}\,\mrT^{d}_{220}\,\xphs 
         + 4\,\frac{1}{\lww}\,\mrT^{d}_{213} - \mrT^{d}_{226} ) \Bigr]\,\xphs \Bigr\}\,\frac{1}{c^2}
\nl &\times& \cfun{ - \mhs}{ - \mws}{ - \mws}{\mh}{\mw}{\mh}
\nl &+&
         \frac{1}{2048}
         \,\Bigl\{ 16\,c^2\,s\mrW_{69} + \Bigl[ ( 2\,\mrT^{d}_{216} 
         - \mrT^{d}_{220}\,\xphs ) \Bigr]\,\frac{\xphs}{\lww} \Bigr\}\,\frac{1}{c^2}
         \,\bfun{ - \mhs}{\mh}{\mh}
\nl &-&
         \frac{1}{6144}
         \,\Bigl\{ 2\,\frac{1}{\lww}\,\mrT^{d}_{202} 
         + \Bigl[ 9\,\mrT^{d}_{218}\,c^4\,\xphs - ( \frac{\xphs}{\lww}\,\mrT^{d}_{228} 
         + 2\,\mrT^{d}_{207} ) \Bigr]\,c^2 \Bigr\}\,\frac{1}{c^8}
\nl &\times& \cfun{ - \mhs}{ - \mws}{ - \mws}{\mz}{\mw}{\mz}
\nl &-&
         \frac{1}{6144}
         \,\Bigl\{ 2\,\frac{1}{\lww}\,\mrT^{d}_{203} + \Bigl[ 48\,c^4\,\xphs 
         - (  - \frac{\xphs}{\lww}\,\mrT^{d}_{227} 
         + \mrT^{d}_{198} ) \Bigr]\,c^2 \Bigr\}\,\frac{1}{c^6}
         \,\bfun{ - \mhs}{\mz}{\mz}
\nl &+&
         \frac{1}{6144}
         \,\Bigl\{ \Bigl[ 144\,c^2\,\xphs + ( \frac{\xphs}{\lww}\,\mrT^{d}_{212}\,\xphs 
         + 2\,\mrT^{d}_{210} ) \Bigr]\,c^2 
         + \Bigl[ ( 2\,\mrT^{d}_{201} 
         + \mrT^{d}_{204}\,\xphs ) \Bigr]\,\frac{1}{\lww} \Bigr\}\,\frac{1}{c^4}
\nl &\times& \bfun{ - \mhs}{\mw}{\mw}
\nl &+&
         \frac{1}{2048}
         \,( 16\,c^4\,\xphs - \mrT^{d}_{205} )\,\frac{1}{c^4}
         \,\LR
\nl &-&
         \frac{1}{6144}
         \,( \frac{\xphs}{\lww}\,\mrT^{d}_{212}\,\xphq 
         + 2\,\frac{\xphs}{\lww}\,\mrT^{d}_{224}\,\xphs - 2\,\mrT^{d}_{216}\,\mrV_{0} 
         + \mrT^{d}_{221}\,\xphq )\,\frac{1}{c^2}
\nl &\times& \cfun{ - \mhs}{ - \mws}{ - \mws}{\mw}{\mh}{\mw}
\nl &+&
         \frac{1}{768}
         \,( \mrT^{d}_{200}\,\xphs - 2\,\mrT^{d}_{214} )\,\frac{1}{c^2}
         \,\cfunf{ - \mhs}{ - \mws}{ - \mws}{\mw}{0}{\mw}
\nl &-&
         \frac{1}{6144}
         \,( \mrT^{d}_{222} - \mrT^{d}_{225}\,c^2\,\xphs )\,\frac{1}{c^4}
\eqas
\bqas
\mcT^{\nfact}_{\ssD\,;\,\PW\PW}(\aAZ) &=&
       -
         \frac{1}{384}
         \,s\,c
         \,\mrX_{23}
     -
         \frac{1}{256}
         \,s\,c\,\xphs
         \,\mrX_{22}
\nl &-&
         \frac{1}{768}
         \,\frac{s}{c}
         \,\mrX_{24}
     +
         \frac{1}{16}
         \,\frac{c^3}{s}\,\mrV_{39}
         \,\mrX_{25}
     +
         \frac{1}{64}
         \,\frac{s}{c}\,\mrT^{d}_{88}\,\xphs
         \,\afun{\mw}
\nl &+&
         \frac{1}{8}
         \,\frac{1}{\lww}\,\frac{s}{c^3}\,\mrT^{d}_{88}
         \,\bfun{ - \mhs}{0}{\mz}
\nl &-&
         \frac{1}{32}
         \,\mrV_{36}\,s\,c\,\xphs
         \,\afun{\mh}
\nl &+&
         \frac{1}{256}
         \,\Bigl[ 48\,c^2\,\xphs + (  - \frac{1}{\lww}\,\mrT^{d}_{240}\,\mrV_{28} 
         + \mrT^{d}_{237} ) \Bigr]\,\frac{s}{c}\,\xphs
         \,\cfun{ - \mhs}{ - \mws}{ - \mws}{\mh}{\mw}{\mh}
\nl &+&
         \frac{1}{96}
         \,\Bigl[ 12\,s^2\,c^2\,\xphq - ( \mrT^{d}_{258}\,\xphs
         - 2\,\mrT^{d}_{259} ) \Bigr]\,\frac{s}{c}
         \,\cfunf{ - \mhs}{ - \mws}{ - \mws}{\mw}{0}{\mw}
\nl &+&
         \frac{1}{16}
         \,\Bigl[ 2\,( c^2\,\xphs - \mrT^{d}_{232} )\,c^2 
         - ( \frac{1}{\lww}\,\mrT^{d}_{88} - \mrV_{9} ) \Bigr]\,\frac{s}{c^3}
         \,\bfun{ - \mws}{0}{\mw}
\nl &-&
         \frac{1}{768}
         \,\Bigl\{ 2\,\frac{1}{\lww}\,\mrT^{d}_{246} 
         + \Bigl[ 6\,( c^2\,\xphs + 2\,\mrT^{d}_{229} )\,c^2\,\xphs 
         + ( \frac{\xphs}{\lww}\,\mrT^{d}_{243} + \mrT^{d}_{255} ) \Bigr]\,c^2
         \Bigr\}\,\frac{s}{c^5}
\nl &\times& \bfun{ - \mhs}{\mz}{\mz}
\nl &-&
         \frac{1}{768}
         \,\Bigl\{ 2\,\frac{1}{\lww}\,\mrT^{d}_{247} 
         + \Bigl[ 3\,\mrT^{d}_{257}\,c^4\,\xphs + ( \frac{\xphs}{\lww}\,\mrT^{d}_{245} 
         + 2\,\mrT^{d}_{250} ) \Bigr]\,c^2 \Bigr\}\,\frac{s}{c^7}
\nl &\times& \cfun{ - \mhs}{ - \mws}{ - \mws}{\mz}{\mw}{\mz}
\nl &+&
         \frac{1}{384}
         \,\Bigl\{ \frac{1}{\lww}\,\mrT^{d}_{248} 
         - \Bigl[ \frac{\xphs}{\lww}\,\mrT^{d}_{252}\,c^2 + 2\,( 3\,\mrT^{d}_{233}\,\xphs 
         - \mrT^{d}_{253} + 12\,\mrV_{9} ) \Bigr]\,c^2 \Bigr\}\,
         \frac{s}{c^5}
\nl &\times& \bfun{ - \mws}{\mw}{\mz}
\nl &-&
         \frac{1}{384}
         \,\Bigl\{ 12\,\mrV_{40}\,c^2\,\xphs + \Bigl[ ( 4\,\frac{\xphs}{\lww}\,\mrT^{d}_{239} 
         - \frac{\xphs}{\lww}\,\mrT^{d}_{240}\,\xphs + 2\,\mrT^{d}_{237} ) \Bigr] 
         \Bigr\}\,\frac{s}{c})
\nl &\times& \bfun{ - \mws}{\mw}{\mh}
\nl &-&
         \frac{1}{768}
         \,\Bigl\{ \Bigl[ 12\,c^2\,\xphq - ( \frac{\xphs}{\lww}\,\mrT^{d}_{240}\,\xphs 
         + 24\,\mrT^{d}_{88}\,\xphs - 2\,\mrT^{d}_{251} ) \Bigr]\,c^2 
         - ( \mrT^{d}_{244}\,\xphs + 2\,\mrT^{d}_{249} )\,\frac{1}{\lww} \Bigr\}\,\frac{s}{c^3}
\nl &\times& \bfun{ - \mhs}{\mw}{\mw}
\nl &+&
         \frac{1}{768}
         \,\Bigl\{ \Bigl[ 96\,c^6\,\xphq + ( 2\,\mrT^{d}_{254} 
         - \mrT^{d}_{256}\,\xphs ) \Bigr]\,c^2 - ( \mrT^{d}_{242}\,\xphs 
         + 2\,\mrT^{d}_{249} )\,\frac{1}{\lww} \Bigr\}\,\frac{s}{c^5}
\nl &\times& \cfun{ - \mhs}{ - \mws}{ - \mws}{\mw}{\mz}{\mw}
\nl &+&
         \frac{1}{32}
         \,\Bigl\{ 2\,\Bigl[ \mrT^{d}_{8} + ( c^2\,\xphs 
         - 2\,\mrT^{d}_{230} )\,c^2\,\xphs \Bigr]\,c^2 + ( \frac{1}{\lww}\,\mrT^{d}_{231} 
         - \mrV_{9} ) \Bigr\}\,\frac{s}{c^5}
\nl &\times& \cfun{ - \mhs}{ - \mws}{ - \mws}{\mz}{\mw}{0}
\nl &+&
         \frac{1}{32}
         \,\Bigl\{ 2\,\Bigl[ \mrT^{d}_{8} 
         + ( c^2\,\xphs - 2\,\mrT^{d}_{230} )\,c^2\,\xphs \Bigr]\,c^2 
         + ( \frac{1}{\lww}\,\mrT^{d}_{231} - \mrV_{9} ) \Bigr\}\,\frac{s}{c^5}
\nl &\times& \cfun{ - \mhs}{ - \mws}{ - \mws}{0}{\mw}{\mz}
\nl &+&
         \frac{1}{256}
         \,( 1 - c^2\,\xphs )\,\frac{s}{c^3}
\nl &-&
         \frac{1}{256}
         \,( 12\,s^2\,c^4\,\xphq + 3\,\mrT^{d}_{234} 
         + 8\,\mrT^{d}_{235}\,c^2\,\xphs )\,\frac{1}{s\,c^3}
         \,\LR
\nl &-&
         \frac{1}{768}
         \,( 2\,\frac{\xphs}{\lww}\,\mrT^{d}_{236}\,\xphs 
         + \frac{\xphs}{\lww}\,\mrT^{d}_{240}\,\xphq + \mrT^{d}_{237}\,\mrV_{38}\,\xphs 
         - 4\,\mrT^{d}_{238} )\,\frac{s}{c}
\nl &\times& \cfun{ - \mhs}{ - \mws}{ - \mws}{\mw}{\mh}{\mw}
\nl &+&
         \frac{1}{256}
         \,( \frac{1}{\lww}\,\mrT^{d}_{240}\,\mrV_{0} + 6\,\mrV_{40}\,c^2 )\,\frac{s}{c}\,\xphs
         \,\bfun{ - \mhs}{\mh}{\mh}
\nl &-&
         \frac{1}{192}
         \,( 3\,\mrT^{d}_{100}\,\xphs - 4\,\mrT^{d}_{241} )\,\frac{s}{c^3}
         \,\afun{\mz}
\eqas
\bqas
\mcT^{\nfact}_{\ssD\,;\,\PW\PW}(\aAA) &=&
         \frac{1}{32}
         \,s^2
         \,\mrX_{27}
     -
         \frac{1}{64}
         \,s^2\,\xphs
         \,\mrX_{26}
     +
         \frac{1}{8}
         \,\mrV_{39}\,c^2
         \,\mrX_{25}
\nl &-&
         \frac{1}{16}
         \,s^2\,\xphs
         \,\afun{\mw}
\nl &-&
         \frac{1}{32}
         \,\frac{s^2}{c^2}\,\mrT^{d}_{7}\,\mrV_{36}
         \,\afun{\mz}
\nl &-&
         \frac{1}{64}
         \,\frac{s^2}{c^2}\,\mrT^{d}_{269}
\nl &-&
         \frac{1}{8}
         \,\mrV_{0}\,s^2\,\xphs
         \,\cfun{ - \mhs}{ - \mws}{ - \mws}{0}{\mw}{0}
\nl &-&
         \frac{1}{32}
         \,\mrV_{36}\,s^2\,\xphs
         \,\afun{\mh}
\nl &+&
         \frac{1}{8}
         \,\mrV_{39}\,c^2
         \,\ssdCZ^{(4)}_{c}
\nl &-&
         \frac{3}{128}
         \,\mrW_{73}\,s^2\,\xphs
         \,\bfun{ - \mhs}{\mh}{\mh}
\nl &+&
         \frac{3}{64}
         \,\mrW_{74}\,s^2\,\xphs
         \,\cfun{ - \mhs}{ - \mws}{ - \mws}{\mh}{\mw}{\mh}
\nl &+&
         \frac{1}{32}
         \,\mrW_{75}\,s^2
         \,\bfun{ - \mws}{\mw}{\mh}
\nl &+&
         \frac{1}{64}
         \,\mrW_{76}\,s^2
         \,\cfun{ - \mhs}{ - \mws}{ - \mws}{\mw}{\mh}{\mw}
\nl &-&
         \frac{1}{64}
         \,\Bigl\{ c^2\,s\mrW_{72} + \Bigl[ ( 2\,\mrT^{d}_{77} 
         + \mrT^{d}_{260}\,\xphs ) \Bigr]\,\frac{1}{\lww} \Bigr\}\,\frac{s^2}{c^2}
         \,\bfun{ - \mhs}{\mw}{\mw}
\nl &+&
         \frac{1}{8}
         \,\Bigl\{ s^2\,\xphq 
         + \Bigl[ ( 2\,\mrT^{d}_{107} - \mrT^{d}_{270}\,\xphs ) \Bigr] \Bigr\}\,s^2
         \,\cfunf{ - \mhs}{ - \mws}{ - \mws}{\mw}{0}{\mw}
\nl &+&
         \frac{1}{128}
         \,\Bigl\{ 4\,\frac{1}{\lww}\,\mrT^{d}_{261} + \Bigl[ \mrV_{2}\,c^2\,\xphs 
         + 2\,(  - \frac{\xphs}{\lww}\,\mrT^{d}_{195} 
         + \mrT^{d}_{77} ) \Bigr]\,c^2 \Bigr\}\,\frac{s^2}{c^4}
         \,\bfun{ - \mhs}{\mz}{\mz}
\nl &-&
         \frac{1}{32}
         \,\Bigl\{ \frac{1}{\lww}\,\mrT^{d}_{262} - \Bigl[ \frac{\xphs}{\lww}\,c^2 
         + ( \xphs - 2\,\mrT^{d}_{264} ) \Bigr]\,c^2 \Bigr\}\,\frac{s^2}{c^4}
         \,\bfun{ - \mws}{\mw}{\mz}
\nl &+&
         \frac{1}{64}
         \,\Bigl\{ 2\,\frac{1}{\lww}\,\mrT^{d}_{263} 
         + \Bigl[ \frac{\xphs}{\lww}\,\mrT^{d}_{196} 
         + ( 11\,c^2\,\xphs + 2\,\mrT^{d}_{268} )\,c^2 \Bigr]\,s^2\,c^2 \Bigr\}\,\frac{1}{c^6}
\nl &\times& \cfun{ - \mhs}{ - \mws}{ - \mws}{\mz}{\mw}{\mz}
\nl &+&
         \frac{1}{64}
         \,\Bigl\{ \Bigl[ 8\,c^4\,\xphq + ( \mrT^{d}_{265}\,\xphs
         - 2\,\mrT^{d}_{266} ) \Bigr]\,c^2 
         + ( \xphs + 2\,\mrT^{d}_{77} )\,\frac{1}{\lww} \Bigr\}\,\frac{s^2}{c^4}
\nl &\times& \cfun{ - \mhs}{ - \mws}{ - \mws}{\mw}{\mz}{\mw}
\nl &+&
         \frac{1}{4}
         \,( \xphs - 2\,c^2 )\,s^2
         \,\bfun{ - \mws}{0}{\mw}
\nl &-&
         \frac{1}{64}
         \,( 3\,s^2\,c^2\,\xphq - 22\,\mrT^{d}_{29}\,c^2\,\xphs
       - \mrT^{d}_{267} )\,\frac{1}{c^2}
         \,\LR
\eqas
\bqas
\mcT^{\nfact}_{\ssD\,;\,\PW\PW}(\aZZ) &=&
         \frac{1}{64}
         \,c^2
         \,\mrX_{29}
     -
         \frac{1}{128}
         \,c^2\,\xphs
         \,\mrX_{28}
     +
         \frac{1}{64}
         \,s^2
         \,\mrX_{20}
\nl &+&
         \frac{1}{2}
         \,s^2\,c^2
         \,\bfun{ - \mws}{0}{\mw}
\nl &+&
         \frac{1}{32}
         \,\frac{s^2}{c^2}\,\mrT^{d}_{7}\,\mrV_{36}
         \,\afun{\mz}
\nl &-&
         \frac{1}{32}
         \,\mrT^{d}_{195}\,\xphs
         \,\afun{\mw}
\nl &-&
         \frac{1}{32}
         \,\mrV_{36}\,c^2\,\xphs
         \,\afun{\mh}
\nl &+&
         \frac{3}{128}
         \,\Bigl[ 8\,c^2\,\xphs + ( \frac{1}{\lww}\,\mrT^{d}_{195}\,\mrV_{28} 
         - \mrT^{d}_{290} ) \Bigr]\,\xphs
         \,\cfun{ - \mhs}{ - \mws}{ - \mws}{\mh}{\mw}{\mh}
\nl &+&
         \frac{1}{16}
         \,\Bigl[ 2\,c^2\,\xphq - ( \mrT^{d}_{195}\,\xphs + 2\,\mrT^{d}_{287} ) \Bigr]\,s^2
         \,\cfunf{ - \mhs}{ - \mws}{ - \mws}{\mw}{0}{\mw}
\nl &+&
         \frac{1}{128}
         \,\Bigl[ \mrT^{d}_{282} 
         - 2\,( 3\,c^2\,\xphs + 2\,\mrT^{d}_{275} )\,c^2\,\xphs \Bigr]\,\frac{1}{c^2}
         \,\LR
\nl &-&
         \frac{1}{64}
         \,\Bigl[ 2\,\mrV_{40}\,c^2\,\xphs + ( \frac{\xphs}{\lww}\,\mrT^{d}_{195}\,\xphs 
         - 4\,\frac{\xphs}{\lww}\,\mrT^{d}_{288} - 2\,\mrT^{d}_{290} ) \Bigr]
         \,\bfun{ - \mws}{\mw}{\mh}
\nl &+&
         \frac{1}{128}
         \,\Bigl\{ 2\,\frac{1}{\lww}\,\mrT^{d}_{277} 
         - \Bigl[ ( c^2\,\xphs - 2\,\mrT^{d}_{286} )\,c^2\,\xphs 
         + ( \frac{\xphs}{\lww}\,\mrT^{d}_{196} + \mrT^{d}_{285} ) \Bigr]\,c^2 \Bigr\}\,
         \frac{1}{c^4}
         \,\bfun{ - \mhs}{\mz}{\mz}
\nl &+&
         \frac{1}{128}
         \,\Bigl\{ 2\,\frac{1}{\lww}\,\mrT^{d}_{278} 
         + \Bigl[ ( 16\,c^4\,\xphs + \mrT^{d}_{281} )\,c^2\,\xphs 
         - (  - \frac{\xphs}{\lww}\,\mrT^{d}_{292} 
         + 2\,\mrT^{d}_{284} ) \Bigr]\,c^2 \Bigr\}\,\frac{1}{c^6}
\nl &\times& \cfun{ - \mhs}{ - \mws}{ - \mws}{\mz}{\mw}{\mz}
\nl &-&
         \frac{1}{64}
         \,\Bigl\{ \frac{1}{\lww}\,\mrT^{d}_{279} 
         - \Bigl[ \frac{\xphs}{\lww}\,\mrT^{d}_{195}\,c^2 
         - 2\,( \mrT^{d}_{272}\,\xphs - \mrT^{d}_{283} ) \Bigr]\,c^2 \Bigr\}\,\frac{1}{c^4}
         \,\bfun{ - \mws}{\mw}{\mz}
\nl &-&
         \frac{1}{128}
         \,\Bigl\{ \Bigl[ 2\,c^2\,\xphq 
         + ( \mrT^{d}_{195}\,\mrW_{77}\,\xphs + 2\,\mrT^{d}_{289} ) \Bigr]\,c^2 
         - ( \mrT^{d}_{271}\,\xphs + 2\,\mrT^{d}_{274} )\,\frac{1}{\lww} \Bigr\}\,\frac{1}{c^2}
\nl &\times& \bfun{ - \mhs}{\mw}{\mw}
\nl &+&
         \frac{1}{128}
         \,\Bigl\{ \Bigl[ 16\,c^6\,\xphq + ( 2\,\mrT^{d}_{273} 
         - \mrT^{d}_{280}\,\xphs ) \Bigr]\,c^2 - ( \mrT^{d}_{7}\,\xphs
         + 2\,\mrT^{d}_{274} )\,\frac{1}{\lww} \Bigr\}\,\frac{1}{c^4}
\nl &\times& \cfun{ - \mhs}{ - \mws}{ - \mws}{\mw}{\mz}{\mw}
\nl &+&
         \frac{1}{128}
         \,( \frac{\xphs}{\lww}\,\mrT^{d}_{195}\,\xphq 
         + 2\,\frac{\xphs}{\lww}\,\mrT^{d}_{291}\,\xphs - 4\,\mrT^{d}_{289} 
         + \mrT^{d}_{290}\,\mrV_{38}\,\xphs )
\nl &\times& \cfun{ - \mhs}{ - \mws}{ - \mws}{\mw}{\mh}{\mw}
\nl &-&
         \frac{3}{128}
         \,( \frac{1}{\lww}\,\mrT^{d}_{195}\,\mrV_{0} - \mrV_{40}\,c^2 )\,\xphs
         \,\bfun{ - \mhs}{\mh}{\mh}
\nl &-&
         \frac{1}{128}
         \,( \mrT^{d}_{153}\,c^2\,\xphs + \mrT^{d}_{276} )\,\frac{1}{c^2}
\eqas
\bqas
\mcT^{\nfact}_{\ssD\,;\,\PW\PW}(\ren) &=&
         \frac{1}{32}
         \,\mrX_{30}      
\eqas
\bqas
\mcT^{\nfact}_{\ssP\,;\,\PW\PW}(\atp) &=&
         \frac{3}{32}
         \,\mrW_{78}\,\xpts\,\xpbs
         \,\afun{\mb}
     -
         \frac{3}{32}
         \,\mrW_{78}\,\xptq
         \,\afun{\mt}
     +
         \frac{3}{32}
         \,\mrW_{79}\,\xpts
\nl &-&
         \frac{3}{64}
         \,\mrW_{80}\,\xpts
         \,\bfun{ - \mws}{\mt}{\mb}
\nl &+&
         \frac{3}{64}
         \,\mrW_{81}\,\xpts
         \,\bfun{ - \mhs}{\mt}{\mt}
\nl &-&
         \frac{3}{64}
         \,\mrW_{82}\,\xpts
         \,\cfun{ - \mhs}{ - \mws}{ - \mws}{\mt}{\mb}{\mt}
\eqas
\bqas
\mcT^{\nfact}_{\ssP\,;\,\PW\PW}(\abp) &=&
         \frac{3}{32}
         \,\mrW_{78}\,\xpts\,\xpbs
         \,\afun{\mb}
     -
         \frac{3}{32}
         \,\mrW_{78}\,\xptq
         \,\afun{\mt}
     +
         \frac{3}{32}
         \,\mrW_{79}\,\xpts
\nl &-&
         \frac{3}{64}
         \,\mrW_{80}\,\xpts
         \,\bfun{ - \mws}{\mt}{\mb}
\nl &+&
         \frac{3}{64}
         \,\mrW_{81}\,\xpts
         \,\bfun{ - \mhs}{\mt}{\mt}
\nl &-&
         \frac{3}{64}
         \,\mrW_{82}\,\xpts
         \,\cfun{ - \mhs}{ - \mws}{ - \mws}{\mt}{\mb}{\mt}
\eqas
\bqas
\mcT^{\nfact}_{\ssP\,;\,\PW\PW}(\atBW) &=&
         \frac{3}{16}
         \,\frac{\xpts}{\lww}\,c\,\xpbs
         \,\bfun{ - \mhs}{\mb}{\mb}
\nl &+&
         \frac{3}{32}
         \,\frac{\xpts}{\lww}\,\mrW_{87}\,c
         \,\cfun{ - \mhs}{ - \mws}{ - \mws}{\mt}{\mb}{\mt}
\nl &-&
         \frac{3}{64}
         \,\mrW_{83}\,c\,\xpts
\nl &-&
         \frac{3}{64}
         \,\mrW_{83}\,c\,\xpts\,\xpbs
         \,\afun{\mb}
\nl &+&
         \frac{3}{64}
         \,\mrW_{83}\,c\,\xptq
         \,\afun{\mt}
\nl &+&
         \frac{3}{64}
         \,\mrW_{84}\,c\,\xpts
         \,\bfun{ - \mhs}{\mt}{\mt}
\nl &-&
         \frac{3}{32}
         \,\mrW_{85}\,c\,\xpts\,\xpbs
         \,\cfun{ - \mhs}{ - \mws}{ - \mws}{\mb}{\mt}{\mb}
\nl &-&
         \frac{3}{64}
         \,\mrW_{86}\,c\,\xpts
         \,\bfun{ - \mws}{\mt}{\mb}
\eqas
\bqas
\mcT^{\nfact}_{\ssP\,;\,\PW\PW}(\atWB) &=&
         \frac{3}{16}
         \,\frac{\xpts}{\lww}\,s\,\xpbs
         \,\bfun{ - \mhs}{\mb}{\mb}
\nl &+&
         \frac{3}{32}
         \,\frac{\xpts}{\lww}\,\mrW_{87}\,s
         \,\cfun{ - \mhs}{ - \mws}{ - \mws}{\mt}{\mb}{\mt}
\nl &-&
         \frac{3}{64}
         \,\mrW_{83}\,s\,\xpts
\nl &-&
         \frac{3}{64}
         \,\mrW_{83}\,s\,\xpts\,\xpbs
         \,\afun{\mb}
\nl &+&
         \frac{3}{64}
         \,\mrW_{83}\,s\,\xptq
         \,\afun{\mt}
\nl &+&
         \frac{3}{64}
         \,\mrW_{84}\,s\,\xpts
         \,\bfun{ - \mhs}{\mt}{\mt}
\nl &-&
         \frac{3}{32}
         \,\mrW_{85}\,s\,\xpts\,\xpbs
         \,\cfun{ - \mhs}{ - \mws}{ - \mws}{\mb}{\mt}{\mb}
\nl &-&
         \frac{3}{64}
         \,\mrW_{86}\,s\,\xpts
         \,\bfun{ - \mws}{\mt}{\mb}
\eqas
\bqas
\mcT^{\nfact}_{\ssP\,;\,\PW\PW}(\abBW) &=&
       -
         \frac{3}{64}
         \,\mrW_{88}\,c\,\xpbs
         \,\afun{\mb}
\nl &+&
         \frac{3}{64}
         \,\mrW_{88}\,c\,\xpts
         \,\afun{\mt}
\nl &-&
         \frac{3}{64}
         \,\mrW_{89}\,c
\nl &-&
         \frac{3}{64}
         \,\mrW_{90}\,c\,\xpts
         \,\bfun{ - \mhs}{\mt}{\mt}
\nl &+&
         \frac{3}{64}
         \,\mrW_{91}\,c
         \,\bfun{ - \mws}{\mt}{\mb}
\nl &+&
         \frac{3}{64}
         \,\mrW_{92}\,c\,\xpts
         \,\cfun{ - \mhs}{ - \mws}{ - \mws}{\mt}{\mb}{\mt}
\nl &-&
         \frac{3}{64}
         \,\mrW_{93}\,c\,\xpbs
         \,\cfun{ - \mhs}{ - \mws}{ - \mws}{\mb}{\mt}{\mb}
\nl &-&
         \frac{3}{64}
         \,\mrW_{94}\,c\,\xpbs
         \,\bfun{ - \mhs}{\mb}{\mb}
\eqas
\bqas
\mcT^{\nfact}_{\ssP\,;\,\PW\PW}(\abWB) &=&
       -
         \frac{3}{64}
         \,\mrW_{88}\,s\,\xpbs
         \,\afun{\mb}
\nl &+&
         \frac{3}{64}
         \,\mrW_{88}\,s\,\xpts
         \,\afun{\mt}
\nl &-&
         \frac{3}{64}
         \,\mrW_{89}\,s
\nl &-&
         \frac{3}{64}
         \,\mrW_{90}\,s\,\xpts
         \,\bfun{ - \mhs}{\mt}{\mt}
\nl &+&
         \frac{3}{64}
         \,\mrW_{91}\,s
         \,\bfun{ - \mws}{\mt}{\mb}
\nl &+&
         \frac{3}{64}
         \,\mrW_{92}\,s\,\xpts
         \,\cfun{ - \mhs}{ - \mws}{ - \mws}{\mt}{\mb}{\mt}
\nl &-&
         \frac{3}{64}
         \,\mrW_{93}\,s\,\xpbs
         \,\cfun{ - \mhs}{ - \mws}{ - \mws}{\mb}{\mt}{\mb}
\nl &-&
         \frac{3}{64}
         \,\mrW_{94}\,s\,\xpbs
         \,\bfun{ - \mhs}{\mb}{\mb}
\eqas
\bqas
\mcT^{\nfact}_{\ssP\,;\,\PW\PW}(\apB) &=&
         \frac{3}{16}
         \,\mrW_{78}
         \,\afun{\mw}
\nl &+&
         \frac{3}{16}
         \,\mrW_{79}
\nl &+&
         \frac{3}{16}
         \,\mrW_{95}
         \,\afun{\mh}
\nl &+&
         \frac{3}{32}
         \,\mrW_{96}
         \,\bfun{ - \mhs}{\mh}{\mh}
\nl &-&
         \frac{3}{32}
         \,\mrW_{97}
         \,\bfun{ - \mws}{\mw}{\mh}
\nl &-&
         \frac{3}{32}
         \,\mrW_{98}
         \,\cfun{ - \mhs}{ - \mws}{ - \mws}{\mh}{\mw}{\mh}
\eqas
\bqas
\mcT^{\nfact}_{\ssP\,;\,\PW\PW}(\apBox) &=&
       -
         \frac{5}{64}
         \,\frac{s^2}{c^4}\,\mrT^{d}_{309}\,\mrW_{78}
         \,\afun{\mz}
\nl &+&
         \frac{1}{64}
         \,\frac{1}{c^2}\,\mrT^{d}_{297}\,\mrW_{78}
         \,\afun{\mw}
\nl &-&
         \frac{5}{2}
         \,\frac{1}{\lww}\,s^2
         \,\bfun{ - \mws}{0}{\mw}
\nl &-&
         \frac{3}{64}
         \,\mrW_{95}
         \,\afun{\mh}
\nl &-&
         \frac{1}{16}
         \,\mrW_{102}
         \,\bfun{ - \mws}{\mw}{\mh}
\nl &+&
         \frac{1}{256}
         \,\mrW_{103}
         \,\bfun{ - \mhs}{\mh}{\mh}
\nl &+&
         \frac{1}{256}
         \,\mrW_{104}
         \,\cfun{ - \mhs}{ - \mws}{ - \mws}{\mw}{\mh}{\mw}
\nl &+&
         \frac{1}{256}
         \,\mrW_{105}
         \,\cfun{ - \mhs}{ - \mws}{ - \mws}{\mh}{\mw}{\mh}
\nl &+&
         \frac{1}{64}
         \,\Bigl[ 4\,c^2\,s\mrW_{101} + ( \frac{1}{\lww}\,\mrT^{d}_{298} 
         + \mrT^{d}_{299}\,\mrV_{9} ) \Bigr]\,\frac{1}{c^2}
\nl &-&
         \frac{1}{256}
         \,\Bigl\{ 5\,c^2\,s\mrW_{99}\,\xphs 
         + \Bigl[ (  - 12\,\frac{\xphs}{\lwws}\,\mrT^{d}_{308} 
         - \frac{\xphs}{\lww}\,\mrT^{d}_{306} - 24\,\frac{1}{\lwws}\,\mrT^{d}_{293} 
\nl &+& 22\,\frac{1}{\lww}\,\mrT^{d}_{55} + 2\,\mrT^{d}_{296}\,\mrV_{9} 
         + \mrT^{d}_{307} ) \Bigr] \Bigr\}\,\frac{1}{c^2}
         \,\bfun{ - \mhs}{\mw}{\mw}
\nl &-&
         \frac{1}{128}
         \,\Bigl\{ 7\,c^4\,s\mrW_{100} - \Bigl[ ( 12\,\frac{1}{\lwws}\,\mrT^{d}_{304} 
         - \frac{1}{\lww}\,\mrT^{d}_{305} + 5\,\mrT^{d}_{295}\,\mrV_{9} ) \Bigr] \Bigr\}\,\frac{1}{c^4}
         \,\bfun{ - \mws}{\mw}{\mz}
\nl &+&
         \frac{1}{256}
         \,\Bigl\{ 7\,\Bigl[ ( 12\,\frac{\xphs}{\lwws}\,\mrT^{d}_{195} 
         + \mrW_{95} ) \Bigr]\,c^2 + 2\,\Bigl[ (  - 12\,\frac{1}{\lwws}\,\mrT^{d}_{302} 
         - \frac{1}{\lww}\,\mrT^{d}_{301} + 5\,\mrT^{d}_{58}\,\mrV_{9} ) \Bigr] \Bigr\}\,\frac{1}{c^4}
\nl &\times& \bfun{ - \mhs}{\mz}{\mz}
\nl &-&
         \frac{1}{256}
         \,\Bigl\{ 7\,\Bigl[ ( 12\,\frac{\xphs}{\lwws}\,\mrT^{d}_{196} 
         + \frac{\xphs}{\lww}\,\mrT^{d}_{312} - \mrT^{d}_{311} ) \Bigr]\,c^2 
\nl &-& 2\,\Bigl[ (  - 12\,\frac{1}{\lwws}\,\mrT^{d}_{300} 
         - \frac{1}{\lww}\,\mrT^{d}_{303} + 5\,\mrT^{d}_{294}\,\mrV_{9} ) \Bigr] \Bigr\}\,\frac{1}{c^6}
         \,\cfun{ - \mhs}{ - \mws}{ - \mws}{\mz}{\mw}{\mz}
\nl &-&
         \frac{1}{256}
         \,( 84\,\frac{\xphs}{\lwws} + 7\,\frac{\xphs}{\lww}\,\mrT^{d}_{8} 
         + 24\,\frac{1}{\lwws}\,\mrT^{d}_{293} - 2\,\frac{1}{\lww}\,\mrT^{d}_{310} 
         - 10\,\mrT^{d}_{294}\,\mrV_{9} - \mrT^{d}_{313} )\,\frac{1}{c^4}
\nl &\times& \cfun{ - \mhs}{ - \mws}{ - \mws}{\mw}{\mz}{\mw}
\eqas
\bqas
\mcT^{\nfact}_{\ssP\,;\,\PW\PW}(\apD) &=&
       -
         \frac{1}{4}
         \,\frac{1}{\lww}\,\mrT^{d}_{12}
         \,\bfun{ - \mws}{0}{\mw}
\nl &+&
         \frac{1}{512}
         \,\mrW_{107}
         \,\cfun{ - \mhs}{ - \mws}{ - \mws}{\mw}{\mh}{\mw}
\nl &-&
         \frac{1}{128}
         \,\mrW_{108}
         \,\bfun{ - \mhs}{\mh}{\mh}
\nl &+&
         \frac{1}{256}
         \,\mrW_{109}
         \,\afun{\mh}
\nl &-&
         \frac{1}{128}
         \,\mrW_{110}
         \,\cfun{ - \mhs}{ - \mws}{ - \mws}{\mh}{\mw}{\mh}
\nl &-&
         \frac{1}{512}
         \,\mrW_{111}
         \,\bfun{ - \mws}{\mw}{\mh}
\nl &+&
         \frac{1}{256}
         \,\Bigl[ c^2\,s\mrW_{95} + 2\,( \frac{1}{\lww}\,\mrT^{d}_{323} 
         - \mrT^{d}_{325}\,\mrV_{9} ) \Bigr]\,\frac{1}{c^4}
         \,\afun{\mz}
\nl &+&
         \frac{1}{512}
         \,\Bigl[ \mrT^{d}_{331}\,c^2\,s\mrW_{100} - 2\,( 12\,\frac{1}{\lwws}\,\mrT^{d}_{329} 
         - \frac{1}{\lww}\,\mrT^{d}_{330} + \mrT^{d}_{327}\,\mrV_{9} ) \Bigr]\,\frac{1}{c^4}
         \,\bfun{ - \mws}{\mw}{\mz}
\nl &-&
         \frac{1}{128}
         \,\Bigl\{ 2\,c^2\,s\mrW_{106} + \Bigl[ ( \frac{1}{\lww}\,\mrT^{d}_{321} 
         + 5\,\mrT^{d}_{315}\,\mrV_{9} ) \Bigr] \Bigr\}\,
         \frac{1}{c^2}
\nl &+&
         \frac{1}{512}
         \,\Bigl\{ 3\,c^2\,s\mrW_{99}\,\xphs - \Bigl[ ( 60\,\frac{\xphs}{\lwws} 
         + \frac{\xphs}{\lww}\,\mrT^{d}_{318} + 48\,\frac{1}{\lwws}\,\mrT^{d}_{316} 
         - 8\,\frac{1}{\lww}\,\mrT^{d}_{229}
\nl &-& 8\,\mrT^{d}_{314}\,\mrV_{9} - \mrT^{d}_{319} ) \Bigr] \Bigr\}\,\frac{1}{c^2}
         \,\bfun{ - \mhs}{\mw}{\mw}
\nl &-&
         \frac{1}{256}
         \,\Bigl\{ 2\,\Bigl[ ( 12\,\frac{\xphs}{\lwws}\,\mrT^{d}_{195} + \mrW_{95} ) \Bigr]\,c^2 
         + \Bigl[ (  - 12\,\frac{1}{\lwws}\,\mrT^{d}_{324} 
         - \frac{1}{\lww}\,\mrT^{d}_{322} 
         + 5\,\mrT^{d}_{58}\,\mrV_{9} ) \Bigr] \Bigr\}\,\frac{1}{c^4}
\nl &\times& \bfun{ - \mhs}{\mz}{\mz}
\nl &+&
         \frac{1}{256}
         \,\Bigl\{ 2\,\Bigl[ ( 12\,\frac{\xphs}{\lwws}\,\mrT^{d}_{196} 
         + \frac{\xphs}{\lww}\,\mrT^{d}_{312} - \mrT^{d}_{332} ) \Bigr]\,c^2 
         - \Bigl[ (  - 12\,\frac{1}{\lwws}\,\mrT^{d}_{326} 
         - \frac{1}{\lww}\,\mrT^{d}_{328} 
\nl &+& 5\,\mrT^{d}_{294}\,\mrV_{9} ) \Bigr] \Bigr\}\,\frac{1}{c^6}
         \,\cfun{ - \mhs}{ - \mws}{ - \mws}{\mz}{\mw}{\mz}
\nl &+&
         \frac{1}{512}
         \,( 60\,\frac{\xphs}{\lwws} + 5\,\frac{\xphs}{\lww}\,\mrT^{d}_{8} 
         + 48\,\frac{1}{\lwws}\,\mrT^{d}_{316} - 8\,\frac{1}{\lww}\,\mrT^{d}_{317} 
         - 8\,\mrT^{d}_{294}\,\mrV_{9} - \mrT^{d}_{333} )\,\frac{1}{c^4}
\nl &\times& \cfun{ - \mhs}{ - \mws}{ - \mws}{\mw}{\mz}{\mw}
\nl &+&
         \frac{1}{128}
         \,(  - \frac{1}{\lww}\,\mrT^{d}_{320} + 5\,\mrT^{d}_{144}\,\mrV_{9} )\,\frac{1}{c^2}
         \,\afun{\mw}
\eqas
\bqas
\mcT^{\nfact}_{\ssP\,;\,\PW\PW}(\aAZ) &=&
       -
         \frac{1}{16}
         \,s\,c
         \,\mrX_{31}
       -
         \frac{1}{8}
         \,\frac{c^3}{s}
         \,\mrX_{25}
\nl &-&
         \frac{3}{8}
         \,\frac{\xphs}{\lww}\,\mrV_{0}\,s\,c
         \,\cfun{ - \mhs}{ - \mws}{ - \mws}{\mh}{\mw}{\mh}
\nl &-&
         \frac{1}{8}
         \,\frac{1}{s}\,c^3
         \,\ssdCZ^{(4)}_{c}
\nl &-&
         \frac{1}{4}
         \,\mrV_{0}\,s^3\,c
         \,\cfunf{ - \mhs}{ - \mws}{ - \mws}{\mw}{0}{\mw}
\nl &-&
         \frac{3}{64}
         \,\mrW_{112}\,s\,c\,\xphs
         \,\bfun{ - \mhs}{\mh}{\mh}
\nl &+&
         \frac{1}{64}
         \,\mrW_{113}\,s\,c
         \,\bfun{ - \mws}{\mw}{\mh}
\nl &+&
         \frac{1}{64}
         \,\mrW_{114}\,s\,c
         \,\cfun{ - \mhs}{ - \mws}{ - \mws}{\mw}{\mh}{\mw}
\nl &-&
         \frac{1}{32}
         \,\mrW_{115}\,s\,c
         \,\afun{\mh}
\nl &-&
         \frac{1}{128}
         \,\Bigl[ 12\,\frac{\xphs}{\lwws}\,s^2\,c^2 + \frac{\xphs}{lww}\,\mrT^{d}_{351}\,c^2 
         + ( 24\,\frac{1}{\lwws}\,\mrT^{d}_{352} 
         - 2\,\frac{1}{\lww}\,\mrT^{d}_{356} + \mrT^{d}_{346} 
\nl &+& 2\,\mrT^{d}_{350}\,\mrV_{9} 
         + 24\,\mrV_{92} ) \Bigr]\,\frac{s}{c^3}
         \,\bfun{ - \mws}{\mw}{\mz}
\nl &+&
         \frac{1}{64}
         \,\Bigl[ \frac{\xphs}{\lww}\,s^2\,c^2 - (  - 2\,\frac{1}{\lww}\,\mrT^{d}_{353} 
         + \mrT^{d}_{341} + 2\,\mrT^{d}_{349}\,\mrV_{9} ) \Bigr]\,\frac{s}{c^3}
         \,\afun{\mz}
\nl &-&
         \frac{1}{16}
         \,\Bigl[ (  - 18\,\frac{1}{\lwws}\,\mrT^{d}_{339}
         + \mrT^{d}_{98}\,\mrV_{9} ) + ( 6\,\frac{1}{\lww}\,\mrT^{d}_{336} 
         + 5\,\mrT^{d}_{7}\,c^2 )\,c^2 \Bigr]\,\frac{s}{c^5}
\nl &\times& \cfun{ - \mhs}{ - \mws}{ - \mws}{\mz}{\mw}{\mz}
\nl &-&
         \frac{1}{128}
         \,\Bigl\{ 4\,c^2\,\xphs - \Bigl[ ( 12\,\frac{\xphs}{\lwws}\,\mrT^{d}_{54} 
         + \frac{\xphs}{\lww}\,\mrT^{d}_{354} - 24\,\frac{1}{\lwws}\,\mrT^{d}_{355} 
         + 6\,\frac{1}{\lww}\,\mrT^{d}_{334} 
\nl &+& 10\,\mrT^{d}_{88}\,\mrV_{9} 
         - \mrT^{d}_{335} ) \Bigr] \Bigr\}\,\frac{s}{c}
         \,\bfun{ - \mhs}{\mw}{\mw}
\nl &+&
         \frac{1}{128}
         \,\Bigl\{ 32\,c^6\,\xphs - 12\,\frac{\xphs}{\lwws}\,s^2 
         - \Bigl[ ( 3\,\frac{\xphs}{\lww}\,\mrT^{d}_{343} 
         - 24\,\frac{1}{\lwws}\,\mrT^{d}_{355} + 2\,\frac{1}{\lww}\,\mrT^{d}_{344}
\nl &+& 10\,\mrT^{d}_{88}\,\mrV_{9} - \mrT^{d}_{340} ) \Bigr] \Bigr\}\,\frac{s}{c^3}
         \,\cfun{ - \mhs}{ - \mws}{ - \mws}{\mw}{\mz}{\mw}
\nl &-&
         \frac{1}{64}
         \,\Bigl\{ \Bigl[ c^2\,\xphs - 2\,(  - 6\,\frac{1}{\lww}\,\mrT^{d}_{7} 
         + \mrT^{d}_{2} ) \Bigr]\,c^2 + 4\,(  - 18\,\frac{1}{\lwws}\,\mrT^{d}_{100} 
         + \mrT^{d}_{7}\,\mrV_{9} ) \Bigr\}\,\frac{s}{c^3}
\nl&\times& \bfun{ - \mhs}{\mz}{\mz}
\nl &+&
         \frac{1}{32}
         \,\Bigl\{  - 2\,\Bigl[ 2\,( 2 - c^2\,\xphs )\,c^2 
         + ( \frac{1}{\lww}\,\mrT^{d}_{88} - 3\,\mrV_{9} ) \Bigr]\,c^2
         + 3\,( \frac{1}{\lwws}\,\mrT^{d}_{231} - \mrV_{92} ) \Bigr\}\,\frac{s}{c^5}
\nl &\times& \cfun{ - \mhs}{ - \mws}{ - \mws}{\mz}{\mw}{0}
\nl &+&
         \frac{1}{32}
         \,\Bigl\{  - 2\,\Bigl[ 2\,( 2 - c^2\,\xphs )\,c^2 
         + ( \frac{1}{\lww}\,\mrT^{d}_{88} - 3\,\mrV_{9} ) \Bigr]\,c^2
         + 3\,( \frac{1}{\lwws}\,\mrT^{d}_{231} - \mrV_{92} ) \Bigr\}\,\frac{s}{c^5}
\nl&\times& \cfun{ - \mhs}{ - \mws}{ - \mws}{0}{\mw}{\mz}
\nl &-&
         \frac{1}{16}
         \,\Bigl\{  - \Bigl[ ( \frac{\xphs}{\lww}\,\mrT^{d}_{2}
         + \mrT^{d}_{342} )\,c^2 - ( \frac{1}{\lww}\,\mrT^{d}_{338} 
         + 3\,\mrV_{9} ) \Bigr]\,c^2 + 3\,( \frac{1}{\lwws}\,\mrT^{d}_{88} 
         - \mrV_{92} ) \Bigr\}\,\frac{s}{c^3}
\nl &\times&          \,\bfun{ - \mws}{0}{\mw}
\nl &-&
         \frac{1}{32}
         \,( 1 + 2\,\frac{1}{\lww}\,\mrT^{d}_{348} + 2\,\mrT^{d}_{110}\,\mrV_{9} )\,\frac{s}{c}
\nl &+&
         \frac{1}{32}
         \,( s\mrW_{78} + 12\,\frac{1}{\lwws}\,\mrT^{d}_{88} )\,\frac{s}{c^3}
         \,\bfun{ - \mhs}{0}{\mz}
\nl &-&
         \frac{1}{32}
         \,( 3\,s^2\,c^2\,\xphs + 2\,\mrT^{d}_{337} )\,\frac{1}{s\,c}
         \,\LR
\nl &+&
         \frac{1}{32}
         \,(  - 2\,\frac{1}{\lww}\,\mrT^{d}_{347} + \mrT^{d}_{88} 
         + 2\,\mrT^{d}_{345}\,\mrV_{9} )\,\frac{s}{c}
         \,\afun{\mw}
\eqas
\bqas
\mcT^{\nfact}_{\ssP\,;\,\PW\PW}(\aAA) &=&
       -
         \frac{1}{4}
         \,c^2
         \,\mrX_{25}
     -
         \frac{1}{16}
         \,s^2
         \,\mrX_{32}
\nl &+&
         \frac{1}{4}
         \,s^2\,\xphs
         \,\cfun{ - \mhs}{ - \mws}{ - \mws}{0}{\mw}{0}
\nl &-&
         \frac{3}{8}
         \,\frac{\xphs}{\lww}\,\mrV_{0}\,s^2
         \,\cfun{ - \mhs}{ - \mws}{ - \mws}{\mh}{\mw}{\mh}
\nl &-&
         \frac{1}{4}
         \,\mrV_{0}\,s^4
         \,\cfunf{ - \mhs}{ - \mws}{ - \mws}{\mw}{0}{\mw}
\nl &-&
         \frac{1}{16}
         \,\mrW_{79}\,s^2
\nl &-&
         \frac{3}{64}
         \,\mrW_{112}\,s^2\,\xphs
         \,\bfun{ - \mhs}{\mh}{\mh}
\nl &+&
         \frac{1}{64}
         \,\mrW_{113}\,s^2
         \,\bfun{ - \mws}{\mw}{\mh}
\nl &+&
         \frac{1}{64}
         \,\mrW_{114}\,s^2
         \,\cfun{ - \mhs}{ - \mws}{ - \mws}{\mw}{\mh}{\mw}
\nl &-&
         \frac{1}{32}
         \,\mrW_{115}\,s^2
         \,\afun{\mh}
\nl &+&
         \frac{1}{16}
         \,\mrW_{117}\,s^2
         \,\afun{\mw}
\nl &+&
         \frac{1}{8}
         \,\Bigl[ 5\,s^2\,c^4 + 6\,\frac{1}{\lww}\,\mrT^{d}_{358}\,c^2 
         + (  - 18\,\frac{1}{\lwws}\,\mrT^{d}_{361} + \mrT^{d}_{359}\,\mrV_{9} ) \Bigr]\,\frac{1}{c^4}
\nl &\times& \cfun{ - \mhs}{ - \mws}{ - \mws}{\mz}{\mw}{\mz}
\nl &+&
         \frac{1}{64}
         \,\Bigl[ 8\,\frac{\xphs}{\lww}\,s^2\,c^2 + ( 12\,\frac{1}{\lwws}\,\mrT^{d}_{364} 
         - \frac{1}{\lww}\,\mrT^{d}_{366} 
         + \mrT^{d}_{363}\,\mrV_{9} + 4\,\mrT^{d}_{367} ) \Bigr]\,\frac{s^2}{c^2}
\nl &\times& \bfun{ - \mws}{\mw}{\mz}
\nl &+&
         \frac{1}{64}
         \,\Bigl\{ c^2\,s\mrW_{118} - 8\,\Bigl[ ( 18\,\frac{1}{\lwws}\,\mrT^{d}_{7} 
         - \mrV_{9} ) \Bigr] \Bigr\}\,\frac{s^2}{c^2}
         \,\bfun{ - \mhs}{\mz}{\mz}
\nl &-&
         \frac{1}{64}
         \,\Bigl\{ c^2\,s\mrW_{116} + \Bigl[ ( 12\,\frac{1}{\lwws}\,\mrT^{d}_{55} 
         + \frac{1}{\lww}\,\mrT^{d}_{357} - \mrT^{d}_{334}\,\mrV_{9} ) \Bigr] \Bigr\}\,\frac{s^2}{c^2}
         \,\bfun{ - \mhs}{\mw}{\mw}
\nl &+&
         \frac{1}{8}
         \,\Bigl\{ \frac{\xphs}{\lww}\,c^2 + \Bigl[ ( 4\,\frac{1}{\lww}\,\mrT^{d}_{106} 
         + \mrT^{d}_{362} ) \Bigr] \Bigr\}\,s^2
         \,\bfun{ - \mws}{0}{\mw}
\nl &+&
         \frac{1}{64}
         \,\Bigl\{ 8\,\frac{\xphs}{\lww}\,s^2\,c^2 - 16\,\mrV_{0}\,c^6 
         + \Bigl[ ( 12\,\frac{1}{\lwws}\,\mrT^{d}_{55} + \frac{1}{\lww}\,\mrT^{d}_{360} 
         - \mrT^{d}_{88}\,\mrV_{9} ) \Bigr] \Bigr\}\,\frac{s^2}{c^4}
\nl &\times& \cfun{ - \mhs}{ - \mws}{ - \mws}{\mw}{\mz}{\mw}
\nl &-&
         \frac{1}{32}
         \,( 3\,s^2\,\xphs - 16\,\mrT^{d}_{29} )
         \,\LR
\nl &+&
         \frac{1}{32}
         \,(  - \frac{1}{\lww}\,\mrT^{d}_{365} + 2\,\mrT^{d}_{7} + 9\,\mrV_{9} )\,\frac{s^2}{c^2}
         \,\afun{\mz}
\eqas
\bqas
\mcT^{\nfact}_{\ssP\,;\,\PW\PW}(\aZZ) &=&
       -
         \frac{1}{16}
         \,c^2
         \,\mrX_{33}
\nl &-&
         \frac{3}{8}
         \,\frac{\xphs}{\lww}\,\mrV_{0}\,c^2
         \,\cfun{ - \mhs}{ - \mws}{ - \mws}{\mh}{\mw}{\mh}
\nl &-&
         \frac{1}{4}
         \,\mrV_{0}\,s^2\,c^2
         \,\cfunf{ - \mhs}{ - \mws}{ - \mws}{\mw}{0}{\mw}
\nl &-&
         \frac{3}{64}
         \,\mrW_{112}\,c^2\,\xphs
         \,\bfun{ - \mhs}{\mh}{\mh}
\nl &+&
         \frac{1}{64}
         \,\mrW_{113}\,c^2
         \,\bfun{ - \mws}{\mw}{\mh}
\nl &+&
         \frac{1}{64}
         \,\mrW_{114}\,c^2
         \,\cfun{ - \mhs}{ - \mws}{ - \mws}{\mw}{\mh}{\mw}
\nl &-&
         \frac{1}{32}
         \,\mrW_{115}\,c^2
         \,\afun{\mh}
\nl &+&
         \frac{1}{8}
         \,\mrW_{119}\,s^2\,c^2
         \,\bfun{ - \mws}{0}{\mw}
\nl &-&
         \frac{1}{32}
         \,\Bigl[ 2\,c^2 + ( \frac{1}{\lww}\,\mrT^{d}_{378}
       + \mrT^{d}_{375}\,\mrV_{9} ) \Bigr]\,\frac{1}{c^2}
\nl &-&
         \frac{1}{64}
         \,\Bigl[ c^2\,s\mrW_{120} - (  - \frac{\xphs}{\lww}\,\mrT^{d}_{389} 
         - 12\,\frac{1}{\lwws}\,\mrT^{d}_{386} + \frac{1}{\lww}\,\mrT^{d}_{387} 
         + \mrT^{d}_{385}\,\mrV_{9} ) \Bigr]
\nl &\times& \bfun{ - \mhs}{\mw}{\mw}
\nl &-&
         \frac{1}{64}
         \,\Bigl[ c^6\,\xphs + 4\,s^2\,c^4 + (  - 12\,\frac{1}{\lwws}\,\mrT^{d}_{376} 
         - \frac{1}{\lww}\,\mrT^{d}_{370}
         + \mrT^{d}_{369}\,\mrV_{9} ) \Bigr]\,\frac{1}{c^4}
         \,\bfun{ - \mhs}{\mz}{\mz}
\nl &-&
         \frac{1}{64}
         \,\Bigl[ 8\,\frac{\xphs}{\lww}\,s^4\,c^4 - 4\,\mrT^{d}_{390}\,c^2 
         + ( 12\,\frac{1}{\lwws}\,\mrT^{d}_{381} - \frac{1}{\lww}\,\mrT^{d}_{383} 
         + \mrT^{d}_{380}\,\mrV_{9} ) \Bigr]\,\frac{1}{c^4}
\nl &\times& \bfun{ - \mws}{\mw}{\mz}
\nl &-&
         \frac{1}{32}
         \,\Bigl[ 2\,\mrT^{d}_{7}\,c^2 + (  - \frac{1}{\lww}\,\mrT^{d}_{382} 
         + \mrT^{d}_{368}\,\mrV_{9} ) \Bigr]\,\frac{s^2}{c^4}
         \,\afun{\mz}
\nl &-&
         \frac{1}{32}
         \,\Bigl[ 2\,\mrT^{d}_{195}\,c^2 - (  - \frac{1}{\lww}\,\mrT^{d}_{377} 
         + \mrT^{d}_{374}\,\mrV_{9} ) \Bigr]\,
         \frac{1}{c^2}
         \,\afun{\mw}
\nl &-&
         \frac{1}{64}
         \,\Bigl\{ 8\,\frac{\xphs}{\lww}\,s^4 
         + \Bigl[ (  - 12\,\frac{1}{\lwws}\,\mrT^{d}_{386} 
         + \frac{1}{\lww}\,\mrT^{d}_{388} + 9\,\mrT^{d}_{88}\,\mrV_{9} ) \Bigr] 
         - 8\,( 2\,c^4\,\xphs - \mrT^{d}_{384} )\,c^2 \Bigr\}\,\frac{1}{c^2}
\nl &\times& \cfun{ - \mhs}{ - \mws}{ - \mws}{\mw}{\mz}{\mw}
\nl &+&
         \frac{1}{64}
         \,\Bigl\{ \Bigl[ ( 12\,\frac{1}{\lwws}\,\mrT^{d}_{371}
         + \frac{1}{\lww}\,\mrT^{d}_{372} - \mrT^{d}_{373}\,\mrV_{9} ) \Bigr] 
         + 8\,( 2\,c^4\,\xphs + \mrT^{d}_{379} )\,c^4 \Bigr\}\,\frac{1}{c^6}
\nl &\times& \cfun{ - \mhs}{ - \mws}{ - \mws}{\mz}{\mw}{\mz}
\nl &-&
         \frac{1}{32}
         \,( 3\,c^2\,\xphs + 4\,\mrT^{d}_{88} )
         \,\LR
\eqas

\normalsize

\section{Appendix: $\mrdim = 4$ sub-amplitudes \label{LOsub}}

In this Appendix we list the $\mrdim = 4$ sub-amplitudes for $\PH \to \PAZ, \PZZ$
and $\PWW$.
\subsection{$\PH \to \PAZ$}
\footnotesize
\bqa
\mcT^{\PQt}_{\sPHAZ\,;\,\myLO} &=&
         \frac{1}{4}
         \,\frac{\mhs\,\mts}{\mw\,(\mhs-\mzs)}
         \,\frac{s}{c}
         \vtq
       +
         \frac{1}{8}
         \,( 1 - 4\,\frac{\mts}{\mhs-\mzs} )\,\frac{\mhs\,\mts}{\mw}
         \,\frac{s}{c}
         \,\vtq
         \,\cfun{ - \mhs}{0}{ -\mzs}{\mt}{\mt}{\mt}
\nl &-&
         \frac{1}{4}
         \,\frac{\mhs\,\mw}{\mhs-\mzs}
         \,\frac{\mts}{\mhs-\mzs}
         \,\frac{s}{c^3}
         \,\vtq
         \,\bfun{ -\mzs}{\mt}{\mt}
\nl &+&
         \frac{1}{4}
         \,\frac{\mw\,\mhs}{\mhs-\mzs}
         \,\frac{\mts}{\mhs-\mzs}
         \,\frac{s}{c^3}
         \vtq
         \,\bfun{ - \mhs}{\mt}{\mt}
\eqa
\bqa
\mcT^{\PQb}_{\sPHAZ\,;\,\myLO} &=&
         \frac{1}{8}
         \,\frac{\mhs\,\mbs}{\mw\,(\mhs-\mzs)}
         \,\frac{s}{c}
         \,\vbq
       +
         \frac{1}{16}
         \,( 1 - 4\,\frac{\mbs}{\mhs-\mzs} )\,\frac{\mhs\,\mbs}{\mw}
         \,\frac{s}{c}
         \,\vbq
         \,\cfun{ - \mhs}{0}{ -\mzs}{\mb}{\mb}{\mb}
\nl &-&
         \frac{1}{8}
         \,\frac{\mhs\,\mw}{\mhs-\mzs}
         \,\frac{\mbs}{\mhs-\mzs}
         \,\frac{s}{c^3}
         \,\vbq
         \,\bfun{ -\mzs}{\mb}{\mb}
\nl &+&
         \frac{1}{8}
         \,\frac{\mw\,\mhs}{\mhs-\mzs}
         \,\frac{\mbs}{\mhs-\mzs}
         \,\frac{s}{c^3}
         \,\vbq
         \,\bfun{ - \mhs}{\mb}{\mb}
\eqa
\bqa
\mcT^{\PW}_{\sPHAZ\,;\,\myLO} &=&
         \frac{1}{16}
         \,\Bigl[ 2\,( 1 - 6\,c^2 )\,\mw + ( 1 - 2\,c^2 )\,\frac{\mhs}{\mw} \Bigr]
         \,\frac{\mhs}{\mhs-\mzs}
         \,\frac{s}{c}
\nl &-&
         \frac{1}{8}
         \,\Bigl[ 2\,( 1 - 6\,c^2 )\,\frac{\mwc}{\mhs-\mzs} - 2\,( 1 - 4\,c^2 )\,\mw 
         + ( 1 - 2\,c^2 )\,\frac{\mhs\,\mw}{\mhs-\mzs} \Bigr]
         \,\mhs
         \,\frac{s}{c}
\nl &\times&
         \,\cfun{ - \mhs}{0}{ -\mzs}{\mw}{\mw}{\mw}
\nl &-&
         \frac{1}{16}
         \,\Bigl[ 2\,( 1 - 6\,c^2 )\,\frac{\mwc}{\mhs-\mzs} 
         + ( 1 - 2\,c^2 )\,\frac{\mhs\,\mw}{\mhs-\mzs} \Bigr]
         \,\frac{\mhs}{\mhs-\mzs}
         \,\frac{s}{c^3}
         \,\bfun{ -\mzs}{\mw}{\mw}
\nl &+&
         \frac{1}{16}
         \,\Bigl[ 2\,( 1 - 6\,c^2 )\,\frac{\mwc}{\mhs-\mzs} 
         + ( 1 - 2\,c^2 )\,\frac{\mhs\,\mw}{\mhs-\mzs} \Bigr]
         \,\frac{\mhs}{\mhs-\mzs}
         \,\frac{s}{c^3}
         \,\bfun{ - \mhs}{\mw}{\mw}
\eqa

\normalsize
\subsection{$\PH \to \PZZ$}
\footnotesize
\bqa
\mcD^{\PQt}_{\sPHZZ\,;\,\myNLO} &=&
       -
         \frac{3}{64}
         \,\Bigl[ 2\,\LR - ( 1 + \vtqs ) \Bigr]
         \,\frac{1}{c^2}
         \,\frac{\mts}{\mw}
\nl &+&
         \frac{3}{32}
         \,\Bigl[ c^2 + ( 1 + \vtqs )\,\frac{\mws}{\lz} \Bigr]
         \,\frac{1}{c^4}
         \,\frac{\mts}{\mw}
         \,\bfun{ - \mzs}{\mt}{\mt}
\nl &+&
         \frac{3}{32}
         \,\Bigl[ ( 1 + \vtqs ) \Bigr]
         \,\frac{1}{c^4}
         \,\frac{\mw\,\mts}{\lz}
         \,\bfun{ - \mhs}{\mt}{\mt}
\nl &+&
         \frac{3}{128}
         \,\Bigl[  - 4\,( 1 - \vtqs )\,c^4\,\mts + ( 1 - \vtqs )\,c^4\,\mhs
         + 2\,( 1 + \vtqs )\,c^2\,\mws + 4\,( 1 + \vtqs )\,\frac{\mwq}{\lz} \Bigr]
         \,\frac{1}{c^6}
         \,\frac{\mts}{\mw}
\nl &\times& \cfun{ - \mhs}{ - \mzs}{ - \mzs}{\mt}{\mt}{\mt}
\nl &+& \Delta\mcD^{\PQt}_{\sPHZZ\,;\,\myNLO} 
\eqa
\bqa
\mcD^{\PQb}_{\sPHZZ\,;\,\myNLO} &=&
       -
         \frac{3}{64}
         \,\Bigl[ 2\,\LR - ( 1 + \vbqs ) \Bigr]
         \,\frac{1}{c^2}
         \,\frac{\mbs}{\mw}
\nl &+&
         \frac{3}{32}
         \,\Bigl[ c^2 + ( 1 + \vbqs )\,\frac{\mws}{\lz} \Bigr]
         \,\frac{1}{c^4}
         \,\frac{\mbs}{\mw}
         \,\bfun{ - \mzs}{\mb}{\mb}
\nl &+&
         \frac{3}{32}
         \,\Bigl[ ( 1 + \vbqs ) \Bigr]
         \,\frac{1}{c^4}
         \,\frac{\mbs\,\mw}{\lz}
         \,\bfun{ - \mhs}{\mb}{\mb}
\nl &+&
         \frac{3}{128}
         \,\Bigl[ ( 1 - \vbqs )\,c^4\,\mhs - 4\,( 1 - \vbqs )\,\mbs\,c^4 
         + 2\,( 1 + \vbqs )\,c^2\,\mws + 4\,( 1 + \vbqs )\,\frac{\mwq}{\lz} \Bigr]
         \,\frac{1}{c^6}
         \,\frac{\mbs}{\mw}
\nl &\times& \cfun{ - \mhs}{ - \mzs}{ - \mzs}{\mb}{\mb}{\mb}
\nl &+& \Delta\mcD^{\PQb}_{\sPHZZ\,;\,\myNLO} 
\eqa
\bqa
\mcD^{\PW}_{\sPHZZ\,;\,\myNLO} &=&
       -
         \frac{1}{4}
         \,\afun{\mw}
         \,\mw
\nl &+&
         \frac{1}{64}
         \,\Bigl[ 3\,( 1 + 2\,c^2 )\,\LR\,\mws - ( 1 + 2\,c^2 + 32\,c^4 - 24\,s^2\,c^4 )\,\mws 
         - ( 3 - 4\,s^2\,c^2 )\,c^2\,\mhs \Bigr]
         \,\frac{1}{\mw}
         \,\frac{1}{c^4}
\nl &+&
         \frac{1}{32}
         \,\Bigl[ c^2 + 2\,( 1 + 8\,c^2 - 12\,s^2\,c^2 )\,\frac{\mws}{\lz} 
         + ( 1 - 4\,s^2\,c^2 )\,\frac{\mhs}{\lz} \Bigr]
         \,\frac{1}{c^4}
         \,\mw
         \,\bfun{ - \mhs}{\mw}{\mw}
\nl &+&
         \frac{1}{32}
         \,\Bigl[ 4\,( 1 - 2\,c^2 )\,c^2 
         + 2\,( 1 + 8\,c^2 - 12\,s^2\,c^2 )\,\frac{\mws}{\lz} 
         + ( 1 - 4\,s^2\,c^2 )\,\frac{\mhs}{\lz} \Bigr]
         \,\frac{1}{c^4}
         \,\mw
         \,\bfun{ - \mzs}{\mw}{\mw}
\nl &+&
         \frac{1}{64}
         \,( \mws + \frac{\mhq}{\lz}\,c^2 + 2\,\frac{\mws\,\mhs}{\lz} )
         \,\frac{1}{\mw}
         \,\frac{1}{c^4}
         \,\bfun{ - \mhs}{\mz}{\mz}
\nl &+&
         \frac{3}{64}
         \,( 2\,\mws - c^2\,\mhs )
         \,\frac{1}{\mw}
         \,\frac{1}{c^4}
         \,\frac{\mhs}{\lz}
         \,\bfun{ - \mhs}{\mh}{\mh}
\nl &+&
         \frac{1}{32}
         \,( 2\,\mws - \frac{\mhq}{\lz}\,c^2 + 4\,\frac{\mws\,\mhs}{\lz} )
         \,\frac{1}{\mw}
         \,\frac{1}{c^4}
         \,\bfun{ - \mzs}{\mh}{\mz}
\nl &+&
         \frac{1}{32}
         \,\Bigl[ 6\,( 1 - 4\,c^2 - 4\,c^4 )\,c^4\,\mws + ( 1 - 4\,c^2 + 12\,c^4 )\,c^4\,\mhs 
         - 2\,( 1 + 8\,c^2 - 12\,s^2\,c^2 )\,\frac{\mwq}{\lz} 
\nl &-& ( 1 - 4\,s^2\,c^2 )\,\frac{\mws\,\mhs}{\lz} \Bigr]
         \,\frac{1}{c^6}
         \,\mw
         \,\cfun{ - \mhs}{ - \mzs}{ - \mzs}{\mw}{\mw}{\mw}
\nl &+&
         \frac{1}{64}
         \,( 4\,\mwq - 2\,c^2\,\mws\,\mhs - c^4\,\mhq - \frac{\mhvi}{\lz}\,c^4 
         - 2\,\frac{\mws\,\mhq}{\lz}\,c^2 )
         \,\frac{1}{\mw}
         \,\frac{1}{c^6}
\nl &\times& \cfun{ - \mhs}{ - \mzs}{ - \mzs}{\mz}{\mh}{\mz}
\nl &+&
         \frac{3}{64}
         \,(  - c^2\,\mws + \frac{\mhq}{\lz}\,c^4 
         - 4\,\frac{\mws\,\mhs}{\lz}\,c^2 + 4\,\frac{\mwq}{\lz} )
         \,\frac{1}{c^6}
         \,\frac{\mhs}{\mw}
\nl &\times& \cfun{ - \mhs}{ - \mzs}{ - \mzs}{\mh}{\mz}{\mh}
\nl &+& \Delta\mcD^{\PW}_{\sPHZZ\,;\,\myNLO} 
\eqa
\bqa
\Delta\mcD^{\PQt}_{\sPHZZ\,;\,\myNLO} &=&
 \frac{1}{32}\,\frac{\mw}{c^2}\,\lpar 
     \mrW^{(4)}_{\PH\,;\,\PQt} + 2\,\mrW^{(4)}_{\PZ\,;\,\PQt} +
                           4\,\ssdCZ^{(4)}_{c\,;\,\PQt} \rpar
\nl
\Delta\mcD^{\PQb}_{\sPHZZ\,;\,\myNLO} &=&
 \frac{1}{32}\,\frac{\mw}{c^2}\,\lpar 
     \mrW^{(4)}_{\PH\,;\,\PQb} + 2\,\mrW^{(4)}_{\PZ\,;\,\PQb} +
                           4\,\ssdCZ^{(4)}_{c\,;\,\PQb} \rpar
\nl
\Delta\mcD^{\PW}_{\sPHZZ\,;\,\myNLO} &=&
 \frac{1}{32}\,\frac{\mw}{c^2}\,\lpar 
  \mrW^{(4)}_{\PH\,;\,\PW} +
  2\,\mrW^{(4)}_{\PZ\,;\,\PW} -
  \ssdCZ^{(4)}_{\mw\,;\,\PW} -
  2\,\ssdCZ^{(4)}_{g\,;\,\PW} +
  4\,\ssdCZ^{(4)}_{c\,;\,\PW} \rpar
\eqa
\vspace{0.5cm}
\bqa
\mcP^{\PQt}_{\sPHZZ\,;\,\myNLO} &=&
         \frac{3}{64}
         \,\Bigl[ ( 1 + \vtqs ) + ( 1 + \vtqs )\,\frac{\mhs}{\lz} \Bigr]
         \,\frac{1}{c^2}
         \,\frac{\mts}{\mw\,\mhs}
\nl &+&
         \frac{3}{128}
         \,\Bigl[ ( 1 + \vtqs )\,c^2 + 12\,( 1 + \vtqs )\,\frac{\mws\,\mhs}{\lzs} 
         + ( 7 - \vtqs )\,\frac{\mhs}{\lz}\,c^2 \Bigr]
         \,\frac{1}{c^4}
         \,\frac{\mts}{\mw\,\mhs}
         \,\bfun{ - \mhs}{\mt}{\mt}
\nl &+&
         \frac{3}{128}
         \,\Bigl[ ( 1 + \vtqs )\,c^2 
         + 12\,( 1 + \vtqs )\,\frac{\mws\,\mhs}{\lzs} 
         + ( 7 - \vtqs )\,\frac{\mhs}{\lz}\,c^2 \Bigr]
         \,\frac{1}{c^4}
         \,\frac{\mts}{\mw\,\mhs}
         \,\bfun{ - \mzs}{\mt}{\mt}
\nl &+&
         \frac{3}{128}
         \,\Bigl[ 2\,( 1 - \vtqs )\,c^4\,\mhs - ( 1 + \vtqs )\,c^2\,\mws 
         + 4\,( 1 + \vtqs )\,c^4\,\mts 
         + 4\,( 1 + \vtqs )\,\frac{\mhs\,\mts}{\lz}\,c^4 
\nl &+& 12\,( 1 + \vtqs )\,\frac{\mwq\,\mhs}{\lzs} 
         + ( 9 + \vtqs )\,\frac{\mws\,\mhs}{\lz}\,c^2 \Bigr]
         \,\frac{1}{c^6}
         \,\frac{\mts}{\mw\,\mhs}
\nl &\times& \cfun{ - \mhs}{ - \mzs}{ - \mzs}{\mt}{\mt}{\mt}
\eqa
\bqa
\mcP^{\PQb}_{\sPHZZ\,;\,\myNLO} &=&
         \frac{3}{64}
         \,\Bigl[ ( 1 + \vbqs ) + ( 1 + \vbqs )\,\frac{\mhs}{\lz} \Bigr]
         \,\frac{1}{c^2}
         \,\frac{\mbs}{\mw\,\mhs}
\nl &+&
         \frac{3}{128}
         \,\Bigl[ ( 1 + \vbqs )\,c^2 
         + 12\,( 1 + \vbqs )\,\frac{\mws\,\mhs}{\lzs} 
         + ( 7 - \vbqs )\,\frac{\mhs}{\lz}\,c^2 \Bigr]
         \,\frac{1}{c^4}
         \,\frac{\mbs}{\mw\,\mhs}
         \,\bfun{ - \mhs}{\mb}{\mb}
\nl &+&
         \frac{3}{128}
         \,\Bigl[ ( 1 + \vbqs )\,c^2 
         + 12\,( 1 + \vbqs )\,\frac{\mws\,\mhs}{\lzs} 
         + ( 7 - \vbqs )\,\frac{\mhs}{\lz}\,c^2 \Bigr]
         \,\frac{1}{c^4}
         \,\frac{\mbs}{\mw\,\mhs}
         \,\bfun{ - \mzs}{\mb}{\mb}
\nl &+&
         \frac{3}{128}
         \,\Bigl[ 2\,( 1 - \vbqs )\,c^4\,\mhs - ( 1 + \vbqs )\,c^2\,\mws 
         + 4\,( 1 + \vbqs )\,\mbs\,c^4 
         + 12\,( 1 + \vbqs )\,\frac{\mwq\,\mhs}{\lzs} 
\nl &+& 4\,( 1 + \vbqs )\,\frac{\mbs\,\mhs}{\lz}\,c^4 
         + ( 9 + \vbqs )\,\frac{\mws\,\mhs}{\lz}\,c^2 \Bigr]
         \,\frac{1}{c^6}
         \,\frac{\mbs}{\mw\,\mhs}
\nl &\times& \cfun{ - \mhs}{ - \mzs}{ - \mzs}{\mb}{\mb}{\mb}
\eqa
\bqa
\mcP^{\PW}_{\sPHZZ\,;\,\myNLO} &=&
       -
         \frac{1}{64}
         \,\Bigl[ ( 1 + 2\,c^2 + 16\,c^4 - 24\,s^2\,c^4 )\,\mws 
         + ( 1 + 2\,c^2 + 16\,c^4 - 24\,s^2\,c^4 )\,\frac{\mws\,\mhs}{\lz} 
\nl &+& ( 3 - 4\,s^2\,c^2 )\,c^2\,\mhs 
         + ( 3 - 4\,s^2\,c^2 )\,\frac{\mhq}{\lz}\,c^2 \Bigr]
         \,\frac{1}{c^4}
         \,\frac{1}{\mw\,\mhs}
\nl &+&
         \frac{1}{64}
         \,( \mws - c^2\,\mhs 
         + \frac{\mhq}{\lz}\,c^2 - \frac{\mws\,\mhs}{\lz} )
         \,\frac{1}{c^2}
         \,\frac{1}{\mwc}
         \,\afun{\mh}
\nl &+&
         \frac{1}{64}
         \,( \mws - c^2\,\mhs 
         + \frac{\mhq}{\lz}\,c^2 - \frac{\mws\,\mhs}{\lz} )
         \,\frac{1}{c^4}
         \,\frac{1}{\mw\,\mhs}
         \,\afun{\mz}
\nl &+&
         \frac{1}{128}
         \,\Bigl[ 2\,( 1 + 8\,c^2 - 12\,s^2\,c^2 )\,c^2\,\mws 
         + 24\,( 1 + 8\,c^2 - 12\,s^2\,c^2 )\,\frac{\mwq\,\mhs}{\lzs} 
\nl &+& ( 1 - 4\,s^2\,c^2 )\,c^2\,\mhs 
         - ( 1 - 4\,s^2\,c^2 )\,\frac{\mhq}{\lz}\,c^2 
         + 12\,( 1 - 4\,s^2\,c^2 )\,\frac{\mws\,\mhq}{\lzs} 
\nl &+& 2\,( 15 - 8\,c^2 + 12\,s^2\,c^2 )\,\frac{\mws\,\mhs}{\lz}\,c^2 \Bigr]
         \,\frac{1}{c^4}
         \,\frac{1}{\mw\,\mhs}
         \,\bfun{ - \mhs}{\mw}{\mw}
\nl &+&
         \frac{1}{128}
         \,\Bigl[ 2\,( 1 + 8\,c^2 - 12\,s^2\,c^2 )\,c^2\,\mws 
         + 24\,( 1 + 8\,c^2 - 12\,s^2\,c^2 )\,\frac{\mwq\,\mhs}{\lzs} 
\nl &+& ( 1 - 4\,s^2\,c^2 )\,c^2\,\mhs 
         - ( 1 - 4\,s^2\,c^2 )\,\frac{\mhq}{\lz}\,c^2 
         + 12\,( 1 - 4\,s^2\,c^2 )\,\frac{\mws\,\mhq}{\lzs} 
\nl &+& 2\,( 15 - 8\,c^2 + 12\,s^2\,c^2 )\,\frac{\mws\,\mhs}{\lz}\,c^2 \Bigr]
         \,\frac{1}{c^4}
         \,\frac{1}{\mw\,\mhs}
         \,\bfun{ - \mzs}{\mw}{\mw}
\nl &+&
         \frac{1}{256}
         \,( 4\,\mwq - c^4\,\mhq + \frac{\mhvi}{\lz}\,c^4 
         + 12\,\frac{\mws\,\mhvi}{\lzs}\,c^2 
         + 28\,\frac{\mwq\,\mhs}{\lz} 
         + 24\,\frac{\mwq\,\mhq}{\lzs} )
         \,\frac{1}{c^4}
         \,\frac{1}{\mwc\,\mhs}
\nl &\times& \bfun{ - \mhs}{\mz}{\mz}
\nl &+&
         \frac{1}{128}
         \,( 4\,c^2\,\mws - c^4\,\mhs + \frac{\mhq}{\lz}\,c^4
         - 4\,\frac{\mws\,\mhs}{\lz}\,c^2 
         - 12\,\frac{\mws\,\mhq}{\lzs}\,c^2 
         + 16\,\frac{\mwq}{\lz} 
\nl &+& 48\,\frac{\mwq\,\mhs}{\lzs} )
         \,\frac{1}{c^4}
         \,\frac{1}{\mwc}
         \,\bfun{ - \mzs}{\mh}{\mz}
\nl &+&
         \frac{3}{256}
         \,( c^4 - \frac{\mhs}{\lz}\,c^4 
         - 12\,\frac{\mws\,\mhs}{\lzs}\,c^2 
         + 24\,\frac{\mwq}{\lzs} )
         \,\frac{1}{c^4}
         \,\frac{\mhs}{\mwc}
         \,\bfun{ - \mhs}{\mh}{\mh}
\nl &+&
         \frac{1}{128}
         \,\Bigl[ ( 1 - 8\,c^2 - 44\,c^4 + 112\,c^6 )\,c^2\,\mhs 
         + 2\,( 1 - 8\,c^2 + 28\,c^4 - 48\,c^6 )\,c^2\,\mws 
\nl &-& 24\,( 1 + 8\,c^2 - 12\,s^2\,c^2 )\,\frac{\mwq\,\mhs}{\lzs} 
         - ( 1 - 12\,c^4 + 16\,c^6 )\,\frac{\mhq}{\lz}\,c^2 
         - 12\,( 1 - 4\,s^2\,c^2 )\,\frac{\mws\,\mhq}{\lzs} 
\nl &-& 2\,( 17 - 4\,c^4 + 48\,c^6 )\,\frac{\mws\,\mhs}{\lz}\,c^2 \Bigr]
         \,\frac{1}{c^6}
         \,\frac{\mw}{\mhs}
         \,\cfun{ - \mhs}{ - \mzs}{ - \mzs}{\mw}{\mw}{\mw}
\nl &+&
         \frac{1}{256}
         \,( 8\,\mwq + 2\,c^2\,\mws\,\mhs - c^4\,\mhq 
         + \frac{\mhvi}{\lz}\,c^4 + 6\,\frac{\mws\,\mhq}{\lz}\,c^2 
         + 12\,\frac{\mws\,\mhvi}{\lzs}\,c^2 
         + 40\,\frac{\mwq\,\mhs}{\lz} 
\nl &+& 24\,\frac{\mwq\,\mhq}{\lzs} )
         \,\frac{1}{c^4}
         \,\frac{1}{\mwc}
         \,\cfun{ - \mhs}{ - \mzs}{ - \mzs}{\mz}{\mh}{\mz}
\nl &+&
         \frac{3}{256}
         \,( 4\,c^4\,\mws - c^6\,\mhs + \frac{\mhq}{\lz}\,c^6
         - 4\,\frac{\mws\,\mhs}{\lz}\,c^4 
         + 12\,\frac{\mws\,\mhq}{\lzs}\,c^4 
         + 8\,\frac{\mwq}{\lz}\,c^2 
         - 48\,\frac{\mwq\,\mhs}{\lzs}\,c^2 
\nl &+& 48\,\frac{\mwvi}{\lzs} )
         \,\frac{1}{c^6}
         \,\frac{\mhs}{\mwc}
         \,\cfun{ - \mhs}{ - \mzs}{ - \mzs}{\mh}{\mz}{\mh}
\eqa

\normalsize
\subsection{$\PH \to \PWW$}
\footnotesize
\bqa
\mcD^{\PQq}_{\sPHWW\,;\,\myNLO} &=&
         \frac{3}{32}
         \,( \mbs + \mts - \LR\,\mbs - \LR\,\mts )
         \,\frac{1}{\mw}
\nl &+&
         \frac{3}{16}
         \,( \mbs - \mts + \mws )
         \,\frac{1}{\mw}
         \,\frac{\mts}{\lw}
         \,\bfun{ - \mhs}{\mt}{\mt}
\nl &+&
         \frac{3}{16}
         \,(  - \mbs + \mts + \mws )
         \,\frac{1}{\mw}
         \,\frac{\mbs}{\lw}
         \,\bfun{ - \mhs}{\mb}{\mb}
\nl &-&
         \frac{3}{32}
         \,( \mbs + \mts - 2\,\frac{\mbq}{\lw} + 4\,\frac{\mts\,\mbs}{\lw} 
         - 2\,\frac{\mtq}{\lw} + 2\,\frac{\mws\,\mbs}{\lw} + 2\,\frac{\mws\,\mts}{\lw} )
         \,\frac{1}{\mw}
         \,\bfun{ - \mws}{\mt}{\mb}
\nl &-&
         \frac{3}{32}
         \,(  - \mbs + \mts + \mws + 2\,\frac{\mbq}{\lw} - 4\,\frac{\mts\,\mbs}{\lw} 
         + 2\,\frac{\mtq}{\lw} - 4\,\frac{\mws\,\mbs}{\lw} 
         + 4\,\frac{\mws\,\mts}{\lw} + 2\,\frac{\mwq}{\lw} )
         \,\frac{\mbs}{\mw}
\nl &\times& \cfun{ - \mhs}{ - \mws}{ - \mws}{\mb}{\mt}{\mb}
\nl &-&
         \frac{3}{32}
         \,( \mbs - \mts + \mws + 2\,\frac{\mbq}{\lw} - 4\,\frac{\mts\,\mbs}{\lw} 
         + 2\,\frac{\mtq}{\lw} + 4\,\frac{\mws\,\mbs}{\lw} 
         - 4\,\frac{\mws\,\mts}{\lw} + 2\,\frac{\mwq}{\lw} )
         \,\frac{\mts}{\mw}
\nl &\times& \cfun{ - \mhs}{ - \mws}{ - \mws}{\mt}{\mb}{\mt}
\nl &+& \Delta\mcD^{\PQq}_{\sPHWW\,;\,\myNLO} 
\eqa
\bqa
\mcD^{\PW}_{\sPHWW\,;\,\myNLO} &=&
       -
         \frac{1}{4}
         \,\afun{\mw}
         \,\mw
\nl &-&
         \frac{1}{64}
         \,\Bigl[ 3\,c^2\,\mhs - 3\,( 1 + 2\,c^2 )\,\LR\,\mws + ( 1 + 34\,c^2 )\,\mws \Bigr]
         \,\frac{1}{\mw}
         \,\frac{1}{c^2}
\nl &-&
         \frac{1}{64}
         \,\Bigl[ c^2 - 2\,( 1 + 6\,c^2 - 16\,c^4 )\,\frac{\mws}{\lw} 
         - (  - c^2 + s^2 )\,\frac{\mhs}{\lw}\,c^2 \Bigr]
         \,\frac{1}{c^4}
         \,\mw
         \,\bfun{ - \mhs}{\mz}{\mz}
\nl &-&
         \frac{1}{64}
         \,\Bigl[ 2\,c^2\,\mws + \frac{\mhq}{\lw}\,c^2 
         + ( 1 + 2\,c^2 )\,\frac{\mws\,\mhs}{\lw} + 2\,( 1 + 8\,c^2 )\,\frac{\mwq}{\lw} \Bigr]
         \,\frac{1}{\mw}
         \,\frac{1}{c^2}
         \,\bfun{ - \mhs}{\mw}{\mw}
\nl &+&
         \frac{1}{32}
         \,\Bigl[ \frac{\mhs}{\lw}\,c^4 + 2\,( 1 - 3\,c^2 )\,c^2 
         - ( 1 + 5\,c^2 - 24\,c^4 )\,\frac{\mws}{\lw} \Bigr]
         \,\frac{1}{c^4}
         \,\mw
         \,\bfun{ - \mws}{\mw}{\mz}
\nl &+&
         \frac{3}{64}
         \,( \mhs - 2\,\mws )
         \,\frac{1}{\mw}
         \,\frac{\mhs}{\lw}
         \,\bfun{ - \mhs}{\mh}{\mh}
\nl &-&
         \frac{1}{32}
         \,(  - 2\,\mws + \frac{\mhq}{\lw} - 4\,\frac{\mws\,\mhs}{\lw} )
         \,\frac{1}{\mw}
         \,\bfun{ - \mws}{\mw}{\mh}
\nl &-&
         \frac{1}{64}
         \,\Bigl[ 9\,c^6\,\mhs - 2\,( 1 + 4\,c^2 - 28\,c^4 + 32\,c^6 )\,\frac{\mwq}{\lw} 
         + 2\,( 2 - 7\,c^2 - 16\,c^4 )\,c^2\,\mws 
\nl &-& ( c^2 - s^2 )^2\,\frac{\mws\,\mhs}{\lw}\,c^2 \Bigr]
         \,\frac{1}{c^6}
         \,\mw
         \,\cfun{ - \mhs}{ - \mws}{ - \mws}{\mz}{\mw}{\mz}
\nl &+&
         \frac{1}{64}
         \,\Bigl[ \frac{\mws\,\mhs}{\lw} 
         - 2\,( 1 - 22\,c^2 + 8\,s^2\,c^2 )\,c^2\,\mws 
         + 2\,( 1 + 8\,c^2 )\,\frac{\mwq}{\lw} 
\nl &-& ( 1 + 8\,c^4 )\,c^2\,\mhs \Bigr]
         \,\frac{1}{c^4}
         \,\mw
         \,\cfun{ - \mhs}{ - \mws}{ - \mws}{\mw}{\mz}{\mw}
\nl &+&
         \frac{1}{64}
         \,( \mhq + 2\,\mws\,\mhs - 4\,\mwq + \frac{\mhvi}{\lw} + 2\,\frac{\mws\,\mhq}{\lw} )
         \,\frac{1}{\mw}
         \,\cfun{ - \mhs}{ - \mws}{ - \mws}{\mw}{\mh}{\mw}
\nl &+&
         \frac{3}{64}
         \,(  - \mws + \frac{\mhq}{\lw} - 4\,\frac{\mws\,\mhs}{\lw} + 4\,\frac{\mwq}{\lw} )
         \,\frac{\mhs}{\mw}
         \,\cfun{ - \mhs}{ - \mws}{ - \mws}{\mh}{\mw}{\mh}
\nl &-&
         \frac{1}{8}
         \,( \mhs - 2\,\mws )
         \,s^2\,\mw
         \,\cfunf{ - \mhs}{ - \mws}{ - \mws}{\mw}{0}{\mw}
\nl &+& \Delta\mcD^{\PW}_{\sPHWW\,;\,\myNLO} 
\eqa
\bqa
\mcP^{\PQq}_{\sPHWW\,;\,\myNLO} &=&
         \frac{3}{32}
         \,( \mbs + \mts + \frac{\mhs\,\mbs}{\lw} + \frac{\mhs\,\mts}{\lw} )
         \,\frac{1}{\mw\,\mhs}
\nl &+&
         \frac{3}{32}
         \,( \mbs - \mts - \frac{\mhs\,\mbs}{\lw} + \frac{\mhs\,\mts}{\lw} )
         \,\frac{1}{\mws}
         \,\frac{\mbs}{\mw\,\mhs}
         \,\afun{\mb}
\nl &-&
         \frac{3}{32}
         \,( \mbs - \mts - \frac{\mhs\,\mbs}{\lw} + \frac{\mhs\,\mts}{\lw} )
         \,\frac{1}{\mws}
         \,\frac{\mts}{\mw\,\mhs}
         \,\afun{\mt}
\nl &+&
         \frac{3}{64}
         \,(  - \mbs + \mts + \mws + \frac{\mhs\,\mbs}{\lw} - \frac{\mhs\,\mts}{\lw} 
         + 3\,\frac{\mws\,\mhs}{\lw} + 12\,\frac{\mws\,\mhs\,\mbs}{\lws} 
\nl &-& 12\,\frac{\mws\,\mhs\,\mts}{\lws} + 12\,\frac{\mwq\,\mhs}{\lws} )
         \,\frac{1}{\mws}
         \,\frac{\mts}{\mw\,\mhs}
         \,\bfun{ - \mhs}{\mt}{\mt}
\nl &+&
         \frac{3}{64}
         \,( \mbs - \mts + \mws - \frac{\mhs\,\mbs}{\lw} 
         + \frac{\mhs\,\mts}{\lw} + 3\,\frac{\mws\,\mhs}{\lw} 
         - 12\,\frac{\mws\,\mhs\,\mbs}{\lws} + 12\,\frac{\mws\,\mhs\,\mts}{\lws} 
\nl &+& 12\,\frac{\mwq\,\mhs}{\lws} )
         \,\frac{1}{\mws}
         \,\frac{\mbs}{\mw\,\mhs}
         \,\bfun{ - \mhs}{\mb}{\mb}
\nl &-&
         \frac{3}{64}
         \,(  - \mbq + 2\,\mts\,\mbs - \mtq + \mws\,\mbs + \mws\,\mts + \frac{\mhs\,\mbq}{\lw} 
         - 2\,\frac{\mhs\,\mts\,\mbs}{\lw} + \frac{\mhs\,\mtq}{\lw} 
\nl &+& 3\,\frac{\mws\,\mhs\,\mbs}{\lw} - 12\,\frac{\mws\,\mhs\,\mbq}{\lws} 
         + 3\,\frac{\mws\,\mhs\,\mts}{\lw} + 24\,\frac{\mws\,\mhs\,\mts\,\mbs}{\lws} 
\nl &-& 12\,\frac{\mws\,\mhs\,\mtq}{\lws} + 12\,\frac{\mwq\,\mhs\,\mbs}{\lws} 
         + 12\,\frac{\mwq\,\mhs\,\mts}{\lws} )
         \,\frac{1}{\mwc\,\mhs}
         \,\bfun{ - \mws}{\mt}{\mb}
\nl &+&
         \frac{3}{64}
         \,( \mbq - 2\,\mts\,\mbs + \mtq - 2\,\mws\,\mbs - 2\,\mws\,\mts
         + \mwq - \frac{\mhs\,\mbq}{\lw} + 2\,\frac{\mhs\,\mts\,\mbs}{\lw}
         - \frac{\mhs\,\mtq}{\lw} 
\nl &-& 10\,\frac{\mws\,\mhs\,\mbs}{\lw} 
         - 12\,\frac{\mws\,\mhs\,\mbq}{\lws} + 6\,\frac{\mws\,\mhs\,\mts}{\lw} 
         + 24\,\frac{\mws\,\mhs\,\mts\,\mbs}{\lws} - 12\,\frac{\mws\,\mhs\,\mtq}{\lws} 
\nl &-& 5\,\frac{\mwq\,\mhs}{\lw} - 24\,\frac{\mwq\,\mhs\,\mbs}{\lws} 
         + 24\,\frac{\mwq\,\mhs\,\mts}{\lws} 
\nl &-& 12\,\frac{\mwvi\,\mhs}{\lws} )
         \,\frac{1}{\mws}
         \,\frac{\mts}{\mw\,\mhs}
         \,\cfun{ - \mhs}{ - \mws}{ - \mws}{\mt}{\mb}{\mt}
\nl &+&
         \frac{3}{64}
         \,( \mbq - 2\,\mts\,\mbs + \mtq - 2\,\mws\,\mbs - 2\,\mws\,\mts
         + \mwq - \frac{\mhs\,\mbq}{\lw} + 2\,\frac{\mhs\,\mts\,\mbs}{\lw}
         - \frac{\mhs\,\mtq}{\lw} 
\nl &+& 6\,\frac{\mws\,\mhs\,\mbs}{\lw} 
         - 12\,\frac{\mws\,\mhs\,\mbq}{\lws} - 10\,\frac{\mws\,\mhs\,\mts}{\lw}
         + 24\,\frac{\mws\,\mhs\,\mts\,\mbs}{\lws} - 12\,\frac{\mws\,\mhs\,\mtq}{\lws} 
\nl &-& 5\,\frac{\mwq\,\mhs}{\lw} + 24\,\frac{\mwq\,\mhs\,\mbs}{\lws} 
         - 24\,\frac{\mwq\,\mhs\,\mts}{\lws} - 12\,\frac{\mwvi\,\mhs}{\lws} )
         \,\frac{1}{\mws}
         \,\frac{\mbs}{\mw\,\mhs}
\nl &\times& \cfun{ - \mhs}{ - \mws}{ - \mws}{\mb}{\mt}{\mb}
\eqa
\bqa
\Delta\mcD^{\PQq}_{\sPHWW\,;\,\myNLO} &=&
 \frac{1}{32}\,\mw\,\lpar
  \mrW^{(4)}_{\PH\,;\,\PQt,\PQb} +
  2\,\mrW^{(4)}_{\PW\,;\,\PQt,\PQb} -
  \ssdCZ^{(4)}_{\mw\,;\,\PQt,\PQb} -
  2\,\ssdCZ^{(4)}_{g\,;\,\PQt,\PQb} \rpar
\nl
\Delta\mcD^{\PW}_{\sPHWW\,;\,\myNLO} &=&
 \frac{1}{32}\,\mw\,\lpar
  \mrW^{(4)}_{\PH\,;\,\PW} +
  2\,\mrW^{(4)}_{\PW\,;\,\PW} -
  \ssdCZ^{(4)}_{\mw\,;\,\PW} -
  2\,\ssdCZ^{(4)}_{g\,;\,\PW} \rpar
\eqa
\vspace{0.5cm}
\bqa
\mcP^{\PW}_{\sPHWW\,;\,\myNLO} &=&
       -
         \frac{1}{64}
         \,\Bigl[ 3\,c^2\,\mhs + 3\,\frac{\mhq}{\lw}\,c^2 + ( 1 + 18\,c^2 )\,\mws 
         + ( 1 + 18\,c^2 )\,\frac{\mws\,\mhs}{\lw} \Bigr]
         \,\frac{1}{c^2}
         \,\frac{1}{\mw\,\mhs}
\nl &+&
         \frac{1}{64}
         \,\Bigl[ c^2\,\mhs - \frac{\mhq}{\lw}\,c^2 + ( 1 - 2\,c^2 )\,\mws
       - ( 1 - 2\,c^2 )\,\frac{\mws\,\mhs}{\lw} \Bigr]
         \,\frac{1}{c^2}
         \,\frac{1}{\mw\,\mhs}
         \,\afun{\mw}
\nl &-&
         \frac{1}{64}
         \,\Bigl[ ( 9 - 8\,s^2 ) - ( 9 - 8\,s^2 )\,\frac{\mhs}{\lw} \Bigr]
         \,\frac{\mw}{\mhs}
         \,\frac{s^2}{c^4}
         \,\afun{\mz}
\nl &-&
         \frac{1}{64}
         \,( \mhs - \mws - \frac{\mhq}{\lw} + \frac{\mws\,\mhs}{\lw} )
         \,\frac{1}{\mwc}
         \,\afun{\mh}
\nl &+&
         \frac{1}{2}
         \,\frac{\mw}{\lw}
         \,s^2
         \,\bfun{ - \mws}{0}{\mw}
\nl &-&
         \frac{1}{256}
         \,\Bigl[ c^2\,\mhs - \frac{\mhq}{\lw}\,c^2 - 2\,( 1 - 8\,c^2 )\,\frac{\mws\,\mhs}{\lw} 
         - 24\,( 1 + 6\,c^2 - 16\,c^4 )\,\frac{\mwq\,\mhs}{\lws} 
         + 2\,( 1 + 8\,c^2 )\,\mws 
\nl &-& 12\,(  - c^2 + s^2 )\,\frac{\mws\,\mhq}{\lws}\,c^2 \Bigr]
         \,\frac{1}{c^4}
         \,\frac{1}{\mw\,\mhs}
         \,\bfun{ - \mhs}{\mz}{\mz}
\nl &+&
         \frac{1}{256}
         \,\Bigl[ c^2\,\mhq - \frac{\mhvi}{\lw}\,c^2 - 12\,\frac{\mws\,\mhvi}{\lws}\,c^2 
         + 2\,( 1 - 12\,c^2 )\,\mwq + ( 1 - 2\,c^2 )\,\mws\,\mhs 
\nl &-& ( 1 - 2\,c^2 )\,\frac{\mws\,\mhq}{\lw} 
         - 12\,( 1 + 2\,c^2 )\,\frac{\mwq\,\mhq}{\lws} 
         - 24\,( 1 + 8\,c^2 )\,\frac{\mwvi\,\mhs}{\lws} 
\nl &-& 2\,( 1 + 20\,c^2 )\,\frac{\mwq\,\mhs}{\lw} \Bigr]
         \,\frac{1}{c^2}
         \,\frac{1}{\mwc\,\mhs}
         \,\bfun{ - \mhs}{\mw}{\mw}
\nl &+&
         \frac{1}{128}
         \,\Bigl[ c^4\,\mhs - \frac{\mhq}{\lw}\,c^4 
         + 12\,\frac{\mws\,\mhq}{\lws}\,c^4 - ( 1 + 5\,c^2 - 24\,c^4 )\,\mws 
\nl &-& 12\,( 1 + 5\,c^2 - 24\,c^4 )\,\frac{\mwq\,\mhs}{\lws} 
         + ( 1 + 21\,c^2 - 72\,c^4 + 64\,c^6 )\,\frac{\mws\,\mhs}{\lw} \Bigr]
         \,\frac{1}{c^4}
         \,\frac{1}{\mw\,\mhs}
\nl &\times& \bfun{ - \mws}{\mw}{\mz}
\nl &-&
         \frac{3}{256}
         \,( 1 - \frac{\mhs}{\lw} - 12\,\frac{\mws\,\mhs}{\lws} + 24\,\frac{\mwq}{\lws} )
         \,\frac{\mhs}{\mwc}
         \,\bfun{ - \mhs}{\mh}{\mh}
\nl &-&
         \frac{1}{128}
         \,( \mhs - 4\,\mws - \frac{\mhq}{\lw} + 4\,\frac{\mws\,\mhs}{\lw} 
         + 12\,\frac{\mws\,\mhq}{\lws} - 16\,\frac{\mwq}{\lw} - 48\,\frac{\mwq\,\mhs}{\lws} )
         \,\frac{1}{\mwc}
\nl &\times& \bfun{ - \mws}{\mw}{\mh}
\nl &+&
         \frac{1}{256}
         \,\Bigl[ 12\,\frac{\mws\,\mhq}{\lws} 
         + 2\,( 1 - 36\,c^2 + 96\,c^4 )\,\frac{\mws\,\mhs}{\lw} 
         + ( 1 + 4\,c^2 )\,\frac{\mhq}{\lw} 
\nl &-& 2\,( 1 + 4\,c^2 - 32\,c^4 )\,\mws + 24\,( 1 + 8\,c^2 )\,\frac{\mwq\,\mhs}{\lws} 
\nl &-& (  - 2\,c^2 + 63\,c^4 + s^4 )\,\mhs \Bigr]
         \,\frac{1}{c^4}
         \,\frac{\mw}{\mhs}
         \,\cfun{ - \mhs}{ - \mws}{ - \mws}{\mw}{\mz}{\mw}
\nl &+&
         \frac{1}{256}
         \,\Bigl[ 2\,( 1 - 12\,c^2 + 8\,c^4 + 64\,c^6 )\,\frac{\mws\,\mhs}{\lw} 
         - 2\,( 1 + 4\,c^2 - 32\,c^4 )\,\mws 
\nl &+& 24\,( 1 + 4\,c^2 
         - 28\,c^4 + 32\,c^6 )\,\frac{\mwq\,\mhs}{\lws} 
         + (  - 2\,c^2 + 7\,c^4 + s^4 )\,\frac{\mhq}{\lw}\,c^2 
         - (  - 2\,c^2 + 63\,c^4 + s^4 )\,c^2\,\mhs 
\nl &+& 12\,( c^4 - 2\,s^2\,c^2 + s^4 )\,\frac{\mws\,\mhq}{\lws}\,c^2 \Bigr]
         \,\frac{1}{c^6}
         \,\frac{\mw}{\mhs}
         \,\cfun{ - \mhs}{ - \mws}{ - \mws}{\mz}{\mw}{\mz}
\nl &-&
         \frac{3}{256}
         \,( \mhs - 4\,\mws - \frac{\mhq}{\lw} + 4\,\frac{\mws\,\mhs}{\lw} 
         - 12\,\frac{\mws\,\mhq}{\lws} - 8\,\frac{\mwq}{\lw} 
         + 48\,\frac{\mwq\,\mhs}{\lws} 
\nl &-& 48\,\frac{\mwvi}{\lws} )
         \,\frac{\mhs}{\mwc}
         \,\cfun{ - \mhs}{ - \mws}{ - \mws}{\mh}{\mw}{\mh}
\nl &-&
         \frac{1}{256}
         \,( \mhq - 2\,\mws\,\mhs - 8\,\mwq - \frac{\mhvi}{\lw} - 6\,\frac{\mws\,\mhq}{\lw} 
         - 12\,\frac{\mws\,\mhvi}{\lws} - 40\,\frac{\mwq\,\mhs}{\lw} 
\nl &-& 24\,\frac{\mwq\,\mhq}{\lws} )
         \,\frac{1}{\mwc}
         \,\cfun{ - \mhs}{ - \mws}{ - \mws}{\mw}{\mh}{\mw}
\eqa

\normalsize

\section{Appendix: Corrections for the $\PW$ mass \label{DMW}}

In this Appendix we present the full list of corrections for $\mw$ in the $\alpha\,$-scheme,
as given in \eqn{premw}. In this Appendix we use $s= \sth$ and $c= \cth$ where $\cths$ is defined 
in \eqn{defscipsb}. Furthermore, in this Appendix, ratios of masses are defined according to
\bq
\xph = \frac{\mh}{\mz}, \quad x_{\Pf} = \frac{M_{\Pf}}{\mz}
\eq
\etc We introduce the following polynomials:

\scriptsize
\bqas
\mrQ^{a}_{0}= 23 - 4\,c^2 &\qquad&
\mrQ^{a}_{1}= 77 - 12\,c^2
\eqas


\vspace{0.4cm}
\[
\begin{array}{lll}
\mrQ^{b}_{0}= 21 - 4\,c^2 \;&
\mrQ^{b}_{1}= 65 - 12\,c^2 \;&
\mrQ^{b}_{2}= 109 - 6\,\mrQ^{a}_{0}\,c^2 \\
\mrQ^{b}_{3}= 5 - 2\,c^2 \;&
\mrQ^{b}_{4}= 62 - \mrQ^{a}_{1}\,c^2 \;&  \\
\end{array}
\]


\vspace{0.4cm}
\[
\begin{array}{lll}
\mrQ^{c}_{0}= 2 - c^2 \;&
\mrQ^{c}_{1}= 12\,c^3 + 29\,s \;&
\mrQ^{c}_{2}= 19 - \mrQ^{b}_{0}\,c^2 \\
\mrQ^{c}_{3}= 75 - 16\,c^2 \;&
\mrQ^{c}_{4}= 22 - \mrQ^{b}_{0}\,c^2 \;&
\mrQ^{c}_{5}= 52 - \mrQ^{b}_{1}\,c^2 \\
\mrQ^{c}_{6}= 5 - 2\,c^2 \;&
\mrQ^{c}_{7}= 32 - \mrQ^{b}_{2}\,c^2 \;&
\mrQ^{c}_{8}= 7 - \mrQ^{b}_{3}\,c^2 \\
\mrQ^{c}_{9}= 1 + 2\,\mrQ^{b}_{4}\,c^2 \;& \\
\end{array}
\]


\vspace{0.4cm}
\[
\begin{array}{lll}
\mrQ^{d}_{0}= 1 - 2\,c^2 \;&
\mrQ^{d}_{1}= 119 - 128\,\mrQ^{c}_{0}\,c^2 \;&
\mrQ^{d}_{2}= 29 - \mrQ^{c}_{1}\,s \\

\mrQ^{d}_{3}= 43 - 6\,\mrQ^{c}_{2}\,c^2 \;&
\mrQ^{d}_{4}= 15 - \mrQ^{c}_{3}\,c^2 \;&
\mrQ^{d}_{5}= 61 - 12\,\mrQ^{c}_{4}\,c^2 \\

\mrQ^{d}_{6}= 2 + 3\,\mrQ^{c}_{5}\,c^2 \;&
\mrQ^{d}_{7}= 1 - s^2 \;&
\mrQ^{d}_{8}= 1 - s^2\,c^2 \\

\mrQ^{d}_{9}= 5 - 4\,c^2 \;&
\mrQ^{d}_{10}= 7 - 4\,c^2 \;&
\mrQ^{d}_{11}= 9 - 8\,c^2 \\

\mrQ^{d}_{12}= 13 - 8\,c^2 \;&
\mrQ^{d}_{13}= 19 + 4\,\mrQ^{c}_{6}\,c^2 \;&
\mrQ^{d}_{14}= 1 - 2\,\mrQ^{c}_{7}\,c^2 \\

\mrQ^{d}_{15}= 7 - \mrQ^{c}_{8}\,c^2 \;&
\mrQ^{d}_{16}= 7 - \mrQ^{c}_{9}\,c^2 \;& \\
\end{array}
\]


\vspace{0.4cm}
\[
\begin{array}{lll}
\mrQ^{e}_{0}= 16\,\mrQ^{d}_{0}\,s - \mrQ^{d}_{1}\,c \;&
\mrQ^{e}_{1}= 1 + 16\,c^2 - 4\,\mrQ^{d}_{2}\,s\,c \;&
\mrQ^{e}_{2}= 1 + 62\,\mrQ^{d}_{0}\,s\,c - 2\,\mrQ^{d}_{3}\,c^2 \\

\mrQ^{e}_{3}= 1 + \mrQ^{d}_{4}\,c^2 \;&
\mrQ^{e}_{4}= 9 - 8\,s^2 \;&
\mrQ^{e}_{5}= 20\,\mrQ^{d}_{0}\,s + \mrQ^{d}_{5}\,c \\

\mrQ^{e}_{6}= 59\,\mrQ^{d}_{0}\,s + \mrQ^{d}_{6}\,c \;&
\mrQ^{e}_{7}= 5 - 8\,c^2 \;&
\mrQ^{e}_{8}= 9 - 128\,\mrQ^{d}_{7}\,s^2 \\

\mrQ^{e}_{9}= 3\,\mrQ^{d}_{0}\,c + 32\,\mrQ^{d}_{8}\,s \;&
\mrQ^{e}_{10}= 4\,\mrQ^{d}_{0}\,s - \mrQ^{d}_{9}\,c \;&
\mrQ^{e}_{11}= 4\,\mrQ^{d}_{0}\,s - \mrQ^{d}_{10}\,c \\

\mrQ^{e}_{12}= 4\,\mrQ^{d}_{0}\,s - 3\,\mrQ^{d}_{11}\,c \;&
\mrQ^{e}_{13}= 8\,\mrQ^{d}_{0}\,s - \mrQ^{d}_{12}\,c \;&
\mrQ^{e}_{14}= 1 - 2\,c^2 \\

\mrQ^{e}_{15}= 1 - 2\,\mrQ^{d}_{13}\,c^2 \;&
\mrQ^{e}_{16}= 12 - \mrQ^{d}_{14}\,s\,c - 6\,\mrQ^{d}_{15}\,c^2 \;&
\mrQ^{e}_{17}= 12 - 6\,\mrQ^{d}_{15}\,c^2 + \mrQ^{d}_{16}\,s\,c \\
\end{array}
\]

\normalsize

With their help we derive the correction factors:

\footnotesize

\bqa
\Delta^{(4)}_{\Pl}\,\mw &=&
       - \frac{1}{24}\,\bfun{-\mzs}{0}{0}\,\frac{1}{s^2}\,\frac{c^2}{s^2-c^2}
       + \frac{1}{24}\,\bfunp{0}{0}{\mle}\,\frac{1}{s^2-c^2}\,\xplq
\nl
{}&+& \frac{1}{24}\,\afun{\mle}\,\frac{1}{s^2\,c^2}\,\xplq
       + \frac{1}{24}\,\Bigl[ \xpls - \vles\,c^2\,\xpls - 8\,s^4\,c^2 \Bigr]\,
            \afun{\mle}\,\frac{1}{s^2}\,\frac{1}{s^2-c^2}
\nl
{}&+& \frac{1}{24}\,\Bigl[ \xplq + c^2\,\xpls - 2\,s\,c^3 \Bigr]\,
                \bfun{-\mws}{0}{\mle}\,\frac{1}{s^2\,c^2}
\nl
{}&+& \frac{1}{144}\,\Bigl[ 6\,c\,\xpls + \mrQ^{e}_{0} + (1 - 6\,\xpls)\,\vles\,c + 3\,(
      \vqus + \vqds)\,c \Bigr]\,\frac{1}{s^2}\,\frac{c}{s^2-c^2}
\nl
{}&-& \frac{1}{48}\,\Bigl[ (1 - 4\,\xpls) + (1 + 2\,\xpls)\,\vles \Bigr]\,
            \bfun{-\mzs}{\mle}{\mle}\,\frac{1}{s^2}\,\frac{c^2}{s^2-c^2}
\eqa


\bqa
\Delta^{(4)}_{\PQq}\,\mw &=&
       - \frac{1}{8}\,\Bigl[ \xpds - \xpus \Bigr]\,\afun{\mqu}\,\frac{1}{s^2\,c^2}\,\xpus
       + \frac{1}{8}\,( \xpus - \xpds)^2\,\bfunp{0}{\mqu}{\mqd}\,\frac{1}{s^2-c^2}
\nl
{}&+& \frac{1}{72}\,\Bigl[ 9\,\xpus - 9\,\vqus\,c^2\,\xpus + 18\,s^2\,\xpds - 32\,s^4\,c^2
       + 18\,\frac{\xpdq}{\xpus-\xpds}\,s^2 \Bigr]\,\afun{\mqu}\,\frac{1}{s^2}\,\frac{1}{s^2-c^2}
\nl
{}&-& \frac{1}{8}\,\Bigl[ c^2 - (\xpds - \xpus) \Bigr]\,\afun{\mqd}\,\frac{1}{s^2\,c^2}\,\xpds
       - \frac{1}{8}\,\Bigl[ \vqus\,\xpus + \vqds\,\xpds - (\xpds + \xpus) \Bigr]\,
           \frac{1}{s^2}\,\frac{c^2}{s^2-c^2}
\nl
{}&-& \frac{1}{72}\,\Bigl[ 9\,\vqds\,c^2\,\xpds + 8\,s^4\,c^2 + 18\,\frac{\xpdq}{\xpus-\xpds}\,s^2 \Bigr]\,
            \afun{\mqd}\,\frac{1}{s^2}\,\frac{1}{s^2-c^2}
\nl
{}&-& \frac{1}{8}\,\Bigl[ 2\,s\,c^3 - (\xpds - \xpus)^2 - (\xpds + \xpus)\,c^2 \Bigr]\,
         \bfun{\mws}{\mqu}{\mqd}\,\frac{1}{s^2\,c^2}
\nl
{}&-& \frac{1}{16}\,\Bigl[ (1 - 4\,\xpds) + (1 + 2\,\xpds)\,\vqds \Bigr]\,
         \bfun{\mzs}{\mqd}{\mqd}\,\frac{1}{s^2}\,\frac{c^2}{s^2-c^2}
\nl
{}&-& \frac{1}{16}\,\Bigl[ (1 - 4\,\xpus) + (1 + 2\,\xpus)\,\vqus \Bigr]\,
         \bfun{-\mzs}{\mqu}{\mqu}\,\frac{1}{s^2}\,\frac{c^2}{s^2-c^2}
\eqa


\bqa
\Delta^{(4)}_{\PB}\,\mw &=&
        \bfun{-\mws}{0}{\mw}\,c^2
       - \frac{1}{6}\,\bfunp{0}{0}{\mw}\,\frac{s^2\,c^4}{s^2-c^2}
\nl
{}&+& \frac{1}{48}\,\afun{\mw}\,\frac{1}{s^2}\,\xphs
       - \frac{1}{48}\,\mrQ^{e}_{1}\,\bfun{-\mws}{\mw}{\mz}\,\frac{1}{s^2\,c^2}
\nl
{}&-& \frac{1}{48}\,\mrQ^{e}_{4}\,\bfunp{0}{\mw}{\mz}\,\frac{1}{s^2-c^2}\,s^4
       - \frac{1}{48}\,\mrQ^{e}_{5}\,\bfun{-\mzs}{\mw}{\mw}\,\frac{1}{s^2}\,\frac{c}{s^2-c^2}
\nl
{}&-& \frac{1}{48}\,\Bigl[ 12 - 4\,\xphs + \xphq \Bigr]\,
           \bfun{-\mzs}{\mh}{\mz}\,\frac{1}{s^2}\,\frac{c^2}{s^2-c^2}
\nl
{}&-& \frac{1}{48}\,\Bigl[ \xphq - 4\,c^2\,\xphs + 12\,s\,c^3 \Bigr]\,
           \bfun{-\mws}{\mw}{\mh}\,\frac{1}{s^2\,c^2}
\nl
{}&-& \frac{1}{48}\,\Bigl[ \xphq - 2\,c^2\,\xphs + c^4 \Bigr]\,
           \bfunp{0}{\mw}{\mh}\,\frac{1}{s^2-c^2}
\nl
{}&+& \frac{1}{48}\,\Bigl[ 3\,c^2\,\xphs - 8\,c^4 - s^2\,\xphq - 8\,\frac{c^6}{\xphs - c^2} \Bigr]\,
           \afun{\mh}\,\frac{1}{c^2}\,\frac{1}{s^2-c^2}
\nl
{}&+& \frac{1}{48}\,\Bigl[ c^4\,\xphs - \mrQ^{e}_{3} \Bigr]\,
        \afun{\mz}\,\frac{1}{s^2\,c^2}\,\frac{1}{s^2-c^2}
\nl
{}&+& \frac{1}{48}\,\Bigl[ 8\,\frac{s^2\,c^4}{\xphs - c^2} + \mrQ^{e}_{2} \Bigr]\,
           \afun{\mw}\,\frac{1}{s^2}\,\frac{1}{s^2-c^2}
\nl
{}&+& \frac{1}{36}\,\Bigl[ \mrQ^{e}_{6} - (100 - 9\,\Delta_g)\,s^2\,c \Bigr]\,\frac{1}{s^2}\,
           \frac{c}{s^2-c^2}
\eqa


\bqa
\Delta^{(6)}_{\Pl}\,\mw &=&
  - \frac{1}{48}\,\Bigl[ 8\,\aplo - \apD \Bigr]\,\bfun{-\mzs}{0}{0}\,\frac{1}{s^2}\,\frac{c^2}{s^2-c^2}
\nl
{}&-& \frac{1}{144}\,\Bigl[ 9\,\xpls\,\apD - 8\,\mrQ^{e}_{7}\,\apD + 48\,(\aplt + \aplo
       + \apl)\,s^2 + 72\,(\aplo - \apl)\,\xpls \Bigr]\,\frac{1}{s^2}\,\frac{c^2}{s^2-c^2}\,\LR
\nl
{}&+& \frac{1}{48}\,\Bigl[ c\,\xpls\,\apD - 8\,\vle\,c\,\xpls\,\aplV + 8\,\vles\,c\,\xpls\,
      \aplt + 8\,s^2\,c^3\,\apD + 64\,s^4\,c\,\aplt + \mrQ^{e}_{11}\,\vle\,\xpls\,\apD
\nl
{}&+& 8\,(\aplo - \apl)\,c\,\xpls \Bigr]\,\afun{\mle}\,\frac{1}{s^2}\,\frac{c}{s^2-c^2}
\nl
{}&+& \frac{1}{288}\,\Bigl[ 6\,c^2\,\xpls\,\apD - \mrQ^{e}_{8}\,c^2\,\apD - 32\,\mrQ^{e}_{9}
      \,s\,\aplt - \mrQ^{e}_{10}\,\vqd\,c\,\apD - \mrQ^{e}_{13}\,\vqu\,c\,\apD 
\nl
{}&+& 8\,(1 - 
      6\,\xpls)\,\vle\,c^2\,\aplV - 8\,(1 - 6\,\xpls)\,\vles\,c^2\,\aplt 
\nl
{}&-& (1 - 6\,
      \xpls)\,\mrQ^{e}_{11}\,\vle\,c\,\apD + 8\,(122\,\aplt + \aplo + \apl)\,s^2 + 
      48\,(\aplo - \apl)\,c^2\,\xpls 
\nl
{}&-& 24\,(\vqus + \vqds)\,c^2\,\aplt \Bigr]\,
      \frac{1}{s^2}\,\frac{1}{s^2-c^2}
\nl
{}&+& \frac{1}{6}\,(\aplt - \aplo - \apl)\,\frac{1}{s^2}
       + \frac{1}{288}\,\Bigl[ \mrQ^{e}_{10}\,\vqd + \mrQ^{e}_{13}\,\vqu \Bigr]\,\frac{1}{s^2}\,
      \frac{c}{s^2-c^2}\,\myNG\,\apD
\nl
{}&+& \frac{1}{96}\,\Bigl[ \mrQ^{e}_{11}\,\vle\,\apD + 2\,\mrQ^{e}_{12}\,\vle\,\xpls\,\apD + (1
       - 4\,\xpls)\,c\,\apD - 8\,(1 + 2\,\xpls)\,\vle\,c\,\aplV 
\nl
{}&+& 8\,(1 + 2\,\xpls)\,
      \vles\,c\,\aplt + 8\,(\aplo - \apl)\,(1 - 4\,\xpls)\,c \Bigr]\,
         \bfun{-\mzs}{\mle}{\mle}\,\frac{1}{s^2}\,\frac{c}{s^2-c^2}
\eqa


\bqa
\Delta^{(6)}_{\PQq}\,\mw &=&
        \frac{1}{3}\,\Bigl[ c\,\apqt - 3\,s\,\xpds \Bigr]\,\frac{1}{s}
\nl
{}&+& \frac{1}{144}\,\Bigl[ 9\,c\,\xpds\,\apD - 72\,\vqd\,c\,\xpds\,\apdV + 72\,\vqds\,c\,\xpds
      \,\apqt + 8\,s^2\,c^3\,\apD + 64\,s^4\,c\,\apqt + 3\,\mrQ^{e}_{10}\,\vqd\,\xpds\,
      \apD 
\nl
{}&+& 72\,(\apqo - \apd)\,c\,\xpds \Bigr]\,\afun{\mqd}\,\frac{1}{s^2}\,
      \frac{c}{s^2-c^2}
\nl
{}&+& \frac{1}{144}\,\Bigl[ 9\,c\,\xpus\,\apD - 72\,\vqu\,c\,\xpus\,\apuV + 72\,\vqus\,c\,\xpus
      \,\apqt + 32\,s^2\,c^3\,\apD + 256\,s^4\,c\,\apqt + 3\,\mrQ^{e}_{13}\,\vqu\,\xpus
      \,\apD 
\nl
{}&-& 72\,(\apqo - \apu)\,c\,\xpus \Bigr]\,\afun{\mqu}\,\frac{1}{s^2}\,
      \frac{c}{s^2-c^2}
\nl
{}&+& \frac{1}{48}\,\Bigl[ 24\,\vqus\,c\,\xpus\,\apqt + 24\,\vqds\,c\,\xpds\,\apqt + 
            \mrQ^{e}_{10}\,\vqd\,\xpds\,\apD + \mrQ^{e}_{13}\,\vqu\,\xpus\,\apD 
\nl
{}&+& 4\,(1 - 6\,\xpds)\,\vqd\,c\,\apdV + 4\,(1 - 6\,\xpus)\,\vqu\,c\,\apuV 
\nl
{}&+& 4\,(\apdA + \apuA)\,c + 24
      \,(\apqo - \apd)\,c\,\xpds - 24\,(\apqo - \apu)\,c\,\xpus + 3\,(\xpds + 
      \xpus)\,c\,\apD \Bigr]\,\frac{1}{s^2}\,\frac{c}{s^2-c^2}
\nl
{}&-& \frac{1}{288}\,\Bigl[ \mrQ^{e}_{10}\,\vqd + \mrQ^{e}_{13}\,\vqu \Bigr]\,\frac{1}{s^2}\,
      \frac{c}{s^2-c^2}\,\myNG\,\apD
\nl
{}&+& \frac{1}{96}\,\Bigl[ 3\,(1 - 4\,\xpds)\,c\,\apD - 24\,(1 + 2\,\xpds)\,\vqd\,c\,\apdV
       + 24\,(1 + 2\,\xpds)\,\vqds\,c\,\apqt + (1 + 2\,\xpds)\,\mrQ^{e}_{10}\,\vqd\,
      \apD 
\nl
{}&+& 24\,(\apqo - \apd)\,(1 - 4\,\xpds)\,c \Bigr]\,
         \bfun{-\mzs}{\mqd}{\mqd}\,\frac{1}{s^2}\,\frac{c}{s^2-c^2}
\nl
{}&+& \frac{1}{96}\,\Bigl[ 3\,(1 - 4\,\xpus)\,c\,\apD - 24\,(1 + 2\,\xpus)\,\vqu\,c\,\apuV
       + 24\,(1 + 2\,\xpus)\,\vqus\,c\,\apqt + (1 + 2\,\xpus)\,\mrQ^{e}_{13}\,\vqu\,
      \apD 
\nl
{}&-& 24\,(\apqo - \apu)\,(1 - 4\,\xpus)\,c \Bigr]\,
         \bfun{-\mzs}{\mqu}{\mqu}\,\frac{1}{s^2}\,\frac{c}{s^2-c^2}
\nl
{}&-& \frac{1}{48}\,\Bigl[ 16\,(3\,\apqt - \apqo + \apd - 2\,\apu)\,s^2 + 72\,(
      \apqo - \apd)\,\xpds - 72\,(\apqo - \apu)\,\xpus 
\nl
{}&+& 9\,(\xpds + \xpus)\,
      \apD \Bigr]\,\frac{1}{s^2}\,\frac{c^2}{s^2-c^2}\,\LR
\eqa


\bqa
\Delta^{(6)}_{\PB}\,\mw &=&
       - \frac{1}{48}\,\mrQ^{e}_{1}\,\bfun{-\mws}{\mw}{\mz}\,\frac{1}{s^2\,c^2}\,\apD
\nl
{}&-& \frac{1}{48}\,\mrQ^{e}_{4}\,\bfunp{0}{\mw}{\mz}\,\frac{s^4}{s^2-c^2}\,\apD
       - \frac{1}{48}\,\mrQ^{e}_{5}\,\bfun{-\mzs}{\mw}{\mw}\,\frac{1}{s^2}\,\frac{c}{s^2-c^2}\,\apD
\nl
{}&-& \frac{1}{12}\,\mrQ^{e}_{14}\,\bfunp{0}{0}{\mw}\,\frac{c^4}{s^2-c^2}\,\apD
       - \frac{1}{48}\,\Bigl[ \apD 
\nl
{}&+& 2\,\apBox \Bigr]\,\Bigl[ 12 - 4\,\xphs + \xphq \Bigr]\,
          \bfun{-\mzs}{\mh}{\mz}\,\frac{1}{s^2}\,\frac{c^2}{s^2-c^2}
\nl
{}&+& \frac{1}{96}\,\Bigl[ 33\,\apD + 8\,\apBox \Bigr]\,\afun{\mz}\,\frac{1}{s^2}
       - \frac{1}{96}\,\Bigl[ 4\,\xphs\,\apBox + 15\,c^2\,\apD \Bigr]\,\afun{\mh}\,\frac{1}{s^2\,c^2}\,\xphs
\nl
{}&-& \frac{1}{24}\,\Bigl[ \xphq - 4\,c^2\,\xphs + 12\,s\,c^3 \Bigr]\,
          \bfun{-\mws}{\mw}{\mh}\,\frac{1}{s^2\,c^2}\,\apBox
\nl
{}&-& \frac{1}{24}\,\Bigl[ \xphq - 2\,c^2\,\xphs + c^4 \Bigr]\,
          \bfunp{0}{\mw}{\mh}\,\frac{1}{s^2-c^2}\,\apBox
\nl
{}&+& \frac{1}{48}\,\Bigl[ 2\,c\,\xphs\,\apBox - (39\,\apD - 4\,\apBox)\,s^3 + (51\,
      \apD - 4\,\apBox)\,s \Bigr]\,\afun{\mw}\,\frac{1}{s^2\,c}
\nl
{}&+& \frac{1}{72}\,\Bigl[ 3\,s\,c^3\,\xphs\,\apD - 3\,\mrQ^{e}_{17}\,\apD + (19\,\apD + 
      4\,\apBox)\,s^3\,c - (182\,\apD + 6\,\apBox - 9\,\Delta_g\,\apD)\,s^3\,c^3 \Bigr]\,
      \frac{1}{s^3\,c}\,\frac{1}{s^2-c^2}
\nl
{}&-& \frac{1}{96}\,\Bigl[ 32\,s^2\,c^2\,\apBox + 32\,\frac{s^2\,c^4}{\xphs - c^2}
      \,\apBox + 2\,(\apD + 2\,\apBox)\,c^2\,\xphq 
\nl
{}&-& 3\,(5\,\apD + 4\,\apBox
      )\,s^2\,\xphs \Bigr]\,\afun{\mh}\,\frac{1}{s^2}\,\frac{1}{s^2-c^2}
\nl
{}&+& \frac{1}{48}\,\Bigl[ 16\,\frac{s^3\,c^5}{\xphs - c^2}\,\apBox - \mrQ^{e}_{16}
      \,\apD - 2\,(15\,\apD + \apBox)\,s^3\,c^3 \Bigr]\,\afun{\mw}\,\frac{1}{s^3\,c}\,\frac{1}{s^2-c^2}
\nl
{}&-& \frac{1}{96}\,\Bigl[ 2\,\mrQ^{e}_{15}\,\apD - 2\,(\apD + 2\,\apBox)\,c^4\,\xphs + 
      (139\,\apD + 8\,\apBox)\,s^2\,c^2 \Bigr]\,\afun{\mz}\,\frac{1}{s^2\,c^2}\,
      \frac{1}{s^2-c^2}
\nl
{}&-& \frac{1}{96}\,\Bigl[ (25 + 9\,\xphs)\,\apD - 4\,(5\,\apD - 11\,\apBox)\,s^2 \Bigr]\,
         \frac{1}{s^2}\,\frac{c^2}{s^2-c^2}\,\LR
\nl
{}&+& \frac{1}{36}\,\Bigl[ (\apD - 2\,\apBox)\,c - (33\,\apD + 2\,\apBox)\,s^3 + 
      (51\,\apD + 2\,\apBox)\,s \Bigr]\,\frac{1}{s^2\,c}
\eqa


\normalsize

\section{Appendix: $\ssT$ parameter \label{Tpar}}

In this Appendix we present explicit results for the $\ssT$ parameter of \eqn{PTparsix}. For
simplicity we only include PTG operators in loops. We have introduced $s = \sth, c= \cth$,
$c_2 = {\hat c}_{_{2\,\theta}}$ and
\bq
\alpha\,\ssT = \frac{\alpha}{\pi}\,\frac{t}{s^2\,c^2\,c_2}
\eq

\footnotesize
\vspace{0.5cm}
\bei
\item {\underline{$\mrdim = 4$ component}}
\eei


\bqa
t^{(4)} &=&
        \frac{5}{4}\,c^2\,c_2
       - \frac{1}{4}\,\sum_{\gen}\,\afun{\mqu}\,\frac{\xpus\,\xpds}{\xpus-\xpds}\,c_2
\nl
{}&+& \frac{1}{4}\,\sum_{\gen}\,\afun{\mqd}\,\frac{\xpus\,\xpds}{\xpus-\xpds}\,c_2
       - \frac{1}{24}\,\bfunp{0}{0}{\mle}\,c_2\,\sum_{\gen}\,\xplq
\nl
{}&+& \frac{1}{6}\,\bfunp{0}{0}{\mw}\,s^2\,c^4\,c_2
       - \frac{1}{48}\,(1 - \xphs)^2\,\bfunp{0}{\mh}{\mz}\,c_2
\nl
{}&+& \frac{1}{48}\,(5 + 4\,c_2)\,\bfunp{0}{\mw}{\mz}\,s^4\,c_2
       - \frac{1}{8}\,\sum_{\gen}\,(\xpds - \xpus)^2\,\bfunp{0}{\mqu}{\mqd}\,c_2
\nl
{}&+& \frac{1}{24}\,\sum_{\gen}\,(3\,\xpds + 3\,\xpus + \xpls)\,c_2
       + \frac{1}{48}\,(\xphs - c^2)\,(\xphs - c^2)\,\bfunp{0}{\mw}{\mh}\,c_2
\nl
{}&-& \frac{1}{6}\,\Bigr[  - c^2 - \frac{c^4}{\xphs - c^2} 
           + \frac{\xphs}{\xphs-1} \Bigr]\,\afun{\mh}\,c_2
\nl
{}&+& \frac{1}{24}\,\Bigr[ 4\,\frac{s^2}{\xphs-1} - (9 + 17\,c_2 + 4\,c_2^2) \Bigr]\,
           \afun{\mz}\,\frac{c_2}{s^2}
\nl
{}&-& \frac{1}{12}\,\Bigr[ 2\,\frac{s^2\,c^2}{\xphs - c^2} - 5\,(2 + c_2) \Bigr]\,
           \afun{\mw}\,\frac{c_2}{s^2}\,c^2
\eqa


\vspace{0.5cm}
\bei
\item {\underline{$\mrdim = 6$ component}}
\eei


\bqa
t^{(6)} &=&
       - \frac{1}{6}\,\bfunp{0}{0}{\mle}\,c_2\,\sum_{\gen}\,\xplq\,\aplt
       - \frac{1}{12}\,\bfunp{0}{0}{\mw}\,c^6\,c_2\,\apD
\nl
{}&+& \frac{1}{96}\,(5 + 4\,c_2)\,\bfunp{0}{\mw}{\mz}\,s^4\,c_2\,\apD
       - \frac{1}{96}\,(\apD + 4\,\apBox)\,(1 - \xphs)^2\,\bfunp{0}{\mh}{\mz}\,c_2
\nl
{}&-& \frac{1}{2}\,\sum_{\gen}\,(\xpds - \xpus)^2\,\bfunp{0}{\mqu}{\mqd}\,c_2\,\apqt
       - \frac{1}{96}\,(\xphs - c^2)\,(\xphs - c^2)\,(\apD - 4\,\apBox)\,\bfunp{0}{\mw}{\mh}\,c_2
\nl
{}&+& \frac{1}{16}\,\sum_{\gen}\,\Bigr[  - 3\,\apD + 16\,\frac{\xpds}{\xpus-\xpds}\,\apqt + 8\,(2
      \,\apqt - 3\,\apqo + 3\,\apd) \Bigr]\,\afun{\mqd}\,c_2\,\xpds
\nl
{}&-& \frac{1}{16}\,\sum_{\gen}\,\Bigr[ \apD + 8\,(\aplo - \apl) \Bigr]\,\afun{\mle}\,c_2\,\xpls
\nl
{}&-& \frac{1}{16}\,\sum_{\gen}\,\Bigr[ 3\,\apD + 16\,\frac{\xpds}{\xpus-\xpds}\,\apqt - 24\,(
      \apqo - \apu) \Bigr]\,\afun{\mqu}\,c_2\,\xpus
\nl
{}&-& \frac{1}{96}\,\Bigr[ 9\,\xphs\,\apD + 8\,(\apD - 4\,\apBox)\,c^2 
           + 8\,(\apD - 4\,\apBox)\,\frac{c^4}{\xphs - c^2} 
           + 8\,(\apD + 4\,\apBox)\,\frac{\xphs}{\xphs-1} \Bigr]\,\afun{\mh}\,c_2
\nl
{}&-& \frac{1}{24}\,\Bigr[ c_2\,\xphs\,\apD - 2\,c^4\,\apBox + 2\,s^4\,\apD - 2\,(\apD - 
              \apBox)\,s^2\,c^2 + 2\,(\apD + \apBox)\,c_2 \Bigr]
\nl
{}&-& \frac{1}{32}\,\Bigr[ 6\,s^2\,\xphs\,\apD - 3\,(7 + \xphs)\,\apD - 8\,(5\,\apD - 3\,
              \apBox)\,s^4 + 2\,(31\,\apD - 6\,\apBox)\,s^2 \Bigr]\,\LR
\nl
{}&-& \frac{1}{24}\,\Bigr[ 5\,(1 - 3\,c_2 - c_2^2)\,\apD 
            - 2\,(\apD - 4\,\apBox)\,\frac{s^2\,c^2}{\xphs - c^2}\,c_2 
\nl
{}&+& (\apD - \apBox)\,s^2\,c^2 - (21\,\apD - \apBox)\,s^4 \Bigr]\,
            \afun{\mw}\,\frac{c^2}{s^2}
\nl
{}&+& \frac{1}{192}\,\Bigr[ (51 - 152\,c_2 - 3\,c_2^2 - 16\,c_2^3)\,\apD + 16\,(\apD + 4\,
            \apBox)\,\frac{s^2}{\xphs-1}\,c_2 + 2\,(11\,\apD - 4\,\apBox)\,s^2\,c^2
\nl
{}&-& 2\,(113\,\apD - 4\,\apBox)\,s^4 \Bigr]\,\afun{\mz}\,\frac{1}{s^2}
\nl
{}&-& \frac{1}{48}\,\sum_{\gen}\,
            \Bigr[  - 24\,(\apqt - 3\,\apqo + 3\,\apd)\,\xpds - 24\,(\apqt + 3\,
            \apqo - 3\,\apu)\,\xpus - 8\,(\aplt - 3\,\aplo + 3\,\apl)\,\xpls 
\nl
{}&+& 3\,(3\,\xpds + 3\,\xpus + \xpls)\,\apD \Bigr]\,c_2
\nl
{}&+& \frac{1}{16}\,\sum_{\gen}\,
            \Bigr[ 24\,(\apqo - \apd)\,\xpds - 24\,(\apqo - \apu)\,\xpus + 8\,(
            \aplo - \apl)\,\xpls + (3\,\xpds + 3\,\xpus + \xpls)\,\apD \Bigr]\,\LR\,c_2
\eqa


\normalsize




\clearpage
\bibliographystyle{atlasnote}
\bibliography{HEFTF_arXiv}{}

\end{document}